\documentclass{puthesis}

\usepackage{cite}

\usepackage{graphicx}                       
\usepackage{amsmath}
\usepackage{amssymb}
\usepackage{amsthm}
\usepackage{dsfont}       

\usepackage{color} 

\usepackage[format=plain, font=small, labelfont=bf, justification = raggedright]{caption}
\usepackage{array}  
\usepackage{overpic}
\usepackage{placeins} 
\usepackage{fancyhdr}

\usepackage{empheq} 

\usepackage{appendix} 

\newcommand{\eq}[1]{(\ref{#1})}
\newcommand{\Eq}[1]{Eq.~\eq{#1}}
\newcommand{\Eqs}[1]{Eqs.~\eq{#1}}
\newcommand{\Fig}[1]{Fig.~\ref{#1}}
\newcommand{\Sec}[1]{Sec.~\ref{#1}}

\renewcommand{\Ref}[1]{Ref.~\cite{#1}}

\newcommand{\Refs}[1]{Refs.~\cite{#1}}
\newcommand{\App}[1]{Appendix~\ref{#1}}

\newcommand{\Ch}[1]{Ch.~\ref{#1}}

\newcommand{\eg}{{e.g., }}
\newcommand{\ie}{{i.e., }}

\newcommand{\mc}[1]{\mathcal{#1}}

\newcommand{\msf}[1]{\mathsf{#1}}

\newcommand{\mbb}[1]{\mathbb{#1}}

\newcommand{\bra}[1]{\langle#1 |}
\newcommand{\ket}[1]{|#1 \rangle}
\newcommand{\braket}[2]{\langle#1 |  #2 \rangle}
\newcommand{\com}[2]{\left[#1, #2\right]}

\newcommand{\oper}[1]{\smash{\hat{#1}}}
\newcommand{\unit}[1]{\smash{\check{#1}}}
\newcommand{\fourier}[1]{\smash{\widetilde{#1}}}

\newcommand{\pd}[1]{\partial_{#1}}

\newcommand{\nablaNEW}{\widetilde{\nabla}}
\newcommand{\dd}{\mathrm{d}}

\DeclareMathOperator{\airyA}{Ai}
\DeclareMathOperator{\airyB}{Bi}
\DeclareMathOperator{\pearcey}{Pe}

\newcommand{\Sp}[1]{\text{Sp}\left(#1,\R \right)}

\newcommand{\Vect}[1]{{\boldsymbol{\rm #1}}}
\newcommand{\VectOp}[1]{\oper{\Vect{#1}}}

\newcommand{\Mat}[1]{\msf{#1}}
\newcommand{\IMat}[1]{\Mat{I}_{#1}}
\newcommand{\OMat}[1]{\Mat{0}_{#1}}
\newcommand{\JMat}[1]{\Mat{J}_{#1}}
\newcommand{\TpMat}{\Mat{T}_+}
\newcommand{\TmMat}{\Mat{T}_-}
\newcommand{\dubdot}{\text{\large :}}
\newcommand{\tripdot}{ \text{\large $\therefore$}}
\newcommand{\Tr}{\text{tr}}

\newcommand{\Symb}[1]{\mc{#1}}
\newcommand{\Weyl}{\mbb{W}}
\newcommand{\WeylInv}{\mbb{W}^{-1}}
\newcommand{\QoSymb}{\Symb{D}_\text{c}}

\newcommand{\R}{\mbb{R}}

\newcommand{\IdentOp}{\oper{\mathds{1}}}
\newcommand{\deltaQO}{\Delta_\text{QO}}

\newcommand{\cont}[1]{\mc{C}_{#1}}
\newcommand{\curv}{\mc{K}}
\newcommand{\aber}{a}
\newcommand{\MTnorm}{\mc{N}_\Vect{t}}
\newcommand{\env}{\phi}

\newcommand{\NIMT}[1]{\mbb{N}^{}_{#1}}
\newcommand{\PMT}[1]{\mbb{P}^{}_{#1}}
\newcommand{\MT}{\mbb{M}}

\renewcommand{\Re}{\textrm{Re}}
\renewcommand{\Im}{\textrm{Im}}

\theoremstyle{definition}
\newtheorem{definition}{Definition}

\theoremstyle{plain}
\newtheorem*{corollary}{Corollary}

\theoremstyle{plain}
\newtheorem{theorem}{Theorem}

\newcommand{\nullFrac}{\vphantom{\frac{}{}}}

\newcommand{\Stroke}[1]{\text{\ooalign{ $#1$\cr \hidewidth\raise.225ex \hbox{$-\mkern.5mu$}\cr}}}

\usepackage{subfiles} 

\title{Metaplectic geometrical optics}
\author{Nicolas Alexander Lopez}
\adviser{Ilya Y. Dodin}

\abstract{%
Ray optics is an intuitive and computationally efficient model for wave propagation through nonuniform media. It is therefore the standard choice for multiphysics simulations that involve wave physics in some capacity, such as designing and analyzing nuclear fusion experiments. However, the underlying geometrical-optics (GO) approximation of ray optics breaks down at caustics such as cutoffs and focal points, erroneously predicting the wave intensity to be infinite and thereby limiting the predictive capabilities of these codes. Full-wave modeling can be used instead, but the added computational cost brings its own set of tradeoffs. Developing cheaper, more efficient caustic remedies has therefore been an active area of research for the past few decades.

In this thesis, I present a new ray-based approach called `metaplectic geometrical optics' (MGO) that can be applied to any linear wave equation. Instead of evolving waves in the usual $\Vect{x}$ (coordinate) or $\Vect{k}$ (spectral) representation, MGO uses a mixed $\Vect{X} \equiv \Mat{A}\Vect{x} + \Mat{B}\Vect{k}$ representation. By continuously adjusting the matrix coefficients $\Mat{A}$ and $\Mat{B}$ along the rays via sequenced metaplectic transforms (MTs) of the wavefield, corresponding to symplectic transformations of the ray phase space, one can ensure that GO remains valid in the $\Vect{X}$ coordinates without caustic singularities. The caustic-free result is then mapped back onto the original $\Vect{x}$ space using metaplectic transforms, as demonstrated and verified on a number of examples.

Besides outlining the basic theory of MGO, this thesis also presents specialized fast algorithms for MGO. These algorithms focus on the MT, which is a unitary integral mapping used in MGO that can be considered a generalization of the Fourier transform. First, a discrete representation of the MT is developed that can be computed in linear time [$O(N_p)$ for $N_p$ sample points] when evaluated in the near-identity limit; finite MTs can then be implemented as successive applications of $K \gg 1$ near-identity MTs. Second, an algorithm based on Gauss--Freud quadrature is developed for efficiently computing finite MTs along their steepest-descent curves, which may be useful in catastrophe-optics applications beyond MGO. These algorithms lay the foundations for the development of an MGO-based ray-tracing code.
}

\begin{document}
	
\chapter{Introduction}
\label{ch:Intro}

\section{Background}

Electromagnetic (EM) waves are widely used in plasma applications, whether that be the radiofrequency (RF) waves used to heat and drive current in a tokamak plasma~\cite{Stix92,Fisch87,Prater08,Freidberg10,Wesson11,Farina14,Poli15,Lopez18a} or the high-power lasers used to compress a fuel pellet~\cite{Marinak01,Lindl04,Hohenberger15,Craxton15,Robey18,Kritcher21}. Accurately modeling how EM waves propagate in plasma is therefore of upmost importance. Full-wave modeling, that is, directly solving Maxwell's equations with appropriate source terms, can be computationally expensive. Instead, ray optics is often used to quickly calculate the wave amplitude along the characteristic rays that illuminate the region of interest~\cite{Tracy14, Kravtsov90}. The obtained amplitude profile can then be used as a source term in calculations of macroscopic plasma equilibrium~\cite{Prater08,Poli18a}. Indeed, the speed of ray-tracing codes makes them invaluable in multiphysics simulation packages that are increasingly used to design and optimize future experimental campaigns~\cite{Poli18a,Poli15,Poli16,Lopez18a}, such as the TRANSP~\cite{Hawryluk81} or the HYDRA~\cite{Marinak01} code. 

Unfortunately, the ray-tracing formalism, based on the geometrical-optics (GO) approximation~\cite{Kravtsov90,Tracy14}, breaks down at mode-conversion regions and at caustics, such as cutoffs or focal points~\cite{Kravtsov93,Berry80b}. This is a particular impediment to fusion research since the wave behavior at such regions is often precisely the quantity being optimized. One example is initializing spherical tokamak plasmas~\cite{Peng86,Peng00,Ono15a} via electron cyclotron resonance heating~\cite{Erckmann94,Prater04}, where the time-evolution of caustic surfaces directly defines the window of operation~\cite{Lopez18a}; another example is driving current in overdense plasmas via mode conversion to the electron Bernstein wave~\cite{Ram00,Shiraiwa06,Uchijima15,Seltzman17,Lopez18b,Laqua07,Preinhaelter73,Hansen85,Mjolhus84,Laqua03,Shevchenko07}, where the field structure near the EM wave cutoffs must be precisely resolved to obtain accurate mode-conversion efficiency estimates. Reduced modeling of these processes requires a more advanced machinery than traditional GO. There has been much recent progress to rigorously incorporate mode conversion into the GO framework, notably the normal-form approach~\cite{Tracy93,Tracy01,Tracy03a,Tracy07,Jaun07,Richardson08} and the extended GO (XGO) framework~\cite{Ruiz15a,Ruiz15b,Ruiz17a} with its beam-tracing generalization~\cite{Dodin19,Yanagihara19a,Yanagihara19b,Yanagihara21a,Yanagihara21b}. Comparatively less work has been dedicated to caustics.

Loosely speaking, the validity of GO is determined by the condition $kL \gg 1$, where $k$ is the local wavenumber and $L$ is the smallest scale among those that characterize the local properties of the medium, the wave envelope, and $k$ itself. However, even for initially smooth fields, GO often predicts the appearance of caustics where $kL \to 0$; at caustics, GO erroneously predicts an infinite wave amplitude. Examples of simple caustics include cutoffs (where $k \to 0$) and focal points (where $L \to 0$). Equivalently, caustics can be understood as surfaces across which the number of rays arriving at a given point changes abruptly~\cite{Kravtsov93}. The general properties of such surfaces have long been known from catastrophe theory, which provides a classification wherein only a finite number of caustic types are possible for a given number of spatial dimensions~\cite{Berry76,Berry80b}. This result underlies modern research into caustics~\cite{Hobbs07,Borghi16,Zannotti17a,EspindolaRamos19}, as general properties of a given caustic type can be inferred by studying a particular case. Still, practical calculations continue to rely on directly solving wave equations~\cite{Wright09,Shiraiwa10,Myatt17}, which is computationally expensive. It would be advantageous to find a more efficient way to calculate these caustic structures. In particular, the question whether caustics can be modeled by somehow extending GO has been attracting attention for a long time.

In some cases, this problem is solved by locally reducing a given wave equation to a simpler one with a known solution, such as Airy's equation~\cite{Kravtsov93,Ludwig66,Berry72}. (This method can be generalized for the entire hierarchy of catastrophes via `etalon integrals', as I shall discuss in later chapters.) In other cases, cutoffs are modeled as discrete interfaces, such as in specular-reflection or perfect-conductor approximations~\cite{Jackson75,Born99,Lopez18b}. However, such approaches assume that the spatial structure of the caustic is known \textit{a~priori}, which is not often the case. A more fundamental description of caustics was developed by Maslov~\cite{Maslov81,Ziolkowski84,Thomson85} based on geometrical properties of GO solutions in the ray phase space. By occasionally rotating the phase space by $\pi/2$ using the Fourier transform in one or more spatial variables, one can remove a caustic and locally reinstate GO. However, the exact moment for performing the FT is only loosely specified. This makes Maslov's method cumbersome to implement in codes, since it would require supervision by the user or numerical modeler. More recent methods~\cite{Littlejohn85,Littlejohn86b,Kay94a,Zor96,Madhusoodanan98,Alonso97b,Alonso99} remedy this issue by replacing the occasional FT of Maslov's theory with a metaplectic transformation (MT) applied continually along a ray. However, these works either introduced additional free parameters~\cite{Kay94a,Zor96,Madhusoodanan98,Alonso97b,Alonso99} or made overly restrictive assumptions on the class of solutions sought (\eg wavepackets)~\cite{Littlejohn85,Littlejohn86b,Kay94a,Zor96,Madhusoodanan98}. They also tended to simultaneously under- and over-emphasize the use of rays by expressing the wavefield as an integral taken over all points along every ray (rather than only the rays that actually arrive at a given observation point), whose integrand is determined entirely by the phase-space ray geometry (only true for scalar diffraction-free waves).

In this thesis I propose a new ray-tracing framework called metaplectic geometrical optics (MGO)~\cite{Lopez19,Lopez20,Lopez21a,Lopez21b,Donnelly21,Lopez22}. MGO is formulated using the Weyl symbol calculus~\cite{Tracy14}, and as such, it is applicable for any caustic structure in any linear scalar wave equation with minimal assumptions on the type of solutions seeked, including integro-differential equations that arise in kinetic treatments of plasma waves. This robustness is highlighted by the success of MGO in modeling a variety of caustics in a variety of different wave equations, as will be shown in later chapters. The other aforementioned methods do not possess such generality. Hence, this approach promises to be useful for a wide variety of applications, such as in optics and in plasma physics.

I also propose several fast specialized algorithms for numerically solving the MGO equations, specifically targeting how to compute the MT in various limits. In recent years, many fast algorithms have been developed to compute the MT for one-dimensional (1-D) and 2-D fields~\cite{Ozaktas96,Hennelly05b,Healy10,Koc10a,Ding12,Pei16,Sun18a}. Many of them are reviewed in \Ref{Healy18}. These algorithms often involve the fast Fourier transform, since the MT has a simple spectral representation; as such, they often scale as $O(N \log_2 N)$, where $N$ is the number of sample points. Despite this multitude, however, there also exist applications for which suitable MT algorithms have yet to be designed. One example is the continuous MT the MGO framework applies along a ray; the corresponding MTs will be near-identity. Since the existing algorithms treat the MT as an integral transform, they are not optimal for computing the MT in this limit, and instead, a differential representation would be advantageous. In this thesis I develop a general pseudo-differential form of the MT that readily yields asymptotic \textit{differential} representations of the MT in the near-identity limit, which are easy to compute numerically. Specifically, they can be computed in linear $O(N)$ time, which is potentially a significant speedup compared to other MT algorithms. Finite MTs can be computed iteratively as successive applications of $K \gg 1$ near-identity MTs, which preserves the $O(N)$ scaling. I therefore expect this algorithm to be useful in a broad range of applications beyond just the MGO modeling of EM waves.

Similarly, by default, MGO yields an integral representation of the wavefield, which can be approximated analytically to some extent but in general must be evaluated numerically. Unfortunately, the integrands in MGO are highly oscillatory, so standard integration methods are insufficient~\cite{Deano17}. Special numerical algorithms tailored to MGO are needed. Hence, in this thesis I also develop a novel quadrature rule for calculating MGO integrals based on numerical steepest-descent integration~\cite{Deano09}. This algorithm emerges naturally from the MGO framework in that MGO integrals always contain saddlepoints that correspond to the ray contributions to the wavefield. The quadrature rule is then successfully benchmarked on a class of examples in which the MGO integral contains a single isolated saddlepoint of various degeneracy, physically representing a wavefield either far from a caustic or at the critical point of a cuspoid-type caustic, and on two examples of more immediate relevance, namely, the propagation and reflection of an EM wave off an isolated cutoff and bounded between a pair of cutoffs in $1$-D. The numerical MGO solutions agrees with the exact results amazingly well, laying the foundations for the development of a future MGO-based ray-tracing code.


\section{Outline}

This thesis is organized as follows: In \Ch{ch:GO} I introduce the standard geometrical-optics approximation for scalar wavefields. I then summarize how caustic singularities arise in GO and examine the five stable caustics that can occur in 3-D within the context of a paraxial wave propagating in uniform medium. I conclude this chapter by discussing how certain unifying principles naturally lead to practical algorithms to avoid ray caustics in simulations, including MGO but also other popular methods such as etalon integrals and Maslov transformations. I overview these various methods, beginning with the most crude methods and culminating in an preview discussion of MGO.

In \Ch{ch:MT} I introduce the mathematical machinery necessary to develop the phase-space rotation scheme of MGO, namely, linear symplectic transformations of the ray phase space and their corresponding metaplectic operators that act on the wavefields. I also derive their representations in certain convenient basis choices that will be useful in later chapters, such as when ray patterns are `quasiuniform'. I also discuss these two classes of transformations in the near-identity limit, deriving a novel pseudo-differential representation of the MT in the process, and in the limit when the transformations are orthogonal in addition to symplectic (orthosymplectic), as will be the case for any practical MGO implementation in a code. I conclude this chapter with a detailed discussion of the MT in two familiar physics contexts to aid the reader develop further intuition: first, in the familiar setting of elementary quantum mechanics in which the MT acts as the time propagator for the quantum harmonic oscillator problem; second, in the familiar setting of paraxial optics in which the MT acts as the spatial evolution operator for the diffracted (transverse) field.

In \Ch{ch:NIMT} I propose two fast, linear-time algorithms specifically optimized to calculate the MT in the near-identity limit (the so-called `NIMT'). The first algorithm is called the `local' algorithm, and is based on defining the NIMT via truncated Taylor expansions to the desired order. I describe an iterative algorithm to perform cumulative MTs which are not near-identity, and then discuss the computational complexity and stability of such an algorithm, augmented with numerical demonstrations. The second algorithm is called the `unitary' algorithm, and is based on instead defining the NIMT via truncated Pad\'e expansions to the desired order. I then assess the computational complexity and stability of the iterated Pad\'e-based algorithm and validate the analysis with a series of numerical demonstrations.

In \Ch{ch:GF} I propose a novel quadrature to compute the finite MTs of catastrophe wavefields that will necessarily appear in MGO. To do this, I first introduce the basic idea of steepest-descent integration and of Gaussian quadrature for numerical integration. I then proceed to derive the new quadrature rule. I conclude the chapter with a number of benchmarking examples.

In \Ch{ch:MGO} I present the main result of this thesis, namely, the development of a new ray-based caustic-removal scheme called metaplectic geometrical optics (MGO). MGO is a general framework that does not assume a specific caustic structure nor a specific wave equation, and thereby promises to become a new paradigm in caustic modeling. The discussion of this chapter is built upon the ideas of caustics and metaplectic transforms that were discussed at length in previous chapters. I start by introducing the idea that phase-space rotations can remove caustics in the familiar example of Airy's problem, in which an EM wave propagates towards and subsequently from an isolated cutoff. I then develop the GO formalism in arbitrarily rotated phase space; choosing the rotations properly while the rays propagate then leads to the MGO formalism, as I next discuss. I then compare the MGO formalism with two other related formalisms: first, I show that MGO reduces analytically to the standard GO formalism when evaluated far away from caustics; second, I show how MGO can be considered as a delta-windowed semiclassical integral expression with no \textit{a priori} assumptions on the class of tangent-space wavefields sought, \eg wavepackets or delta-shaped fields as commonly assumed in other published semiclassical methods.

Lastly, in \Ch{ch:Ex} I demonstrate the MGO formalism in a series of examples to highlight the variety of caustics and types of wave equations that MGO can accomodate. I begin this chapter with a step-by-step summary list of the MGO procedure that outlines the algebraic steps of the forthcoming examples and might also serve as outline for an MGO-based ray-tracing code. Then, I show that MGO can exactly describe simple plane wave propagation (as one would desire if MGO is to be a robust simulation tool) by deriving the MGO solution to a unidirectional wave equation in $1$-D. Next, I show that MGO can accurately describe the wavefield behavior near a single isolated fold caustic (Airy's equation) in $1$-D, and bounded within a pair of fold caustics (Weber's equation) in $1$-D. I then extend the Airy result from $1$-D to $2$-D as well, as a means of gently introducing the added complexities that multiple dimensions bring to the MGO-modeling table. I then conclude this chapter with the most sophisticated example - an isolated cusp caustic in $2$-D. (I anticipate this example constitutes the boundary to what is feasible analytically, and more complicated MGO examples can only be done numerically.) In all examples, the MGO solution remains finite at the caustic, unlike the standard GO solution, and accurately reproduces the exact wavefield throughout the entire space.

In all chapters, the main results are summarized in the final sections and auxiliary calculations are presented in appendices.


\section{Contributing publications}

The following publications contributed to the body of work and original results presented in this thesis:
\begin{itemize}
	\item{N.~A.~Lopez and I.~Y.~Dodin, \emph{Pseudo-differential representation of the metaplectic transform and its application to fast algorithms}, J.~Opt.~Soc.~Am.~A \textbf{36}, 1846 (2019)}
	%
	\item{N.~A.~Lopez and I.~Y.~Dodin, \emph{Restoring geometrical optics near caustics using sequenced metaplectic transforms}, New J.~Phys.~\textbf{22}, 083078 (2020)}
	%
	\item{N.~A.~Lopez and I.~Y.~Dodin, \emph{Metaplectic geometrical optics for modeling caustics in uniform and nonuniform media}, J.~Opt.~\textbf{23}, 025601 (2021)}
	%
	\item{N.~A.~Lopez and I.~Y.~Dodin, \emph{Exactly unitary discrete representations of the metaplectic transform for linear-time algorithms}, J.~Opt.~Soc.~Am.~A \textbf{38}, 634 (2021)}
	%
	\item{Sean M.~Donnelly, Nicolas A.~Lopez, and I.~Y.~Dodin, \emph{Steepest-descent algorithm for simulating plasma-wave caustics via metaplectic geometrical optics}, Phys.~Rev.~E \textbf{104}, 025304 (2021)}
	%
	\item{N.~A.~Lopez and I.~Y.~Dodin, \emph{Metaplectic geometrical optics for ray-based modeling of caustics: Theory and algorithms}, Phys.~Plasmas \textbf{29}, 052111 (2022)}
\end{itemize}

\chapter{Geometrical optics for scalar waves}
\label{ch:GO}

\section{Introduction}
\label{sec:2_intro}

In this section, I introduce the standard geometrical-optics approximation for scalar wavefields. I then summarize how caustic singularities arise in GO and examine the five stable caustics that can occur in three dimensions ($3$-D) within the context of a paraxial wave propagating in uniform medium. I conclude this chapter with an overview of the various methods used by researchers to remove caustic singularities from ray-tracing codes, beginning with the most crude methods and culminating in an preview discussion of MGO. The presentation is based on discussions presented in \Refs{Lopez20, Lopez21a, Donnelly21, Lopez22}. Note that the following procedure can be generalized for multicomponent vector waves propagating in arbitrarily curved spaces if desired; see \Ref{Dodin19} for details.

\section{The geometrical-optics approximation}
\label{sec:2_GO}

Let $\psi(\Vect{x})$ be a scalar stationary wavefield in a plasma described by an $N$-dimensional ($N$-D) Euclidean coordinate system with coordinates $\Vect{x}$ ($\Vect{x}$-space). Neglecting nonlinear effects, the governing wave equation for $\psi(\Vect{x})$ is most generally written in the following integral form:
\begin{equation}
	\int \dd \Vect{x}' \, D(\Vect{x}, \Vect{x}') \psi(\Vect{x}') = 0 
	,
	\label{eq:2_intWAVE}
\end{equation}

\noindent where $D(\Vect{x},\Vect{x}')$ is some dispersion kernel. In particular, differential wave equations (partial or ordinary) have dispersion kernels that consist of delta functions and their derivatives. For example, the Helmholtz equation has 
\begin{equation}
	D(\Vect{x},\Vect{x}') = \nabla'^2 \delta(\Vect{x}' - \Vect{x}) + n^2(\Vect{x}') \delta(\Vect{x}' - \Vect{x})
	,
	\label{eq:2_HelmD}
\end{equation} 

\noindent where $\nabla'$ is the gradient with respect to $\Vect{x}'$ and $n(\Vect{x})$ is a spatially varying index of refraction. More generally, the kernel $D$ can be a smooth function, as is the case for waves in warm plasma, for example~\cite{Tracy14,Stix92}. (A review of the general theory of linear dispersion can be found in \Ref{Dodin17d}.) It is not necessary for my purposes to specify this function; let me simply state that $D$ encodes all information about the linear medium where the wave propagates, whatever that medium may be. (Specific examples will be discussed in \Ch{ch:Ex}.)

Let me introduce%
\footnote{Here, I use the bra-ket notation that is standard in quantum mechanics and optics~\cite{Shankar94,Stoler81}.} %
a Hilbert space of state vectors $\ket{\psi}$ such that $\psi(\Vect{x})$ is the projection of a given $\ket{\psi}$ onto the eigenbasis $\{ \ket{\Vect{x}} \}$ of the coordinate operator $\VectOp{x}$. I adopt the usual normalization such that
\begin{equation}
	\int \dd \Vect{x} \, \ket{\Vect{x}} \bra{\Vect{x}} = \IdentOp 
	,
\end{equation}

\noindent where $\IdentOp$ is the identity operator. Then,
\begin{equation}
	\psi(\Vect{x}) \doteq \braket{\Vect{x}}{\psi} 
	,
\end{equation}

\noindent where the symbol $\doteq$ denotes definitions. I define $\VectOp{x}$ through its action on the Hilbert space as $\VectOp{x}\ket{\Vect{x}'} = \Vect{x}' \ket{\Vect{x}'}$, or equivalently, through its matrix elements 
\begin{equation}
	\bra{\Vect{x}}\VectOp{x} \ket{\Vect{x}'} = \Vect{x}' \delta(\Vect{x}' - \Vect{x}) 
	.
\end{equation}

\noindent The canonically conjugate momentum operator $\VectOp{k}$ is similarly defined through its matrix elements as 
\begin{equation}
	\bra{\Vect{x}}\VectOp{k} \ket{\Vect{x}'} = i\nabla' \delta(\Vect{x}' - \Vect{x}) 
	.
\end{equation}

Let me further define the dispersion operator $\oper{D}$ through its matrix elements $\bra{\Vect{x}}\oper{D} \ket{\Vect{x}'} = D(\Vect{x},\Vect{x'})$. Then, \Eq{eq:2_intWAVE} can be represented as
\begin{equation}
	\oper{D}(\VectOp{x}, \VectOp{k}) \ket{\psi} = \ket{0} 
	,
	\label{eq:2_hilbertWAVE}
\end{equation}

\noindent with \Eq{eq:2_intWAVE} being simply the projection of \Eq{eq:2_hilbertWAVE} onto the coordinate eigenbasis. (Here, $\ket{0}$ is the null vector.) Note that $\oper{D}$ is expressed as a function of $\VectOp{x}$ and $\VectOp{k}$. When the dispersion kernel $D(\Vect{x},\Vect{x}')$ describes a local differential wave equation, the construction of $\oper{D}(\VectOp{x},\VectOp{k})$ is trivial. For example, the aforementioned Helmholtz equation [\Eq{eq:2_HelmD}] has 
\begin{equation}
	\oper{D}(\VectOp{x},\VectOp{k}) = -\VectOp{k}^2 + n^2(\VectOp{x})
	.
\end{equation}

\noindent However, constructing $\oper{D}(\VectOp{x},\VectOp{k})$ for integro-differential wave equations requires a pseudo-differential representation of $D(\Vect{x},\Vect{x}')$. Such a representation can be obtained using the Wigner--Weyl symbol calculus, which I shall introduce momentarily.

GO is the asymptotic model of \Eq{eq:2_hilbertWAVE} for the short-wavelength limit. Loosely speaking, I require $\lambda \ll L$ with $\lambda$ being the local wavelength and $L$ the smallest scale among those that characterize the local properties of the
medium, the wave envelope, and $\lambda$ itself. In this limit, the wavefield can be partitioned into a rapidly varying phase and a slowly varying envelope. Following \Ref{Dodin19}, I define the envelope state vector $\ket{\env}$ via the unitary transformation
\begin{equation}
	\ket{\env} \doteq \exp\left[-i \theta(\VectOp{x}) \right] \ket{\psi} 
	,
	\label{eq:2_phiENV}
\end{equation}

\noindent where $\theta(\VectOp{x})$ is a hermitian operator representing the phase of $\psi(\Vect{x})$. Under this transformation, \Eq{eq:2_hilbertWAVE} becomes
\begin{equation}
	\exp\left[-i \theta(\VectOp{x}) \right]
	\oper{D}(\VectOp{x}, \VectOp{k}) \,
	\exp\left[i \theta(\VectOp{x}) \right]
	\ket{\env} = \ket{0} 
	.
	\label{eq:2_hilbertENV}
\end{equation}

I shall now approximate the envelope dispersion operator of \Eq{eq:2_hilbertENV} in the GO limit. This is readily accomplished using the Wigner--Weyl symbol calculus, which provides a mapping between functions and operators~\cite{McDonald88a} and is reviewed in \App{app:2_WeylRev}. With the Weyl symbol calculus, approximating operators becomes as easy as approximating functions: one performs a Wigner--Weyl transform (WWT) to obtain the operator's Weyl symbol, approximates the symbol in the desired limit using, say, familiar Taylor expansions, then performs an inverse WWT to obtain the correspondingly approximated operator. Indeed, I perform the GO approximation~\cite{Dodin19}
\begin{align}
	&\Weyl
	\left\{
		\exp\left[-i \theta(\VectOp{x}) \right]
		\oper{D}(\VectOp{x}, \VectOp{k}) \,
		\exp\left[i \theta(\VectOp{x}) \right]
	\right\}
	\nonumber\\
	&=
	\int \dd \Vect{s} \, \exp\left(i\Vect{k}^\intercal \Vect{s} \right) 
	\bra{\Vect{x}-\Vect{s}/2} 
		\exp\left[-i \theta(\VectOp{x}) \right]
		\oper{D}(\VectOp{x}, \VectOp{k}) \,
		\exp\left[i \theta(\VectOp{x}) \right] 
	\ket{\Vect{x} + \Vect{s}/2}
	\nonumber\\
	&=
	\int \dd \Vect{s} \, 
	\exp\left[
		i\Vect{k}^\intercal \Vect{s}
		+ i \theta(\Vect{x} + \Vect{s}/2)
		- i \theta(\Vect{x}-\Vect{s}/2)
	\right]
	\bra{\Vect{x}-\Vect{s}/2} 
		\oper{D}(\VectOp{x}, \VectOp{k}) \,
	\ket{\Vect{x} + \Vect{s}/2}
	\nonumber\\
	&\approx
	\int \dd \Vect{s} \, 
	\exp\left\{
		i \left[
			\Vect{k}
			+ \pd{\Vect{x}} \theta(\Vect{x})
		\right]^\intercal \Vect{s}
	\right\}
	\bra{\Vect{x}-\Vect{s}/2} 
		\oper{D}(\VectOp{x}, \VectOp{k}) \,
	\ket{\Vect{x} + \Vect{s}/2}
	\nonumber\\
	&= \Symb{D}\left[
		\Vect{x},
		\Vect{k} + \pd{\Vect{x}} \theta(\Vect{x})
	\right]
	\nonumber\\
	&\approx \Symb{D}\left[
		\Vect{x},
		\pd{\Vect{x}} \theta(\Vect{x})
	\right]
	+ \Vect{k}^\intercal \pd{\Vect{k}} \Symb{D}\left[
		\Vect{x},
		\pd{\Vect{x}} \theta(\Vect{x})
	\right]
	.
\end{align}

\noindent Hence, \Eq{eq:2_hilbertENV} becomes
\begin{equation}
	\left\{ 
		\Symb{D}\left[\VectOp{x}, \Vect{k}(\VectOp{x}) \right]
		+ \Vect{v}(\VectOp{x})^\intercal \VectOp{k}
		- \frac{i}{2} \pd{\Vect{x}} \cdot \Vect{v}(\VectOp{x})
	\right\} \ket{\env} = \ket{0}
	,
	\label{eq:2_approxENV}
\end{equation}

\noindent where 
\begin{subequations}%
	\label{eq:2_GOdefs}%
	\begin{align}
		\Symb{D}(\Vect{z}) &\doteq \Weyl \left[\oper{D}(\VectOp{z}) \right]
		, \\
		\label{eq:2_pGRADtheta}
		\Vect{k}(\Vect{x}) &\doteq \pd{\Vect{x}} \theta(\Vect{x}) 
		, \\
		\Vect{v}(\Vect{x}) &\doteq \left.\pd{{\Vect{k}}} \Symb{D}(\Vect{x}, \Vect{k}) \right|_{\Vect{k}=\Vect{k}(\Vect{x})}
	\end{align}%
\end{subequations}%

\noindent are interpreted respectively as the local dispersion function, the local wavevector, and the local group velocity. Importantly, note that $\Vect{k}(\Vect{x})$ is irrotational, meaning that
\begin{equation}
	\pd{x_\ell} k_m = \pd{x_m} k_\ell 
	,
	\quad 
	\ell,m = 1, \ldots, N .
	\label{eq:2_curlp}
\end{equation}

Projecting \Eq{eq:2_approxENV} onto $\Vect{x}$-space then yields the GO equations,
\begin{subequations}
	\label{eq:2_GO}
	\begin{gather}
		\label{eq:2_GOdisp}
		\Symb{D}\left[\Vect{x}, \Vect{k}(\Vect{x}) \right] = 0 
		, \\
		\label{eq:2_GOenv}
		\Vect{v}(\Vect{x})^\intercal \pd{{\Vect{x}}} \env(\Vect{x}) 
		+ \frac{1}{2} \left[\pd{\Vect{x}} \cdot \Vect{v}(\Vect{x}) \right] \env(\Vect{x}) = 0 
		,
	\end{gather}%
\end{subequations}

\noindent where $\env(\Vect{x}) \doteq \braket{\Vect{x}}{\env}$. Specifically, \Eq{eq:2_GOdisp} represents a local dispersion relation, and \Eq{eq:2_GOenv} represents an envelope equation. For simplicity, I neglect dissipation such that $\oper{D}$ is Hermitian and consequently, both $\Symb{D}$ and $\Vect{v}$ are real. Then, \Eq{eq:2_GOenv} can also be cast as a conservation relation:
\begin{equation}
	\pd{\Vect{x}} \cdot \left[|\env(\Vect{x})|^2 \Vect{v}(\Vect{x})  \right]
	= 0
	,
	\label{eq:2_actionCONS}
\end{equation}

\noindent where the quantity within square brackets is recognized as the wave action flux (or the wave energy flux, which for stationary waves is the same up to a constant factor)~\cite{Whitham74}.

The local dispersion relation \eq{eq:2_GOdisp} naturally resides within the $2N$-D phase space with coordinates $\Vect{z}$; \Eq{eq:2_GOdisp} implicitly defines a $(2N-1)$-D volume within this phase space that describes the local momentum of the wavefield at a given point $\Vect{x}$ in configuration space. For coherent wavefields that have a single wavevector $\Vect{k}(\Vect{x})$ (or a finite superposition of such wavevectors) at each point given by \Eq{eq:2_pGRADtheta}, then $\Vect{k}$ is actually restricted to an $N$-D surface contained within the $(2N-1)$-D volume defined by \Eq{eq:2_GOdisp}. (The specific $N$-D surface is dictated by initial conditions.) This $N$-D surface is called the ray manifold, and by resulting from a gradient lift [\Eq{eq:2_pGRADtheta}] it is a Lagrangian manifold~\cite{Tracy14,Arnold89}. This class of manifolds possesses certain special properties that will be useful when I develop MGO in \Ch{ch:MGO}.

The ray manifold is a central object in GO and MGO. It is therefore useful for practical purposes to have an explicit construction of it, rather than relying on the formal construction described in the preceding paragraph. This explicit construction is provided by the ray (Hamilton's) equations~\cite{Tracy14}
\begin{equation}
	\pd{\xi} \Vect{x} = 
	\pd{\Vect{k}} \Symb{D}(\Vect{x}, \Vect{k})
	, \quad
	\pd{\xi} \Vect{k} =
	- \pd{\Vect{x}} \Symb{D}(\Vect{x}, \Vect{k})
	,
	\label{eq:2_GOrays}
\end{equation}

\noindent subject to the constraints that the initial conditions satisfy both the local dispersion relation \eq{eq:2_GOdisp} and the gradient lift \eq{eq:2_pGRADtheta}. The family of solution trajectories $(\Vect{x}(\xi), \Vect{k}(\xi))$ for a corresponding family of initial conditions $(\Vect{x}(0), \Vect{k}(0))$ then trace out the ray manifold.

Since the ray manifold is $N$-D, let me introduce a set of $N$-D coordinates $\Vect{\tau}$ such that it can be parameterized as $(\Vect{x}(\Vect{\tau}),\Vect{k}(\Vect{\tau}))$. I shall choose $\tau_1 = \xi$ as a `longitudinal' coordinate along each ray and the remaining $\Vect{\tau}_\perp \doteq (\tau_2,\ldots \tau_N)$ as `transverse' coordinates that describe the different initial conditions of each ray. (For example, $\tau_1$ can be the time variable or one of the spatial coordinates; then $\Vect{\tau}_\perp$ are the remaining spatial coordinates.) Alternatively, let me parameterize the ray manifold as the zero set of $N$ independent functions~\cite{Tracy14} 
\begin{equation}
	\Vect{\Symb{M}}(\Vect{z}) \doteq (\Symb{M}_1(\Vect{z}), \ldots \Symb{M}_N(\Vect{z}))
	, \quad
	\Symb{M}_1 \equiv \Symb{D}
	.
\end{equation}

\noindent Then, the remaining $\tau_2, \ldots, \tau_N$ can be chosen as the coordinates generated by $\Symb{M}_2, \ldots, \Symb{M}_N$ in the same sense that evolution by $\xi$ is generated by $\Symb{D}$ in \Eq{eq:2_GOrays}.

It is important to note that the rays $\Vect{z}(\Vect{\tau})$ can never cross in phase space since \Eq{eq:2_GOrays} is a first-order autonomous system; however, their projections onto $\Vect{x}$-space, $\Vect{x}(\Vect{\tau})$, have no such restriction. This is problematic for the GO model [\Eqs{eq:2_GO}] that is constructed in $\Vect{x}$-space using $\Vect{x}(\Vect{\tau})$. To see why, let me integrate \Eq{eq:2_GOenv} along a ray. The first term in \Eq{eq:2_GOenv} is clearly the directional derivative of $\env$ along the ray trajectory. Following \Ref{Kravtsov90}, the second term is simplified upon noting that
\begin{equation}
	\pd{\Vect{x}} \cdot \Vect{v}(\Vect{\tau}) = \Tr \left[ \pd{\Vect{x}} \Vect{v}(\Vect{\tau}) \right] 
	,
\end{equation}

\noindent where $\Tr$ denotes the matrix trace and $\Vect{v}(\Vect{\tau}) \doteq \Vect{v}\left[ \Vect{x}(\Vect{\tau}) \right]$. Then, since
\begin{align}
	\pd{\Vect{x}} \Vect{v}(\Vect{\tau}) &= \pd{\Vect{\tau}} \Vect{v}(\Vect{\tau})
	\left[\pd{\Vect{\tau}} \Vect{x}(\Vect{\tau}) \right]^{-1} 
	= \pd{\Vect{\tau}} \left[\pd{{\tau_1}}\Vect{x}(\Vect{\tau}) \right]
	\left[\pd{\Vect{\tau}} \Vect{x}(\Vect{\tau}) \right]^{-1}
	= \pd{\tau_1} \left[\pd{\Vect{\tau}} \Vect{x}(\Vect{\tau}) \right]
	\left[\pd{\Vect{\tau}} \Vect{x}(\Vect{\tau}) \right]^{-1}
\end{align}

\noindent by the chain rule, Jacobi's formula for the derivative of the matrix determinant implies that
\begin{equation}
	\Tr\left[\pd{\Vect{x}} \Vect{v}(\Vect{\tau}) \right]
	= \pd{\tau_1} \log j(\Vect{\tau}) 
	,
\end{equation}

\noindent where 
\begin{equation}
	j(\Vect{\tau}) \doteq \det \pd{\Vect{\tau}} \Vect{x}(\Vect{\tau})
	\label{eq:2_jacDEF}
\end{equation}

\noindent is the Jacobian determinant of the ray evolution in $\Vect{x}$-space, and $\log(x)$ is the natural logarithm. 

Hence, \Eq{eq:2_GOenv} is written along a ray as
\begin{equation}
	\pd{\tau_1} \env(\Vect{\tau}) 
	+ \frac{1}{2}\left[ \pd{\tau_1} \log j(\Vect{\tau}) \right]
	\env(\Vect{\tau}) = 0 
	.
\end{equation}

\noindent The solution is readily obtained to be
\begin{equation}
	\env(\Vect{\tau}) = \env_0(\Vect{\tau}_\perp) 
	\sqrt{ \frac{j_0(\Vect{\tau}_\perp)}{j(\Vect{\tau})} }
	,
	\label{eq:2_rayENV}
\end{equation}

\noindent where $\env_0(\Vect{\tau}_\perp) \doteq \env(0, \Vect{\tau}_\perp)$ and $j_0(\Vect{\tau}_\perp) \doteq j(0, \Vect{\tau}_\perp)$ are set by initial conditions. Intuitively, since the matrix determinant equals the (signed) volume spanned by the constituent column (or row) vectors, \Eq{eq:2_rayENV} states that $|\env|^2 |\Vect{v}| \dd A$ is constant along a ray, where $\dd A$ is an infinitesimal cross-sectional area of a ray family. This is consistent with action conservation \eq{eq:2_actionCONS} for an infinitesimal `ray tube' volume centered on a specific ray. 

Finally, having determined $\env$ and $\theta$ from integrating the rays, the full field $\psi$ is constructed by summing over all rays that arrive at a given $\Vect{x}$, that is,
\begin{equation}
	\psi(\Vect{x})
	= \sum_{\Vect{t} \in \Vect{\tau}(\Vect{x})}
	\env(\Vect{t}) \exp[i\theta(\Vect{t})]
	\equiv
	\sum_{\Vect{t} \in \Vect{\tau}(\Vect{x})}
	\env_0(\Vect{t}_\perp) 
	\sqrt{
		\frac{j_0(\Vect{t}_\perp)}{j(\Vect{t})}
	} \exp\left(i \int \Vect{k}^\intercal \dd \Vect{x} \right)
	,
	\label{eq:2_GOsol}
\end{equation}

\noindent where $\Vect{\tau}(\Vect{x})$ is the formal function inverse of $\Vect{x}(\Vect{\tau})$ and is generally multi-valued (corresponding to the multiple rays whose interference pattern determines $\psi$), and the phase integral is performed along a ray. Clearly though, the GO field \eq{eq:2_GOsol} diverges where the ray Jacobian vanishes, \ie
\begin{equation}
	j(\Vect{t}) = 0
	.
	\label{eq:2_caustic}
\end{equation}

\noindent Equivalently, since $\pd{\Vect{k}} \Vect{\Symb{M}}\left[\Vect{x}, \nabla \theta(\Vect{x}) \right] = \pd{\Vect{\tau}} \Vect{x}(\Vect{\tau}) $ when $\Vect{\tau}$ is generated by $\Vect{\Symb{M}}$, \ie when $\pd{\tau_j} \Vect{x}(\Vect{\tau}) = \pd{\Vect{k}} \Symb{M}_j\left[\Vect{x}, \nabla \theta(\Vect{x}) \right]$, the condition \eq{eq:2_caustic} can also be written as
\begin{equation}
	\det \pd{\Vect{k}} 
	\Vect{\Symb{M}}\left[\Vect{x}, \nabla \theta(\Vect{x}) \right]
	= 0
	.
\end{equation}

\noindent Such locations are called `caustics'. The accurate modeling of $\psi$ in the neighborhood of caustics is the primary goal of this work.


\section{Caustics and catastrophes of geometrical optics}

\begin{figure}
	\centering
	\includegraphics[width=0.5\linewidth]{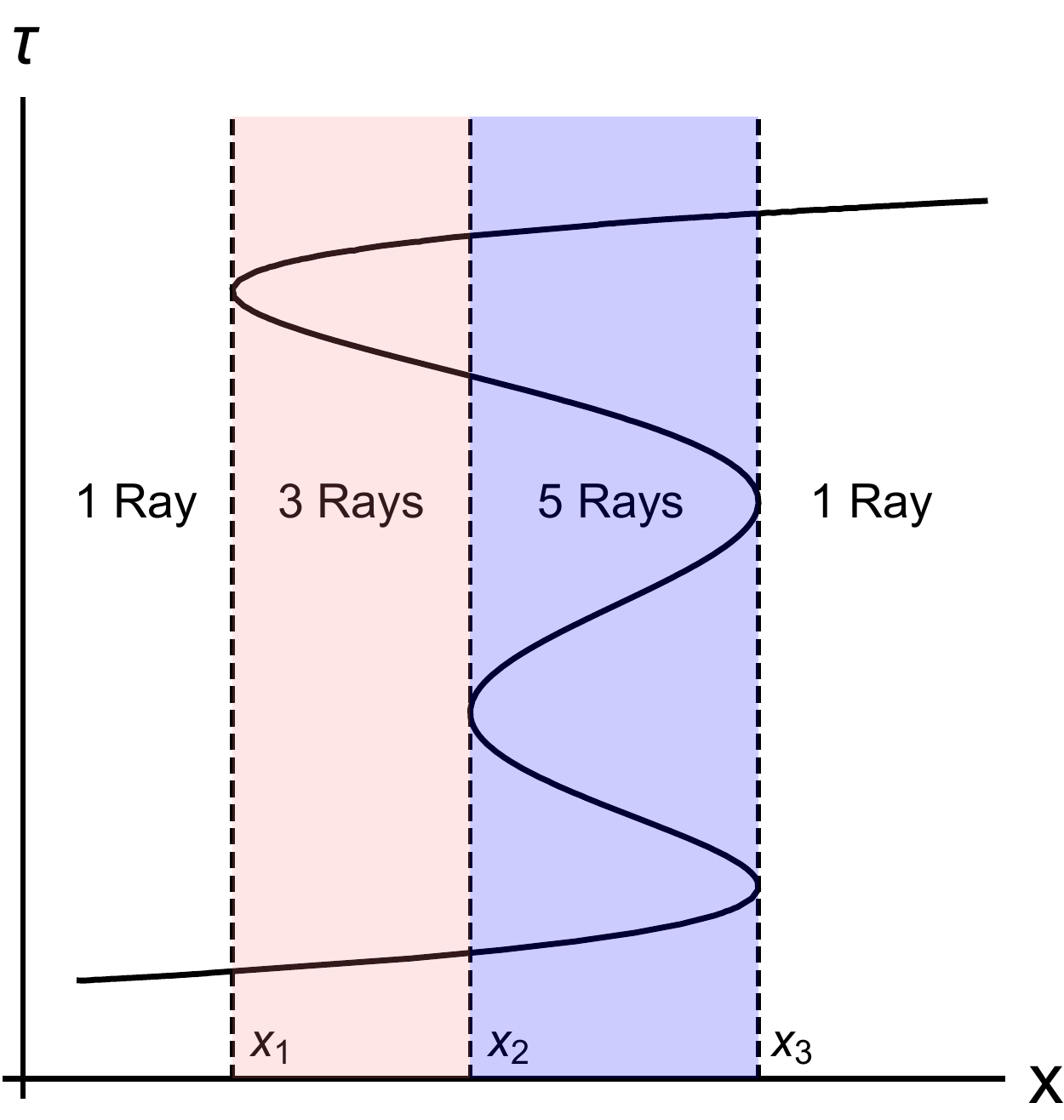}
	\caption{For a continuous function $\Vect{x}(\Vect{\tau})$, $\det \pd{\Vect{\tau}} \Vect{x} = 0$ typically occurs at boundaries between regions where $\Vect{x}(\Vect{\tau}) = \Vect{x}_0$ has differing numbers of roots. This is illustrated above in $1$-D, where $x'(\tau) = 0$ at $x_1$, $x_2$, and $x_3$. An exception is when $\det \pd{\Vect{\tau}} \Vect{x} = 0$ corresponds to a degenerate saddlepoint (an inflection point in $1$-D); however, these structures are not stable to small perturbations, so they do not correspond to any realistic 1-D physical system (although they can occur stably in higher dimensions).}
	\label{fig:2_caustPROJ}
\end{figure}

To better understand where and why caustics occur, let me consider the extended `ray parameter' space $(\Vect{x}, \Vect{\tau})$. In this space, the ray trajectories are represented by the graph $\Vect{\tau} = \Vect{\tau}(\Vect{x})$, which is obtained by a formal inversion of $\Vect{x}(\Vect{\tau})$. The condition $j(\Vect{\tau}) = 0$ has a geometric interpretation in this space: $j(\Vect{\tau}) = 0$ where the projection of $\Vect{\tau}(\Vect{x})$ onto $\Vect{x}$-space becomes singular. As can be seen from \Fig{fig:2_caustPROJ}, this condition importantly implies that caustics do not occur every time rays cross in $\Vect{x}$-space, but rather, when the number of rays crossing in $\Vect{x}$-space changes abruptly. In this sense, caustics appear as topological boundaries.

A powerful mathematical result concerning caustics can be obtained by the following observations. First, consider a given wavefield $\psi$ in terms of its Fourier transform $\fourier{\psi}$:
\begin{equation}
	\psi(\Vect{x})
	= \int \dd \Vect{k} \,
	\fourier{\env}(\Vect{k})
	\exp\left[
		i \fourier{\theta}(\Vect{k} )
		+ i \Vect{x}^\intercal \Vect{k}
	\right]
	.
	\label{eq:2_FT}
\end{equation}

\noindent The rays that underlie $\psi$ in $\Vect{x}$-space are determined by the stationarity condition
\begin{equation}
	\Vect{x} = - \pd{\Vect{k}} \fourier{\theta}(\Vect{k})
	,
	\label{eq:2_FTphase}
\end{equation}

\noindent and caustics occur where stationary points coalesce, \ie where
\begin{equation}
	\left.
		\det \pd{\Vect{k}\Vect{k}}^2 \fourier{\theta}(\Vect{k})
	\right|_{\Vect{x} = \pd{\Vect{k}} \fourier{\theta}(\Vect{k})}
	= 0
	.
\end{equation}

\noindent One can readily verify this definition coincides with that provided by \Eq{eq:2_caustic} by noting that the solution to \Eq{eq:2_FTphase} lies on the same ray manifold $\Vect{z}(\Vect{\tau})$ as \Eq{eq:2_pGRADtheta} (as I shall show explicitly in \Ch{ch:MGO}):
\begin{equation}
	\left.
		\det \pd{\Vect{k}\Vect{k}}^2 \fourier{\theta}(\Vect{k})
	\right|_{\Vect{x} = \pd{\Vect{k}} \fourier{\theta}(\Vect{k})}
	=
	\pd{\Vect{k}} \Vect{x}\left[ \Vect{k}(\Vect{\tau}) \right]
	=
	\pd{\Vect{\tau}} \Vect{x} 
	\left[
		\pd{\Vect{\tau}} \Vect{k}
	\right]^{-1}
	.
\end{equation}

\begin{figure}
	\centering
	\includegraphics[width=0.7\linewidth]{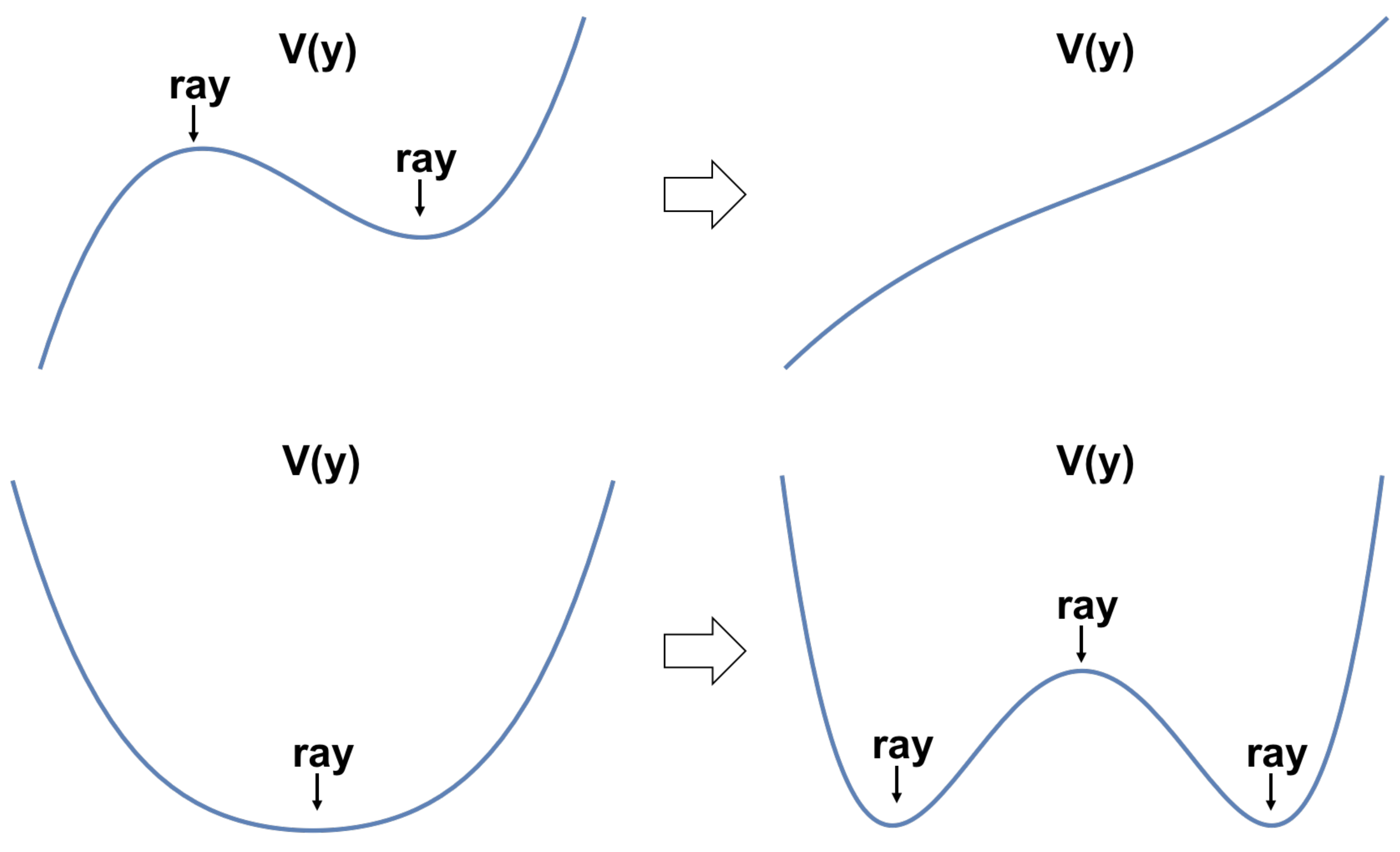}
	\caption{Under certain smooth parameter changes, the number of equilibria (rays) for a given potential function $V(y)$ can change number. Catastrophe theory studies how such changes occur for generic conditions.}
	\label{fig:2_equilibrium}
\end{figure}

Hence, rays can be considered the `equilibria' (with respect to $\Vect{k}$) of a parameterized `potential' function
\begin{equation}
	V(\Vect{x}, \Vect{k})
	\doteq
	\fourier{\theta}(\Vect{k} )
	+ \Vect{x}^\intercal \Vect{k}
	,
	\label{eq:2_rayPOTENTIAL}
\end{equation}

\noindent and (by the observations of \Fig{fig:2_caustPROJ}) caustics correspond to the values of $\Vect{x}$ for which the number of equilibria change (or `bifurcate').%
\footnote{One can also use the more obvious potential function $V(\Vect{x}, \Vect{k}) = \theta(\Vect{x}) - \Vect{x}^\intercal \Vect{k}$, considered as a function of $\Vect{x}$ and parameterized by $\Vect{k}$ [and whose equilibria correspondingly satisfy \Eq{eq:2_pGRADtheta}], to obtain a similar conclusion regarding $\Vect{k}$-space caustics.} %
This situation is illustrated in \Fig{fig:2_equilibrium}. It turns out that there exists an entire field of mathematics dedicated to studying such changes of equilibria: \textit{catastrophe theory}~\cite{Berry80b,Kravtsov93, Poston96}. Some basic results of catastrophe theory (taken as \textit{definitions} in lieu of formal proofs) are listed in \App{sec:2_catastropheREV} for completeness; for my purposes, however, it suffices to simply state that catastrophe theory provides a classification system for caustics based on their codimension, that is, the minimum number of spatial dimensions in which they can be observed. For example, the simplest type of caustic is the fold caustic, which occurs when a wave encounters a cutoff. The fold caustic has codimension $1$, so it can be observed in $N$-D systems with $N \ge 1$. On the other hand, the cusp caustic, which occurs at a focal point, has codimension $2$ and can thus only be observed for $N \ge 2$. This supports the intuition that cutoffs are well-described by $1$-D models like Airy's equation, but foci are inherently $2$-D.

\begin{table*}
	\centering
	\begin{tabular}{| c | c | c | c | c |}
		\multicolumn{1}{c}{\textbf{Name}} & \multicolumn{1}{c}{$\alpha$} & \multicolumn{1}{c}{$m$} & \multicolumn{1}{c}{$M$} & \multicolumn{1}{c}{$f_\alpha(\Vect{\kappa}, \Vect{y})$} \\
		\hline
		No caustic & $A_1$ & $0$ & $1$ 
		& $\kappa_1^2$ \\
		Fold & $A_2$ & $1$ & $1$ 
		& $\kappa_1^3 + y_1 \kappa_1$ \\
		Cusp & $A_3$ & $2$ & $1$ 
		& $\kappa_1^4 + y_2 \kappa_1^2 + y_1 \kappa_1$ \\
		Swallowtail & $A_4$ & $3$ & $1$ 
		& $\kappa_1^5 + y_3 \kappa_1^3 + y_2 \kappa_1^2 + y_1 \kappa_1$ \\
		Hyperbolic umbilic & $D_4^+$ & $3$ & $2$ 
		& $\kappa_1^3 + \kappa_2^3 + y_3 \kappa_1 \kappa_2 + y_2 \kappa_2 + y_1 \kappa_1$ \\[1mm]
		Elliptic umbilic & $D_4^-$ & $3$ & $2$ 
		& $\kappa_1^3 - 3 \kappa_1 \kappa_2^2 + y_3(\kappa_1^2 + \kappa_2^2) + y_2 \kappa_2 + y_1 \kappa_1$ \\[1mm]
		\hline
	\end{tabular}
	\caption{A complete list of the normal-form generators $f_\alpha(\Vect{\kappa}, \Vect{y})$ for caustics with codimension $m \le 3$~\cite{Berry80b, Arnold83, Kravtsov93}. (In the language of \App{sec:2_catastropheREV}, these are the unique \textit{universal unfoldings} of a given codimension.) The interference patterns that correspond to these caustics can be viewed in their entirety when the number of spatial dimensions $N = 3$. For each caustic, $\alpha$ is the Arnold label~\cite{Arnold83}, $m$ is the codimension, and $M$ is the corank.}
	\label{tab:2_caustic}
\end{table*}

There are three main advantages to using the catastrophe classification system to study caustics: \textbf{(i)} Only `structurally stable' caustics that are robust under small perturbations are included. These are the caustics that are most physically relevant, since a structurally unstable caustic will be destroyed by any imperfections in the experimental setup (which are of course unavoidable). For example, an EM wave propagating in an unmagnetized cold plasma with a linear density profile $n(x)$ will have a cutoff at some $x_c$. If the density is perturbed from $n(x)$ to $\tilde{n}(x)$ by some global motion of the plasma, the cutoff location will shift from $x_c$ to $\tilde{x}_c$, but will generally not disappear; hence, the cutoff (fold caustic) is `structurally stable'. \textbf{(ii)} There are only a finite number of distinct caustic types that are stable in a given number of dimensions. For example, only six different caustics can occur in $3$-D (including `no caustic'; see Table~\ref{tab:2_caustic}). \textbf{(iii)} General properties of a given caustic type can be determined by studying a single member in detail, often chosen to be the `simplest' member (the so-called `normal-form generator' of the caustic class; see below). These three results from catastrophe theory greatly reduce the work required to validate any new method in catastrophe optics; indeed, a new method for modeling caustics need only be tested on two different nontrivial caustics to be fully viable in $2$-D, or on five different nontrivial caustics for $3$-D.

As suggested by \Eq{eq:2_FT}, for my purposes I shall be concerned with the standard integrals of catastrophe theory, which are integral representations for caustic wavefields and take the general form
\begin{equation}
	I_\alpha(\Vect{y}) \doteq 
	\int \dd \Vect{\kappa} \,
	\exp\left[
		i f_\alpha(\Vect{\kappa}, \Vect{y})
	\right]
	,
	\label{eq:2_Ialpha}
\end{equation}

\noindent where $\Vect{y}$ is an $m$-D collection of `external' (or `control') variables, $\Vect{\kappa}$ is an $M$-D collection of `internal' (or `state') variables, and $\alpha$ labels the type of caustic. The integers $m \le N$ and $M \le N$ are the `codimension' and `corank' of the caustic, respectively, and the function $f_\alpha(\Vect{\kappa}, \Vect{y})$ is the normal-form generator for a type-$\alpha$ caustic. For example, the fold caustic, also called the $A_2$ caustic in Arnold's nomenclature~\cite{Arnold83}, has $m = 1$, $M = 1$, and
\begin{equation}
	f_{A_2}(\kappa_1, y_1)
	= \kappa_1^3 + y_1 \kappa_1
	.
\end{equation}

\noindent The corresponding $I_{A_2}(y_1)$ is proportional to the Airy function $\airyA(y_1/ \sqrt[3]{3})$~\cite{Olver10a}. See Table~\ref{tab:2_caustic} for more examples. Note that if $M < N$ such that only a subset of the integration variables of \Eq{eq:2_FT} are included in the standard integral $I_\alpha$, then by the splitting lemma of catastrophe theory (which is simply an extension of the Morse lemma)~\cite{Berry80b,Poston96}, the remaining $N - M$ integrals contained in \Eq{eq:2_FT} are decoupled and involve phase functions that are quadratic at most, and thereby trivially integrated. 


\section{Case study: caustics in paraxial propagation}
\label{sec:2_parax}

Before discussing how to accurately model caustics, it is instructive to review the different types of caustics that can occur in some detail. As Table~\ref{tab:2_caustic} shows, it turns out that only five unique caustics can occur for $3$-D systems: the fold ($A_2$), the cusp ($A_3$), the swallowtail ($A_4$), the hyperbolic umbilic ($D_4^+$), and the elliptic umbilic ($D_4^-$) catastrophe functions. Note that the labels within parenthesis correspond to the commonly adopted Arnold classification~\cite{Arnold75}. And as those labels suggest, the first three caustics in the list are members of a larger family of caustics called the `cuspoids' (represented by the label $A_{m+1}$ for $m \ge 1$), while the final two are members of the `umbilic' family (represented by the label $D_{m+1}^\pm$ for $m\ge 3$). Recall that $m$ is intuitively the minimum number of dimensions required to view the corresponding caustic in its entirety, and also that $m+1$ is the number of rays involved in creating the caustic pattern. For example, the fold caustic (which corresponds to a cutoff) can occur entirely in $1$-D and involves two interfering rays (the incoming and reflected rays).

To see how these caustics can occur in practice, let me consider a wavefield propagating paraxially in an ($N + 1$)-D uniform medium according to the wave equation
\begin{equation}
	4 \pi i \pd{z} \psi(\Vect{x}, z)
	+ \lambda \pd{\Vect{x}}^2 \psi(\Vect{x}, z) 
	= 0
	,
	\label{eq:2_paraxial}
\end{equation}

\noindent where $z$ is the direction of propagation, $\Vect{x}$ are the $N$-D coordinates transverse to $z$ (\ie the optical axis corresponds to $\Vect{x} = \Vect{0}$), and $\lambda$ is the wavelength. The formal solution to \Eq{eq:2_paraxial} is readily obtained:
\begin{subequations}
	\begin{equation}
		\psi(\Vect{x}, z) = 
		\exp\left(
			i \frac{\lambda z}{4 \pi} 
			\pd{\Vect{x}}^2
		\right) \psi(\Vect{x},0)
		,
		\label{eq:2_freePMT}
	\end{equation}

	\noindent or equivalently,
	\begin{equation}
		\psi(\Vect{x}, z) =
		\int \dd \Vect{y} \,
		\frac{
			\psi(\Vect{y}, 0) 
		}{ 
			\left( i \lambda z \right)^{N/2}
		}
		\exp\left( 
			i \pi\frac{
				\|\Vect{y} - \Vect{x}\|^2
			}{\lambda z} 
		\right)
		.
		\label{eq:2_freeMT}
	\end{equation}
\end{subequations}

\noindent One recognizes \Eq{eq:2_freeMT} as the well-known Fresnel diffraction integral~\cite{Born99}, but it is also an example of a metaplectic transform (\Ch{ch:MT}), which feature prominently in MGO. The GO rays for \Eq{eq:2_paraxial} solve the local dispersion relation
\begin{equation}
	\Symb{D}(\Vect{x},z, \Vect{k}, k_z)
	= \frac{4\pi k_z}{\lambda} + \Vect{k}^\intercal \Vect{k}
	= 0
	.
\end{equation}

\noindent Hence, they are given explicitly as
\begin{align}
	\Vect{x}(z, \Vect{x}_0) &= \Vect{x}_0 + \frac{\lambda z}{2\pi} \Vect{k}_0(\Vect{x}_0)
	, \quad
	\Vect{k}(z, \Vect{x}_0) = \Vect{k}_0(\Vect{x}_0)
	,
	\label{eq:2_paraxRAYS}
\end{align}

\noindent where I have chosen to set $\Vect{\tau} = (z, \Vect{x}_0)$.


\subsection{Cuspoid caustics}
\label{sec:2_cuspoid}

Consider first the case $N = 1$ (so $2$-D including the longitudinal dimension). An $A_{m+1}$-type cuspoid caustic can be generated by choosing the following initial conditions for $\psi$ (up to an arbitrary constant factor):
\begin{equation}
	\psi(x, 0)
	= \exp\left( i \, \frac{x^{m+2}}{\ell^{m+2}} + i \sum_{j = 1}^m a_j \frac{x^j}{\ell^j} \right)
	,
	\label{eq:2_cuspoidINIT}
\end{equation}

\noindent which corresponds to
\begin{equation}
	k_{x,0}(x_0) = (m+2)\frac{x_0^{m+1}}{\ell^{m+2}} + \sum_{j = 1}^m j a_j \frac{x_0^{j-1}}{\ell^j}
	.
	\label{eq:2_cuspoidINITk}
\end{equation}

\noindent (Here $\{a_j\}$ are constant parameters, \eg lens aberrations, and $\ell$ determines the characteristic length.) Then, for $m = 1$, \Eq{eq:2_freeMT} leads to
\begin{equation}
	\psi(x, z) =
	\frac{\ell}{\sqrt{ i \lambda z} } \,
	\exp \left(
		i \frac{\pi x^2}{\lambda z} 
		+ i \frac{2 \pi^2 \ell^3 x}{3 \lambda^2 z^2}
		+ i \frac{2 \pi^3 \ell^6}{27 \lambda^3 z^3}
		- i a_1 \frac{\pi \ell^2}{3 \lambda z}
	\right)
	A_2\left(
		a_1 
		- \frac{2 \pi \ell x}{\lambda z} 
		- \frac{\pi^2 \ell^4}{3 \lambda^2 z^2}
	\right)
	,
	\label{eq:2_cuspoidSOLm1}
\end{equation}

\noindent and for $m > 1$, \Eq{eq:2_freeMT} leads to
\begin{equation}
	\psi(x, z) =
	\frac{\ell}{\sqrt{ i \lambda z} } \,
	\exp \left(
		i \frac{\pi x^2}{\lambda z} 
	\right)
	A_{m+1}\left(
		a_1 - \frac{2\pi \ell x}{\lambda z}
		,
		a_2 + \frac{\pi \ell^2}{\lambda z}
		,
		a_3
		, \ldots
		,
		a_m
	\right)
	,
	\label{eq:2_cuspoidSOL}
\end{equation}

\noindent where the $A_{m+1}$ `catastrophe integral' is defined as
\begin{equation}
	A_{m+1}\left(
		a_1
		, \ldots
		,
		a_m
	\right)
	\doteq
	\int \dd y \,
	\exp\left(
		i y^{m+2} + i \sum_{j = 1}^m a_j y^j 
	\right)
	.
\end{equation}

\noindent Note that \Eq{eq:2_cuspoidSOLm1} can also be written in terms of the Airy function by using the relation
\begin{equation}
	A_2\left(
		a_1 
	\right)
	=
	\frac{2\pi}{3^{1/3}}
	\airyA
	\left(
		\frac{a_1}{3^{1/3}}
	\right)
	.
\end{equation}

\begin{figure}
	\centering
	\includegraphics[width=0.7\linewidth,trim={3mm 22mm 13mm 21mm},clip]{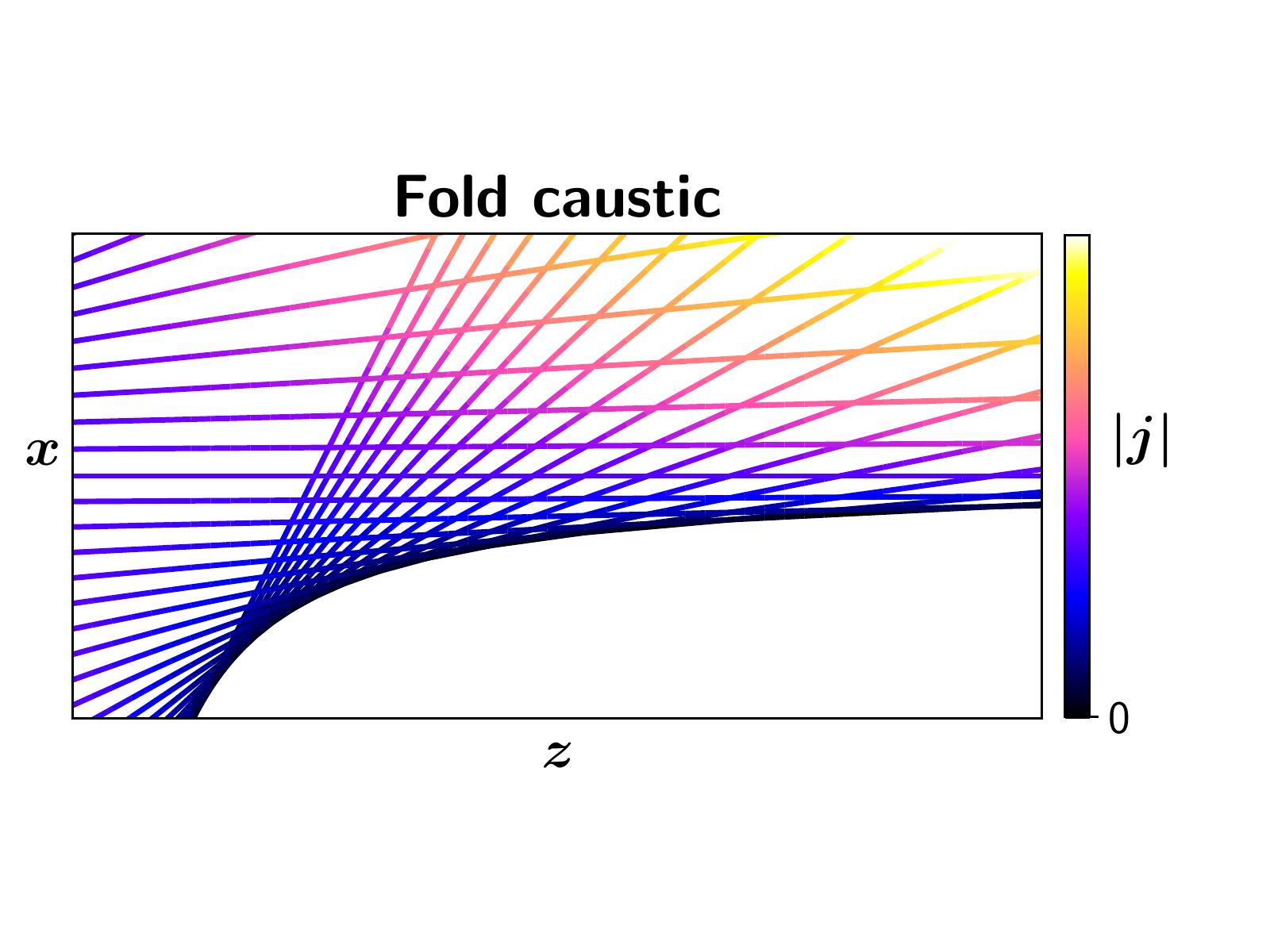}
	\caption{Ray trajectories for the fold caustic obtained via \Eq{eq:2_rayFOLD} with $\lambda = 2\pi$, $\ell = \sqrt[3]{3}$, and $a_1 = 0$. The color shows the magnitude of the Jacobian $j(\Vect{\tau})$ defined by \Eq{eq:2_jacDEF}. The caustic occurs where $j = 0$ (black curve).}
	\label{fig:2_fold}
\end{figure}

The simplest caustic of the cuspoid family is the $A_2$ fold caustic. The field near a fold caustic is given by \Eq{eq:2_cuspoidSOLm1}, and the underlying ray trajectories are given by \Eqs{eq:2_paraxRAYS} and \eq{eq:2_cuspoidINITk}; namely,
\begin{equation}
	x(z, x_0)
	= x_0
	+ \frac{\lambda z}{2\pi \ell }
	\left(
		\frac{3 x_0^2}{\ell^2} + a_1
	\right)
	.
	\label{eq:2_rayFOLD}
\end{equation}

\noindent The fold caustic occurs where \Eq{eq:2_caustic} is satisfied, or equivalently, where $\pd{x_0}x(z,x_0) = 0$. This ultimately yields the caustic curve
\begin{equation}
	x_c(z)
	=
	a_1\frac{\lambda z}{2 \pi \ell}
	- \frac{\pi \ell^3}{6 \lambda z}
	.
	\label{eq:2_causticFOLD}
\end{equation}

\noindent The ray pattern \eq{eq:2_rayFOLD} is shown in \Fig{fig:2_fold}, in which the caustic curve \eq{eq:2_causticFOLD} is clearly visible.

Let me next consider the $A_3$ cusp caustic. The field near a cusp caustic is given by \Eq{eq:2_cuspoidSOL} with $m = 2$, and the underlying ray trajectories are given as
\begin{equation}
	x(z, x_0)
	= x_0
	+ \frac{\lambda z}{2\pi \ell}
	\left(
		\frac{4 x_0^3}{\ell^3} 
		+ a_1
		+ a_2 \frac{2 x_0}{\ell}
	\right)
	.
	\label{eq:2_rayCUSP}
\end{equation}

\noindent The cusp caustic can be shown to occur along the curve
\begin{equation}
	x_c(z) = 
	a_1\frac{ \lambda z}{2 \pi \ell} 
	\pm
	\frac{
	    \sqrt{- 6 \lambda z (\pi \ell^2 + a_2 \lambda z )^3}
	}{
		9\pi \ell \lambda  z
	}
	.
	\label{eq:2_causticCUSP}
\end{equation}

\begin{figure}
	\centering
	\includegraphics[width=0.7\linewidth,trim={3mm 22mm 13mm 21mm},clip]{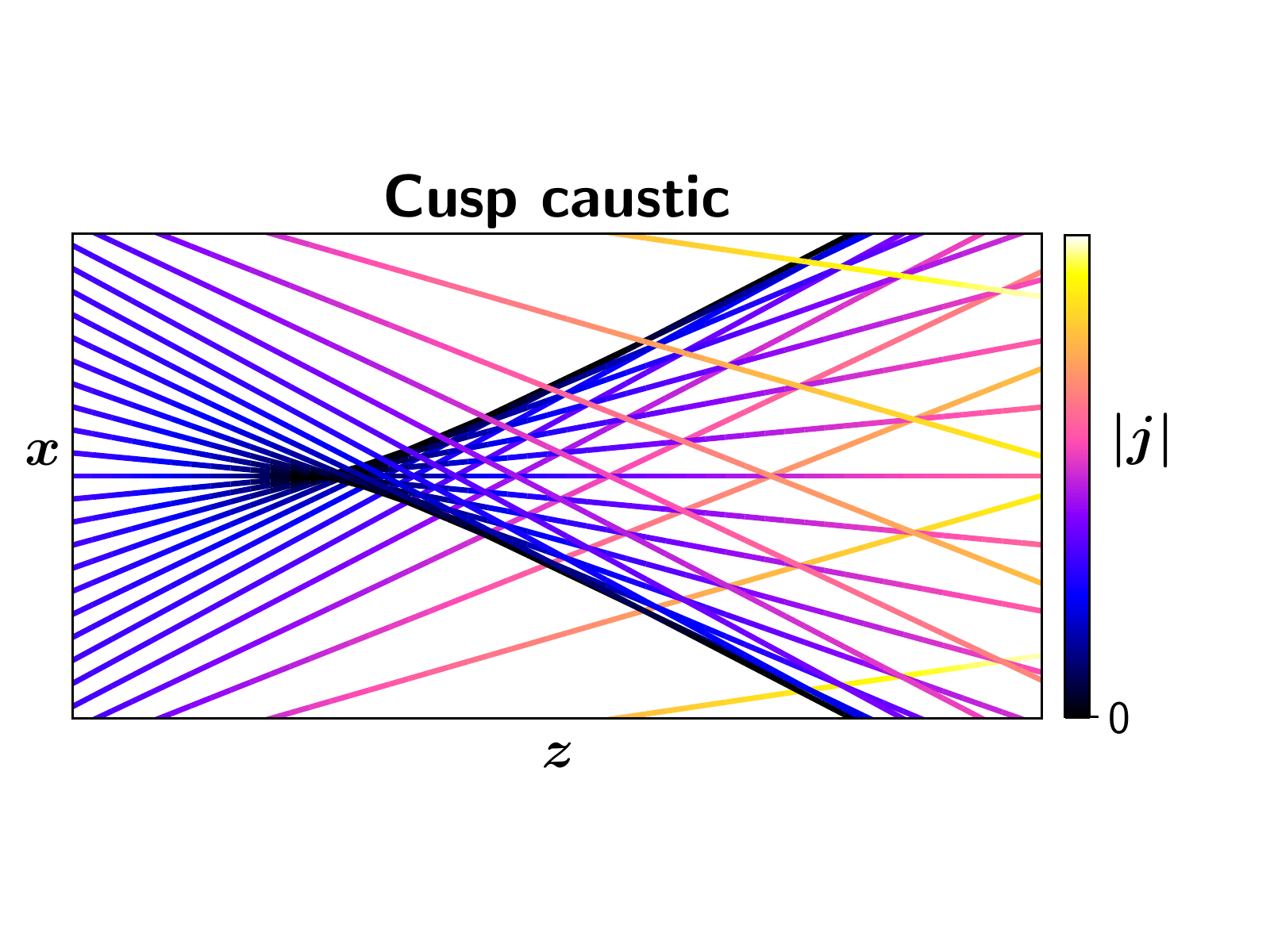}
	\caption{Same as \Fig{fig:2_fold} but for the cusp caustic \eq{eq:2_rayCUSP} with $\lambda = 2\pi$, $\ell = \sqrt{2}$, $a_1 = 0$, and $a_2 = -2$.}
	\label{fig:2_cusp}
\end{figure}

\noindent The ray pattern \eq{eq:2_rayCUSP} is shown in \Fig{fig:2_cusp}, in which the caustic curve \eq{eq:2_causticCUSP} is clearly visible.

Let me next consider the final stable cuspoid in $3$-D, the $A_4$ swallowtail caustic. Here, I shall choose to generate the swallowtail via the mathematically simpler approach of having a high-order $1$-D aberration, rather than with a low-order $2$-D aberration as would more commonly occur in practice. Correspondingly, the field near a swallowtail caustic is given by \Eq{eq:2_cuspoidSOL} with $m = 3$, and the underlying ray trajectories are given as
\begin{equation}
	x(z, x_0)
	= x_0
	+ \frac{\lambda z}{2\pi \ell}
	\left(
		\frac{5 x_0^4}{\ell^4} 
		+ a_1
		+ a_2 \frac{2 x_0}{\ell}
		+ a_3 \frac{3 x_0^2}{\ell^2}
	\right)
	.
	\label{eq:2_raySWTAIL}
\end{equation}

\noindent The cusp caustic can be shown to occur along the parametric curve
\begin{align}
	x_c(\zeta)
	=
	\frac{a_1 \lambda z_c(\zeta)}{2 \pi \ell}
	- \frac{3 \lambda z_c(\zeta)}{2 \pi \ell} (a_3 + 5 \zeta^2) \zeta^2
	, \quad
	z_c(\zeta)
	= - \frac{\pi \ell^2}
	{
		\lambda \left[
			a_2 + \zeta (3 a_3 + 10 \zeta^2)
		\right]
	}
	,
	\label{eq:2_causticSWTAIL}
\end{align}

\begin{figure}
	\centering
	\includegraphics[width=0.7\linewidth,trim={3mm 22mm 13mm 21mm},clip]{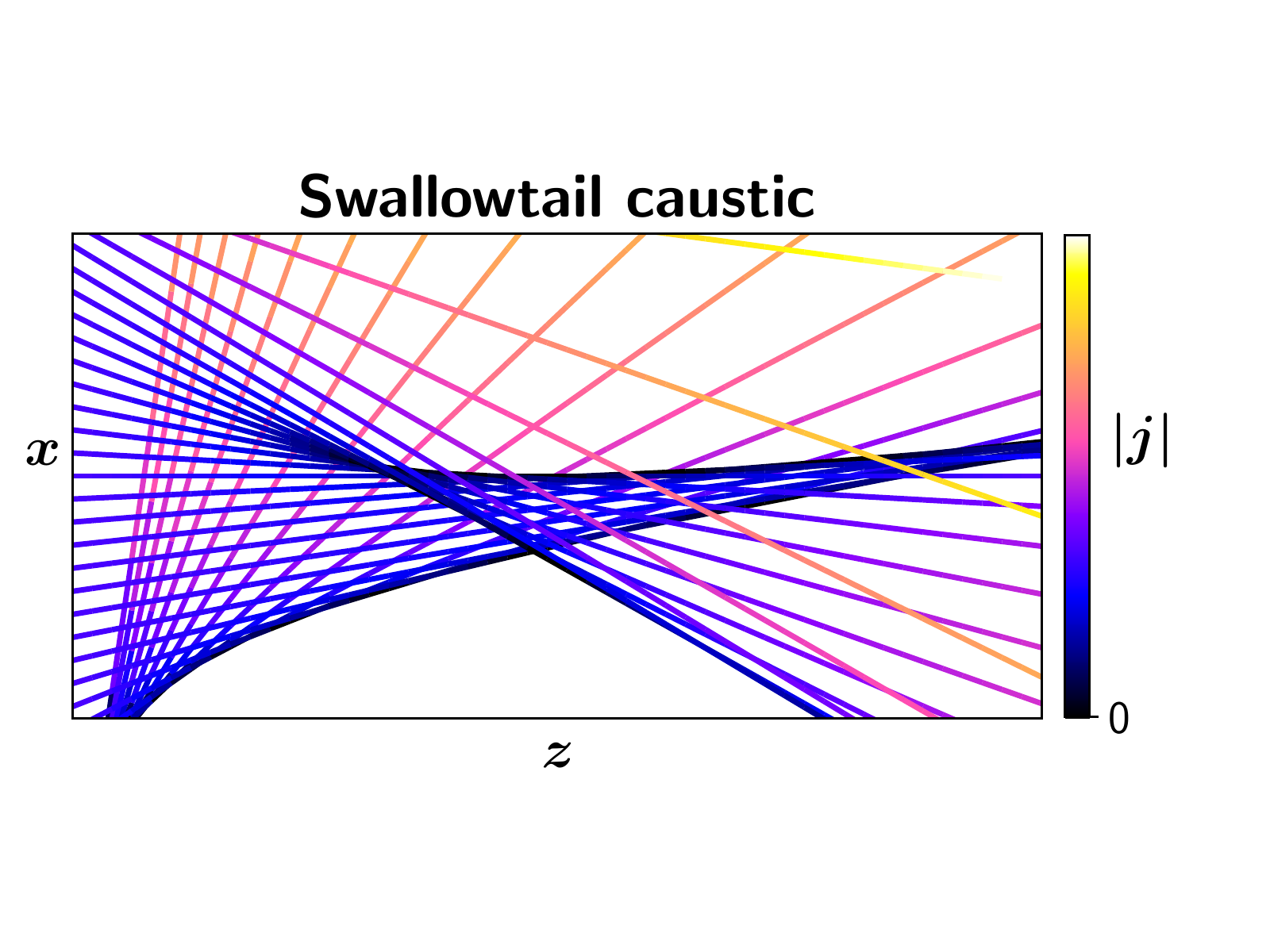}
	\caption{Same as \Fig{fig:2_fold} but for the swallowtail caustic \eq{eq:2_raySWTAIL} with $\lambda = 2\pi$, $\ell = \sqrt[5]{5}$, $a_1 = 0$, $a_2 = -1$, and $a_3 = -1$.}
	\label{fig:2_swtail}
\end{figure}

\noindent where $\zeta \in (-\infty, \infty)$ is a $1$-D parameterization of the caustic curve in the ($x$,$z$) plane. The ray pattern \eq{eq:2_raySWTAIL} is shown in \Fig{fig:2_swtail} for parameters specifically chosen to have the eponymous swallowtail section of the caustic curve \eq{eq:2_causticSWTAIL} appear in the longitudinal plane.


\subsection{Umbilic caustics}

Consider now $N = 2$ (so the total number of spatial dimensions is three). In addition to the cuspoids, a new class of caustics, the umbilics, are now possible. A wavefield containing any caustic from the $D_{m+1}^\pm$ umbilic series can be generated by choosing
\begin{equation}
	\psi(\Vect{x}, 0)
	=
	\exp\left(
		i \frac{y^m}{\ell^m}
		\pm i \frac{x^2 y}{\ell^3}
		+ i a_1 \frac{x}{\ell}
		+ i a_2 \frac{y}{\ell}
		+ i a_3 \frac{x^2}{\ell^2}
		+ i \sum_{j = 4}^m a_j \frac{y^{j - 2}}{\ell^{j-2}}
	\right)
	.
	\label{eq:2_umbilicINIT}
\end{equation}

\noindent The corresponding solution \eq{eq:2_freeMT} for $m \ge 4$ is
\begin{align}
	\psi(\Vect{x}, z) =
	\frac{\ell^2 }{i \lambda z}
	\exp \left(
		i \frac{\pi \Vect{x}^\intercal \Vect{x}}{\lambda z} 
	\right)
	D_{m+1}^\pm
	\left(
		a_1 - \frac{2\pi \ell x}{\lambda z}
		,
		a_2 - \frac{2\pi \ell y}{\lambda z}
		,
		a_3 + \frac{\pi \ell^2 }{\lambda z}
		, 
		a_4 + \frac{\pi \ell^2 }{\lambda z}
		,
		a_5
		, \ldots
		,
		a_m
	\right)
	,
	\label{eq:2_umbilicSOL}
\end{align}

\noindent while for $m = 3$ is given as
\begin{align}
	\psi(\Vect{x}, z) &=
	\frac{\ell^2 }{i \lambda z}
	\exp \left(
		i \frac{\pi \Vect{x}^\intercal \Vect{x}}{\lambda z} 
		-
		i a_2 \frac{\pi \ell^2}{3 \lambda z}
		+ i \frac{2 \pi^2 \ell^3}{3 \lambda^2 z^2} y
		+ i \frac{2 \pi^3 \ell^6}{27 \lambda^3 z^3}
	\right)
	\nonumber\\
	&\hspace{4mm}\times
	D_{4}^\pm
	\left(
		a_1 - \frac{2\pi \ell x}{\lambda z}
		,
		a_2 - \frac{2\pi \ell y}{\lambda z} - \frac{\pi^2 \ell^4 }{3 \lambda^2 z^2}
		,
		a_3 + \frac{3 \mp 1 }{3 \lambda z} \pi \ell^2
	\right)
	,
	\label{eq:2_umbilicSOLm3}
\end{align}
    
\noindent where the $D_{m+1}^\pm$ catastrophe integral is defined as
\begin{align}
	D_{m+1}^\pm
	\left(
		a_1
		, \ldots
		,
		a_m
	\right)
	\doteq 
	\int \dd U \, \dd V
	\exp\left[
		i V^m
		\pm i U^2 V
		+ i a_1 U
		+ i a_2 V
		+ i a_3 U^2 
		+ i \sum_{j = 4}^m a_j V^{j - 2}
	\right]
	.
\end{align} 

\noindent Note that the initial condition \eq{eq:2_umbilicINIT} generates rays having
\begin{equation}
	k_{x,0}(\Vect{x}_0) =
	\pm 2 \frac{x_0 y_0}{\ell^3}
	+ \frac{a_1}{\ell}
	+ 2 a_3 \frac{x_0}{\ell^2}
	, \quad
	k_{y,0}(\Vect{x}_0) = m\frac{y_0^{m-1}}{\ell^m}
	\pm \frac{x_0^2}{\ell^3}
	+ \frac{a_2}{\ell}
	+ \sum_{j = 4}^m (j - 2) a_j \frac{y_0^{j - 3}}{\ell^{j-2}}
	.
	\label{eq:2_umbilicINITk}
\end{equation}

The only umbilics that occur stably in $3$-D are the $D_4^+$ hyperbolic and the $D_4^-$ elliptic umbilic caustics. The field near these caustics are given by \Eq{eq:2_umbilicSOLm3}. The underlying ray trajectories for the hyperbolic umbilic are given as
\begin{align}
	x(z; \Vect{x}_0) =
	x_0 +
	\frac{\lambda z}{2\pi \ell}
	\left(
		a_1
		+ 2 a_3 \frac{x_0}{\ell}
		+ 2 \frac{x_0 y_0}{\ell^2} 
	\right)
	, \quad
	y(z; \Vect{x}_0) = 
	y_0 +
	\frac{\lambda z}{2 \pi \ell}
	\left(
		a_2
		+ \frac{3y_0^2 + x_0^2}{\ell^2}
	\right)
	,
	\label{eq:2_rayHUMB}
\end{align}

\noindent and the caustic in the $(x,y)$ transverse plane at fixed propagation distance $z$ is given by the parametric curve 
\begin{subequations}
	\label{eq:2_causticHUmb}
	\begin{align}
		x_c(\zeta)
		&= 
		a_1 \frac{\lambda z}{2\pi \ell}
		- \frac{\sqrt{3} \lambda z}{4 \pi \ell}\left( 
			a_3 + \frac{2\pi \ell^2}{3\lambda z}
		\right)^2
		\sinh(\zeta)
		\left[
			\cosh(\zeta) \pm 1
		\right]
		,\\
		y_c(\zeta)
		&=
		a_2 \frac{\lambda z}{2\pi \ell}
		- \frac{\pi \ell^3}{6 \lambda z}
		+ \frac{3 \lambda z}{4\pi \ell}
		\left( 
			a_3 + \frac{2\pi \ell^2}{3\lambda z}
		\right)^2
		\cosh(\zeta)
		\left[
			\cosh(\zeta) \mp 1
		\right]
		.
	\end{align}
\end{subequations}

\begin{figure}
	\centering
	\includegraphics[width=0.6\linewidth,trim={29mm 5mm 11mm 4mm}, clip]{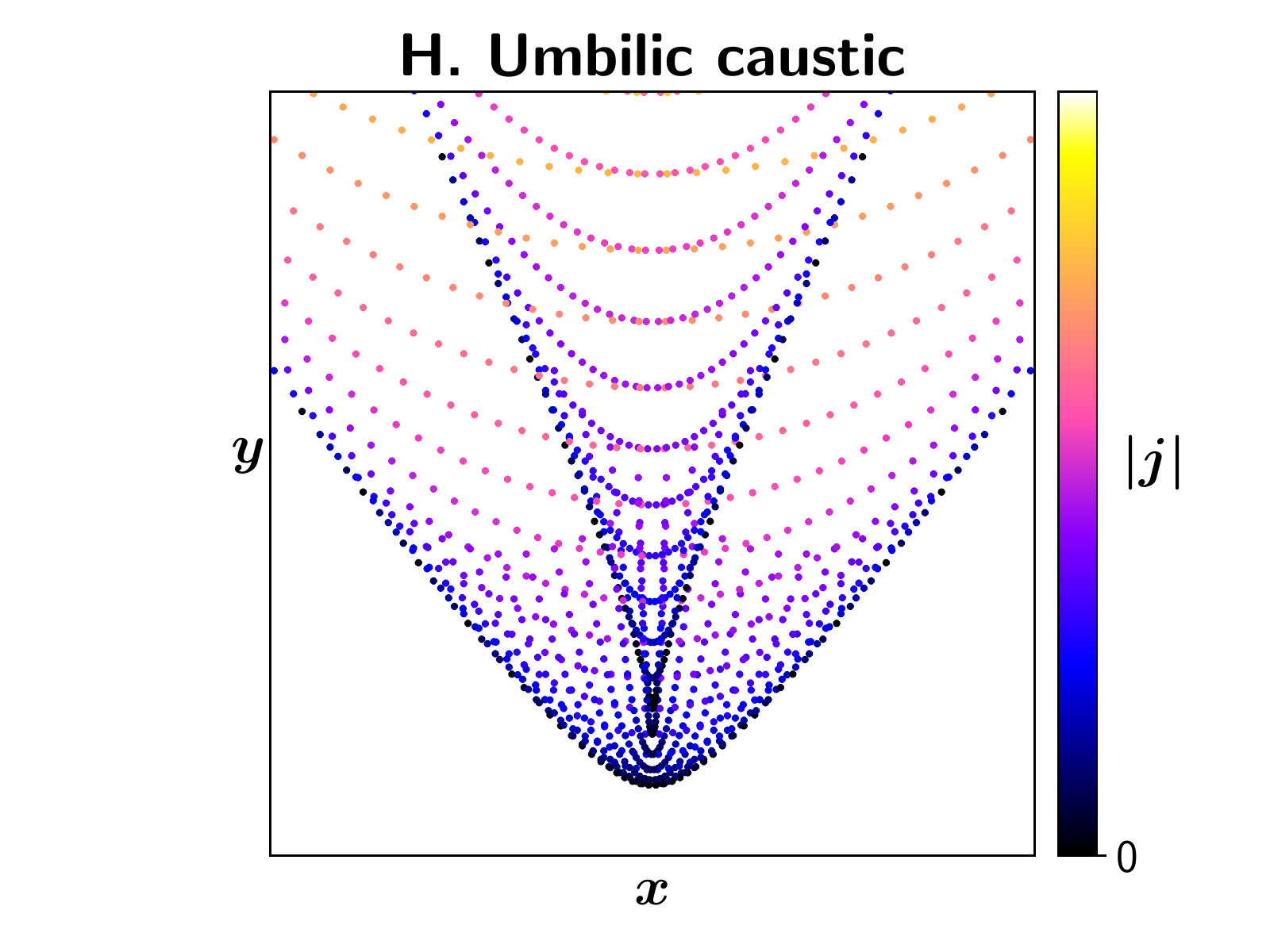}
	\caption{Intersection plot of the ray trajectories for the hyperbolic umbilic caustic obtained via \Eq{eq:2_rayHUMB} with $\lambda = 4\pi$, $\ell = 1$, $a_1 = 0$, $a_2 = -1.3$, and $a_3 = -1$ through the plane $z = 1$.}
	\label{fig:2_HUmb}
\end{figure}

\noindent Note that the hyperbolic umbilic actually contains two separate caustic curves (indicated by the $\pm$ terms above): a fold curve (top sign) and a cusp curve (bottom sign). These two curves lie in a squid-like orientation with the cusp residing within the bow of the fold. This is readily observed in \Fig{fig:2_HUmb}, which presents an intersection plot for the rays \eq{eq:2_rayHUMB} with respect to the plane $z = 1$. As each dot represents a single traversing ray, the $D_4^+$ caustic \eq{eq:2_causticHUmb} manifests as a visible increase in the ray density.

Similarly, the ray trajectories for the elliptic umbilic are given as
\begin{align}
	x(z; \Vect{x}_0) =
	x_0 +
	\frac{\lambda z}{2\pi \ell}
	\left(
		a_1
		+ 2 a_3 \frac{x_0}{\ell}
		- 2 \frac{x_0 y_0}{\ell^2} 
	\right)
	, \quad
	y(z; \Vect{x}_0) =
	y_0 +
	\frac{\lambda z}{2\pi \ell}
	\left(
		a_2
		+ \frac{3 y_0^2 - x_0^2}{\ell^2}
	\right)
	,
	\label{eq:2_rayEUMB}
\end{align}

\noindent with the caustic at a fixed distance $z$ given by the parametric curve
\begin{subequations}
	\label{eq:2_causticEUmb}
	\begin{align}
		x_c(\zeta)
		&= 
		a_1 \frac{\lambda z}{2\pi \ell}
		+ \frac{\sqrt{3} \lambda z}{2 \pi \ell}\left( 
			a_3 + \frac{4\pi \ell^2}{3\lambda z}
		\right)^2
		\sin^2\left(\frac{\zeta}{2} \right)
		\sin(\zeta)
		, \\
		y_c(\zeta)
		&=
		a_2 \frac{\lambda z}{2\pi \ell}
		- \frac{\pi \ell^3}{6 \lambda z}
		+ \frac{3 \lambda z}{2\pi \ell}
		\left( 
			a_3 + \frac{4\pi \ell^2}{3\lambda z}
		\right)^2
		\cos^2\left(\frac{\zeta}{2} \right)
		\cos(\zeta)
		.
	\end{align}
\end{subequations}

\begin{figure}
	\centering
	\includegraphics[width=0.6\linewidth,trim={29mm 5mm 11mm 4mm}, clip]{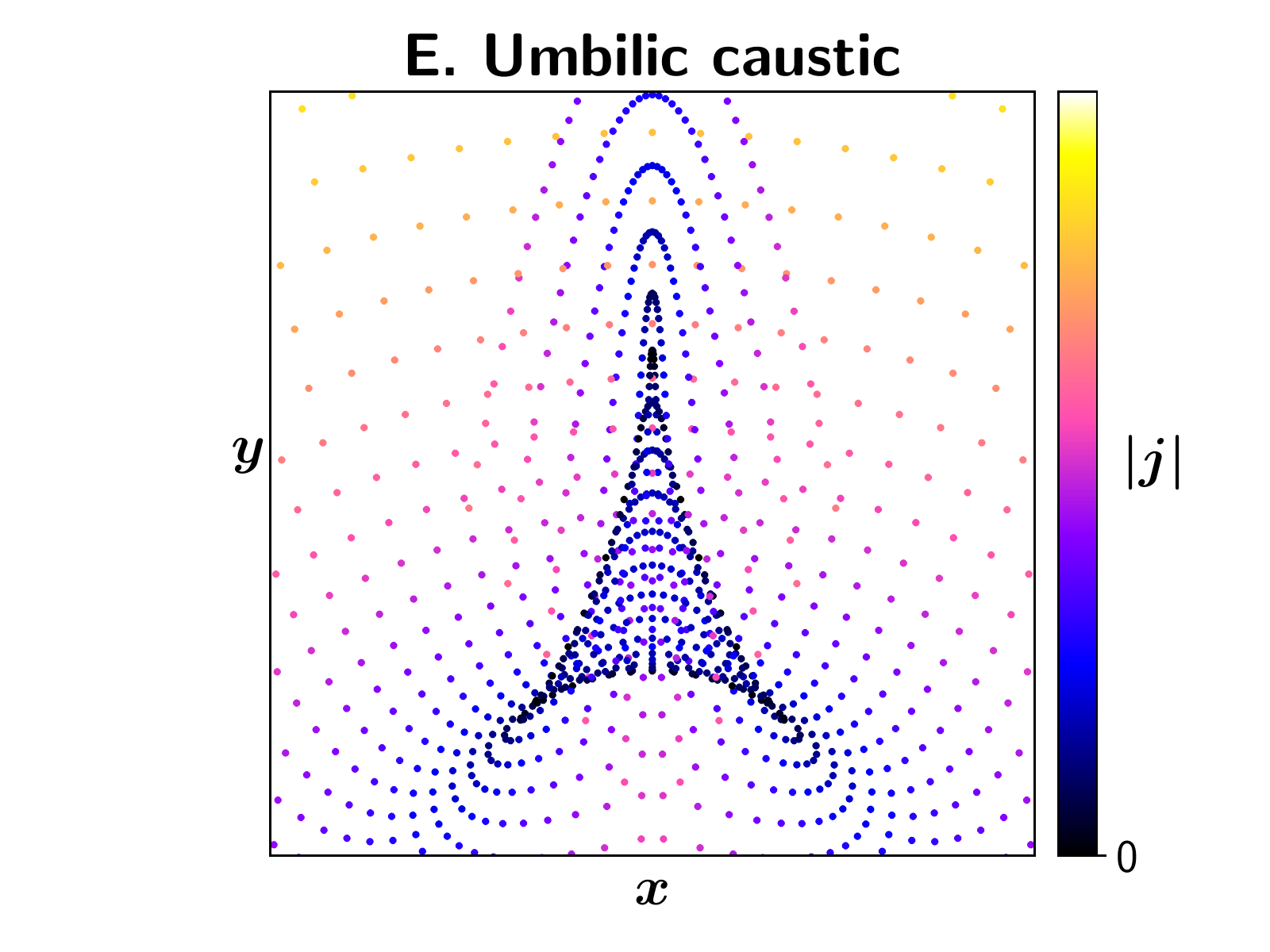}
	\caption{Same as \Fig{fig:2_HUmb} but for the elliptic umbilic caustic \eq{eq:2_rayEUMB} with $\lambda = 4\pi$, $\ell = 1$, $a_1 = 0$, $a_2 = -0.7$, and $a_3 = -2.1$.}
	\label{fig:2_EUmb}
\end{figure}

\noindent Figure \ref{fig:2_EUmb} presents an intersection plot of the ray trajectories \eq{eq:2_rayEUMB} that generate the $D_4^-$ caustic with respect to the plane $z = 1$. The characteristic tricorn shape of the caustic curve \eq{eq:2_causticEUmb} is readily observed by the visible increase in the ray density.


\section{Remedies to avoid caustics in ray-tracing simulations}

Having identified how caustics arise in GO (the vanishing of the ray Jacobian), and having discussed the standard classification of caustics at length, let me now begin to consider how caustic singularities might be removed from ray-tracing simulations.


\begin{figure}
	\centering
	\includegraphics[width=\linewidth]{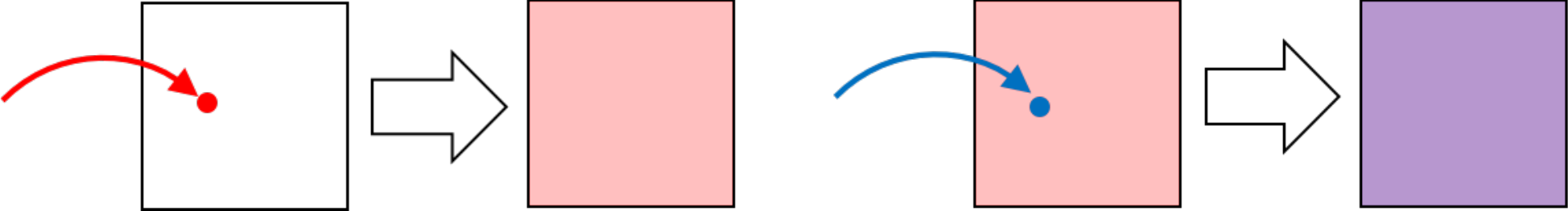}
	\caption{In many artificial intensity regularization schemes, the intensity of a ray is `smeared' over a simulation cell and thereby maintained finite even at caustics. When multiple rays intersect the same simulation cell, their contributions are simply added (just as adding red and blue makes purple).}
	\label{fig:2_smear}
\end{figure}

\subsection{Artificial intensity limiters}

Since caustics are singularities of the GO envelope, a simple, crude, yet surprisingly common remedy is to cap the wave intensity at some maximum value artificially. This is most often done by `smearing' the envelope over a simulation cell, as depicted in \Fig{fig:2_smear}. Then, since a realistic simulation has a finite discretization and only uses a finite number of rays, the intensity will remain finite as well. There is a logic to this approach, albeit flawed: as discussed following \Eq{eq:2_rayENV}, the GO envelope satisfies a flux conservation of the form
\begin{equation}
	\int \env^2 \, \Vect{v} \cdot \dd \Vect{a} = \textrm{constant} ,
	\label{eq:2_conserved}
\end{equation}

\noindent where $\Vect{v}$ is the ray (group) velocity and $\dd \Vect{a}$ is the cross-sectional area of a bundle of neighboring rays. Although this conservation breaks down at caustics because $\dd \Vect{a}$ becomes ill-defined, nevertheless researchers often leverage \Eq{eq:2_conserved} to forgo the explicit evolution of $\env$ to save computational resources. They later attempt to reconstruct $\env$ by using some form of heuristic but dimensionally correct analogue of \Eq{eq:2_conserved}~\cite{Kaiser00, Colaitis18}. A review of such methods is provided in \Ref{Colaitis21}; for my purposes it suffices to state that these methods express $\env$ as
\begin{equation}
	\env^2 \propto \frac{\textrm{num. rays}}{\textrm{mesh volume}}
	.
	\label{eq:2_causticHEURISTIC}
\end{equation}

\noindent Hence, such `fixes' have unpredictable behavior at caustics due to the explicit dependence of $\env$ on simulation parameters. This can lead to spurious effects if one is not careful. Additionally, the simulation grid is usually determined by plasma transport processes, which vary on scales much larger than the typical wavelength; hence, such intensity estimators are unable to resolve small-scale features that might be necessary to resolve important physics such as mode-conversion and parametric decay instabilities. These meshes also typically move and deform as the plasma evolves in time, which can erroneously cause the caustic intensities calculated via \Eq{eq:2_causticHEURISTIC} to change due to local compression of the mesh even if the plasma optical properties, \eg locations of density cutoffs, are not changing appreciably.

Moreover, these schemes typically add intensities, rather than fields, when multiple rays intersect. Although this approach can reproduce certain large-scale behavior, it fails to reproduce the correct linear behavior of interfering waves by neglecting the final term in the following expression:
\begin{equation}
	|\psi_1 + \psi_2|^2 =
	|\psi_1|^2 + |\psi_2|^2 + 2 |\psi_1| |\psi_2| \cos \vartheta
	,
\end{equation}

\noindent where $\vartheta$ is the relative phase between the two interfering wavefields $\psi_1$ and $\psi_2$. The interference term is responsible for establishing the beat pattern between overlapping wavefields; its improper calculation therefore prevents the proper calculation of important nonlinear phenomena such as cross-beam energy transfer (CBET) that are integral to hohlraum performance~\cite{Michel09}. Considerable effort must be made to introduce CBET back into the equations `by hand'~\cite{Colaitis18}, which may not be necessary if the linear modeling is improved. A theory-based model for caustics is needed.


\subsection{Etalon integrals and uniform approximations}

As mentioned, catastrophe theory provides a complete list of stable caustics that can occur in a given number of dimensions. However, these standard caustics typically govern only the local behavior of a given caustic wavefield. The global behavior can sometimes be modeled using the method of `uniform approximation'~\cite{Olver10a,Chester57,Ludwig66,Berry80b,Kravtsov93}, also called `etalon integrals'~\cite{Colaitis19a,Colaitis19b,Colaitis21}, in which a given wavefield is asymptotically matched to one of these local normal forms. However, this method relies on \textbf{(i)} the caustic type being known beforehand or somehow guessed reliably (which is fine for interpretive but not predictive simulations) and \textbf{(ii)} only a single caustic being present. Indeed, the elementary catastrophes mentioned here often combine to form `caustic networks', an example being an EM wave focused on a cutoff producing a fold-cusp network.%
\footnote{One might notice that the fold-cusp network of a focused beam reflecting off a cutoff is actually just a $2$-D section of the codimension-$3$ hyperbolic umbilic caustic~\cite{Lopez21DPPb}; hence, often one can equivalently think of caustic networks as being `organized', \ie lower-dimensional slices of, by a higher-dimensional single caustic, although the complicated nature of higher-dimensional caustics often prevents such `direct' analysis from being practical.} %
It might be possible to infer basic properties of such caustic networks from the constituent members, but complete understanding can only be achieved by considering the caustic network as a whole, which is very difficult to do.

Moreover, simulation parameter scans `invalidate' the catastrophe classification. By this, I mean that each parameter being scanned increases the `effective' dimensionality of the simulation such that higher-codimension caustics can now appear stably in lower (physical) dimensional simulations. For example, a $2$-D simulation with one scan parameter might start having swallowtail and umbilic caustics appearing. (I actually used this fact previously in \Sec{sec:2_cuspoid} to create the swallowtail caustic in $1$-D.) Hence, attempting to use etalon integrals in parameter scans quickly becomes impractical because too many caustic types need to be remembered (\ie stored in memory) and checked against. In addition, the underlying logic of catastrophe theory, namely the interest in describing `generic' behavior, does not hold in simulations in which researchers can create non-generic, unrealistic behavior with ease. For example, a perfect aberration-free lens cannot ever be made in reality, but one can just as well include a perfect lens in a ray-tracing code. The resulting perfect-lens caustic that will be observed in the simulation is not included in any catastrophe classification because it has infinite codimension and is thus formally unstable. Etalon integrals will fail for this situation. Hence, a method of modeling caustics that does not rely on catastrophe theory is needed.


\subsection{Phase-space rotations}

Recall from \Fig{fig:2_caustPROJ} that caustics occur when the inverse ray map $\Vect{\tau}(\Vect{x})$ has a singular projection onto $\Vect{x}$-space (\ie a vertical tangent plane). Since $\Vect{\tau}$ are coordinates on the dispersion manifold, the same geometric interpretation of caustics must hold in the ray phase space as well. Indeed, one can readily verify that
\begin{equation}
	\det \pd{\Vect{x}} \Vect{k}(\Vect{\tau})
	= \frac{\det \pd{\Vect{\tau}} \Vect{k}(\Vect{\tau})}
	{j(\Vect{\tau})} 
	.
\end{equation}

\noindent Hence, $j(\Vect{\tau}) = 0$ where the dispersion manifold has a singular projection onto $\Vect{x}$-space as well, \ie where%
\begin{equation}
	\det \pd{\Vect{x}}\Vect{k} 
	\equiv \det \pd{\Vect{x} \Vect{x}} \theta
	\to \infty
	.
	\label{eq:2_causticSING}
\end{equation}

\noindent Formulating caustics as projection singularities in phase space is advantageous because \textbf{(i)} it is a `caustic-agnostic' description in that it does not rely on catastrophe theory, \textbf{(ii)} it highlights the arbitrariness in the initial choice to project \Eq{eq:2_hilbertWAVE} onto $\Vect{x}$-space, and \textbf{(iii)} it can be used to create a practical caustic-removal scheme for ray-tracing simulations. Indeed, a caustic generally occurs wherever the dispersion manifold has a singular projection onto the chosen projection plane (not necessarily $\Vect{x}$-space), so they can be removed by rotating the projection plane.


\subsubsection{Maslov's method for caustic removal}

A popular paradigm for performing such phase-space rotations is Maslov's method~\cite{Maslov81,Ziolkowski84}. This method takes advantage of the fact that for a well-behaved dispersion manifold, there always exists some mixed $(\Vect{x}, \Vect{k})$ coordinates that locally have no caustics. A caustic that appears in $\Vect{x}$-space is thus absent in one of these mixed representations, and vice versa. By repeatedly switching between such representations as caustics are approached, one can construct a GO framework that does not produce singularities along a given ray.

\begin{figure}
	\centering
	\includegraphics[width=0.6\linewidth]{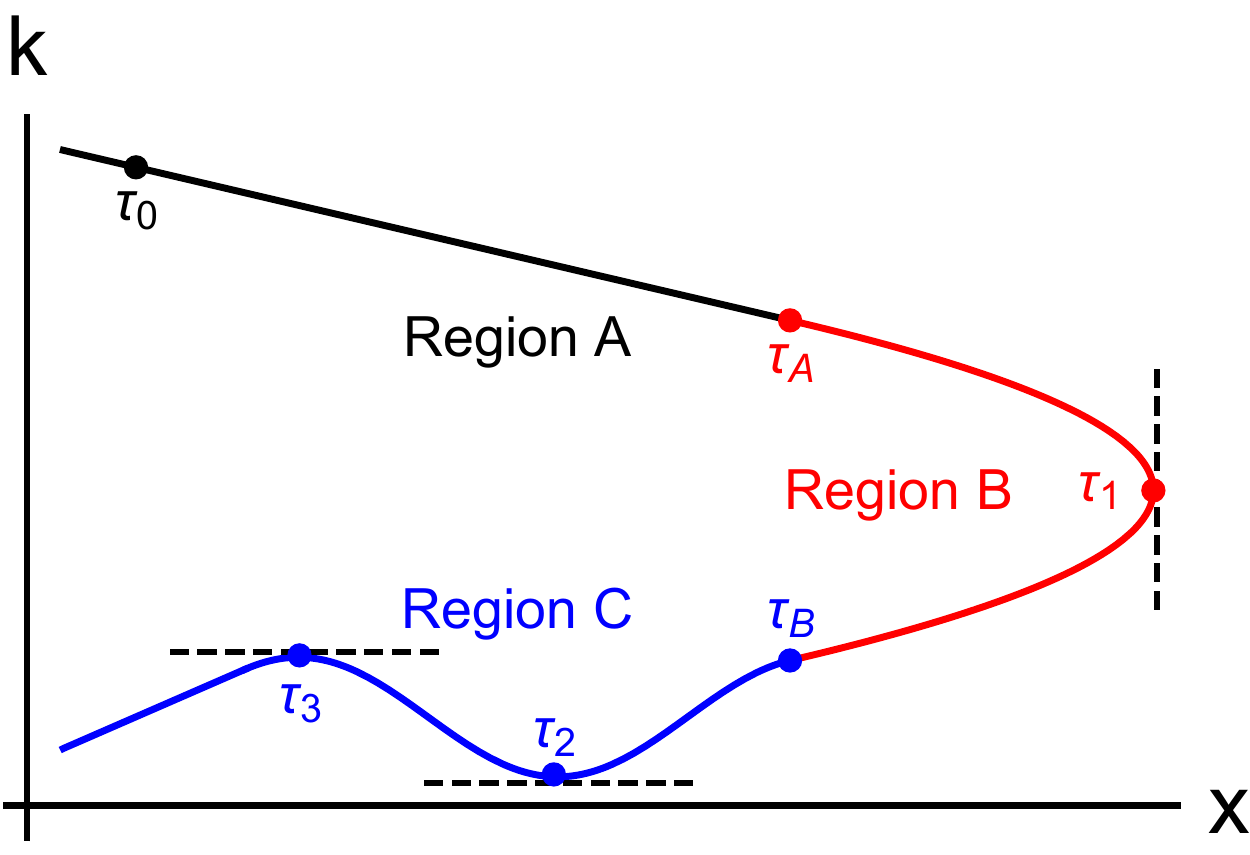}
	\caption{A $1$-D dispersion manifold with coordinate $\tau$ that exhibits a $x$-space caustic at $\tau = \tau_1$ and two $k$-space caustics at $\tau = \tau_2$ and $\tau = \tau_3$. Region A ($\tau \le \tau_A$) is far from the caustics, so both the $x$-space and $k$-space GO solutions are well-behaved. However, region B ($\tau_A < \tau < \tau_B$) is close to the $x$-space caustic, so the $x$-space GO solution is singular while the $k$-space GO solution is well-behaved. Similarly, region C ($\tau \ge \tau_B$) is close to the $k$-space caustics, so the $x$-space GO solution is well-behaved while the $k$-space GO solution is singular.}
	\label{fig:2_Maslov}
\end{figure}

To illustrate this method qualitatively, let me consider the $1$-D dispersion manifold shown in \Fig{fig:2_Maslov}. An $x$-space caustic occurs at $x(\tau_1)$, while $k$-space caustics occur at $k(\tau_2)$ and $k(\tau_3)$. Consider a wave initially located at $x(\tau_0)$. In region A, that is, for $\tau$ between $\tau_0$ and some $\tau_a$ to be specified momentarily, $x$-space can be used as the projection plane. Consequently, the wave envelope is evolved using \Eq{eq:2_GOenv}. Near the $x$-space caustic at $x(\tau_1)$, however, \Eq{eq:2_GOenv} breaks down and cannot be used. Instead, the projection plane should be switched from $x$-space to $k$-space prior to encountering the caustic at $\tau_1$. The switching location, $\tau_a$, must be far enough from $x$-space and $k$-space caustics such that GO is accurate in both representations near $\tau_a$, but is otherwise arbitrary. 

Next, the wavefield $\psi(x)$ is transformed to its $k$-space representation $\Psi(k)$ at $\tau_a$. This is achieved using the Fourier transform (FT) subsequently evaluated via the stationary phase approximation (SPA)~\cite{Olver10a}. Indeed, since
\begin{equation}
	\hspace{-1.2mm}
	\Psi(k) = \int \frac{\dd x}{\sqrt{2\pi}} \, \psi(x) \exp\left(-i kx \right) = \int \frac{\dd x}{\sqrt{2\pi}} \, \phi(x) \exp\left[ i \theta(x) - i kx \right]
	,
	\label{eq:2_FTeik}
\end{equation}

\noindent the phase of the FT integrand is stationary where $\pd{x} \theta(x) = k$, which is satisfied along the dispersion manifold by definition. When $\tau_a$ is chosen sufficiently far from both $x$-space caustics (such that $\phi$ is not singular) and $k$-space caustics (such that $\pd{x}^2 \theta \equiv \pd{x} k$ is nonzero), the SPA of \Eq{eq:2_FTeik} is
\begin{equation}
	\Psi\left[ k(\tau_a) \right] = \frac{\psi\left[ x(\tau_a) \right]}{\sqrt{\pd{x} k(\tau_a) }} \exp\left[ i\frac{\pi}{4} - i k(\tau_a) x(\tau_a) \right]
	.
\end{equation}

\noindent Thus, the SPA has the important role in Maslov's method of localizing the FT to become a pointwise mapping from $\psi\left[x(\tau_a)\right]$ to $\Psi\left[k(\tau_a)\right]$. 

Being absent from $k$-space caustics, $\Psi(k)$ is evolved in region B from $\tau_a$ to $\tau_b$ using GO formulated in $k$-space. I shall derive the GO equations in various projection planes, including $\Vect{k}$-space, in \Ch{ch:MGO}; for the moment, let me simply note that the $\Vect{k}$-space GO equations are not obtained by projecting \Eq{eq:2_approxENV} onto $\{\ket{\Vect{k}} \}$, but instead, more sophisticated machinery must be introduced. After propagating $\Psi(k)$ through region B, the projection plane must be switched back to $x$-space to avoid the $k$-space caustic at $k(\tau_2)$. This is accomplished by using the inverse FT evaluated via SPA. Since there are no remaining $x$-space caustics, $\psi(x)$ can be evolved using $x$-space GO for all subsequent $\tau > \tau_b$.


\subsubsection{Metaplectic geometrical optics}

Maslov's method has been very successful for the theoretical analysis of caustics. There are two issues, however. First is that 
there are often multiple valid mixed $(\Vect{x}, \Vect{k})$ representations, and Maslov's methods provides no general rule for choosing one. This is not a problem in $1$-D since in that case the alternative is uniquely specified (if not $x$-space, then $k$-space), but in higher dimensions this is no longer true. For example, in $2$-D if $(x,y)$ is a bad set of coordinates, at least one of either $(x, k_y)$, $(k_x, y)$, or $(k_x, k_y)$ will be a good set of coordinates, but it is not clear \textit{a priori} which one(s) that will be. The second issue is the lack of rigorous criteria for choosing when to switch between these various mixed representations. Ultimately for a code, this selection must be performed using an external module that supervises the envelope evaluation, detects when a caustic is becoming `close' using some \textit{ad hoc} cost function, then triggers a switch in representation~\cite{Jaun07}, or it must be done via interpolation formulas between the desired representations that have been constructed by hand for the specific problem of interest~\cite{Knudson85, Knudson86}. A framework that could proceed unsupervised would be more desirable.

\begin{figure}
	\centering
	\includegraphics[width=0.85\linewidth,trim={0mm 0mm 0mm 0mm},clip]{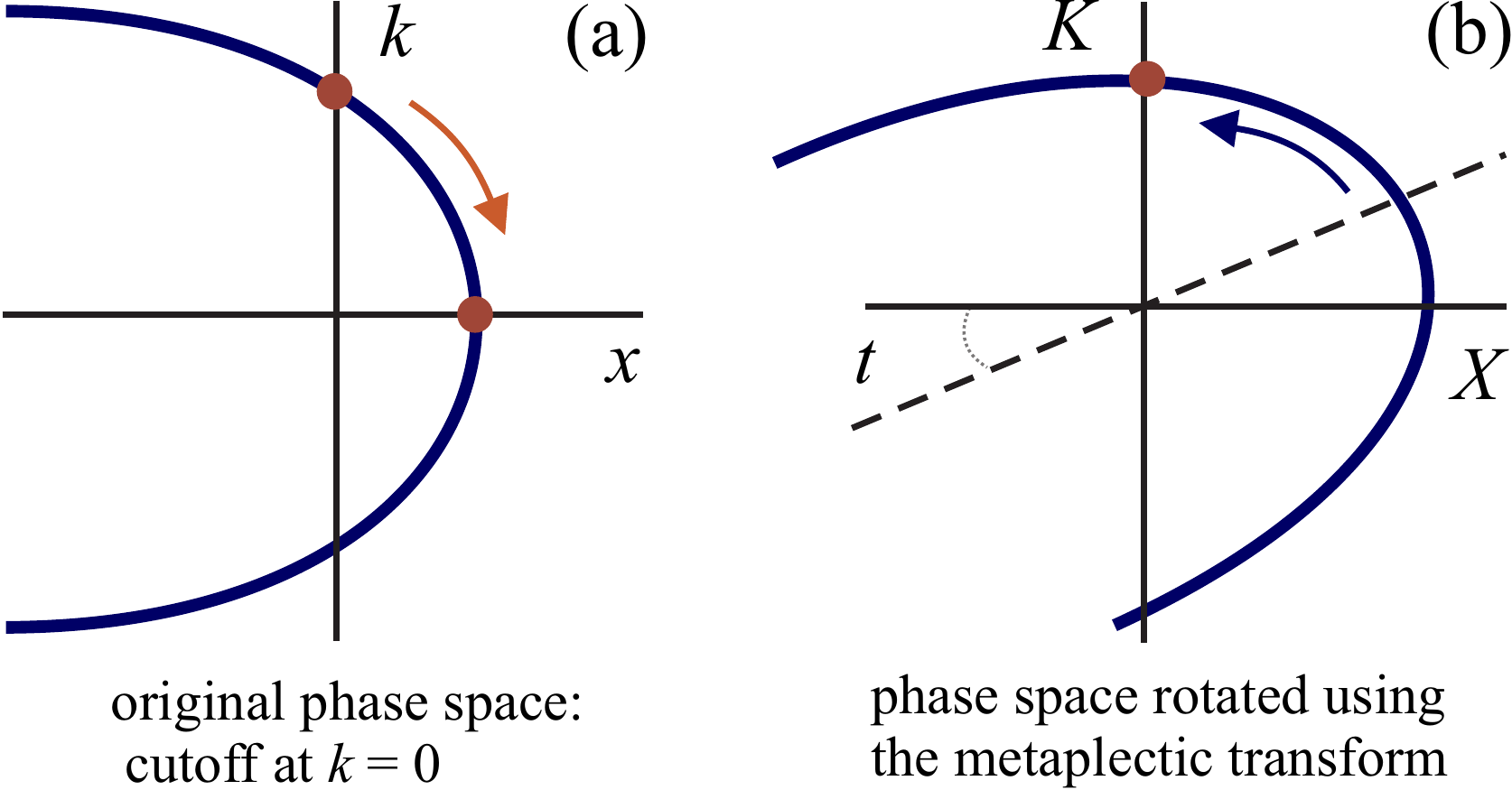}
	\caption{In the original phase space (a), rays travel along the blue curve and encounter the cutoff where $\dd k/\dd x \to \infty$. After rotation by some angle $t$ (b), $\dd K/\dd X$ is finite at the former cutoff, and the envelope remains finite at this location, i.e., the caustic is avoided.}
	\label{fig:2_caustROT}
\end{figure}

In this thesis, I propose an alternative method called metaplectic geometrical optics (MGO) that solves these issues. In short, rather than \textit{sometimes} switching between $\Vect{x}$-space and some mixed $(\Vect{x}, \Vect{k})$-space as in Maslov's method, I propose to \textit{always} switch between $\Vect{x}$-space and the local tangent plane of the dispersion manifold at a desired query point $\Vect{z}(\Vect{\tau})$, as illustrated in \Fig{fig:2_caustROT}. The desired rotation is uniquely specified and each point on the dispersion manifold is treated equally, so there is no need to arbitrarily designate specific points as `switching' points. Also, there will never be a caustic near $\Vect{z}(\Vect{\tau})$ by definition. For these reasons, my approach should be relatively easy to implement in a code. Before discussing MGO, however, I must first introduce the necessary mathematics for performing such arbitrary phase-space rotations. These are the metaplectic operators, which are the subject of the next chapter.


\section{Summary}

In summary, the geometrical-optics (GO) approximation that underlies ray-tracing codes provides a short-wavelength asymptotic solution to a broad class of wave equations, including integro-differential equations, using the Weyl calculus. However, the GO approximation breaks down at caustics, which are formally understood as the locations in $\Vect{x}$-space where the ray-map Jacobian determinant vanishes, \ie where $\det \pd{\Vect{\tau}} \Vect{x} = 0$. One can show that this condition is equivalent to the condition that the number of equilibria for the potential function $V(\Vect{k}) = \fourier{\theta}(\Vect{k} ) + \Vect{x}^\intercal \Vect{k}$ change number, where $\fourier{\theta}(\Vect{k} )$ is the Legendre transform of the wavefield phase $\theta(\Vect{x})$, \ie $\fourier{\theta}$ is the phase of the stationary-phase-approximated FT of $\psi$ [\Eq{eq:2_FT}]. As such, caustics can be described by catastrophe theory, an observation that forms the foundation for common methods of removing caustic singularities in ray-tracing codes. However, these methods often require knowing the caustic type \textit{a priori}, and they do not scale well with the number of simulation parameters (as these increase the effective codimension). 

Caustics can be equivalently understood as locations where the ray trajectories in phase space have an ill-defined projection onto $\Vect{x}$-space. Unlike the description based on catastrophe theory, this definition has no reference to the specific caustic type and as such, potentially offers an avenue for developing a `caustic-agnostic' means of removing caustics in ray-tracing codes. In the remainder of this thesis, I shall develop such a method by showing how the use of phase-space rotation operators (so-called metaplectic operators) can improve the projection properties of the rays and thereby remove caustics.


\begin{subappendices}

\section*{Appendix}

\section{Review of the Wigner--Weyl symbol calculus}
\label{app:2_WeylRev}

In the following, I shall only consider the mapping between scalar functions and scalar operators; matrix-valued functions and matrix-valued operators can be transformed elementwise using the scalar formulae. Also, I assume all functions are square integrable, and all operators are Hilbert--Schmidt normalizable for the purposes of presenting formal theorems.%
\footnote{For proofs, see \Ref{Pool66}. For some extensions to non-Euclidean coordinates, see also the Supplementary Material in \Ref{Dodin19}.} %
The Wigner--Weyl transform (WWT) (denoted $\Weyl$) maps a given operator $\oper{A}(\VectOp{z})$ to a corresponding phase-space function $\Symb{A}(\Vect{z})$ (called the Weyl symbol of $\oper{A}$) as
\begin{equation}
	\Symb{A}(\Vect{z})
	= \Weyl\left[ \oper{A}(\VectOp{z}) \right]
	\doteq
	\int \dd \Vect{\zeta} \,
	\frac{
		\exp\left(
			i\Vect{\zeta}^\intercal \JMat{2N} \Vect{z} 
		\right)
	}{(2\pi)^N} \, \Tr \left[
		\exp\left(
			-i\Vect{\zeta}^\intercal \JMat{2N} \VectOp{z}
		\right)
		\oper{A}(\VectOp{z})
	\right] 
	,
	\label{eq:2_WWT}
\end{equation}

\noindent where $\Tr$ is the matrix trace and the integral is taken over phase space. I have also introduced the fiducial symplectic matrix
\begin{equation}
	\JMat{2N} \doteq 
	\begin{pmatrix}
		\OMat{N} & \IMat{N} \\
		-\IMat{N} & \OMat{N}
	\end{pmatrix}
	,
	\label{eq:2_jMAT}
\end{equation}

\noindent with $\OMat{N}$ and $\IMat{N}$ being respectively the $N$-D null and identity matrices. (Here, $^\intercal$ denotes the matrix transpose, which also denotes the scalar dot product for vectors, \ie $\Vect{a}^\intercal \Vect{b} \equiv \Vect{a} \cdot \Vect{b}$, since I assume all vectors are column-oriented unless explicitly transposed.) The inverse WWT maps a phase-space function $\Symb{A}$ to an operator $\oper{A}$ as
\begin{equation}
	\oper{A}(\VectOp{z})
	=
	\WeylInv \left[ \Symb{A}(\Vect{z}) \right]
	\doteq
	\int \frac{\dd \Vect{z}' \, \dd \Vect{\zeta}}{(2\pi)^{2N}} \, 
	\Symb{A}(\Vect{z}') 
	\exp\left(
		-i\Vect{\zeta}^\intercal \JMat{2N} \Vect{z}'
	\right)
	\exp\left(
		i\Vect{\zeta}^\intercal \JMat{2N} \VectOp{z}
	\right) 
	,
	\label{eq:2_WWTinv}
\end{equation}

\noindent where both integrals are taken over phase space. Note that if the $\Vect{x}$-space matrix elements of $\oper{A}$ are known, the WWT [\Eq{eq:2_WWT}] can also be represented as
\begin{equation}
	\Weyl\left[ \oper{A}(\VectOp{z}) \right]
	\doteq \int \dd \Vect{s} \, \exp\left(i\Vect{k}^\intercal \Vect{s} \right) 
	\bra{\Vect{x}-\Vect{s}/2} \oper{A} \ket{\Vect{x} + \Vect{s}/2} 
	,
	\label{eq:2_WWTkernel}
\end{equation}

\noindent where the integral is now taken over $\Vect{x}$-space. Also note that \Eq{eq:2_WWTkernel} is derived from \Eq{eq:2_WWT} using the $\Vect{x}$-space matrix elements~\cite{Littlejohn86a}
\begin{equation}
	\bra{\Vect{x}} \exp\left[ -i\left(\Vect{z}'\right)^\intercal \JMat{2N} \VectOp{z} \right] \ket{\Vect{x}''} 
	= \exp\left[\frac{i}{2}(\Vect{k}')^\intercal (\Vect{x}+\Vect{x}'') \right]
	\delta(\Vect{x} - \Vect{x}' - \Vect{x}'' ) 
	.
\end{equation}

The WWT preserves hermiticity, \ie 
\begin{equation}
	\Weyl\left[\oper{A}^\dagger \right] = \Symb{A}^*
	,
\end{equation}

\noindent and it preserves locality, \ie 
\begin{align}
	\Weyl \left[ \alpha \oper{A} + \beta \oper{B} \right] &= \alpha \Weyl \left[\oper{A} \right] + \beta \Weyl \left[\oper{B} \right]\, , \\
        \| \oper{A} \|_\text{HS} &= \left(2\pi\right)^{-N} \| \Weyl \left[\oper{A} \right] \|_{L_2} \, ,
\end{align}

\noindent such that two operators that are close approximations of each other map to two functions that are also close approximations of each other, and vice versa. Both of these properties make the Weyl symbol calculus an attractive means to approximate wave equations. Note that I have introduced $\| \cdot \|_\text{HS}$ as the Hilbert--Schmidt norm on the space of operators, defined as
\begin{equation}
    \|\oper{A} \|_\text{HS} \doteq \Tr\left( \oper{A}^\dagger \oper{A} \right) \, ,
\end{equation}

\noindent and $\| \cdot \|_{L_2}$ as the $L_2$ norm on the space of functions.

The WWT of the product of two operators can be concisely represented as the so-called Moyal product $\star$ of their symbols:
\begin{equation}
	\Weyl[\oper{A}\oper{B}] = \Symb{A}(\Vect{z}) \star \Symb{B}(\Vect{z}) .
	\label{eq:2_WWTstar}
\end{equation}

\noindent This (non-commutative) product is given explicitly as
\begin{equation}
	\Symb{A}(\Vect{z}) \star \Symb{B}(\Vect{z}) 
	= \left. 
		\sum_{s = 0}^\infty 
		\frac{\left( \frac{i}{2} \pd{\Vect{z}}^\intercal \, \JMat{2N} \, \pd{\Vect{\zeta}} \right)^{s}}
		{s!}  
		\Symb{A}(\Vect{z}) \Symb{B}(\Vect{\zeta}) 
	\right|_{\Vect{\zeta} = \Vect{z}}
	.
	\label{eq:2_moyalSERIES}
\end{equation}

\noindent These rules can be used to compute the following relevant WWT pairs:
\begin{subequations}
	\begin{align}
		f(\Vect{x}) 
		&\Longleftrightarrow f(\VectOp{x}) 
		, \\
		f(\Vect{k}) 
		&\Longleftrightarrow f(\VectOp{k})
		, \\
		\Vect{k}^\intercal \Vect{v}(\Vect{x}) 
		&\Longleftrightarrow 
		\frac{
			\Vect{v}(\VectOp{x})^\intercal \VectOp{k} 
			+ \VectOp{k}^\intercal\Vect{v}(\VectOp{x})
		}{2}
		, \\
		\Vect{k}^\intercal \Mat{M}(\Vect{x}) \Vect{k} 
		&\Longleftrightarrow
		\frac{
			\Mat{M}(\VectOp{x}) \dubdot \VectOp{k} \VectOp{k} 
			+ 2 \VectOp{k}^\intercal \Mat{M}(\VectOp{x}) \VectOp{k}
			+ \VectOp{k} \VectOp{k} \dubdot  \Mat{M}(\VectOp{x})
		}{4}
		.
	\end{align}
\end{subequations}

\noindent where $\dubdot$ denotes the double contraction.

Lastly, let me note that the WWT of the two-point correlation function $\psi(\Vect{x})\psi^*(\Vect{x}')$, \ie
\begin{equation}
	W_\psi(\Vect{z}) \doteq 
	\int \frac{\dd \Vect{s}}{(2\pi)^N} \, 
	\exp\left( i\Vect{k}^\intercal \Vect{s} \right) \,
	\psi
	\left(
		\Vect{x} 
		- \frac{\Vect{s}}{2}
	\right)
	\psi^*
	\left(
		\Vect{x} 
		+ \frac{\Vect{s}}{2}
	\right)
	,
	\label{eq:2_wigFUNC}
\end{equation}

\noindent acts as a phase-space (quasi-)distribution function for the field intensity, satisfying~\cite{Case08}
\begin{equation}
	|\psi(\Vect{x})|^2 = \int \dd \Vect{k} \, W_\psi(\Vect{z} )
	, \quad
	|\fourier{\psi}(\Vect{k})|^2 = \int \dd \Vect{x} \, W_\psi(\Vect{z}),
\end{equation}

\noindent where $\fourier{\psi}$ is the Fourier transform of $\psi$. The function $W_\psi$ is commonly known as the Wigner function, and its application to classical waves are discussed in detail in \Refs{Alonso11,Ruiz17t,Dodin22}, among others.


\section{Basic results from catastrophe theory}
\label{sec:2_catastropheREV}

(The following definitions and results are taken from \Ref{Poston96}.) Let me begin with some basic definitions:
\begin{definition}
	For any smooth function $f: \R^N \mapsto \R$, the \textit{$k$-jet} $j^k f$ is the Taylor expansion of $f$ up to order $k$. I denote truncation to order $k$ as $j^k f \equiv \overline{f}^k$
\end{definition}

\begin{definition}
	A function $f$ is \textit{$k$-determinate} at $\Vect{x} = \Vect{0}$ if any smooth function $f(\Vect{x}) + g(\Vect{x})$ can be expressed as $f(\Vect{y}(\Vect{x}))$ for some smooth reversible coordinate change $\Vect{y}: \R^N \mapsto \R^N$, where $g(\Vect{x})$ is order $k+1$ at $\Vect{x} = \Vect{0}$. 
\end{definition}

\begin{corollary}
	Any $k$-determinate function is also $k'$-determinate for all $k' \ge k$.
\end{corollary}

\begin{definition}
	A function $f$ is \textit{strongly $k$-determinate} at $\Vect{x} = \Vect{0}$ if $f$ is $k$-determinate at $\Vect{x} = \Vect{0}$ and $\Vect{y}(\Vect{x})$ can be chosen such that the Jacobian matrix at $\Vect{x} = \Vect{0}$ is the identity, that is, $\pd{x_j} y_i(\Vect{0}) = \delta_{ij}$.
\end{definition}

\begin{corollary}
	Clearly, if $f$ is (strongly) $k$-determinate, then $j^k f$ is also (strongly) $k$-determinate.
\end{corollary}

\begin{definition}
	The \textit{determinancy} of $f$ at $\Vect{x} = \Vect{0}$, denoted $\sigma(f)$, is the lowest $k$ for which $f$ is $k$-determinate at $\Vect{x} = \Vect{0}$. If there is no such finite $k$, then I write $\sigma(f) = \infty$ and say $f$ is \textit{indeterminate}.
\end{definition}

Let me also introduce the following polynomial vector spaces:
\begin{definition}
	~
\end{definition}
\begin{itemize}
	\item{$E_N^k$ is the polynomial vector space consisting of all $N$-variate polynomials of degree $\in [0,k]$.}
	\item{$J_N^k$ is the subspace of $E_N^k$ with no constant term, \ie polynomials of degree $\in [1,k]$.}
	\item{$I_N^k$ is the subspace of $J_N^k$ with no linear term, \ie polynomials of degree $\in [2,k]$.}
	\item{$M_N^k$ is the subspace of $I_N^k$ with only homogeneous polynomials of degree $k$.}
\end{itemize}

\begin{definition}
	I denote by $\Delta_k(f)$ the subspace of $E_N^k$ spanned by all polynomials of the form $\overline{P_E \cdot j^k(\pd{x_i}f)}^k$, where $P_E \in E_N^k$.
\end{definition}

\begin{corollary}
	It follows from truncation that $\Delta_k(f) = \Delta_k(j^{k+1}f)$.
\end{corollary}

\noindent A convenient choice for $P_E \in E_N^k$ are the basis monomials $x_1^{\alpha_1} \ldots x_N^{\alpha_N}$ for $\sum \alpha_j \in [0,k]$.

\begin{definition}
	I denote by $\overline{J_N^k \cdot \Delta_k(f)}^k$ the subspace of $J_N^k$ spanned by all polynomials of the form $\overline{P_J \cdot j^k(\pd{x_i}f)}^k$, where $P_J \in J_N^k$.
\end{definition}

\begin{corollary}
	It follows from truncation that $\overline{J_N^k \cdot \Delta_k(f)}^k = \overline{J_N^k \cdot \Delta_{k-1}(f)}^k$.
\end{corollary}

\noindent A convenient choice for $P_J \in J_N^k$ are the basis monomials $x_1^{\alpha_1} \ldots x_N^{\alpha_N}$ for $\sum \alpha_j \in [1,k]$.

\begin{definition}
	I denote by $\overline{I_N^k \cdot \Delta_k(f)}^k$ the subspace of $I_N^k$ spanned by all polynomials of the form $\overline{P_I \cdot j^k(\pd{x_i}f)}^k$, where $P_I \in I_N^k$.
\end{definition}

\begin{corollary}
	It follows from truncation that $\overline{I_N^k \cdot \Delta_k(f)}^k = \overline{I_N^k \cdot \Delta_{k-2}(f)}^k$.
\end{corollary}

\noindent A convenient choice for $P_I \in I_N^k$ are the basis monomials $x_1^{\alpha_1} \ldots x_N^{\alpha_N}$ for $\sum \alpha_j \in [2,k]$.

I can now state some main theorems of catastrophe theory:
\begin{theorem}
	$f$ is strongly $k$-determinate if and only if $M_N^{k+1} \subseteq \overline{I_N^{k+1} \cdot \Delta_{k+1}(f)}^{k+1}$.
\end{theorem}

\noindent The converse statement is as follows: \textit{if $M_N^{k+1} \nsubseteq \overline{I_N^{k+1} \cdot \Delta_{k+1}(f)}^{k+1}$ for a given $k$, then $f$ is not strongly $k$-determinate}.

\begin{theorem}
	If $f$ is $k$-determinate, then $M_N^{k+1} \subseteq \overline{J_N^{k+1} \cdot \Delta_{k+1}(f)}^{k+1}$.
	\label{th:noKdet}
\end{theorem}

\noindent The converse statements are as follow: (i) \textit{If $M_N^{k+1} \subseteq \overline{J_N^{k+1} \cdot \Delta_{k+1}(f)}^{k+1}$ for a given $k$, then $f$ may or may not be $k$-determinate}; (ii) \textit{If $M_N^{k+1} \nsubseteq \overline{J_N^{k+1} \cdot \Delta_{k+1}(f)}^{k+1}$ for a given $k$, then $f$ is not $k$-determinate}.

\begin{corollary}
	If $f$ is $k$-determinate, then $f$ is strongly $(k+1)$-determinate, although $f$ may not be strongly $k$-determinate. 
\end{corollary}

\begin{corollary}
	If the lowest $k$ for which $f$ is strongly $k$-determinate is $k_0$, then $f$ may be $(k_0 - 1)$-determinate, but it cannot be $(k_0 - m)$-determinate for $m > 1$.
\end{corollary}

\begin{theorem}
	$f$ is $k$-determinate if and only if $M_N^{k+1} \subseteq \overline{J_N^{k+1} \cdot \Delta_{k+1}(f + P_M)}^{k+1}$ for all $P_M \in M_N^{k+1}$.
\end{theorem}

\noindent Theorem \ref{th:noKdet} constitutes the particular case when $P_M$ is the zero polynomial.

\vspace{3mm}
Lastly, let me state some definitions and theorems that underlie the `standard' catastrophe optics normal forms (such as those listed in Table~\ref{tab:2_caustic}).

\begin{definition}
	Let $f$ be $k$-determinate for some $k$ at $\Vect{x} = \Vect{0}$. The \textit{codimension} of $f$ at $\Vect{x} = \Vect{0}$, denoted $\textrm{cod}(f)$, is the codimension of $\Delta_{k}(f)$ in $J_N^k$. If $f$ is indeterminate, then the codimension of $f$ is infinite.
\end{definition}

\noindent In other words, the codimension of $f$ is the number of `missing' basis elements of $J_N^k$ in $\Delta_{k}(f)$.

\begin{definition}
	Let $f$ have finite codimension. An $d$-dimensional \textit{unfolding} ($d$-unfolding) of $f$ at $\Vect{x} = \Vect{0}$ is a function $F: \R^{N+d} \mapsto \R$, given symbolically as $F(\Vect{x}, \Vect{t}) = F(x_1, \ldots, x_N, t_1, \ldots, t_d)$, such that $F(\Vect{x}, \Vect{0}) = f(\Vect{x})$. 
\end{definition}

\noindent In this context, $f$ is often called the \textit{germ} of the unfolding $F$.

\begin{definition}
	An $d$-unfolding $F$ of $f$ is called \textit{versal} if all other unfoldings of $f$ can be induced from it via smooth coordinate transformation plus the addition of a shear function. $F$ is \textit{universal} if it is both versal and $d$ is minimal, that is, $d = \textrm{cod}(f)$.
\end{definition}

\begin{corollary}
	Two universal unfoldings $F$ and $\bar{F}$ of $f$ are equivalent, in that they can be obtained from each other via smooth coordinate transformation.
\end{corollary}

\begin{definition}
	Let $F$ be a $d$-unfolding of $f$. I define the subspace $V^k(F)$ of  $J_N^k$ as the span of the vector set $\{ \pd{t_j}[j^k F(\Vect{x}, \Vect{0}) ] \}$.
\end{definition}

\begin{theorem}
	Let $f$ be $k$-determinate. An $d$-unfolding $F$ of $f$ is versal if and only if $V^k(F)$ and $\Delta_k(f)$ are transverse subspaces of $J_N^k$, that is, $V^k(F)$ and $\Delta_k(f)$ together span $J_N^k$.
\end{theorem}

\begin{corollary}
	If $f$ is $k$-determinate, then a universal unfolding $F$ for $f$ can be constructed by choosing a co-basis $\{p_1, \ldots, p_d \}$ for $\Delta_k(f)$ in $J_N^k$ and setting $F(\Vect{x}, \Vect{t}) = f(\Vect{x}) + t_1 \, p_1(\Vect{x}) + \ldots + t_d  \, p_d(\Vect{x})$.
\end{corollary}

\vspace{3mm}
Finally, I conclude with the definition of structural stability of a function.
\begin{definition}
	A function $F(\Vect{x}, \Vect{t})$ is structurally stable with respect to arbitrary perturbations of the form $g(\Vect{x})$ if it is a versal unfolding of a $k$-determinate germ $f(\Vect{x})$.
\end{definition}

\noindent The intuition of this result is as follows: $k$-determinancy of the germ $f$ kills high-order ($>k$) perturbations to $F$ via smooth coordinate transformations, while the unfolding terms kill low-order ($\le k$) perturbations via shifts in $t_1$, \ldots $t_d$. 

It turns out (via Thom's theorem~\cite{Poston96}) that there are only a finite number of structurally stable universal unfoldings with codimension $ \textrm{cod}(f) \le 5$, up to equivalency. Table~\ref{tab:2_caustic} provides the complete list for $ \textrm{cod}(f) \le 3$. I shall not prove this fact as it is quite involved to do so; instead, let me note the intuition of this result: functions with $\textrm{cod}(f)> 6$ generically include trivariate cubic functions, because a trivariate quadratic function contains $6$ coefficients ($x^2$, $xy$, $xz$, $y^2$, $yz$, and $z^2$) that can each be made zero by appropriately choosing the $6$ unfolding terms. Unlike bivariate cubics, which are classified via the types of umbilic points for surfaces in $\R^3$, trivariate cubics can not be systematically classified because there are `conserved quantities' (ratios of roots typically) that can not be altered via smooth coordinate transformations; hence, such functions are structurally unstable because any perturbation that alters a conserved quantity cannot be undone by a smooth reparameterization of the function. As \Ref{Poston96} states (p.~120), `\textit{only for $r \le 5$ is it typical for an $r$-parameter family of functions to be stable, although even for $r>5$ is may be}'. The $A_m$ cuspoid and $D_m^\pm$ umbilic series introduced in \Sec{sec:2_parax} are examples of stable caustics with arbitrarily large codimension, but in general, describing such caustics requires more advanced machinery (so-called \textit{moduli}) that are beyond the scope of this brief review.

\end{subappendices}

\chapter{Linear symplectic and metaplectic transforms}
\label{ch:MT}

\section{Introduction}

In this chapter I introduce the mathematical machinery necessary to develop the phase-space rotation scheme of MGO, namely, linear symplectic transformations of the ray phase space and their corresponding metaplectic operators that act on the wavefields. I also derive their representations in certain convenient basis choices that will be useful in later chapters, such as when ray patterns are `quasiuniform'. I also discuss these two classes of transformations in the near-identity limit, deriving a novel pseudo-differential representation of the MT in the process, and in the limit when the transformations are orthogonal in addition to symplectic (orthosymplectic), as will be the case for any practical MGO implementation in a code. I conclude this chapter with a detailed discussion of the MT in two familiar physics contexts to aid the reader develop further intuition: first, in the familiar setting of elementary quantum mechanics in which the MT acts as the time propagator for the quantum harmonic oscillator problem; second, in the familiar setting of paraxial optics in which the MT acts as the spatial evolution operator for the diffracted (transverse) field. The text in this chapter is based on discussions published previously in \Refs{Lopez19,Lopez20,Lopez21a,Lopez21b,Lopez22}.

\section{Linear symplectic transformations}


\subsection{Basic definitions and identities}

Before discussing phase-space rotations, let us first discuss more general phase-space transformations. Suppose I want to transform the original phase-space coordinates $\Vect{z} \doteq (\Vect{x}, \Vect{k})$ to some new set of coordinates $\Stroke{\Vect{Z}}(\Vect{z}) \doteq (\Vect{X}(\Vect{z}), \Vect{K}(\Vect{z}))$. A special class of transformations, \textit{symplectic} transformations, are those which leave the ray equations of motion \eq{eq:2_GOrays} structurally unchanged~\cite{Goldstein02}. This feature makes such transformations belong to a sort of equivalency class on phase space, in that any physical result will be invariant with respect to one's choice of symplectically related coordinates.

To develop this idea more quantitatively, I first note that the ray equations of motion \eq{eq:2_GOrays} can be written more compactly as
\begin{equation}
	\pd{\xi} \Vect{z}
	= \JMat{2N} \, \pd{\Vect{z}} \Symb{D}(\Vect{z})
	,
	\label{eq:3_EOM}
\end{equation}

\noindent where $\JMat{2N}$ is defined in \Eq{eq:2_jMAT}. Hence, I can directly compute the evolution of $\Stroke{\Vect{Z}}$:
\begin{align}
	\pd{\xi} \Stroke{\Vect{Z}}
	= \left(\pd{\Vect{z}} \Stroke{\Vect{Z}} \right) \pd{\xi} \Vect{z}
	= \left(\pd{\Vect{z}} \Stroke{\Vect{Z}} \right) 
	\JMat{2N} \, \pd{\Vect{z}} \fourier{\Symb{D}}\left[ \Stroke{\Vect{Z}}(\Vect{z}) \right]
	=\left(\pd{\Vect{z}} \Stroke{\Vect{Z}} \right) 
	\JMat{2N} \, 
	\left(\pd{\Vect{z}} \Stroke{\Vect{Z}} \right)^\intercal
	\pd{\Stroke{\Vect{Z}}} \fourier{\Symb{D}}\left( \Stroke{\Vect{Z}} \right)
	,
	\label{eq:3_transEOM}
\end{align}

\noindent where $\fourier{\Symb{D}}\left( \Stroke{\Vect{Z}} \right)$ is simply the original dispersion function $\Symb{D}$ expressed in the new coordinates $\Stroke{\Vect{Z}}$, \ie
\begin{equation}
	\fourier{\Symb{D}}\left( \Stroke{\Vect{Z}} \right)
	=
	\Symb{D}\left[ \Vect{z}(\Stroke{\Vect{Z}}) \right]
	.
\end{equation}

\noindent Hence, \Eq{eq:3_transEOM} will have the same structural form as \Eq{eq:3_EOM}, meaning that the transformation $\Vect{z} \to \Stroke{\Vect{Z}}$ is symplectic, if the Jacobian matrix satisfies
\begin{equation}
	\left(\pd{\Vect{z}} \Stroke{\Vect{Z}} \right) 
	\JMat{2N} \, 
	\left(\pd{\Vect{z}} \Stroke{\Vect{Z}} \right)^\intercal
	= \JMat{2N}
	.
\end{equation}

Of particular interest for developing MGO will be linear symplectic transformations. These take the form
\begin{equation}
	\Stroke{\Vect{Z}} = \Mat{S} \Vect{z}
	,
	\label{eq:3_Strans}
\end{equation}

\noindent where $\Mat{S}$ is a constant symplectic matrix that satisfies
\begin{equation}
	\Mat{S} \JMat{2N} \Mat{S}^\intercal
	= \JMat{2N}
	.
	\label{eq:3_symplecDEF}
\end{equation}

\noindent Note that any symplectic matrix is invertible, since \Eq{eq:3_symplecDEF} implies that
\begin{equation}
	\left( \det \Mat{S} \right)^2 = 1
	.
\end{equation}

\noindent Hence, \Eq{eq:3_symplecDEF} also implies that
\begin{equation}
	\Mat{S}^{-1}
	= 
	- \JMat{2N} \Mat{S}^\intercal \JMat{2N}
	,
	\label{eq:3_Sinv}
\end{equation}

\noindent where I have used the fact that $\JMat{2N} \JMat{2N} = - \IMat{2N}$. One can then show that if $\Mat{S}$ is symplectic, both $\Mat{S}^\intercal$ and $\Mat{S}^{-1}$ are as well, and that the product $\Mat{S}_1 \Mat{S}_2$ is symplectic if both $\Mat{S}_1$ and $\Mat{S}_2$ are. Also note that both $\IMat{2N}$ and $\JMat{2N}$ are symplectic.

It is useful to adopt an $N \times N$ block decomposition for $\Mat{S}$ of the form
\begin{equation}
	\Mat{S}
	=
	\begin{pmatrix}
		\Mat{A} & \Mat{B} \\
		\Mat{C} & \Mat{D}
	\end{pmatrix}
	.
	\label{eq:3_blockS}
\end{equation}

\noindent Note that \Eq{eq:3_Sinv} implies the block-decomposition of $\Mat{S}^{-1}$ to be
\begin{equation}
	\Mat{S}^{-1}
	=
	\begin{pmatrix}
		\Mat{D}^\intercal & - \Mat{B}^\intercal \\
		-\Mat{C}^\intercal & \Mat{A}^\intercal
	\end{pmatrix}
	.
	\label{eq:3_blockSinv}
\end{equation}

\noindent Computing \Eq{eq:3_symplecDEF} blockwise then yields the following relations between the submatrices $\Mat{A}$, $\Mat{B}$, $\Mat{C}$, and $\Mat{D}$:%
\begin{subequations}%
	\label{eq:3_symplecABCD}%
	\begin{align}
		\label{eq:3_symplec1}
		\Mat{A}\Mat{B}^\intercal - \Mat{B}\Mat{A}^\intercal &= \OMat{N} 
		, \\
		\label{eq:3_symplec2}
		\Mat{A}\Mat{D}^\intercal - \Mat{B}\Mat{C}^\intercal &= \IMat{N}
		, \\
		\label{eq:3_symplec3}
		\Mat{C}\Mat{D}^\intercal - \Mat{D}\Mat{C}^\intercal &= \OMat{N}
		.
	\end{align}
	Similarly, \Eq{eq:3_Sinv} along with the symplecticity of $\Mat{S}^{-1}$ implies the further relations
	\begin{align}
		\label{eq:3_symplec4}
		\Mat{B}^\intercal\Mat{D} - \Mat{D}^\intercal\Mat{B} &= \OMat{N}
		, \\
		\label{eq:3_symplec5}
		\Mat{A}^\intercal\Mat{D} - \Mat{C}^\intercal\Mat{B} &= \IMat{N}
		, \\
		\label{eq:3_symplec6}
		\Mat{A}^\intercal\Mat{C} - \Mat{C}^\intercal\Mat{A} &= \OMat{N}
		.
	\end{align}
\end{subequations}

\noindent Equations \eq{eq:3_symplecABCD} are sometimes called the Luneburg relations in the optics literature~\cite{Luneburg64}. Also, note that at $N=1$, \Eqs{eq:3_symplec1}, \eq{eq:3_symplec3}, \eq{eq:3_symplec4}, and \eq{eq:3_symplec6} are satisfied automatically, and \Eqs{eq:3_symplec2} and \eq{eq:3_symplec5} are equivalent to $\det \Mat{S} = 1$; hence, a $2 \times 2$ matrix is symplectic if and only if it has unit determinant.


\subsection{Symplectic submatrices in the SVD basis}
\label{sec:3_SVD}

It will be necessary in developing MGO to consider submatrices $\Mat{A}$, $\Mat{B}$, $\Mat{C}$, $\Mat{D}$ in the diagonalizing basis of $\Mat{B}$. Suppose that $\Mat{B}$ has rank $\rho$ and corank $\varsigma = N - \rho$. A singular-value decomposition of $\Mat{B}$ can be performed as%
\footnote{A less restrictive decomposition is used in \Ref{Littlejohn86a} to derive results analogous to those presented in \Sec{sec:3_SVD}. However, my use of the singular-value decomposition is more practical due to the plethora of efficient algorithms for its computation~\cite{Press07}.}%
\begin{equation}
	\Mat{B} = \Mat{L}_\textrm{s} \, \widetilde{\Mat{B}} \, \Mat{R}_\textrm{s}^\intercal ,
	\label{eq:3_bSVD}
\end{equation}

\noindent where $\widetilde{\Mat{B}}$ is a diagonal matrix given by
\begin{equation}
	\widetilde{\Mat{B}}
	=
	\begin{pmatrix}
		\Mat{\Lambda}_{\rho \rho} & \OMat{ \rho \varsigma} \\
		\OMat{\varsigma \rho} & \OMat{\varsigma \varsigma}
	\end{pmatrix}
	.
\end{equation}

\noindent The subscript $_{mn}$ is used to indicate a matrix sub-block is size $m \times n$. Correspondingly, $\Mat{\Lambda}_{\rho \rho}$ is a diagonal matrix containing all nonzero singular values of $\Mat{B}$ and hence has $\det \Mat{\Lambda}_{\rho \rho} \neq 0$ by definition. The matrices $\Mat{L}_\textrm{s}$ and $\Mat{R}_\textrm{s}$ are both orthogonal and can be written in terms of the left and right singular vectors $\{ \unit{\Vect{\ell}}_j \}$ and $\{ \unit{\Vect{r}}_j \}$ of $\Mat{B}$ as
\begin{equation}
	\Mat{L}_\textrm{s} 
	=
	\begin{pmatrix}
		\uparrow &  & \uparrow \\
		\unit{\Vect{\ell}}_1 & \ldots & \unit{\Vect{\ell}}_N \\
		\downarrow &  & \downarrow
	\end{pmatrix}
	,
	\quad
	\Mat{R}_\textrm{s} 
	= 
	\begin{pmatrix}
		\uparrow &  & \uparrow \\
		\unit{\Vect{r}}_1 & \ldots & \unit{\Vect{r}}_N \\
		\downarrow &  & \downarrow
	\end{pmatrix}
	.
\end{equation}

\noindent (The arrows emphasize that the constituent vectors are oriented columnwise.) Note that these vectors are mutually orthonormal:
\begin{equation}
	\unit{\Vect{\ell}}_j^\intercal \unit{\Vect{\ell}}_k = \delta_{jk}
	,
	\quad
	\unit{\Vect{r}}_j^\intercal \unit{\Vect{r}}_k = \delta_{jk} 
	.
\end{equation} 

\noindent I then introduce the matrix projections
\begin{align}
	\widetilde{\Mat{A}} 
	&\doteq \Mat{L}_\textrm{s}^\intercal \Mat{A} \Mat{R}_\textrm{s}
	\doteq
	\begin{pmatrix}
		\Mat{a}_{\rho \rho} & \Mat{a}_{\rho \varsigma} \\
		\Mat{a}_{\varsigma \rho} & \Mat{a}_{\varsigma \varsigma}
	\end{pmatrix}
	, \quad
	\widetilde{\Mat{C}} 
	\doteq \Mat{L}_\textrm{s}^\intercal \Mat{C} \Mat{R}_\textrm{s}
	\doteq
	\begin{pmatrix}
		\Mat{c}_{\rho \rho} & \Mat{c}_{\rho \varsigma} \\
		\Mat{c}_{\varsigma \rho} & \Mat{c}_{\varsigma \varsigma}
	\end{pmatrix}
	, \quad
	\widetilde{\Mat{D}} 
	\doteq \Mat{L}_\textrm{s}^\intercal \Mat{D} \Mat{R}_\textrm{s}
	\doteq
	\begin{pmatrix}
		\Mat{d}_{\rho \rho} & \Mat{d}_{\rho \varsigma} \\
		\Mat{d}_{\varsigma \rho} & \Mat{d}_{\varsigma \varsigma}
	\end{pmatrix}
	.
\end{align}

\noindent The goal shall be to determine the identities that the various submatrices of $\widetilde{\Mat{A}}$, $\widetilde{\Mat{C}}$, and $\widetilde{\Mat{D}}$ satisfy.

To do this, note that the orthogonality of $\Mat{L}$ and $\Mat{R}$ means that $\widetilde{\Mat{A}}$, $\widetilde{\Mat{B}}$, $\widetilde{\Mat{C}}$, and $\widetilde{\Mat{D}}$ also satisfy \Eqs{eq:3_symplecABCD}. Explicitly, \Eq{eq:3_symplec1} reads
\begin{equation}
	\begin{pmatrix}
		\Mat{a}_{\rho \rho} \Mat{\Lambda}_{\rho \rho} - \left(\Mat{a}_{\rho \rho} \Mat{\Lambda}_{\rho \rho} \right)^\intercal 
		& -\left( \Mat{a}_{\varsigma \rho} \Mat{\Lambda}_{\rho \rho} \right)^\intercal \\
		\Mat{a}_{\varsigma \rho} \Mat{\Lambda}_{\rho \rho} & \OMat{\varsigma \varsigma}
	\end{pmatrix}
	= \OMat{N}
	.
\end{equation}

\noindent Since $\Mat{\Lambda}_{\rho \rho} \neq \OMat{\rho \rho}$, this implies that
\begin{equation}
	\Mat{a}_{\varsigma \rho} = \OMat{\varsigma \rho}
	,
	\quad
	\Mat{a}_{\rho \rho} \Mat{\Lambda}_{\rho \rho}
	= 
	\left(
		\Mat{a}_{\rho \rho} \Mat{\Lambda}_{\rho \rho}
	\right)^\intercal 
	.
	\label{eq:3_aSUBsymm}
\end{equation}

\noindent Similarly, \Eq{eq:3_symplec4} is written explicitly as
\begin{equation}
	\begin{pmatrix}
		\Mat{\Lambda}_{\rho \rho} \Mat{d}_{\rho \rho} - \left( \Mat{\Lambda}_{\rho \rho} \Mat{d}_{\rho \rho} \right)^\intercal
		& \Mat{\Lambda}_{\rho \rho} \Mat{d}_{\rho \varsigma}
		\\
		- \left(\Mat{\Lambda}_{\rho \rho} \Mat{d}_{\rho \varsigma} \right)^\intercal
		&
		\OMat{\varsigma \varsigma}
	\end{pmatrix}
	= 
	\OMat{N} .
\end{equation}

\noindent This yields analogous constraints on $\widetilde{\Mat{D}}$, namely
\begin{equation}
	\Mat{d}_{\rho \varsigma} = \OMat{\rho \varsigma}
	,
	\quad
	\Mat{\Lambda}_{\rho \rho} \Mat{d}_{\rho \rho} 
	= 
	\left( 
		\Mat{\Lambda}_{\rho \rho} \Mat{d}_{\rho \rho} 
	\right)^\intercal .
	\label{eq:3_dSUBsymm}
\end{equation}

Similarly, \Eq{eq:3_symplec2} is explicitly written as
\begin{equation}
	\begin{pmatrix}
		\Mat{a}_{\rho \rho} \Mat{d}_{\rho \rho}^\intercal - \Mat{\Lambda}_{\rho \rho} \Mat{c}_{\rho \rho}^\intercal
		&
		\Mat{a}_{\rho \rho} \Mat{d}_{\varsigma \rho}^\intercal + \Mat{a}_{\rho \varsigma} \Mat{d}_{\varsigma \varsigma}^\intercal - \Mat{\Lambda}_{\rho \rho} \Mat{c}_{\varsigma \rho}^\intercal
		\\
		\OMat{\varsigma \rho}
		&
		\Mat{a}_{\varsigma \varsigma} \Mat{d}_{\varsigma \varsigma}^\intercal
	\end{pmatrix}
	=
	\IMat{N} .
\end{equation}

\noindent Since the matrix inverse is unique, I therefore obtain
\begin{equation}
	\Mat{d}_{\varsigma \varsigma} 
	=
	\Mat{a}_{\varsigma \varsigma}^{-\intercal}
	\label{eq:3_dINV}
	,
\end{equation}

\noindent which means that $\Mat{a}_{\varsigma \varsigma}$ is invertible. Since $\Mat{\Lambda}_{\rho \rho}$ is invertible, I also obtain
\begin{align}
	\Mat{c}_{\rho \rho} 
	= \Mat{d}_{\rho \rho} \Mat{a}_{\rho \rho}^\intercal \Mat{\Lambda}_{\rho \rho}^{-1} 
	- \Mat{\Lambda}_{\rho \rho}^{-1}
	,
	\quad
	\Mat{c}_{\varsigma \rho} 
	= \Mat{d}_{\varsigma \rho} \Mat{a}_{\rho \rho}^\intercal \Mat{\Lambda}_{\rho \rho}^{-1} 
	+ \Mat{a}_{\varsigma \varsigma}^{-\intercal} \Mat{a}_{\rho \varsigma}^\intercal \Mat{\Lambda}_{\rho \rho}^{-1}
	,
\end{align}

\noindent where I have used \Eq{eq:3_dINV}. Consequently, \Eq{eq:3_symplec5} is greatly simplified; it reads
\begin{equation}
	\begin{pmatrix}
		\IMat{\rho}
		&
		\OMat{\rho \varsigma}
		\\
		\Mat{a}_{\rho \varsigma}^\intercal \Mat{d}_{\rho \rho}
		+
		\Mat{a}_{\varsigma \varsigma}^\intercal \Mat{d}_{\varsigma \rho}
		-
		\Mat{c}_{\rho \varsigma}^\intercal \Mat{\Lambda}_{\rho \rho}
		&
		\IMat{\varsigma}
	\end{pmatrix}
	=
	\IMat{N} .
\end{equation}

\noindent I therefore obtain
\begin{equation}
	\Mat{c}_{\rho \varsigma} = \Mat{\Lambda}_{\rho \rho}^{-1} \Mat{d}_{\rho \rho}^\intercal \Mat{a}_{\rho \varsigma} + \Mat{\Lambda}_{\rho \rho}^{-1} \Mat{d}_{\varsigma \rho}^\intercal \Mat{a}_{\varsigma \varsigma}
	.
\end{equation}

\noindent Finally, since $\Mat{a}_{\varsigma \varsigma}$ is invertible, both \Eq{eq:3_symplec3} and \Eq{eq:3_symplec6} take the form
\begin{equation}
	\begin{pmatrix}
		\OMat{\rho \rho}
		&
		\OMat{\rho \varsigma} 
		\\
		\OMat{\varsigma \rho}
		&
		\Mat{n}
		-
		\Mat{n}^\intercal
	\end{pmatrix}
	=
	\OMat{N} ,
\end{equation}

\noindent where 
\begin{equation}
	\Mat{n}
	\doteq
	\Mat{a}_{\varsigma \varsigma}^\intercal \Mat{d}_{\varsigma \rho} \Mat{\Lambda}_{\rho \rho}^{-1} \Mat{a}_{\rho \varsigma}
	- 
	\Mat{a}_{\varsigma \varsigma}^\intercal \Mat{c}_{\varsigma \varsigma} .
\end{equation}

\noindent I therefore require $\Mat{c}_{\varsigma \varsigma}$ to satisfy
\begin{equation}
	\Mat{a}_{\varsigma \varsigma}^\intercal \Mat{d}_{\varsigma \rho} \Mat{\Lambda}_{\rho \rho}^{-1} \Mat{a}_{\rho \varsigma}
	- 
	\Mat{a}_{\varsigma \varsigma}^\intercal \Mat{c}_{\varsigma \varsigma}
	=
	\left(
		\Mat{a}_{\varsigma \varsigma}^\intercal \Mat{d}_{\varsigma \rho} \Mat{\Lambda}_{\rho \rho}^{-1} \Mat{a}_{\rho \varsigma}
		- 
		\Mat{a}_{\varsigma \varsigma}^\intercal \Mat{c}_{\varsigma \varsigma}
	\right)^\intercal .
	\label{eq:3_cSUBsymm}
\end{equation}

In summary, the SVD projections of $\Mat{A}$, $\Mat{C}$, and $\Mat{D}$ take the form
\begin{equation}
	\widetilde{\Mat{A}} 
	=
	\begin{pmatrix}
		\Mat{a}_{\rho \rho} & \Mat{a}_{\rho \varsigma} \\
		\OMat{\varsigma \rho} & \Mat{a}_{\varsigma \varsigma}
	\end{pmatrix}
	,
	\quad
	\widetilde{\Mat{C}}
	=
	\begin{pmatrix}
		\Mat{c}_{\rho \rho} & \Mat{c}_{\rho \varsigma} \\
		\Mat{c}_{\varsigma \rho} & \Mat{c}_{\varsigma \varsigma}
	\end{pmatrix}
	, 
	\quad
	\widetilde{\Mat{D}} 
	=
	\begin{pmatrix}
		\Mat{d}_{\rho \rho} & \OMat{\rho \varsigma} \\
		\Mat{d}_{\varsigma \rho} & \Mat{a}_{\varsigma \varsigma}^{-\intercal}
	\end{pmatrix}
	.
\end{equation}

\noindent This result shall be used later in this chapter and also in \Ch{ch:MGO} to derive MGO formulas that are valid for use on any type of linear wave equation.


\subsection{Near-identity symplectic matrices}

It will be necessary in \Ch{ch:NIMT} to computing the quantities $\det{\Mat{A}}$, $\Mat{A}^{-1}$, $\Mat{A}^{-1}\Mat{B}$, and $\Mat{C} \Mat{A}^{-1}$ when $\Mat{S}$ is near-identity, that is, $\Mat{S} \approx \IMat{2N}$. It will therefore be useful to derive approximate asymptotic formulas for these quantities which help calculate them more efficiently. In particular, I will show that calculating the lowest-order terms of such asymptotic representations will not require any explicit matrix multiplications.

Generally speaking, the near-identity behavior of a group is governed by its Lie algebra. For the group of $2N\times 2N$ real symplectic matrices, denoted $\Sp{2N}$, the Lie algebra is the space of all $2N\times2N$ real Hamiltonian matrices~\cite{Dragt05}. Note that a matrix $\Mat{H}$ is Hamiltonian if and only if 
\begin{equation}
	\JMat{2N} \Mat{H} 
	= - \Mat{H}^\intercal \JMat{2N} 
	\equiv \left( \JMat{2N} \Mat{H}\right)^\intercal
	.
	\label{eq:3_hamDEF}
\end{equation}

\noindent By the connectivity of $\Sp{2N}$ and the polar decomposition, any symplectic matrix $\Mat{S}$ can be parameterized as~\cite{Littlejohn86a, Dragt82, Hall15}
\begin{equation}
	\Mat{S} = \exp\left( \epsilon \Mat{H}_s \right) \exp\left( \epsilon \Mat{H}_a \right)
	,
	\label{eq:3_polarS}
\end{equation}

\noindent where $\Mat{H}_s$ and $\Mat{H}_a$ are symmetric and antisymmetric Hamiltonian matrices, respectively. The formal small parameter $\epsilon$ has been introduced to aid with ordering the forthcoming expansions when $\Mat{H}_s$ and $\Mat{H}_a$ are small. Note that if $\Mat{H}$ is Hamiltonian, then $\Mat{H}^\intercal$ also is; hence, $\Mat{H}_s$ and $\Mat{H}_a$ can be uniquely represented as~\cite{Dragt82}
\begin{equation}
	\Mat{H}_s = \frac{\Mat{H} + \Mat{H}^\intercal}{2} \, , \quad \Mat{H}_a = \frac{\Mat{H} - \Mat{H}^\intercal}{2}
	.
	\label{eq:3_HDecomp}
\end{equation}

\noindent In this sense, $\Mat{S}$ is parameterized by a single Hamiltonian matrix $\Mat{H}$.

Let us consider the case when $\Mat{S}$ is near-identity, meaning $\Mat{H}$ is close to $\OMat{2N}$. Expanding \Eq{eq:3_polarS} in $\epsilon$ yields
\begin{equation}
	\Mat{S} = \IMat{2N} + \epsilon \Mat{H} + \frac{\epsilon^2}{4}\left(2\Mat{H} \Mat{H} - \Mat{H} \Mat{H}^\intercal + \Mat{H}^\intercal \Mat{H} \right) + O(\epsilon^3) 
	.
	\label{eq:3_SMATexpand}
\end{equation}

\noindent Since any Hamiltonian matrix can be block-decomposed as
\begin{equation}
	\Mat{H} = \begin{pmatrix}
		\Mat{V}^\intercal & \Mat{W}\\
		-\Mat{U} & -\Mat{V}
	\end{pmatrix} = \Mat{J}
	\begin{pmatrix}
		\Mat{U} & \Mat{V}\\
		\Mat{V}^\intercal & \Mat{W}
	\end{pmatrix} 
	,
	\label{eq:3_Hblock}
\end{equation}

\noindent with $\Mat{U}$ and $\Mat{W}$ being symmetric matrices, I obtain the following expansions from \Eq{eq:3_SMATexpand}:
\begin{subequations}
	\begin{align}
		\Mat{A} &\approx \IMat{N} + \epsilon\Mat{V}^\intercal + \epsilon^2 \frac{\Mat{V}_s \Mat{V}^\intercal - \Mat{V}^\intercal \Mat{V}_a+ \TmMat \Mat{U} - \Mat{W} \TpMat}{2} 
		,\\
		\Mat{B} &\approx \epsilon\Mat{W} + \epsilon^2 \frac{\Mat{V}_s\Mat{W} - \Mat{W} \, \Mat{V}_a + \TmMat \Mat{V} + \Mat{V}^\intercal \TpMat}{2} 
		,\\
		\Mat{C} &\approx -\epsilon\Mat{U} + \epsilon^2 \frac{\Mat{V}_s \Mat{U} + \Mat{U} \, \Mat{V}_a - \TmMat \Mat{V}^\intercal + \Mat{V} \, \TpMat}{2} 
		,\\
		\Mat{D} &\approx \IMat{N}-\epsilon\Mat{V} + \epsilon^2 \frac{\Mat{V}_s \Mat{V} + \Mat{V} \, \Mat{V}_a - \TmMat \Mat{W} - \Mat{U} \TpMat}{2} 
		,
	\end{align}
\end{subequations}

\noindent where
\begin{subequations}
	\begin{align}
		\Mat{V}_s &\doteq \frac{1}{2}\left(\Mat{V} + \Mat{V}^\intercal \right) \, , \quad \Mat{V}_a \doteq \frac{1}{2}\left(\Mat{V} - \Mat{V}^\intercal \right) 
		,\\
		\TpMat &\doteq \frac{1}{2}\left(\Mat{U} + \Mat{W} \right) \, , \quad \TmMat \doteq \frac{1}{2}\left(\Mat{U} - \Mat{W} \right) 
		.
	\end{align}
\end{subequations}

One can show that
\begin{equation}
	\Mat{A}^{-1} \approx \IMat{N} - \epsilon \Mat{V}^\intercal + \epsilon^2 \frac{\Mat{V}^\intercal \Mat{V}_s - \Mat{V}_a\Mat{V}^\intercal - \TmMat \Mat{U} + \Mat{W} \TpMat}{2}
	\label{eq:3_AInvExpans}
\end{equation}

\noindent satisfies both $\Mat{A}^{-1} \Mat{A} = \IMat{N}$ and $\Mat{A} \Mat{A}^{-1} = \IMat{N}$ to $O(\epsilon^3)$. By direct multiplication one also obtains
\begin{align}
	\Mat{A}^{-1}\Mat{B} \approx \epsilon\Mat{W} + \epsilon^2\left(\Mat{V}_a\Mat{W} + \Mat{V}^\intercal\TmMat \right)_s 
	,\quad
	\Mat{C} \Mat{A}^{-1} \approx -\epsilon\Mat{U} + \epsilon^2\left(\Mat{V}_s \Mat{U} + \Mat{V}\TpMat \right)_s
	,
	\label{eq:3_invPRODexpand}
\end{align}

\noindent where the subscript $_s$ denotes the symmetric part. Notably, the expansions of both $\Mat{A}^{-1} \Mat{B}$ and $\Mat{C} \Mat{A}^{-1}$ are symmetric at each order of $\epsilon$, as required by \Eqs{eq:3_symplec1} and \eq{eq:3_symplec6}. Finally, let us approximate $\det{\Mat{A}}$ as
\begin{equation}
	\det{\Mat{A}} \approx \det\left(\IMat{N} + \epsilon \Mat{M} \right) 
	,\quad
	\Mat{M} \doteq \Mat{V}^\intercal + \epsilon \frac{\Mat{V}_s \Mat{V}^\intercal - \Mat{V}^\intercal \Mat{V}_a+ \TmMat \Mat{U} - \Mat{W} \TpMat}{2} 
	.
	\label{eq:3_CharPoly}
\end{equation}

\noindent Up to the factor $\epsilon^N$, the right-hand side of \Eq{eq:3_CharPoly} is simply the characteristic polynomial of $-\Mat{M}$. Using, for example, Faddeev--LeVerrier's method leads to
\begin{align}
	\det{\Mat{A}} \approx
	1 + \epsilon \, \Tr \left( \Mat{V} \right)
	+ \epsilon^2 \frac{ \left[ \Tr \left( \Mat{V} \right) \right]^2 
	+ \Tr \left( \TmMat \Mat{U} - \Mat{W} \TpMat \right)}{2} 
	.
	\label{eq:3_detAexpand}
\end{align}


\subsection{Linear orthosymplectic transformations}

Lastly, it will be useful for MGO to also consider linear \textit{orthosymplectic} transformations, that is, linear transformations governed by a matrix $\Mat{S}$ that is both symplectic and orthogonal. The latter condition requires
\begin{equation}
	\Mat{S}^\intercal = \Mat{S}^{-1}
	.
	\label{eq:3_orthoDEF}
\end{equation}

\noindent By taking the transpose of \Eq{eq:3_blockS} and comparing with \Eq{eq:3_blockSinv}, it is easy to observe that the orthogonality condition \eq{eq:3_orthoDEF} implies $\Mat{D} = \Mat{A}$ and $\Mat{C} = - \Mat{B}$. Hence, an orthosymplectic matrix has the block decomposition
\begin{equation}
	\Mat{S}
	=
	\begin{pmatrix}
		\Mat{A} & \Mat{B} \\
		- \Mat{B} & \Mat{A}
	\end{pmatrix}
	.
	\label{eq:3_blockSortho}
\end{equation}

\noindent One can then readily substitute this block decomposition into the symplectic relations \eq{eq:3_symplecABCD} to obtain their orthosymplectic analogues:
\begin{subequations}%
	\label{eq:3_symplecABCDortho}
	\begin{align}%
		\Mat{A} \Mat{B}^\intercal - \Mat{B} \Mat{A}^\intercal &= \OMat{N}
		, \\
		\Mat{A} \Mat{A}^\intercal + \Mat{B} \Mat{B}^\intercal &= \IMat{N} 
		, \\
		\Mat{B}^\intercal \Mat{A} - \Mat{A}^\intercal \Mat{B} &= \OMat{N}
		, \\
		\Mat{A}^\intercal \Mat{A} + \Mat{B}^\intercal \Mat{B} &= \IMat{N}
		.
	\end{align}%
\end{subequations}%

By inspection, one can also show from the results of \Sec{sec:3_SVD} that $\Mat{A}$ has a block-diagonal structure in the diagonalizing basis of $\Mat{B}$:
\begin{equation}
	\widetilde{\Mat{A}}
	=
	\begin{pmatrix}
		\Mat{a}_{\rho \rho} & \OMat{\rho \varsigma} \\
		\OMat{\varsigma \rho} & \Mat{a}_{\varsigma \varsigma}
	\end{pmatrix}
	,
	\label{eq:3_aSVDortho}
\end{equation}

\noindent where $\Mat{a}_{\varsigma \varsigma}$ is orthogonal, that is,
\begin{equation}
	\Mat{a}_{\varsigma \varsigma} = \Mat{a}_{\varsigma \varsigma}^{-\intercal}
	,
\end{equation}

\noindent (which implies that $\det \Mat{a}_{\varsigma \varsigma} = \pm 1$), and $\Mat{a}_{\rho \rho}$ obeys the symmetry
\begin{equation}
	\Mat{a}_{\rho \rho}\Mat{\Lambda}_{\rho \rho} 
	= 
	\left( 
		\Mat{a}_{\rho \rho}\Mat{\Lambda}_{\rho \rho}
	\right)^\intercal
	, \quad
	\Mat{\Lambda}_{\rho \rho} \Mat{a}_{\rho \rho} 
	= 
	\left( 
		\Mat{\Lambda}_{\rho \rho} \Mat{a}_{\rho \rho} 
	\right)^\intercal
	, \quad
	\Mat{a}_{\rho \rho} \Mat{a}_{\rho \rho}^\intercal
	= \Mat{a}_{\rho \rho}^\intercal \Mat{a}_{\rho \rho}
	= \IMat{\rho} - \Mat{\Lambda}_{\rho \rho}\Mat{\Lambda}_{\rho \rho}
	.
\end{equation}

Lastly, note that when $\Mat{S}$ is orthosymplectic, its near-identity behavior [\Eq{eq:3_polarS}] is governed solely by an antisymmetric Hamiltonian matrix, \ie
\begin{equation}
	\Mat{S} = \exp\left( \epsilon \Mat{H}_a \right)
	,
\end{equation}

\noindent where the antisymmetric Hamiltonian matrix $\Mat{H}_a$ has the block decomposition
\begin{equation}
	\Mat{H}_a = \begin{pmatrix}
		- \Mat{V} & \Mat{W}\\
		-\Mat{W} & -\Mat{V}
	\end{pmatrix}
	,
\end{equation}

\noindent with $\Mat{W}$ being symmetric and $\Mat{V}$ being antisymmetric. Note that this latter condition implies that $\Tr(\Mat{V}) = 0$ and that $V = 0$ for $1$-D problems.

\noindent Hence, I compute the expansions
\begin{subequations}
	\begin{align}
		\Mat{A} &\approx \IMat{N} - \epsilon\Mat{V} + \epsilon^2 \frac{\Mat{V} \, \Mat{V} - \Mat{W} \, \Mat{W}}{2} 
		,\\
		\Mat{B} &\approx \epsilon\Mat{W} - \epsilon^2 \frac{ \Mat{W} \, \Mat{V} + \Mat{V} \, \Mat{W}}{2} 
		.
	\end{align}
\end{subequations}

\noindent Additionally, I obtain
\begin{equation}
	\Mat{A}^{-1} \approx \IMat{N} + \epsilon \Mat{V} + \epsilon^2 \frac{ \Mat{V}\, \Mat{V} + \Mat{W} \, \Mat{W}}{2}
	,
	\label{eq:3_AInvExpansORTHO}
\end{equation}

\noindent meaning I can calculate
\begin{align}
	\Mat{A}^{-1}\Mat{B} \approx \epsilon\Mat{W} + \epsilon^2 \frac{ \Mat{V} \, \Mat{W} - \Mat{W} \, \Mat{V} }{2}
	,\quad
	\Mat{B} \Mat{A}^{-1} \approx \epsilon\Mat{W} - \epsilon^2 \frac{ \Mat{V} \, \Mat{W} - \Mat{W} \, \Mat{V} }{2}
	,
\end{align}

\noindent along with
\begin{align}
	\det{\Mat{A}} \approx
	1
	- \epsilon^2 \frac{ \Tr \left( \Mat{W} \Mat{W} \right)}{2} 
	.
\end{align}


\section{Unitary metaplectic transforms}


\subsection{Definition, derivation, and integral form}
\label{sec:3_MT}

Recall from \Ch{ch:GO} that GO is constructed in a Hilbert space consisting of state vectors $\ket{\psi}$. Linear symplectic phase-space transformations are special because they have an exact operator analogue in a Hilbert space; these are the unitary metaplectic operators~\cite{Moshinsky71,Littlejohn86a}, denoted as $\oper{M}(\Mat{S})$. Specifically, if $\VectOp{z} \doteq (\VectOp{x}, \VectOp{k})$ denotes the original phase-space position and momentum operators, then one can obtain the transformed operators $\Stroke{\VectOp{Z}} \doteq (\VectOp{X}, \VectOp{K})$ corresponding to the linear symplectic transformation of \Eq{eq:3_Strans} through the action of $\oper{M}(\Mat{S})$ as
\begin{equation}
	\Stroke{\VectOp{Z}} = 
	\oper{M}^\dagger(\Mat{S}) \VectOp{z} \oper{M}(\Mat{S})
	\equiv
	\Mat{S} \VectOp{z}
	,
	\label{eq:3_operTRANS}
\end{equation}

\noindent or equivalently in terms of the constituent operators,
\begin{subequations}
	\begin{align}
		\label{eq:3_XtransOPER}
		\VectOp{X} &= \oper{M}^\dagger(\Mat{S}) \VectOp{x} \oper{M}(\Mat{S})
		\equiv
		\Mat{A} \VectOp{x} + \Mat{B} \VectOp{k}
		, \\
		\label{eq:3_KtransOPER}
		\VectOp{K} &= \oper{M}^\dagger(\Mat{S}) \VectOp{k} \oper{M}(\Mat{S})
		\equiv
		\Mat{C} \VectOp{x} + \Mat{D} \VectOp{k}
		.
	\end{align}
\end{subequations}

In addition to generating the mapping $\VectOp{z} \mapsto \Stroke{\VectOp{Z}}$, the metaplectic operator also induces a mapping between the \textit{projections} of $\ket{\psi}$ onto the original $\Vect{x}$-space and onto the transformed $\Vect{X}$-space. Recall from \Ch{ch:GO} that the former is defined as $\psi(\Vect{y}) \doteq \braket{\Vect{x}(\Vect{y})}{\psi}$, where $\ket{\Vect{x}(\Vect{y})}$ is the eigenvector of $\VectOp{x}$ corresponding to the eigenvalue $\Vect{y}$. (This is a generalized notation compared to what was used in \Ch{ch:GO}, for reasons that will be apparent shortly.) Likewise, the projection onto $\Vect{X}$-space is obtained as $\Psi(\Vect{y}) \doteq \braket{\Vect{X}(\Vect{y})}{\psi}$, where $\ket{\Vect{X}(\Vect{y})}$ is the eigenvector of $\VectOp{X}$ corresponding to the eigenvalue $\Vect{y}$. The $\Vect{X}$-space basis elements are generated via direct action of $\oper{M}(\Mat{S})$ as%
\footnote{Analogous to the Schr\"odinger versus Heisenberg representations of time evolution, there exists in the general case a distinction between whether $\oper{M}$ transforms the wavefunction (`active' representation) or transforms the projection basis (`passive' representation). In my discussion, I assume the passive representation.}%
\begin{equation}
	\ket{\Vect{X}(\Vect{y})} = \oper{M}^\dagger(\Mat{S}) \ket{\Vect{x}(\Vect{y})}
	. 
\end{equation}

\noindent The induced mapping between the $\psi$ and $\Psi$ can therefore be calculated in the usual manner:
\begin{align}
	\Psi(\Vect{Y}) = \int \dd \Vect{y} \, U(\Vect{Y},\Vect{y}) \psi(\Vect{y}) 
	,\quad
	U(\Vect{Y},\Vect{y}) = \braket{\Vect{X}(\Vect{Y})}{\Vect{x}(\Vect{y})} = \bra{\Vect{x}(\Vect{Y})}\oper{M} \ket{\Vect{x}(\Vect{y})} 
	.
\end{align}

To calculate $U$, consider \Eq{eq:3_XtransOPER} and apply $\bra{\Vect{X}(\Vect{Y})}$ from the left and $\ket{\Vect{x}(\Vect{y})}$ from the right. Using the eigenvalue relations along with
\begin{equation}
	\bra{\Vect{X}(\Vect{Y})} \VectOp{k} \ket{\Vect{x}(\Vect{y})} = \left[\bra{\Vect{x}(\Vect{y})} \VectOp{k} \ket{\Vect{X}(\Vect{Y})}\right]^* = i\partial_\Vect{y} U(\Vect{Y},\Vect{y})
\end{equation}

\noindent leads to a differential equation~\cite{Littlejohn86a,Moshinsky71}
\begin{equation}
	\Vect{Y} \, U(\Vect{Y},\Vect{y}) = \left(\Mat{A} \Vect{y} + i \Mat{B} \pd{\Vect{y}} \right) U(\Vect{Y},\Vect{y})  
	,
\end{equation}

\noindent which can be solved to yield
\begin{equation}
	U(\Vect{Y},\Vect{y}) = f(\Vect{Y}) \, 
	\exp\left( 
		\frac{i}{2} \Vect{y}^\intercal \Mat{B}^{-1}\Mat{A}\Vect{y} 
		- i \Vect{y}^\intercal \Mat{B}^{-1} \Vect{Y} 
	\right)
	,
	\label{eq:3_xSOL}
\end{equation}

\noindent where $f(\Vect{Y})$ is an unknown function. Doing the same with \Eq{eq:3_KtransOPER} leads to
\begin{equation}
	\pd{\Vect{Y}} U(\Vect{Y},\Vect{y}) = \left(i\Mat{C} \Vect{y} - \Mat{D} \pd{\Vect{y}}\right) U(\Vect{Y},\Vect{y}) 
	.
	\label{eq:3_pde2}
\end{equation}

\noindent Using \Eqs{eq:3_symplecABCD}, \eq{eq:3_xSOL}, and \eq{eq:3_pde2} determines $f(\Vect{Y})$ up to a multiplicative constant:
\begin{equation}
	f(\Vect{Y}) = \alpha \, 
	\exp\left( 
		\frac{i}{2} \Vect{Y}^\intercal \Mat{D}\Mat{B}^{-1}\Vect{Y} 
	\right)
	.
	\label{eq:3_MTf}
\end{equation}

\begin{figure}
	\centering
	\includegraphics[width=0.6\linewidth,trim={2mm 5mm 13mm 2mm},clip]{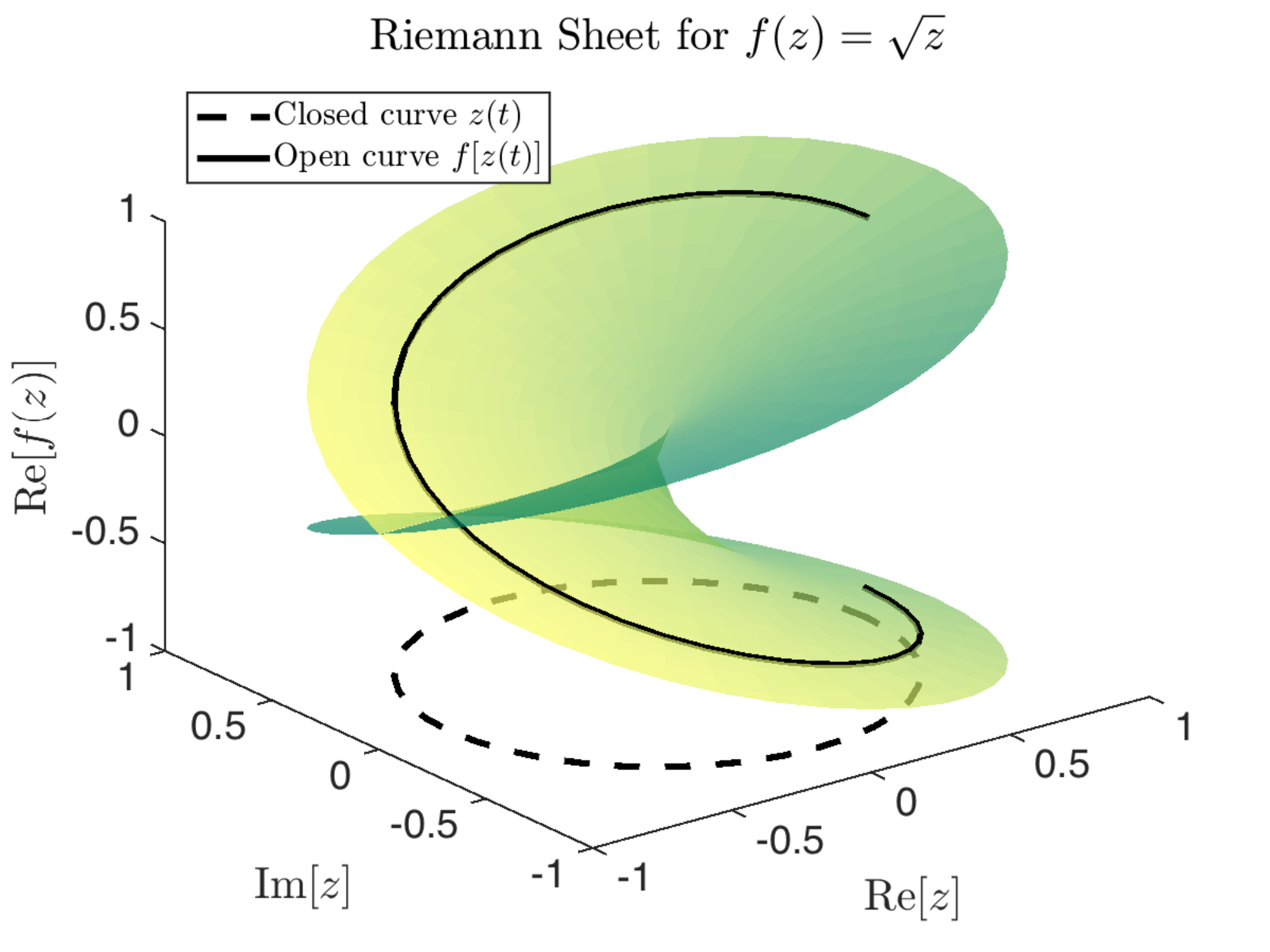}
	\caption{The Riemann sheet (colored) of the function $f(z) \doteq \sqrt{z}$ illustrates the relationship between the family $\{\oper{M}_t\}$ (more generally, the metaplectic group) and the family of all phase-space rotations (more generally, the symplectic group) for all~$t$. As depicted in the figure, $f$ maps a closed curve on the $z$ plane (dashed) to a closed curve on the Riemann sheet only if the winding number is even. Likewise, it takes two rotation periods for $\oper{M}_t$ to return to its original value $\oper{M}_0 = \IdentOp$. In a more general formulation, the metaplectic group forms a double-cover of the symplectic group. For details, see \Ref{Littlejohn86a}.}
	\label{fig:3_riemann}
\end{figure}

Normalization determines the constant $\alpha$ up to a phase. The phase requires more involved analysis to determine, and the result is not unique: there exist two possible phases which differ by $\pi$. This ambiguity is required to ensure that the metaplectic operators form a group, but results in a one-to-two correspondence between the symplectic and the metaplectic groups~\cite{Littlejohn86a}. In other words, changing the overall sign of a metaplectic operator does not change the resulting phase space transformation. As shall be discussed in \Sec{sec:3_SecMETQHO} (and also related to the general Bohr-Sommerfeld rule~\cite{Shankar94}), the sign ambiguity becomes important when one considers a family of transformations parameterized by some path variable $t$. A closed trajectory in the space of symplectic matrices, $\Mat{S}_t$, results in a closed trajectory in the space of metaplectic operators only for even winding numbers. In contrast, for odd winding numbers $\oper{M}_t$ changes sign, just like the function $f(z) \doteq \sqrt{z}$ changes sign each time $z$ encircles the origin in the complex plane~\cite{Littlejohn86a} (see \Fig{fig:3_riemann}).

Including the phase and sign ambiguity, the final result for the transformation is~\cite{Collins70,Moshinsky71,Littlejohn86a}
\begin{equation}
	\Psi(\Vect{X}) = \pm
	\frac{ 
		\exp\left( \frac{i}{2}\Vect{X}^\intercal \Mat{D} \Mat{B}^{-1} \Vect{X} \right)
	}{
		(2\pi i)^{\frac{N}{2}} \sqrt{\det{\Mat{B}}}
	}
	 \int \dd\Vect{x} \,
	 \exp\left(
	 	\frac{i}{2}\Vect{x}^\intercal \Mat{B}^{-1}\Mat{A} \Vect{x} - i\Vect{x}^\intercal \Mat{B}^{-1} \Vect{X}
	\right) \, \psi(\Vect{x}) 
	 ,
	\label{eq:3_MTint}
\end{equation}

\noindent where $\Mat{B}^{-1}\Mat{A}$ and $\Mat{D} \Mat{B}^{-1}$ are symmetric due to \Eqs{eq:3_symplec1} and \eq{eq:3_symplec4}. Equation \eq{eq:3_MTint} defines $\Psi(\Vect{X})$ as the metaplectic transform (MT) of $\psi(\Vect{x})$. In writing \Eq{eq:3_MTint}, I have dropped the $\Vect{y}$ and $\Vect{Y}$ notation in favor of $\Vect{x}$ and $\Vect{X}$, as there is no longer any risk of ambiguity, and my branch cut convention restricts all complex phases to the interval $(-\pi,\pi]$. Similarly, the inverse MT is given as
\begin{equation}
	\psi(\Vect{x}) = \pm
	\frac{ 
		\exp\left(- \frac{i}{2} \Vect{x}^\intercal \Mat{B}^{-1}\Mat{A} \Vect{x} \right)
	}{
		(- 2\pi i)^{\frac{N}{2}} \sqrt{\det{\Mat{B}}}
	} \int \dd \Vect{X} \,
	\exp\left(
		- \frac{i}{2}\Vect{X}^\intercal \Mat{D} \Mat{B}^{-1} \Vect{X} 
		+ i\Vect{x}^\intercal \Mat{B}^{-1} \Vect{X}
	\right) \, \Psi(\Vect{X})
	,
	\label{eq:3_invMTint}
\end{equation}

\noindent with complex phases restricted to the interval $[-\pi, \pi)$.


\subsection{Singular metaplectic transforms}

The standard MT integral given by \Eq{eq:3_MTint} requires $\Mat{B}$ to be invertible. When $\det \Mat{B} = 0$, one might think that the MT diverges, but this is actually not the case. The MT kernel $U(\Vect{X}, \Vect{x}; \Mat{S})$ [\Eqs{eq:3_xSOL} and \eq{eq:3_MTf}] that corresponds to a symplectic matrix $\Mat{S}$ with $\det \Mat{B} = 0$ can be considered as a limiting case~\cite{Littlejohn86a}
\begin{equation}
	U(\Vect{X}, \Vect{x}; \Mat{S}) = \lim_{\varepsilon \to 0} U(\Vect{X}, \Vect{x}; \Mat{S}_\varepsilon) 
	,
	\label{eq:3_MTlim}
\end{equation}

\noindent where $\Mat{S}_\varepsilon$ is a symplectic matrix that has $\Mat{S}$ as a limit at $\varepsilon \to 0$. For example, I can adopt
\begin{equation}
	\Mat{S}_\varepsilon 
	= 
	\begin{pmatrix}
		\Mat{A} & \Mat{B} + \varepsilon \Mat{A} \\
		\Mat{C} & \Mat{D} + \varepsilon \Mat{C}
	\end{pmatrix} ,
\end{equation}

\noindent whose symplecticity (to all orders in $\varepsilon$) can be readily verified by definition \eq{eq:3_symplecDEF}.

To show that $\det \left(\Mat{B} + \varepsilon \Mat{A} \right) \neq 0$ and subsequently compute the limit in \Eq{eq:3_MTlim}, it is useful to perform a singular value decomposition (SVD) of $\Mat{B}$. Using results from \Sec{sec:3_SVD} I compute
\begin{align}
	\det
	\left(
		\Mat{B} + \varepsilon \Mat{A}
	\right)
	=
	\det
	\begin{pmatrix}
		\Mat{\Lambda}_{\rho \rho} + \varepsilon \Mat{a}_{\rho \rho} 
		& 
		\varepsilon \Mat{a}_{\rho \varsigma} 
		\\
		\OMat{\varsigma \rho} 
		& 
		\varepsilon \Mat{a}_{\varsigma \varsigma}
	\end{pmatrix}
	=
	\det 
	\left( 
		\Mat{\Lambda}_{\rho \rho} + \varepsilon \Mat{a}_{\rho \rho} 
	\right)
	\det
	\left( 
		\varepsilon \Mat{a}_{\varsigma \varsigma}
	\right)
	\approx
	\varepsilon^\varsigma
	\det \Mat{\Lambda}_{\rho \rho} \,
	\det \Mat{a}_{\varsigma \varsigma}
	,
	\label{eq:3_detBmod}
\end{align}

\noindent where I have used $\det \Mat{L}_\textrm{s} = \det \Mat{R}_\textrm{s} = 1$. Since $\det \Mat{a}_{\varsigma \varsigma} \ne 0$ and $\det \Mat{\Lambda}_{\rho \rho} \ne 0$ by definition, $\det(\Mat{B} + \varepsilon \Mat{A})$ is nonzero for finite $\varepsilon$. (I adopt the convention that $0 \times 0$ matrices have unit determinant.) 

Similarly, I must compute the terms within the phase factors in $U(\Vect{x}, \Vect{X}; \Mat{S}_\varepsilon)$ to leading order in $\varepsilon$. For reference, this phase factor is given explicitly in the SVD basis as
\begin{align}
	\widetilde{G}(\Vect{x}, \Vect{X})
	\doteq
	\frac{1}{2}
	\Vect{X}^\intercal
	\Mat{L}_\textrm{s}
	\left(
		\widetilde{\Mat{D}} + \varepsilon \widetilde{\Mat{C}}
	\right)
	\left(
		\widetilde{\Mat{B}} + \varepsilon \widetilde{\Mat{A}} 
	\right)^{-1}
	\Mat{L}_\textrm{s}^\intercal
	\Vect{X}
	-
	\Vect{x}^\intercal
	\Mat{R}_\textrm{s}
	\left(
		\widetilde{\Mat{B}} + \varepsilon \widetilde{\Mat{A}}
	\right)^{-1}
	\left(
		\Mat{L}_\textrm{s}^\intercal \Vect{X}
		+\frac{1}{2} \widetilde{\Mat{A}} \,
		\Mat{R}_\textrm{s}^\intercal \Vect{x}
	\right)
	.
\end{align}

\noindent The first step is to approximate the matrix inverse term to leading order in $\varepsilon$. Let us adopt the parameterization
\begin{equation}
	\left(
		\widetilde{\Mat{B}}
		+ \varepsilon \widetilde{\Mat{A}}
	\right)^{-1}
	=
	\begin{pmatrix}
		\Mat{m}_{\rho \rho} & \Mat{m}_{\rho \varsigma} \\
		\Mat{m}_{\varsigma \rho} & \Mat{m}_{\varsigma \varsigma}
	\end{pmatrix}
	.
\end{equation}

\noindent Then, since 
\begin{equation}
	\begin{pmatrix}
		\Mat{m}_{\rho \rho}
		\left(
			\Mat{\Lambda}_{\rho \rho} 
			+ \varepsilon \Mat{a}_{\rho \rho}
		\right)
		&
		\varepsilon
		\left(
			\Mat{m}_{\rho \rho} \Mat{a}_{\rho \varsigma} 
			+ \Mat{m}_{\rho \varsigma} \Mat{a}_{\varsigma \varsigma}
		\right)
		\\[1mm]
		\Mat{m}_{\varsigma \rho} 
		\left(
			\Mat{\Lambda}_{\rho \rho} 
			+ \varepsilon \Mat{a}_{\rho \rho}
		\right)
		&
		\varepsilon
		\left(
			\Mat{m}_{\varsigma \rho} \Mat{a}_{\rho \varsigma}
			+ \Mat{m}_{\varsigma \varsigma} \Mat{a}_{\varsigma \varsigma}
		\right)
	\end{pmatrix}
	= \IMat{N}
	,
\end{equation}

\noindent I require
\begin{subequations}
	\label{eq:mEQs}
	\begin{align}
		\Mat{m}_{\varsigma \rho} 
		\left(
			\Mat{\Lambda}_{\rho \rho} 
			+ \varepsilon \Mat{a}_{\rho \rho}
		\right)
		&= \OMat{\varsigma \rho}
		, 
		\\
		\varepsilon
		\left(
			\Mat{m}_{\rho \rho} \Mat{a}_{\rho \varsigma} 
			+ \Mat{m}_{\rho \varsigma} \Mat{a}_{\varsigma \varsigma}
		\right)
		&= \OMat{\rho \varsigma}
		,
		\\
		\varepsilon
		\left(
			\Mat{m}_{\varsigma \rho} \Mat{a}_{\rho \varsigma}
			+ \Mat{m}_{\varsigma \varsigma} \Mat{a}_{\varsigma \varsigma}
		\right)
		&= \IMat{\varsigma}
		,
		\\
		\Mat{m}_{\rho \rho}
		\left(
			\Mat{\Lambda}_{\rho \rho} 
			+ \varepsilon \Mat{a}_{\rho \rho}
		\right)
		&= \IMat{\rho}
		.
	\end{align}
\end{subequations}

\noindent Solving \Eqs{eq:mEQs} in sequence yields
\begin{subequations}
	\begin{align}
		\Mat{m}_{\varsigma \rho} 
		= \OMat{\varsigma \rho}
		,
		\quad
		\Mat{m}_{\rho \varsigma} 
		= - \Mat{m}_{\rho \rho} \Mat{a}_{\rho \varsigma} \Mat{a}_{\varsigma \varsigma}^{-1}
		,
		\quad
		\Mat{m}_{\varsigma \varsigma}
		= \varepsilon^{-1} \Mat{a}_{\varsigma \varsigma}^{-1}
		,
		\quad
		\Mat{m}_{\rho \rho}
		= (\Mat{\Lambda}_{\rho \rho} + \varepsilon \Mat{a}_{\rho \rho})^{-1}
		\approx \Mat{\Lambda}_{\rho \rho}^{-1}
		,
	\end{align}
\end{subequations}

\noindent where I have used the fact that $\Mat{a}_{\varsigma \varsigma}$ is invertible. Hence, I obtain the following leading-order approximations:
\begin{align}
	\left(
		\widetilde{\Mat{D}} 
		+ \varepsilon \widetilde{\Mat{C}}
	\right)
	\left(
		\widetilde{\Mat{B}} 
		+ \varepsilon \widetilde{\Mat{A}}
	\right)^{-1}
	&\approx
	\begin{pmatrix}
		\Mat{d}_{\rho \rho} \Mat{\Lambda}_{\rho \rho}^{-1}
		&
		\Mat{\Lambda}_{\rho \rho}^{-1} \Mat{d}_{\varsigma \rho}^\intercal
		\\[1mm]
		\Mat{d}_{\varsigma \rho} \Mat{\Lambda}_{\rho \rho}^{-1}
		&
		\varepsilon^{-1} \Mat{a}_{\varsigma \varsigma}^{-\intercal} \Mat{a}_{\varsigma \varsigma}^{-1} + \Mat{c}_{\varsigma \varsigma} \Mat{a}_{\varsigma \varsigma}^{-1} - \Mat{d}_{\varsigma \rho} \Mat{\Lambda}_{\rho \rho}^{-1} \Mat{a}_{\rho \varsigma} \Mat{a}_{\varsigma \varsigma}^{-1}
	\end{pmatrix}
	,
	\\
	\left(
		\widetilde{\Mat{B}} 
		+ \varepsilon \widetilde{\Mat{A}}
	\right)^{-1}
	\widetilde{\Mat{A}}
	&\approx
	\begin{pmatrix}
		\Mat{\Lambda}_{\rho \rho}^{-1} \Mat{a}_{\rho \rho}
		&
		\OMat{\rho \varsigma}
		\\[1mm]
		\OMat{\varsigma \rho}
		&
		\varepsilon^{-1} \IMat{\varsigma}
	\end{pmatrix}
	.
\end{align}

Hence, I obtain to leading order in $\varepsilon$
\begin{align}
	U(\Vect{X}, \Vect{x}; \Mat{S}_\varepsilon)
	\approx 
	\pm \frac
	{ 
		\varepsilon^{-\varsigma/2} 
		\exp
		\left[
			i \, g(\Vect{x}_\rho, \Vect{X})
		\right]
	}
	{
		(2 \pi i)^{N/2}
		\sqrt
		{
			\det \Mat{\Lambda}_{\rho \rho}
			\det \Mat{a}_{\varsigma \varsigma}
		}
	}
	\exp
	\left(
		\frac{i}{2\varepsilon}
		\left|
			\Vect{x}_\varsigma - \Mat{a}_{\varsigma \varsigma}^{-1} \Vect{X}_\varsigma
		\right|^2
	\right)
	,
	\label{eq:3_MTepsilon}
\end{align}

\noindent where I have defined 
\begin{align}
	g(\Vect{x}_\rho, \Vect{X})
	\doteq
	\frac{1}{2} \Vect{x}_\rho^\intercal \, \Mat{\Lambda}_{\rho \rho}^{-1} \Mat{a}_{\rho \rho} \, \Vect{x}_\rho
	- \Vect{x}_\rho^\intercal \, \Mat{M}_1 \, \Mat{L}_\textrm{s}^\intercal \Vect{X}
	+ \frac{1}{2} \Vect{X}^\intercal \, \Mat{L}_\textrm{s} \Mat{M}_2 \Mat{L}_\textrm{s}^\intercal \, \Vect{X}
	,
\end{align}

\noindent along with the matrices 
\begin{align}
	\Mat{M}_1
	\doteq
	\begin{pmatrix}
		\Mat{\Lambda}_{\rho \rho}^{-1} 
		& 
		-\Mat{\Lambda}_{\rho \rho}^{-1} \Mat{a}_{\rho \varsigma} \Mat{a}_{\varsigma \varsigma}^{-1} 
	\end{pmatrix}
	, \quad
	\Mat{M}_2
	\doteq
	\begin{pmatrix}
		\Mat{d}_{\rho \rho} \, \Mat{\Lambda}_{\rho \rho}^{-1}
		&
		\Mat{\Lambda}_{\rho \rho}^{-1} \Mat{d}_{\varsigma \rho}^\intercal
		\\
		\Mat{d}_{\varsigma \rho} \Mat{\Lambda}_{\rho \rho}^{-1}
		&
		\Mat{c}_{\varsigma \varsigma} \Mat{a}_{\varsigma \varsigma}^{-1} - \Mat{d}_{\varsigma \rho} \Mat{\Lambda}_{\rho \rho}^{-1} \Mat{a}_{\rho \varsigma} \Mat{a}_{\varsigma \varsigma}^{-1}
	\end{pmatrix}
	,
	\label{eq:3_M12}
\end{align}

\noindent and the vector decompositions
\begin{equation}
	\Mat{R}_\textrm{s}^\intercal\Vect{x}
	=
	\begin{pmatrix}
		\Vect{x}_\rho \\[1mm]
		\Vect{x}_\varsigma
	\end{pmatrix}
	, \quad
	\Mat{L}_\textrm{s}^\intercal \Vect{X}
	=
	\begin{pmatrix}
		\Vect{X}_\rho \\[1mm]
		\Vect{X}_\varsigma
	\end{pmatrix}
	.
	\label{eq:3_Qproj}
\end{equation}

\noindent Note that $\Mat{M}_2$ and $\Mat{\Lambda}_{\rho \rho}^{-1} \Mat{a}_{\rho \rho}$ are symmetric [\Eqs{eq:3_aSUBsymm}, \eq{eq:3_dSUBsymm}, and \eq{eq:3_cSUBsymm}]. Also note that $\Mat{M}_1$ is size $\rho \times N$, $\Mat{M}_2$ is size $N \times N$, and any vector $\Vect{\nu}_m$ is size $m \times 1$.

Then, using \Eq{eq:3_MTlim} along with
\begin{align}
	\lim_{\varepsilon \to 0} 
	\varepsilon^{-\varsigma/2}
	\exp
	\left(
		\frac{i}{2 \varepsilon}
		\left| 
			\Vect{x}_\varsigma
			- \Mat{a}_{\varsigma \varsigma}^{-1} \Vect{X}_\varsigma
		\right|^2
	\right)
	=
	(2\pi i)^{\varsigma/2} \,
	\delta
	\left(
		\Vect{x}_\varsigma
		- \Mat{a}_{\varsigma \varsigma}^{-1} \Vect{X}_\varsigma
	\right) ,
\end{align}

\noindent I obtain the limit of the MT kernel at $\det \Mat{B} \to 0$:
\begin{align}
	U(\Vect{X}, \Vect{x})
	= 
	\pm \frac
	{ 
		\exp
		\left[
			i \, g(\Vect{x}_\rho, \Vect{X})
		\right]
		\,
		\delta 
		\left( 
			\Vect{x}_\varsigma - \Mat{a}_{\varsigma \varsigma}^{-1} \Vect{X}_\varsigma
		\right)
	}
	{
		(2 \pi i)^{\rho/2} 
		\sqrt
		{
			\det \Mat{\Lambda}_{\rho \rho}
			\det \Mat{a}_{\varsigma \varsigma}
		}
	}
	,
	\label{eq:3_METkernB}
\end{align}

\noindent where, for brevity, I no longer mention the dependence of $M$ on the symplectic matrix explicitly.

Following straightforward delta-function manipulations, I also obtain the inverse MT kernel when $\det \Mat{B} = 0$:
\begin{equation}
	U^{-1}(\Vect{x}, \Vect{X})
	= 
	\pm \frac
	{
		\exp
		\left[
			- i \,
			\widetilde{g}(\Vect{X}_\rho, \Vect{x})
		\right]
		\, 
		\delta
		\left(
			\Vect{X}_\varsigma - \Mat{a}_{\varsigma \varsigma} \Vect{x}_\varsigma
		\right)
	}
	{
		(-2\pi i)^{\rho/2}
		\sqrt
		{
			\det \Mat{\Lambda}_{\rho \rho}
			\det \Mat{a}_{\varsigma \varsigma}^{-1}
		}
	}
	,
	\label{eq:3_invMETkernB}
\end{equation}

\noindent where I have defined 
\begin{align}
	\widetilde{g}(\Vect{X}_\rho, \Vect{x})
	\doteq
	\frac{1}{2} \Vect{X}_\rho^\intercal \, \Mat{d}_{\rho \rho} \Mat{\Lambda}_{\rho \rho}^{-1} \, \Vect{X}_\rho
	- \Vect{X}_\rho^\intercal \, \Mat{M}_3 \Mat{R}_\textrm{s}^\intercal \Vect{x}
	+ \frac{1}{2} \Vect{x}^\intercal \, \Mat{R}_\textrm{s} \Mat{M}_4 \Mat{R}_\textrm{s}^\intercal \, \Vect{x}
	,
\end{align}

\noindent along with the matrices
\begin{align}
	\Mat{M}_3 
	\doteq
	\begin{pmatrix}
		\Mat{\Lambda}_{\rho \rho}^{-1}
		&
		- \Mat{\Lambda}_{\rho \rho}^{-1} \Mat{d}_{\varsigma \rho}^\intercal \Mat{a}_{\varsigma \varsigma}
	\end{pmatrix}
	,
	\quad
	\Mat{M}_4
	\doteq
	\begin{pmatrix}
		\Mat{\Lambda}_{\rho \rho}^{-1} \Mat{a}_{\rho \rho}
		& \Mat{\Lambda}_{\rho \rho}^{-1} \Mat{a}_{\rho \varsigma}
		\\
		\Mat{a}_{\rho \varsigma}^\intercal \Mat{\Lambda}_{\rho \rho}^{-1}
		&
		\Mat{a}_{\varsigma \varsigma}^\intercal \Mat{c}_{\varsigma \varsigma} - \Mat{a}_{\varsigma \varsigma}^\intercal \Mat{d}_{\varsigma \rho} \Mat{\Lambda}_{\rho \rho}^{-1} \Mat{a}_{\rho \varsigma}
	\end{pmatrix}
	.
\end{align}

\noindent Note that $\Mat{M}_4$ and $\Mat{d}_{\rho \rho} \Mat{\Lambda}_{\rho \rho}^{-1}$ are symmetric [\Eqs{eq:3_aSUBsymm}, \eq{eq:3_dSUBsymm}, and \eq{eq:3_cSUBsymm}]. Also, $\Mat{M}_3$ is size $\rho \times N$ while $\Mat{M}_4$ is size $N \times N$. Lastly, I choose the following branch-cut convention: $\text{arg}(i) = \pi/2$ and $\text{arg}(\det \Mat{\Lambda}_{\rho \rho} \det\Mat{a}_{\varsigma \varsigma}) \in (-\pi, \pi]$ in \Eq{eq:3_METkernB}; $\text{arg}(-i) = -\pi/2$ and $\text{arg}(\det \Mat{\Lambda}_{\rho \rho} \det\Mat{a}_{\varsigma \varsigma}^{-1}) \in [-\pi, \pi)$ in \Eq{eq:3_invMETkernB}.


\subsection{Pseudo-differential representation of the metaplectic transform}
\label{sec:3_PMT}

Here, I develop a pseudo-differential representation of \Eq{eq:3_MTint}. (The prefix `pseudo' refers to the fact that this representation will formally involve an infinite series of derivative operators.) This representation is particularly useful when $\Mat{A}^{-1}\Mat{B}$ is small (or similarly, when $\Mat{S} \approx \IMat{2N}$), because then the MT can be approximated by a finite-order differential transform, which might easier to evaluate than the integral transform of \Eq{eq:3_MTint}. Specifically, I proceed as follows. Using the substitution $\Vect{u} \doteq \Vect{x} - \Mat{A}^{-1} \Vect{X}$, \Eq{eq:3_MTint} can be re-written as
\begin{align}
	\Psi(\Vect{X}) &= 
	\pm 
	\frac{
		\exp\left( \frac{i}{2} \Vect{X}^\intercal \Mat{C} \Mat{A}^{-1} \Vect{X} \right)
	}{
		(2\pi i)^{\frac{N}{2}} \sqrt{ \det{\Mat{B}} }
	}
	\int \dd\Vect{u} \, 
	\exp\left[
		\frac{i}{2} \Vect{u}^\intercal \left( \Mat{A}^{-1} \Mat{B} \right)^{-1} \Vect{u}
	\right] \, 
	\psi(\Mat{A}^{-1} \Vect{X} + \Vect{u}) 
	.
	\label{eq:3_NIMTtrans}
\end{align}

\noindent In the following, I shall assume that $\Mat{A}^{-1} \Mat{B}$ is small. This assumption is not strictly necessary, since the final result is convergent for all values of $\| \Mat{A}^{-1} \Mat{B} \|$ and thereby possesses a natural analytic continuation; however, it aides intuition in the forthcoming derivation.

Since $\Mat{A}^{-1} \Mat{B}$ is a symmetric matrix, by the spectral theorem it can be diagonalized. Let us enumerate with subscripts $j \in \{1,\ldots,N \}$ vector components with respect to the diagonalizing basis of $\Mat{A}^{-1} \Mat{B}$. Then,
\begin{align}
	\int \dd\Vect{u} \, 
	\exp\left[
		\frac{i}{2} \Vect{u}^\intercal \left( \Mat{A}^{-1} \Mat{B} \right)^{-1} \Vect{u}
	\right] \, 
	\psi(\Mat{A}^{-1} \Vect{X} + \Vect{u}) 
	= \int 
	\prod_{j = 1}^N \dd u_j \,
	\exp\left(i \frac{u_j^2}{2 \lambda_j } \right) \,
	\psi(\Mat{A}^{-1}\Vect{X} + \Vect{u}) 
	,
\end{align}

\noindent where $\lambda_j$ is the $j$-th eigenvalue of $\Mat{A}^{-1} \Mat{B}$. Since each $\lambda_j^{-1}$ is assumed large, only small values of $u$ will contribute to the integral of \Eq{eq:3_NIMTtrans}. Therefore, I can expand the function $\psi(\Mat{A}^{-1}\Vect{X} + \Vect{u})$ around $\Vect{u} = 0$ as
\begin{align}
	\psi(\Mat{A}^{-1}\Vect{X} + \Vect{u})
	= \sum_{n_1 = 0}^\infty \ldots \sum_{n_N = 0}^\infty 
	\frac{u_1^{n_1}}{n_1!}\ldots \frac{u_N^{n_N}}{n_N!} \, 
	\psi^{(n_1, \ldots , n_N)} (\Mat{A}^{-1}\Vect{X} ) \, ,
\end{align}

\noindent with the shorthand notation 
\begin{equation}
	\psi^{(n_1,\ldots , n_N)} (\Mat{A}^{-1}\Vect{X}) 
	\doteq 
	\left. 
		\frac{\partial^{n_1+ \ldots +n_N}}{\partial x_1^{n_1} \ldots \partial x_N^{n_N}} 
		\psi(\Vect{x}) 
	\right|_{\Vect{x} = \Mat{A}^{-1}\Vect{X}}
\end{equation}

\noindent denoting the derivatives of $\psi$ along the eigenvectors of $\Mat{A}^{-1}\Mat{B}$. (Here, I assume that $\psi$ is smooth, but I shall revisit this assumption below.) The integral therefore becomes

\begin{align}
	\int \dd\Vect{u} \, 
	\exp\left[
		\frac{i}{2} \Vect{u}^\intercal \left( \Mat{A}^{-1} \Mat{B} \right)^{-1} \Vect{u}
	\right] \, 
	\psi(\Mat{A}^{-1} \Vect{X} + \Vect{u})
	&\sim \sum_{n_1 = 0}^\infty \ldots \sum_{n_N = 0}^\infty 
	\frac{
		\psi^{(2n_1, \ldots , 2n_N)} (\Mat{A}^{-1}\Vect{X}) 
	}{
		\left(2n_1\right)! \ldots \left(2n_N\right)!
	} 
	\nonumber\\
	&\hspace{25mm}\times
	\prod_{j = 1}^N
	\int \dd u_j \,
	u_j^{2n_j}
	\exp\left(i \frac{u_j^2}{2 \lambda_j } \right)
	,
\end{align}

\noindent where the summation has been restricted to even integers by parity considerations, since all integrals with odd powers of $u_j$ are identically zero.

Let us introduce a dummy multiplicative variable $s$, which will eventually be taken to unity. Then, since
\begin{equation}
	\pd{s}^{n_j} \, \exp\left(
		i s \frac{u_j^2}{2 \lambda_j }
	\right) 
	= 
	\left(
		-2 i \lambda_j 
	\right)^{-n_j} u_j^{2n_j} 
	\exp\left(
		i s \frac{u_j^2}{2 \lambda_j }
	\right)  
	,
\end{equation}

\noindent I obtain
\begin{align}
	\int \dd u_j \,
	u_j^{2n_j}
	\exp\left(i \frac{u_j^2}{2 \lambda_j } \right)
	&= \frac{(2 \lambda_j)^{n_j}}{i^{n_j}} \left.
		\pd{s}^{n_j} \int_{-\infty}^{\infty} \dd u_j \, \exp\left(i s \frac{u_j^2}{2 \lambda_j } \right)
	\right|_{s=1}\nonumber\\
	&= \frac{(2 \lambda_j)^{n_j}}{i^{n_j}} 
	\left.
		\pd{s}^{n_j} \left(\sqrt{2 \pi i \lambda_j }~s^{-\frac{1}{2}}\right)
	\right|_{s=1}\nonumber\\
	&= \frac{(2 \lambda_j)^{n_j}}{i^{n_j}}  
	\sqrt{2 \pi i \lambda_j} \, 
	\frac{
		\Gamma\left(\frac{1}{2}\right)
	}{
		\Gamma\left(\frac{1}{2}-n_j\right)
	} 
	,
\end{align}

\noindent where the first line invokes Leibniz's rule, the final equality follows from the binomial theorem~\cite{Olver10a}, and $\Gamma(z)$ is the gamma function~\cite{Olver10a}. Hence, I obtain the asymptotic representation
\begin{align}
	&\int \dd\Vect{u} \, 
	\exp\left[
		\frac{i}{2} \Vect{u}^\intercal \left( \Mat{A}^{-1} \Mat{B} \right)^{-1} \Vect{u}
	\right] \, 
	\psi(\Mat{A}^{-1} \Vect{X} + \Vect{u})
	\nonumber\\
	&\sim 
	(2 \pi i)^{N/2} \sqrt{\det \Mat{A}^{-1}\Mat{B} }
	\prod_{j = 1}^N
	\sum_{n_j = 0}^\infty 
	\frac{(2 \lambda_j)^{n_j}}{i^{n_j}}   
	\frac{
		\Gamma\left(\frac{1}{2}\right)
		\psi^{(2n_1, \ldots , 2n_N)} (\Mat{A}^{-1}\Vect{X})
	}{
		\Gamma \left(2n_j + 1\right) \Gamma\left(\frac{1}{2}-n_j\right)
	}  
	.
	\label{eq:3_PMTunsimple}
\end{align}

Finally, using well-known properties of the gamma function yields the pseudo-differential representation of the MT:
\begin{subequations}
	\begin{align}
		\label{eq:3_PMTseries}
		\Psi(\Vect{X}) &= \pm
		\frac{
			\exp\left(
				\frac{i}{2} \Vect{X}^\intercal \Mat{C} \Mat{A}^{-1} \Vect{X}
			\right)
		}{
			\sqrt{\det{\Mat{A}}}
		} \, 
		\sum_{n_1=0}^\infty \ldots \sum_{n_N=0}^\infty 
		\frac{
			\left(i \lambda_1/2 \right)^{n_1} \ldots \left(i \lambda_N/2 \right)^{n_N}
		}{n_1! \ldots n_N!}
		\psi^{(2n_1, \ldots , 2n_N)} \left(\Mat{A}^{-1}\Vect{X}\right) 
		,
	\end{align}

	\noindent or symbolically,
	\begin{equation}
		\Psi(\Vect{X}) = \pm\left. 
			\frac{
				\exp\left(
					\frac{i}{2} \Vect{X}^\intercal \Mat{C} \Mat{A}^{-1} \Vect{X}
				\right)
			}{
				\sqrt{\det{\Mat{A}}}
			} 
			\, \exp\left(
				\frac{i}{2} \, \Mat{A}^{-1} \Mat{B} \dubdot \pd{\Vect{x} \Vect{x}}^2 
			\right) 
			\psi(\Vect{x} )
		\right|_{\Vect{x} = \Mat{A}^{-1}\Vect{X}} 
		,
		\label{eq:3_PMT}
	\end{equation}
\end{subequations}

\noindent where the notation $\dubdot$ denotes the double dot product, \ie $\Mat{A}^{-1} \Mat{B}:\pd{\Vect{x} \Vect{x}}^2 = (\Mat{A}^{-1} \Mat{B})_{ab}(\pd{x_a})(\pd{x_b})$ summed over common indices. I can also express \Eq{eq:3_PMT} in an equivalent vector form $\ket{\Psi} = \oper{M}(\Mat{S}) \ket{\psi}$, where $\oper{M}(\Mat{S})$ is the manifestly-unitary MT operator given as
\begin{equation}
	\oper{M}(\Mat{S}) \doteq \pm \sqrt{\det \Mat{A}^{-1}} \, 
	\oper{D}_{\Mat{A}} \, 
	\exp\left(
		\frac{i}{2} \VectOp{x}^\intercal \Mat{A}^\intercal \Mat{C} \VectOp{x}
	\right) \, 
	\exp\left(
		- \frac{i}{2} \VectOp{p}^\intercal \Mat{A}^{-1} \Mat{B} \VectOp{p}
	\right)
	,
	\label{eq:3_operMTold}
\end{equation}

\noindent where $\oper{D}_\Mat{A}$ is the non-unitary inverse dilation operator [not to be confused with the dispersion operator in \Eq{eq:2_hilbertWAVE}] defined via its effect in the spatial representation, $\bra{\Vect{x}}\oper{D}_{\Mat{A}} \ket{\psi} \doteq \psi(\Mat{A}^{-1}\Vect{x})$. This dilation action can also be affected using multidimensional quantum squeeze operators~\cite{Scully12} as
\begin{equation}
	\sqrt{\det \Mat{A}^{-1}}
	\oper{D}_{\Mat{A}}
	=
	\exp\left[ 
		i \frac{
			\VectOp{x}^\intercal \left(\log \Mat{A}^{-\intercal}\right) \VectOp{k} 
			+ \VectOp{k}^\intercal \left(\log \Mat{A}^{-1}\right) \VectOp{x}
		}{2}
	\right] 
	,
    \label{eq:3_dilOPER}
\end{equation}

\noindent as verified in \App{app:3_squeeze}. Hence, \Eq{eq:3_operMTold} can be written in the manifestly unitary form
\begin{align}
	\oper{M}(\Mat{S}) 
	&= \pm 
	\exp\left[
		i \frac{
			\VectOp{x}^\intercal \left(\log \Mat{A}^{-\intercal}\right) \VectOp{k} 
			+ \VectOp{p}^\intercal \left(\log \Mat{A}^{-1}\right) \VectOp{q}
		}{2}
	\right] 
	\exp\left(
		\frac{i}{2} \VectOp{x}^\intercal \Mat{A}^\intercal \Mat{C} \VectOp{x}
	\right)
	\, 
	\exp\left(
		- \frac{i}{2} \VectOp{k}^\intercal \Mat{A}^{-1} \Mat{B} \VectOp{k}
	\right)
	.
	\label{eq:3_operMT}
\end{align}

The three operator exponentials that enter \Eq{eq:3_operMT} can be understood as the individual MTs induced by the following block matrices:
\begin{align}
	\Mat{G} 
	\doteq
	\begin{pmatrix}
		\Mat{A} & \OMat{N}\\
		\OMat{N} & \Mat{A}^{-\intercal}
	\end{pmatrix}
	, \quad
	\Mat{L}
	\doteq
	\begin{pmatrix}
		\IMat{N} & \OMat{N} \\
		\Mat{A}^\intercal \Mat{C} & \IMat{N}
	\end{pmatrix}
	, \quad
	\Mat{U}
	\doteq
	\begin{pmatrix}
		\IMat{N} & \Mat{A}^{-1}\Mat{B} \\
		\OMat{N} & \IMat{N}
	\end{pmatrix}
	.
\end{align}

\noindent In this sense, \Eq{eq:3_operMT} is the decomposition of the MT induced by the following decomposition of the corresponding symplectic matrix $\Mat{S}$:
\begin{equation}
	\Mat{S}
	= \Mat{G} \Mat{L} \Mat{U}
	,
	\label{eq:3_decompS}
\end{equation}

\noindent which can be viewed as a modified LDU decomposition. By recognizing $\Mat{G}$, $\Mat{L}$, and $\Mat{U}$ as ray-transfer matrices~\cite{Kogelnik66} (see also \Sec{sec:3_EXparax}), a physical interpretation to \Eq{eq:3_operMT} is readily obtained: $\oper{M}(\Mat{S})$ represents the action of a paraxial optical system consisting of propagation in uniform media $\Mat{U}$ (generally anisotropic), followed by a thin lens $\Mat{L}$ (generally asymmetric), followed by magnification $\Mat{G}$.%
\footnote{A similar decomposition, along with the resulting pseudo-differential representation of the MT, appears in Ch.~9 of \Ref{Wolf79}, suggesting that \Eq{eq:3_operMT} may have been known prior to my publication of it in \Ref{Lopez19}. However, it appears that \Ref{Lopez19} was the first to recognize the potential of \Eq{eq:3_operMT} for use in fast MT algorithms, as further discussed in \Ch{ch:NIMT}.}

Although the above derivation assumes smooth $\psi$, the final result can be understood more generally, which is why the asymptotic relation has been replaced with an exact equality. As shown in \App{app:3_equivalence}, the operator \eq{eq:3_operMT} has exactly the same kernel as the original integral MT \eq{eq:3_MTint} and exists on the space of all functions which have a well-defined Fourier transform, \ie smoothness of $\psi$ is not required. In this sense, \Eq{eq:3_PMT} should be understood not as a symbolic representation of the series \eq{eq:3_PMTseries} (whose convergence depends on details of $\psi$) but rather as a symbolic representation of the integral MT \eq{eq:3_MTint}. This new representation is advantageous because it is compact, and it facilitates asymptotic expansions of the MT to any order of $\Mat{A}^{-1} \Mat{B}$, which is particularly useful when $\Mat{S}$ is near-identity.


\subsection{Metaplectic transforms of some example functions}


\subsubsection{Gaussian}

Let us denote by $\MT$ the MT operation. I consider the multidimensional function
\begin{equation}
	\psi(\Vect{x}) = \exp\left(\Vect{x}^\intercal \Mat{M} \Vect{x} + \Vect{x}^\intercal \Vect{v} \right)
\end{equation}

\noindent for some $\Mat{M}$ and $\Vect{v}$. Then, one computes
\begin{align}
	\MT\left[ \psi(\Vect{x}) \right] 
	&= \pm
	\frac{ 
		\exp\left(
			\frac{i}{2}\Vect{X}^\intercal \Mat{D} \Mat{B}^{-1} \Vect{X}
		\right)
	}{
		(2\pi i)^{\frac{N}{2}} \sqrt{\det{\Mat{B}}}
	} 
	\int \dd\Vect{x} \, 
	\exp \left[
		\frac{i}{2}\Vect{x}^\intercal \left(\Mat{B}^{-1}\Mat{A} -2i \Mat{M}\right) \Vect{x} + i\Vect{x}^\intercal \left(-\Mat{B}^{-1} \Vect{X} - i \Vect{v}\right)
	\right] 
	\nonumber\\
	&=
	\pm \frac{ 
		\exp\left[
			\frac{i}{2} \Vect{v}^\intercal \left(
				\Mat{B}^{-1}\Mat{A} 
				-2i \Mat{M} 
			\right)^{-1} \Vect{v}
		\right]
	}
	{
		\sqrt{\det\left(\Mat{A} -2i \Mat{B}\Mat{M}\right)}
	}
	\nonumber\\
	&\hspace{4mm}\times
	\exp\left\{
		\frac{i}{2}\Vect{X}^\intercal \left[
			\Mat{D}\Mat{B}^{-1} 
			- \left(
				\Mat{A}\Mat{B}^\intercal 
				- 2i\Mat{B}\Mat{M}\Mat{B}^\intercal 
			\right)^{-1} 
		\right]\Vect{X} 
		+ \Vect{v}^\intercal \left[
			\left(
				\Mat{A} 
				- 2i\Mat{B}\Mat{M}
			\right)^{-1} 
		\right]_\textrm{s} \Vect{X}
	\right\}
	,
\end{align}

\noindent where $[~]_\textrm{s}$ denotes the symmetric part. In particular, for a 1-D chirped Gaussian $\psi(x) = \exp\left[(i-1)x^2 \right]$, then
\begin{equation}
	\MT\left\{ \exp\left[(i-1)x^2 \right] \right\} = 
	\pm \frac{
		\exp\left\{
			\frac{i}{2}\left[ 
				\frac{D}{B} 
				- \frac{1}{AB + (2+2i)B^2} 
			\right] X^2
		\right\}
	}{
		\sqrt{A + (2 + 2i)B}
	}
	.
\end{equation}

\noindent Also, for a 1-D squeezed coherent state $\psi(x) = \exp\left[ -\alpha(x - x_0)^2 + ik_0 x \right]$, then
\begin{equation}
	\MT\left\{ \exp\left[ -\alpha(x - x_0)^2 + ik_0 x \right] \right\} 
	= \pm \frac{
		\exp\left[
			\frac{iD}{2B}X^2 
			- \frac{i}{2}\frac{\left(X/B + 2i \alpha x_0 - k_0 \right)^2}{A/B + 2i \alpha} -\alpha x_0^2 
		\right]
	}
	{
		\sqrt{A + 2i\alpha B}
	}
	.
\end{equation}


\subsubsection{Hermite function}

I consider the 1-D function
\begin{equation}
	\psi_m(x) = \frac{1}{\sqrt{2^m m! \sqrt{\pi}}} \exp\left(-\frac{x^2}{2} \right) H_m(x) 
	.
\end{equation}

\noindent The Hermite polynomials $H_m(x)$ have the exponential generator\cite{Olver10a}
\begin{equation}
	\exp\left( 2zx - z^2 \right) = \sum_{m=0}^\infty \frac{z^m}{m!}H_m(x) 
	,
\end{equation}

\noindent which implies that the Hermite functions have the exponential generator:
\begin{equation}
	\exp\left(2zx - z^2 - \frac{x^2}{2} \right) = \sum_{m=0}^\infty \sqrt{\frac{2^m \sqrt{\pi}}{m!}} z^m \psi_m(x) 
	.
	\label{eq:3_HermFuncGen}
\end{equation}

\noindent Applying the MT (with respect to $x$) to both sides yields the relation
\begin{equation}
	\MT\left[
		\exp\left( 2zx - z^2 - \frac{x^2}{2} \right)
	\right] = \sum_{m=0}^\infty \sqrt{\frac{2^m \sqrt{\pi}}{m!}} z^m \MT\left[ \psi_m(x) \right] \, .
	\label{eq:3_MetRotSum}
\end{equation}

I therefore proceed to compute
\begin{align}
	\MT \left[
		\exp\left( 2zx - z^2 - \frac{x^2}{2} \right) 
	\right] 
	&=
	\frac{
		\exp\left( \frac{iD}{2B}X^2 -z^2 \right)
	}{
		\sqrt{2\pi  i B}
	} \int \dd x \,
	\exp\left(
		\frac{i}{2}\frac{A + i B}{B} x^2 
		- i \frac{Q + 2 i B z}{B} x
	\right)
	\nonumber\\
	&=
	\frac{1}{\sqrt{A + iB}} \exp\left[\frac{iD}{2B}X^2 -\frac{iB}{2A + 2iB}\left(\frac{Q + 2 i B z}{B} \right)^2 - z^2 \right]
	\nonumber\\
	&= \frac{
		\exp\left(-i\phi/2 \right)
	}{
		\left(A^2 + B^2\right)^{1/4}
	} \exp\left[
		2 \left(ze^{-i\phi} \right) \left(\frac{X}{\sqrt{A^2 + B^2}}\right) 
		- \left(ze^{-i\phi}\right)^2 
		\right.\nonumber\\
		&\left.\hspace{50mm}
		- \left( \frac{1 - iAC - i DB}{A^2 + B^2} \right)\frac{X^2}{2}
	\right]
	,
\end{align}

\noindent where $\phi = \text{Arg}\left(A + i B \right)$. Applying \Eq{eq:3_HermFuncGen} to the right-hand side of this expression, I can therefore identify
\begin{align}
	\MT \left[
		\exp\left( 2zx - z^2 - \frac{x^2}{2} \right)
	\right] &=
	\sum_{m = 0}^\infty \sqrt{\frac{2^m \sqrt{\pi}}{m!}} 
	z^m e^{-i \left(m + 1/2 \right)\phi} 
	\psi_m\left( \frac{X}{\sqrt{A^2 + B^2}} \right) 
	\nonumber\\
	&\hspace{50mm}\times
	\frac{
		\exp\left( i\frac{AC + DB}{2A^2 + 2B^2}X^2 \right)
	}{
		\left(A^2 + B^2\right)^{1/4}
	}
	.
\end{align}

\noindent Comparing with \Eq{eq:3_MetRotSum}, and using the completeness of the monomial basis, I conclude that
\begin{equation}
	\MT\left[ 
		\psi_m(x) 
	\right] = 
	\frac{
		\exp\left[ i\frac{AC + DB}{2A^2 + 2B^2}X^2 -i \left(m + \frac{1}{2} \right)\phi \right]
	}{
		\left(A^2 + B^2\right)^{1/4}
	}
	\, \psi_m\left( \frac{X}{\sqrt{A^2 + B^2}} \right) 
	.
\end{equation}

\noindent In particular, for phase space rotations, \ie $A = D = \cos(\theta)$, $B = -C = \sin(\theta)$, $AC + BD = 0$, $A^2 + B^2 = 1$, and $\phi = \theta$, thus yielding an eigenvalue relationship.


\subsubsection{Airy function}

I consider the $1$-D function
\begin{equation}
	\psi(x) = \frac{1}{\sqrt{2\pi i}} \exp\left( \frac{i}{3}x^3 - i h x \right) 
	,
\end{equation}

\noindent which is proportional to the well-known Fourier transform of the Airy function (shifted by $h$). I compute
\begin{align}
	\MT\left[ \psi(x) \right] &= 
	\pm\frac{ 
		\exp\left( \frac{i D}{2B}X^2 \right)
	}{
		2\pi i \sqrt{B}
	} \int \dd x \,
	\exp\left[
		\frac{i}{3}x^3 
		+ \frac{iA}{2B}x^2 
		- i \left(\frac{X}{B} + h\right) x
	\right]
	\nonumber\\
	&=
	\pm\frac{ 
		\exp\left[
			\frac{i D}{2B}X^2
			+ \frac{iA^3}{12B^3} 
			+ \frac{iA}{2B^2}X 
			+ i\frac{A}{2B}h
		\right]
	}{
		2\pi i \sqrt{B}
	} 
	\nonumber\\
	&\hspace{50mm}\times
	\int \dd x' \, 
	\exp\left[
		\frac{i}{3}\left(x' \right)^3 - i\left(\frac{A^2}{4B^2} + \frac{X}{B} + h\right)x'
	\right]
	.
\end{align}

\noindent where I have made the variable substitution $x' = x + \frac{A}{2B}$. Finally, I use Airy's integral,
\begin{equation}
	2\pi \airyA(x) =  \int \dd y \, e^{\frac{i}{3} y^3 + i x y} 
	,
\end{equation}

\noindent to conclude that
\begin{equation}
	\MT\left[ \psi(x) \right] = \pm 
	\frac{1}{i \sqrt{B}} 
	\exp\left(
		\frac{i D}{2B}X^2 
		+ \frac{iA}{2B^2}X 
		+ \frac{iA^3}{12B^3} 
		+ \frac{iA}{2B}h 
	\right)
	\airyA\left(- \frac{X}{B} - \frac{A^2}{4B^2} - h\right) 
	.
\end{equation}

\noindent Hence, the MT of the Airy function corresponding to $\Mat{S}$ is given by the above equation with symplectic matrix $-\Mat{S}\Mat{J}$ (Since $\Mat{J}$ effects a Fourier transform). Explicitly, this is
\begin{equation}
	\MT\left[Ai(x) \right] = 
	\pm \frac{1}{\sqrt{A}} 
	\exp\left(
		\frac{i C}{2A}X^2 
		+ \frac{iB}{2A^2}X 
		- \frac{iB^3}{12A^3} 
		-\frac{iB}{2A} h
	\right) 
	Ai\left(\frac{X}{A} - \frac{B^2}{4A^2} - h\right) 
	.
\end{equation}


\section{Case study: time evolution of the quantum harmonic oscillator}
\label{sec:3_SecMETQHO}

To better understand what the MT is, let us consider an elementary problem from quantum mechanics, namely, the quantum harmonic oscillator (QHO). The QHO is described by the Schr\"odinger equation
\begin{equation}
	i\pd{t} \ket{\psi_t} = \oper{H}\ket{\psi_t} 
	,
	\quad
	\oper{H} \doteq (\oper{k}^2 + \oper{x}^2 )/2 
	.
	\label{eq:3_QHO}
\end{equation}

\noindent Equation \eq{eq:3_QHO} has the solution $\ket{\psi_t} = \oper{M}_t \ket{\psi_0}$, where $\ket{\psi_0}$ is an initial wavefunction and the propagator $\oper{M}_t$ is a unitary operator given by
\begin{equation}
	\oper{M}_t = \exp\left(-i\oper{H}t\right) 
	.
	\label{eq:3_metOPERATOR}
\end{equation}

An interesting property of $\oper{M}_t$ is revealed by switching from the Schr\"odinger representation to the Heisenberg representation, in which the wavefunction is fixed but $\oper{x}$ and $\oper{k}$ evolve in time as governed by~\cite{Shankar94}
\begin{subequations}
	\label{eq:3_HeisEQ}
	\begin{align}
		\pd{t} (\oper{M}_t^\dagger \oper{x} \oper{M}_t ) &= i \oper{M}_t^\dagger \left[\oper{H},\oper{x} \right] \oper{M}_t = \oper{M}_t^\dagger \oper{k} \oper{M}_t 
		,\\
		\pd{t} (\oper{M}_t^\dagger \oper{k} \oper{M}_t ) &= i \oper{M}_t^\dagger \left[\oper{H},\oper{k} \right] \oper{M}_t = -\oper{M}_t^\dagger \oper{x} \oper{M}_t 
		.
	\end{align}
\end{subequations}

\noindent The coordinate and momentum operators of the QHO are seen to satisfy the same Hamilton's equations that describe a classical harmonic oscillator~\cite{Goldstein02}. The solution to \Eqs{eq:3_HeisEQ} is therefore
\begin{equation}
	\oper{X} = +\cos(t) \oper{x} + \sin(t) \oper{k} \, ,
	\quad
	\oper{K} = -\sin(t) \oper{x} + \cos(t) \oper{k} \, ,
	\label{eq:3_HeisSOL}
\end{equation}

\noindent where I introduced [analogous to \Eq{eq:3_operTRANS}]
\begin{equation}
	\oper{X} \doteq \oper{M}_t^\dagger \oper{x} \oper{M}_t 
	,
	\quad
	\oper{K} \doteq \oper{M}_t^\dagger \oper{k} \oper{M}_t 
	.
\end{equation}

\noindent Equations \eq{eq:3_HeisSOL} can be considered as a mapping $(\oper{x},\oper{k}) \mapsto (\oper{X},\oper{k})$ which is a phase-space rotation by angle $t$. The unitary propagator $\oper{M}_t$ that effects this rotation is called a metaplectic operator.%
\footnote{Here, $\oper{M}_t$ also acts as the \textit{fractional Fourier transform} operator, up to a phase.}

Finally, let us notice the following. As is well-known, the eigenvalues of the QHO Hamiltonian are~\cite{Shankar94}
\begin{equation}
	\oper{H}\ket{n} = (n + 1/2)\ket{n} 
	,
	\label{eq:3_QHOmet}
\end{equation}

\noindent with $n$ an integer and $\ket{n}$ the $n$-th eigenstate of $\oper{H}$; hence, the specific MT considered in \Eq{eq:3_metOPERATOR} can also be represented as
\begin{equation}
	\oper{M}_t = \exp\left(-it/2 \right) \sum_{n=0}^\infty \exp\left(-int \right)\ket{n}\bra{n} 
	.
	\label{eq:3_metDIAG}
\end{equation}

\noindent A notable aspect of this formula is that it takes not one but \textit{two} rotation periods ($t=4\pi$) for $\oper{M}_t$ to return to its original value $\oper{M}_0 = \IdentOp$. More generally, $\oper{M}_{2\pi n} = \IdentOp$ for even $n$ yet $\oper{M}_{2\pi n} = -\IdentOp$ for odd $n$. Hence, the same identity transformation on phase space [governed by \Eq{eq:3_HeisSOL}] can be effected by two distinct metaplectic operators, $\pm \IdentOp$. This double-valuedness also holds for arbitrary rotation angles, and is in fact a general property of the MT, as was discussed in \Sec{sec:3_MT} and illustrated in \Fig{fig:3_riemann}.


\section{Case study: paraxial propagation}
\label{sec:3_EXparax}


\subsection{General expression and special cases}

As in \Ch{ch:GO}, let us consider an $(N + 1)$-D system where the optical axis is aligned with $z$ and the transverse $N$-D coordinates are denoted by $\Vect{x}$. Suppose a wavefield $E(\Vect{x}, z, t)$ propagates according to the Helmholtz-type equation
\begin{equation}
	\pd{z}^2 E(\Vect{x}, z, t) 
	+ \pd{\Vect{x}}^2 E(\Vect{x}, z, t) 
	- \frac{\epsilon(\Vect{x},z)}{c^2} \pd{t}^2 E(\Vect{x}, z, t) = 0
	,
	\label{eq:3_waveEQ}
\end{equation}

\noindent where $\epsilon(\Vect{x},z)$ is the medium dielectric function, assumed to be time-independent. Let us partition the wavefield as
\begin{equation}
	E(\Vect{x}, z, t) = \psi(\Vect{x},z) \exp(i k z - i \omega t)
	.
	\label{eq:3_paraxE}
\end{equation}

\noindent Then, \Eq{eq:3_waveEQ} becomes
\begin{equation}
	\pd{z}^2 \psi(\Vect{x}, z)
	+ 2 i k \, \pd{z} \psi(\Vect{x}, z)
	- k^2 \psi(\Vect{x}, z) 
	+ \pd{\Vect{x}}^2 \psi(\Vect{x}, z) 
	+ \frac{\omega^2}{c^2} \epsilon(\Vect{x},z) \psi(\Vect{x}, z) = 0
	.
\end{equation}

\noindent If I assume that $\psi$ varies slowly with respect to $z$ compared to $\exp(i k z)$, then the first term is negligible and I arrive at the standard paraxial equation:
\begin{equation}
	2 i k \, \pd{z} \psi(\Vect{x}, z)
	+ \pd{\Vect{x}}^2 \psi(\Vect{x}, z) 
	+ \left[
		\frac{\omega^2}{c^2} \epsilon(\Vect{x},z) 
		- k^2
	\right]	
	\psi(\Vect{x}, z) = 0
	.
	\label{eq:3_paraxial}
\end{equation}

Equation \eq{eq:3_paraxial} has a Schrodinger form; if I assume that $\epsilon(\Vect{x}, z)$ is independent of $z$, then the solution can formally be written as
\begin{equation}
	\psi(\Vect{x}, z) = 
	\exp\left\{
		i \frac{z}{2 k} 
		\left[
			\pd{\Vect{x}}^2
			+ \frac{\omega^2}{c^2} \epsilon(\Vect{x}) 
			- k^2
		\right]
	\right\} \psi(\Vect{x},0)
	.
	\label{eq:3_paraxSOL}
\end{equation}

\noindent Let us now consider some special cases.


\subsubsection{Uniform medium}

First, let us consider the case when the medium is uniform and 
\begin{equation}
	\epsilon(\Vect{x}) = \epsilon_0
\end{equation}

\noindent for some constant $\epsilon_0 > 0$. Equation \eq{eq:3_paraxSOL} accordingly becomes
\begin{align}
	\psi(\Vect{x}, z) &= 
	\exp\left[ i \frac{z}{2 k} 
		\left( \frac{\omega^2}{c^2} \epsilon_0 - k^2\right)
	\right]
	\exp\left(
		i \frac{z}{2 k} 
		\pd{\Vect{x}}^2
	\right) \psi(\Vect{x},0)
	\nonumber\\
	&\equiv
	\exp\left[ i \frac{z}{2 k} 
		\left( \frac{\omega^2}{c^2} \epsilon_0 - k^2\right)
	\right]
	\oper{M}(\Mat{S}_\textrm{prop})
	\psi(\Vect{x},0)
	,
	\label{eq:3_freePMT}
\end{align}

\noindent where I have used \Eq{eq:3_PMT} to introduce the metaplectic operator $\oper{M}(\Mat{S}_\textrm{prop})$ corresponding to the symplectic matrix
\begin{equation}
	\Mat{S}_\textrm{prop} = 
	\begin{pmatrix}
		\IMat{N} & \frac{z}{k} \IMat{N} \\
		\OMat{N} & \IMat{N}
	\end{pmatrix}
	,
	\label{eq:3_freeS}
\end{equation}

\noindent which is the well-known ray-transfer matrix for paraxial propagation in uniform medium~\cite{Kogelnik66}. Instead of the pseudo-differential representation that underlies \Eq{eq:3_freePMT}, I can be equivalently use the integral representation of the MT provided in \Eq{eq:3_MTint} to obtain:
\begin{equation}
	\psi(\Vect{x}, z) =
	\left(
		\frac{k}{2 \pi i z}
	\right)^{N/2}
	\exp \left[
		i \frac{z}{2 k} 
		\left( \frac{\omega^2}{c^2} \epsilon_0 - k^2\right)
	\right]
	\int \dd \Vect{y} \,
	\psi(\Vect{y}, 0) 
	\exp\left[ 
		i \frac{k}{2z} (\Vect{x} - \Vect{y})^\intercal (\Vect{x} - \Vect{y}) 
	\right]
	.
	\label{eq:3_freeMT}
\end{equation}

\noindent Equation \eq{eq:3_freeMT} can be recognized as the well-known Fresnel diffraction formula, a connection which was noted in \Ch{ch:GO}.


\subsubsection{Linearly varying medium}

Let us next consider the case when the medium varies linearly with $\Vect{x}$ such that 
\begin{equation}
	\epsilon = \epsilon_0 + \Vect{\kappa}^\intercal \Vect{x}
\end{equation}

\noindent for some constant scalar $\epsilon_0 > 0$ and constant vector $\Vect{\kappa}$. Equation \eq{eq:3_paraxSOL} accordingly becomes
\begin{equation}
	\psi(\Vect{x}, z)
	= 
	\exp\left[ i \frac{z}{2 k} 
		\left( \frac{\omega^2}{c^2} \epsilon_0 - k^2\right)
	\right]
	\exp\left(
		i \frac{z}{2 k} 
		\pd{\Vect{x}}^2
		+ i \frac{z}{2 k} \frac{\omega^2}{c^2} \Vect{\kappa}^\intercal \Vect{x}
	\right) \psi(\Vect{x},0)
	.
	\label{eq:3_linearPMT}
\end{equation}

\noindent Using well-known operator manipulations, namely the BCH formula, the exponential operator can be recast as
\begin{align}
	\exp\left(
		i \frac{z}{2 k} 
		\pd{\Vect{x}}^2
		+ i \frac{z}{2 k} \frac{\omega^2}{c^2} \Vect{\kappa}^\intercal \Vect{x}
	\right)
	=
	\exp\left( i \frac{z^3}{48 k^3} \frac{\omega^4}{c^4}\Vect{\kappa}^\intercal \Vect{\kappa}\right) 
	\oper{T}(\Vect{\zeta}_\textit{lin}) \oper{M}(\Mat{S}_\textrm{prop})
	,
	\label{eq:3_inhomMT}
\end{align}

\noindent where I have introduced the ($\Vect{x}$-space representation) Heisenberg operator $\oper{T}(\Vect{\zeta}_0)$ that effects a phase-space translation by $\Vect{\zeta}_0 = (\Vect{x}_0, \Vect{k}_0)$ as
\begin{equation}
	\oper{T}(\Vect{\zeta}_0) \doteq \exp(i \Vect{k}_0^\intercal \Vect{x} - \Vect{x}_0^\intercal \pd{\Vect{x}} )
	\equiv
	\exp\left(
		-\frac{i}{2} \Vect{k}_0^\intercal \Vect{x}_0
	\right)
	\exp(i \Vect{k}_0 \Vect{x})
	\exp( - \Vect{x}_0^\intercal \pd{\Vect{x}} )
	,
\end{equation}

\noindent and I have defined the specific displacement vector for paraxial propagation in linearly-varying medium:
\begin{equation}
	\Vect{\zeta}_\textrm{lin} \doteq
	\begin{pmatrix}
		\frac{z^2}{4 k^2} \frac{\omega^2}{c^2} \Vect{\kappa} \\
		\frac{z}{2 k} \frac{\omega^2}{c^2} \Vect{\kappa}
	\end{pmatrix}
	.
\end{equation}

\noindent Note that \Eq{eq:3_inhomMT} represents an \textit{inhomogeneous} metaplectic operator rather than a standard metaplectic operator, due to having linear terms in $\Vect{x}$ and $\pd{\Vect{x}}$. An inhomogeneous metaplectic operator has the general form $\exp(i \gamma) \oper{T}(\Vect{\zeta}_0) \oper{M}(\Mat{S})$ for some scalar $\gamma$, phase-space displacement vector $\Vect{\zeta}_0$, and symplectic matrix $\Mat{S}$~\cite{Littlejohn86a}. These operators form a group (the \textit{inhomogeneous} metaplectic group), just as the standard metaplectic operators form a group.

Hence, \Eq{eq:3_linearPMT} can also be written as
\begin{equation}
	\psi(\Vect{x}, z)
	= 
	\exp\left[ i \frac{z}{2 k} 
		\left( \frac{\omega^2}{c^2} \epsilon_0 
			+ \frac{\omega^2}{c^2} \Vect{\kappa}^\intercal \Vect{x}
			- k^2
		\right)
		- i \frac{z^3}{24 k^3} \frac{\omega^4}{c^4}
		\Vect{\kappa}^\intercal \Vect{\kappa}
	\right]
	\Psi\left(
		\Vect{x} - \frac{z^2}{4 k^2} \frac{\omega^2}{c^2} \Vect{\kappa}
		, z
	\right)
	,
	\label{eq:3_linearMT}
\end{equation}

\noindent where I have used the fact that $\exp\left(- \Vect{v}^\intercal \pd{\Vect{x}} \right)$ is simply a translation operator (specifically, translate $\Vect{x}$ by $\Vect{v}$), and I have defined $\Psi(\Vect{x}, z)$ as the Fresnel transform of $\psi(\Vect{x}, 0)$:
\begin{equation}
	\Psi(\Vect{x},z)
	\doteq
	\left(
		\frac{k}{2 \pi i z}
	\right)^{N/2}
	\int \dd \Vect{y} \,
	\psi(\Vect{y}, 0) 
	\exp\left[ 
		i \frac{k}{2z} (\Vect{x} - \Vect{y})^\intercal (\Vect{x} - \Vect{y})
	\right]
	.
	\label{eq:3_linPSI}
\end{equation}

\noindent Note that the inhomogeneous linear symplectic transformation of the transverse ray phase space that underlies \Eq{eq:3_linearMT} is readily computed to be
\begin{equation}
	\Vect{\zeta}(z) = \Mat{S}_\textrm{prop} \, \Vect{\zeta}(0) + \Vect{\zeta}_\textrm{lin}
	.
\end{equation}


\subsubsection{Cascaded media}

As just discussed, paraxial propagation in uniform media is governed by the symplectic matrix $\Mat{S}_\textrm{prop}$ given in \Eq{eq:3_freeS}. A more general paraxial optical setup will be governed by an arbitrary symplectic matrix $\Mat{S}$, and the composition (cascade) of multiple paraxial optical setups will be governed by the product of the constituent governing symplectic matrices, \ie $\Mat{S}_\textrm{tot} = \prod \Mat{S}_j$ with $\Mat{S}_j$ being the governing symplectic matrix of the $j$-th paraxial optical setup in the cascade.

For example, the action of a thin ideal spherical lens in the paraxial limit is governed by the symplectic matrix
\begin{equation}
	\Mat{S}_\textrm{lens}
	=
	\begin{pmatrix}
		\IMat{N} & \OMat{N} \\
		-\frac{k}{f} \, \IMat{N} & \IMat{N}
	\end{pmatrix}
	,
\end{equation}

\noindent where $f$ is the focal distance and $k$ is the wavenumber introduced in \Eq{eq:3_paraxE}. Hence, propagation in free space over a distance $d$ to a thin lens of focal length $f$ followed by free-space propagation over a distance $d'$ is governed by
\begin{equation}
	\Mat{S}_\textrm{p-l-p}
	=
	\Mat{S}_\textrm{prop}(d') \, \Mat{S}_\textrm{lens} \, \Mat{S}_\textrm{prop}(d)
	= 
	\begin{pmatrix}
		(1 - \frac{d'}{f}) \, \IMat{N} & [\frac{d'}{k} + \frac{d}{k}( 1 - \frac{d'}{f})] \, \IMat{N} \\
		-\frac{k}{f} \, \IMat{N} & (1 - \frac{d}{f}) \, \IMat{N}
	\end{pmatrix}
	.
	\label{eq:3_plpS}
\end{equation}

\noindent The corresponding integral MT is given via \Eq{eq:3_MTint} as
\begin{align}
	\psi(\Vect{x},d + d'; f) = &\pm 
	\frac{1}{ (2 \pi i)^{N/2} }
	\left[\frac{k f}{ f d' + d(f - d')} \right]^{N/2}
	\,
	\exp\left[
		i \frac{
			k (f - d) \, \Vect{x}^\intercal \Vect{x}
		}{
			2 f d' + 2 d(f - d')
		} 
	\right]
	\nonumber\\
	&\times
	\,
	\int \dd \Vect{y} \,
	\psi(\Vect{y},0) 
	\exp\left[ 
		i \frac{
			k (f - d') \, \Vect{y}^\intercal \Vect{y}
		}{
			2 f d' + 2 d(f - d')
		} 
		- i \frac{
			k f \, \Vect{x}^\intercal \Vect{y} 
		}{
			f d' + d(f - d')
		}
	\right]
	.
	\label{eq:3_plpMT}
\end{align}

\noindent Well-known special cases, such as the optical Fourier transformer ($d = d' = f$), can readily be derived from \Eq{eq:3_plpMT}.


\subsection{Connection to far-field diffraction theory}

As noted in the previous section, the MT corresponding to paraxial propagation in a uniform medium is equivalent to the Fresnel diffraction formula. Indeed, introducing the wavelength $\lambda = 2\pi/k$ and setting it such that the phase detuning vanishes, \Eq{eq:3_freeMT} can be written as
\begin{equation}
	\psi(\Vect{x},z)
	=
	\int \frac{
		\dd \Vect{y}
	}{ 
		( i \lambda z)^{N/2} 
	} \,
	\psi(\Vect{y},0) 
	\exp\left[ i \frac{\pi }{\lambda z} (\Vect{x} - \Vect{y})^\intercal (\Vect{x} - \Vect{y}) \right]
	.
	\label{eq:3_Fresnel}
\end{equation}

\noindent This is the Fresnel diffraction formula~\cite{Born99}. I emphasize that no approximation regarding the propagation length (`far-field' approximation) has been made. It is well known that in the `far-field' limit, the Fresnel diffraction formula reduces to the Fraunhofer diffraction formula, in which the diffracted field is the Fourier transform of the initial field. 

From a matrix perspective, this far-field limit is somewhat odd because it requires the block-triangular matrix $\Mat{S}_\textrm{prop}$ [\Eq{eq:3_freeS}] that underlies Fresnel diffraction to become proportional to the block-antidiagonal matrix $\JMat{2N}$ that generates the FT. In other words, matrix elements that are exactly zero must be approximated as being non-zero in a transparent manner, and it is not obvious how to do this \textit{a priori}. Fortunately, it becomes clear how this limit manifests after performing a horizontal polar decomposition of $\Mat{S}_\textrm{prop}$. Recall that the horizontal polar decomposition of a symplectic matrix can be written as~\cite{Qin19}
\begin{equation}
	\Mat{S}
	=
	\begin{pmatrix}
		\Mat{A} & \Mat{B} \\
		\Mat{C} & \Mat{D}
	\end{pmatrix}
	=
	\begin{pmatrix}
		\IMat{N} & \OMat{N} \\
		( \Mat{C} \Mat{A}^\intercal + \Mat{D} \Mat{B}^\intercal ) \Mat{A}_0^{-1} \Mat{A}_0^{-1} & \IMat{N}
	\end{pmatrix}
	\begin{pmatrix}
		\Mat{A}_0 & \OMat{N} \\
		\OMat{N} & \Mat{A}_0^{-1}
	\end{pmatrix}
	\begin{pmatrix}
		\Mat{A}_0^{-1} \Mat{A} & \Mat{A}_0^{-1} \Mat{B} \\
		- \Mat{A}_0^{-1} \Mat{B} & \Mat{A}_0^{-1} \Mat{A}
	\end{pmatrix}
	,
\end{equation}

\noindent where $\Mat{A}_0 \doteq \sqrt{\Mat{A} \Mat{A}^\intercal + \Mat{B} \Mat{B}^\intercal}$, with the square root corresponding to the unique positive square root. It is also useful to first non-dimensionalize $\Mat{S}_\textrm{prop}$ as
\begin{equation}
	\Mat{S}_\textrm{prop} = 
	\begin{pmatrix}
		W \, \IMat{N} & \OMat{N} \\
		\OMat{N} & \frac{1}{W} \, \IMat{N}
	\end{pmatrix}
	\fourier{\Mat{S}}_\textrm{prop}
	\begin{pmatrix}
		\frac{1}{W} \, \IMat{N} & \OMat{N} \\
		\OMat{N} & W \, \IMat{N}
	\end{pmatrix}
	,
	\quad
	\fourier{\Mat{S}}_\textrm{prop}
	=
	\begin{pmatrix}
		\IMat{N} & \frac{1}{\mc{F}} \, \IMat{N} \\
		\OMat{N} & \IMat{N}
	\end{pmatrix}
	.
\end{equation}

\noindent Here, $W$ corresponds to a typical scale length of $\Vect{x}$, or equivalently, of $\psi(\Vect{x},0)$, \eg the width of an aperture. Also, I have introduced the Fresnel number, defined as
\begin{equation}
	\mc{F} = \frac{2 \pi W^2}{\lambda z}
	.
\end{equation}

\noindent Performing the horizontal polar decomposition of $\fourier{\Mat{S}}_\textrm{prop}$ then yields
\begin{equation}
	\fourier{\Mat{S}}_\textrm{prop} =
	\begin{pmatrix}
		\IMat{N} & \OMat{N} \\
		\frac{\mc{F} }{1 + \mc{F}^2} \, \IMat{N} & \IMat{N}
	\end{pmatrix}
	\begin{pmatrix}
		\frac{\sqrt{1 + \mc{F}^2}}{\mc{F}} \, \IMat{N} & \OMat{N} \\
		\OMat{N} & \frac{\mc{F}}{\sqrt{1 + \mc{F}^2} } \IMat{N}
	\end{pmatrix}
	\begin{pmatrix}
		\frac{\mc{F}}{\sqrt{1 + \mc{F}^2} } \, \IMat{N} & \frac{1}{\sqrt{1 + \mc{F}^2} } \, \IMat{N} \\
		-\frac{1}{\sqrt{1 + \mc{F}^2} } \, \IMat{N} & \frac{\mc{F}}{\sqrt{1 + \mc{F}^2} } \, \IMat{N}
	\end{pmatrix}
	.
\end{equation}

I next take the limit $\mc{F} \to 0$, which is the traditional definition of `far-field'. In doing this, I must be sure that each matrix factor in the decomposition of $\fourier{\Mat{S}}_\textrm{prop}$ remains symplectic. To lowest order, this is
\begin{align}
	\fourier{\Mat{S}}_\textrm{prop} \approx
	\begin{pmatrix}
		\IMat{N} & \OMat{N} \\
		\mc{F} \, \IMat{N} & \IMat{N}
	\end{pmatrix}
	\begin{pmatrix}
		\frac{1}{\mc{F}} \, \IMat{N} & \OMat{N} \\
		\OMat{N} & \mc{F} \, \IMat{N}
	\end{pmatrix}
	\begin{pmatrix}
		\OMat{N} & \IMat{N} \\
		- \IMat{N} & \OMat{N}
	\end{pmatrix}
	=
	\begin{pmatrix}
		\OMat{N} & \frac{1}{\mc{F}} \, \IMat{N} \\
		- \mc{F} \, \IMat{N} & \IMat{N}
	\end{pmatrix}
	,
\end{align}

\noindent which implies that in the far-field,
\begin{equation}
	\Mat{S}_\textrm{prop} \approx
	\begin{pmatrix}
		\OMat{N} & \frac{z \lambda}{2 \pi} \, \IMat{N} \\
		- \frac{2 \pi }{\lambda z} \, \IMat{N} & \IMat{N}
	\end{pmatrix}
	.
	\label{eq:3_ffS1}
\end{equation}

\noindent Note, however, that I've keep the term $\mc{F} \, \IMat{N}$ in the first matrix factor of $\fourier{\Mat{S}}_\textrm{prop}$ even though it is not lowest order nor is it necessary to maintain symplecticity. Hence, one can equally well make the additional simplification
\begin{align}
	\fourier{\Mat{S}}_\textrm{prop} \approx
	\begin{pmatrix}
		\IMat{N} & \OMat{N} \\
		\OMat{N} & \IMat{N}
	\end{pmatrix}
	\begin{pmatrix}
		\frac{1}{\mc{F}} \, \IMat{N} & \OMat{N} \\
		\OMat{N} & \mc{F} \, \IMat{N}
	\end{pmatrix}
	\begin{pmatrix}
		\OMat{N} & \IMat{N} \\
		- \IMat{N} & \OMat{N}
	\end{pmatrix}
	=
	\begin{pmatrix}
		\OMat{N} & \frac{1}{\mc{F}} \, \IMat{N} \\
		- \mc{F} \, \IMat{N} & \OMat{N}
	\end{pmatrix}
	,
\end{align}

\noindent which yields
\begin{equation}
	\Mat{S}_\textrm{prop} \approx
	\begin{pmatrix}
		\OMat{N} & \frac{\lambda z}{2 \pi} \, \IMat{N} \\
		- \frac{2 \pi}{\lambda z} \, \IMat{N} & \OMat{N}
	\end{pmatrix}
	\label{eq:3_ffS2}
	.
\end{equation}

\noindent Both \Eqs{eq:3_ffS1} and \eq{eq:3_ffS2} are consistent `far-field' approximations, although \Eq{eq:3_ffS1} is slightly more accurate.

The metaplectic transform corresponding to the approximate $\Mat{S}_\textrm{prop}$ of \Eq{eq:3_ffS1} is given by
\begin{equation}
	\psi(\Vect{x},z)
	=
	\exp \left( i \frac{\pi}{\lambda z} \Vect{x}^\intercal \Vect{x} \right)
	\int 
	\frac{
		\dd \Vect{y}
	}{
		(i \lambda z)^{N/2}
	} \,
	\psi(\Vect{y}, 0) \exp \left( - i \frac{2 \pi }{\lambda z} \Vect{x}^\intercal \Vect{y} \right)
	.
	\label{eq:3_ffMT1}
\end{equation}

\noindent Similarly, the metaplectic transform corresponding to the approximate $\Mat{S}_\textrm{prop}$ of \Eq{eq:3_ffS2} is given by
\begin{equation}
	\psi(\Vect{x},z)
	= 
	\int 
	\frac{
		\dd \Vect{y}
	}{
		(i \lambda z)^{N/2}
	} \,
	\psi(\Vect{y},0) \exp \left( - i \frac{2 \pi }{\lambda z} \Vect{x}^\intercal \Vect{y} \right)
	.
	\label{eq:3_ffMT2}
\end{equation}

\noindent Both \Eqs{eq:3_ffMT1} and \eq{eq:3_ffMT2} are referred to as the Fraunhofer diffraction formula in the literature, depending on the reference. As mentioned, both forms are valid and consistent approximations for far-field diffraction, although \Eq{eq:3_ffMT1} is a slightly better approximation by retaining the curvature of the outgoing spherical-wave constituents.


\subsubsection{Cautionary remarks}

One should note that the Fraunhofer diffraction limit requires the presence of a finite-width aperture, even though width $W$ does not appear in the final formula \eq{eq:3_ffMT1}, \eq{eq:3_ffMT2}. Indeed, the existence of $W$ was used to introduce the Fresnel number $\mc{F}$ as the formal small parameter to derive Fraunhofer diffraction from Fresnel diffraction. The underlying mathematical reason for the necessary existence of an aperture is that Fraunhofer propagation does not form a group - it is invertible, but it is not closed under composition (matrix multiplication). Indeed, the composition of any two block-antidiagonal (symplectic) matrices is not itself block anti-diagonal:
\begin{equation}
	\begin{pmatrix}
		\OMat{N} & \Mat{B}_1 \\
		-\Mat{B}_1^{-1} & \OMat{N}
	\end{pmatrix}
	\begin{pmatrix}
		\OMat{N} & \Mat{B}_2 \\
		-\Mat{B}_2^{-1} & \OMat{N}
	\end{pmatrix}
	=
	\begin{pmatrix}
		- \Mat{B}_1 \Mat{B}_2^{-1} & \OMat{N} \\
		\OMat{N} & -\Mat{B}_1^{-1} \Mat{B}_2
	\end{pmatrix}
	\neq 
	\begin{pmatrix}
		\OMat{N} & \Mat{B}_3 \\
		-\Mat{B}_3^{-1} & \OMat{N}
	\end{pmatrix}
	,
\end{equation}

\noindent including Fraunhofer propagation as a special case. Consequently, Fraunhofer diffraction over a distance $d_1$ followed by Fraunhofer diffraction over a distance $d_2$ is not the same as Fraunhofer diffraction over a distance $d_1 + d_2$; the origin of the diffraction pattern \ie the aperture location, therefore has a privileged role in Fraunhofer diffraction.

The same is not true for Fresnel diffraction. Unit upper triangular (symplectic) matrices form a group:
\begin{equation}
	\begin{pmatrix}
		\IMat{N} & \Mat{B}_1 \\
		\OMat{N} & \IMat{N}
	\end{pmatrix}
	\begin{pmatrix}
		\IMat{N} & \Mat{B}_2 \\
		\OMat{N} & \IMat{N}
	\end{pmatrix}
	=
	\begin{pmatrix}
		\IMat{N} & \Mat{B}_3 \\
		\OMat{N} & \IMat{N}
	\end{pmatrix}
	,
	\quad
	\Mat{B}_3 = \Mat{B}_1 + \Mat{B}_2
	,
\end{equation} 

\noindent with unique inverse $\Mat{B}_2 = - \Mat{B}_1$ and identity $\Mat{B} = \OMat{N}$. This implies that the special case of Fresnel diffraction, in which $\Mat{B}_1$ and $\Mat{B}_2$ are diagonal, does as well (since the sum of two diagonal matrices is also diagonal). Hence, Fresnel diffraction over a distance $d_1$ followed by Fresnel diffraction over a distance $d_2$ is the same as Fresnel diffraction over a distance $d_1 + d_2$; the origin has no distinguished role and thus, there is no mathematical need for an aperture. Note also the fact that Fresnel diffraction forms a group means one can use an iterated near-identity metaplectic transform (\Ch{ch:NIMT}) to compute Fresnel diffraction over a finite interval as the composition of Fresnel diffractions over very small intervals.


\section{Summary}

In summary, symplectic transformations in classical mechanics constitute an important class of variable transformations on phase space in that they preserve the structure of Hamilton's equations of motion along with the canonical Poisson bracket. Linear symplectic transformations are particularly important because they have an exact operator analogue for quantum mechanics: the metaplectic operators. In this chapter I presented some preliminary results and basic identities that will be needed to develop MGO in \Ch{ch:MGO}. These include representing symplectic matrices \textbf{(i)} in the particular basis that results from performing a singular-value decomposition (SVD) of its upper-right matrix block, \textbf{(ii)} in the near-identity limit, and \textbf{(iii)} when it is also constrained to also be orthogonal. The corresponding metaplectic transforms that generate the aforementioned list of special cases for linear symplectic transformation are also derived. These results shall then allow us to perform phase-space rotations to remove caustics in the general case, as I shall show in later chapters.


\begin{subappendices}
\section*{Appendix}


\section{Metaplectic transforms in the mixed basis of configuration and coherent states}
\label{app:3_GaussMT}

One can also define the MT with respect to a mixed basis that involves Gaussian coherent states, which has no issues if $\det \Mat{B} = 0$. The Gaussian coherent states, denoted $\ket{\Stroke{\Vect{Z}}_0}$, are defined by their $\Vect{X}$-space representations
\begin{equation}
	\braket{\Vect{X}}{\Stroke{\Vect{Z}}_0}
	=
	\pi^{-N/4}
	\exp
	\left[
		- \frac{|\Vect{X} - \Vect{X}_0|^2}{2}
		+ i \Vect{K}_0^\intercal 
		\left(
			\Vect{X}
			- \frac{\Vect{X}_0}{2}
		\right)
	\right] 
	.
\end{equation}

\noindent The parameters $\Vect{X}_0$ and $\Vect{K}_0$ define the center of $\ket{\Stroke{\Vect{Z}}_0}$ in phase space, that is,
\begin{equation}
	\bra{\Stroke{\Vect{Z}}_0}
	\VectOp{X}
	\ket{\Stroke{\Vect{Z}}_0}
	= \Vect{X}_0
	,
	\quad
	\bra{\Stroke{\Vect{Z}}_0}
	\VectOp{K}
	\ket{\Stroke{\Vect{Z}}_0}
	= \Vect{K}_0
	.
\end{equation}

\noindent The states $\ket{\Stroke{\Vect{Z}}_0}$ also satisfy a completeness relation of the form
\begin{equation}
	\IdentOp = \int \frac{\dd \Vect{X}_0 \dd \Vect{K}_0}{(2\pi)^N} \, \ket{\Stroke{\Vect{Z}}_0}
	\bra{\Stroke{\Vect{Z}}_0} 
	.
\end{equation}

\noindent The completeness of $\{ \ket{\Stroke{\Vect{Z}}_0} \}$ allows one to write the inverse MT of \Eq{eq:3_invMTint} as
\begin{equation}
	\psi(\Vect{x}) 
	=
	\int \dd \Vect{X} \,
	\braket{\Vect{x}}{\Vect{X}}
	\Psi(\Vect{X}) 
	=
	\int 
	\frac{
		\dd \Vect{X} \, \dd \Vect{Q}_0 \, \dd \Vect{K}_0
	}{
		(2\pi)^N
	}
	\braket{\Vect{x}}{\Stroke{\Vect{Z}}_0}
	\braket{\Stroke{\Vect{Z}}_0}{\Vect{X}}
	\Psi(\Vect{X}) 
	.
	\label{eq:3_gaussINT}
\end{equation}

Then, as shown in \Ref{Littlejohn86a}, one can compute
\begin{align}
	&\braket{\Vect{x}}{\Stroke{\Vect{Z}}_0}
	=
	\pm \frac
	{
		\exp
		\left[
			-\frac{1}{2} \Vect{x}^\intercal 
			\left(
				\Mat{D}
				- i \Mat{B}
			\right)^{-1}
			\left(
				\Mat{A}
				+ i \Mat{C}
			\right) \Vect{x}
		\right]
	}
	{
		\pi^{N/4} 
		\sqrt{ \det(\Mat{D} - i \Mat{B} ) }
	}
	\exp
	\left[
		\left(
			\Vect{x}
			- \frac{i}{2} \Mat{B}^\intercal \Vect{\zeta}
		\right)^\intercal
		\left(
			\Mat{D}
			- i \Mat{B}
		\right)^{-1}
		\Vect{\zeta}
		- \frac{1}{2} \Vect{X}_0^\intercal \Vect{\zeta}
	\right]
	,
\end{align}

\noindent where I have introduced the complex vector $\Vect{\zeta} \doteq \Vect{X}_0 + i \Vect{K}_0$. Importantly, the complex matrix $\Mat{D} - i \Mat{B}$ is always invertible~\cite{Littlejohn87}. Then, upon using $\braket{\Stroke{\Vect{Z}}_0}{\Vect{X}} = ( \braket{\Vect{X}}{\Stroke{\Vect{Z}}_0} )^*$, one computes
\begin{align}
	\braket{\Vect{x}}{\Stroke{\Vect{Z}}_0}
	\braket{\Stroke{\Vect{Z}}_0}{\Vect{X}}
	&=
	\pm
	\frac
	{
		\exp
		\left[
			- \frac{1}{2} \Vect{x}^\intercal 
			\left(
				\Mat{D}
				- i \Mat{B}
			\right)^{-1} 
			\left(
				\Mat{A}
				+ i \Mat{C}
			\right)
			\Vect{x}
		\right]
	}
	{
		\pi^{N/2} 
		\sqrt{ \det(\Mat{D} - i \Mat{B} ) }
	}
	\nonumber\\
	&\hspace{4mm}\times
	\exp
	\left[
		\left(
			\Vect{x}
			- \frac{i}{2} \Mat{B}^\intercal \Vect{\zeta}
		\right)^\intercal
		\left(
			\Mat{D}
			- i \Mat{B}
		\right)^{-1} \Vect{\zeta}
		- |\Vect{X}_0|^2
		- \frac{|\Vect{X}|^2}{2}
		+ \Vect{X}^\intercal \Vect{\zeta}^*
	\right]
	.
\end{align}

\noindent Finally, I integrate over $\Vect{X}_0$ in \Eq{eq:3_gaussINT} to obtain
\begin{align}
	U^{-1}(\Vect{x}, \Vect{X})
	=
	\pm
	\int \dd \Vect{K}_0 \ 
	\frac
	{
		\exp
		\left[
			\fourier{g}(\Vect{x}, \Vect{\xi})
			- |\Vect{K}_0|^2
		\right]
	}
	{
		(\sqrt{2} \pi)^N 
		\sqrt{ \det(2 \Mat{D} - i \Mat{B} ) }
	}
	,
	\label{eq:3_gaussMT}
\end{align}

\noindent where I have defined $\Vect{\xi} \doteq \Vect{X} + 2 i \Vect{K}_0$ and 
\begin{align}
	\fourier{g}(\Vect{x}, \Vect{\xi})
	\doteq
	- \frac{1}{2} \Vect{x}^\intercal 
	\left(
		2\Mat{D}
		- i \Mat{B}
	\right)^{-1}
	\left(
		\Mat{A}
		+ 2 i \Mat{C}
	\right)
	\Vect{x}
	+
	\left(
		\Vect{x}
		- \frac{1}{2} \Mat{D}^\intercal \Vect{\xi}
	\right)^\intercal
	\left(
		2 \Mat{D}
		- i \Mat{B}
	\right)^{-1} \Vect{\xi}
	.
\end{align}

\noindent Note that the complex matrix $2 \Mat{D} - i \Mat{B}$ is always invertible~\cite{Littlejohn87}. Also note that $\Vect{K}_0$ cannot be integrated over without inverting $\Mat{B}$.


\section{Verification of the operator MT}
\label{app:3_squeeze}

I can verify that \Eq{eq:3_operMT} is indeed a representation of the MT by verifying that is satisfies \Eq{eq:3_operTRANS}. Let us begin with \Eq{eq:3_XtransOPER}. First, using the fact that $\Vect{x}$ and $\Vect{k}$ satisfy the canonical commutator relation
\begin{equation}
	\com{
		\oper{x}_\ell
	}{
		\oper{k}_m
	}
	=
	i \delta_{\ell m} \IdentOp
	,
	\quad \ell ,m = 1, \ldots, N
	,
	\label{eq:3_xkCOM}
\end{equation}

\noindent I can compute the commutator
\begin{equation}
	\left[
		\frac{
			\VectOp{x}^\intercal \left(\log \Mat{A}^{-\intercal}\right) \VectOp{k} 
			+ \VectOp{k}^\intercal \left(\log \Mat{A}^{-1}\right) \VectOp{x}
		}{
			2 i
		}
		,
		\VectOp{x}
	\right]
	=
	(\log \Mat{A})
	\,
	\VectOp{x}
	.
	\label{eq:3_squeezeQ}
\end{equation}

\noindent Hence, from the Baker--Campbell--Hausdorff (BCH) formula~\cite{Scully12}, it follows by induction that
\begin{align}
	&\exp\left[
		- i \frac{
			\VectOp{x}^\intercal \left(\log \Mat{A}^{-\intercal}\right) \VectOp{k} 
			+ \VectOp{k}^\intercal \left(\log \Mat{A}^{-1}\right) \VectOp{x}
		}{2}
	\right] \,
	\VectOp{x} \,
	\exp\left[
		i \frac{
			\VectOp{x}^\intercal \left(\log \Mat{A}^{-\intercal}\right) \VectOp{k} 
			+ \VectOp{k}^\intercal \left(\log \Mat{A}^{-1}\right) \VectOp{x}
		}{2}
	\right]
	\nonumber\\
	&= 
	\exp\left(\log \Mat{A} \right) \VectOp{x}
	=
	\Mat{A} \VectOp{x}
	.
	\label{eq:3_Q1}
\end{align}

\noindent Since functions of $\VectOp{x}$ commute, I trivially obtain
\begin{equation}
	\exp\left(
		- \frac{i}{2} \VectOp{x}^\intercal \Mat{A}^\intercal \Mat{C} \VectOp{x}
	\right)
	\Mat{A} \VectOp{x} 
	\exp\left(
		\frac{i}{2} \VectOp{x}^\intercal \Mat{A}^\intercal \Mat{C} \VectOp{x}
	\right)
	=
	\Mat{A} \VectOp{x} 
	.
	\label{eq:3_Q2}
\end{equation}

\noindent Next, using \Eq{eq:3_xkCOM} and the fact that $\Mat{A}^{-1} \Mat{B}$ is symmetric [\Eq{eq:3_symplec1}], I compute the commutator
\begin{equation}
	\left[
		\frac{i}{2}
		\VectOp{k}^\intercal \Mat{A}^{-1} \Mat{B} \VectOp{k}
		,
		\Mat{A} \VectOp{x}
		\nullFrac
	\right]
	=
	\Mat{B} \VectOp{k}
	.
	\label{eq:3_chirpQ}
\end{equation}

\noindent Since the right-hand side does not contain $\VectOp{x}$, the BCH series truncates and I obtain
\begin{align}
	\exp\left(
		\frac{i}{2} \VectOp{k}^\intercal \Mat{A}^{-1} \Mat{B} \VectOp{k}
	\right)
	\Mat{A} \VectOp{x} 
	\exp\left(
		- \frac{i}{2} \VectOp{k}^\intercal \Mat{A}^{-1} \Mat{B} \VectOp{k}
	\right)
	=
	\Mat{A} \VectOp{x} + \Mat{B} \VectOp{k}
	.
	\label{eq:3_Q3}
\end{align}

\noindent Combining \Eqs{eq:3_Q1}, \eq{eq:3_Q2}, and \eq{eq:3_Q3} with \Eq{eq:3_operMT} yields \Eq{eq:3_XtransOPER}.

Next, let us consider \Eq{eq:3_KtransOPER}. Analogous to \Eq{eq:3_squeezeQ}, I compute the commutator
\begin{equation}
	\left[
		\frac{
			\VectOp{x}^\intercal \left(\log \Mat{A}^{-\intercal}\right) \VectOp{k} 
			+ \VectOp{k}^\intercal \left(\log \Mat{A}^{-1}\right) \VectOp{x}
		}{2i}
		,
		\VectOp{k}
	\right]
	=
	(\log \Mat{A}^{-\intercal})
	\,
	\VectOp{k}
	.
\end{equation}

\noindent By induction, the BCH formula therefore yields
\begin{align}
	&\exp\left[
		- i \frac{
			\VectOp{x}^\intercal \left(\log \Mat{A}^{-\intercal}\right) \VectOp{k} 
			+ \VectOp{k}^\intercal \left(\log \Mat{A}^{-1}\right) \VectOp{x}
		}{2}
	\right] \,
	\VectOp{k} \,
	\exp\left[
		i \frac{
			\VectOp{x}^\intercal \left(\log \Mat{A}^{-\intercal}\right) \VectOp{k} 
			+ \VectOp{k}^\intercal \left(\log \Mat{A}^{-1}\right) \VectOp{x}
		}{2}
	\right]
	\nonumber\\
	&=
	\exp\left(\log \Mat{A}^{-\intercal} \right) \VectOp{k}
	=
	\Mat{A}^{-\intercal} \VectOp{k}
	.
	\label{eq:3_P1}
\end{align}

\noindent Analogous to \Eq{eq:3_chirpQ}, using the fact that $\Mat{A}^\intercal \Mat{C}$ is symmetric [\Eq{eq:3_symplec6}], I compute the commutator
\begin{equation}
	\left[ 
		-\frac{i}{2} \VectOp{x}^\intercal \Mat{A}^\intercal \Mat{C} \VectOp{x}
		,
		\Mat{A}^{-\intercal} \VectOp{k}
	\right]
	=
	\Mat{C} \VectOp{x}
	.
\end{equation}

\noindent Since the right-hand side does not contain $\VectOp{k}$, the BCH series truncates and I obtain
\begin{align}
	\exp\left(
		- \frac{i}{2} \VectOp{x}^\intercal \Mat{A}^{\intercal} \Mat{C} \VectOp{x}
	\right)
	\Mat{A}^{-\intercal} \VectOp{k} 
	\exp\left(
		\frac{i}{2} \VectOp{x}^\intercal \Mat{A}^{\intercal} \Mat{C} \VectOp{x}
	\right)
	=
	\Mat{A}^{-\intercal} \VectOp{k} + \Mat{C} \VectOp{x}
	.
	\label{eq:3_P2}
\end{align}

\noindent  I then compute the commutator
\begin{equation}
	\com{
		\frac{i}{2}
		\VectOp{k} \Mat{A}^{-1} \Mat{B} \VectOp{k}
	}{
		\Mat{A}^{-\intercal} \VectOp{k} + \Mat{C} \VectOp{x}
	}
	=
	\Mat{C} \Mat{B}^\intercal \Mat{A}^{- \intercal} \VectOp{k}
	.
\end{equation}

\noindent Since the right-hand side does not contain $\VectOp{x}$, the BCH series truncates and I obtain
\begin{align}
	\exp\left(
		\frac{i}{2} \VectOp{k}^\intercal \Mat{A}^{-1} \Mat{B} \VectOp{k}
	\right)
	\left(
		\Mat{A}^{-\intercal} \VectOp{k} 
		+ \Mat{C} \VectOp{x}
	\right)
	\exp\left(
		- \frac{i}{2} \VectOp{k}^\intercal \Mat{A}^{-1} \Mat{B} \VectOp{k}
	\right)
	&=
	\Mat{C} \VectOp{x} 
	+ 
	\left(
		\IMat{m} 
		+ \Mat{C}\Mat{B}^\intercal 
	\right)\Mat{A}^{-\intercal} \VectOp{k}
	\nonumber\\
	&=
	\Mat{C} \VectOp{x} 
	+ \Mat{D} \VectOp{k}
	,
	\label{eq:3_P3}
\end{align}

\noindent after using \Eq{eq:3_symplec2}. Combining \Eqs{eq:3_P1}, \eq{eq:3_P2}, and \eq{eq:3_P3} with \Eq{eq:3_operMT} yields \Eq{eq:3_KtransOPER}.


\section{Deriving the MT from its pseudo-differential representation}
\label{app:3_equivalence}

Here, I show the pseudo-differential representation \eq{eq:3_PMT} leads to the original integral representation \eq{eq:3_MTint} regardless the size of $\|\Mat{A}^{-1} \Mat{B} \|$. This proves that the pseudo-differential representation is in fact exact, even though it was originally derived in \Sec{sec:3_PMT} using an expansion in $\|\Mat{A}^{-1} \Mat{B}\|$. As a starting point, let us rewrite \Eq{eq:3_PMT} as
\begin{align}
	\Psi(\Vect{X}) = 
	\pm 
	\frac{
		\exp\left( \frac{i}{2} \Vect{X}^\intercal \Mat{C} \Mat{A}^{-1} \Vect{X} \right)
	}{
		\sqrt{\det{\Mat{A}}}
	}
	\int \dd \Vect{x}'~ \delta \left(\Vect{x}' - \Mat{A}^{-1}\Vect{X} \right) 
	\exp\left(
		\frac{i}{2} \Mat{A}^{-1} \Mat{B} \dubdot \pd{\Vect{x}'\Vect{x}'}^2
	\right) \, \psi(\Vect{x}') 
	.
	\label{eq:3_DeltaFunc}
\end{align}

\noindent I introduce the Fourier representation of $\psi(\Vect{x})$ as
\begin{align}
	\psi(\Vect{x}) = \frac{1}{(2\pi)^N} \int \dd \Vect{k} \, \fourier{\psi}(\Vect{k}) \, \exp( i\Vect{x}^\intercal \Vect{k} )
	, \quad
	\fourier{\psi}(\Vect{k}) = \int \dd \Vect{x} \, \psi(\Vect{x}) \, \exp( -i\Vect{x}^\intercal \Vect{k} ) 
	,
\end{align}

\noindent which, when substituted into \Eq{eq:3_DeltaFunc}, yields
\begin{align}
	\Psi(\Vect{X}) &= 
	\pm 
	\frac{
		\exp\left( \frac{i}{2} \Vect{X}^\intercal \Mat{C}\Mat{A}^{-1} \Vect{X} \right)
	}{
		(2\pi)^N \sqrt{ \det{\Mat{A}} }
	} 
	\int \dd \Vect{x}' \, \delta \left(\Vect{x}' - \Mat{A}^{-1}\Vect{X} \right)
	\int \dd \Vect{x} \, \psi(\Vect{x}) 
	\nonumber\\
	&\hspace{35mm}\times
	\int \dd \Vect{k} \, 
	\exp\left( 
		-\frac{i}{2} \Vect{k}^\intercal \Mat{A}^{-1} \Mat{B} \Vect{k} 
		+ i(\Vect{x}'-\Vect{x})^\intercal \Vect{k}
	\right) 
	.
\end{align}

\noindent The Gaussian integral can be performed explicitly,
\begin{align}
	\int \dd \Vect{k} \, 
	\exp\left(
		-\frac{i}{2} \Vect{k}^\intercal \Mat{A}^{-1} \Mat{B} \Vect{k} 
		+ i(\Vect{x}'-\Vect{x})^\intercal \Vect{k}
	\right)
	= \frac{
		(-2\pi i)^{N/2}
	}{
		\sqrt{\det{\Mat{A}^{-1} \Mat{B}}}
	} \, 
	\exp\left[
		\frac{i}{2} \left(\Vect{x}' - \Vect{x}\right)^\intercal \Mat{B}^{-1} \Mat{A} \left(\Vect{x}' - \Vect{x}\right) 
	\right]
	,
\end{align}

\noindent with the branch cut chosen to restrict all complex phases to the interval $(-\pi,\pi]$. Then, performing the trivial integration over $\dd \Vect{x}'$ yields \Eq{eq:3_MTint}. Note that neither the smoothness nor even the differentiability of $\psi$ is invoked in the above argument; the existence of the Fourier image of $\psi$ is enough.

\end{subappendices}

\chapter{Fast algorithm for near-identity metaplectic transforms}
\label{ch:NIMT}

\section{Introduction}

A number of numerical algorithms have been proposed which efficiently compute the MT on both 1-dimensional (1-D) and 2-D configuration spaces~\cite{Ozaktas96,Hennelly05b, Healy10, Koc10a, Ding12, Pei16,Sun18a}. Many of them are reviewed in \Ref{Healy18}. Despite this multitude, however, there also exist applications for which suitable MT algorithms have yet to be designed, such as the small-angle rotations that will be used in MGO [\Ch{ch:MGO}]. In this chapter, I propose two algorithms specifically optimized in the near-identity limit to close this gap. The first proposed algorithm is local, but it is not unitary and is unstable as a consequence. The second algorithm is nonlocal but is unitary and stable. In the following, I shall assume that $\Mat{S} \approx \IMat{2N}$, and in particular, that $\Mat{A}^{-1} \Mat{B}$ is small. Note that the material presented in this chapter is based on discussions previously published in \Refs{Lopez19, Lopez21b}.


\section{Local fast near-identity metaplectic transform}

\subsection{Definitions}

For the first near-identity metaplectic transform (NIMT) algorithm, let us recall the pseudo-differential series representation presented in \Eq{eq:3_PMTseries}. When $\Mat{A}^{-1} \Mat{B}$ is small and $\psi$ is smooth enough, \Eq{eq:3_PMTseries} can be approximated with a truncated series. We define the $m$-th order NIMT as the truncation that neglects all terms with $n > m$. Also, to be connected with the identity, we explicitly choose the overall $+$ sign when performing NIMT truncations. Decreasing $m$ will increase the locality of the truncated transformation, because the necessary stencil width to compute the $m$-th order NIMT will decrease. This enables the $m$-th order NIMT to be performed pointwise, as the transformed function evaluated at some point $X = X_0$ depends only on the original function and its first $2m$ derivatives evaluated at the corresponding point $x = x_0(X_0)$. Hence, I shall refer to the algorithm that results from this representation as the `local' NIMT algorithm.

When the order is not specified, the `NIMT' refers solely to the first-order NIMT,
\begin{align}
	\Psi(\Vect{X}) 
	\approx 
	\frac{
		\exp\left(
			\frac{i}{2}\Vect{X}^\intercal \Mat{C}\Mat{A}^{-1} \Vect{X}
		\right)
	}{
		\sqrt{\det{\Mat{A}}}
	}
	\left\{
		\psi(\Mat{A}^{-1}\Vect{X})
		+ \frac{i}{2} \Tr \left[ 
			\Mat{A}^{-1}\Mat{B} \, \pd{\Vect{x} \Vect{x}}^2 
			\psi(\Mat{A}^{-1}\Vect{X})
		\right] 
	\right\} 
	,
	\label{eq:4_NIMTtaylor}
\end{align}

\noindent as it is the lowest-order truncation that remains practical. (The truncation at $m=0$ is too simplified to yield an accurate representation of the MT, regardless the smoothness of $\psi$.) Since matrix operations can be computationally expensive when $N$ is large, low-order approximations for $\det{\Mat{A}}$, $\Mat{A}^{-1}$, $\Mat{A}^{-1}\Mat{B}$, and $\Mat{C} \Mat{A}^{-1}$ for use when $\Mat{S}$ is near-identity were derived in \Ch{ch:MT}. We also provide auxiliary calculations when $\psi(\Vect{x})$ is eikonal in \App{app:4_eikNIMT}.

The pseudo-differential representation of the MT naturally gives rise to an iterative algorithm: successive applications of the NIMT can compute a finite transformation from a sequence of near-identity transformations. To see this, consider the MT of a function $\psi$ that results from a desired symplectic transformation $\fourier{\Mat{S}}$, which may be the result of a single optical operation or a cascade of operations. As the symplectic group is topologically connected, it is always possible to find a smooth trajectory of symplectic matrices $\Mat{S}(t)$ with parameterization $t$ such that $\Mat{S}(0) = \IMat{2N}$ and $\Mat{S}(1) = \fourier{\Mat{S}}$. Note that the path $\Mat{S}(t)$ should also have a compatible winding number with the desired overall sign of $\Mat{M}(\Mat{S})$, as discussed in \Ch{ch:MT}. We then discretize $\Mat{S}(t)$ to obtain $K$ near-identity symplectic matrices as the single-step iterates, namely,
\begin{equation}
	\Mat{S}_j \doteq 
	\Mat{S}\left( \frac{j}{K} \right) \, 
	\Mat{S}^{-1}\left( \frac{j-1}{K} \right)
	, \quad
	j = 1, \ldots, K
	,
	\label{eq:4_NIsymp}
\end{equation}

\noindent where we have assumed a uniform step size $\Delta t \doteq 1/K$ for simplicity. Then, since we can decompose $\fourier{\Mat{S}}$ as
\begin{equation}
	\fourier{\Mat{S}} = \Mat{S}_K \ldots \Mat{S}_1
	,
\end{equation}

\noindent we can compute $\oper{M}(\fourier{\Mat{S}})$ via the iteration scheme
\begin{equation}
	\oper{M}(\fourier{\Mat{S}})
	=
	\oper{M}(\Mat{S}_K) \ldots \oper{M}(\Mat{S}_1) 
	,
	\label{eq:4_iterMT}
\end{equation}

\noindent where each $\oper{M}(\Mat{S}_j)$ is near-identity. Hence, an arbitrary MT can be approximately computed via the iterated NIMT algorithm as
\begin{equation}
	\oper{M}(\fourier{\Mat{S}})
	\approx
	\oper{N}(\Mat{S}_K) \ldots \oper{N}(\Mat{S}_1)
	\label{eq:4_iterNIMT}
	,
\end{equation}

\noindent where $\oper{N}$ denotes the appropriate NIMT operator. Note that the discretization of $\Mat{S}_t$ by itself does not incur any errors, so the accuracy of \Eq{eq:4_iterNIMT} depends solely on the `composition error' of the NIMT, which in turn depends on the truncation order and the discretization scheme used for the Hessian matrix. This is the well-known `loss of additivity' that many discrete MTs experience~\cite{Zhao15}. Note also that our approach is advantageous because it is independent of the dimensionality $N$. One only needs to adjust the size of $\Mat{S}$ when changing from, say, a $1$-D application to a $3$-D application. This is not true for other numerical MT algorithms in the literature, which can only handle up to $2$-D and are explicitly different depending on whether $\Mat{S}$ is `separable' or `nonseparable'~\cite{Koc10a,Ding12,Pei16}. Such restrictions do not arise with the iterated NIMT.


\subsection{Runtime estimate}

Let us estimate the computational efficiency of the iterated NIMT. We should first emphasize that although the NIMT appears to require interpolation, this is not strictly necessary. Suppose that $\psi(\Vect{x})$ is only known on a discrete set of points $\{\Vect{x}_k\}$. The discretization of $\psi(\Vect{x})$ can be used to inform the discretization of $\Psi(\Vect{X})$ by evaluating the NIMT only at the corresponding points $\{\Vect{X}_k \doteq \Mat{A}\Vect{x}_k \}$. No interpolation is required, unless, one needs to evaluate $\Psi(\Vect{X})$ off-grid. In that case, either the discrete set $\{\psi(\Vect{x}_k)\}$ must be interpolated and transformed, or the discrete set $\{\Psi(\Vect{X}_k) \}$ must be interpolated. For this reason, and because interpolation efficiency is highly implementation-specific, we do not account for interpolation in our runtime estimate.

From \Eq{eq:4_NIMTtaylor}, evaluating $\Psi(\Vect{X} )$ at $N_p$ discrete points requires only $O(N_p N^3)$ floating-point operations (FLOPs), since each evaluation includes a matrix multiplication that scales as $N^3$ (with $N$ being the dimensionality)~\cite{Trefethen97}. Thus, the NIMT always scales linearly with the number of sample points, independent of dimensionality. The iterated NIMT remains `fast' with respect to the number of sample points, since the FLOP count scales as $O(N_p N^3K)$, with $K$ the number of iterations. The linear scaling is faster than many of the other MT algorithms found in the literature~\cite{Healy18}, which scale as $O(N_p \log N_p)$.


\subsection{Stability}

Although the iterated NIMT scales faster than other published MT algorithms, it may not be as stable. Intuitively, one would expect that refining the discretization of $\Mat{S}_t$ would increase the accuracy of the iterated NIMT since the magnitude of $\|\Mat{A}^{-1}_j\Mat{B}_j\|$ for each successive $j$-th application of the NIMT would decrease. As the magnitude of $\|\Mat{A}^{-1}_j \Mat{B}_j\|$ decreases, however, the number of iterations required to generate a fixed final transformation increases. Careful analysis is needed to determine if the truncation errors of the iterated NIMT accumulate coherently, which we accomplish by estimating the parameter regimes in which the iterated NIMT is non-unitary. For simplicity, the forthcoming analysis is restricted to $1$-D.

Let us consider how the pseudo-differential MT (PMT) and the iterated NIMT transform the single-parameter family of exponential functions $\psi_\kappa(x) \doteq \exp(\kappa x)$, with $\kappa$ being complex. Generally speaking, we define an MT algorithm as \textit{stable}, or non-magnifying, if the norms of the transformed function $\Psi_\kappa(X)$ and the original function $\psi_\kappa(x)$ satisfy $\|\Psi_\kappa(X)\| \le \| \psi_\kappa(x)\|$; conversely, we define an MT algorithm as \textit{unstable}, or magnifying, if $\|\Psi_\kappa(X)\| > \| \psi_\kappa(x)\|$. Unitarity corresponds to a strict equality. The ratio $\|\Psi_\kappa(X)\|/\| \psi_\kappa(x)\|$ is referred to as the \textit{magnification factor}. Additionally, we define an MT algorithm as either \textit{$L^2$-stable} or, respectively, \textit{$L^2$-unitary} if the algorithm is stable or unitary along the entire imaginary $\kappa$ axis. This is because any $L^2$ function can be expanded into Fourier modes; thus, an $L^2$-unitary MT algorithm will be exactly unitary for any $L^2$ function. In our analysis, we shall only consider the class of function norms where $\|e^{ig(X)}f(X/A) \| = \sqrt{A} \, \|f(x) \|$ for $g(X)$ real, an example of which being the $L^2$ norm.

Since $\psi_\kappa'(x) = \kappa \psi_\kappa (x)$, the PMT of $\psi_\kappa(x)$ is
\begin{equation}
	\PMT{\Mat{S}} \left[\psi_\kappa(x) \right] 
	= 
	\frac{
		\exp\left( i\frac{C}{2A}X^2 \right)
	}{
		\sqrt{A}
	} \,
	\exp\left(i\frac{B}{2A}\kappa^2 \right) \, \psi_\kappa\left(\frac{X}{A} \right) 
	,
\end{equation}

\noindent where $\PMT{\Mat{S}}$ is the PMT for symplectic matrix $\Mat{S}$. Let us define the rescaled variable $w \doteq \kappa \sqrt{B/A}$. Then, the PMT is stable when
\begin{equation}
	\left| \exp\left( \frac{i}{2}w^2 \right) \right| \le 1 
	.
	\label{eq:4_PMTstabCond}
\end{equation}

\begin{figure}
	\centering
	\includegraphics[width=0.6\linewidth,trim={4mm 4mm 3mm 3mm},clip]{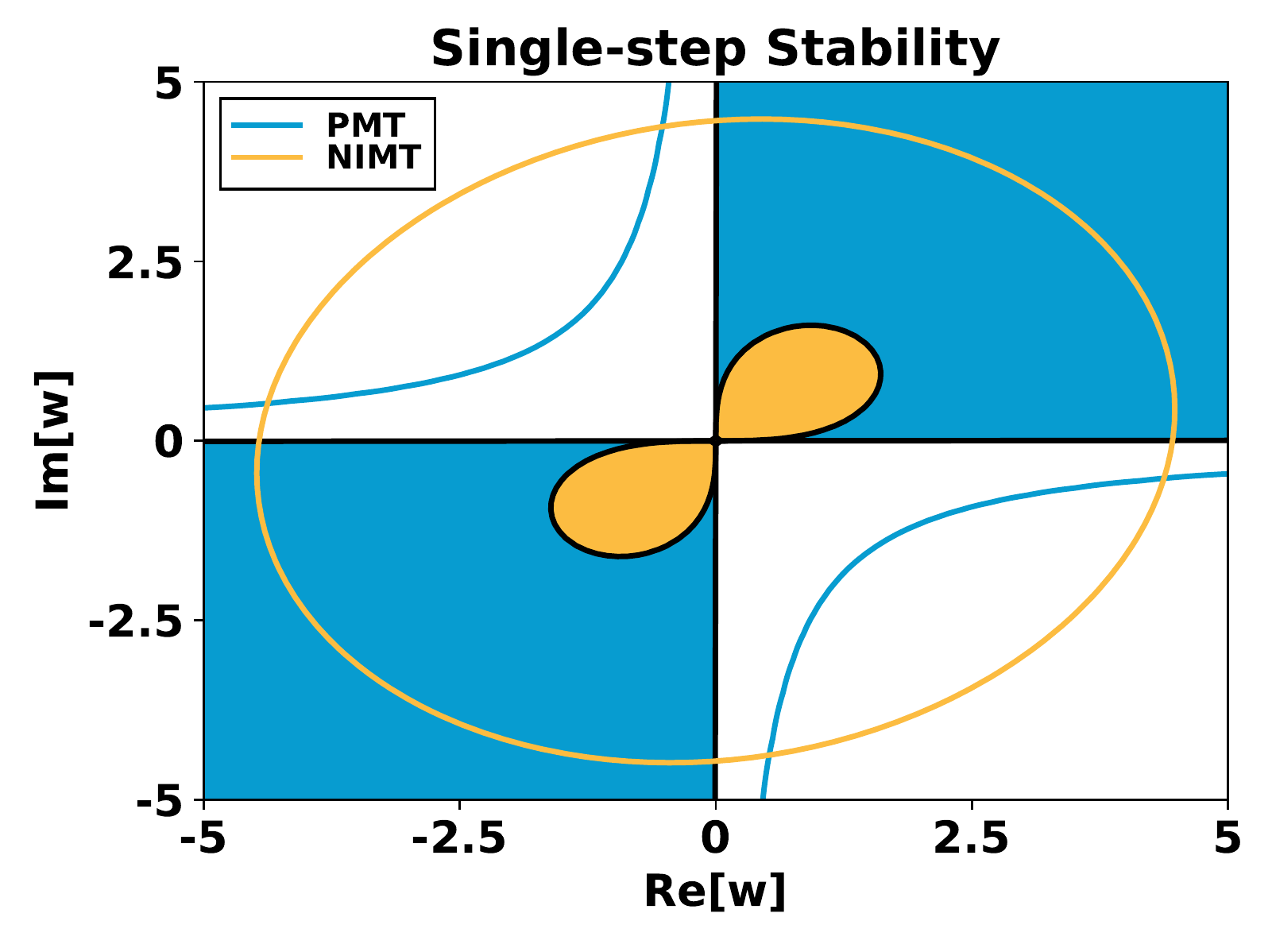}
	\caption{Stability diagrams for the PMT [blue shaded, from \Eq{eq:4_PMTstabCond}] and the NIMT [orange shaded, from \Eq{eq:4_NIMTstabCond}]. The solid lines of each color mark the contour of the respective algorithm at which the magnification factor equals 10.}
	\label{fig:4_1StepStab}
\end{figure}

\noindent This region of the complex $w$ plane is shown in \Fig{fig:4_1StepStab}. The PMT is stable within the first and third quadrants of the complex plane, and is unitary along the real and imaginary $w$ axes. Hence, the PMT is $L^2$-unitary. Interestingly, the PMT is not unitary on its entire domain. This is because the domain of the PMT includes both square-integrable functions and functions where the integral of \Eq{eq:3_MTint} does not converge, such as $\exp(x)$. The cost of this expanded domain is the loss of global unitarity, albeit for functions whose $L^2$ norms are undefined.

We proceed to analyze the NIMT. Applied once, the NIMT of $\psi_\kappa(x)$ is
\begin{equation}
	\NIMT{\Mat{S}} \left[\psi_\kappa(x) \right] = 
	\frac{1+\frac{iB}{2A} \, \kappa^2}{\sqrt{A}} \, \exp\left(\frac{i C}{2A}X^2 \right) 
	\psi_\kappa\left(\frac{X}{A} \right) 
	.
\end{equation}

\noindent Reintroducing $w$, the NIMT is stable where
\begin{equation}
	\left|1+\frac{i}{2}w^2 \right| \le 1 
	,
	\label{eq:4_NIMTstabCond}
\end{equation}

\noindent which is shown alongside the stability region of the PMT in \Fig{fig:4_1StepStab}. This region makes up only a small subset of the stability region of the PMT. Notably, the NIMT is no longer $L^2$-stable; as such, square-integrable functions will be magnified. There are three ways to minimize the magnification: (i) reduce the step size \eg let $B/A \lesssim 1/2\kappa_i$ with $\kappa_i$ the largest Fourier mode number; (ii) apply a low-pass filter to remove fast-growing Fourier modes; or (iii) increase the truncation order. However, we show shortly that increasing the truncation order of the NIMT increases its vulnerability to numerical noise, so only (i) and (ii) are recommended.

Let us now assess how subsequent iterations of the NIMT affect its stability. We first observe that each iteration of the NIMT adds the overall phase $X^2C/2A$ that contributes to the derivatives of subsequent NIMT iterations. This sequence will very quickly become unwieldy as the iteration number increases. To achieve an analytical estimate of the iterated NIMT stability, we shall therefore neglect the contributions of the phase to all derivatives. This is consistent with the near-identity limit, where $C/A$ is vanishingly small. In this approximation, the norm of the $K$-iterated NIMT is
\begin{align}
	\left\| 
		\NIMT{\Mat{S}_K}\left\{
			\ldots 
			\NIMT{\Mat{S}_{1}}\left[
				\psi_\kappa(x) \nullFrac 
			\right] 
		\right\}
	\right\|
	\approx \left|
		\prod_{n=1}^K \frac{1+\frac{iB_n}{2A_n} 
		\frac{\kappa^2}{\prod_{j=1}^{n-1} A^2_j}}{\sqrt{A_n}} 
	\right| 
	\left\| 
		\psi_\kappa\left(\frac{X}{\prod_{n=1}^{K} A_n} \right)
	\right\| 
	,
	\label{eq:4_iterNIMTerr}
\end{align}

\noindent where for $n = 1$, we define $\prod_{j=1}^{n-1} A^2_j = 1$. When the iteration is uniform, \ie $A_n = A$ and $B_n = B$, \Eq{eq:4_iterNIMTerr} simplifies to
\begin{align}
	\left\| 
		\NIMT{\Mat{S}_K}\left\{
			\ldots 
			\NIMT{\Mat{S}_{1}}\left[
				\psi_\kappa(q) \nullFrac 
			\right] 
		\right\}
	\right\|
	= \left| 
		A^{-\frac{K}{2}}\prod_{n=0}^{K-1} \left(1 + \frac{i w^2}{2} A^{-2n} \right) 
	\right| \, 
	\left\| 
		\psi_\kappa\left(\frac{X}{A^K} \right)
	\right\| 
	,
	\label{eq:4_iterNIMTerrUNIFORM}
\end{align}

\noindent where I have reintroduced $w \doteq \kappa \sqrt{B/A}$. Note also that the iteration range has been shifted from $n \in [1, K]$ to $n \in [0, K-1]$ in proceeding from \Eq{eq:4_iterNIMTerr} to \Eq{eq:4_iterNIMTerrUNIFORM}. Hence, the $K$-iterated NIMT is stable where
\begin{equation}
	\left| 
		\left(-\frac{iw^2}{2};A^{-2} \right)_K 
	\right| \le 1 \, ,
	\label{eq:4_IterStab}
\end{equation}

\noindent where $(a;q)_K \doteq \prod_{n=0}^{K-1} \left(1 - aq^n \right)$ is the $q$-Pochhammer symbol~\cite{Olver10a}, \ie the $q$-analog of the rising factorial.

\begin{figure}
	\centering
	\begin{overpic}[width=0.43\linewidth,trim={4mm 4mm 3mm 3mm},clip]{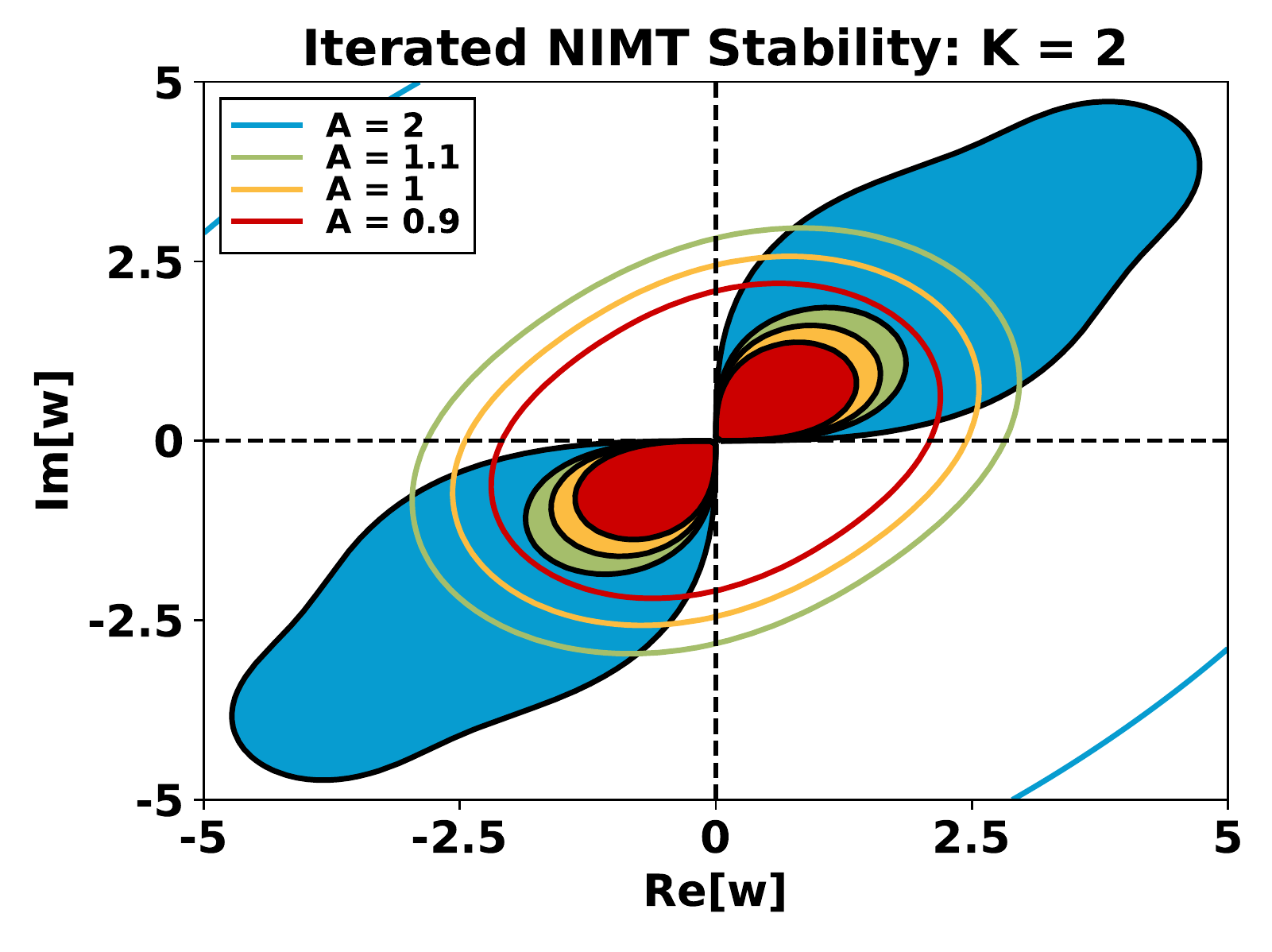}
		\put(87,12){\textbf{\large(a)}}
	\end{overpic}
	\hspace{3mm}
	\begin{overpic}[width=0.43\linewidth,trim={4mm 4mm 3mm 3mm},clip]{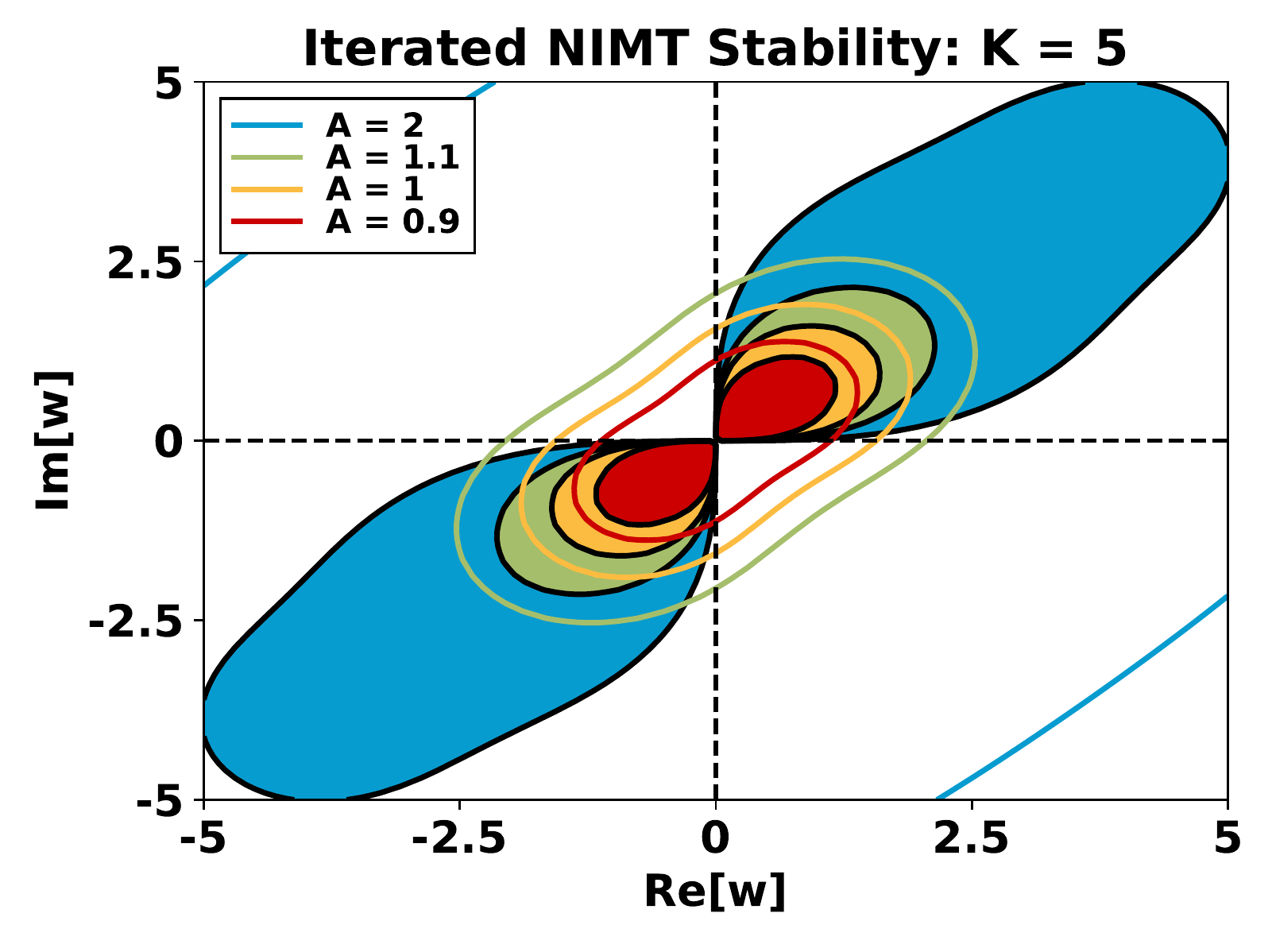}
		\put(87,12){\textbf{\large(b)}}
	\end{overpic}

	\vspace{3mm}
	\begin{overpic}[width=0.43\linewidth,trim={4mm 4mm 3mm 3mm},clip]{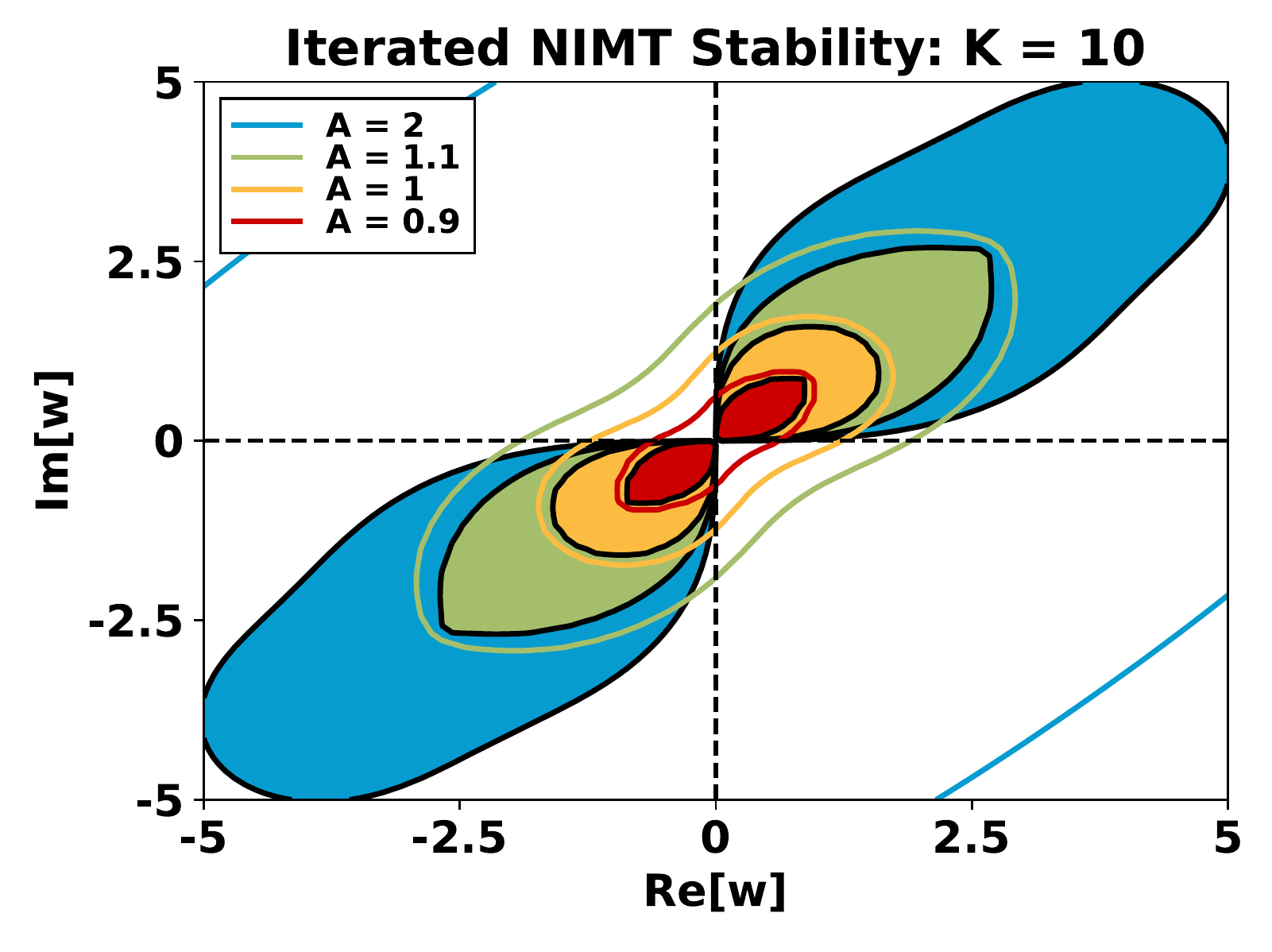}
		\put(87,12){\textbf{\large(c)}}
	\end{overpic}
	\hspace{3mm}
	\begin{overpic}[width=0.43\linewidth,trim={4mm 4mm 3mm 3mm},clip]{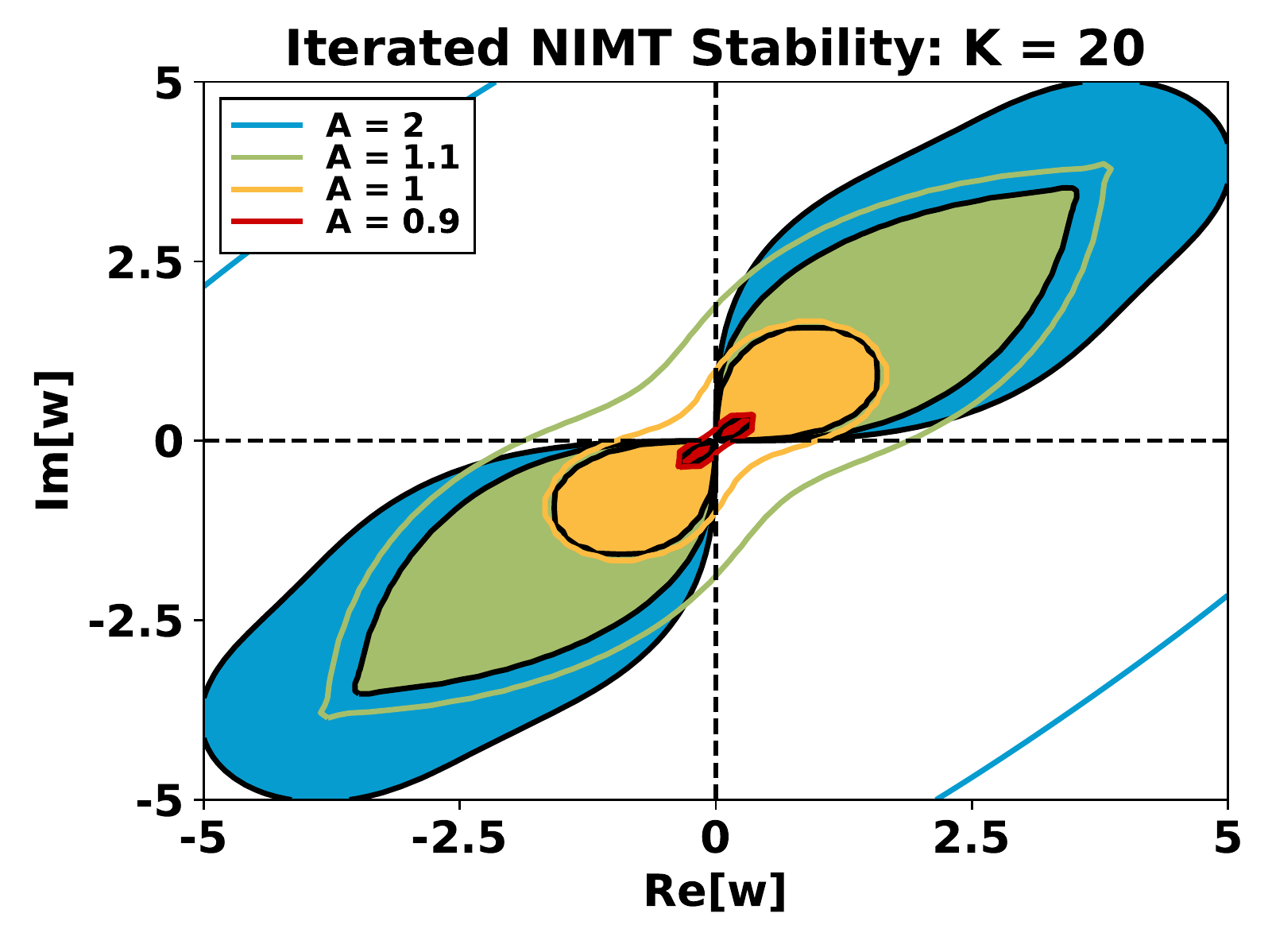}
		\put(87,12){\textbf{\large(d)}}
	\end{overpic}
	\caption{Stability diagrams for the iterated NIMT at various values of the iteration number $K$ and $A$, as computed via \Eq{eq:4_IterStab}. For each color, the shaded region is the region of stability for the respective value of $A$, while the solid line labels the contour at which the magnification factor equals 10. Note that the near-identity limit corresponds to $A \approx 1$.}
	\label{fig:4_IterNIMTstab}
\end{figure}

Figure~\ref{fig:4_IterNIMTstab} shows the stability region at four different iteration numbers: $K = 2$, $K = 5$, $K = 10$, and $K = 20$. As \Eq{eq:4_IterStab} indicates, the stability of the iterated NIMT now explicitly depends on $A$, so each subplot of \Fig{fig:4_IterNIMTstab} includes stability diagrams for $A = 0.9$, $A = 1$, $A = 1.1$, and $A = 2$. These values were chosen to emphasize the near-identity behavior of the iterated NIMT, when $A \approx 1$. There are two notable observations. First, the stability region for $A=1$ is independent of $K$. For other values of $A$, the stability region changes significantly with $K$, decreasing for $A < 1$ and increasing for $A > 1$. Second, the sensitivity of the iterated NIMT increases with $K$, as seen by considering the rate at which the $A = 1.1$ and $A = 0.9$ contours separate. Consequently, a step size $B/A$ that is initially stable, but with $A < 1$, will become quickly and increasingly unstable as the NIMT is iterated. This introduces an interesting tradeoff consideration when computing a finite transformation: is it better to use a coarse discretization with a large step size but few iterations, or a fine discretization with a small step size but many iterations? The answer depends largely on implementation specifics; we find in the following subsection that a fine discretization is preferable for our chosen example, but this is not necessarily indicative of a general principle.

\begin{figure}
	\centering
	\begin{overpic}[width=0.44\linewidth,trim={13mm 7mm 4mm 6mm},clip]{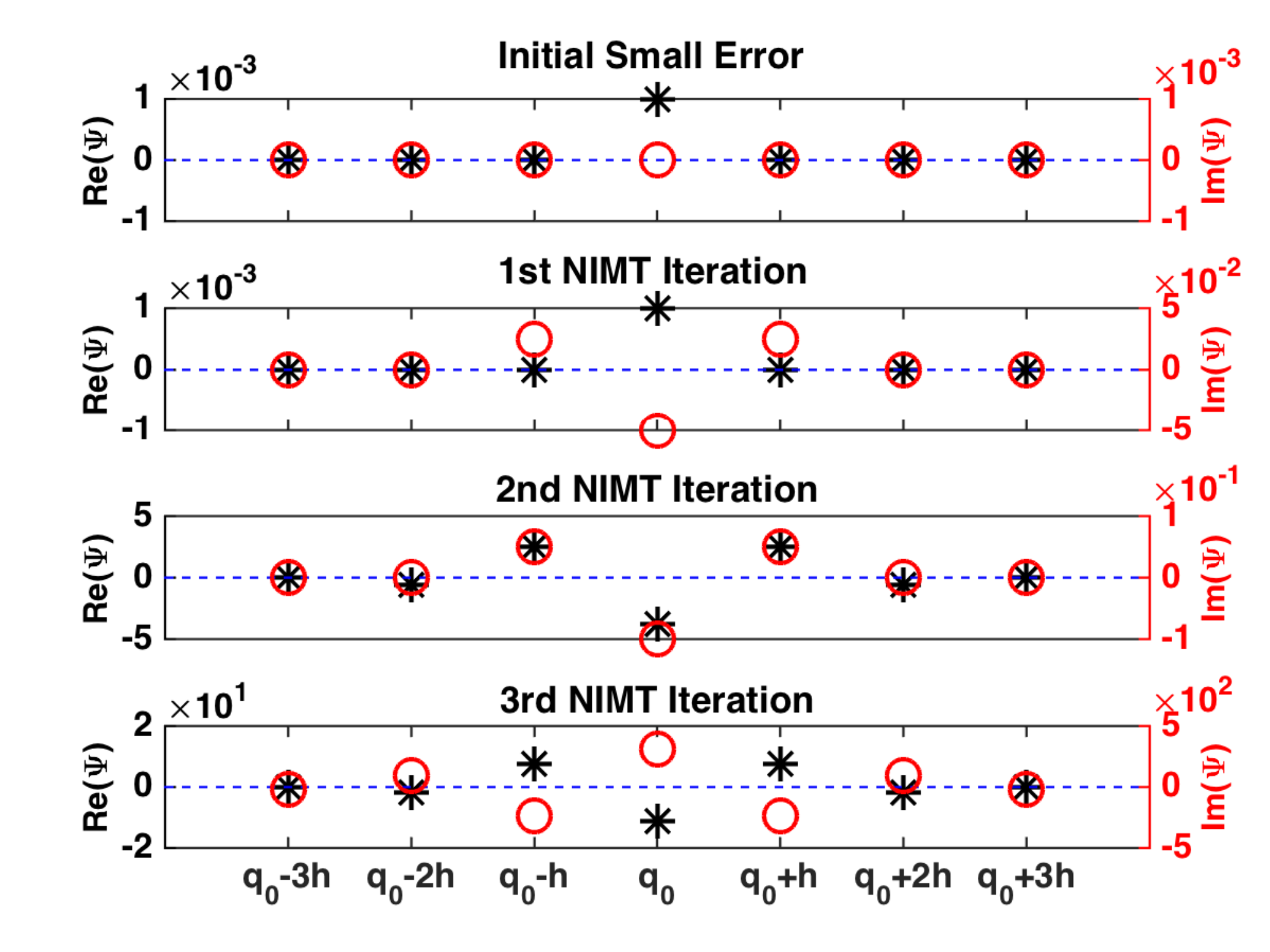}
		\put(85,12){\textbf{\large(a)}}
	\end{overpic}
	\hspace{2mm}
	\begin{overpic}[width=0.44\linewidth,trim={7mm 4mm 15mm 3mm},clip]{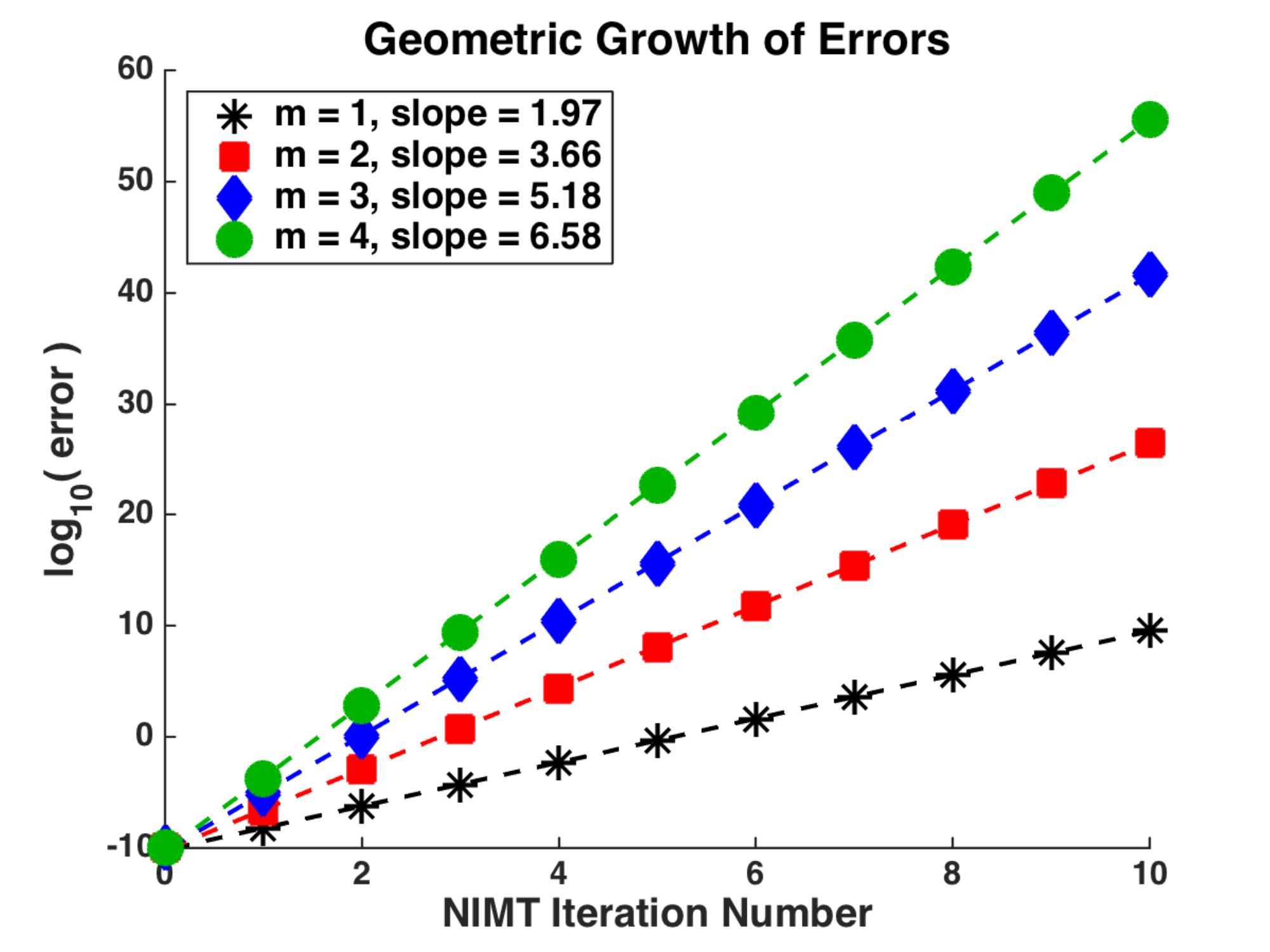}
		\put(85,12){\textbf{\large(b)}}
	\end{overpic}
	\caption{\textbf{(a)} The d-instability increases an initial error of $10^{-3}$ by multiple orders of magnitude after only three iterations. The error also propagates away from the initial location, with width proportional to the iteration number. \textbf{(b)} Error from the d-instability as the NIMT truncation number $m$ is varied. The error is defined as $\| \boldsymbol{\Psi}\|$. Notable parameters are $A = D = 1$, $B = 1/2$, $C = 0$, $h = 0.1$, and initial error $10^{-10}$. Using \Eq{eq:4_growthRATE} [\Eq{eq:4_incGAMMA}] to estimate the growth rate and then taking the base-10 logarithm, the slopes for $m = 1$, $m = 2$, $m = 3$, and $m = 4$ are estimated respectively as $2.000$ ($2.004$), $3.699$ ($3.708$), $5.222$ ($5.235$), and $6.620$ ($6.637$). These values agree well with those obtained by a best fit line (dashed line).}
	\label{fig:4_noiseINSTAB}
\end{figure}

Although we shall not dwell much on implementation details, we must make one cautionary remark regarding the finite-difference scheme used to discretize the NIMT. Because discrete differentiation is poorly conditioned, any noise in the original function $\psi(x)$ will be magnified when its derivatives are computed. Since derivatives are computed with each iteration of the NIMT, this noise will grow geometrically. We call this instability the \textit{d-instability} (with `d' standing for discretization). As shown in \Fig{fig:4_noiseINSTAB}, it is particularly disastrous for iterated NIMTs with large truncation order. 

A basic description of the d-instability is afforded by the transformation of a constant function. Suppose one attempts to transform a function that is identically zero everywhere except at a single grid point, where the function is erroneously non-zero by some unspecified noise source. When the grid spacing $h$ is uniform, the growth rate of the d-instability, $\gamma$, can be estimated analytically. Let $\Delta_{k}$ be a $k$-th order finite-difference matrix such that $h^{-k} \Delta_k \Vect{f}$ equals the $k$-th discrete derivative of $\Vect{f}$. In this specific test problem, any non-zero norm is due to noise; hence, the error of the $K$-th iterated, $m$-th order NIMT is bounded with the triangle inequality as
\begin{equation}
	\|\boldsymbol{\Psi} \| 
	\le 
	\left[
		\frac{1}{\sqrt{|A|}} \sum_{n=0}^m \frac{\|\Delta_{2n} \|}{h^{2n}} \frac{|B/2A|^n}{n!}
	\right]^K \|\boldsymbol{\psi}\| 
	,
	\label{eq:4_noiseINEQ}
\end{equation}

\noindent where $\boldsymbol{\psi}$ and $\boldsymbol{\Psi}$ are the discretized versions of $\psi(x)$ and $\Psi(X)$ respectively, and $\|\Delta_{2n} \|$ is the subordinate matrix norm of $\Delta_{2n}$. There is a freedom to choose the norm with which \Eq{eq:4_noiseINEQ} is evaluated; we choose the $\infty$-norm, denoted $\|\ldots \|_\infty$, as it yields the readily-evaluated matrix row norm as its subordinate~\cite{Trefethen97}. Considering only the leading order in $1/h$, $\gamma$ is estimated as
\begin{equation}
	\gamma \approx \frac{1}{\sqrt{|A|}} \, \frac{\|\Delta_{2m} \|_\infty}{h^{2m}} \, \frac{|B/2A|^m}{m!} 
	.
	\label{eq:4_growthRATE}
\end{equation}

Equation \eq{eq:4_growthRATE} has been purposefully separated: for increasing $m$, the factor $\|\Delta_{2m} \|/h^{2m}$ increases while the factor $|B/2A|^m/m!$ decreases. In fact, as defined, $\gamma \to 0$ as $m \to \infty$ for any reasonable class of $\Delta_{2m}$; this does not mean the d-instability disappears for high truncation orders, but rather that the d-instability is not dominated by the leading order in $1/h$ when $m$ is large. Instead, a subset of intermediate-order terms dominate, which are not included in \Eq{eq:4_growthRATE}. For central finite-difference scheme with homogeneous boundary conditions, $\|\Delta_{2n} \|_\infty = 2^{2n}$~\cite{Abramowitz70}, and the growth rate is uniformly estimated to be
\begin{equation}
	\gamma 
	\approx \frac{1}{\sqrt{|A|}} \, 
	\exp\left(
		\left| \frac{2B}{Ah^2} \right| 
	\right) \, 
	\frac{
		\Gamma\left(m+1, \left| \frac{2B}{Ah^2} \right| \right)
	}{
		\Gamma\left(m+1 \right)
	} 
	,
	\label{eq:4_incGAMMA}
\end{equation}

\noindent where $\Gamma\left(s,x \right)$ is the incomplete gamma function~\cite{Olver10a}. Notably, $\Gamma\left(s,x \right) \to \Gamma\left(s\right)$ as $s \to \infty$. Equation \eq{eq:4_growthRATE} is sufficient for error estimation of low truncation order schemes; for large $m$, however, \Eq{eq:4_incGAMMA} should be used instead.

Thus far, our discussion of the d-instability has been contingent on a maliciously designed initial condition. Such a specific state will not likely arise in practical applications; nevertheless, local d-instabilities can certainly arise. For example, consider the NIMT of a function $\psi(x)$ that asymptotes to $0$ at the domain edge. Near the domain edge, $\psi(x)$ is nearly constant, but a source of error, interpolation or otherwise, will inevitably cause at least one data point to deviate. The local d-instability will then grow rapidly, and will propagate inward from the domain edge until the transformed function is entirely dominated by noise. Since the d-instability growth rate scales with truncation order, using low $m$ schemes will minimize its deleterious effects. Marginally smoothing the input data before taking derivatives will also delay its onset.


\subsection{Examples}

To demonstrate the iterated NIMT, let us consider once again the time evolution of the $1$-D QHO, introduced in \Ch{ch:MT}. There, the symplectic matrix $\Mat{S}$ that governs time evolution can be expressed as
\begin{equation}
	\Mat{S}_t = 
	\begin{pmatrix}
		\cos{t} & \sin{t}\\
		-\sin{t} & \cos{t}
	\end{pmatrix}
	,
	\label{eq:4_SEvo}
\end{equation}

\noindent where we have added the index $t$ to emphasize the dependence on time. This matrix can be represented as $\Mat{S}_t = (\Mat{S}_{\Delta t})^K$, where $\Mat{S}_{\Delta t}$ evolves the system by $\Delta t \ll 1$ and $K = t/\Delta t$. The scalar functions $A_{\Delta t}$, $B_{\Delta t}$, $C_{\Delta t}$, and $D_{\Delta t}$ to be used in the iterated NIMT are
\begin{equation}
	A_{\Delta t} = D_{\Delta t} = \cos(\Delta t) 
	, \quad 
	B_{\Delta t} = -C_{\Delta t} = \sin(\Delta t) 
	.
\end{equation}

For visualization, it is useful to re-introduce the \textit{Wigner function} (\Ch{ch:GO}) that corresponds to $\psi(x)$, defined as~\cite{Wigner32}:
\begin{equation}
	W_\psi(x,k) = \frac{1}{2\pi}\int \limits_{-\infty}^\infty 
	\dd y
	\, \psi\left(x-\frac{y}{2}\right)\psi^*\left(x+\frac{y}{2}\right)
	\exp\left( iky \right) 
	.
	\label{eq:4_wigFUNC}
\end{equation}

\noindent As shown in \Refs{deGosson06,Littlejohn86a,Lohmann93} (see also \Ch{ch:MGO}), the Wigner function of the metaplectic image $\Psi$ is simply the Wigner function of the original $\psi$ correspondingly rotated. This is also readily understood from the physical meaning of $W_\psi$. Specifically, if $\psi$ is a wave field, then $W_\psi$ can be interpreted as the phase space quasiprobability distribution function of the wave quanta. The prefix `quasi' marks the fact that $W_\psi$ is not positive-definite unless averaged over a phase space volume of size $\Delta x \, \Delta k \gtrsim 2\pi$~\cite{Cartwright76, OConnell81}; nonetheless, $W_\psi$ is always real by definition, even for complex $\psi$.

\begin{figure}[t!]
	\centering
	\begin{overpic}[width=0.32\linewidth,trim={-4mm 7mm 4mm 8mm},clip]{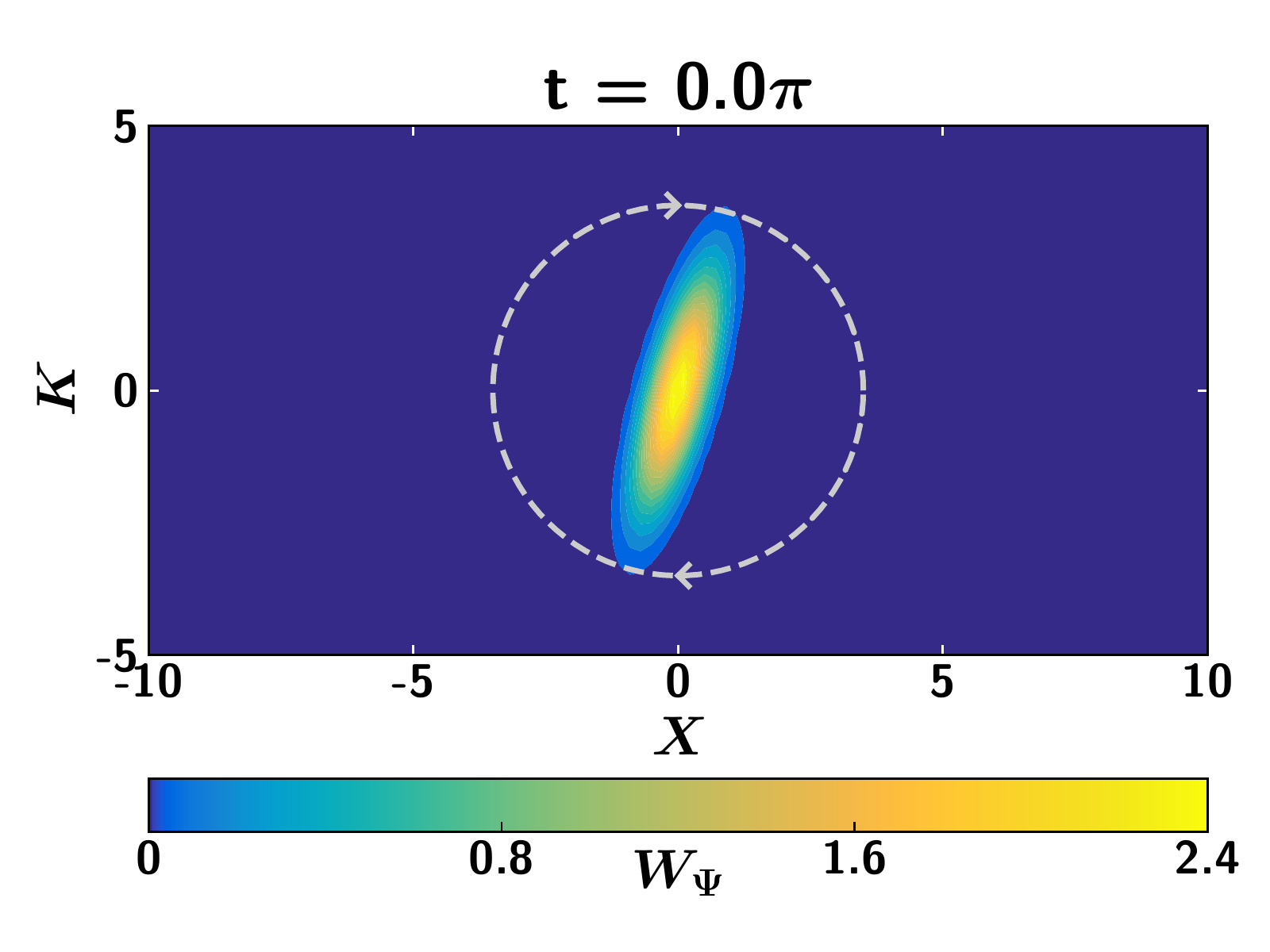}
		\put(85,25){\textbf{\color{white} \large(a)}}
	\end{overpic}
	\begin{overpic}[width=0.32\linewidth,trim={-4mm 7mm 4mm 8mm},clip]{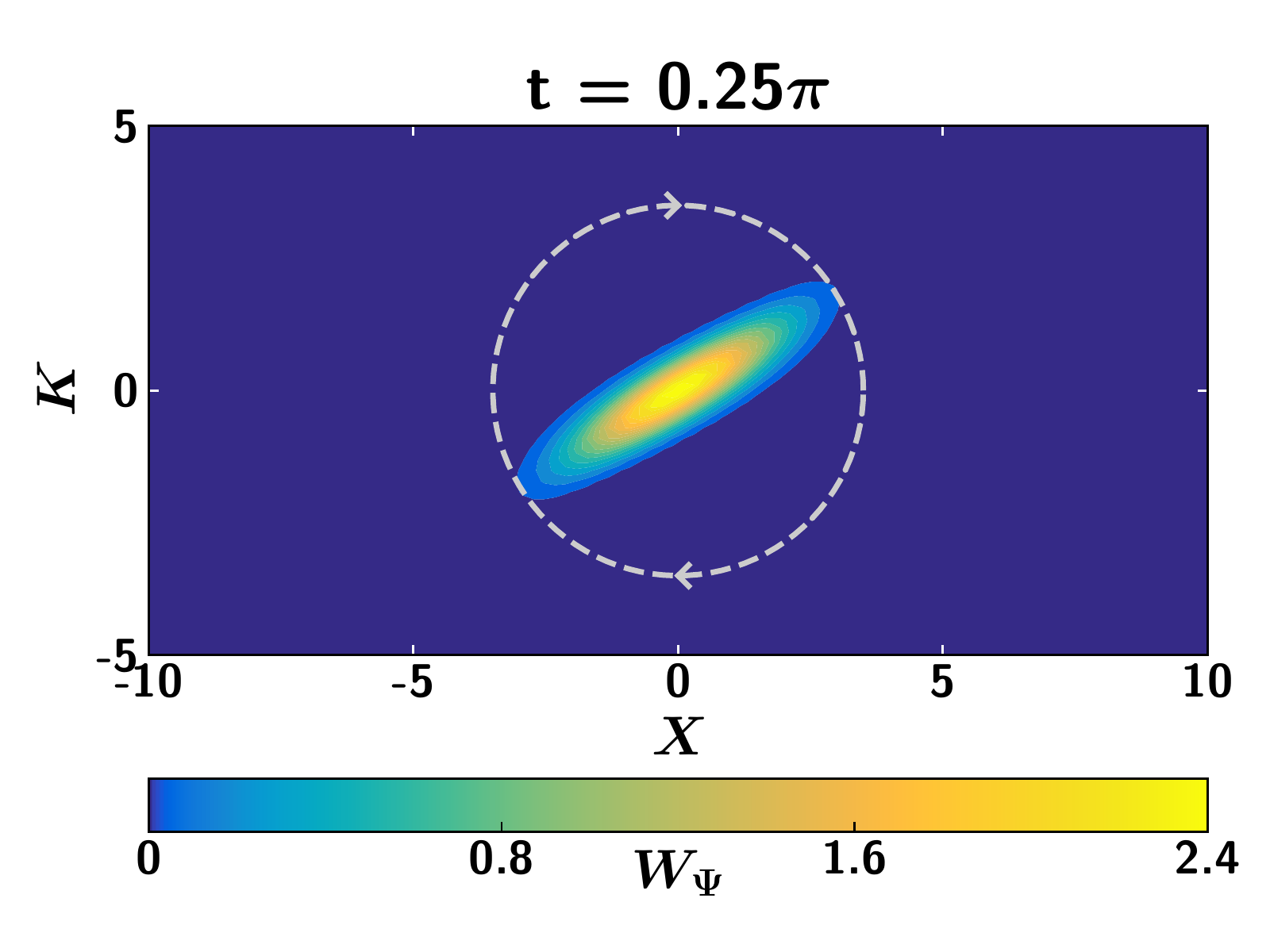}
		\put(85,25){\textbf{\color{white} \large(b)}}
	\end{overpic}
	\begin{overpic}[width=0.32\linewidth,trim={-4mm 7mm 4mm 8mm},clip]{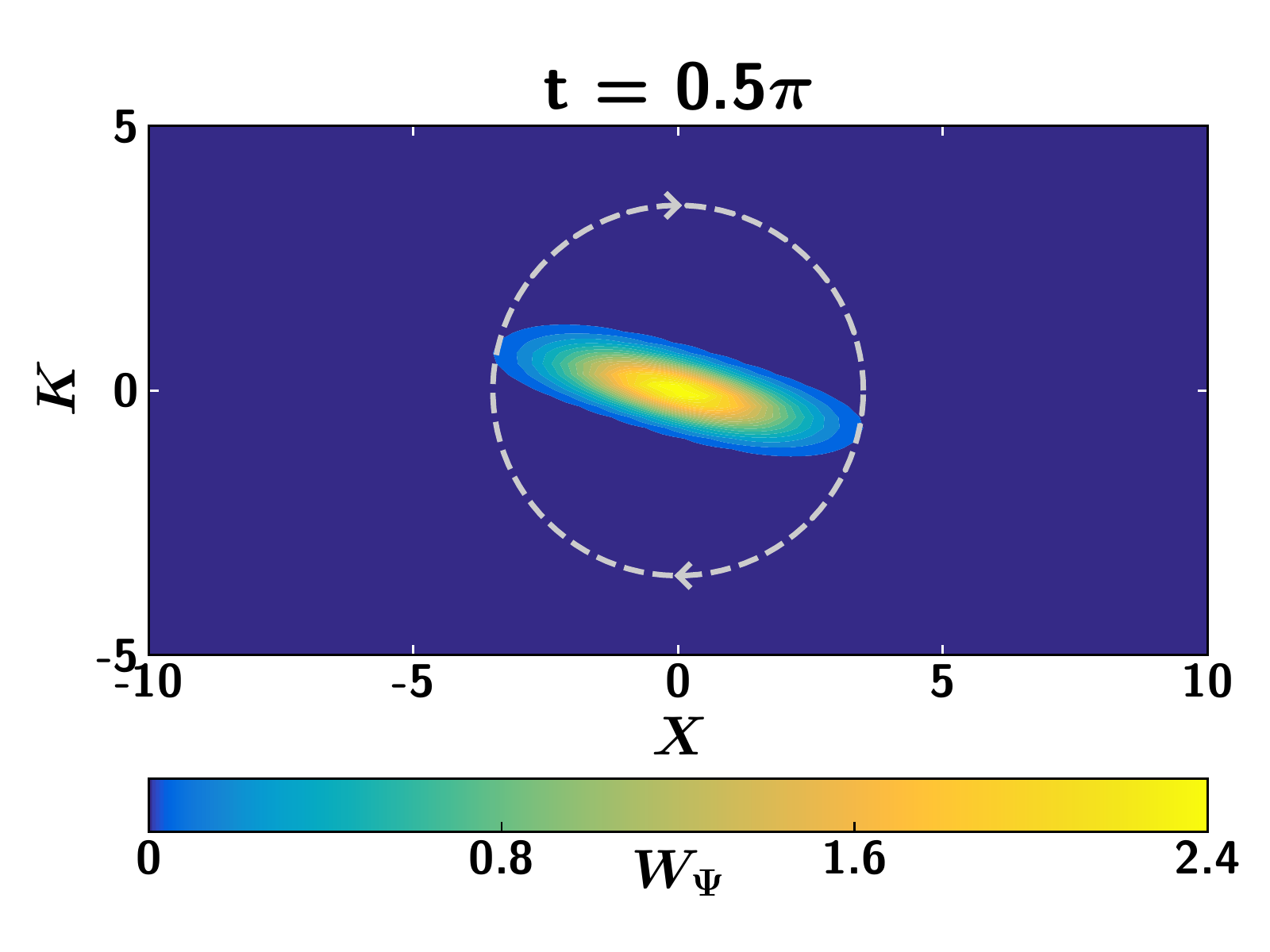}
		\put(85,25){\textbf{\color{white} \large(c)}}
	\end{overpic}

	\begin{overpic}[width=0.32\linewidth,trim={-4mm 7mm 4mm 12mm},clip]{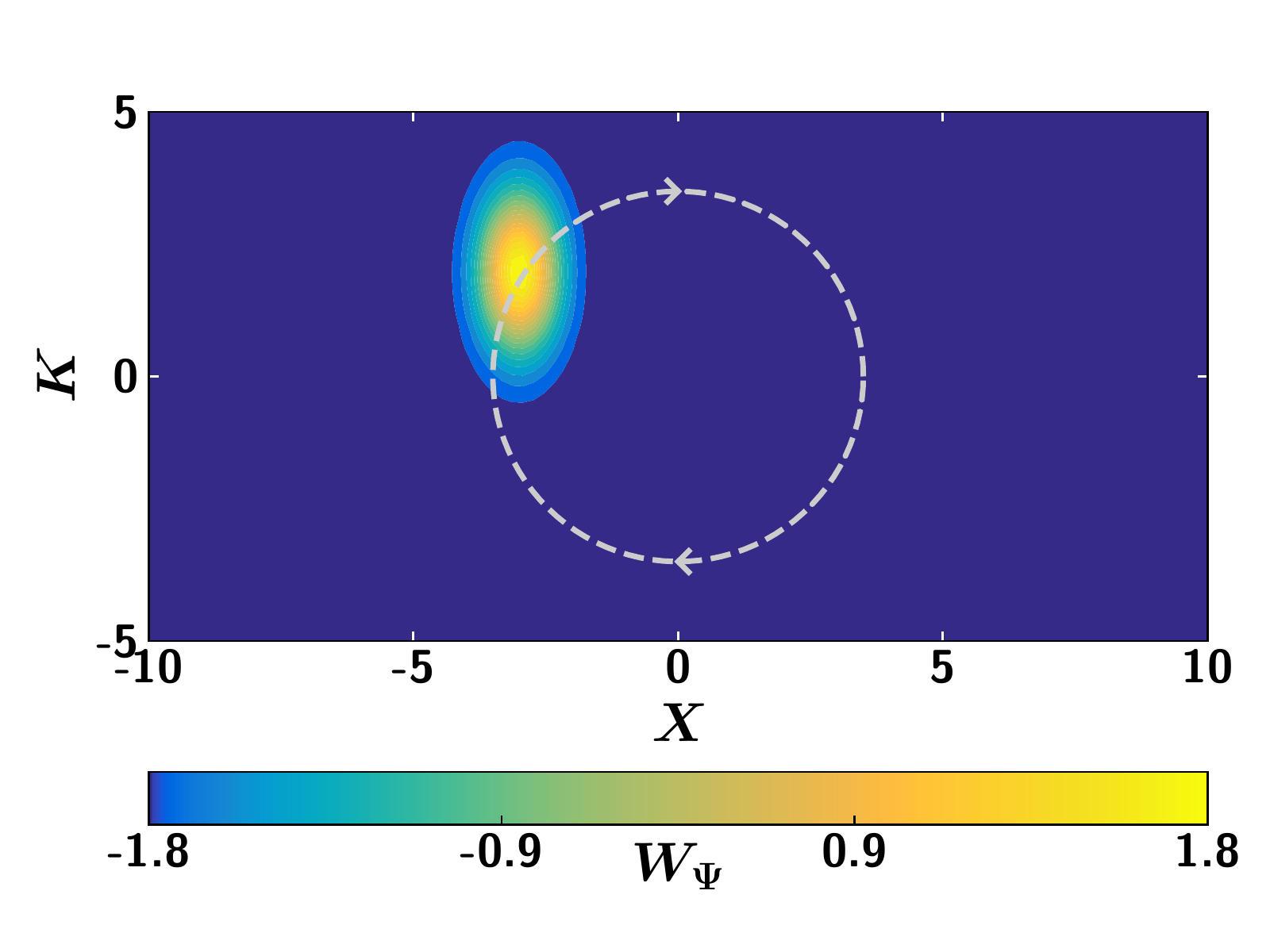}
		\put(85,25){\textbf{\color{white} \large(d)}}
	\end{overpic}
	\begin{overpic}[width=0.32\linewidth,trim={-4mm 7mm 4mm 12mm},clip]{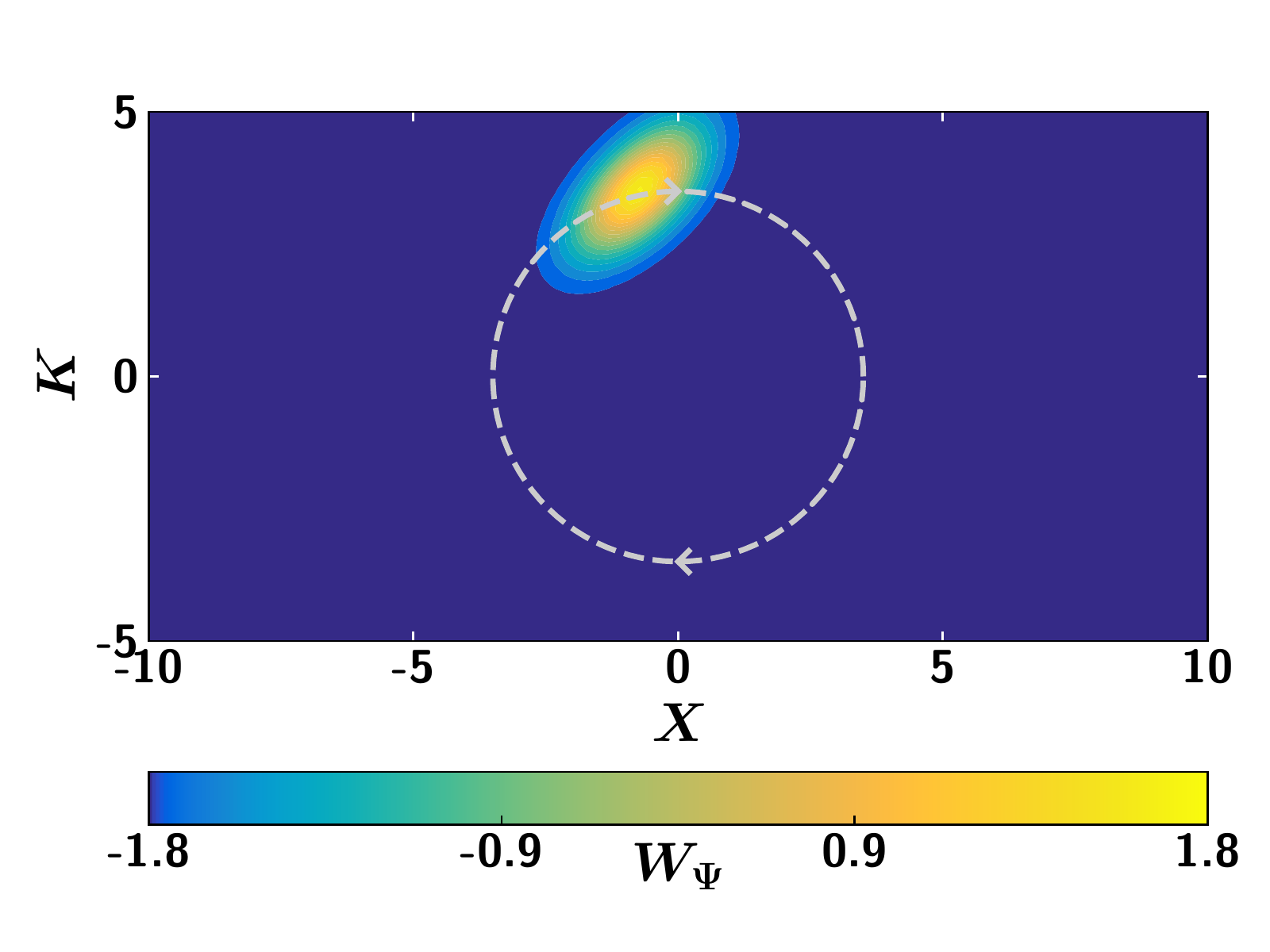}
		\put(85,25){\textbf{\color{white} \large(e)}}
	\end{overpic}
	\begin{overpic}[width=0.32\linewidth,trim={-4mm 7mm 4mm 12mm},clip]{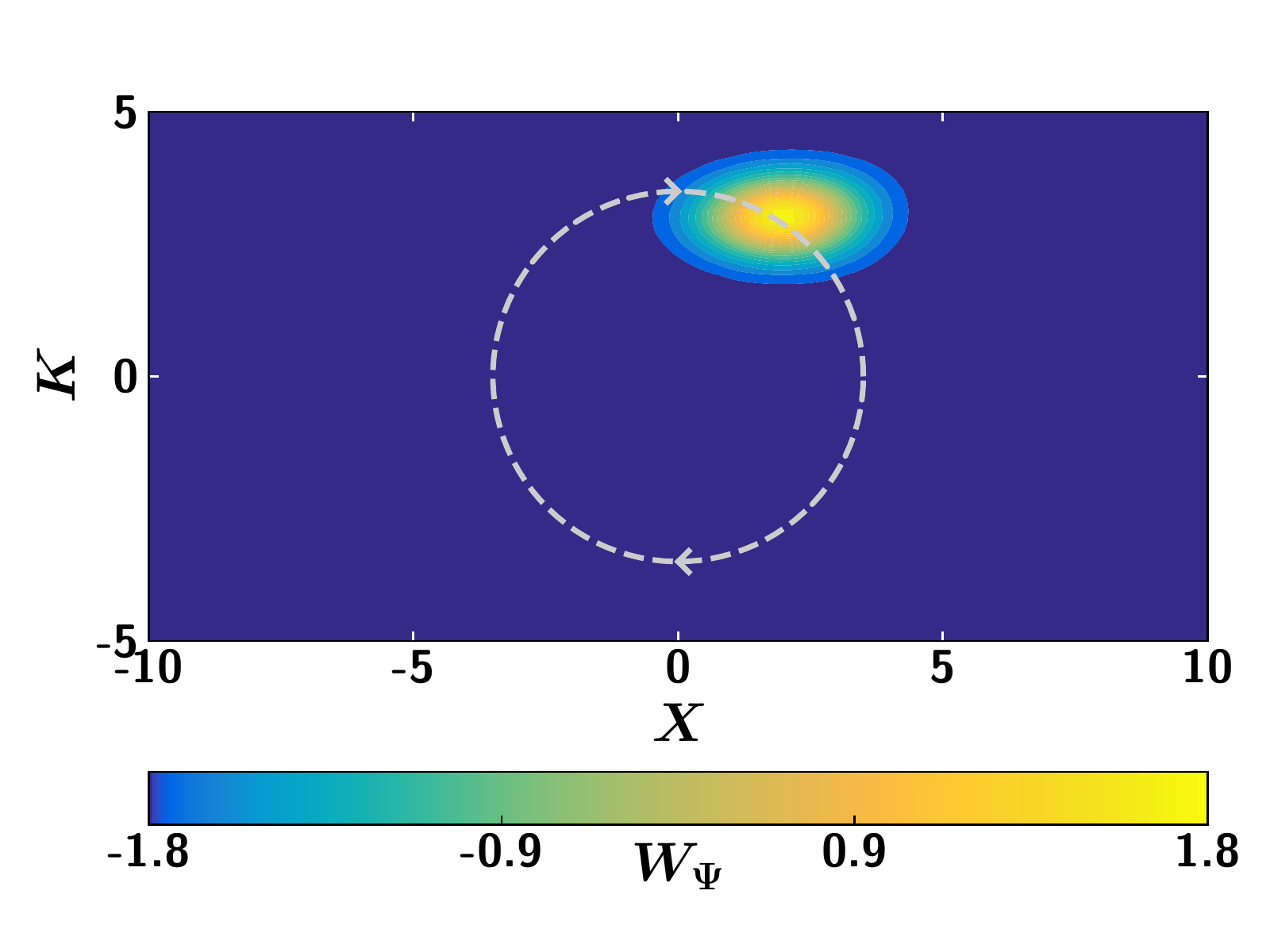}
		\put(85,25){\textbf{\color{white} \large(f)}}
	\end{overpic}

	\begin{overpic}[width=0.32\linewidth,trim={-4mm 7mm 4mm 12mm},clip]{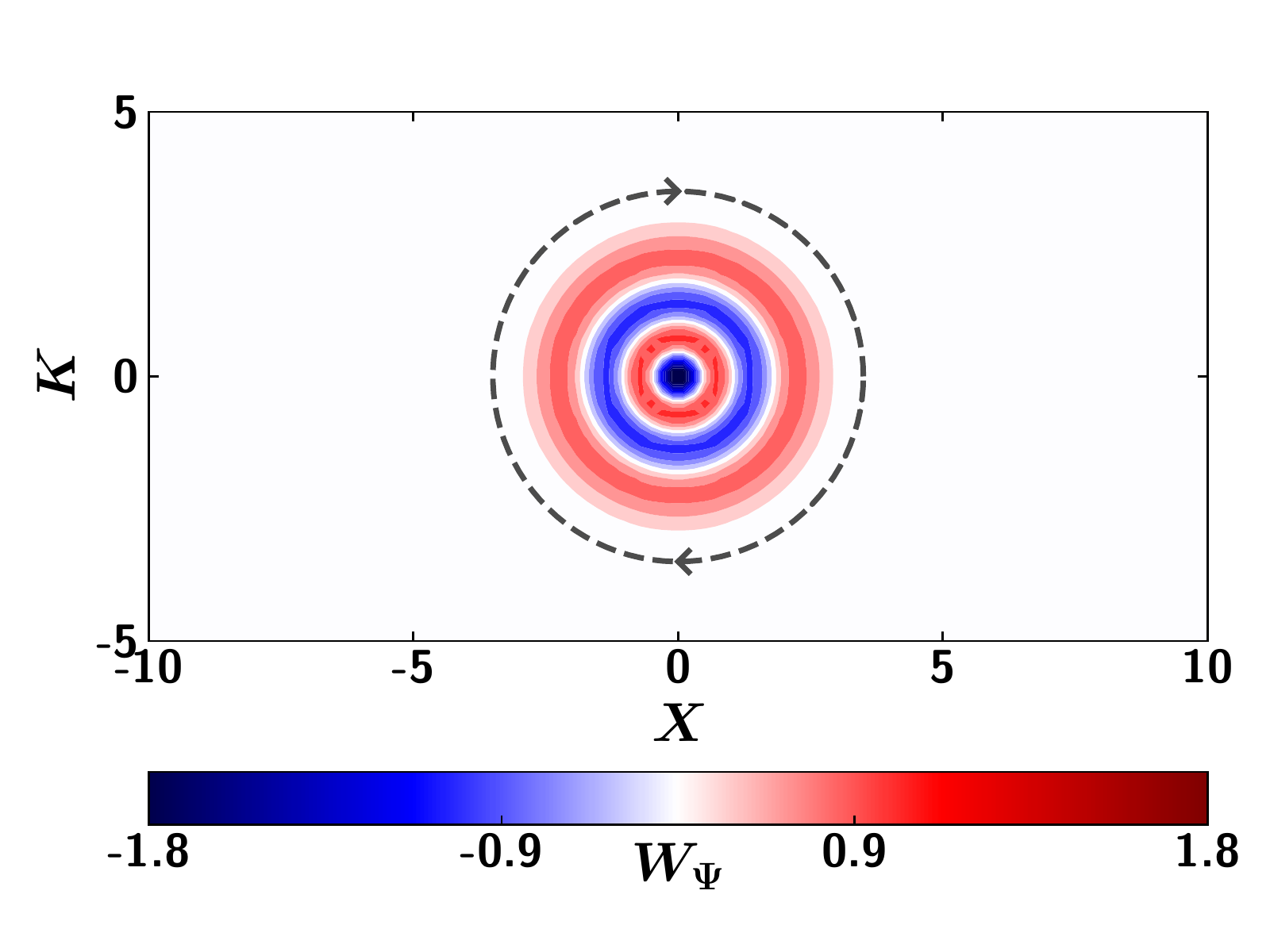}
		\put(85,25){\textbf{\color{black} \large(g)}}
	\end{overpic}
	\begin{overpic}[width=0.32\linewidth,trim={-4mm 7mm 4mm 12mm},clip]{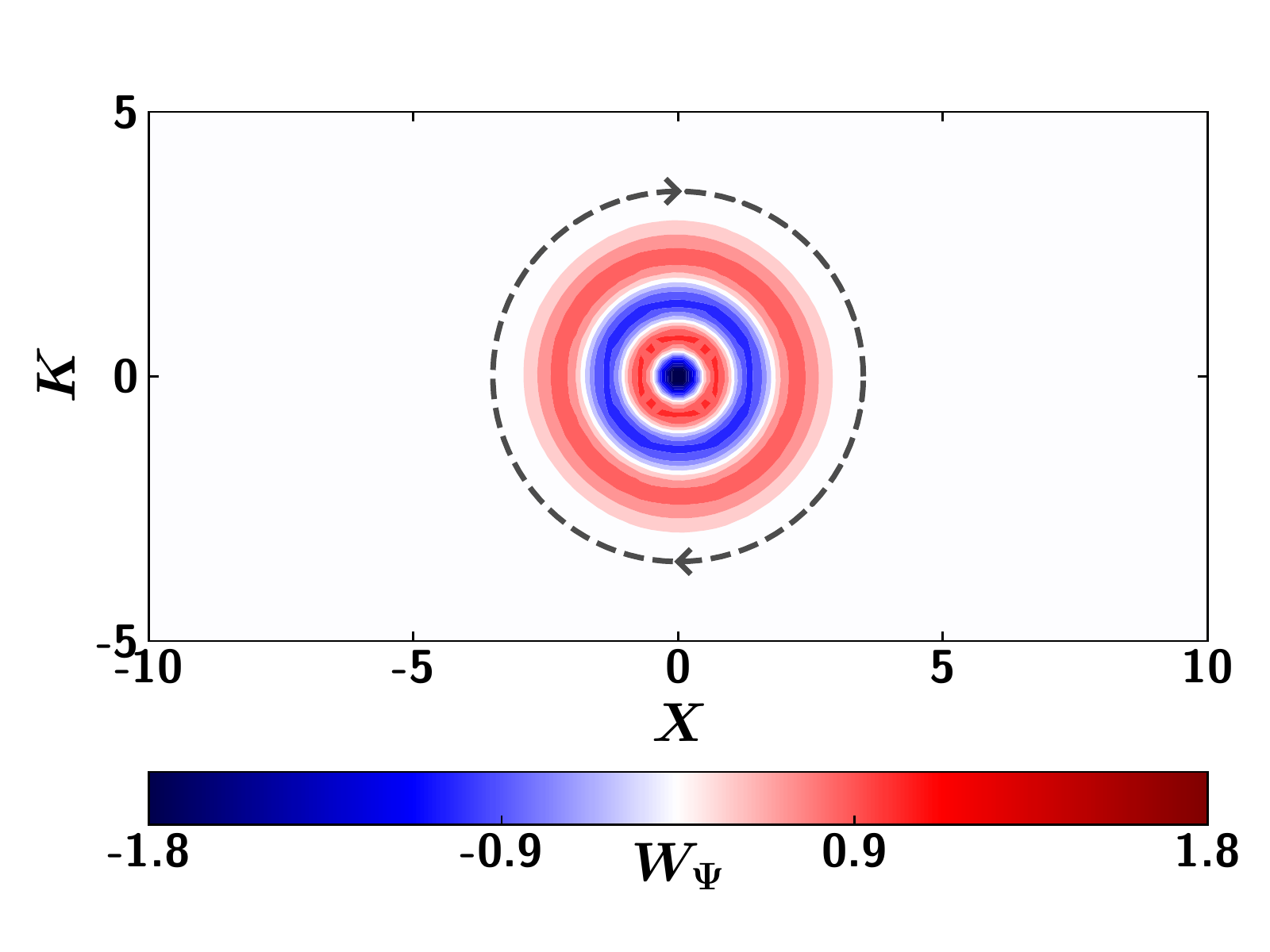}
		\put(85,25){\textbf{\color{black} \large(h)}}
	\end{overpic}
	\begin{overpic}[width=0.32\linewidth,trim={-4mm 7mm 4mm 12mm},clip]{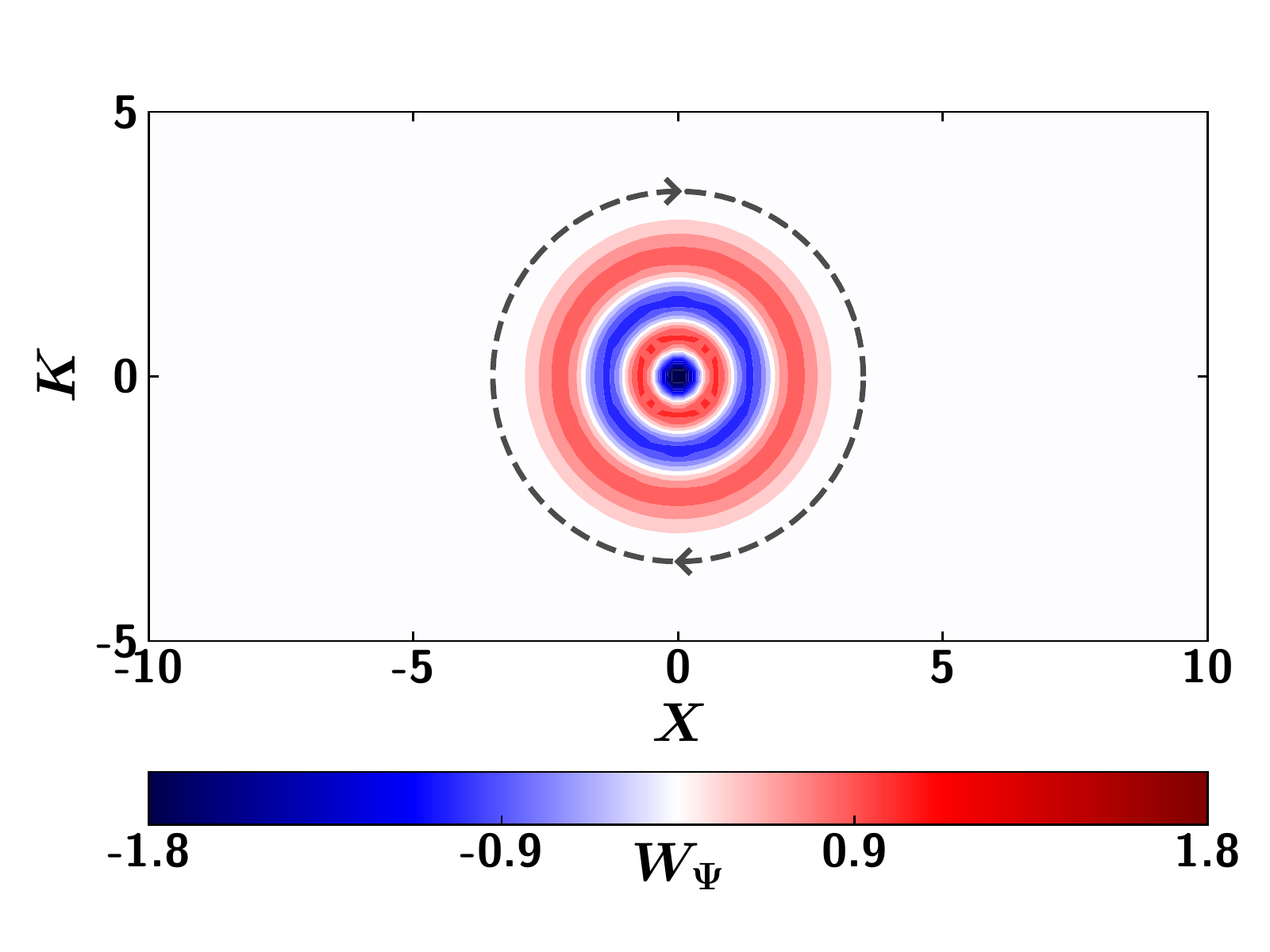}
		\put(85,25){\textbf{\color{black} \large(i)}}
	\end{overpic}
	\caption{\textbf{(a)-(c)} Time evolution of the Wigner function $W_\psi$ for the QHO with initial wavefunction $\psi(x) = \exp[(i-1)x^2]$. The wavefunction is evolved using the iterated NIMT with $\Mat{S}_t$ given by \Eq{eq:4_SEvo} and a step size of $\pi/2000$. $W_\psi$ is subsequently calculated via \Eq{eq:4_wigFUNC}. Shown are snapshots at (a) $t = 0$, (b) $t = \pi/4$, and (c) $t = \pi/2$. \textbf{(d)-(f)} Same as (a)-(c), but with $\psi(x) = \exp[-(x+3)^2+2ix]$. \textbf{(g)-(i)} Same as (a)-(c), but with $\psi(x) = (48\sqrt{\pi})^{-1/2} \exp(-x^2/2 ) H_3(x)$. As expected, the Wigner functions are rotated by the MT along classical trajectories, shown in grey. In the third example, $W_\psi$ is rotationally-symmetric, so it is preserved by the MT.}
	\label{fig:4_FT}
\end{figure}

For our example, we consider the time evolution of three initial states: (i) a chirped Gaussian profile, $\psi(x) = \exp\left[(i-1)x^2\right]$, which is relevant for bit-flip operations in chirp-modulated communication~\cite{Kaminsky05}, (ii) a squeezed coherent state, $\psi(x) = \exp[-(x+3)^2+2ix ]$, which is relevant for high-sensitivity detectors~\cite{Xiao87}, and (iii) the QHO eigenstate corresponding to $n = 3$, namely, $\psi(x) = (48\sqrt{\pi})^{-1/2} \exp)-x^2/2)H_3(x)$, where $H_n(x)$ is the $n$-th degree Hermite polynomial. For these choices, the exact metaplectic image of $\psi(x)$ can be found explicitly from the examples presented in \Ch{ch:MT}, which facilitates benchmarking of our algorithm. They are given respectively by
\begin{subequations}
	\begin{align}
		\label{GAUSS}
		\Psi_t(X) &= \pm \frac{\exp\left[i\frac{(2+2i)\sin(2t) + \cos(2t) - 1}{(8+8i)\sin^2(t)+2\sin(2t)} X^2 \right] }{\sqrt{\cos(t) + (2+2i)\sin(t)}} 
		, \\
		\label{SQCOH}
		\Psi_t(X) &= \pm \frac{\exp\left\{ \frac{i}{2}\cot\left(t \right)X^2 - i\frac{\left[\csc\left(t \right) X - 2 - 6i \right]^2}{2\cot\left(t \right) + 4i} - 9\right\}}{ \sqrt{\cos\left( t \right) +2i \sin\left( t \right)} }
		,\\
		\label{HERM3}
		\Psi_t(X) &= (48\sqrt{\pi})^{-1/2}\exp\left(-\frac{X^2 + 7it}{2}\right) H_3(X) 
		.
	\end{align}
\end{subequations}

\noindent The overall sign is chosen based on the winding number of $\Mat{S}_t$ as discussed in \Ch{ch:MT}: an odd winding number corresponds to the $-$ sign, while an even winding number corresponds to the $+$ sign. Each of these three example functions is evolved in time using the iterated NIMT with a uniform step size of $\Delta t = \pi/2000$, and the resulting Wigner functions are shown in \Fig{fig:4_FT}. Here, $\Psi_t(X)$ is discretized on an equally-spaced grid ranging from $[-10,10]$, and in the final example, the second-order NIMT was used in place of the first-order NIMT. 

As the NIMT is sequentially applied, \Fig{fig:4_FT} shows that the resultant Wigner functions indeed rotate in phase space as expected. (In the third example, $W_\psi$ is rotationally-symmetric, so it is preserved by the MT.) This shows that the iterated NIMT can indeed perform finite transformations with high accuracy. For computing the Fourier transform, which corresponds to $t=\pi/2$, the iterated NIMT is robust to changes in the step size; discretizing the trajectory into $10^2$, $10^3$, and $10^4$ steps all yielded well-behaved solutions. The same is not true for changes in grid resolution, nor in changes of truncation order. Indeed, $\Psi_t(X)$ quickly succumbed to amplified noise when (i) a Chebyshev-spaced grid was used in place of the equally-spaced grid, and (ii) the truncation order was increased beyond third-order.

\begin{figure}
	\centering
	\begin{overpic}[width=0.32\linewidth,trim={3mm 5mm 8mm 3mm},clip]{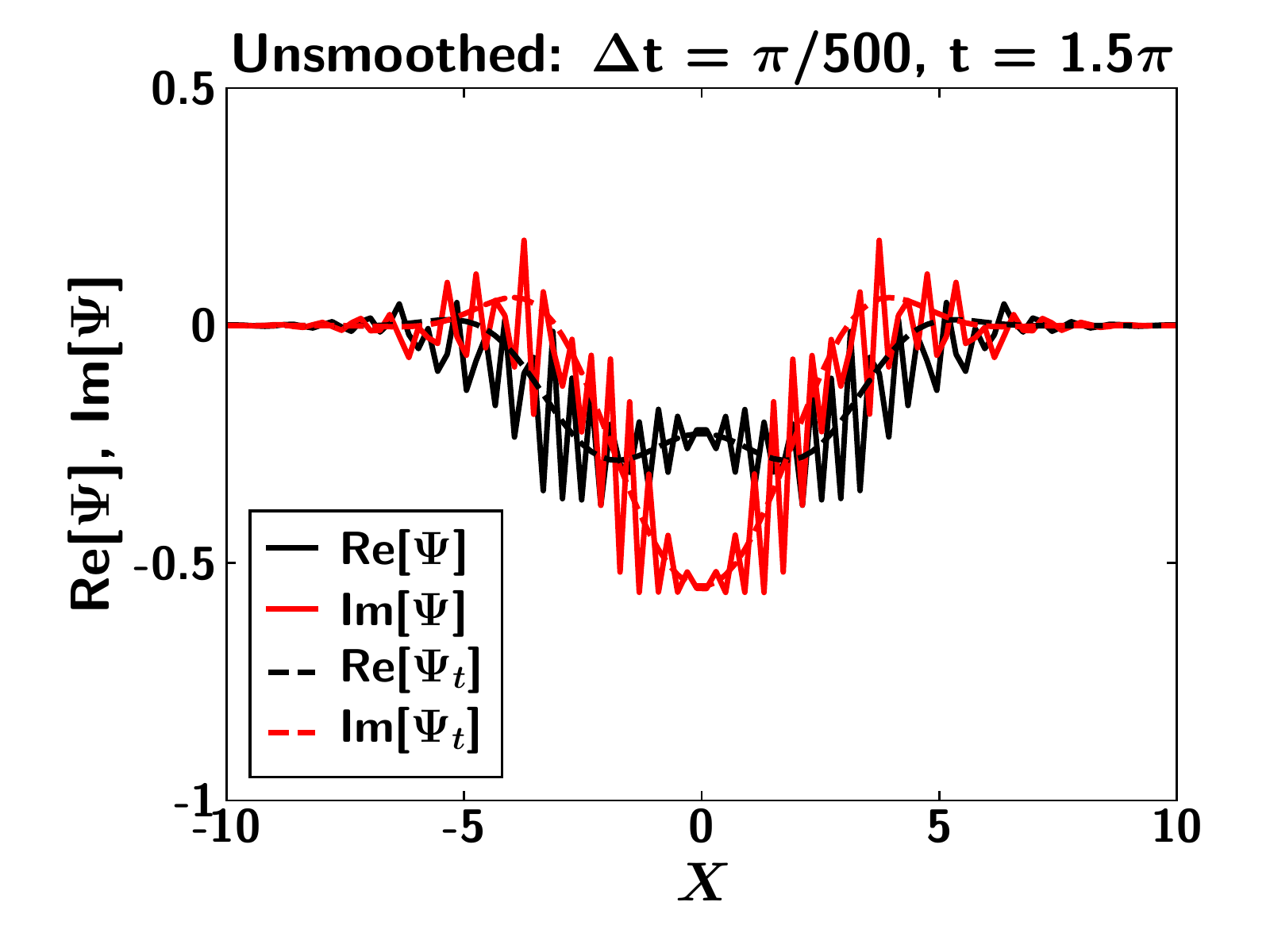}
		\put(85,14){\textbf{\large(a)}}
	\end{overpic}
	\begin{overpic}[width=0.32\linewidth,trim={3mm 5mm 8mm 3mm},clip]{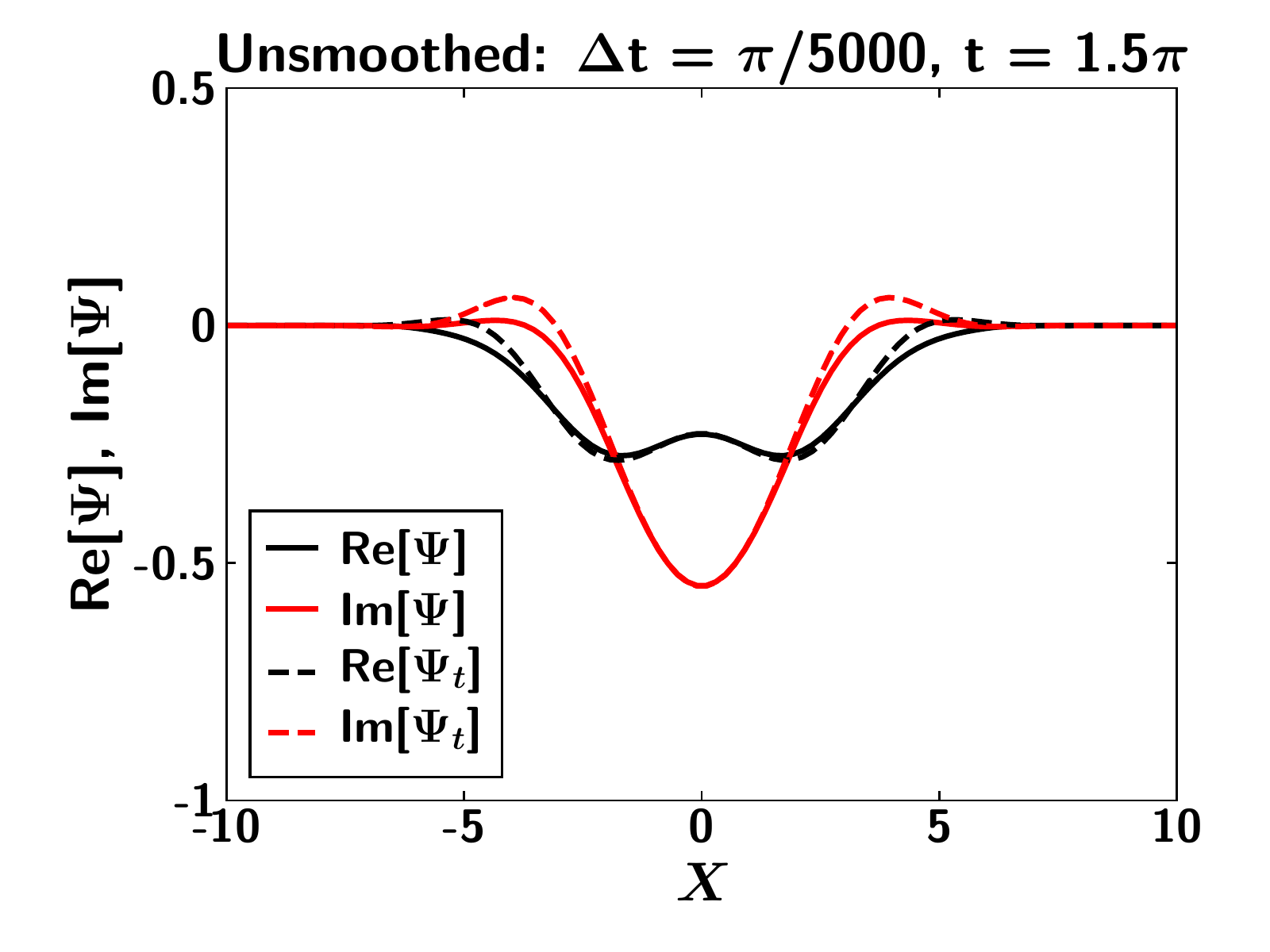}
		\put(85,14){\textbf{\large(b)}}
	\end{overpic}
	\begin{overpic}[width=0.32\linewidth,trim={3mm 5mm 8mm 3mm},clip]{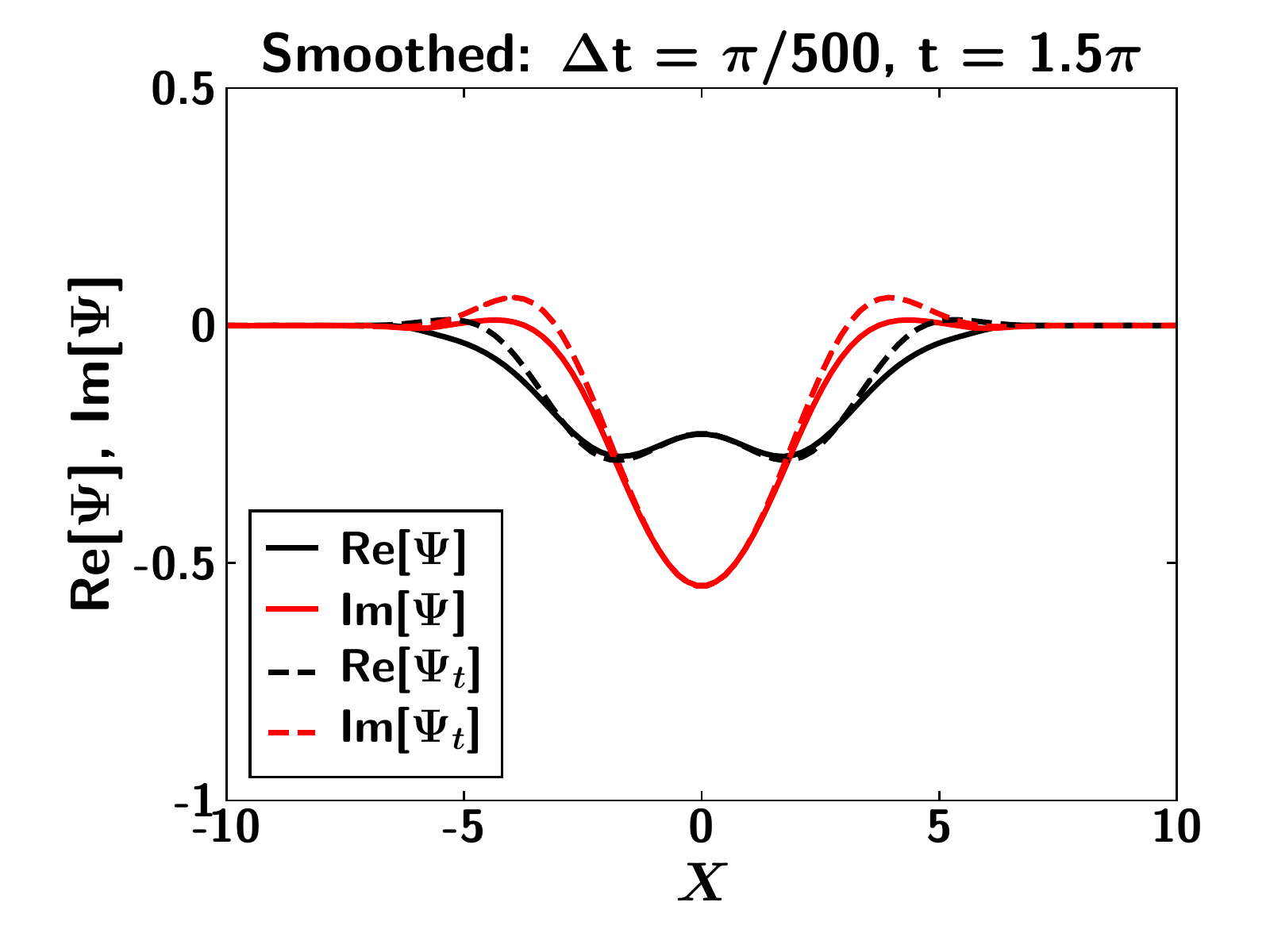}
		\put(85,14){\textbf{\large(c)}}
	\end{overpic}
	\caption{Time evolution of $\psi(x) = \exp[(i-1)x^2]$ using the iterated NIMT for different choices of $\Delta t$, with and without smoothing. The final time is $t = 3\pi/2$ for all three figures. \textbf{(a)} Result of applying the iterated NIMT at a relatively large step size of $\Delta t = \pi/500$ without smoothing. In this case, high-frequency noise is amplified and quickly dominates the signal. \textbf{(b)} Same as (a) but with a smaller step size of $\Delta t = \pi/5000$. \textbf{(c)} Same as (a) but with low-pass filtering. As is clearly seen, low-pass filtering suppresses noise amplification.}
	\label{fig:4_disruption}
\end{figure}

Recall from \Fig{fig:4_IterNIMTstab} that the iterated NIMT is typically a magnifying transformation whose magnification factor depends in a complicated manner on both the path discretization and the input function. For our chosen examples, the magnification is reduced by refining the discretization of the path $\Mat{S}_t$ (\Fig{fig:4_disruption}). When a step size of $\pi/500$ is used, $\Psi_t(X)$ quickly disrupts and becomes dominated by noise. However, refining the discretization by a factor of $10$ avoids the numerical instability and leads to a well-behaved solution. 

We reiterate that the magnification of the NIMT is not reduced for \textit{every} input function by refining the discretization; a rigorous profiling should be performed to determine how the magnification scales with path discretization when using the iterated NIMT in a new application. Alternatively, since the magnification scales with Fourier mode number, occasionally smoothing the signal between NIMT iterations will suppress high-frequency growth. This approach is shown in the final column of \Fig{fig:4_disruption}, where a third-degree Savitzky--Golay filter~\cite{Savitzky64} with a window size of $5$ is applied every $50$ iterations.


\section{Unitary fast near-identity metaplectic transform}

\subsection{Derivation}

As I just showed, using a Taylor expansion to define the NIMT leads to a loss of unitarity and a correspondingly unstable algorithm. To remedy this, let us instead appeal to the manifestly unitary operator representation of the MT provided by \Eq{eq:3_operMT}. In this form, the MT can be readily discretized. Let us consider the 1-D case ($N = 1$) for simplicity. In this case, \Eq{eq:3_operMT} has the $x$-space representation
\begin{align}
	\oper{M}(\Mat{S}) = \pm 
	\exp\left[ 
		\frac{\log A^{-1} }{2}
		\left(
			x \frac{\dd}{\dd x} 
			+ \frac{\dd}{\dd x} x
		\right)
	\right] 
	\exp\left(
		\frac{i}{2} AC x^2
	\right)
	\, 
	\exp\left(
		\frac{i B }{2A } 
		\frac{\dd^2}{\dd x^2}
	\right)
	.
	\label{eq:4_1DMT}
\end{align}

\noindent Let us consider an $m$-point discretization of $x$-space given by the set $\{ x_j \}$, $j = 1, \ldots, m$. We assume that the set $\{ x_j \}$ is distinct and lexicographically ordered such that $x_j < x_k$ when $j < k$. Then, functions of $x$ are discretized by the values on $\{ x_j \}$, and are represented by vectors of length $m$ as
\begin{equation}
	\psi(x)
	\mapsto
	\Vect{\psi} 
	\doteq
	\begin{pmatrix}
	\vspace{2mm}
		\psi_1 &
		\ldots &
		\psi_m
	\end{pmatrix}^\intercal
	, \quad
	\psi_j \doteq \psi(x_j)
	.
\end{equation}

Similarly, operators are represented as $m \times m$ matrices. In particular, the coordinate operator $x$ is represented by the diagonal matrix
\begin{equation}
	\Mat{x} \doteq
	\begin{pmatrix}
		x_1 & & \\
		& \ddots & \\
		& & x_m
	\end{pmatrix}
	,
\end{equation}

\noindent and analytic functions of $x$ are discretized by the formal replacement $x \mapsto \Mat{x}$. Pseudo-differential operators are represented by functions of finite-difference matrices with certain restrictions. For example, to discretize analytic functions of $\dd/\dd x$, we introduce a family of $m \times m$ finite-difference matrices on $\{ x_i\}$, denoted by the set $\{\delta_\ell \}$ with $\ell \ge 1$ such that $\delta_{\ell} \, \Vect{\psi}$ is a suitable discretization of $\dd^\ell \psi /\dd x^\ell$. We then require $\delta_1$ to be the family generator:
\begin{equation}
	\delta_{\ell} = \delta_1^\ell
	, \quad
	\ell = 1, 2, \ldots
	,
	\label{eq:4_deltaREQ}
\end{equation}

\noindent and also to be skew-Hermitian, so that $\delta_1$ faithfully mimics the purely imaginary eigenspectrum of $\dd/\dd x$. Then, any analytic function of $\dd/ \dd x$ is discretized by the formal replacement $\dd/ \dd x \mapsto \delta_1$.

To discretize analytic functions of $\dd^2 / \dd x^2$, we can simply perform the formal replacement $\dd^2 / \dd x^2 \mapsto \delta_1^2$. However, it is often more convenient to introduce an additional family of $m \times m$ even-order finite-difference matrices on $\{ x_i \}$, denoted by the set $\{\Delta_{2k}\}$ with $k \ge 1$ such that $\Delta_{2k} \, \Vect{\psi}$ is a suitable discretization of $\dd^{2k} \psi /\dd x^{2k}$. We then require $\Delta_2$ to be the family generator:
\begin{equation}
	\Delta_{2k} = \Delta_2^k
	, \quad
	k = 1, 2, \ldots 
	,
	\label{eq:4_DeltaREQ}
\end{equation}

\noindent and also to be Hermitian and negative semi-definite, so that $\Delta_2$ faithfully mimics the negative semi-definite eigenspectrum of $\dd^2 / \dd x^2$. (Note that $\{\Delta_{2k} \}$ need not coincide with $\delta_{\ell}$ at even $\ell$, but clearly, any suitable $\{ \delta_\ell \}$ also constitutes a suitable $\{ \Delta_{2k} \}$.) Then, any analytic function of $\dd^2/ \dd x^2$ is discretized by the formal replacement $\dd^2 / \dd x^2 \mapsto \Delta_2$. In particular, \Eq{eq:4_1DMT} is discretized to yield the discrete MT (dMT), given as
\begin{align}
	\oper{M}(\Mat{S})
	\mapsto
	\Mat{M}(\Mat{S})
	\doteq
	\pm 
	\exp\left(
		\log A^{-1} \,
		\frac{
			\Mat{x} \, \delta_1
			+ \delta_1 \Mat{x}
		}{2}
	\right)
	\exp
	\left(
		\frac{i A C}{2} \Mat{x}^2
	\right)
	\,
	\exp\left(
		\frac{i B}{2A} 
		\Delta_2
	\right)
	.
	\label{eq:4_dMT}
\end{align}

When $\{ x_j \}$ are equally spaced with grid spacing $h$, then suitable choices of $\delta_1$ and $\Delta_2$ are the central-difference matrices. These matrices are of Toeplitz form~\cite{Strang14}; explicitly, a 2nd-order finite-difference scheme has (in the case of $4 \times 4$ matrices with empty spaces denoting null entries)%
\begin{align}
	\delta_1^{(2)}
	= \frac{1}{2h}
	\begin{pmatrix}
		0 &      1 &    &   \\
		-1 & 0 &  1 &  \\
		 &     -1 &  0 & 1 \\
		 &  & -1 & 0
	\end{pmatrix}
	,
	\quad
	\Delta_2^{(2)}
	=
	\frac{1}{h^2}
	\begin{pmatrix}
		-2 &      1 &     &  \\
		1 & -2  &  1 & \\ 
		 &      1 & -2  & 1\\
		 &  & 1 & -2
	\end{pmatrix}
	,
	\label{eq:4_2ndDELTA}
\end{align}

\noindent while a 4th-order finite-difference scheme has (in the case of $5 \times 5$ matrices)%
\begin{align}
	\delta_1^{(4)}
	= \frac{1}{12h}
	\begin{pmatrix}
		0 &      8 &     -1 &    & \\
		-8 & 0 & 8 & -1 & \\
		1 & -8 & 0 &  8 & -1 \\
		 &      1 &     -8 &  0 & 8\\
		 &  & 1 & -8 & 0
	\end{pmatrix}
	, \quad
	\Delta_2^{(4)}
	=
	\frac{1}{12h^2}
	\begin{pmatrix}
		-30 &     16 &     -1 &   &    \\
		16 & -30 & 16 &  -1 & \\ 
		-1 & 16 & -30 &  16 & -1\\
		 &     -1 &     16 & -30 & 16\\
		 &  & -1 & 16 & -30
	\end{pmatrix}
	,
	\label{eq:4_4thDELTA}
\end{align}

\noindent and a 6th-order finite-difference scheme has (in the case of $6 \times 6$ matrices%
\begin{align}
	\delta_1^{(6)}
	= \frac{1}{60h}
	\begin{pmatrix}
		0 &     \hspace{-2.5mm}45 &     \hspace{-2.5mm}-9 &      \hspace{-2.5mm}1 &  \hspace{-2.5mm} &  \hspace{-2.5mm} \\
		-45 & \hspace{-2.5mm}0 & \hspace{-2.5mm}45 & \hspace{-2.5mm}-9 & \hspace{-2.5mm} 1 & \hspace{-2.5mm}\\
		9 & \hspace{-2.5mm}-45 & \hspace{-2.5mm}0 & \hspace{-2.5mm}45 & \hspace{-2.5mm}-9  & \hspace{-2.5mm}1\\
		-1 & \hspace{-2.5mm}9 & \hspace{-2.5mm}-45 & \hspace{-2.5mm}0 & \hspace{-2.5mm}45 & \hspace{-2.5mm}-9\\
		 &     \hspace{-2.5mm}-1 &     \hspace{-2.5mm} 9 &   \hspace{-2.5mm} -45 &  \hspace{-2.5mm}0 & \hspace{-2.5mm}45\\
		 &  \hspace{-2.5mm}& \hspace{-2.5mm}-1 & \hspace{-2.5mm}9 & \hspace{-2.5mm}-45 &\hspace{-2.5mm}0
	\end{pmatrix}
	, \quad
	\Delta_2^{(6)}
	=
	\frac{1}{180h^2}
	\begin{pmatrix}
		-490 & \hspace{-2.5mm}   270 &  \hspace{-2.5mm}  -27 &  \hspace{-2.5mm}    2 &  \hspace{-2.5mm}  &  \hspace{-2.5mm}   \\
		270 &\hspace{-2.5mm} -490 &\hspace{-2.5mm} 270 & \hspace{-2.5mm}-27 & \hspace{-2.5mm}   2 &\hspace{-2.5mm} \\ 
		-27 & \hspace{-2.5mm}270 & \hspace{-2.5mm}-490 & \hspace{-2.5mm}270 &\hspace{-2.5mm}  -27 &\hspace{-2.5mm} 2\\
		2 &\hspace{-2.5mm} -27 & \hspace{-2.5mm}270 &\hspace{-2.5mm} -490 &\hspace{-2.5mm}  270 & \hspace{-2.5mm}-27\\
		 & \hspace{-2.5mm}     2 & \hspace{-2.5mm}   -27 & \hspace{-2.5mm}   270 &\hspace{-2.5mm} -490 & \hspace{-2.5mm}270\\
		 & \hspace{-2.5mm} & \hspace{-2.5mm}2 &\hspace{-2.5mm} -27 & \hspace{-2.5mm}270 & \hspace{-2.5mm}-490 
	\end{pmatrix}
	.
	\label{eq:4_6thDELTA}
\end{align}

\noindent Clearly, each $\delta_1$ is skew-symmetric while each $\Delta_2$ is symmetric, as required; hence, the dMT \eq{eq:4_dMT} is unitary, as we show explicitly below. (In contrast, forward and backward finite-difference matrices do not have the required symmetry for use in the dMT.) In \Sec{sec:4_EXpade}, we shall compare the 2nd-order, 4th-order, and 6th-order dMTs that use \Eqs{eq:4_2ndDELTA}, \eq{eq:4_4thDELTA}, and \eq{eq:4_6thDELTA}, respectively.

Let us now consider simplifications to \Eq{eq:4_dMT} when $\Mat{S}$ is near-identity, that is, $|A| \sim 1$ and $|B| \sim |C| \ll 1$. Instead of a Taylor expansion, we shall now use a [1/1] Pad\'e approximation~\cite{Press07,Olver10a} for the matrix exponential, given explicitly as
\begin{equation}
	\exp(\Mat{H})
	\approx 
	\left(
		\IMat{N} 
		- \frac{\Mat{H}}{2}
	\right)
	\left(
		\IMat{N} 
		+ \frac{\Mat{H}}{2}
	\right)
	, \quad 
	\| \Mat{H} \| \ll 1
	.
	\label{eq:4_11pade}
\end{equation}

\noindent Hence, we approximate $\Mat{M}(\Mat{S}) \approx \Mat{N}(\Mat{S})$, where we have introduced
\begin{align}
	\Mat{N}(\Mat{S}) 
	=
	&\left(
		\IMat{N}
		+ \log A
		\frac{
			\Mat{x} \, \delta_1
			+ \delta_1 \Mat{x}
		}{4}
	\right)^{-1}
	\left(
		\IMat{N}
		- \log A
		\frac{
			\Mat{x} \, \delta_1
			+ \delta_1 \Mat{x}
		}{4}
	\right)
	\nonumber\\
	&\times
	\exp
	\left(
		\frac{i A C}{2} \Mat{x}^2
	\right)
	\left(
		\IMat{N}
		- \frac{i B}{4A} \Delta_2
	\right)^{-1}
	\left(
		\IMat{N}
		+ \frac{i B}{4A} \Delta_2
	\right)
	,
	\label{eq:4_dNIMT}
\end{align}

\noindent and we have chosen the overall $+$ sign in \Eq{eq:4_dMT}. We call \Eq{eq:4_dNIMT} the discrete NIMT (dNIMT). Note that there is no need to approximate $\exp(i A C \Mat{x}^2/2)$ because it is diagonal and therefore trivial to compute. Also note that in fact, any diagonal ([$r/r$] with integer $r$) Pad\'e approximation is suitable for use in the dNIMT (as we show below); we choose the [1/1] approximation for simplicity.


\subsection{Unitarity verification}

Since $\Delta_2$ is Hermitian, the matrix $iB \Delta_2/2A $ is skew-Hermitian, which implies that $\exp(iB \Delta_2/2A)$ is unitary. Likewise, the matrix $\exp(i A C \Mat{x}^2/2)$ is unitary as well. Lastly, since $\Mat{x}$ is Hermitian and $\delta_1$ is skew-Hermitian, the matrix $\Mat{x} \, \delta_1+ \delta_1\Mat{x}$ is skew-Hermitian:
\begin{align}
	( \Mat{x} \, \delta_1+ \delta_1\Mat{x} )^\dagger
	= \delta_1^\dagger \Mat{x} + \Mat{x} \delta_1^\dagger
	= - \left( \Mat{x} \, \delta_1+ \delta_1\Mat{x} \right)
	.
\end{align}

\noindent Consequently, $\exp[- \log{A} ( \Mat{x} \, \delta_1+ \delta_1\Mat{x} )/2 ]$ is unitary. As the product of unitary matrices, we conclude that $\Mat{M}(\Mat{S})$ defined by \Eq{eq:4_dMT} is unitary as well.

Regarding the dNIMT, it is well-known that the [1/1] Pad\'e approximation preserves the unitarity of matrix exponentials by acting as a Cayley transform for the argument of the matrix exponential~\cite{Eves80,Diele98,Iserles01,Golub13,Zhang20,Fu22}. Here, we show this fact explicitly. Suppose that $\Mat{H}$ is a skew-Hermitian matrix. It therefore possesses a complete set of eigenvectors with imaginary eigenvalues, denoted by $\Vect{\lambda}_j$ and $i \lambda_j$ with $\lambda_j$ real. Consequently, the set $\{ \Vect{\lambda}_j \}$ also satisfy the eigenvalue relation
\begin{gather}
	\left(
		\IMat{N}
		- \frac{\Mat{H}}{2}
	\right)^{-1}
	\left(
		\IMat{N}
		+ \frac{\Mat{H}}{2}
	\right)
	\Vect{\lambda}_j
	=
	e^{2 i \theta_j} \Vect{\lambda}_j
	, \quad
	\theta_j 
	\doteq \textrm{arg} 
	\left(
		1 
		+ i \frac{\lambda_j}{2} 
	\right)
	.
\end{gather}

\noindent Hence, the [1/1] Pad\'e approximation of the matrix exponential \eq{eq:4_11pade} for skew-Hermitian $\Mat{H}$ is unitary. Since the Pad\'e approximations included in \Eq{eq:4_dNIMT} are all of this form, we conclude that the dNIMT is unitary as well. (Note that with minor modifications, the above proof also holds for a general diagonal Pad\'e approximation.)


\subsection{Runtime estimate}

Here we estimate the computational complexity of the dMT for a uniform grid. First, let us estimate the cost to construct the matrix factors in $\Mat{M}(\Mat{S})$. Although formally dense, the matrix exponential of a banded matrix is `pseudo-sparse' and essentially banded, since the matrix elements rapidly decrease away from the main diagonal~\cite{Iserles00}. This means that the matrix exponential can be computed (approximately) in linear time, \ie $O(m)$\cite{Benzi07}. Thus, we estimate the construction of the three matrix factors in $\Mat{M}(\Mat{S})$ to be $O(m)$. Alternatively, \Eqs{eq:4_exp1} and \eq{eq:4_exp2} below can be directly approximated in $O(m)$ time by using a low-order `scaling-squaring' method based on Pad\'e~\cite{AlMohy09} or truncated Taylor approximants~\cite{AlMohy11}.

Assuming the matrix factors have been constructed, we next estimate the cost to perform the dMT as
\begin{equation}
	\Vect{\Psi} = \Mat{M}(\Mat{S}) \Vect{\psi}
	.
\end{equation}

\noindent Suppose that the matrix exponentials involving $\Delta_2$ and $\delta_1$ have been approximated by banded matrices with bandwidths of $b_\Delta$ and $b_\delta$ respectively. Since matrix-vector multiplication involving a banded matrix of bandwidth $b$ requires $O(bm)$ operations, computing
\begin{equation}
	\Vect{v}_1 \doteq 
	\exp
	\left(
		\frac{i B}{2 A}
		\Delta_2
	\right)
	\Vect{\psi}
	\label{eq:4_exp1}
\end{equation}

\noindent requires $O(b_\Delta m)$ operations. Similarly, computing
\begin{equation}
	\Vect{v}_2 \doteq 
	\exp
	\left(
		\frac{i A C}{2}
		\Mat{x}^2
	\right)
	\Vect{v}_1
\end{equation}

\noindent requires $O(m)$ operations, and computing 
\begin{equation}
	\Vect{\Psi}
	=
	\exp\left(
		\log A^{-1} \,
		\frac{
			\Mat{x} \, \delta_1
			+ \delta_1 \Mat{x}
		}{2}
	\right)
	\Vect{v}_2
	\label{eq:4_exp2}
\end{equation}

\noindent requires $O(b_\delta m)$ operations. Since these operations are performed in sequence, we conclude that the dMT can be approximately computed in linear time. (Note that the approximation is due to representing the matrix exponentials with banded matrices.)

Let us now present a runtime estimate for the dNIMT. As for the dMT, we restrict the analysis to a uniform grid. We shall also consider only the lowest-order approximations for $\delta_1$ and $\Delta_2$ given by \Eqs{eq:4_2ndDELTA}; the analysis for higher-order approximations is analogous. Since $\Delta_2$ is tridiagonal, computing
\begin{equation}
	\Vect{v}_3
	\doteq
	\left(
		\IMat{N}
		+ \frac{i B}{4A} \Delta_2
	\right) \Vect{\psi}
\end{equation}

\noindent requires $O(3m)$ operations. Next, note that $\Vect{v}_4$, defined as
\begin{equation}
	\Vect{v}_4
	\doteq
	\left(
		\IMat{N}
		- \frac{i B}{4A} \Delta_2
	\right)^{-1} \Vect{v}_3
	,
\end{equation}

\noindent is the solution to the tridiagonal linear system
\begin{equation}
	\left(
		\IMat{N}
		- \frac{i B}{4A} \Delta_2
	\right)
	\Vect{v}_4
	=
	\Vect{v}_3
	,
\end{equation}

\noindent which can be obtained in $O(m)$ computations using a tridiagonal Gaussian elimination algorithm~\cite{Press07}. Next, since $\Mat{x}$ is diagonal, computing
\begin{equation}
	\Vect{v}_5
	\doteq
	\exp \left(
		\frac{i A C}{2} \Mat{x}^2
	\right)
	\Vect{v}_4
\end{equation}

\noindent requires $O(m)$ operations. Next, since $\delta_1$ is tridiagonal, computing
\begin{equation}
	\Vect{v}_6
	\doteq
	\left(
		\IMat{N}
		- \log A
		\frac{
			\Mat{x} \, \delta_1
			+ \delta_1 \Mat{x}
		}{4}
	\right)
	\Vect{v}_5
\end{equation}

\noindent requires $O(3m)$ operations. Lastly, rather than directly computing
\begin{equation}
	\Vect{\Psi}
	\doteq
	\left(
		\IMat{N}
		+ \log A
		\frac{
			\Mat{x} \, \delta_1
			+ \delta_1 \Mat{x}
		}{4}
	\right)^{-1}
	\Vect{v}_6
	,
\end{equation}

\noindent we obtain $\Vect{\Psi}$ by solving the tridiagonal linear system
\begin{equation}
	\left(
		\IMat{N}
		+ \log A
		\frac{
			\Mat{x} \, \delta_1
			+ \delta_1 \Mat{x}
		}{4}
	\right)
	\Vect{\Psi}
	=
	\Vect{v}_6
	,
\end{equation}

\noindent which requires $O(m)$ computations. Thus, by performing these computations in sequence, the dNIMT can be computed in linear time, \ie $O(m)$.


\subsection{Convergence and stability}

Analogous to the analysis of finite-difference approximations~\cite{Suli03}, we can consider the local and global convergence of the iterated dNIMT to the dMT. The local error convergence is determined by the truncation error of the Pad\'e approximation. Indeed, since
\begin{align}
	\exp(\Mat{H})
	-
	\left( \IMat{N} - \frac{\Mat{H}}{2} \right)^{-1}
	\left(\IMat{N} + \frac{\Mat{H}}{2}\right)
	&= O(\Delta t^3)
\end{align}

\noindent for any infinitesimal matrix $\Mat{H}$ that satisfies $\| \Mat{H} \| \sim O(\Delta t)$, it is clear that
\begin{equation}
	\Mat{M}(\Mat{S}) = \Mat{N}(\Mat{S}) + O(\Delta t^3)
	\label{eq:4_localCONVERG}
	.
\end{equation}

\noindent Consequently, the local error between the dNIMT and the dMT converges with a rate of $3$.

To assess the global convergence, let us introduce the sequence of single-step iterates $\{\Mat{m}_j\}$ that are computed during the iterated dNIMT [\Eq{eq:4_iterNIMT}] as
\begin{equation}
	\Mat{m}_{j} = \Mat{N}(\Mat{S}_j) \Mat{m}_{j-1}
	, \quad
	\Mat{m}_0 \doteq \Mat{M}[\Mat{S}(0)]
	,
	\label{eq:4_1STEP}
\end{equation}

\noindent along with the sequence $\{\Mat{M}_j \}$ obtained by the iterating the dMT as
\begin{equation}
	\Mat{M}_j = \Mat{M}(\Mat{S}_j) \Mat{M}_{j - 1}
	, \quad
	\Mat{M}_0 = \Mat{m}_0
	.
\end{equation}

\noindent The local convergence of the dNIMT [\Eq{eq:4_localCONVERG}] implies
\begin{equation}
	\Mat{M}_j 
	= \Mat{N}(\Mat{S}_j) \Mat{M}_{j-1}
	+ O(\Delta t ^3)
	.
	\label{eq:4_1STEPexact}
\end{equation}

\noindent Hence, subtracting \Eq{eq:4_1STEP} from \Eq{eq:4_1STEPexact} yields
\begin{equation}
	\Mat{M}_j 
	- \Mat{m}_j = 
	\Mat{N}(\Mat{S}_j )
	\left\{
		\Mat{M}_{j-1}
		- \Mat{m}_{j-1}
		\nullFrac
	\right\}
	+ O(\Delta t^3)
	.
\end{equation}

\noindent Since $\Mat{N}(\Mat{S}_j)$ is unitary, we can bound the global error as
\begin{equation}
	\left\| 
		\Mat{M}_j 
		- \Mat{m}_j 
	\right\|
	\le
	\left\| 
		\Mat{M}_{j-1}
		- \Mat{m}_{j-1} 
	\right\|
	+ T \Delta t^3
\end{equation}

\noindent for some positive constant $T$. Since $\Mat{m}_0 = \Mat{M}_0$, it follows by induction that
\begin{equation}
	\left\| 
		\Mat{M}_j 
		- \Mat{m}_j 
	\right\| 
	\le j T \Delta t^3
	.
	\label{eq:4_bound}
\end{equation}

\noindent Finally, for a total number of iterations $K = 1/\Delta t$, \Eq{eq:4_bound} implies that
\begin{equation}
	\Mat{M}(\Mat{S}_K) \ldots \Mat{M}(\Mat{S}_1)
	= \Mat{N}(\Mat{S}_K) \ldots \Mat{N}(\Mat{S}_1)
	+ O(\Delta t^2)
	.
\end{equation}

\noindent Hence, the iterated dNIMT converges to the iterated dMT at a rate of $2$.


\subsection{Examples}
\label{sec:4_EXpade}

Here we demonstrate the performance of the dMT in four examples. Specifically, we use the dMT to compute the transformation of a Hermite--Gauss (HG) laser mode, with transverse field profile given as
\begin{equation}
	\psi_n(x) = 
	\frac{H_n(x)}{\sqrt{2^m m! \sqrt{\pi}}} 
	\exp\left(
		-\frac{x^2}{2}
	\right)
	\label{eq:4_Herm}
\end{equation}

\noindent (where $H_n$ is the $n$th Hermite polynomial~\cite{Olver10a}) through the paraxial setups corresponding to the following symplectic matrices:
\begin{subequations}
	\begin{align}
		\label{eq:4_test1}
		\Mat{S}_1
		&=
		\begin{pmatrix}
			1 & 1 \\
			1 & 2
		\end{pmatrix}
		, \\
		\label{eq:4_test2}
		\Mat{S}_2
		&=
		\begin{pmatrix}
			4 & 0 \\
			0 & 0.25
		\end{pmatrix}
		, \\
		\label{eq:4_test3}
		\Mat{S}_3
		&=
		\begin{pmatrix}
			0.5 & 2 \\
			-1 & -2
		\end{pmatrix}
		, \\
		\label{eq:4_test4}
		\Mat{S}_4
		&=
		\frac{1}{\sqrt{2}}
		\begin{pmatrix}
			1 & 1 \\
			-1 & 1
		\end{pmatrix}
		.
	\end{align}
\end{subequations}

\noindent (One such correspondence is discussed following \Eq{eq:3_decompS}, but others exist as well~\cite{Arsenault80b,Nazarathy82,Liu08,Yasir21a,Yasir21b}.) For reference, the exact MT for general $\Mat{S}$ of $\psi_n(x)$ given by \Eq{eq:4_Herm} is
\begin{align}
	\Psi_n(x) =
	\left(A^2 + B^2\right)^{-1/4}
	\psi_n\left( \frac{x}{\sqrt{A^2 + B^2}} \right)
	\exp
	\left[
		i\frac{AC + DB}{2A^2 + 2B^2}x^2
		-i \frac{2n + 1}{2} 
		\tan^{-1}
		\left(
			\frac{B}{A}
		\right)
	\right]
	.
\end{align}

\begin{figure}[t]
	\begin{overpic}[width=0.32\linewidth,trim={3mm 23mm 16mm 26mm},clip]{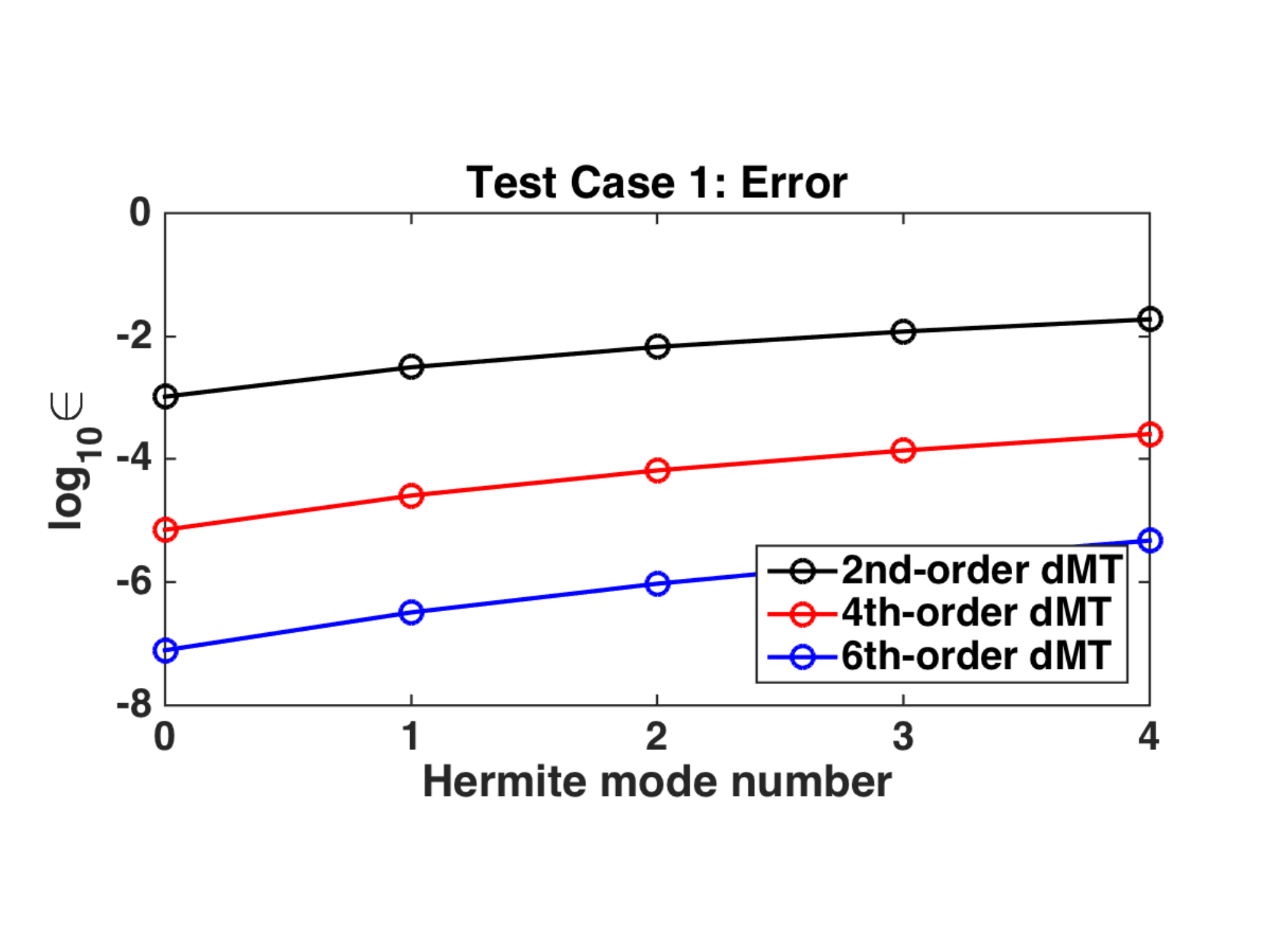}
		\put(15,10){\textbf{\small(a)}}
	\end{overpic}
	\begin{overpic}[width=0.32\linewidth,trim={3mm 23mm 16mm 26mm},clip]{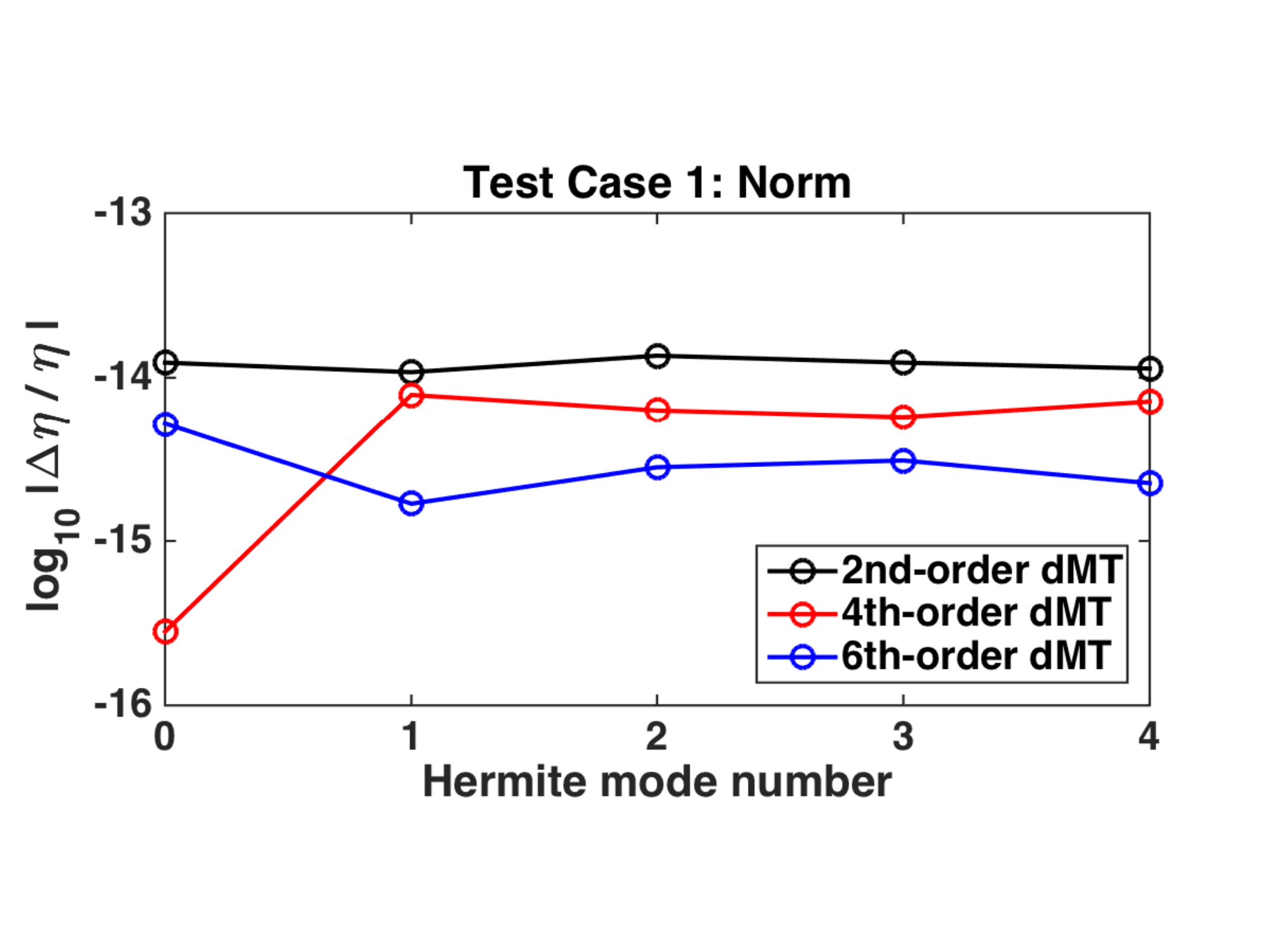}
		\put(15,10){\textbf{\small(b)}}
	\end{overpic}
	\begin{overpic}[width=0.32\linewidth,trim={3mm 23mm 16mm 26mm},clip]{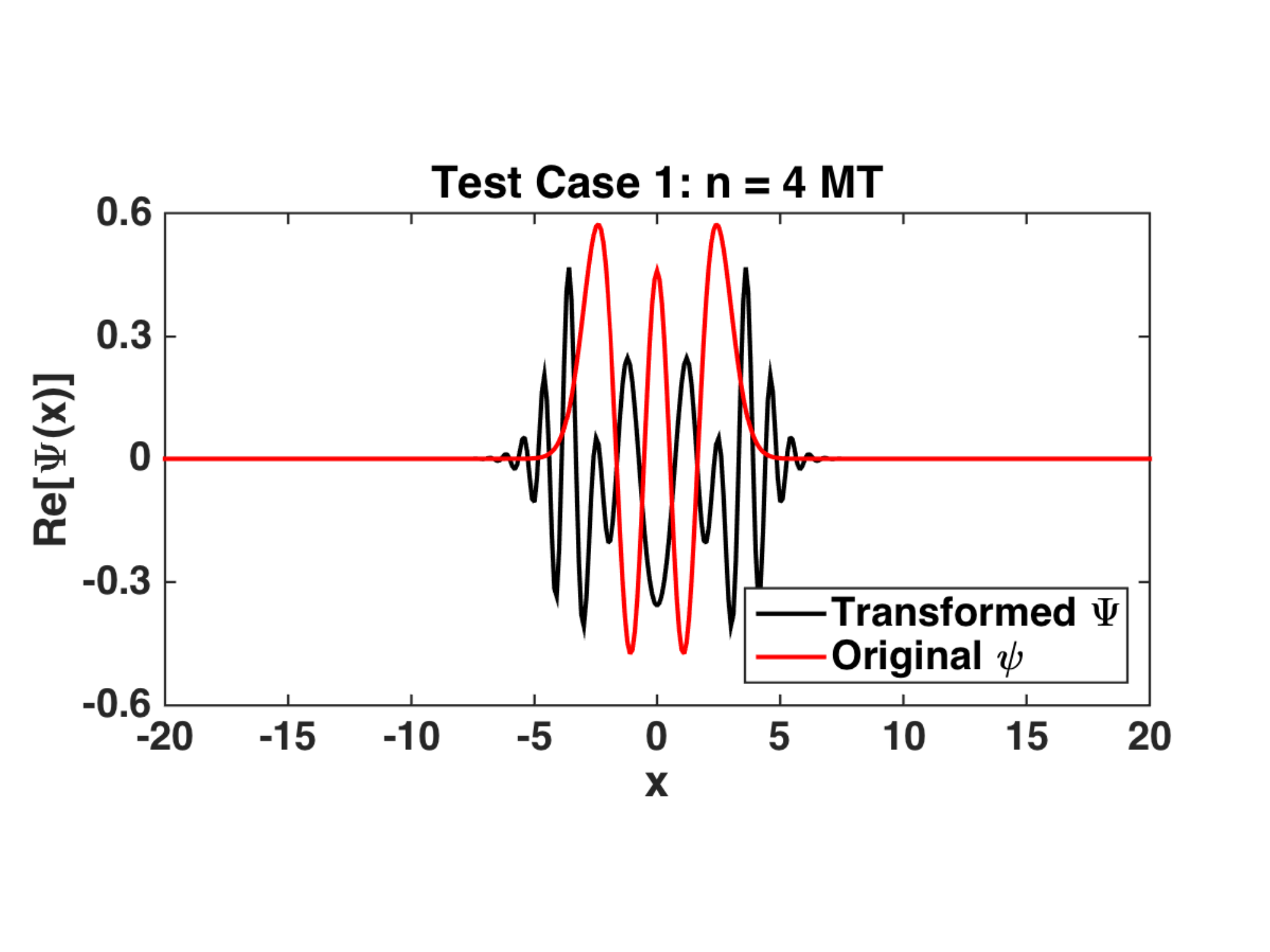}
		\put(15,10){\textbf{\small(c)}}
	\end{overpic}

	\vspace{2mm}
	\begin{overpic}[width=0.32\linewidth,trim={3mm 23mm 16mm 26mm},clip]{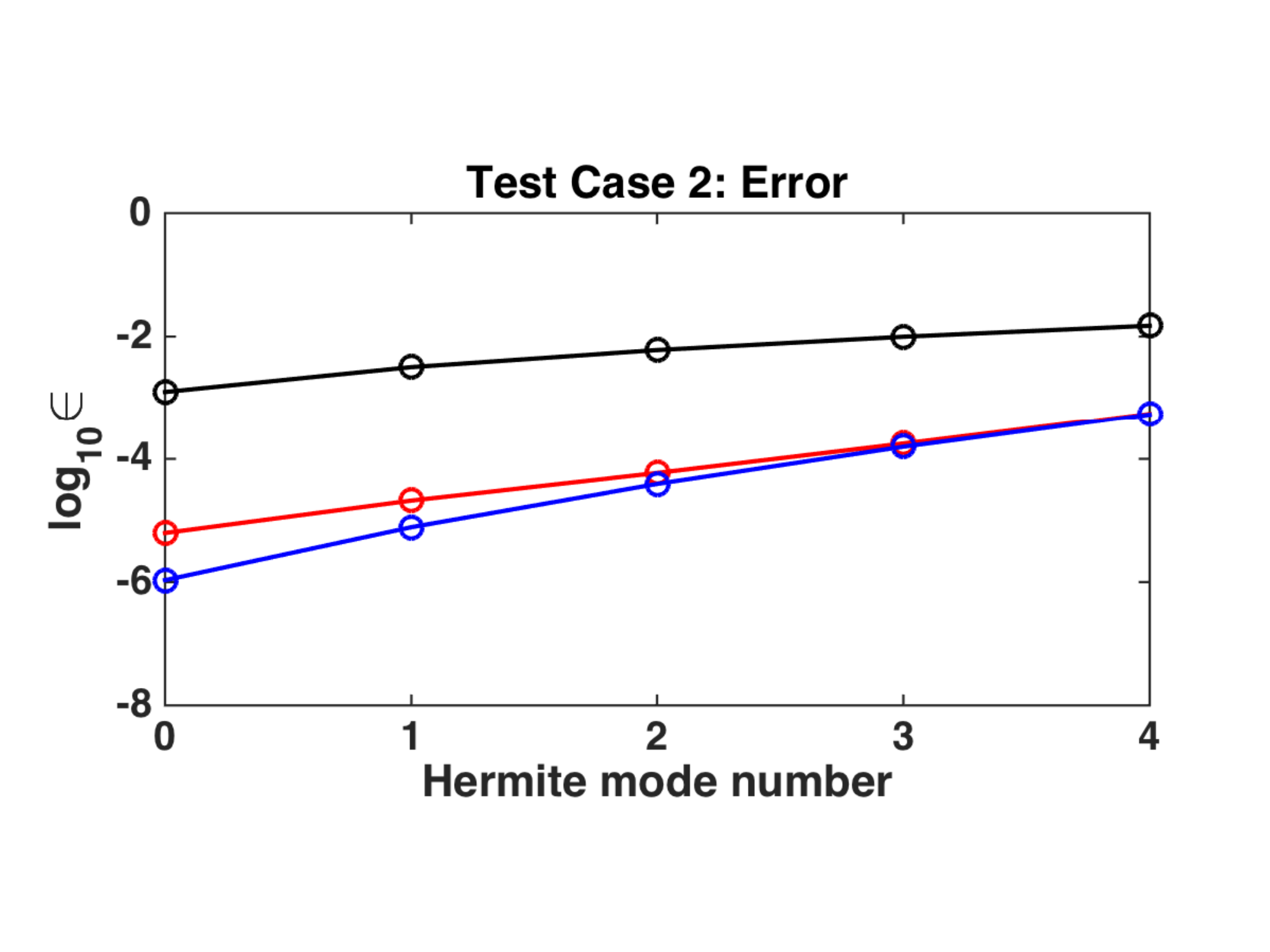}
		\put(15,10){\textbf{\small(d)}}
	\end{overpic}
	\begin{overpic}[width=0.32\linewidth,trim={3mm 23mm 16mm 26mm},clip]{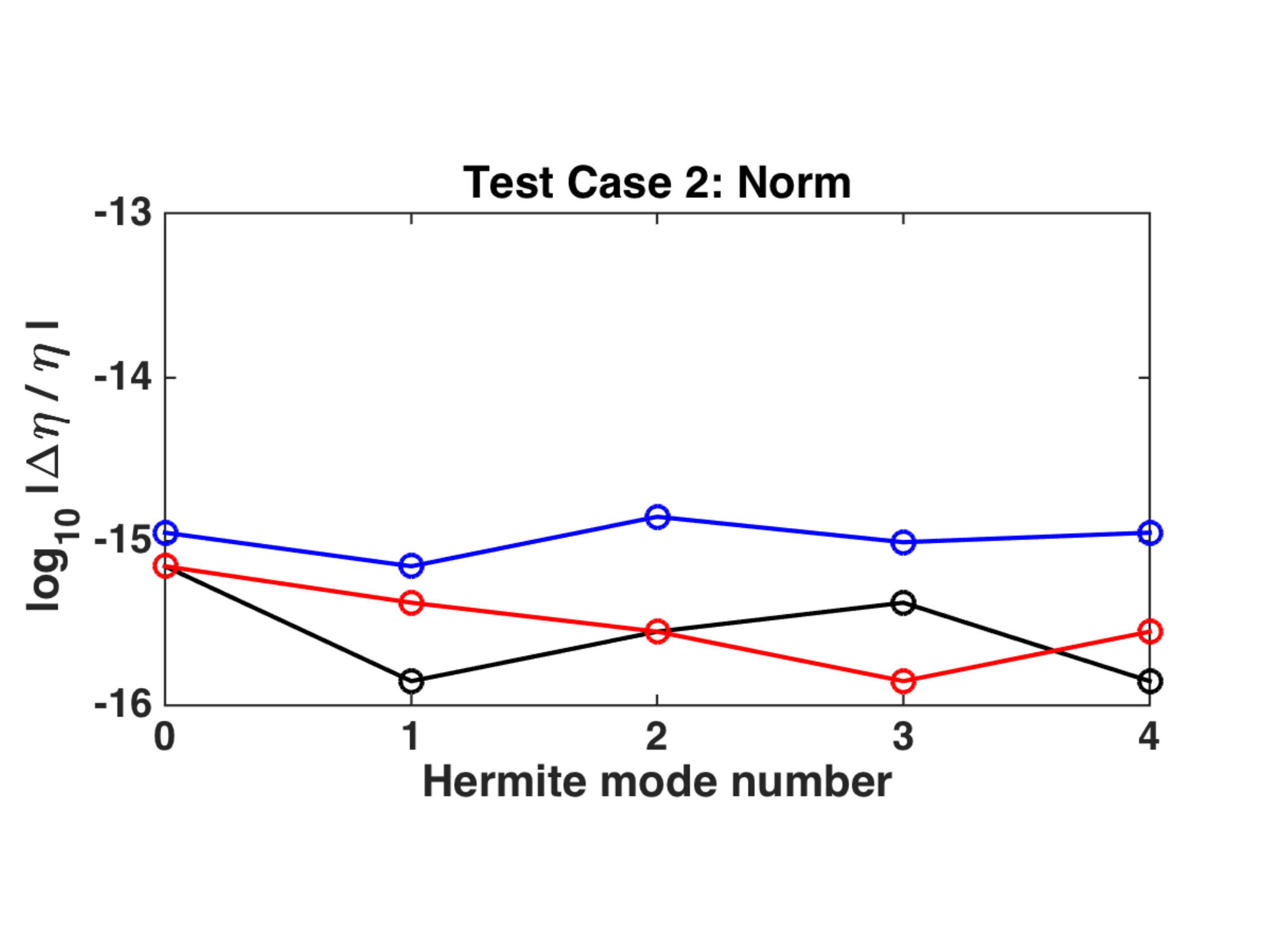}
		\put(15,10){\textbf{\small(e)}}
	\end{overpic}
	\begin{overpic}[width=0.32\linewidth,trim={3mm 23mm 16mm 26mm},clip]{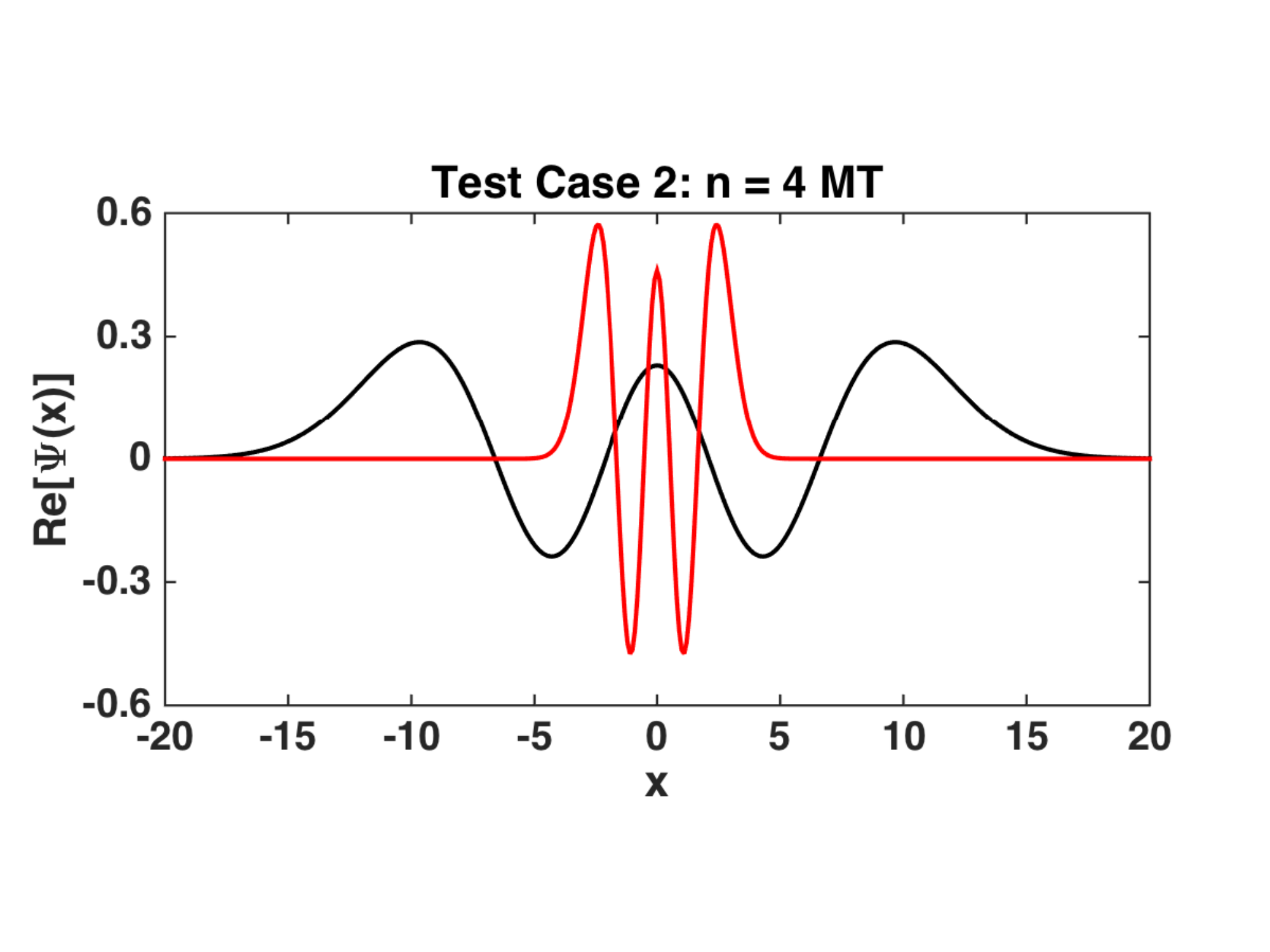}
		\put(15,10){\textbf{\small(f)}}
	\end{overpic}

	\vspace{2mm}
	\begin{overpic}[width=0.32\linewidth,trim={3mm 23mm 16mm 26mm},clip]{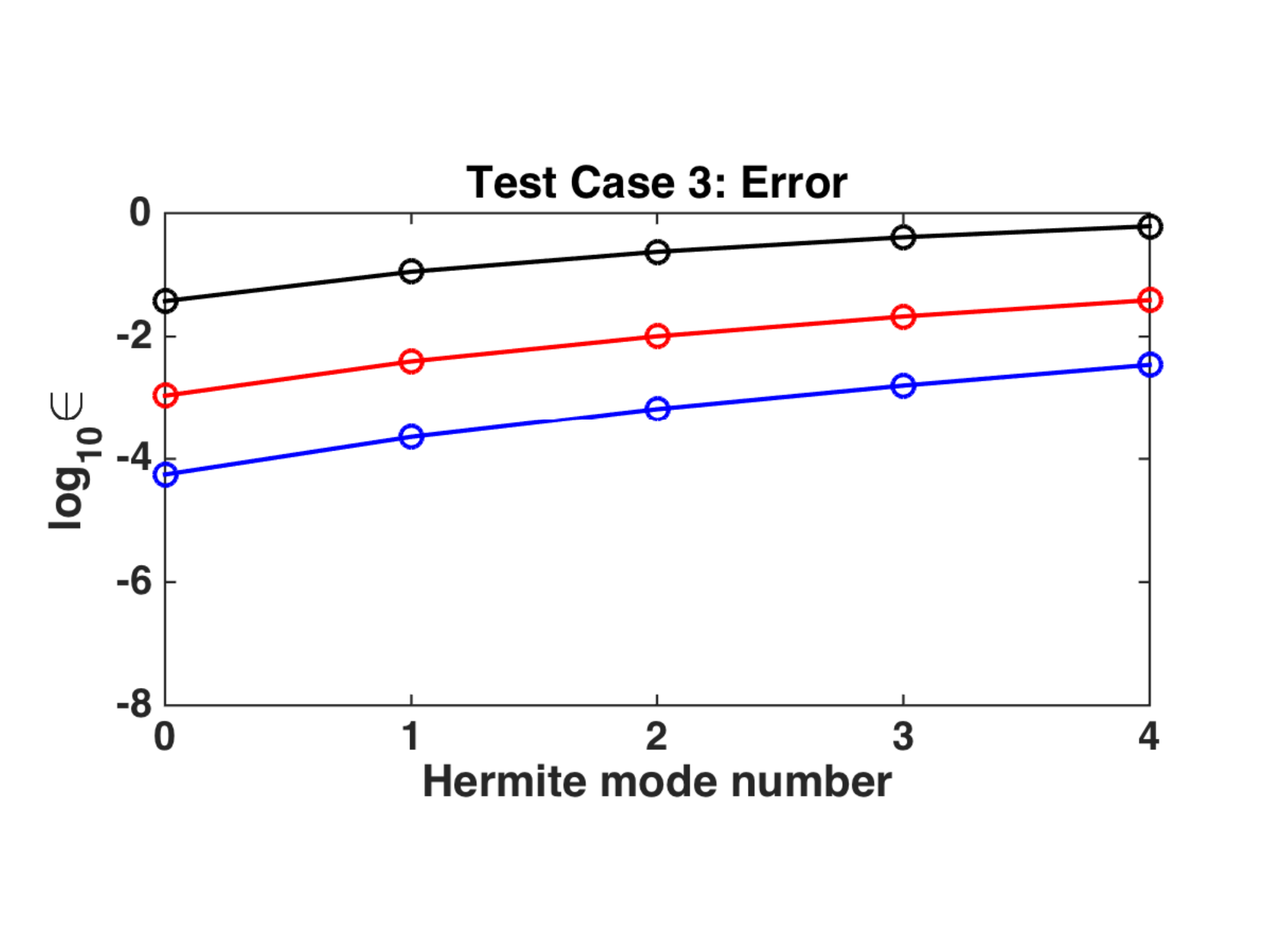}
		\put(15,10){\textbf{\small(g)}}
	\end{overpic}
	\begin{overpic}[width=0.32\linewidth,trim={3mm 23mm 16mm 26mm},clip]{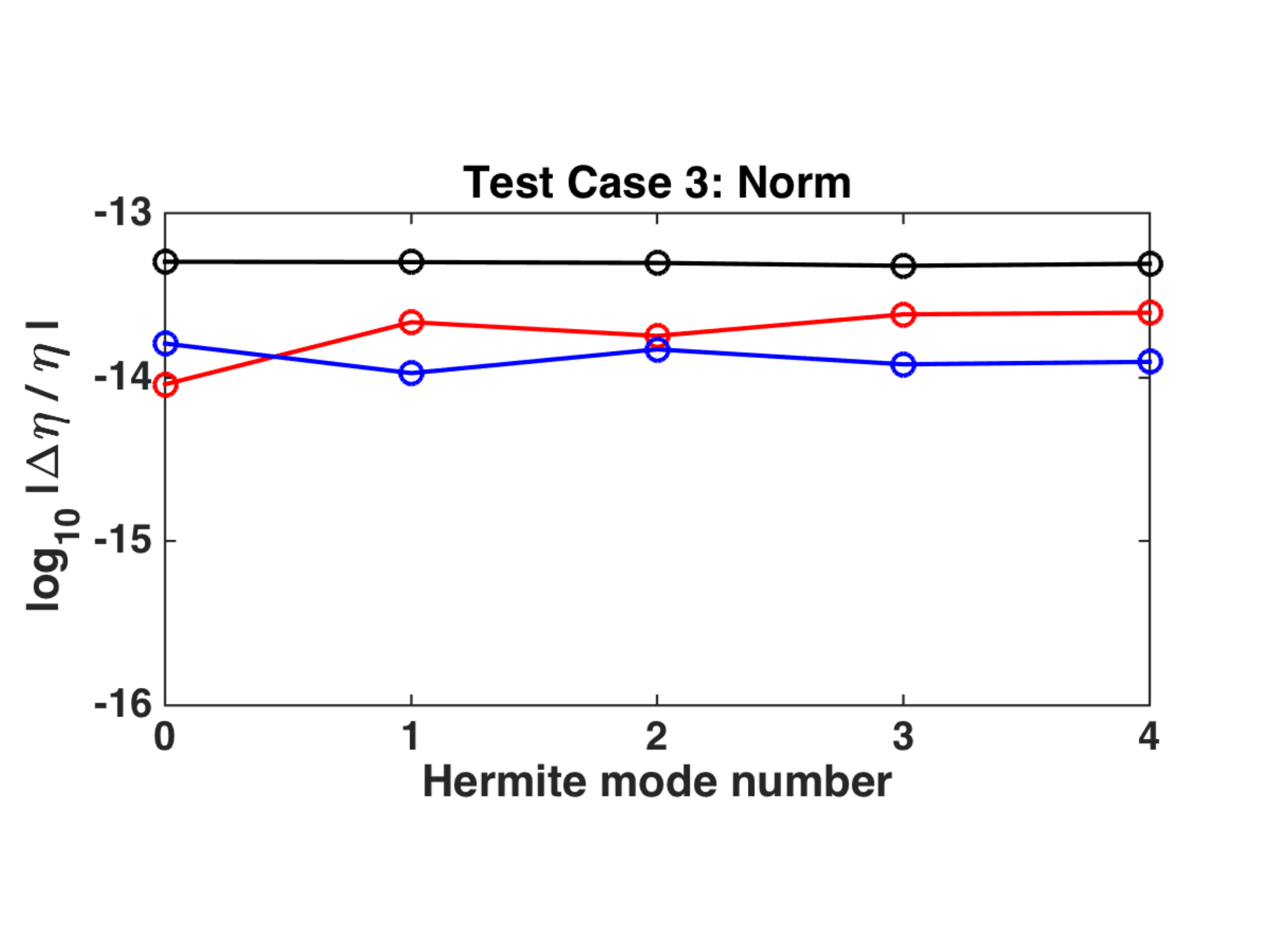}
		\put(15,10){\textbf{\small(h)}}
	\end{overpic}
	\begin{overpic}[width=0.32\linewidth,trim={3mm 23mm 16mm 26mm},clip]{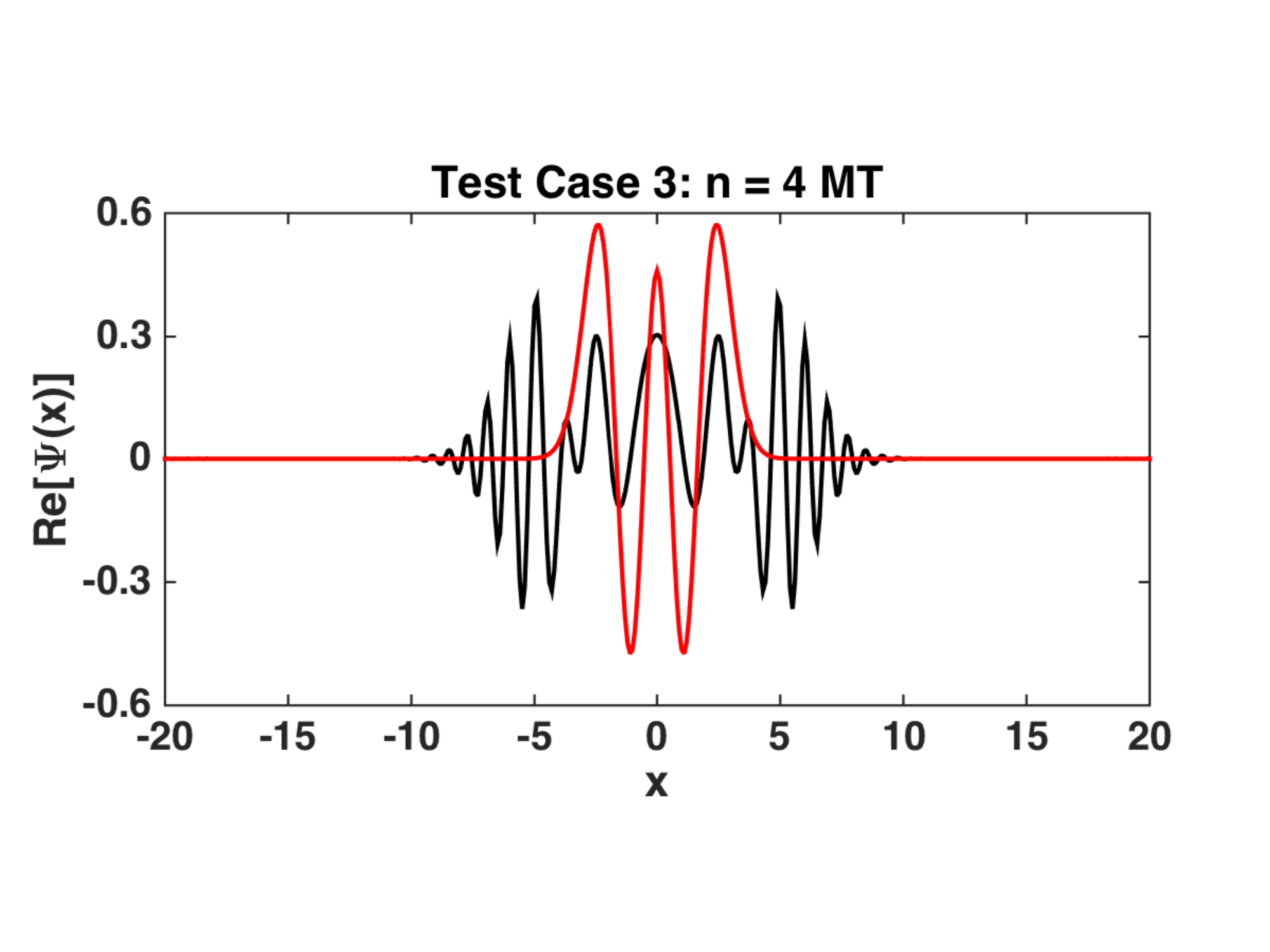}
		\put(15,10){\textbf{\small(i)}}
	\end{overpic}

	\vspace{2mm}
	\begin{overpic}[width=0.32\linewidth,trim={3mm 23mm 16mm 26mm},clip]{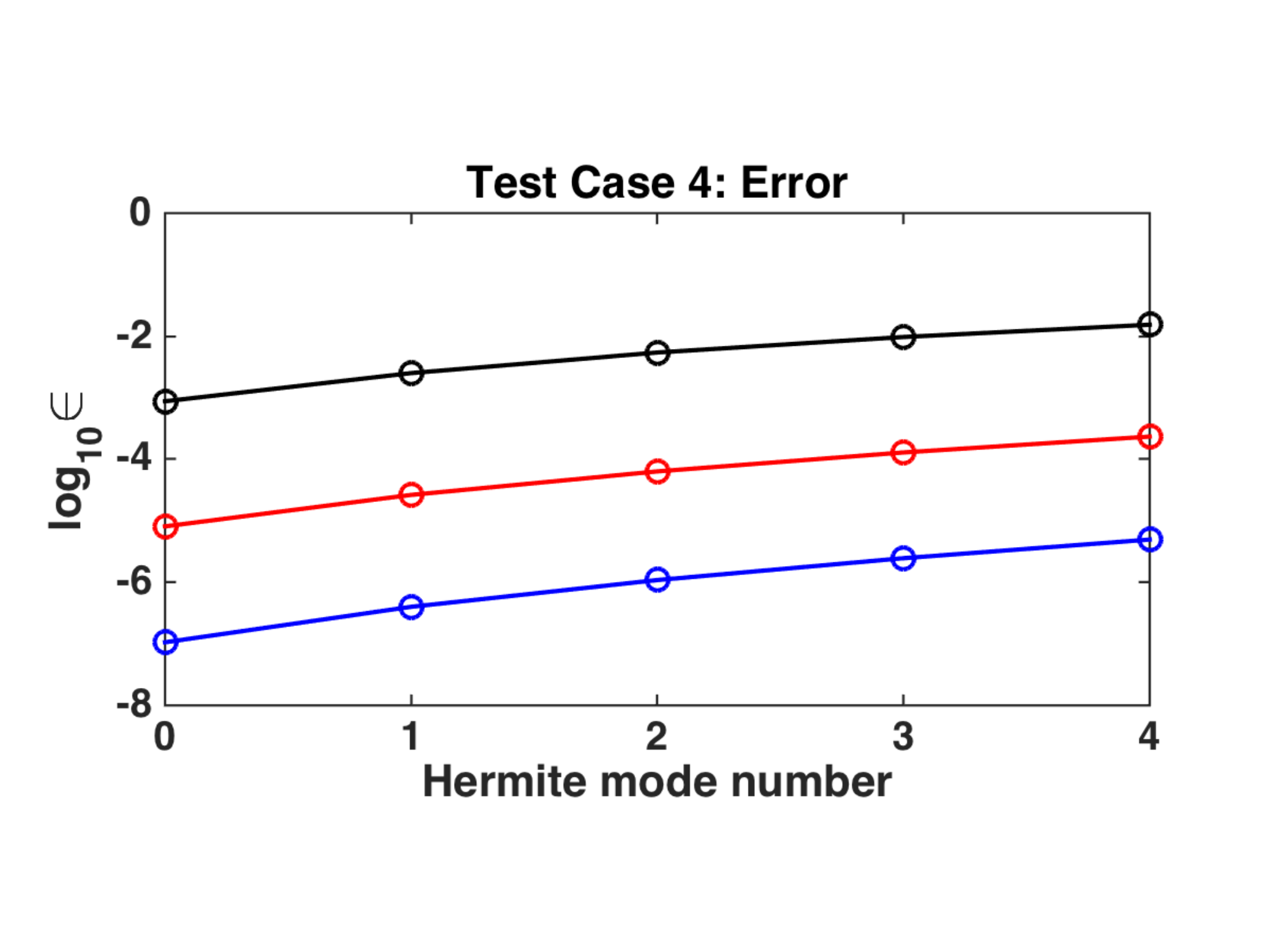}
		\put(15,10){\textbf{\small(j)}}
	\end{overpic}
	\begin{overpic}[width=0.32\linewidth,trim={3mm 23mm 16mm 26mm},clip]{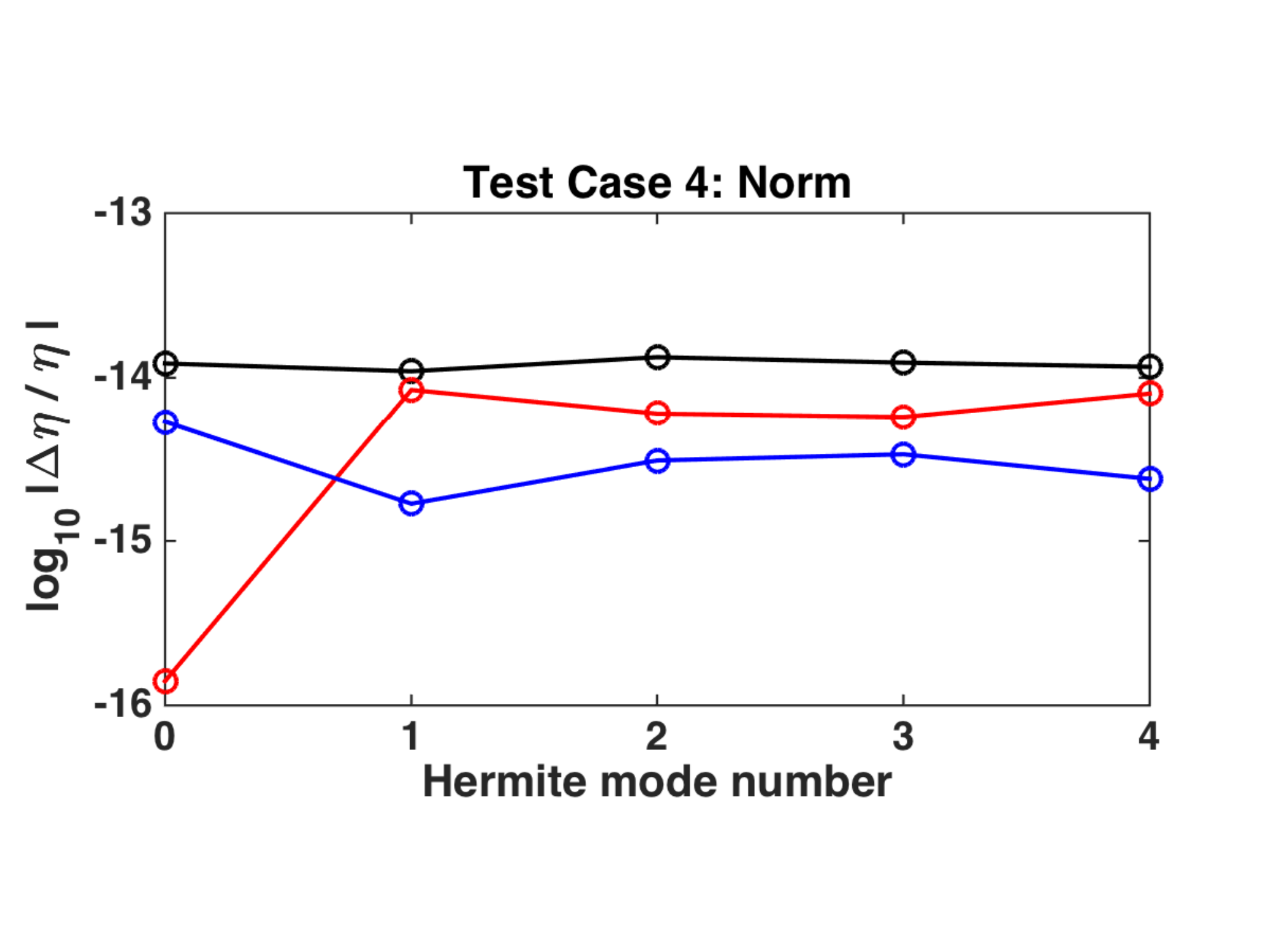}
		\put(15,10){\textbf{\small(k)}}
	\end{overpic}
	\begin{overpic}[width=0.32\linewidth,trim={3mm 23mm 16mm 26mm},clip]{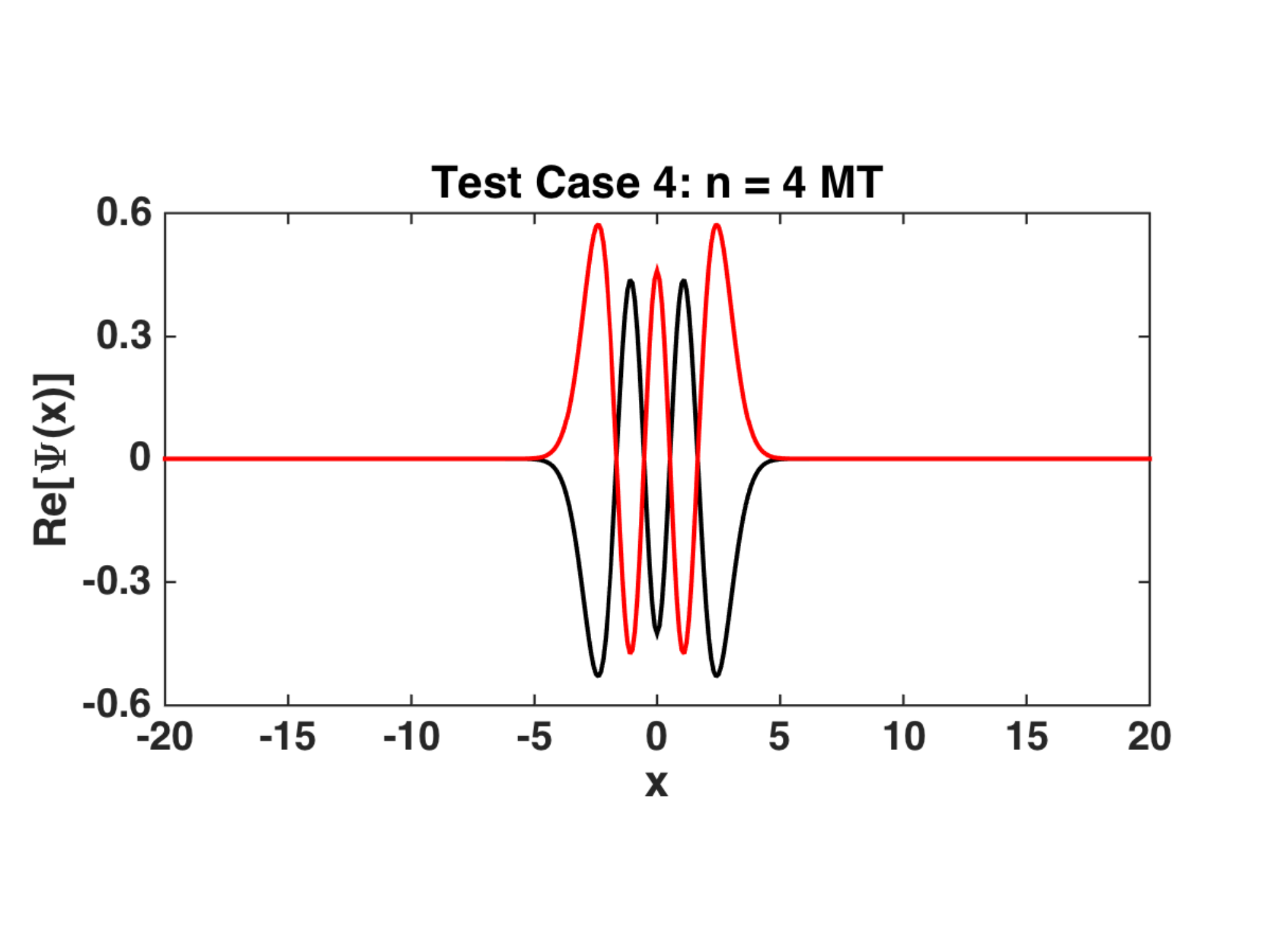}
		\put(15,10){\textbf{\small(l)}}
	\end{overpic}
	\caption{\textbf{(a)} Comparing the error $\epsilon$ [\Eq{eq:4_err}] of the 2nd-order, 4th-order, and 6th-order dMTs applied to the first five HG modes [\Eq{eq:4_Herm}] for the first test case \eq{eq:4_test1}. \textbf{(b)} Same as (a), but for the change in norm $\eta$ [\Eq{eq:4_norm}] rather than $\epsilon$. \textbf{(c)} Comparing the initial and transformed field for the fifth HG mode computed using the 6th-order dMT. For all cases, $x$ is uniformly discretized on the interval $[-20, 20]$ with a step size of $0.1$, so $m = 401$. \textbf{(d) - (f)} Same as (a) - (c), but for the second test case \eq{eq:4_test2}. \textbf{(g) - (i)} Same as (a) - (c), but for the third test case \eq{eq:4_test3}. \textbf{(j) - (l)} Same as (a) - (c), but for the fourth test case \eq{eq:4_test4}. In all these cases, only a single dMT is performed, \ie there are no iterations.}
	\label{fig:4_DMT}
\end{figure}

Figure \ref{fig:4_DMT} shows the error $\epsilon$, defined via the Euclidean $2$-norm as
\begin{equation}
	\epsilon \doteq \frac{\| \Vect{\Psi} - \Vect{\Psi}_\text{exact} \|_2}{\| \Vect{\Psi}_\text{exact} \|_2}
	,
	\label{eq:4_err}
\end{equation}

\noindent along with the change in norm $\eta$, defined as 
\begin{equation}
	\Delta \eta \doteq \| \Vect{\Psi} \|_2 - \eta
	, \quad
	\eta \doteq \| \Vect{\psi} \|_2
	,
	\label{eq:4_norm}
\end{equation}

\noindent as the dMT for the four test cases is applied to the first five HG modes. Figure \ref{fig:4_DMT} also shows a comparison between the real parts of $\psi_4(x)$ and $\Psi_4(x)$ for the four test cases. Overall, the norm is preserved to near machine precision, while the error of the dMT decreases as the order of the dMT is increased. The increase in error as $n$ increases is expected since the length scale of $\psi_n(x)$ decreases with $n$, so the finite-difference error at fixed step size consequently increases with $n$. We should note that in these examples, the matrix exponentials are computed using MATLAB's built-in \texttt{expm} method, which uses the standard Pad\'e-based `scaling-squaring' algorithm~\cite{AlMohy09}.

We next demonstrate the convergence of the iterated dNIMT to the dMT by computing the fourth example \eq{eq:4_test4} via the path
\begin{equation}
	\Mat{S}(t) = 
	\frac{1}{\sqrt{2}}
	\begin{pmatrix}
		\sqrt{2} + (1 - \sqrt{2}) t & t \\
		- t & \frac{2 - t^2}{\sqrt{2} + (1 - \sqrt{2})t}
	\end{pmatrix}
	.
\end{equation}

\begin{figure}
	\centering
	\includegraphics[width=0.6\linewidth,trim={2mm 20mm 16mm 26mm},clip]{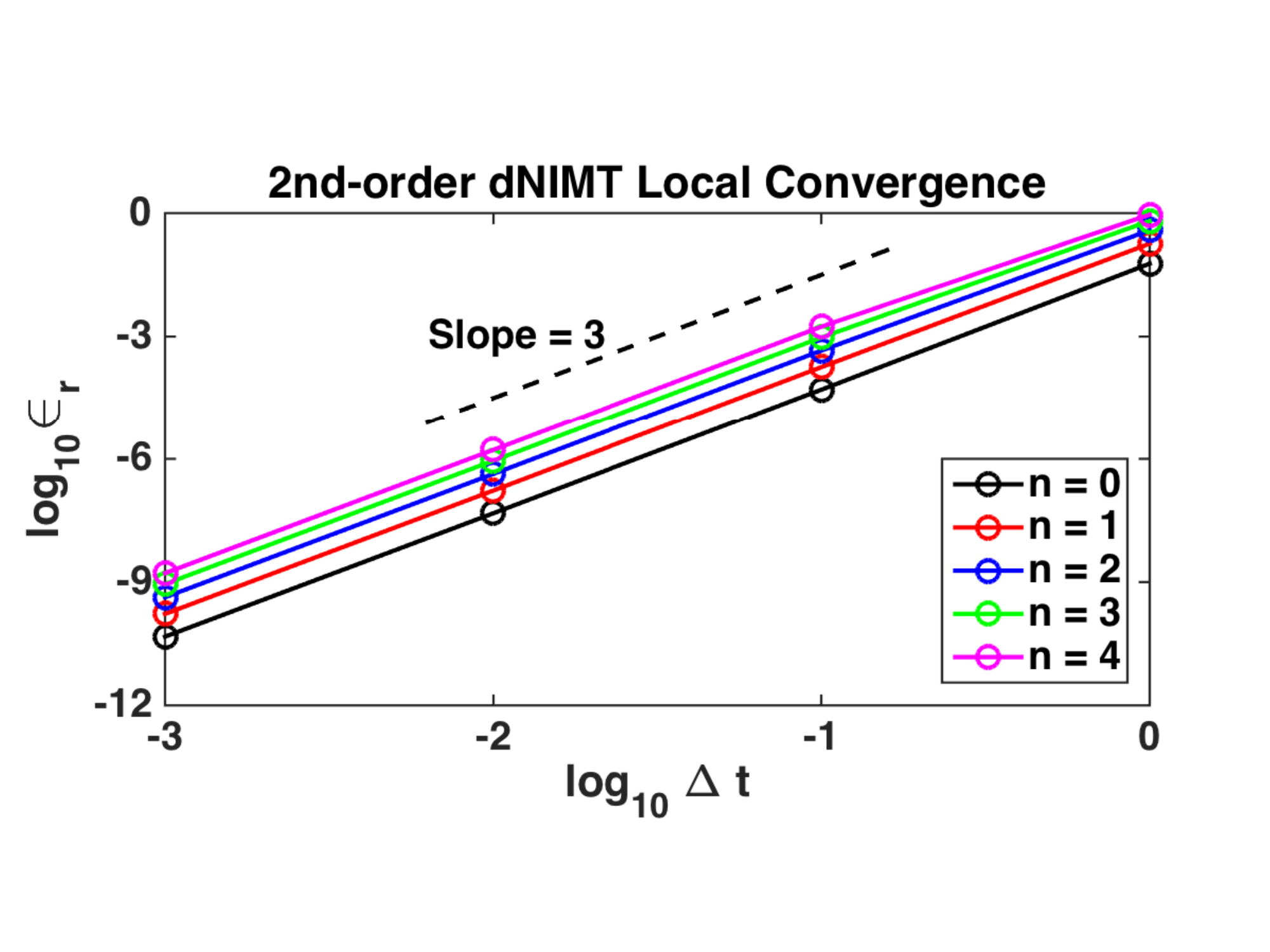}
	\caption{Local (single-step) error convergence of the 2nd-order dNIMT to the dMT for the first five HG modes. The expected convergence rate of $3$ is clearly observed. Although not shown, the 4th-order and the 6th-order dNIMT also exhibit the expected converge rate and in fact, the values of $\epsilon_r$ [\Eq{eq:4_relERR}] are nearly identical to those of the 2nd-order dNIMT.}
	\label{fig:4_localNIMT}
\end{figure}

\noindent One can verify that $\Mat{S}(t)$ is symplectic for all $t \in [0,1]$. Figure~\ref{fig:4_localNIMT} shows the local error convergence of the dNIMT to the dMT, where the relative error $\epsilon_r$ is defined as
\begin{equation}
	\epsilon_r \doteq \frac{\| \Vect{\Psi}_\text{dNIMT} - \Vect{\Psi}_\text{dMT} \|_2}{\| \Vect{\Psi}_\text{dMT} \|_2}
	\label{eq:4_relERR}
	.
\end{equation}

\noindent Clearly, the dNIMT converges to the dMT with a rate of $3$, as expected. Note that $\epsilon_r$ is computed for a single iteration, which means that different values of $\Delta t$ correspond to different final transforms $\Mat{S}(\Delta t)$.

\begin{figure}
	\centering
	\begin{overpic}[width=0.55\linewidth,trim={8mm 21mm 17mm 26mm},clip]{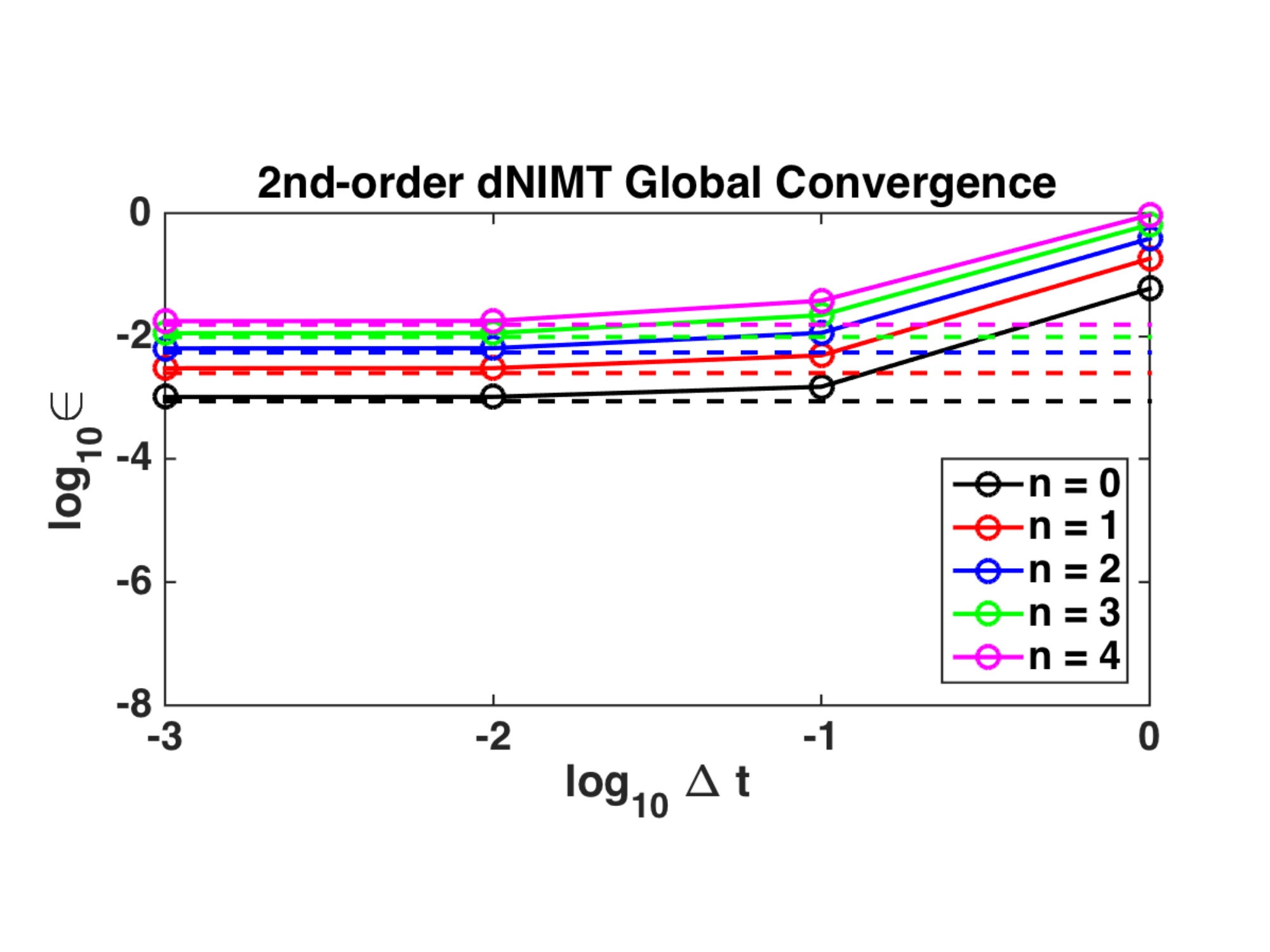}
		\put(15,12){\textbf{\normalsize(a)}}
	\end{overpic}

	\vspace{2mm}
	\begin{overpic}[width=0.55\linewidth,trim={8mm 21mm 17mm 26mm},clip]{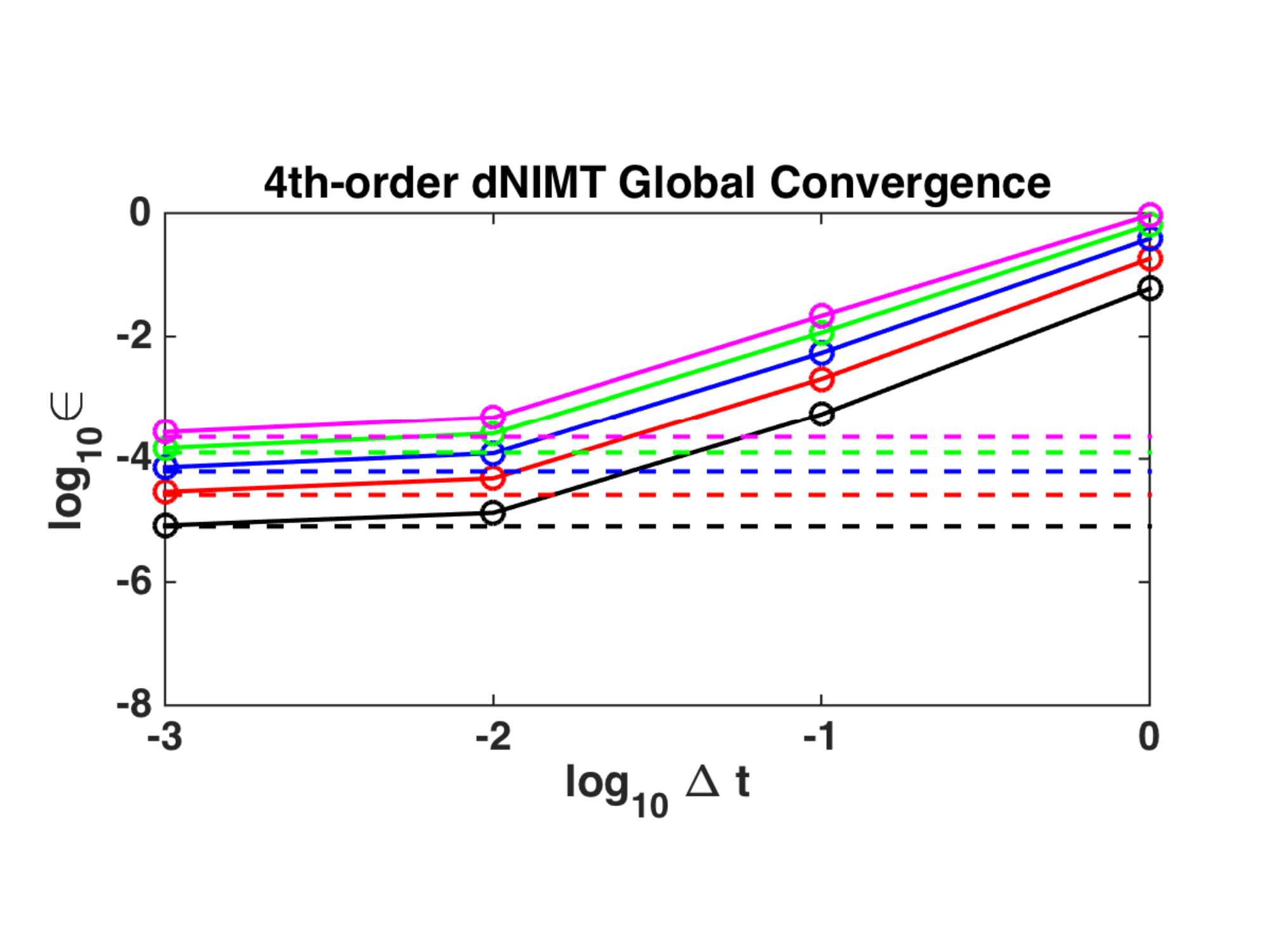}
		\put(15,12){\textbf{\normalsize(b)}}
	\end{overpic}

	\vspace{2mm}
	\begin{overpic}[width=0.55\linewidth,trim={8mm 21mm 17mm 26mm},clip]{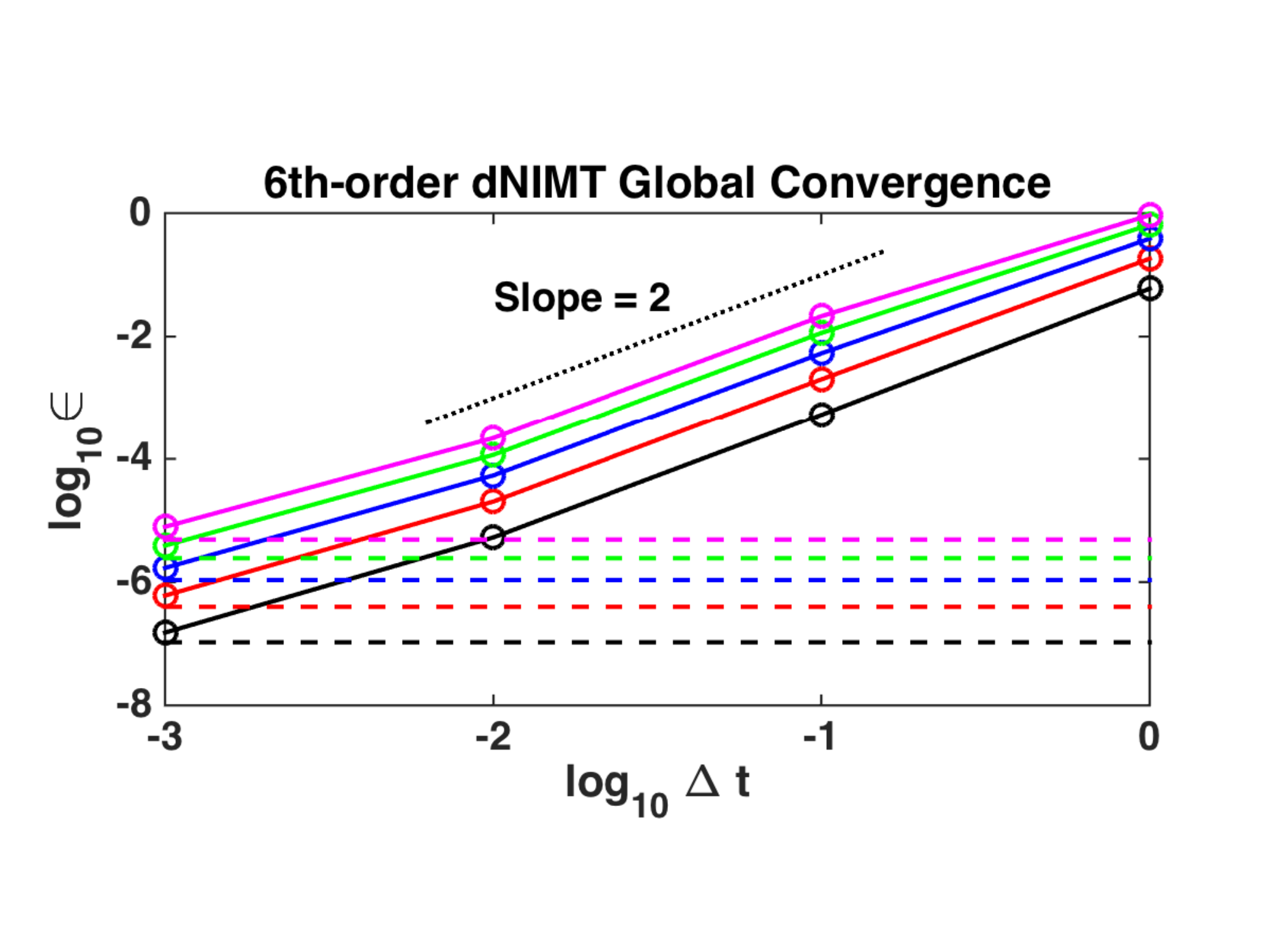}
		\put(15,12){\textbf{\normalsize(c)}}
	\end{overpic}

	\vspace{2mm}
	\begin{overpic}[width=0.55\linewidth,trim={4mm 21mm 17mm 26mm},clip]{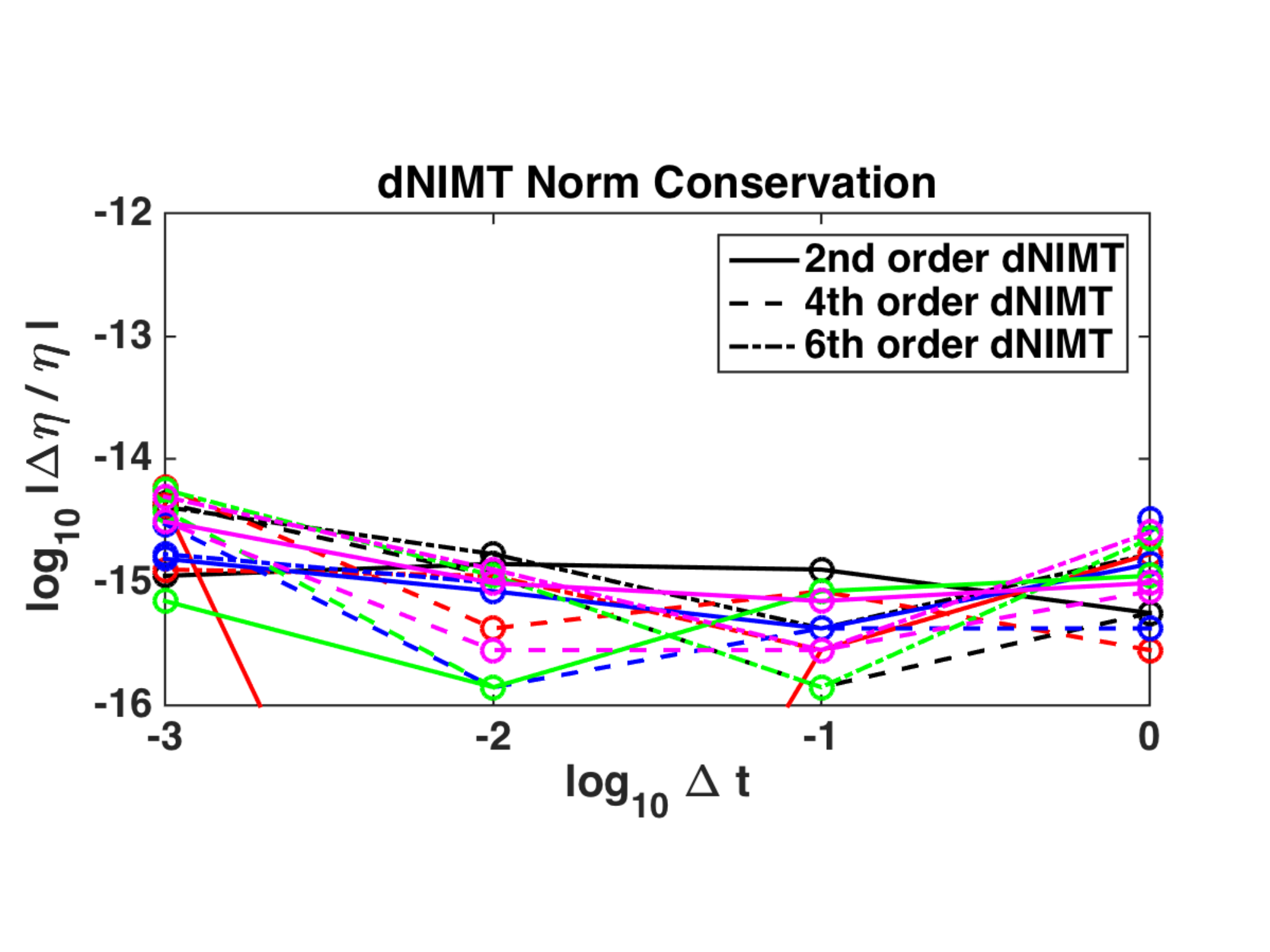}
		\put(15,12){\textbf{\normalsize(d)}}
	\end{overpic}
	\caption{\textbf{(a) - (c)} Global convergence of the 2nd-order, 4th-order, and 6th-order dNIMT to the respective dMT for the first five HG modes. The dashed line shows the intrinsic error of the corresponding dMT. As can be seen, the convergence rate achieves the expected value of $2$ before asymptoting to match the error of the dMT. \textbf{(d)} Global norm conservation of the 2nd-order, 4th-order, and 6th-order dNIMT for the first five HG modes. In all cases, the norm is conserved to near machine precision.}
	\label{fig:4_globalNIMT}
\end{figure}

Figure~\ref{fig:4_globalNIMT} shows the global convergence of the iterated dNIMT to the dMT when computing the final transformation $\Mat{S}(1)$ with the specified step size, along with the norm conservation. The asymptote in the convergence plots marks the intrinsic error between the dMT and the exact result, which arises from the use of finite-difference matrices to perform the spatial derivatives. As expected, the iterated dNIMT converges to the dMT at a rate of $2$ and conserves the norm to near machine precision.


\section{Summary}

In summary, efficiently computing metaplectic transforms in the near-identity limit has been thus far neglected in the computational-optics literature. However, near-identity metaplectic transforms (NIMTs) feature prominently in a variety of physics applications, from evolving certain quantum systems in time to propagating paraxial wavefields to removing caustics via MGO. Indeed, as will be discussed in later chapters, MGO requires NIMTs to be performed along a ray at each timestep of a ray-tracing simulation, so developing fast algorithms for computing the NIMT is crucial.

Above, I proposed two algorithms to close this gap. Both algorithms are based on a pseudo-differential representation of the MT derived in \Ch{ch:MT}. The first algorithm uses truncated Taylor expansions and is given by \Eq{eq:4_NIMTtaylor}; as such, this algorithm is local in that it has a small stencil width when approximated by finite difference matrices. However, the Taylor approximation does not preserve the unitarity of the MT. This means that the local NIMT algorithm can develop instabilities when iterated to compute finite MTs, \ie MTs that are not near-identity. Occasionally smoothing the field while iterating partially remedies the situation in certain cases. The second proposed algorithm is based on using truncated diagonal Pad\'e expansions and is given by \Eq{eq:4_dNIMT}. Unlike the Taylor-based algorithm, this algorithm is exactly unitary, although it is no longer local. As I show in the main text, preserving the unitary property exactly implies that this second algorithm is stable and globally convergent. This algorithm is therefore anticipated to perform superiorly in an MGO-based ray-tracing code.


\begin{subappendices}
\section*{Appendix}

\section{Reducing the PMT to an envelope equation for eikonal functions}
\label{app:4_eikNIMT}

Often, the function $\psi(\Vect{x})$ can be characterized by a rapidly-varying phase $\theta(\Vect{x})$, and a complex envelope $\phi(\Vect{x})$ which varies much slower than $\theta(\Vect{x})$. If such a partition is defined, then we call $\psi(\Vect{x})$ an \textit{eikonal} function. Eikonal solutions to physical systems are frequently sought as a means to develop approximate, reduced models; an example is the WKB approximation for quantum particles~\cite{Heading62a} (see also \Ch{ch:GO}). In reduced models, phase and envelope dynamics are typically governed by separate equations, which often makes it convenient to consider the phase and envelope as separate entities~\cite{Tracy14}. Let us therefore explore how the PMT partitions eikonal functions.

Let $\psi(\Vect{x}) = \phi(\Vect{x}) \exp[ i\theta(\Vect{x})]$, and let $\Vect{k}(\Vect{x}) \doteq \nabla \theta(\Vect{x})$ with component functions $\{k_j (\Vect{x})\}$. Then, by induction
\begin{equation}
	\pd{x_j}^n \, \psi(\Vect{x}) = 
	\exp[i\theta(\Vect{x})]
	\left[
		i k_j(\Vect{x}) 
		+ \pd{x_j} 
	\right]^n \phi(\Vect{x}) 
	.
\end{equation}

\noindent An analogous result is obtained in the case of mixed partial derivatives, which implies that $\nabla$ and $\nablaNEW \doteq i\Vect{k}(\Vect{x}) + \nabla$ have the same commutation relations among their vector components; hence, the phase function effects a formal mapping from a differential operator acting on the full function $\psi(\Vect{x})$ to the differential operator acting solely on the envelope $\phi(\Vect{x})$. For example, see the definition of the envelope dispersion operator in \Ref{Dodin19} and in \Ch{ch:GO}.

For an eikonal function, the PMT is
\begin{equation}
	\Psi(\Vect{X}) = \pm 
	\frac{
		\exp\left[
			\frac{i}{2} \Vect{X}^\intercal \Mat{C} \Mat{A}^{-1} \Vect{X} 
			+ i\theta(\Vect{x})
		\right]
	}{
		\sqrt{\det{\Mat{A}}}
	}\left. 
		\exp\left( \frac{i}{2} \Mat{A}^{-1} \Mat{B} \dubdot \nablaNEW \nablaNEW \right) \phi(\Vect{x}) 
	\right|_{\Vect{x} = \Mat{A}^{-1}\Vect{X}} 
	.
	\label{eq:4_NIMTEik}
\end{equation}

\noindent At least for near-identity transformations, $\Psi$ can also be cast in the eikonal form. Let $\Psi(\Vect{X}) = \Phi(\Vect{X}) \exp[ i\Theta(\Vect{X})]$, then
\begin{align}
	\Phi(\Vect{X}) = 
	\pm 
	\frac{
		\exp\left[ 
			\frac{i}{2} \Vect{X}^\intercal \Mat{C} \Mat{A}^{-1} \Vect{X} 
			+ i\theta(\Vect{x}) 
			- i \Theta(\Vect{X})
		\right]
	}{
		\sqrt{\det{\Mat{A}}}
	} \left. 
		\exp\left(\frac{i}{2} \Mat{A}^{-1} \Mat{B} \dubdot \nablaNEW \nablaNEW \right) \phi(\Vect{x}) 
	\right|_{\Vect{x} = \Mat{A}^{-1}\Vect{X}} 
	.
	\label{eq:4_alphaTRANS}
\end{align}

\noindent Since $\Phi(\Vect{X})$ is generally complex, the definition of $\Theta(\Vect{X})$ is not unique, so choosing it is a matter of convenience (as long as $\Theta$ remains fast compared to $\Phi$). Here, we choose to define $\Theta(\Vect{X})$ such that it is (i) real, (ii) independent of $\phi(\Vect{x})$, and (iii) simplifies the resultant expression for $\Phi(\Vect{X})$ as much as possible. Then, the first-order truncation of \Eq{eq:4_alphaTRANS} yields the eikonal partition
\begin{subequations}
	\label{eq:4_NIET}
	\begin{align}
		\label{eq:4_phaseTRANS}
		\Theta(\Vect{X}) 
		&\approx 
			\theta\left(\Vect{x} \right) 
			+ \frac{1}{2} \Vect{X}^\intercal \Mat{C} \Mat{A}^{-1} \Vect{X}
			\left.
				- \frac{1}{2} \Vect{k}\left(\Vect{x} \right)^\intercal
				\Mat{A}^{-1}\Mat{B} 
				\Vect{k}\left(\Vect{x} \right) 
			\right|_{\Vect{x} = \Mat{A}^{-1}\Vect{X}} 
		,\\
		\Phi(\Vect{X}) 
		&\approx \left.
			\frac{
				\phi\left(\Vect{x} \right) 
				+ \frac{i}{2}\Tr \left\{ \Mat{A}^{-1}\Mat{B} \nullFrac \, \Mat{F}(\Vect{x}) \right\} 
			}{
				\sqrt{\det{\Mat{A}}}
			} 
		\right|_{\Vect{x} = \Mat{A}^{-1}\Vect{X}} 
		,\\
		\Mat{F}(\Vect{x}) 
		&\doteq 
		\nabla\nabla \phi\left(\Vect{x} \right) 
		+ i\Vect{k}\left(\Vect{x} \right) \otimes \nabla \phi\left(\Vect{x} \right)
		+ i\nabla \phi\left(\Vect{x} \right) \otimes \Vect{k}\left(\Vect{x} \right) 
		+ i\phi\left(\Vect{x} \right) \nabla \Vect{k}\left(\Vect{x} \right) 
		,
	\end{align}
\end{subequations}

\noindent where $\otimes$ is the tensor product. If one prefers, additional approximations can be placed on \Eqs{eq:4_NIET} that are consistent with the eikonal ordering ansatz, such as neglecting $\nabla\nabla \phi$ in favor of the terms involving $\Vect{k}$.

Let us also calculate the local wavevector in the new coordinates, $\Vect{K} \doteq \pd{\Vect{X}} \Theta$. From \Eq{eq:4_phaseTRANS} one obtains
\begin{equation}
	\Vect{K}(\Vect{X}) = \Mat{C} \Mat{A}^{-1} \Vect{X} + (\Mat{A}^{-1})^\intercal \Vect{R}(\Mat{A}^{-1}\Vect{X}) 
	,
\label{eq:4_KTrans}
\end{equation}

\noindent where
\begin{equation}
	\Vect{R}(\Vect{x}) \doteq \Vect{k}\left(\Vect{x} \right) - \left[\nabla \Vect{k} \left(\Vect{x} \right) \right] \Mat{A}^{-1}\Mat{B} \, \Vect{k} \left(\Vect{x} \right) 
	.
\end{equation}

\noindent When $\Vect{X}$ is obtained as $\Vect{X} = \Mat{A}\Vect{x} + \Mat{B} \Vect{k}(\Vect{x})$, \Eq{eq:4_KTrans} becomes
\begin{align}
	\Vect{K}(\Vect{x}) &= 
	\Mat{C} \Vect{x} 
	+ \Mat{D} \Vect{k}(\Vect{x}) 
	- (\Mat{A}^{-1})^\intercal 
	\left\{
		\Vect{k}(\Vect{x}) 
		- \Vect{R}\left[
			\Vect{x} + \Mat{A}^{-1}\Mat{B}\Vect{k}(\Vect{x}) 
		\right] \nullFrac 
	\right\} 
	.
	\label{eq:4_KqTrans}
\end{align}

\noindent Assuming that $\epsilon \doteq \|\Mat{A}^{-1}\Mat{B} \|$ is small, then $\Vect{R}\left[\Vect{x} + \Mat{A}^{-1}\Mat{B}\Vect{k}(\Vect{x})\right] \approx \Vect{k}(\Vect{x}) + O(\epsilon^2)$. Substituting this into \Eq{eq:4_KqTrans} yields
\begin{align}
	\Vect{K}(\Vect{x}) = \Mat{C} \Vect{x} + \Mat{D} \Vect{k}(\Vect{x}) + O(\epsilon^2) = \Vect{K} + O(\epsilon^2) 
	,
\end{align}

\noindent where $\Vect{K}$ is defined in \Eq{eq:3_Strans}. This shows that the transform \eq{eq:4_NIET} maps ($\Vect{x}$,$\Vect{k}(\Vect{x})$) to ($\Vect{X}$,$\Vect{K}(\Vect{X})$) with $O(\epsilon^2)$ accuracy, which is consistent with the accuracy of \Eqs{eq:4_NIET}. In this sense, this transform is natural and can be useful for modeling the propagation of eikonal waves, as we shall discuss in \Ch{ch:MGO}.

\end{subappendices}

\chapter{Gauss--Freud quadrature for catastrophe integrals}
\label{ch:GF}

\section{Introduction}

Besides fast algorithms for near-identity metaplectic transforms, it will also be useful in MGO to develop specialized quadrature rules specifically for performing finite MTs of caustic wavefields that are highly oscillatory and thereby difficult to compute using standard numerical methods. In this chapter I present a new algorithm for taking such integrals numerically that is based on the steepest-descent method combined with Gauss--Freud quadrature. The following discussion is based on material published previously in \Ref{Donnelly21}.


\section{Brief review of the steepest-descent method}

Suppose one is interested in computing a standard integral of catastrophe theory (\Ch{ch:GO}), to which the MT of a given caustic wavefield is expected to reduce. As can be seen from \Eq{eq:2_Ialpha} and Table~\ref{tab:2_caustic}, one generally expects $I_\alpha(\Vect{y})$ to involve a highly oscillatory integrand when $\Vect{y}$ and $\Vect{\kappa}$ are both real. These rapid oscillations would make direct evaluation analytically and numerically challenging; however, along the steepest-descent contours oscillatory terms become exponentially decaying terms, reinstating the viability of standard numerical integration methods like Gaussian quadrature (\Sec{sec:5_gaussQUAD}). However, this simplification is contingent on the ability to determine the steepest-descent contours for arbitrary wavefields. Let us therefore characterize the steepest-descent contours of the standard forms $I_\alpha(\Vect{y})$. For simplicity, I restrict attention to $1$-D integrals, \ie $M = 1$ in \Eq{eq:2_Ialpha}.

For integrals of the form
\begin{equation}
	I(\Vect{y}) = \int \dd \kappa \, g(\kappa, \Vect{y}) \exp[i f(\kappa, \Vect{y}) ]
	\label{eq:5_SDMint}
\end{equation}

\noindent [where I have generalized \Eq{eq:2_Ialpha} to include a slowly varying amplitude $g(\kappa, \Vect{y})$], the steepest-descent contours at fixed $\Vect{y}$ are by definition the streamlines of $\nabla \Im(f)$ in the complex $\kappa$ plane $\mbb{C}^1$, where $\nabla \doteq (\pd{\Re(\kappa)},  \pd{\Im(\kappa)} )$. (Here, $\Re$ and $\Im$ denote the real and imaginary parts of a complex function.) When $f$ is analytic in $\kappa$, a more useful definition arises from the Cauchy--Riemann relation~\cite{Rudin87}
\begin{equation}
	i \pd{\Re(\kappa)} f(\kappa, \Vect{y}) = \pd{\Im(\kappa)} f(\kappa, \Vect{y})
	,
	\label{eq:5_cauchy1}
\end{equation}

\noindent which implies that $\nabla \Re(f)$ and $\nabla \Im(f)$ are orthogonal, \ie
\begin{equation}
	\nabla \Re(f) \cdot \nabla \Im(f) = 0
	.
	\label{eq:5_cauchy2}
\end{equation}

\noindent Therefore, the streamlines of $\nabla \Im(f)$ that pass through a given point $\kappa_0$ are also the set of points in $\mbb{C}^1$ that satisfy the implicit equation
\begin{equation}
	\Re\left[ f(\kappa, \Vect{y}) \right] = \Re\left[ f(\kappa_0, \Vect{y}) \right].
\end{equation}

The steepest-descent contours are almost-everywhere smooth curves, although non-differentiable kinks can occur at special points where \Eq{eq:5_cauchy2} is indeterminate, \ie where $\nabla \Re\left(f \right) = \nabla \Im\left(f \right) = \Vect{0}$.%
\footnote{By \Eq{eq:5_cauchy1}, $\nabla \Re\left(f \right)$ and $\nabla \Im\left(f \right)$ always vanish simultaneously.} %
At these points [which are saddlepoints per \Eq{eq:5_cauchy1}%
\footnote{This follows by differentiating \Eq{eq:5_cauchy1} to show that the Hessian matrices for $\Re(f)$ and $\Im(f)$ are symmetric and traceless, thereby possessing two real eigenvalues of opposite sign.}%
], the direction of $\nabla \Im\left(f \right)$ generally changes abruptly, producing the aforementioned kinks that require special parameterization. As shown later, such parameterization can be done by treating the kink as two independent curves that intersect at a finite angle.

Although saddlepoints are `rare' in that they occur at isolated points in $\mbb{C}^1$, they are often of primary interest due to their prominent role in asymptotic wave theory. More specifically, each saddlepoint of $f$ encodes the contribution to $I(\Vect{y})$ from a single corresponding GO ray. Consequently, a saddlepoint $\kappa_s(\Vect{y})$ will be real when $\Vect{y}$ is in the lit region of a caustic, but may be complex when $\Vect{y}$ is in the shadow region. Also, a caustic occurs at the specific values of $\Vect{y}$, denoted $\Vect{y}_c$, such that multiple saddlepoints coalesce and consequently, 
\begin{equation}
	\pd{\kappa}^2 f[\kappa_s(\Vect{y}_c), \Vect{y}_c] = 0
	.
	\label{eq:5_caustic}
\end{equation}


\section{Brief review of Gaussian quadrature for numerical integration }

\label{sec:5_gaussQUAD}

Suppose one wished to compute the integral of some real-valued function $h(\kappa)$ over the real interval $(a,b)$, with both $a$ and $b$ allowed to be infinite. Suppose further that $h(\kappa)$ can be partitioned as 
\begin{equation}
	h(\kappa) = \omega(\kappa) r(\kappa)
	,
	\label{eq:5_GQdecomp}
\end{equation} 

\noindent with $\omega(\kappa)$ positive-definite on $(a,b)$ and $r(\kappa)$ a polynomial of degree $2n - 1$. Then, the following $n$-point quadrature formula holds:
\begin{equation}
	\int_a^b \dd \kappa \, h(\kappa)
	\equiv
	\int_a^b \dd \kappa \, \omega(\kappa) r(\kappa)
	= \sum_{j = 1}^n w_j r(\kappa_j)
	,
	\label{eq:5_gqEXACT}
\end{equation}

\noindent where the quadrature weights $\{ w_j \}$ and nodes $\{ \kappa_j \}$ are determined as follows. 

Let us introduce the inner product
\begin{equation}
	\langle h_1, h_2 \rangle 
	\doteq \int_a^b \dd \kappa \, \omega(\kappa) h_1(\kappa) h_2(\kappa)
	.
	\label{eq:5_innerPROD}
\end{equation}

\noindent Let us also introduce the family of real-valued polynomials $\{p_\ell(\kappa)\}$ (with $\ell$ the polynomial degree) that are orthogonal with respect to \Eq{eq:5_innerPROD}, that is,
\begin{equation}
	\langle p_\ell, p_m \rangle
	= \eta_\ell \, \delta_{\ell m}
	, \quad
	\eta_\ell \doteq \langle p_\ell, p_{\ell} \rangle
	,
\end{equation}

\noindent where $\delta_{\ell m}$ is the Kronecker delta. By performing polynomial division of $r(\kappa)$ by $p_n(\kappa)$ and Lagrange interpolation of the residual, it can be shown~\cite{Gil07} that the quadrature weights $\{w_j \}$ are determined by the formula
\begin{equation}
	w_j = 
	\left\langle \prod_{\substack{\ell = 1 \\ j \neq \ell}}^n
	\frac{\kappa - \kappa_\ell}{\kappa_j - \kappa_\ell}, 1 \right\rangle
	= \frac{\langle p_n(\kappa), (\kappa - \kappa_j)^{-1} \rangle}{p'_n(\kappa_j)}
	,
	\label{eq:5_GQweights}
\end{equation}

\noindent and the quadrature nodes are the $n$ zeros of $p_n(\kappa)$, \ie
\begin{equation}
	\{\kappa_j\} = \{ \kappa ~ | ~ p_n(\kappa) = 0 \}
	.
	\label{eq:5_GQnodes}
\end{equation}

If $h(\kappa)$ cannot be decomposed as \Eq{eq:5_GQdecomp} with polynomial $r(\kappa)$, the corresponding integral can still be approximately computed as
\begin{equation}
	\int_a^b \dd \kappa \, h(\kappa)
	\approx
	\sum_{j = 1}^n w_j \frac{h(\kappa_j)}{\omega(\kappa_j)}
	.
	\label{eq:5_gq}
\end{equation}

\noindent Equation \eq{eq:5_gq} defines the Gaussian quadrature method of numerical integration. The error in using \Eq{eq:5_gq} depends on how `close' $h(\kappa)/\omega(\kappa)$ is to being a $2n-1$ degree polynomial, as determined by the maximum value of $\pd{\kappa}^{2n}(h/\omega)$ over $(a,b)$. Explicitly,~\cite{Suli03}
\begin{equation}
	\left|
		\int_a^b \dd \kappa \, h(\kappa)
		-
		\sum_{j = 1}^n w_j \frac{h(\kappa_j)}{\omega(\kappa_j)}
	\right|
	\le
	\frac{\eta_n}{(2n)!}
	\max_{\zeta \in (a,b)} 
	\left|
		\pd{\kappa}^{2n} \frac{h(\zeta)}{\omega(\zeta)}
	\right|
	.
	\label{eq:5_gqERROR}
\end{equation}

\noindent Note that the right-hand side vanishes when $h(\kappa)/\omega(\kappa)$ is a $2n-1$ degree polynomial, as desired. Also note that \Eq{eq:5_gq} is still valid when $h(\kappa)$ is complex-valued.

Common choices for $\{p_\ell(\kappa) \}$ are the rescaled Legendre polynomials for integrals over finite $(a,b)$ with $\omega(\kappa) = 1$, or the Hermite polynomials for integrals with $(a,b) = (-\infty, +\infty)$ and $\omega(\kappa) = \exp(-\kappa^2)$. For my purposes, though, I will find it more convenient to use the less-common Freud polynomials (\App{sec:5_FreudQUAD}), as I shall now explain.


\section{Gauss--Freud quadrature for catastrophe integrals}


\subsection{Derivation for fixed parameters}

Let us now develop the appropriate Gaussian quadrature rule for MGO. Along the steepest-descent contour that passes through a given saddlepoint at $\kappa = \kappa_0$, denoted $\cont{0}$, \Eq{eq:5_SDMint} takes the general form
\begin{align}
	I(\Vect{y}_0) = 
	\exp 
	\left\{
		i \Re[f(\kappa_0, \Vect{y}_0)] 
		\nullFrac
	\right\}
	\int_{\cont{0}} \dd \kappa \, 
	g(\kappa, \Vect{y}_0)
	\exp
	\left\{
		- \Im
		\left[
			f(\kappa, \Vect{y}_0) 
			\nullFrac
		\right]
	\right\}
	,
\end{align}

\noindent or equivalently,
\begin{align}
	I(\Vect{y}_0) = 
	\exp 
	\left\{
		i \Re[f(\kappa_0, \Vect{y}_0)] 
		\nullFrac
	\right\}
	\int_{-\infty}^\infty \dd l \,
	\kappa'(l)
	g[\kappa(l), \Vect{y}_0]
	\exp
	\left[
		- F(l,\Vect{y}_0)
	\right]
	,
	\label{eq:5_steepI}
\end{align}

\noindent where I have introduced $\kappa(l)$ as a $1$-D parameterization of $\cont{0}$ with $\kappa(0) = \kappa_0$, I have defined
\begin{equation}
	F(l, \Vect{y}_0) \doteq 
	\Im
	\left\{
		f[\kappa(l), \Vect{y}_0] 
		\nullFrac
	\right\}
	,
\end{equation}

\noindent and I have set $\Vect{y} = \Vect{y}_0$ to emphasize that $\Vect{y}$ should be considered a \textit{fixed parameter} for the integration over $\kappa$.%
\footnote{Specifically, $\Vect{y}$ is related to the physical location of the wavefield $\psi(\Vect{x})$ via the ray map $\Vect{x}(\Vect{\tau})$ along with the local coordinate transformation $\Vect{\tau}(\Vect{y})$ needed to place whatever integral expression into standard form.} %
Note that $\kappa_0$ being a saddlepoint implies
\begin{equation}
	\nabla \Im
	\left[
		f(\kappa_0, \Vect{y}_0) 
		\nullFrac
	\right]
	= \Vect{0}
	, \quad
	\pd{l} F(0,\Vect{y}_0)
	= 0 ,
	\label{eq:5_saddleI}
\end{equation}

\noindent and $\mc{C}_0$ being a steepest-descent contour implies
\begin{equation}
	F(l, \Vect{y}_0) \ge F(0, \Vect{y}_0)
	.
	\label{eq:5_Fdecrease}
\end{equation}

Suppose first that $\kappa_0$ is a non-degenerate saddlepoint, which typically occurs when $\Vect{y}_0$ does not coincide with a caustic. This means that
\begin{equation}
	\pd{l}^2 F(0, \Vect{y}_0) > 0
	.
	\label{eq:5_nonDEGEN}
\end{equation}

\noindent Hence, $F(l, \Vect{y}_0)$ is well-approximated around $l = 0$ as
\begin{equation}
	F(l, \Vect{y}_0) \approx F(0, \Vect{y}_0) + \pd{l}^2 F(0, \Vect{y}_0) \, l^2
	.
	\label{eq:5_taylorF}
\end{equation}

\noindent However, this approximation has obvious issues when $\kappa_0$ is a degenerate saddlepoint, which occurs when $\Vect{y}_0$ coincides with a caustic. For this case, although by \Eq{eq:5_caustic},
\begin{equation}
	\pd{l}^2 F(0, \Vect{y}_0) = 0
	,
	\label{eq:5_degen}
\end{equation}

\noindent a quadratic function can still be fit to $F(l, \Vect{y}_0)$ as 
\begin{equation}
	F(l, \Vect{y}_0) \approx F(0, \Vect{y}_0) + s(\Vect{y}_0) \, l^2
	,
	\label{eq:5_secantF}
\end{equation}

\noindent provided the scaling factor $s(\Vect{y}_0)$ is chosen appropriately; I choose to use the finite-difference formula
\begin{align}
	s(\Vect{y}_0) 
	= 
	\left\{
		\begin{array}{lr}
			s_-(\Vect{y}_0), & l \le 0 \\
			s_+(\Vect{y}_0), & l > 0
		\end{array}
	\right.
	, \quad
	s_\pm(\Vect{y}_0) 
	\doteq
	\frac{F(l_\pm, \Vect{y}_0) - F(0, \Vect{y}_0)}{ | l_\pm |^2 }
	,
	\label{eq:5_sDEF}
\end{align}

\noindent where $l_\pm$ satisfy the threshold condition
\begin{equation}
	F(l_\pm, \Vect{y}_0) - F(0, \Vect{y}_0) \ge C_\pm 
	\label{eq:5_kappaL}
\end{equation}

\noindent with $C_\pm$ arbitrary constants. (I use $C_\pm = 1$ for simplicity, but I found varying $C_\pm$ over the range $[0.5, 5]$ produced only $O(10^{-5})$ differences.) Note that I have allowed the possibility for different scaling factors on either side of $l = 0$ in case $\cont{0}$ has a kink at $\kappa_0$ (see \Fig{fig:5_EXcontour}). Importantly, \Eq{eq:5_Fdecrease} implies that $s > 0$. Also note that \Eq{eq:5_secantF} reduces to \Eq{eq:5_taylorF} in the limit $C_\pm \to 0$ when $\kappa_0$ is non-degenerate.

It is worth mentioning here that the quadratic approximation \eq{eq:5_secantF} constitutes a compromise. On one hand, the majority of points in $x$ are not caustics, so it is reasonable to optimize the quadrature rule for quadratic phase functions; on the other hand, the main usefulness of this algorithm for MGO depends heavily on the accuracy in computing $\Upsilon$ at these special points. Hence, I have chosen a scheme that is exact when $F$ is locally quadratic, \ie far from a caustic, and is still well-defined at caustics when $F$ is not locally quadratic, by virtue of the quadratic interpolation.

As a final simplification, let us adopt a piecewise linear approximation to $\cont{0}$ such that%
\footnote{The mapping $\kappa(l)$ being nonlinear is not a problem for Gaussian quadrature \textit{per se}, but it necessitates the inclusion of expensive root-finding steps into the Gaussian quadrature algorithm~\cite{Deano09}}%
\begin{equation}
	\kappa(l) \approx \kappa_0 + |l| \times
	\left\{
		\begin{array}{lr}
			\exp(i \sigma_-), & l \le 0 \\
			\exp(i \sigma_+), & l > 0
		\end{array}
	\right.
	\label{eq:5_unionC0}
\end{equation}

\noindent for suitable rotation angles $\sigma_\pm$, which are allowed to be different in case $\cont{0}$ has a kink at $\kappa_0$. It would be natural to choose $\sigma_\pm$ such that \Eq{eq:5_unionC0} is a tangent-line approximation to $\cont{0}$ at $\kappa_0$; however, this choice cannot be applied to degenerate saddlepoints. Instead, I shall allow \Eq{eq:5_unionC0} to generally describe the secant-line approximations to $\cont{0}$ that underlie \Eq{eq:5_sDEF}, namely,
\begin{align}
	\sigma_\pm 
	= \text{arg}
	\left[
		\kappa(l_\pm) - \kappa_0
	\right]
	=
	\text{sign}
	\left\{
		\frac{
			\Im\left[
				\kappa(l_\pm) - \kappa_0
			\right]
		}{
			\| \kappa(l_\pm) - \kappa_0 \|
		}
	\right\}
	\cos^{-1}
	\left\{
		\frac{
			\Re\left[
				\kappa(l_\pm) - \kappa_0
			\right]
		}{
			\| \kappa(l_\pm) - \kappa_0 \|
		}
	\right\}
	\label{eq:5_sigmaDEF}
	.
\end{align}

\noindent Here, I take the convention that $\text{sign}(0) = 1$; hence \Eq{eq:5_sigmaDEF} restricts $\sigma_\pm$ to lie on the interval $(-\pi, \pi]$.

Inserting \Eqs{eq:5_secantF}, \eq{eq:5_sDEF}, and \eq{eq:5_unionC0} into \Eq{eq:5_steepI} yields
\begin{align}
	\frac{
		I(\Vect{y}_0)
	}{
		\exp[i f(\kappa_0,\Vect{y}_0)]
	}
	&\approx
	\int_0^\infty \dd l 
	\left\{
		\nullFrac
		g[\kappa_0 + l \exp(i \sigma_+), \Vect{y}_0]
		\exp[i \sigma_+ - s_+(\Vect{y}_0) l^2]
		\right.\nonumber\\
		&\hspace{17mm}\left.
		-
		g[\kappa_0 + l \exp(i \sigma_-), \Vect{y}_0]
		\exp[i \sigma_- - s_-(\Vect{y}_0) l^2]
		\nullFrac
	\right\}
	\nonumber\\
	&=
	\int_0^\infty \dd l 
	\left\{
		g
		\left[
			\kappa_0 
			+ \frac{ l \exp(i \sigma_+) }{ \sqrt{s_+(\Vect{y}_0)} }
			, \Vect{y}_0
		\right]
		\frac{
			\exp(i \sigma_+)
		}{
			\sqrt{s_+(\Vect{y}_0)} 
		}
		\right.\nonumber\\
		&\hspace{17mm}\left.
		-
		g
		\left[
			\kappa_0 
			+ \frac{ l \exp(i \sigma_-) }{ \sqrt{s_-(\Vect{y}_0)} }
			, \Vect{y}_0
		\right]
		\frac{
			\exp(i \sigma_-)
		}{
			\sqrt{s_-(\Vect{y}_0)} 
		}
	\right\}
	\exp(- l^2)
	.
\end{align}

\noindent Hence, an appropriate Gaussian quadrature rule for MGO is based on the inner product
\begin{equation}
	\langle h_1, h_2 \rangle
	= \int_0^\infty \dd l \, h_1(l) h_2(l) 
	\exp(-l^2)
	,
	\label{eq:5_FreudIP}
\end{equation}

\noindent for which Freud polynomials are orthogonal (\App{sec:5_FreudQUAD}). Using \Eq{eq:5_gq}, I obtain the quadrature rule for MGO:
\begin{equation}
	I(\Vect{y}_0)
	\approx 
	\sum_{j = 1}^n 
	w_j \exp(l_j^2)
	\left\{
		h\left[
			\kappa_0 
			+ \frac{l_j \exp(i \sigma_+)}{\sqrt{s_+(\Vect{y}_0)}}, \Vect{y}_0
		\right]
		\frac{\exp(i \sigma_+)}{\sqrt{s_+(\Vect{y}_0)}}
		-
		h\left[
			\kappa_0 
			+ \frac{l_j \exp(i \sigma_-)}{\sqrt{s_-(\Vect{y}_0)}}, \Vect{y}_0
		\right]
		\frac{\exp(i \sigma_-)}{\sqrt{s_-(\Vect{y}_0)}}
	\right\}
	,
	\label{eq:5_MGOquad}
\end{equation}

\noindent where $h(\kappa) = g(\kappa, \Vect{y}_0) \exp[i f(\kappa, \Vect{y}_0) ]$, and $\{l_j \}$ are the quadrature nodes. [I use $\{ l_j \}$ rather than $\{ \kappa_j \}$ to be consistent with the notation of \Eq{eq:5_FreudIP}.] Since Gauss--Freud quadrature is somewhat uncommon, a table of the corresponding $\{w_j\}$ and $\{l_j\}$ for various values of $n$ is also provided in \App{sec:5_FreudQUAD}.


\subsection{Angle memory feedback for parameter scans and simulations}
\label{sec:5_feedback}

Let us now allow $\Vect{y}$ to vary in \Eq{eq:5_MGOquad} (rather than being fixed at some $\Vect{y} = \Vect{y}_0$), as will occur when MGO is used to simulate a propagating wave. The steepest-descent topology for $I(\Vect{y})$ will be different for each new value of $\Vect{y}$, with correspondingly new values of $\sigma_\pm$. Repeatedly searching for $\cont{0}$ to compute $\sigma_\pm$ via \Eq{eq:5_sigmaDEF} can be computationally expensive, and merely identifying the correct $\cont{0}$ can be difficult in situations where multiple valid steepest-descent lines exist, as occurs at caustics. Fortunately, the steepest-descent topology of $I(\Vect{y})$ typically evolves smoothly with $\Vect{y}$, which means successive calculations of $\sigma_\pm$ will be correlated. I use this fact to construct a `memory feedback' algorithm to both speed up the time required to calculate the steepest-descent topology of $I(\Vect{y})$ and to correctly identify $\cont{0}$ at caustics.

First, let us initialize the MGO simulation far from a caustic such that $\cont{0}$ is sufficiently simple: I expect the initial angles $\sigma_\pm^{(0)}$ to be approximately given as
\begin{equation}
	\sigma_\pm^{(0)}
	\approx -\frac{\pi}{4} - \frac{\text{arg}[\pd{\kappa}^2f(\kappa_0, \Vect{y}_0)]}{2} \pm \frac{\pi}{2}
\end{equation}

\noindent restricted to the interval $(-\pi, \pi]$. By starting the search for the exact $\cont{0}$ near this value of $\sigma_\pm^{(0)}$, the search time can be reduced. As the simulation progresses, $\cont{0}$ will evolve smoothly; at each new point $\Vect{y}_j$, the search-time for $\cont{0}$ can be reduced by initializing the search near the previously calculated $\sigma_\pm^{(j-1)}$ corresponding to the previous position $\Vect{y}_{j-1}$. Moreover, by restricting the search to \textit{only} consider angles near $\sigma_\pm^{(j-1)}$, \ie restricting $|\sigma_\pm^{(j)} - \sigma_\pm^{(j-i)}| \le \Delta $ for some threshold $\Delta$ (I choose $\Delta = 0.01$), the correct $\cont{0}$ will naturally be identified by analytic continuation, even at caustics.

Thus far I have assumed $1$-D, but the generalization to $N$-D should be straightforward since multivariate analytic functions are analytic in each variable separately; the $N$-D steepest-descent surface is the union of (continuous families of) individual steepest-descent curves for each variable. Hence, it might be possible to evaluate $N$-D integrals as a nested series of $1$-D integrals~\cite{Press07} sequentially evaluated using my $1$-D algorithm along that variable's steepest-descent curve. This approach is best done when the nested integrals are sufficiently simple and the multidimensional weight function factors cleanly into a product of univariate weight functions, which is expected to be true for MGO. If not, though, the more complicated method of $N$-D Gaussian cubature~\cite{Cools97} must be used.

A word of caution: it is possible that this angle-finding scheme may fail to correctly identify the `jump discontinuities' in the contour topology that occur when crossing from one side of a caustic to the other. (The proposed algorithm is designed to handle the case when the simulation propagates into a caustic and stops, since the two examples discussed in Secs.~\ref{sec:5_EM1} and \ref{sec:5_EM2} can be formulated as such.) The proposed algorithm should have no problem when crossing a fold caustic since there will be only two steepest-descent lines emanating from the saddlepoint on either side of the caustic; crossing a cusp point (while remaining on the fold line) may present difficulties however, since there will be three steepest-descent lines emanating from the saddlepoint on either side, from which the correct two (incoming/outgoing) steepest-descent lines must be identified. That said, this is not a failure of the Gauss--Freud algorithm presented in this chapter; rather, the prerequisite contour identification algorithm that I describe here may simply be too primitive for some advanced cases, necessitating further developments. A good example to benchmark a contour-finding algorithm would be the cusp caustic example discussed in \Ch{ch:Ex}, which was outside of the scope of the present thesis work.


\section{Benchmarking results}


\subsection{Isolated saddlepoint}

\begin{figure}
	\centering
	\includegraphics[width=0.32\linewidth,trim={16mm 4mm 26mm 4mm},clip]{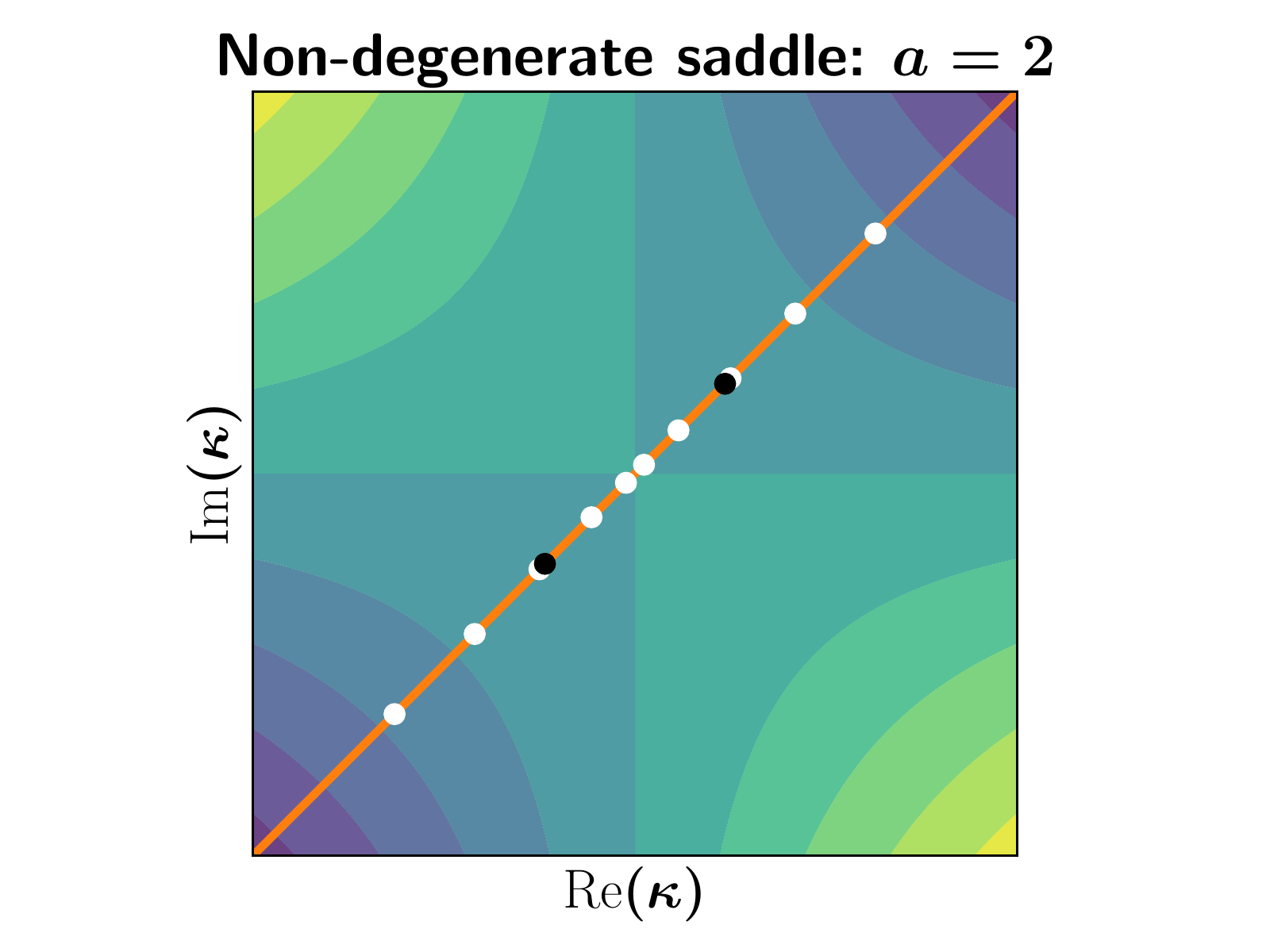}
	\includegraphics[width=0.32\linewidth,trim={16mm 4mm 26mm 4mm},clip]{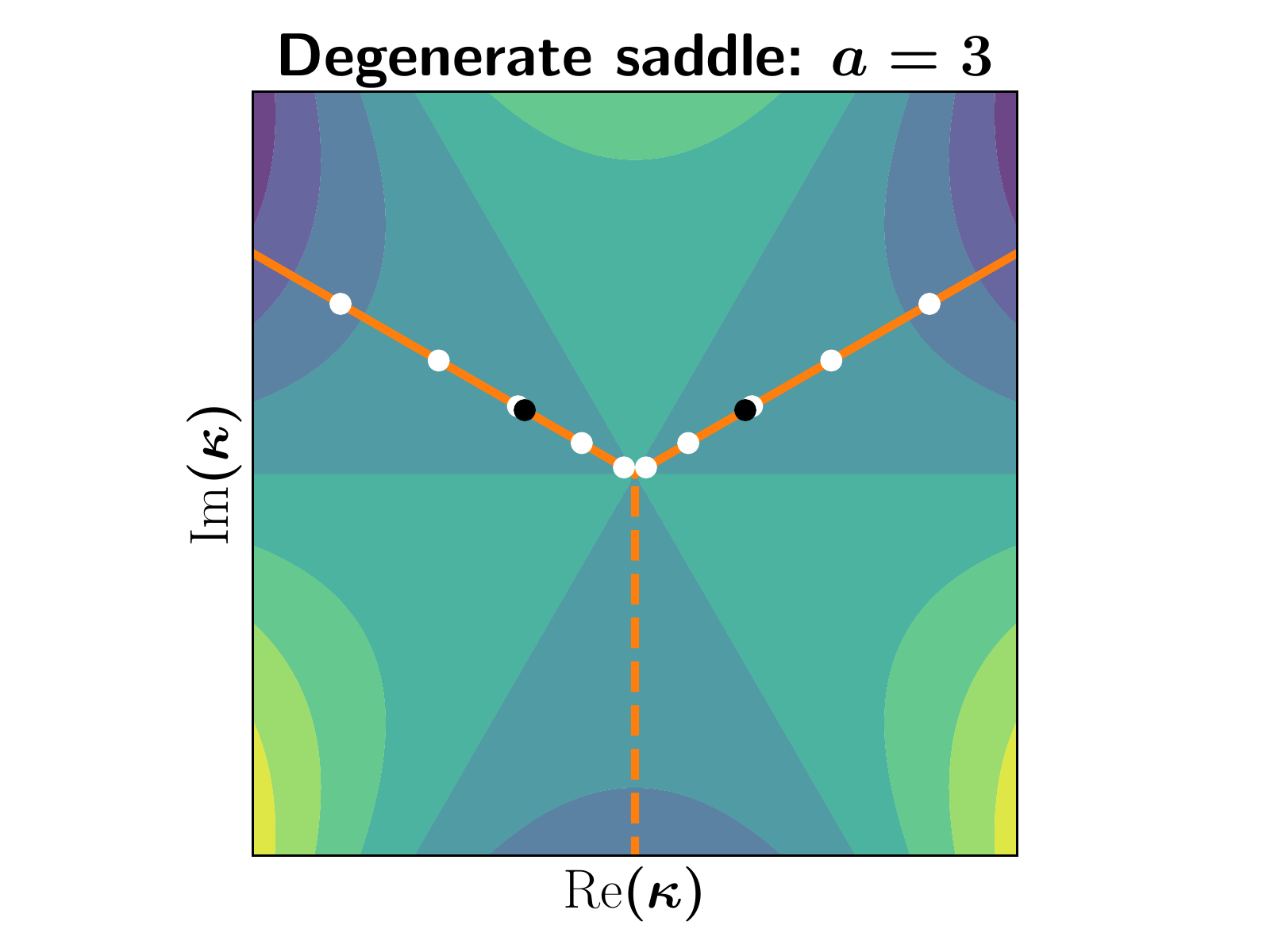}
	\includegraphics[width=0.32\linewidth,trim={16mm 4mm 26mm 4mm},clip]{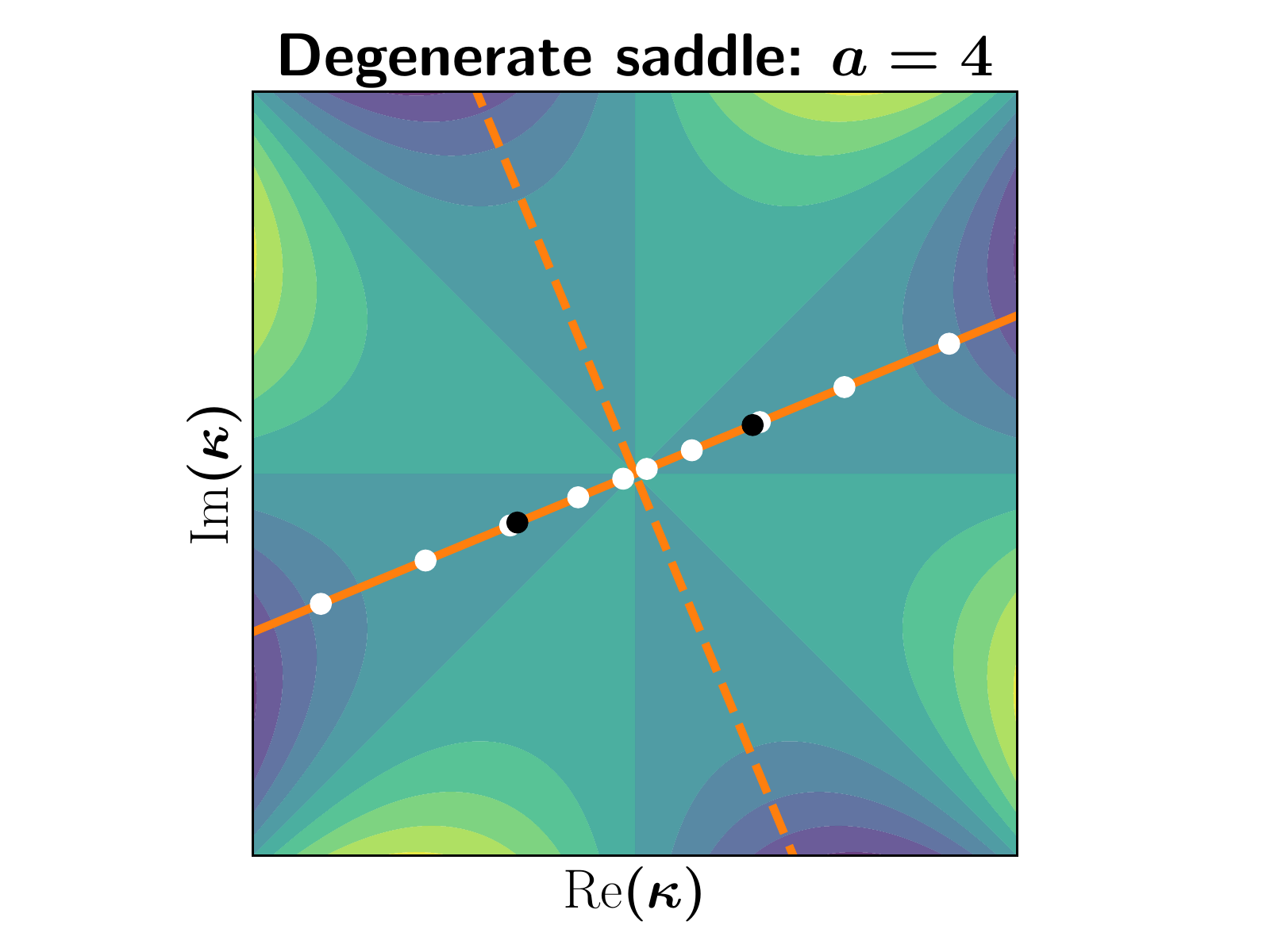}
	\caption{Steepest-descent contours (orange) for the integrand phase function $f(\kappa) \doteq \kappa^a$ of \Eq{eq:5_EXgauss} for various values of the parameter $a$, which characterizes the saddlepoint degeneracy. The background color represents the magnitude of the integrand $-\Im(f)$, with green corresponding to larger values and blue corresponding to smaller values. The order $n = 5$ quadrature nodes are shown as white dots, while the points $\kappa(l_\pm)$ [\Eq{eq:5_kappaL}] that are used to determine the rotation angles $\sigma_\pm$ [\Eq{eq:5_sigmaDEF}] are shown as black dots. Unused steepest-descent contours are shown as dashed orange lines. As can be seen, the steepest-descent contour has a kink when $a$ is odd, which requires $\sigma_+$ and $\sigma_-$ to be calculated separately.}
	\label{fig:5_EXcontour}
\end{figure}

As a first benchmarking of my numerical steepest-descent algorithm \eq{eq:5_MGOquad}, let us consider the numerical evaluation of the following family of integrals:
\begin{align}
	I(a,b) \doteq
	\int_{-\infty}^\infty \dd \kappa \, \kappa^b \exp\left(i \kappa^a\right)
	,
	\quad
	a \ge 2, \,
	b \ge 0
	,
	\label{eq:5_EXgauss}
\end{align}

\noindent whose exact solution is given by
\begin{equation}
	I_\textrm{ex}(a,b)
	=
	\frac{2}{a} \, \Gamma\left( \frac{2 \chi}{\pi} \right)
	\times
	\left\{
		\begin{array}{lr}
			\exp\left( i \chi \right), & \bar{a} = 0, \, \bar{b} = 0\\[1mm]
			0, & \bar{a} = 0, \, \bar{b} = 1\\[1mm]
			\cos \chi, & \bar{a} = 1, \, \bar{b} = 0\\[1mm]
			i \sin \chi, & \bar{a} = 1, \, \bar{b} = 1
		\end{array}
	\right.
	,
\end{equation}

\noindent where I have defined
\begin{equation}
	\chi \doteq \frac{1+b}{2a} \pi
	,
	\quad
	\bar{a} \doteq \textrm{mod}_2(a) 
	,
	\quad
	\bar{b} \doteq \textrm{mod}_2(b) 
	.
\end{equation}

\noindent The family $I(a,b)$ also corresponds to the $A_{a-1}$ `cuspoid' caustic family (Table~\ref{tab:2_caustic}) evaluated at $\Vect{y} = 0$; as such, the integrand of $I(a,b)$ has an isolated saddlepoint at $\kappa = 0$ whose degeneracy is controlled by the value of $a$, with $a = 2$ being non-degenerate. To evaluate $I(a,b)$ via \Eq{eq:5_MGOquad} the scaling factors $s_\pm$ and rotation angles $\sigma_\pm$ are needed; these are given respectively as $s_\pm = 1$ and
\begin{equation}
	\sigma_+ = \frac{\pi}{2a}
	, \quad
	\sigma_- = 
	\left\{
		\begin{array}{lr}
			\sigma_+ - \pi, & \bar{a} = 0 \\
			\pi - \sigma_+, & \bar{a} = 1
		\end{array}
	\right.
	.
	\label{eq:5_EXangles}
\end{equation}

\noindent In particular, \Eq{eq:5_EXangles} implies that the steepest-descent contour has a kink at $\kappa = 0$ when $a$ is odd, which necessitates my partitioning of \Eq{eq:5_MGOquad} into incoming and outgoing branches. This feature is also shown in \Fig{fig:5_EXcontour}.

\begin{figure}
	\centering
	\includegraphics[width=0.6\linewidth,trim={2mm 6mm 4mm 4mm},clip]{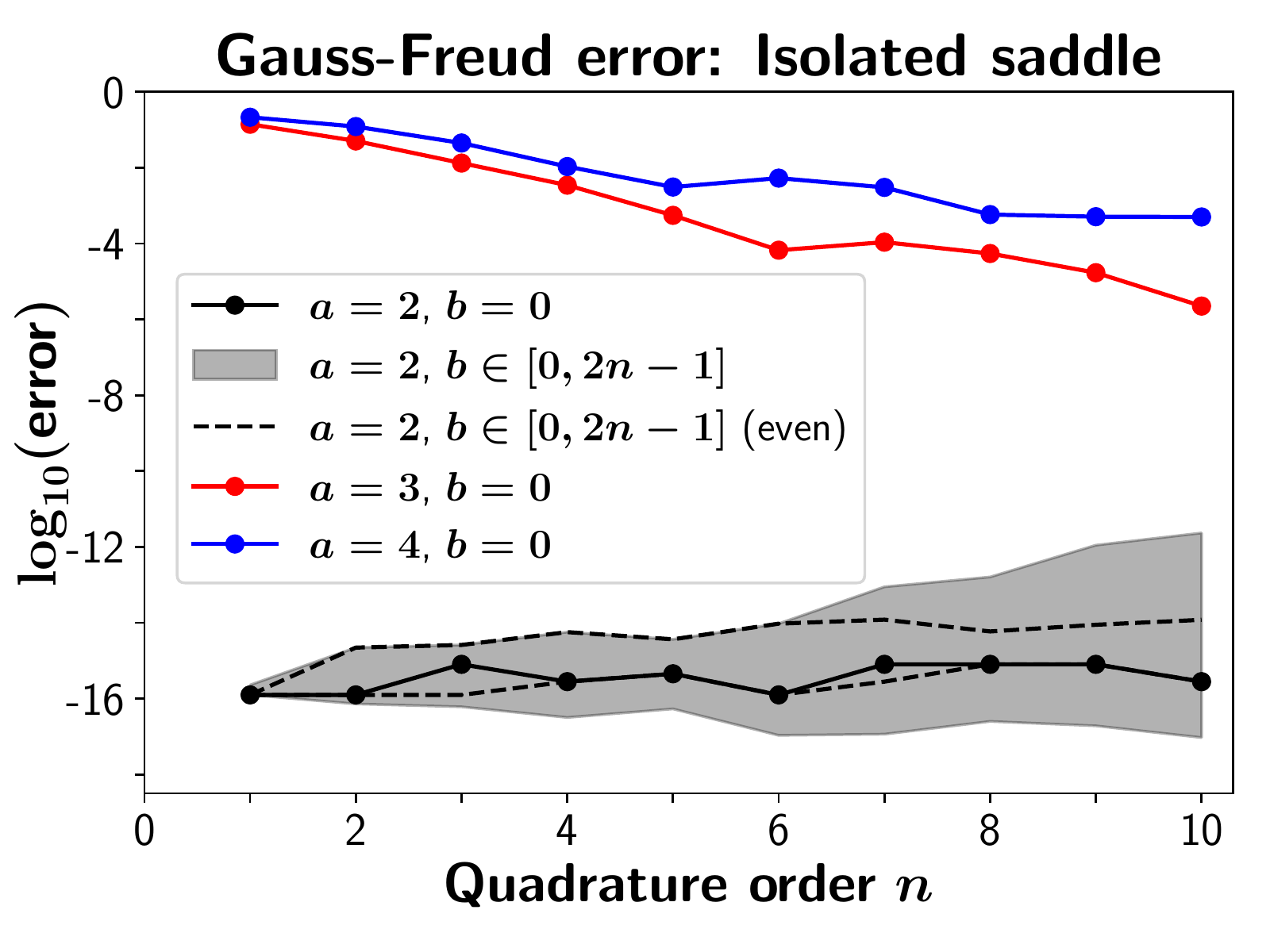}
	\caption{Comparison of the error in computing \Eq{eq:5_EXgauss} using the quadrature rule of \Eq{eq:5_MGOquad} for various values of $a$ and $b$. The error metric used is the relative error when $I(a, b)$ is nonzero and the absolute error otherwise (with the $1$-norm used for both). The shaded gray region marks the range of error over the entire range $b \in [0, 2n-1]$ for which my quadrature rule is expected to be exact, while the dashed black lines bound the region obtained when only even values of $b$ are considered. The precision of the quadrature nodes and weights used is $10^{-15}$. }
	\label{fig:5_EXerr}
\end{figure}

Figure~\ref{fig:5_EXerr} shows the error that results from evaluating $I(a,b)$ via \Eq{eq:5_MGOquad} for quadrature order $n \le 10$. The quadrature weights and nodes used have precision $10^{-15}$ and are listed explicitly in Table~\ref{tab:5_GFnodes}. Note that my quadrature rule was developed to evaluate $I(a,b)$ exactly when $a = 2$ and $b \in [0, 2n - 1]$, and indeed, one observes that the error for these values of $a$ and $b$ remains on the order of the node/weight precision until $n = 6$, beyond which the error is slightly larger than expected. However, this increased error is not due to issues with my quadrature rule \textit{per se}, but rather due to the round-off error that unavoidably accumulates when subtracting large numbers. This conclusion is corroborated by the fact that the increased error is isolated to the cases when $b$ is even and $I(a,b)$ should be identically zero in exact arithmetic by (anti-) symmetry. When $a > 2$, my quadrature rule achieves a respectable accuracy of $10^{-4}$ even at the relatively low quadrature order of $n = 10$, demonstrating the utility of \Eq{eq:5_MGOquad} at caustics and regular points alike.

\subsection{EM wave in unmagnetized plasma slab with linear density profile}
\label{sec:5_EM1}

As a more realistic example, let us consider an EM wave propagating in a stationary unmagnetized plasma slab with a linearly varying density profile. Suppose that the EM wave and all subsequently induced fluctuations have time dependence of the form $\exp(-i \Omega t)$, where $\Omega$ is the wave frequency. Then, after defining $x$ as the direction of inhomogeneity, the electric field of the EM wave can be shown to satisfy~\cite{Stix92}
\begin{equation}
	\pd{x}^2 \psi(x) + \frac{\Omega^2}{c^2}\left[1 - \frac{n(x)}{n_c} \right] \psi(x) = 0
	,
	\label{eq:5_waveEQ}
\end{equation}

\noindent where $c$ is the speed of light in vacuum and $n_c$ is the cutoff density. Let us assume
\begin{equation}
	n(x) = n_c \left(1 + \frac{x}{L_n} \right)
	,
\end{equation}

\noindent where $L_n$ is some constant length scale. Then, \Eq{eq:5_waveEQ} takes the form
\begin{equation}
	\pd{q}^2 \psi(q) - q \psi(q) = 0
	,
	\label{eq:5_eqAIRY}
\end{equation}

\noindent where I have introduced the re-scaled spatial variable
\begin{equation}
	q \doteq x \left( \frac{\Omega^2}{c^2 L_n} \right)^{1/3}
	.
\end{equation}

Equation \eq{eq:5_eqAIRY} is known as Airy's equation, and contains a fold-type $A_2$ caustic at the cutoff location $q = 0$. Assuming that $E(q \to \infty) = 0$, the exact solution is given by the Airy function
\begin{equation}
	\psi_\textrm{ex}(q) = \airyA(q)
	,
	\label{eq:5_exactAIRY}
\end{equation}

\noindent (where the overall constant is set to unity for simplicity) while the MGO solution to \Eq{eq:5_eqAIRY} can be written in the underdense region $q \le 0$ as (\Ch{ch:Ex})
\begin{align}
	\psi_\textrm{MGO}(q) =
	\Upsilon\left(|q|^{1/2} \right) 
	\exp\left(- i \frac{2}{3} |q|^{3/2} \right) 
	+
	\Upsilon\left(-|q|^{1/2} \right) 
	\exp\left(i \frac{2}{3} |q|^{3/2} \right)
	.
	\label{eq:5_mgoAIRY}
\end{align}

\noindent The integral function $\Upsilon$ in \Eq{eq:5_mgoAIRY} has the form
\begin{equation}
	\Upsilon(p) 
	\doteq \frac{1}{2\pi}
	\int_{\cont{0}} \dd \epsilon \,
	\frac{
		\vartheta(p) 
		\exp\left[
			i f(\epsilon, p)
		\right]
	}{
		\left[ 
			\vartheta^4(p) - 8 \vartheta(p) p \epsilon 
		\right]^{1/4}
	}
	,
	\label{eq:5_upsilonAIRY}
\end{equation}

\noindent where the phase function $f$ is given as
\begin{align}
	f(\epsilon,p) 
	\doteq
	\frac{\vartheta^6(p) - \left[\vartheta^4(p) - 8 \vartheta(p) p \epsilon \right]^{3/2}}{96p^3}
	- \frac{\vartheta^3(p)}{8p^2} \epsilon
	+ \frac{\vartheta^2(p)}{4p} \epsilon^2
	,
	\label{eq:5_fAIRY}
\end{align}

\noindent and I have defined $\vartheta(p) \doteq \sqrt{1 + 4 p^2}$. When \Eq{eq:5_upsilonAIRY} is evaluated using the stationary-phase approximation, the standard GO approximation for \Eq{eq:5_eqAIRY} is obtained:
\begin{equation}
	\psi_\textrm{GO}(q) =
	\pi^{-1/2} |q|^{-1/4} 
	\sin\left(\frac{2}{3} |q|^{3/2} + \frac{\pi}{4} \right)
	.
	\label{eq:5_goAIRY}
\end{equation}

\noindent Clearly, the GO solution diverges at the caustic $q = 0$.

\begin{figure}
	\centering
	\includegraphics[width=0.24\linewidth,trim={18mm 4mm 32mm 4mm},clip]{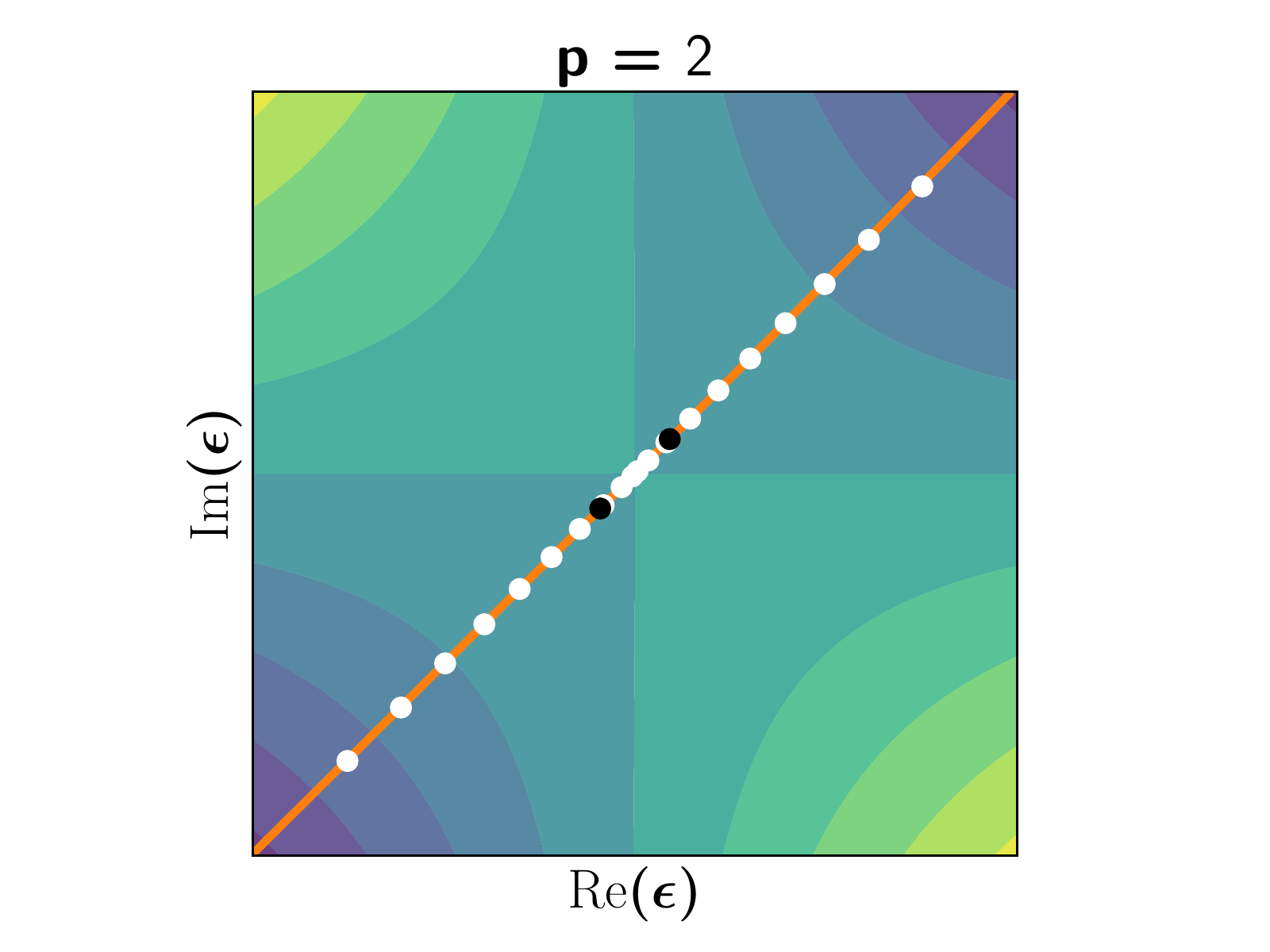}
	\includegraphics[width=0.24\linewidth,trim={18mm 4mm 32mm 4mm},clip]{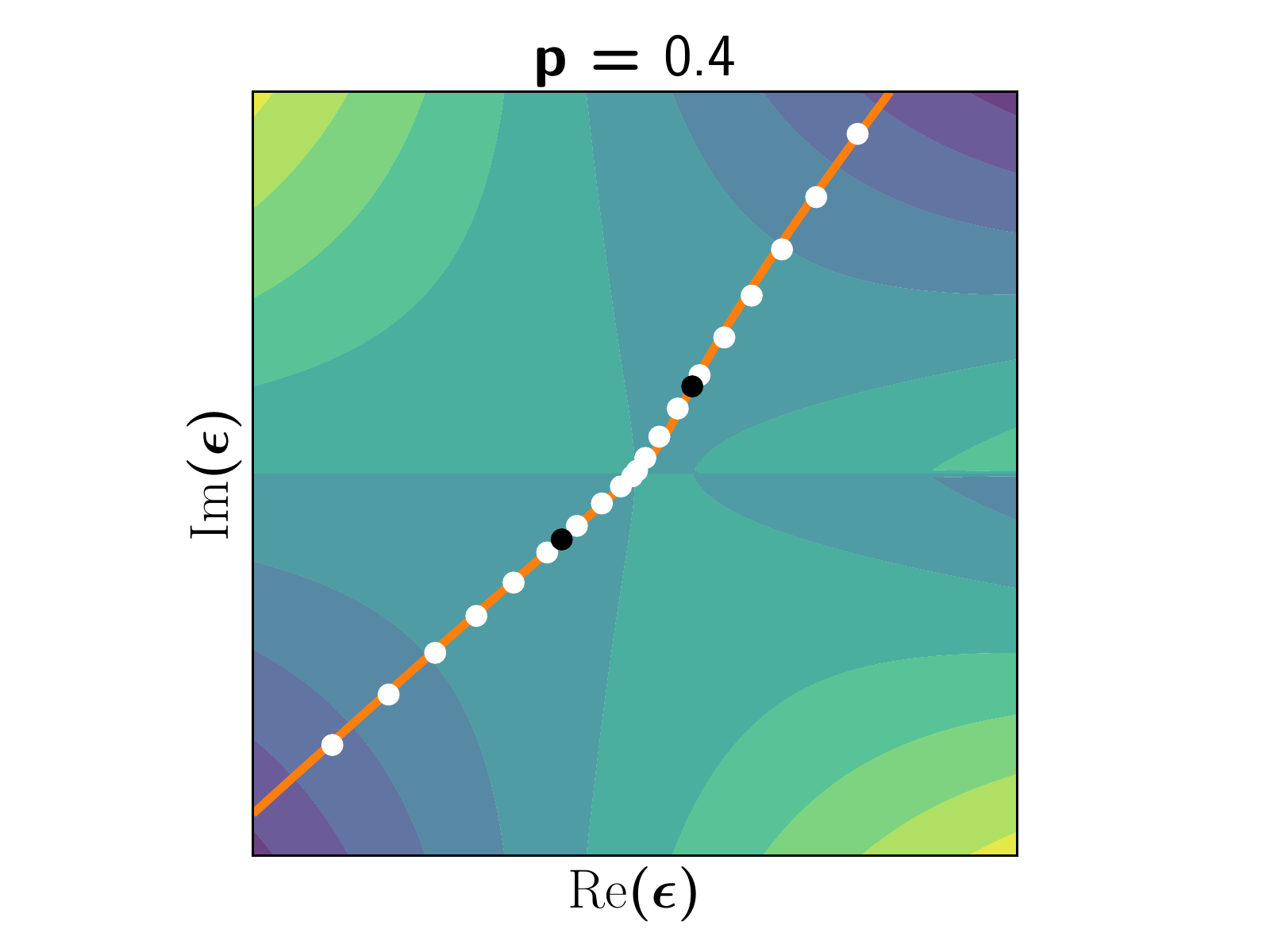}
	\includegraphics[width=0.24\linewidth,trim={18mm 4mm 32mm 4mm},clip]{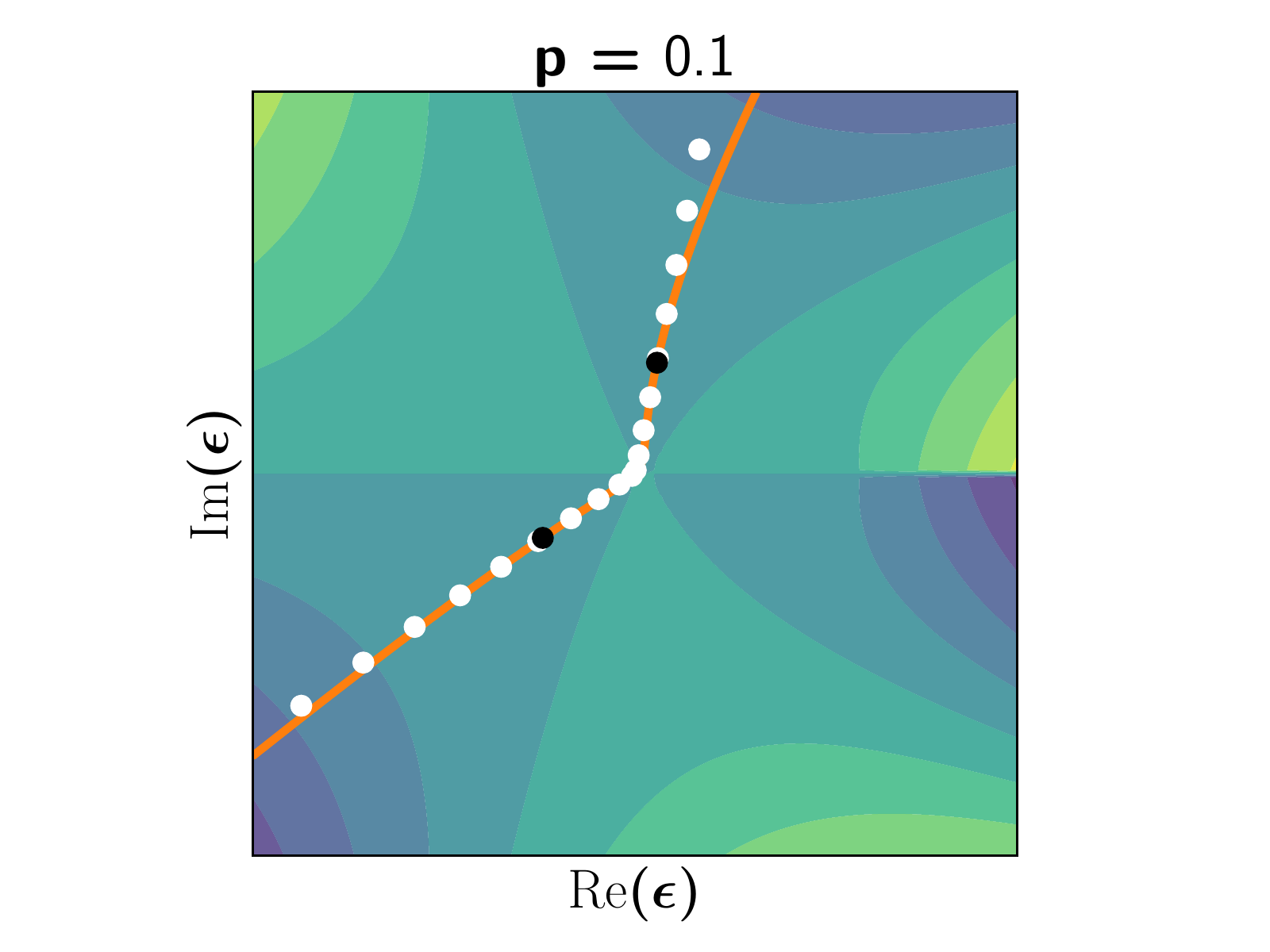}
	\includegraphics[width=0.24\linewidth,trim={18mm 4mm 32mm 4mm},clip]{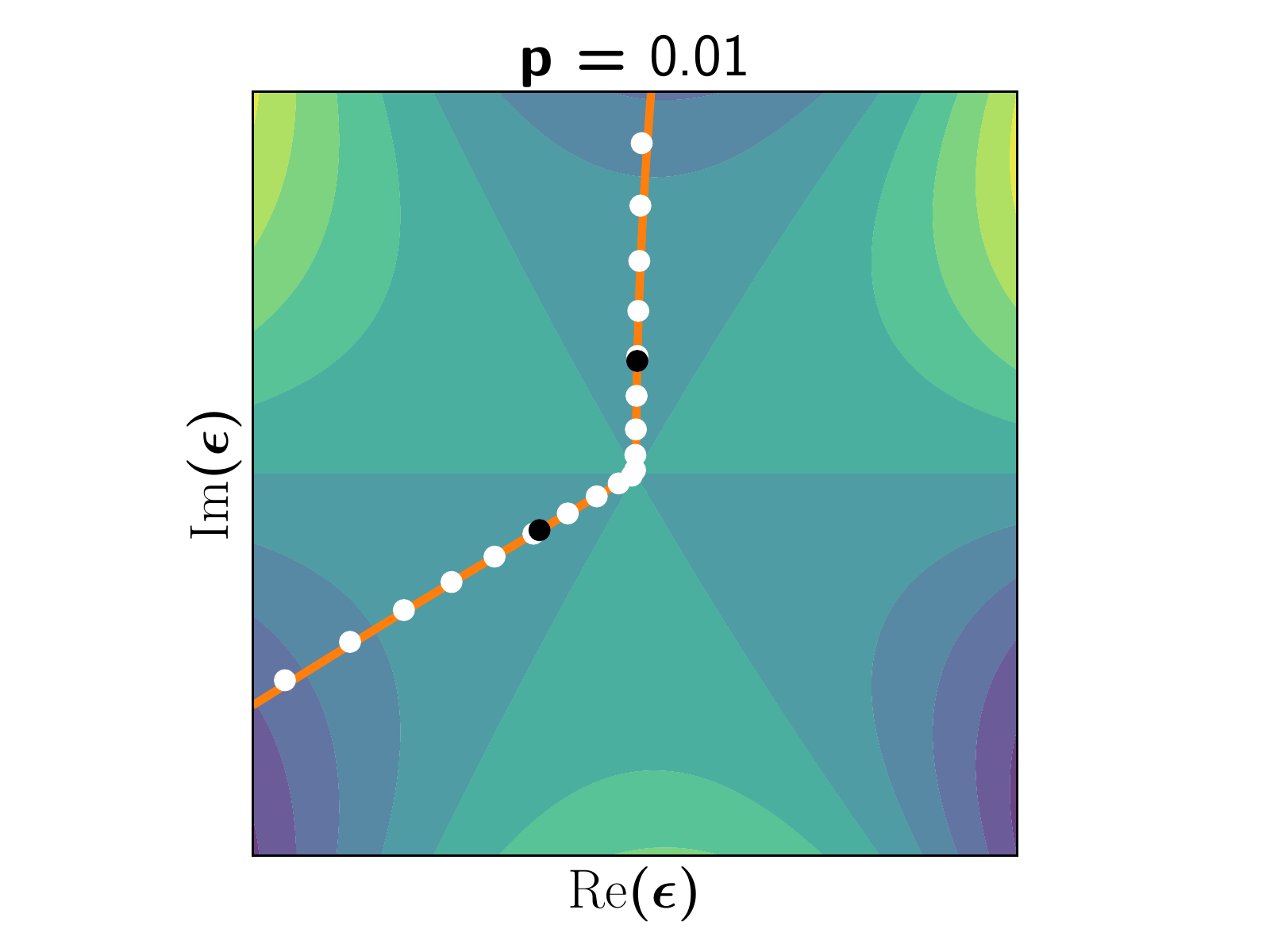}

	\vspace{2mm}
	\includegraphics[width=0.24\linewidth,trim={18mm 4mm 32mm 4mm},clip]{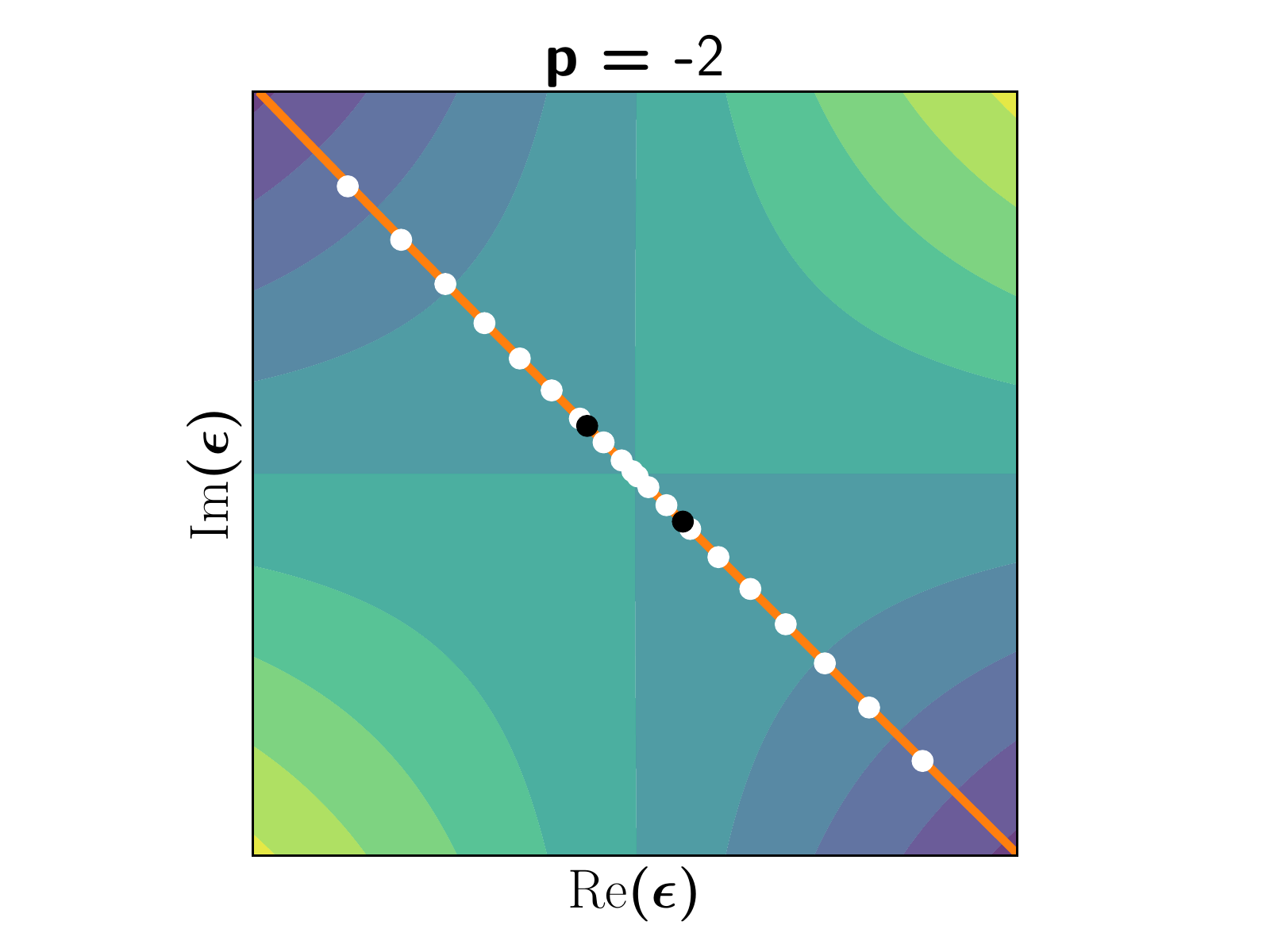}
	\includegraphics[width=0.24\linewidth,trim={18mm 4mm 32mm 4mm},clip]{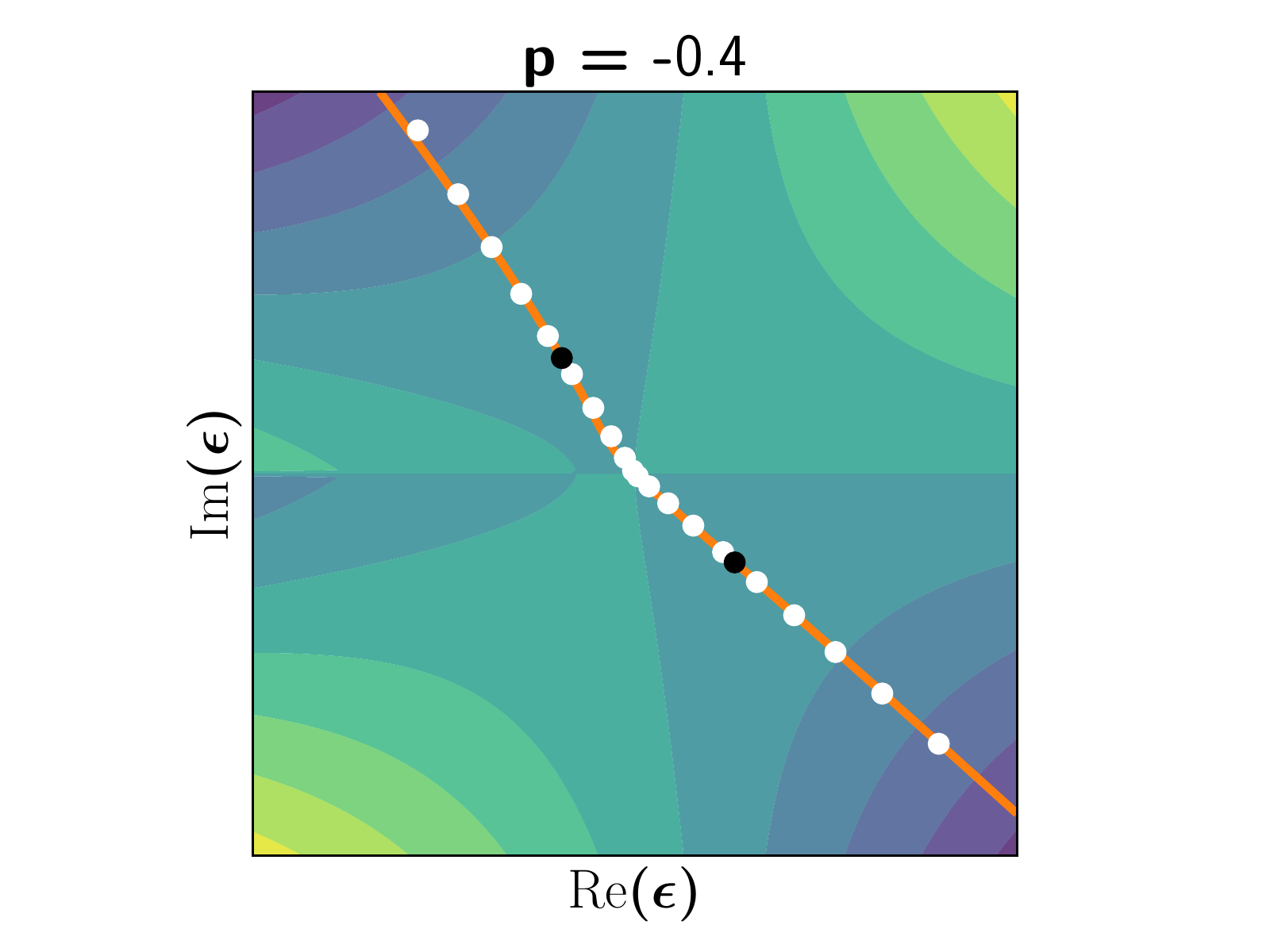}
	\includegraphics[width=0.24\linewidth,trim={18mm 4mm 32mm 4mm},clip]{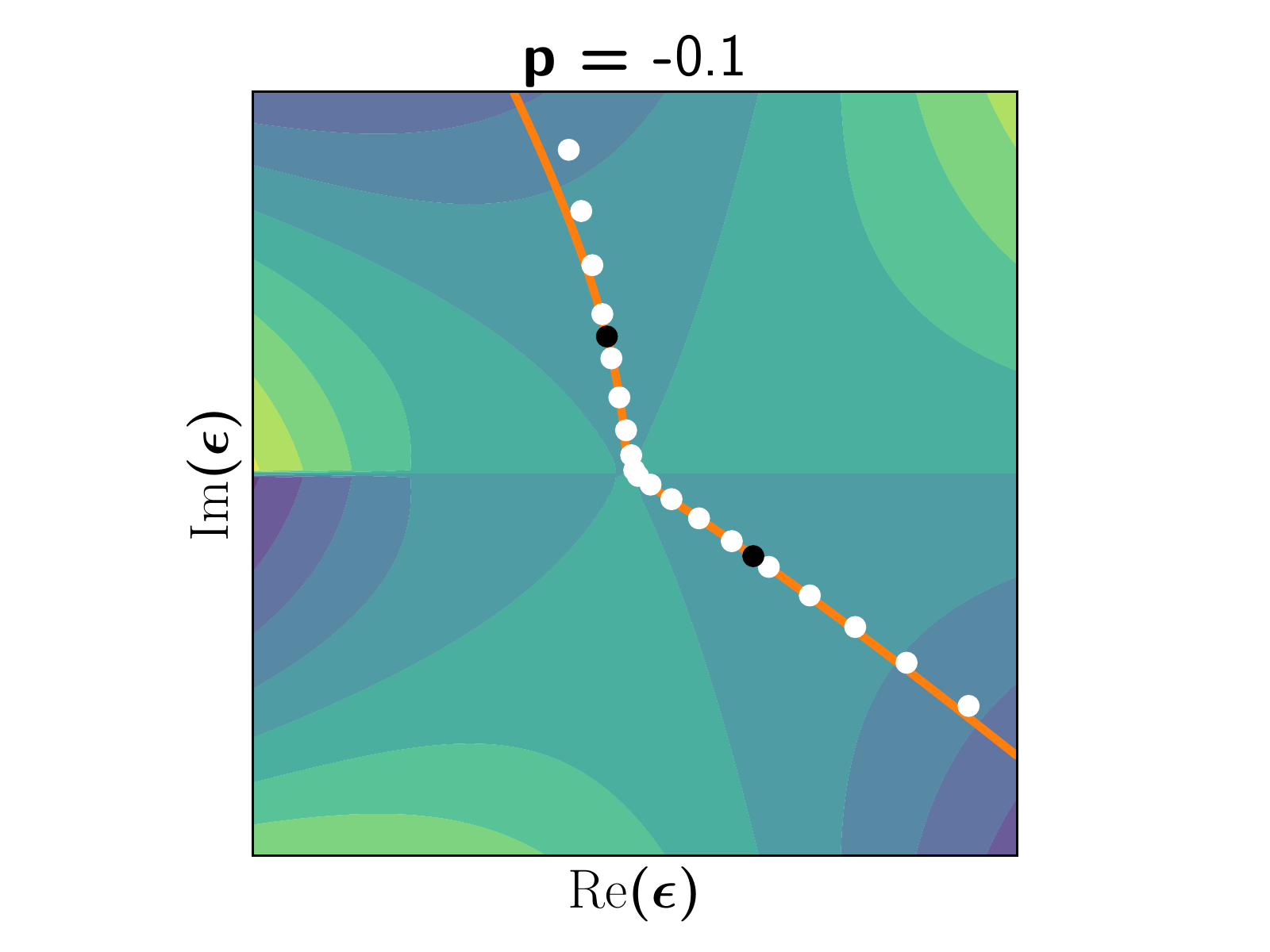}
	\includegraphics[width=0.24\linewidth,trim={18mm 4mm 32mm 4mm},clip]{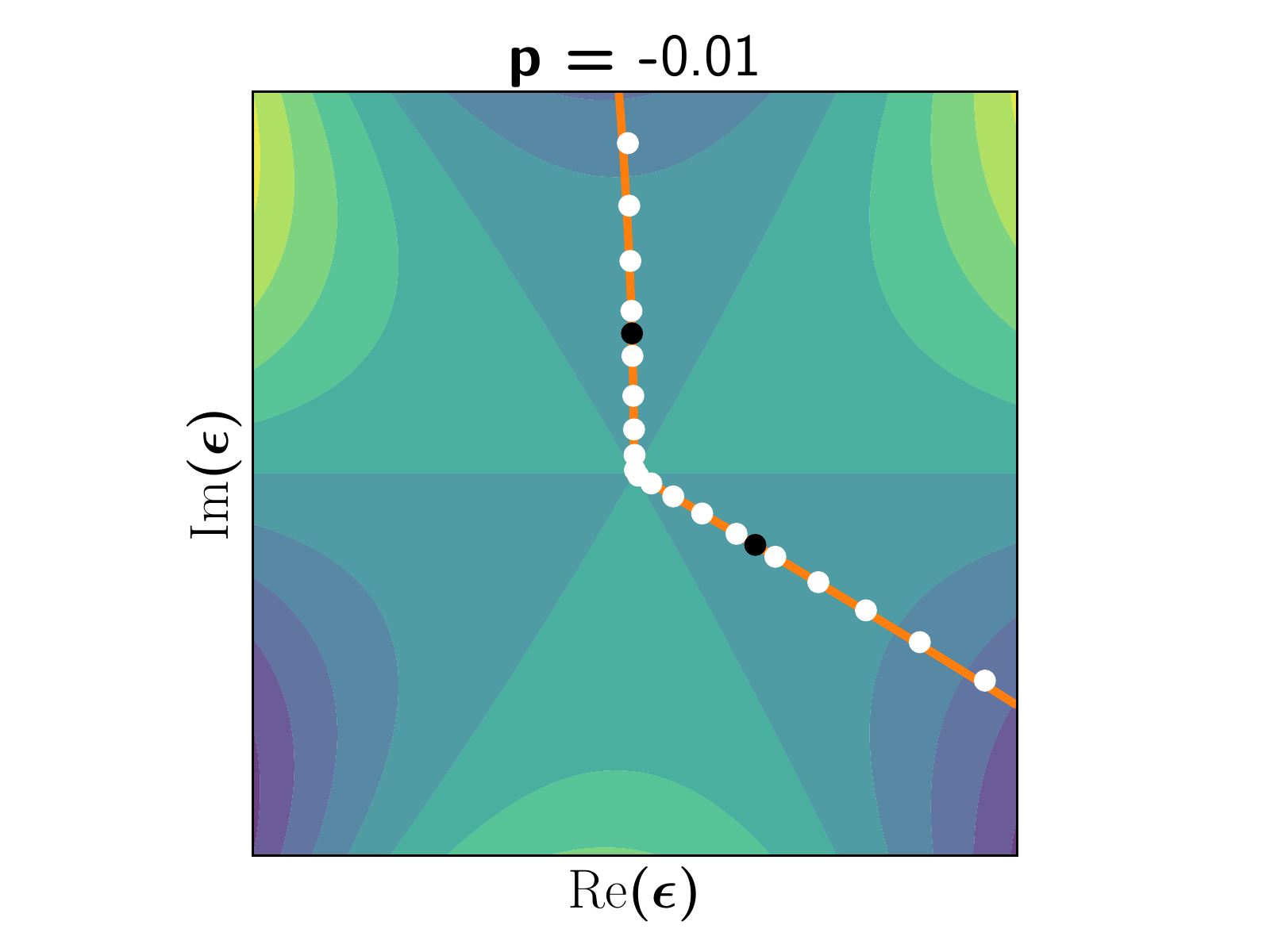}
	\caption{Same as \Fig{fig:5_EXcontour} for the phase function $f(\epsilon, p)$ [\Eq{eq:5_fAIRY}] at various values of $p$. The white dots correspond here to the $n = 10$ quadrature nodes. The steepest-descent contours evolve smoothly with $p$ and ultimately coalesce into a fold-type $A_2$ caustic at $p = 0$.}
	\label{fig:5_MGOcontour}
\end{figure}

\begin{figure}[t]
	\centering
	\begin{overpic}[width=0.6\linewidth,trim={4mm 5mm 4mm 4mm},clip]{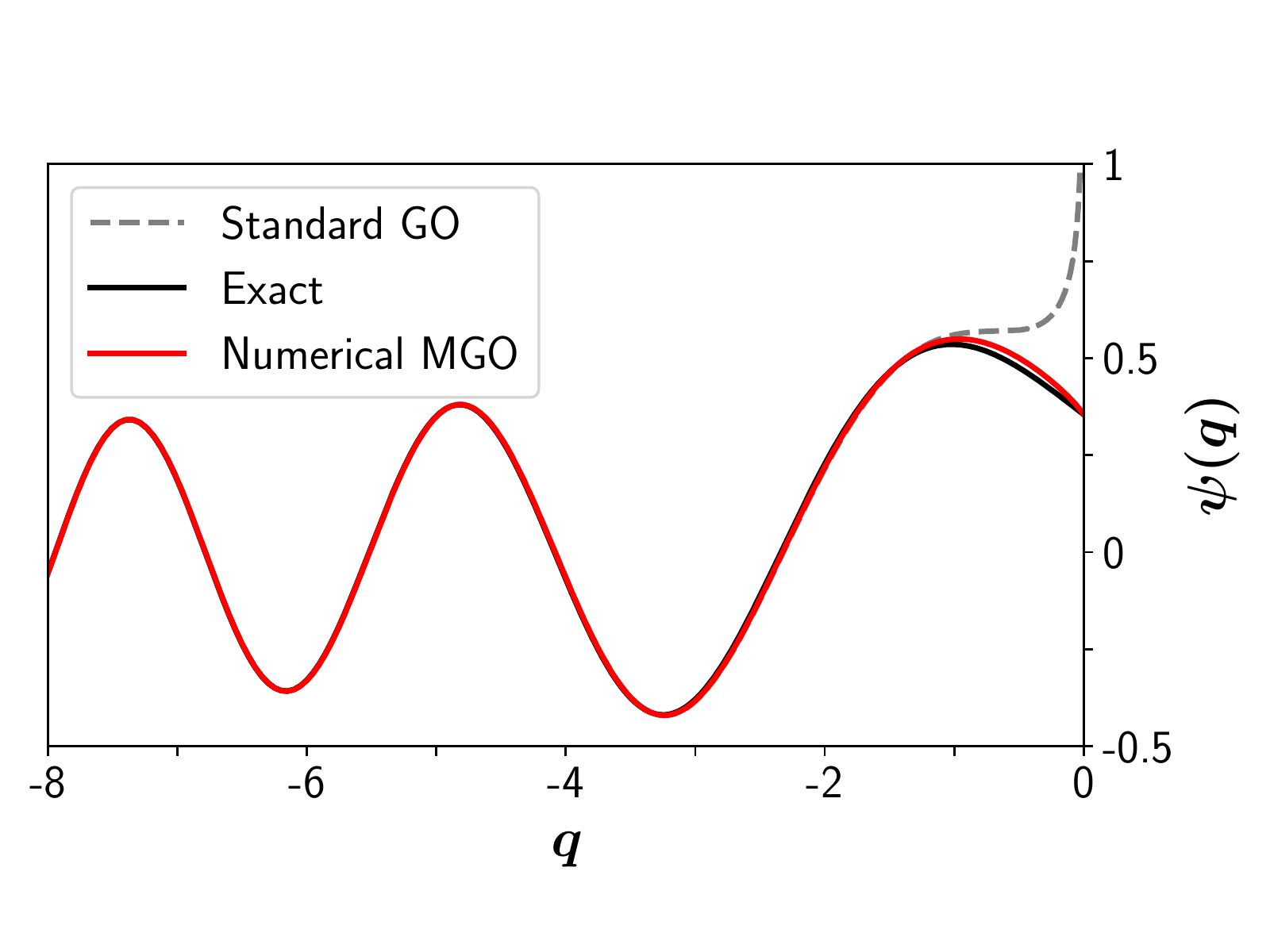}
		\put(5,12){\textbf{\large(a)}}
	\end{overpic}

	\vspace{2mm}
	\hspace{-3mm}\begin{overpic}[width=0.55\linewidth,trim={4mm 22mm 4mm 36mm},clip]{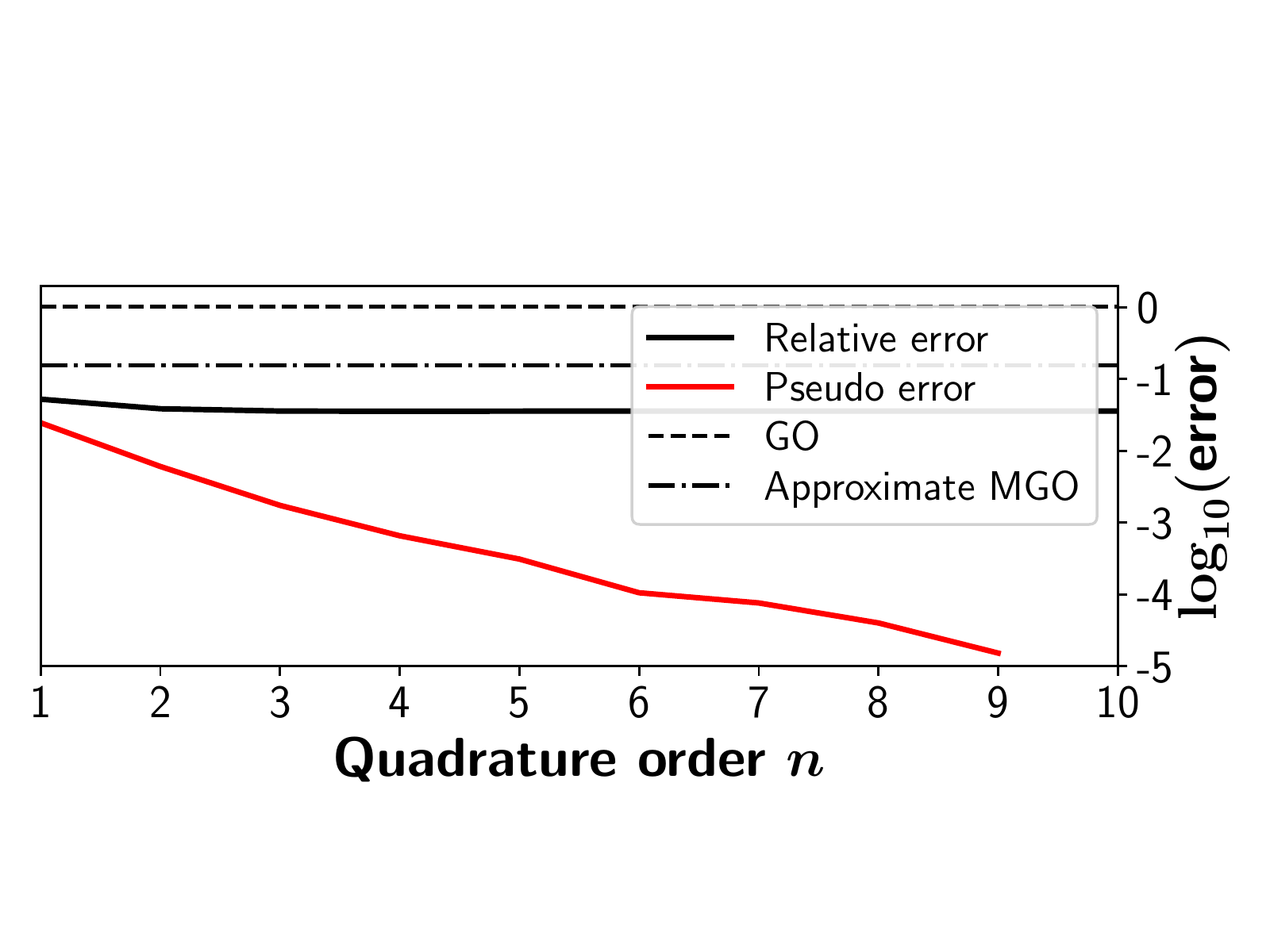}
		\put(5,12){\textbf{\large(b)}}
	\end{overpic}
	\caption{\textbf{(a)} Comparison of the numerical MGO solution (red) with the standard GO solution \eq{eq:5_goAIRY} (dashed gray) and the exact solution \eq{eq:5_exactAIRY} (black) for Airy's equation \eq{eq:5_eqAIRY}. The numerical MGO solution was obtained by applying the quadrature rule of \Eq{eq:5_MGOquad} with order $n = 10$ to \Eqs{eq:5_mgoAIRY} and \eq{eq:5_upsilonAIRY}. The numerical MGO solution displays remarkable agreement with the exact solution compared with the analytical approximations, even near the fold-type caustic at $q = 0$. \textbf{(b)} Error of the numerical MGO solution with respect to a scan over the quadrature order $n$. Note that the `pseudo error' is defined as the relative error with respect to the $n = 10$ solution used in \textbf{(a)}.}
	\label{fig:5_MGOairy}
\end{figure}

Here, I evaluate \Eq{eq:5_upsilonAIRY} numerically via \Eq{eq:5_MGOquad} over the range $q \in [-8, 0]$ using the angle memory feedback algorithm described in \Sec{sec:5_feedback}. Figure~\ref{fig:5_MGOcontour} shows the smooth evolution of steepest-descent curves obtained with the memory feedback algorithm, while \Fig{fig:5_MGOairy} compares the resultant numerical MGO solution with the exact solution \eq{eq:5_exactAIRY} and the GO approximations of \Eqs{eq:5_goAIRY}. As \Fig{fig:5_MGOairy} shows, the numerical MGO solution remains finite at the caustic $q = 0$, whereas the GO solution diverges. It also agrees remarkably well with the exact solution everywhere, even though a relatively low quadrature order of $n = 10$ was used. Moreover, although the relative error with respect to the exact solution does not decrease much after quadrature order $n = 2$, the `pseudo error' (defined as the relative error between the numerical MGO solution for a given $n$ compared with the reference solution $n = 10$) continues to decrease with increasing $n$. This suggests that the numerical MGO algorithm quickly converges to the residual intrinsic error of the MGO theory, at least for this specific example.

\subsection{EM wave in unmagnetized plasma slab with quadratic density profile}
\label{sec:5_EM2}

Similarly, let us take the density profile to be quadratic:
\begin{equation}
	n(x) = n_c \frac{x^2}{L_n^2}
	.
\end{equation}

Then, \Eq{eq:5_waveEQ} takes the form
\begin{equation}
	\pd{q}^2 \psi(x) + (2 \nu + 1 - q^2) \psi(x) = 0
	,
	\label{eq:5_eqQHO}
\end{equation}

\noindent where I have defined the re-scaled coordinate variable $q$ and the mode number $\nu$ as
\begin{equation}
	q \doteq x \sqrt{ \frac{\Omega}{c L_n} }
	, \quad
	\nu = \frac{L_n \Omega}{2 c} - \frac{1}{2}
	.
\end{equation}

Equation \eq{eq:5_eqQHO} is known as the Weber's equation, and contains two fold-type $A_2$ caustics at the cutoff locations $q = \pm R$, where $R \doteq \sqrt{2 \nu + 1}$. Assuming that $E(q \to \pm \infty) = 0$, the exact solution is given by the Airy function
\begin{equation}
	\psi_\textrm{ex}(q) = 
	\frac{\textrm{Ai}(0)}{\sqrt{R}} \, 
	\frac{
		\textrm{D}_{\nu} (\sqrt{2} \, q)
	}{
		\textrm{D}_\nu(\sqrt{2}\,R) 
	}
	,
	\label{eq:5_exactQHO}
\end{equation}

\noindent (where $\textrm{D}_\nu(x)$ is Whittaker's parabolic cylinder function~\cite{Olver10a}) while the MGO solution to \Eq{eq:5_eqQHO} can be written in the underdense region as (\Ch{ch:Ex})
\begin{equation}
	\psi(q) = \text{Im}\left\{  
		\frac{\Upsilon(q) \exp[i \beta(q)] }{\pi (2 R)^{1/3} \sqrt{|q|} } 
	\right\}
	,
	\label{eq:5_MGOqho}
\end{equation}

\noindent where 
\begin{subequations}
	\begin{align}
		\Upsilon(q) 
		&\doteq \int_{\cont{0}} \dd \epsilon \,
		\frac{\exp\left[ i \vartheta(\epsilon, q) \right]}{\left[1 - (\epsilon/R)^2 \right]^{1/4}}
		, \\
		\beta(q) &\doteq
		\frac{R^2}{2} \cos^{-1}\left( \frac{q}{R} \right) 
		- \frac{q}{2} \sqrt{R^2 - q^2}
		+ \frac{\pi}{4} \left[ \text{sgn}(q) + 1 \right]
		, \\
		\vartheta(\epsilon, q) 
		&\doteq
		\frac{\epsilon}{2} \sqrt{R^2 - \epsilon^2}
		+ \frac{R^2}{2} \tan^{-1} \left( \frac{\epsilon}{\sqrt{R^2 - \epsilon^2} } \right)
		- R \epsilon
		- \frac{\epsilon^2}{2 q} \sqrt{R^2 - q^2}
		.
	\end{align}
\end{subequations}

\noindent As before, evaluating the integral function via stationary phase yields the standard GO solution
\begin{equation}
	\psi_\text{GO}(q) 
	= \frac{2^{1/6} \cos\left[ \frac{q}{2}\sqrt{R^2 - q^2} - \frac{R^2 }{2}\cos^{-1}\left(\frac{q}{R} \right)  + \frac{\pi}{4} \right]}{\sqrt{\pi} \, R^{1/3} (R^2 - q^2)^{1/4}}
	.
	\label{eq:5_GOqho}
\end{equation}

\begin{figure}[t!]
	\begin{overpic}[width=0.44\linewidth,trim={6mm 18mm 3mm 23mm},clip]{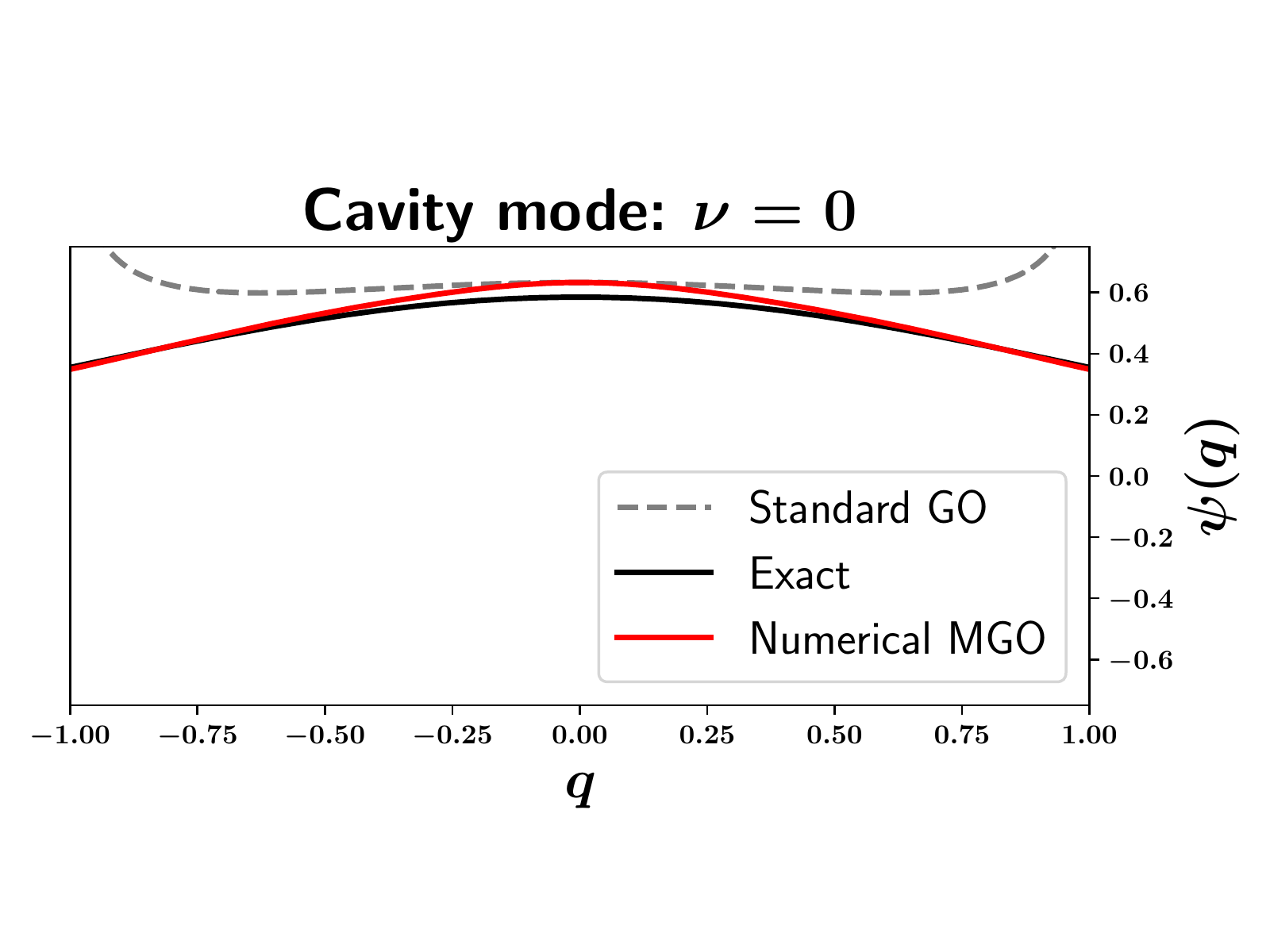}
		\put(5,10){\textbf{\small(a)}}
	\end{overpic}
	\hspace{4mm}
	\begin{overpic}[width=0.44\linewidth,trim={4mm 17mm 3mm 23mm},clip]{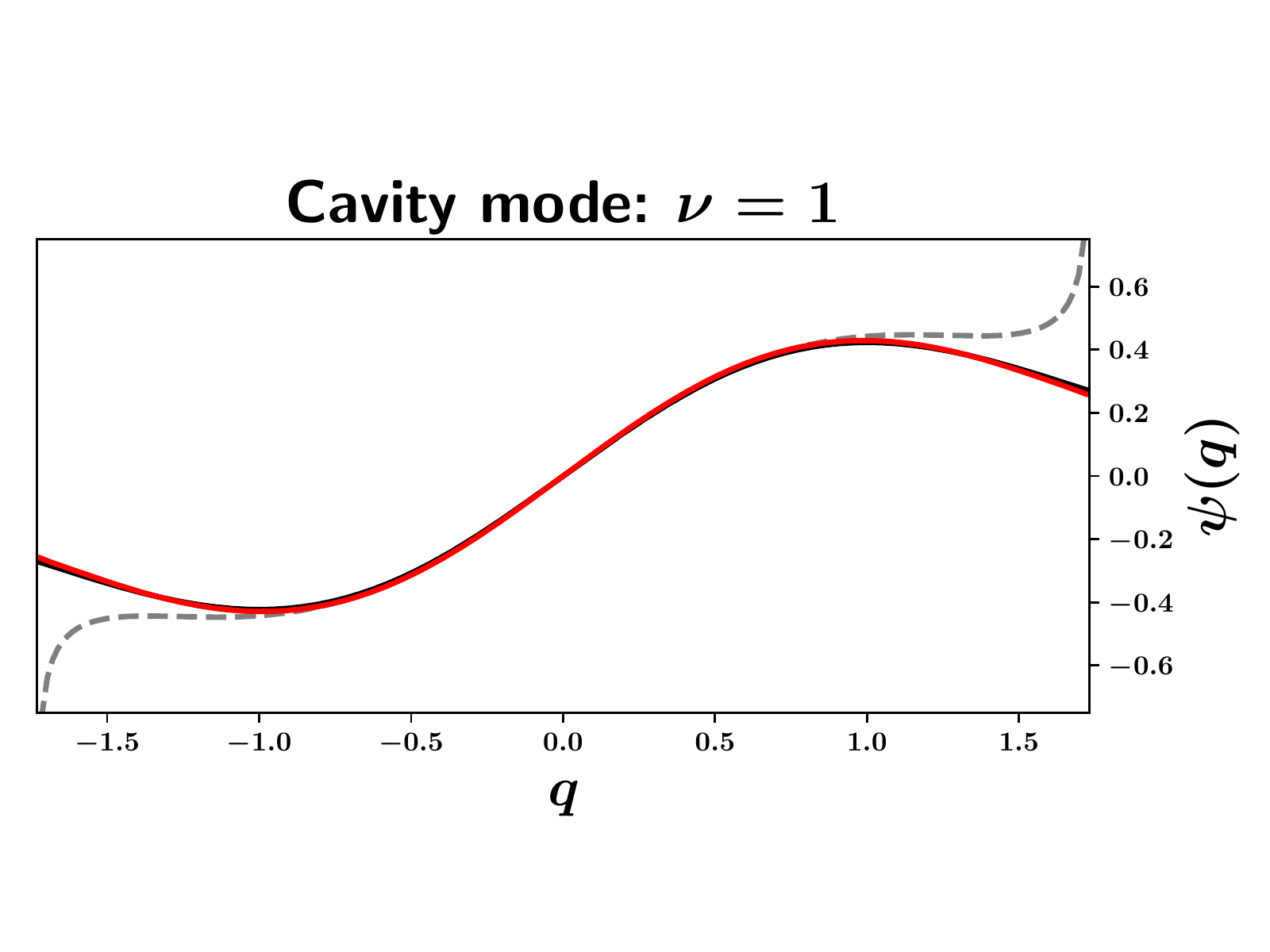}
		\put(5,10){\textbf{\small(b)}}
	\end{overpic}

	\vspace{2mm}
	\begin{overpic}[width=0.44\linewidth,trim={4mm 17mm 3mm 23mm},clip]{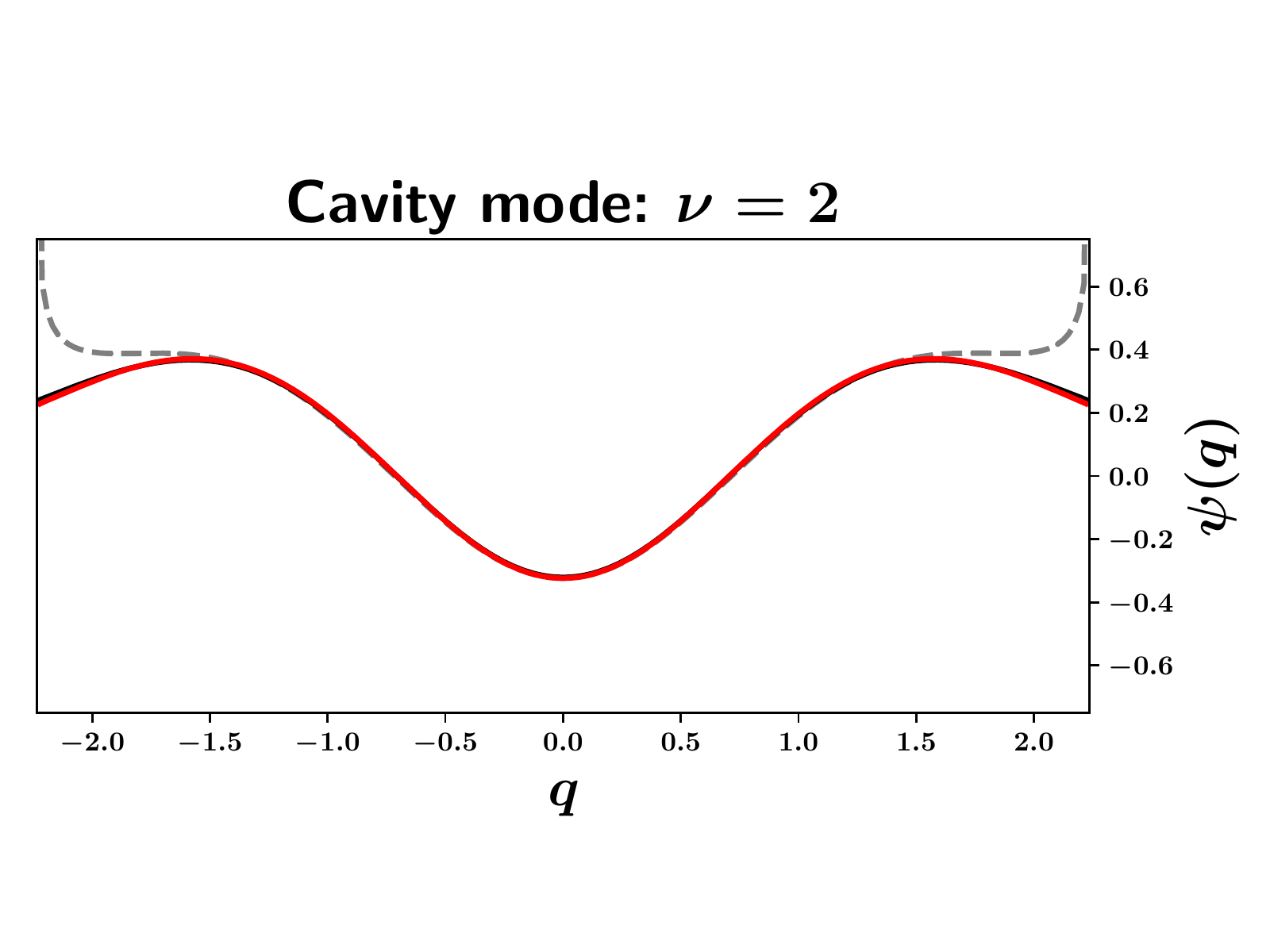}
		\put(5,10){\textbf{\small(c)}}
	\end{overpic}
	\hspace{4mm}
	\begin{overpic}[width=0.44\linewidth,trim={4mm 17mm 3mm 23mm},clip]{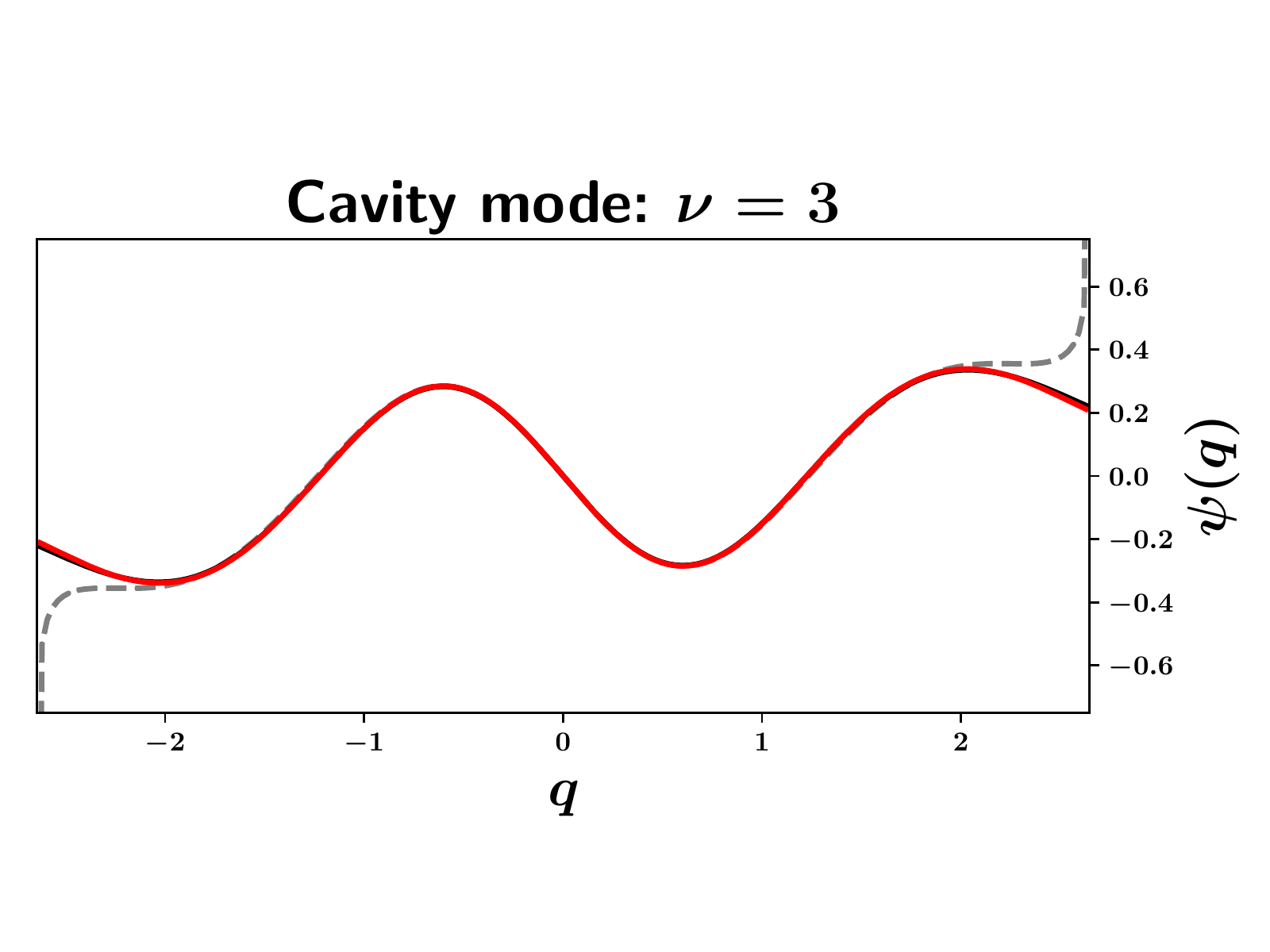}
		\put(5,10){\textbf{\small(d)}}
	\end{overpic}
	
	\vspace{2mm}
	\begin{overpic}[width=0.44\linewidth,trim={4mm 17mm 3mm 23mm},clip]{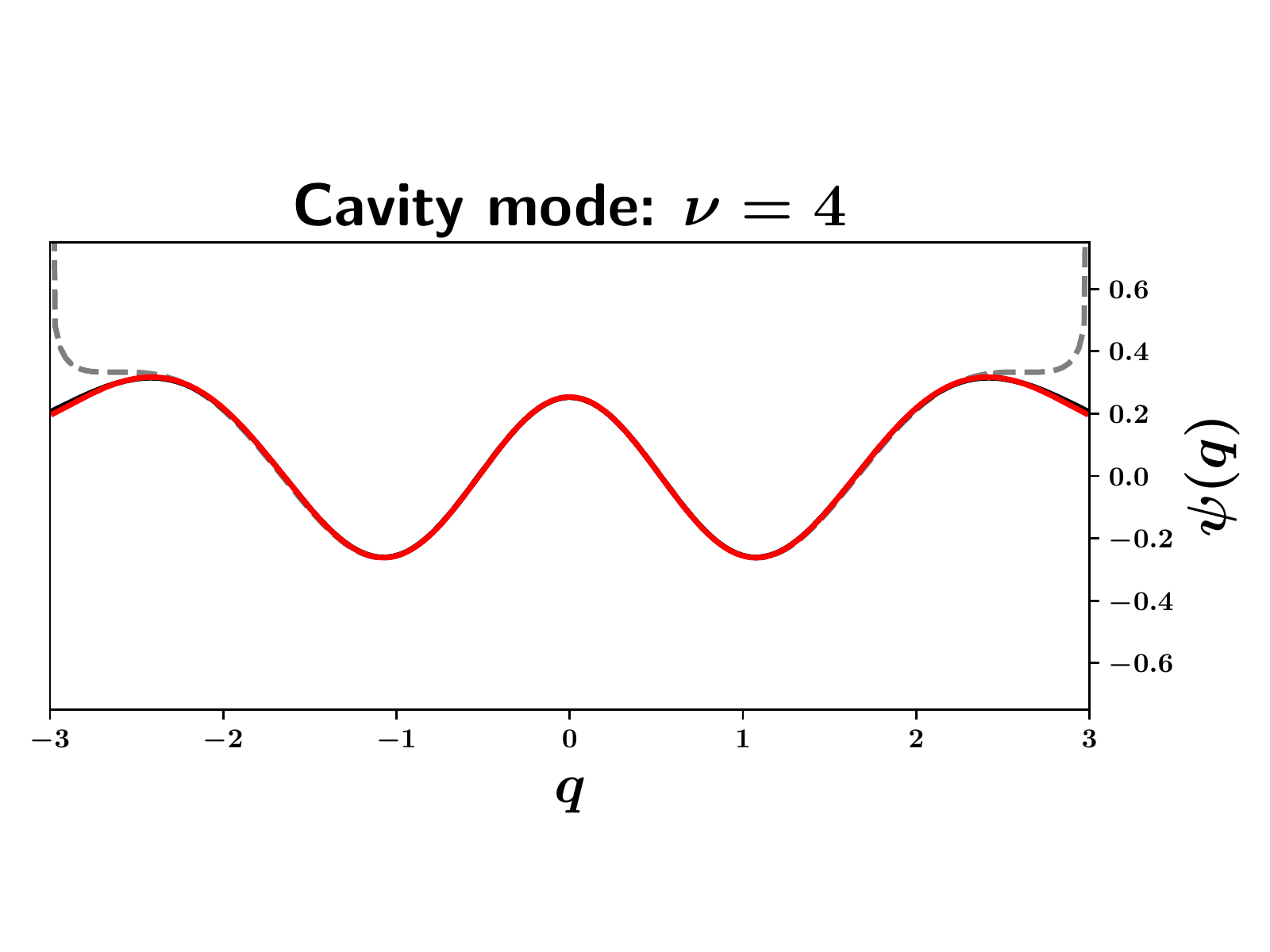}
		\put(5,10){\textbf{\small(e)}}
	\end{overpic}
	\hspace{4mm}
	\begin{overpic}[width=0.44\linewidth,trim={4mm 17mm 3mm 23mm},clip]{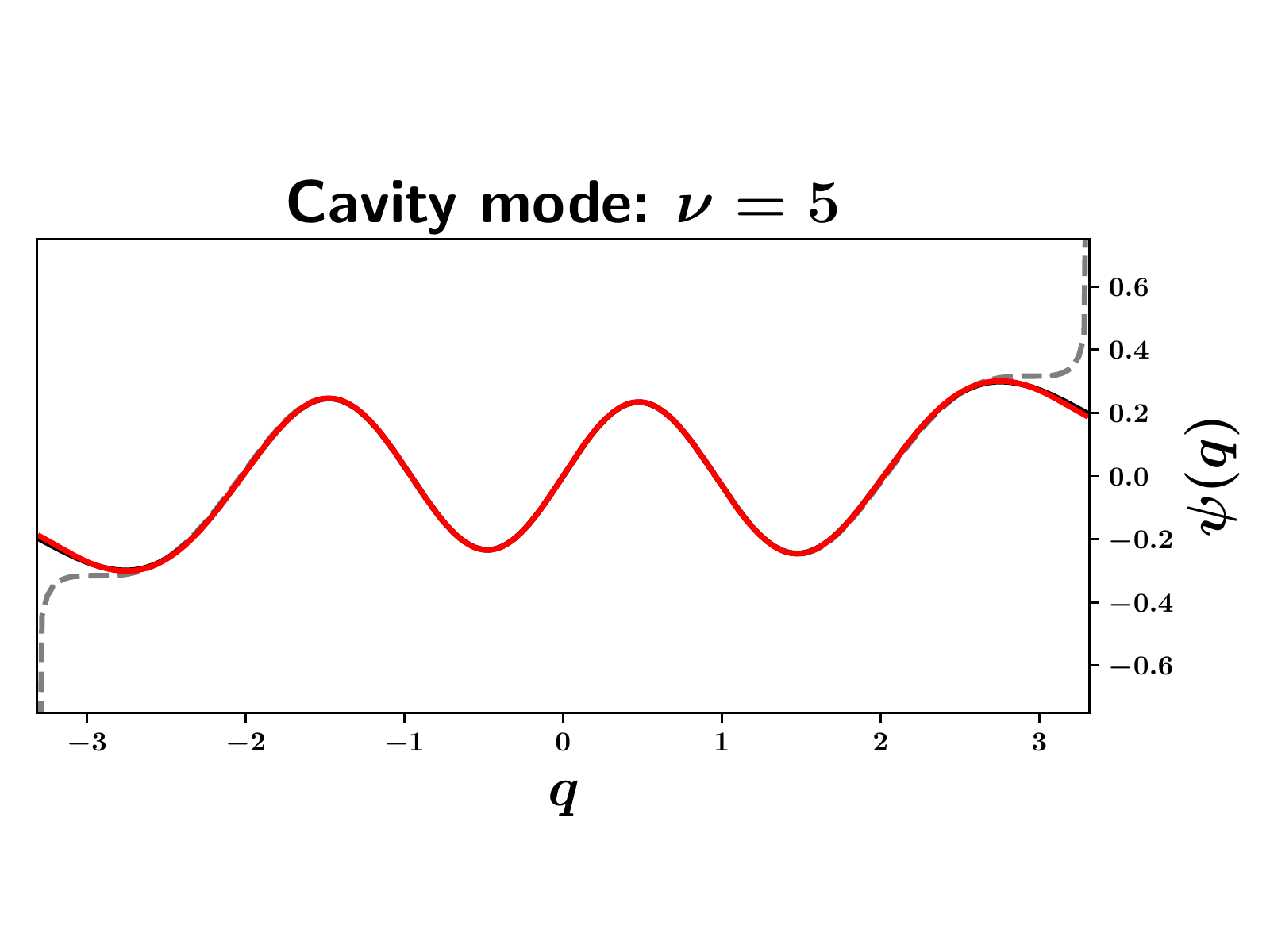}
		\put(5,10){\textbf{\small(f)}}
	\end{overpic}
	
	\vspace{2mm}
	\begin{overpic}[width=0.44\linewidth,trim={4mm 17mm 3mm 23mm},clip]{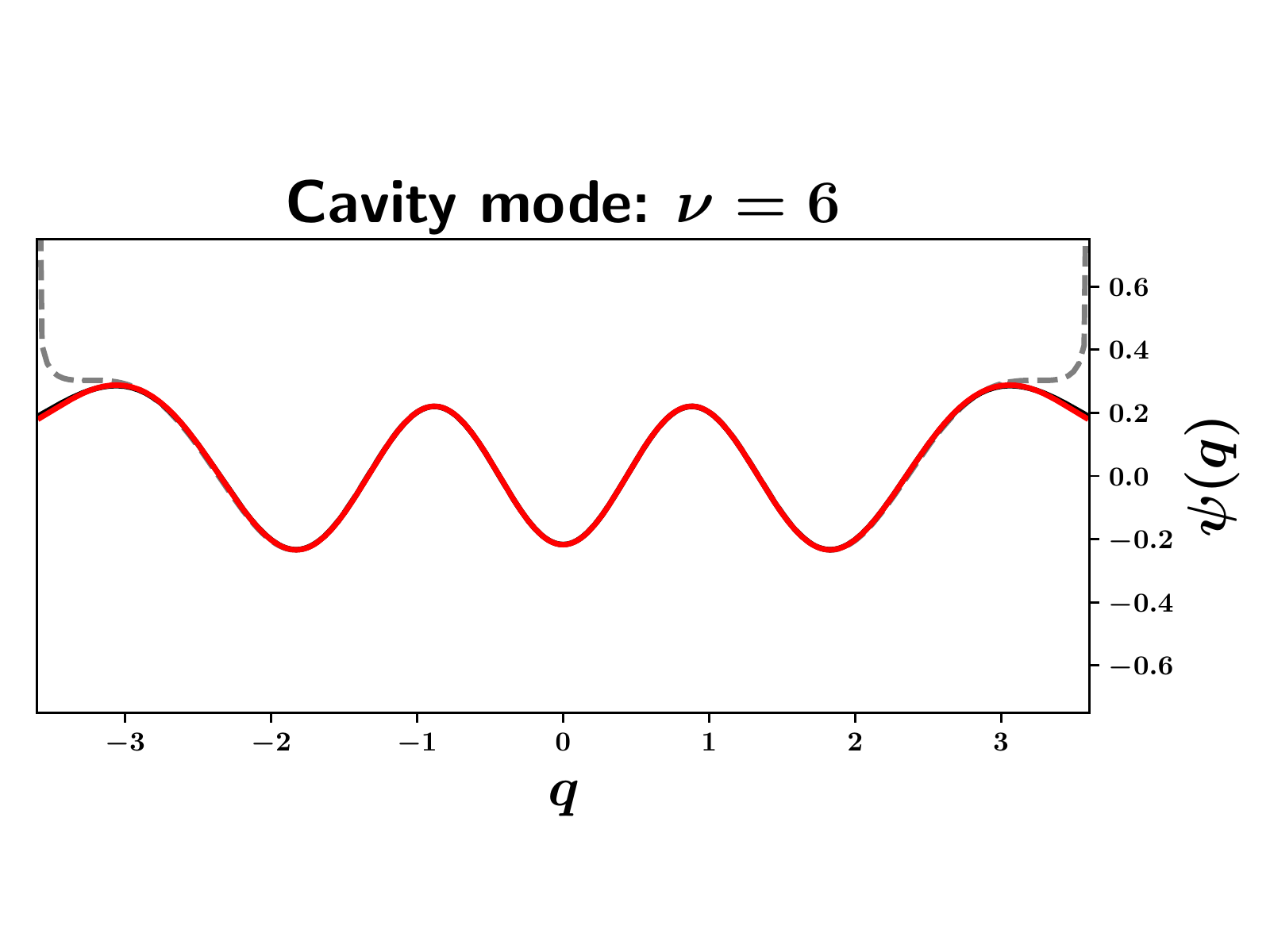}
		\put(5,10){\textbf{\small(g)}}
	\end{overpic}
	\hspace{4mm}
	\begin{overpic}[width=0.44\linewidth,trim={4mm 17mm 3mm 23mm},clip]{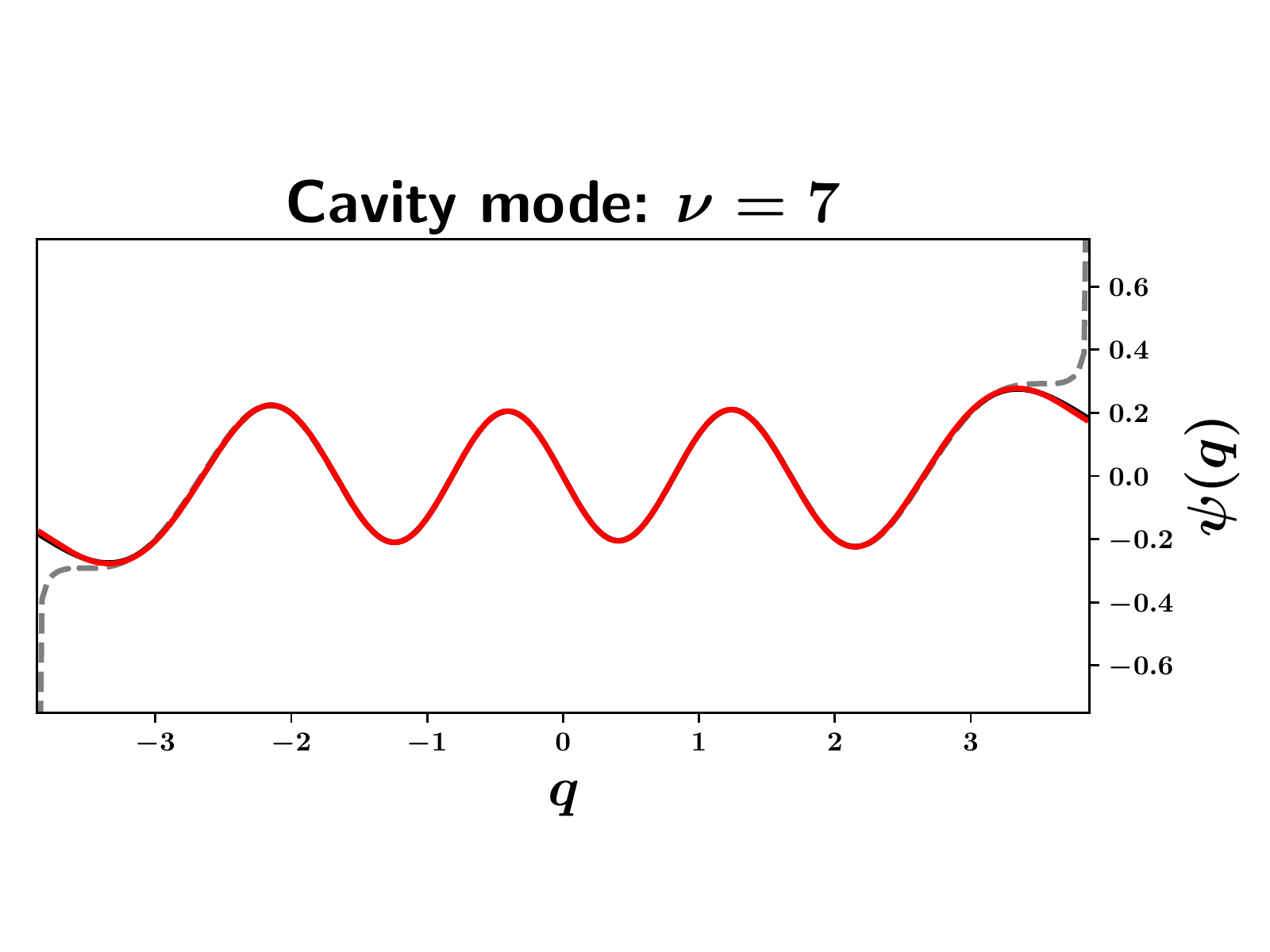}
		\put(5,10){\textbf{\small(h)}}
	\end{overpic}

	\vspace{2mm}
	\begin{overpic}[width=0.44\linewidth,trim={4mm 17mm 3mm 23mm},clip]{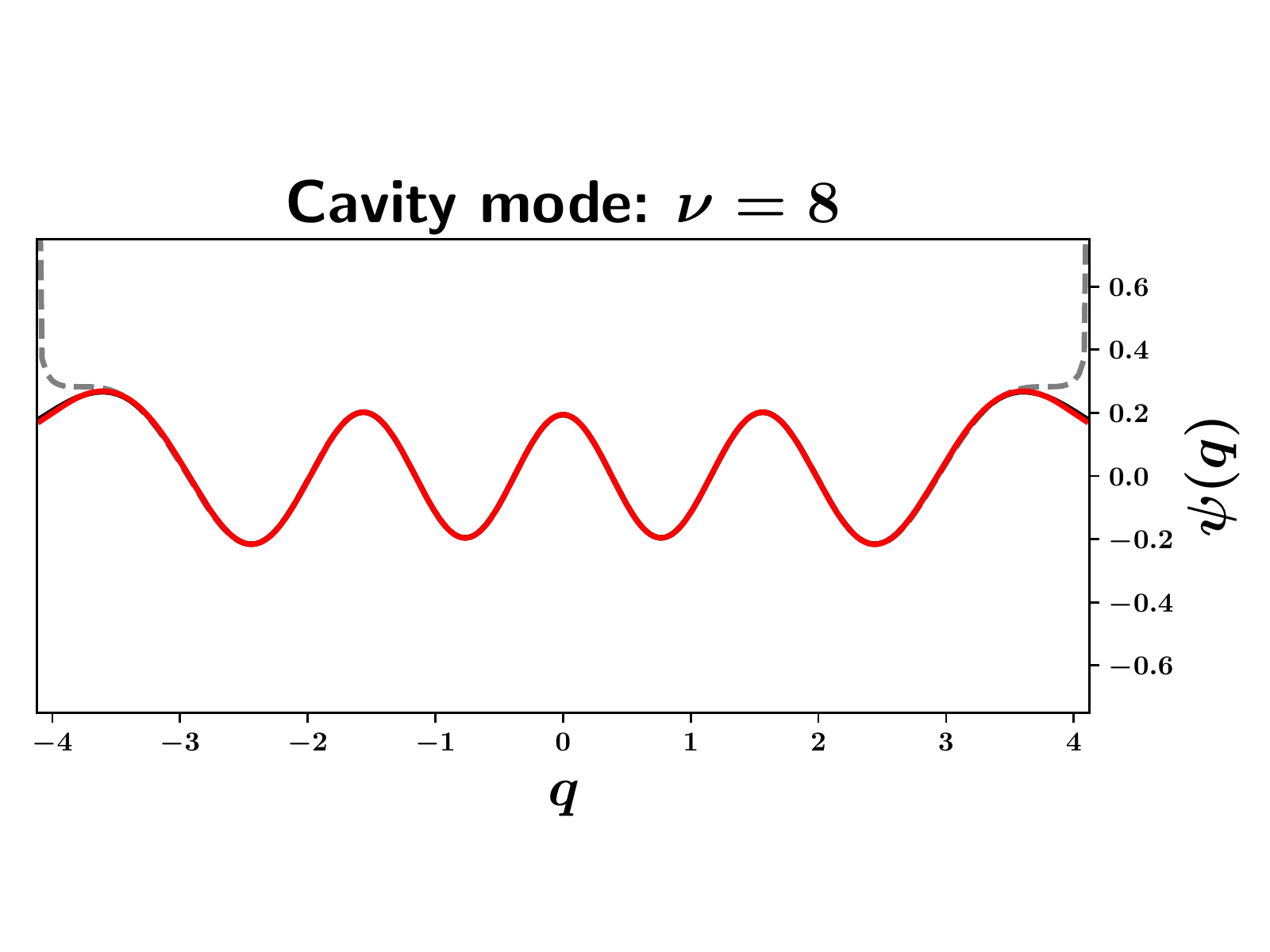}
		\put(5,10){\textbf{\small(i)}}
	\end{overpic}
	\hspace{4mm}
	\begin{overpic}[width=0.44\linewidth,trim={4mm 17mm 3mm 23mm},clip]{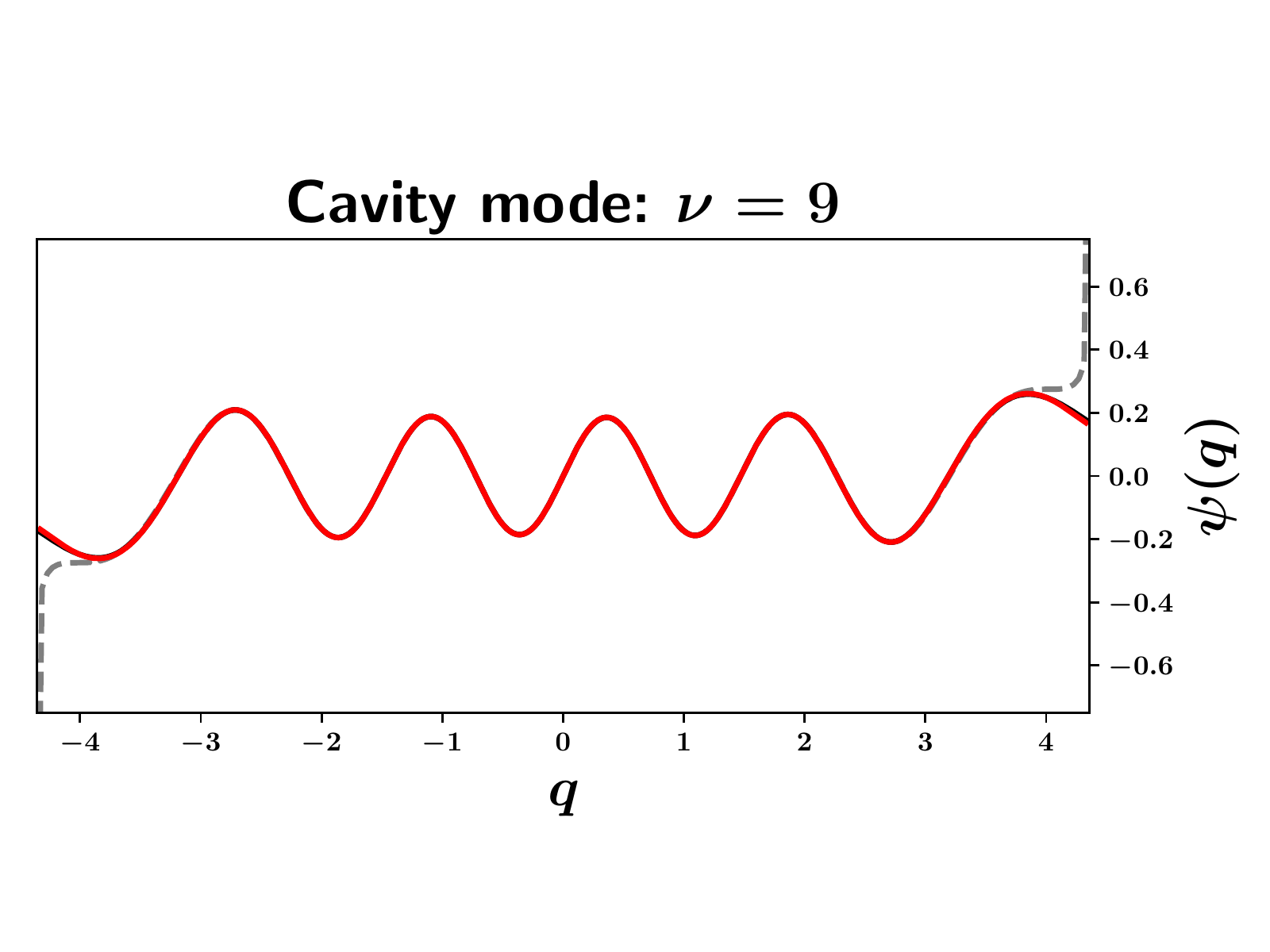}
		\put(5,10){\textbf{\small(j)}}
	\end{overpic}
	\caption{Comparison between the standard GO solution, the exact solution, and the MGO solution \eq{eq:5_MGOqho} computed via \Eq{eq:5_MGOquad} with $n = 2$ for the eigenmodes of a wave bounded within a quadratic cavity potential (\ie the quantum harmonic oscillator problem). The first ten modes are presented (with $\nu$ the mode number). The GO solution diverges at the caustics, but the MGO solution remains finite and agrees well with the exact solution, even though a low quadrature order was used. }
	\label{fig:5_QHO}
\end{figure}

Figure \ref{fig:5_QHO} shows the numerically computed MGO solution using \Eq{eq:5_MGOquad} with quadrature order $n = 2$ and memory feedback to evaluate $\Upsilon$ for the first ten eigenmodes. The agreement between the numerical MGO solution and the exact eigenmodes is remarkable, even for such a low quadrature order. This example, along with the previous one, provide numerical evidence that the quadrature rule \eq{eq:5_MGOquad} converges rapidly with $n$, allowing significant computational savings.


\section{Summary}

In summary, a new algorithm for numerically performing oscillatory integrals of catastrophe theory is developed. The algorithm, given by \Eq{eq:5_MGOquad}, is based on using Gaussian quadrature with Freud polynomials along the steepest-descent curves of the integrand phase. Numerical results show that this algorithm converges rapidly with the number of quadrature points used. As I shall discuss in the next chapter, a key step in MGO is calculating an oscillatory integral along a specified steepest-descent curve; hence, this algorithm is expected to feature naturally within an MGO-based ray-tracing code.


\begin{subappendices}

\section{Gauss--Freud quadrature nodes and weights}
\label{sec:5_FreudQUAD}

\begin{table}[t!]
	\centering
	\resizebox{\columnwidth}{!}{%
	\begin{tabular}{| c | c | c | c | c | c | c |}
		\multicolumn{1}{c}{Order} & \multicolumn{1}{c}{Nodes} & \multicolumn{1}{c}{Weights} & \multicolumn{1}{c}{} & \multicolumn{1}{c}{Order} & \multicolumn{1}{c}{Nodes} & \multicolumn{1}{c}{Weights} \\
		\hline
		\bf{n = 1} & 5.64189583547756 (1) & 8.86226925452758 (1) & & 
			& 5.29786439318514 (2) & 1.34109188453360 (1) \\
		\cline{1-3}
			& 3.00193931060839 (1) & 6.40529179684379 (1) & &
			& 2.67398372167767 (1) & 2.68330754472640 (1) \\
		\bf{n = 2} & 1.25242104533372 (0) & 2.45697745768379 (1) & &
			& 6.16302884182402 (1) & 2.75953397988422 (1) \\
		\cline{1-3}
			& 1.90554149798192 (1) & 4.46029770466658 (1) & &
			& 1.06424631211623 (0) & 1.57448282618790 (1) \\
		\bf{n = 3} & 8.48251867544577 (1) & 3.96468266998335 (1) & &
			\bf{n = 8} & 1.58885586227006 (0) & 4.48141099174625 (2) \\
		& 1.79977657841573 (0) & 4.37288879877644 (2) & &
			& 2.18392115309586 (0) & 5.36793575602526 (3) \\
		\cline{1-3}
			& 1.33776446996068 (1) & 3.25302999756919 (1) & &
			& 2.86313388370808 (0) & 2.02063649132407 (4) \\
		& 6.24324690187190 (1) & 4.21107101852062 (1) & &
			& 3.68600716272440 (0) & 1.19259692659532 (6) \\
		\cline{5-7}
			\bf{n = 4} & 1.34253782564499 (0) & 1.33442500357520 (1) & &
			& 4.49390308011934 (2) & 1.14088970242118 (1) \\
		& 2.26266447701036 (0) & 6.37432348625728 (3) & &
			& 2.28605305560535 (1) & 2.35940791223685 (1) \\
		\cline{1-3}
			& 1.00242151968216 (1) & 2.48406152028443 (1) & &
			& 5.32195844331646 (1) & 2.66425473630253 (1) \\
		& 4.82813966046201 (1) & 3.92331066652399 (1) & &
			& 9.27280745338081 (1) & 1.83251679101663 (1) \\
		\bf{n = 5} & 1.06094982152572 (0) & 2.11418193076057 (1) & &
			\bf{n = 9} & 1.39292385519588 (0) & 7.13440493066916 (2) \\
		& 1.77972941852026 (0) & 3.32466603513439 (2) & &
			& 1.91884309919743 (0) & 1.39814184155604 (2) \\
		& 2.66976035608766 (0) & 8.24853344515628 (4) & &
			& 2.50624783400574 (0) & 1.16385272078519 (3) \\
		\cline{1-3}
			& 7.86006594130979 (2) & 1.96849675488598 (1) & &
			& 3.17269213348124 (0) & 3.05670214897831 (5) \\
		& 3.86739410270631 (1) & 3.49154201525395 (1) & &
			& 3.97889886978978 (0) & 1.23790511337496 (7) \\
		\cline{5-7}
			& 8.66429471682044 (1) & 2.57259520584421 (1) & & 
			& 3.87385243257289 (2) & 9.85520975191087 (2) \\
		\bf{n = 6} & 1.46569804966352 (0) & 7.60131375840058 (2) & & 
			& 1.98233304013083 (1) & 2.08678066608185 (1) \\
		& 2.17270779693900 (0) & 6.85191862513596 (3) & & 
			& 4.65201111814767 (1) & 2.52051688403761 (1) \\
		& 3.03682016932287 (0) & 9.84716452019267 (5) & & 
			& 8.16861885592273 (1) & 1.98684340038387 (1) \\
		\cline{1-3}
			& 6.37164846067008 (2) & 1.60609965149261 (1) & & 
			& 1.23454132402818 (0) & 9.71984227600620 (2) \\
		& 3.18192018888619 (1) & 3.06319808158099 (1) & & 
			\bf{n = 10}& 1.70679814968913 (0) & 2.70244164355446 (2) \\
		& 7.24198989258373 (1) & 2.75527141784905 (1) & & 
			& 2.22994008892494 (0) & 3.80464962249537 (3) \\
		\bf{n = 7} & 1.23803559921509 (0) & 1.20630193130784 (1) & & 
			& 2.80910374689875 (0) & 2.28886243044656 (4) \\
		& 1.83852822027095 (0) & 2.18922863438067 (2) & & 
			& 3.46387241949586 (0) & 4.34534479844469 (6) \\
		& 2.53148815132768 (0) & 1.23644672831056 (3) & & 
			& 4.25536180636608 (0) & 1.24773714817825 (8) \\
		& 3.37345643012458 (0) & 1.10841575911059 (5) & &
			&              &              \\
		\hline
	\end{tabular}
	}
	\caption{Gauss--Freud quadrature nodes and weights for quadrature orders up to $10$. The notation $a$ $(b)$ denotes $a \times 10^{-b}$.}
	\label{tab:5_GFnodes}
\end{table}

The Freud polynomials are the unique family of polynomials that are orthogonal with respect to the inner product
\begin{equation}
	\langle h_1, h_2 \rangle = 
	\int_0^\infty \dd \kappa \, 
	h_1(\kappa) h_2(\kappa) \exp(-\kappa^2)
	.
\end{equation}

\noindent Since the Freud polynomials are uncommon, the corresponding quadrature nodes $\{ \kappa_j \}$ and weights $\{w_j\}$ are not typically provided in standard software. Moreover, the definitions of $\{ \kappa_j \}$ \eq{eq:5_GQnodes} and $\{ w_j \}$ \eq{eq:5_GQweights} are not practical when the functional forms of $\{ p_\ell(\kappa) \}$ are unknown.

In this case, it is better to use the Golub--Welsch algorithm~\cite{Golub69}, which relies on the following eigenvalue relationship that $\{ \kappa_j \}$ and $\{w_j\}$ can be shown to satisfy~\cite{Gil07}:
\begin{equation}
	\Mat{\mc{J}}_n \Vect{\nu}_j = \kappa_j \Vect{\nu}_j
	,
	\quad
	j = 1, \ldots, n .
	\label{eq:5_golubNODES}
\end{equation}

\noindent Here, $\Mat{\mc{J}}_n$ is the symmetric tridiagonal $n \times n$ Jacobi matrix corresponding to the first $n$ members of $\{ p_\ell(\kappa) \}$~\cite{Olver10a}, \ie
\begin{equation}
	\Mat{\mc{J}}_n = 
	\begin{pmatrix}
		a_0        & \sqrt{b_1} &                & \\
		\sqrt{b_1} &        a_1 &         \ddots & \\
		&     \ddots &         \ddots & \sqrt{b_{n-1}} \\
		&            & \sqrt{b_{n-1}} & a_{n-1}
	\end{pmatrix}
	,
\end{equation}

\noindent with $a_\ell$ and $b_\ell$ being the coefficients of the three-term recurrence relation that the \textit{monic} family $\{ \fourier{p}_\ell(\kappa) \}$ satisfy:
\begin{equation}
	\fourier{p}_{\ell + 1}(\kappa) 
	= (\kappa + a_\ell)\fourier{p}_{\ell}(\kappa)
	+ b_\ell \fourier{p}_{\ell - 1}(\kappa) 
	, \quad
	\ell = 0, 1, \ldots
\end{equation}

\noindent subject to the initial conditions
\begin{equation}
	\fourier{p}_{-1}(\kappa) = 0 
	, \quad
	\fourier{p}_0(\kappa) = 1 .
\end{equation}

\noindent There are established algorithms to obtain these coefficients~\cite{Press07,Gautschi16}. The weights are then obtained from the first eigenvector $\Vect{\nu}_1$, which can be normalized such that $\{w_j\}$ are given by its vector components as
\begin{equation}
	\Vect{\nu}_1 = \frac{1}{\sqrt{\langle 1, 1 \rangle}}
	\begin{pmatrix}
		\sqrt{w_1}
		& 
		\ldots
		&
		\sqrt{w_n }
	\end{pmatrix}^\intercal
	, \quad
	\Vect{\nu}_1^\intercal \Vect{\nu}_1 = 1
	.
	\label{eq:5_golubWEIGHTS}
\end{equation}

The resulting list of $\{\kappa_j\}$ and weights $\{w_j\}$ for quadrature orders $n \le 10$ is provided in Table~\ref{tab:5_GFnodes}, adapted from a similar table for $2 \le n \le 20$ presented in \Ref{Steen69}. These values can also be calculated with high precision for arbitrary values of $n$ using the code of \Ref{Gautschi20}; see \Ref{Gautschi21} for more details.

\end{subappendices}

\chapter{Metaplectic geometrical optics}
\label{ch:MGO}

\section{Introduction}

Having discussed at length the idea of caustics and metaplectic transforms, in this chapter I shall present the main result of this thesis, namely, the development of a new ray-based caustic-removal scheme called metaplectic geometrical optics (MGO). MGO is a general framework that does not assume a specific caustic structure nor a specific wave equation, and thereby promises to become a new paradigm in caustic modeling. The following discussion is based on material published previously in \Refs{Lopez19, Lopez20, Lopez21a, Lopez22}.


\section{Phase-space rotation for cutoff removal}

\begin{figure}[t!]
	\centering
	\begin{overpic}[width=0.24\linewidth,trim={2mm 24mm 1mm 17mm},clip]{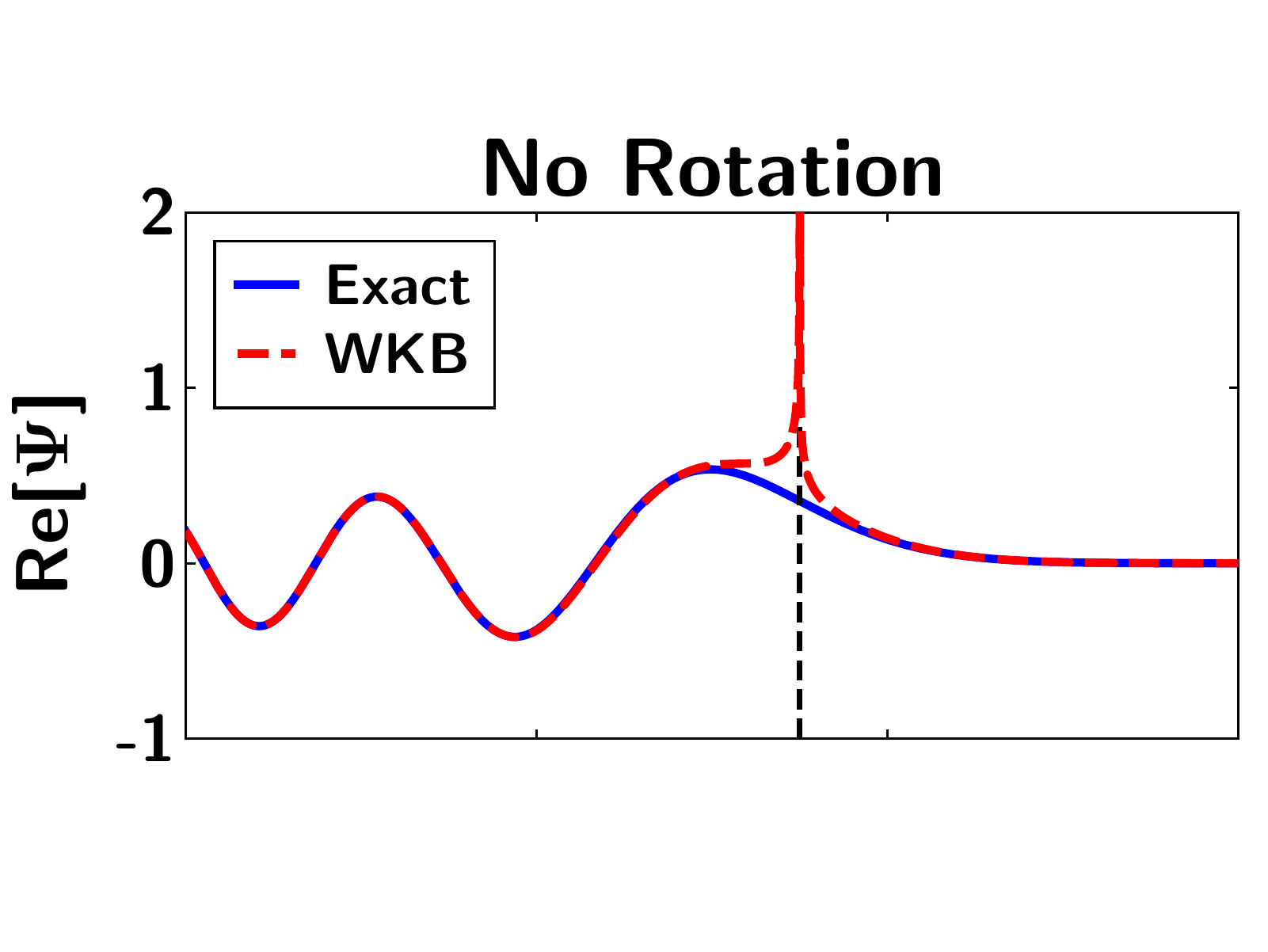}
		\put(85,5){\textbf{\color{black} \large(a)}}
	\end{overpic}
	\begin{overpic}[width=0.24\linewidth,trim={2mm 24mm 1mm 17mm},clip]{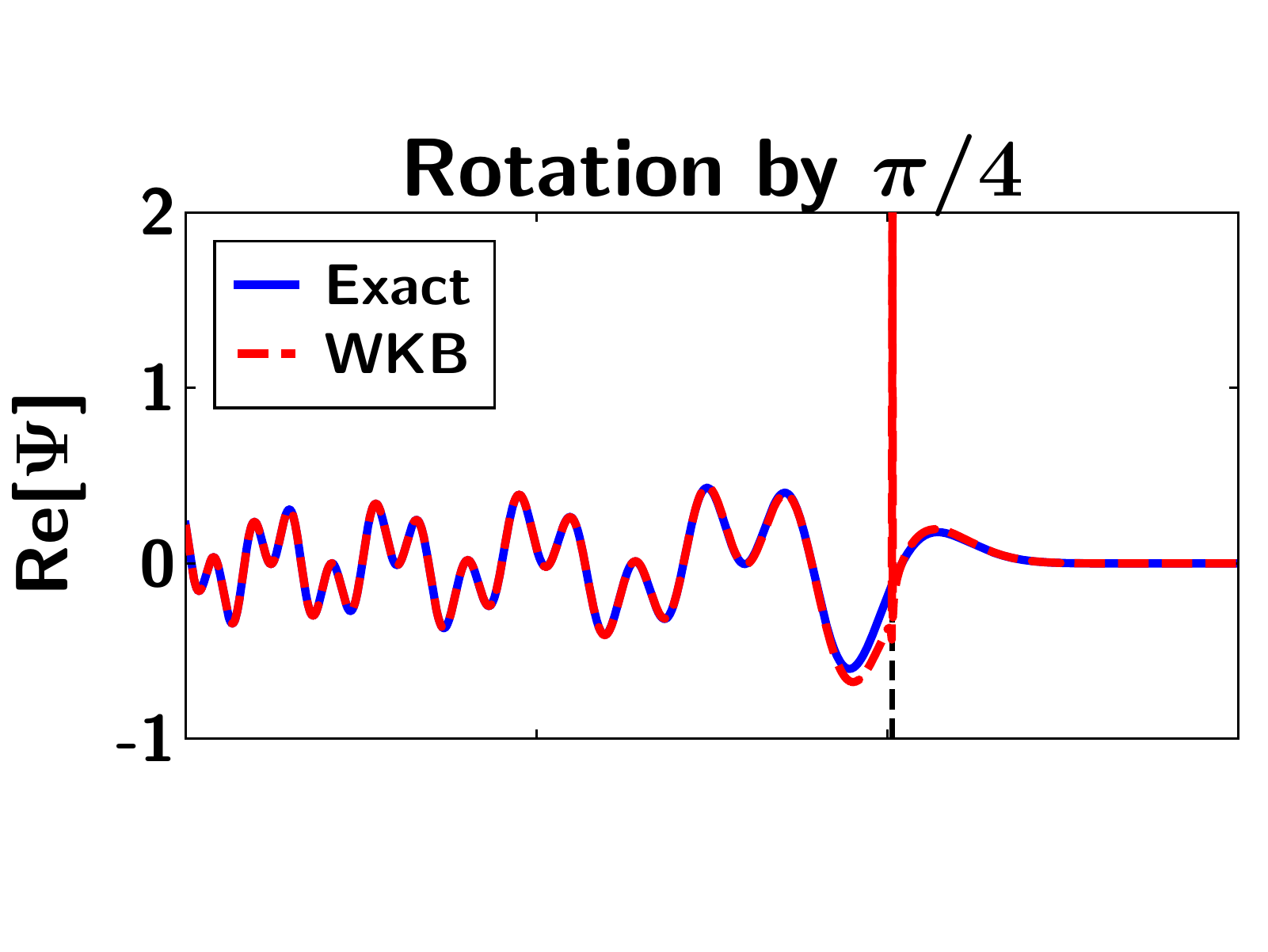}
		\put(85,5){\textbf{\color{black} \large(b)}}
	\end{overpic}
	\begin{overpic}[width=0.24\linewidth,trim={2mm 24mm 1mm 17mm},clip]{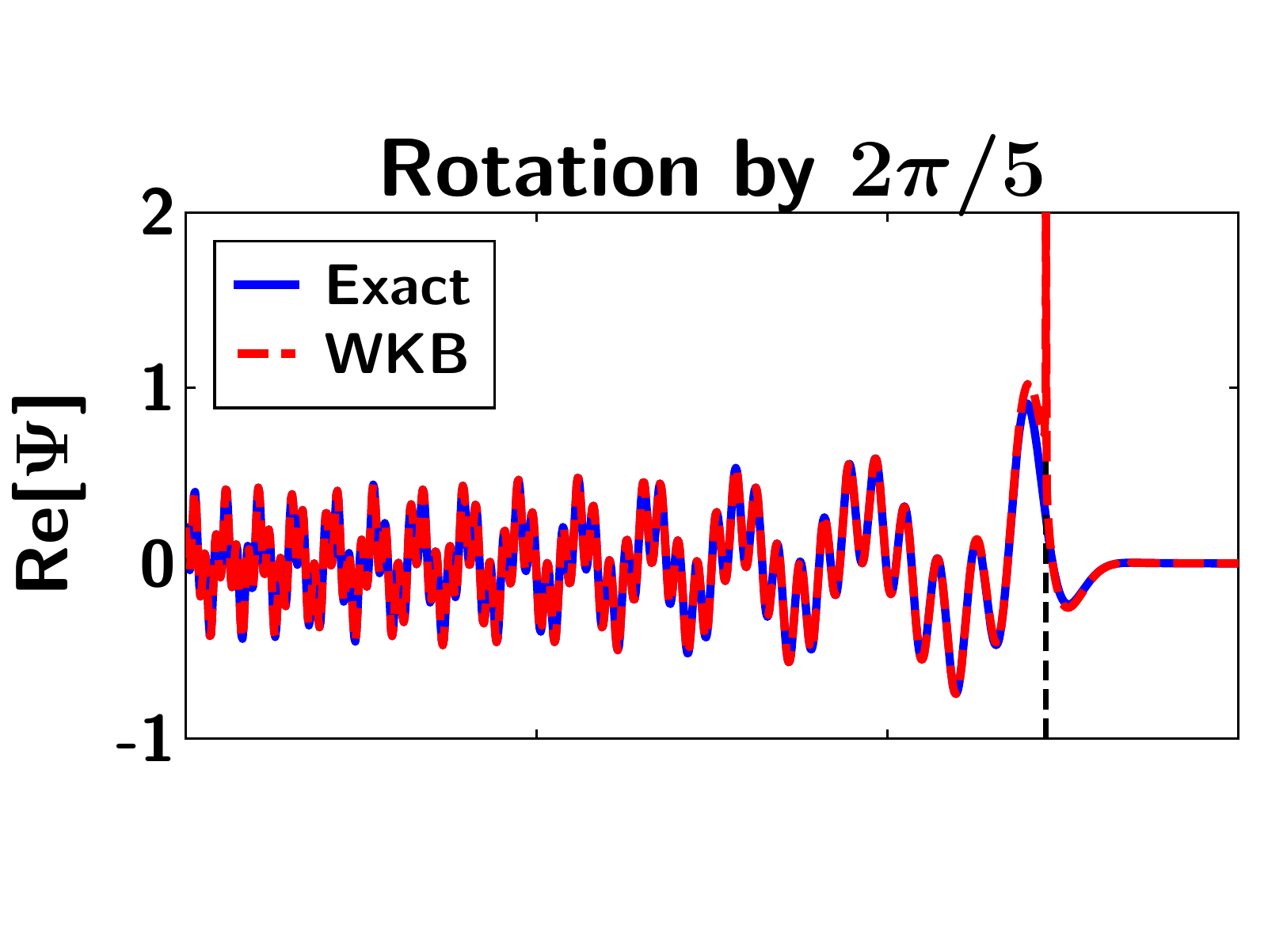}
		\put(85,5){\textbf{\color{black} \large(c)}}
	\end{overpic}
	\begin{overpic}[width=0.24\linewidth,trim={2mm 24mm 1mm 17mm},clip]{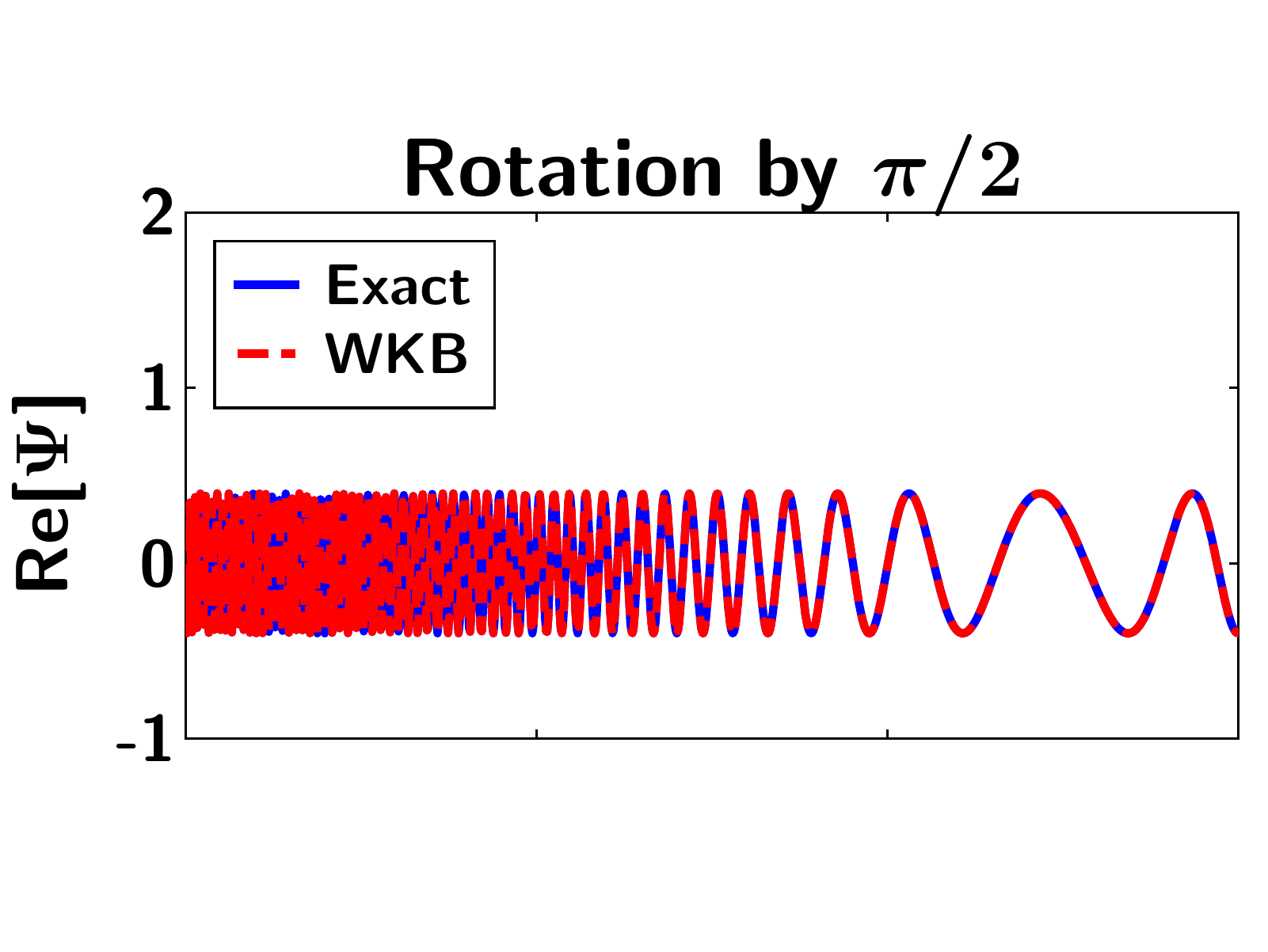}
		\put(85,5){\textbf{\color{black} \large(d)}}
	\end{overpic}

	\begin{overpic}[width=0.24\linewidth,trim={2mm 11mm 3mm 21mm},clip]{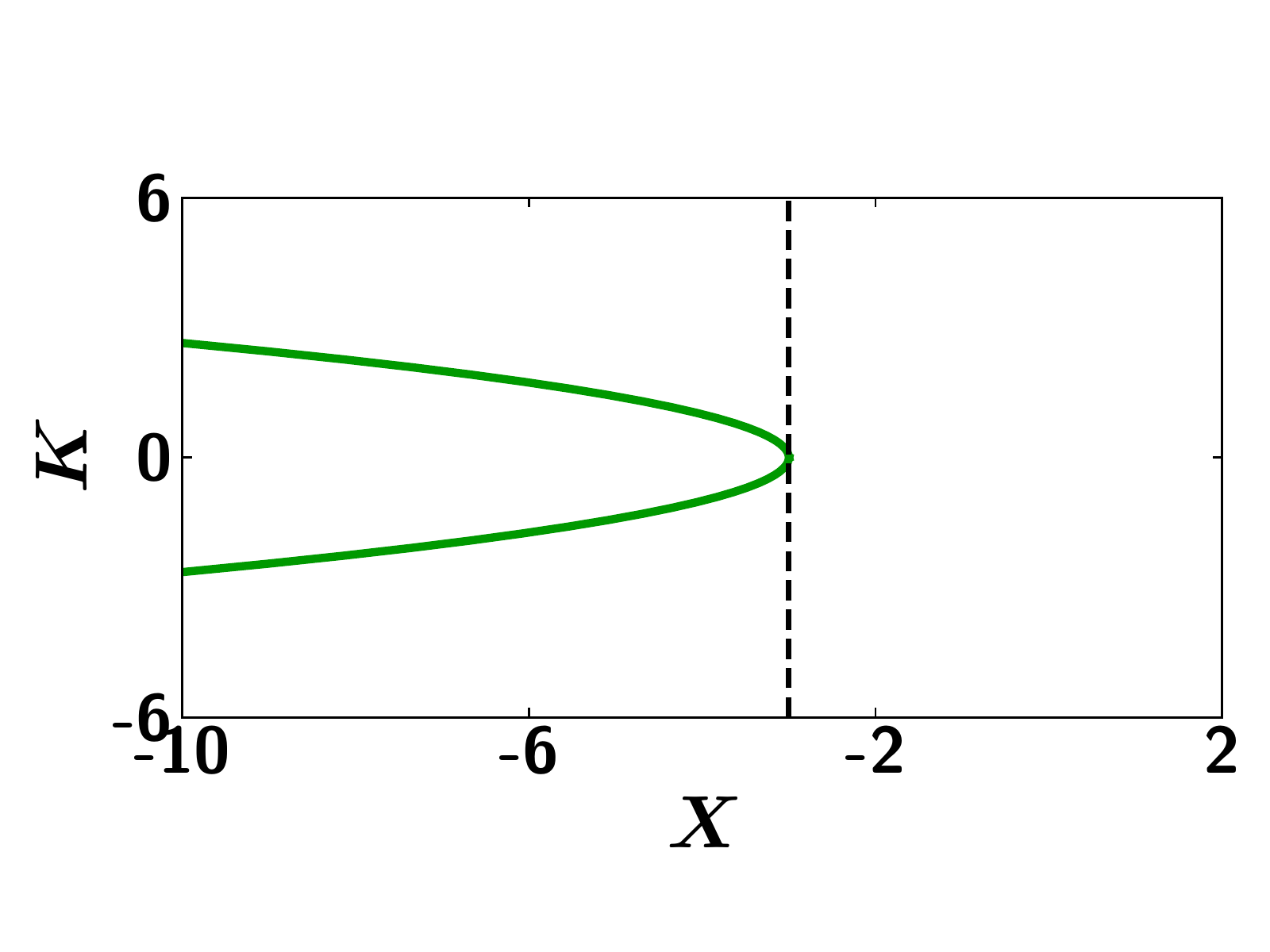}
		\put(85,15){\textbf{\color{black} \large(e)}}
	\end{overpic}
	\begin{overpic}[width=0.24\linewidth,trim={2mm 11mm 3mm 21mm},clip]{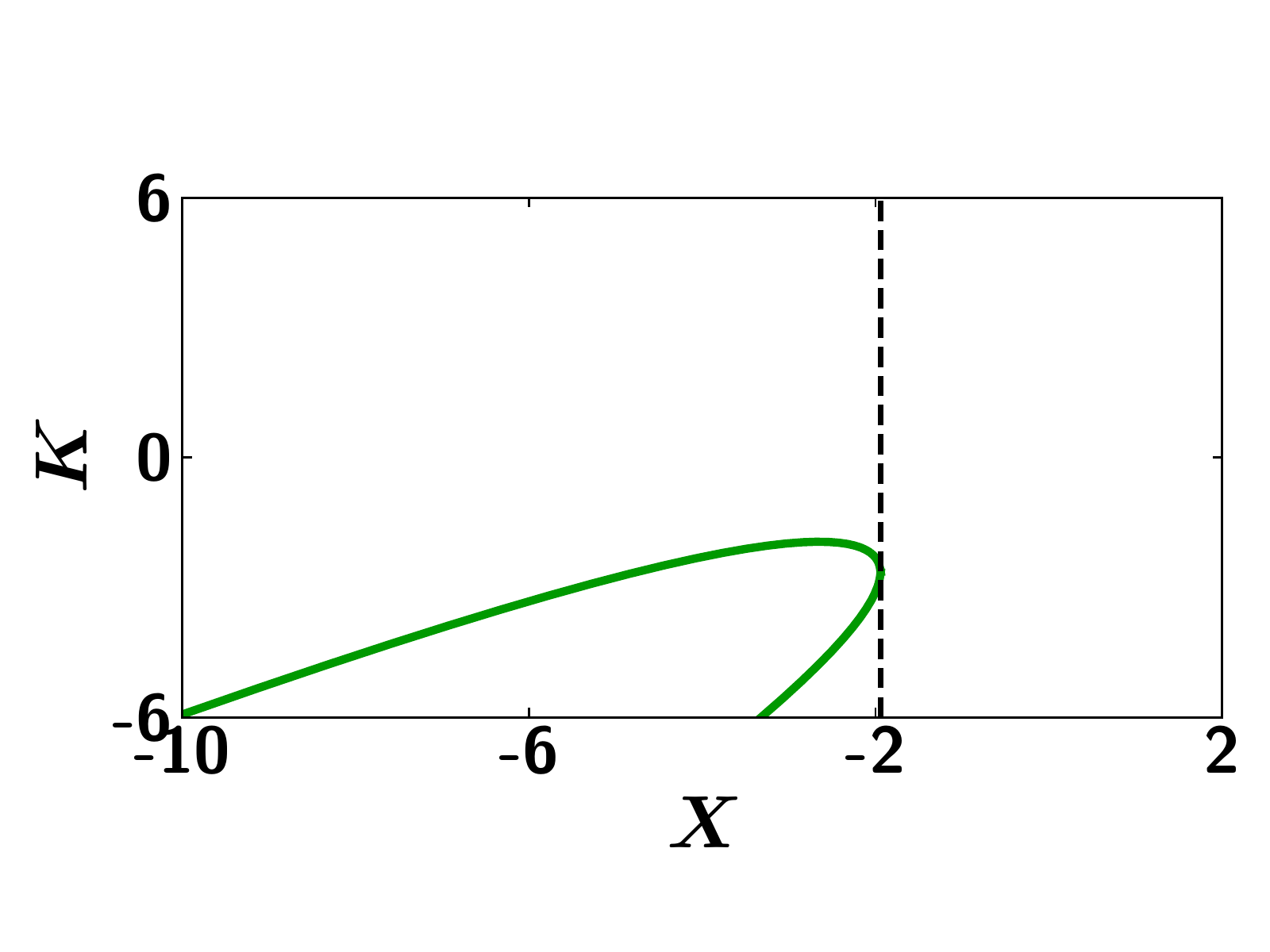}
		\put(85,15){\textbf{\color{black} \large(f)}}
	\end{overpic}
	\begin{overpic}[width=0.24\linewidth,trim={2mm 11mm 3mm 21mm},clip]{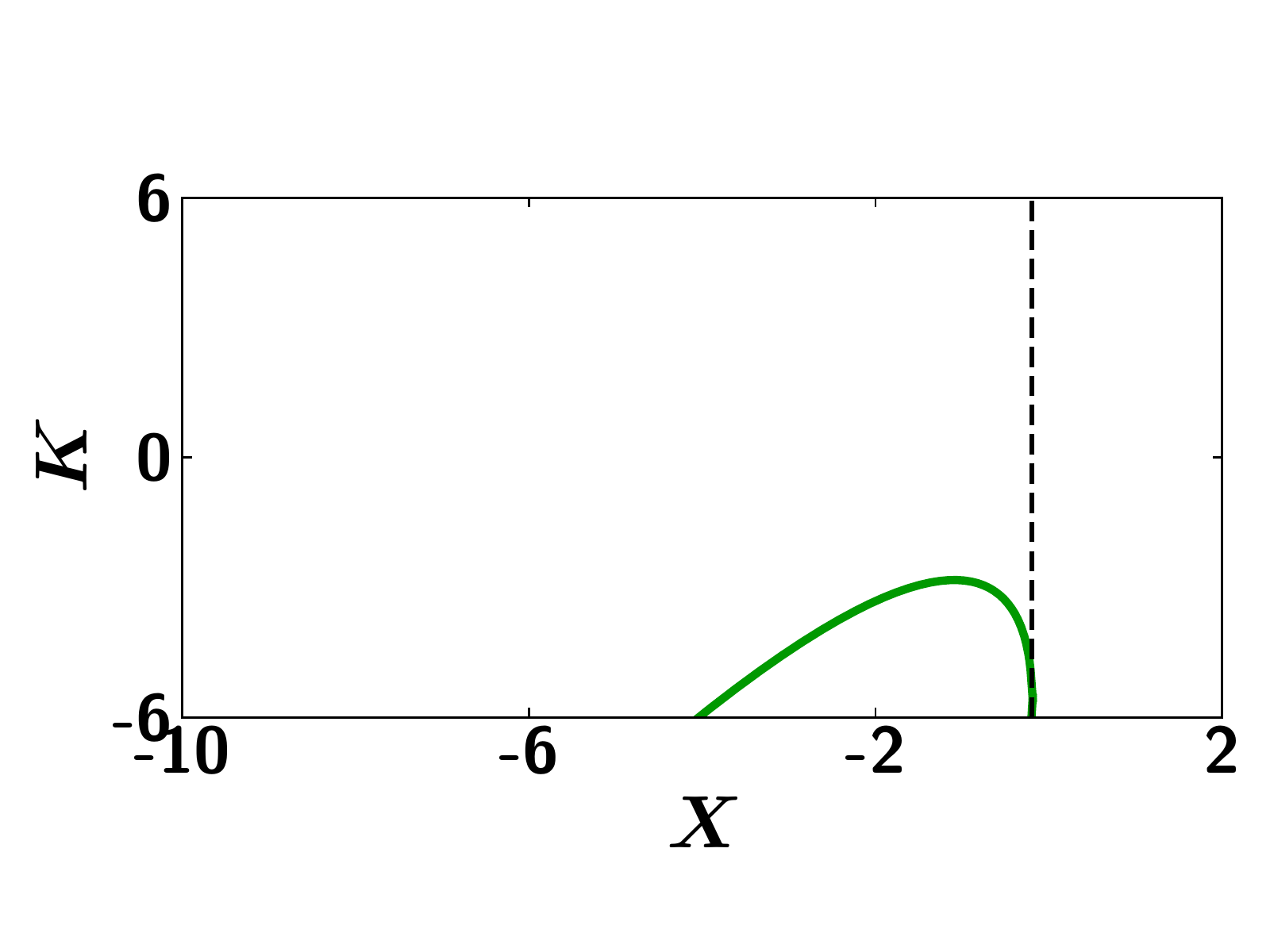}
		\put(85,15){\textbf{\color{black} \large(g)}}
	\end{overpic}
	\begin{overpic}[width=0.24\linewidth,trim={2mm 11mm 3mm 21mm},clip]{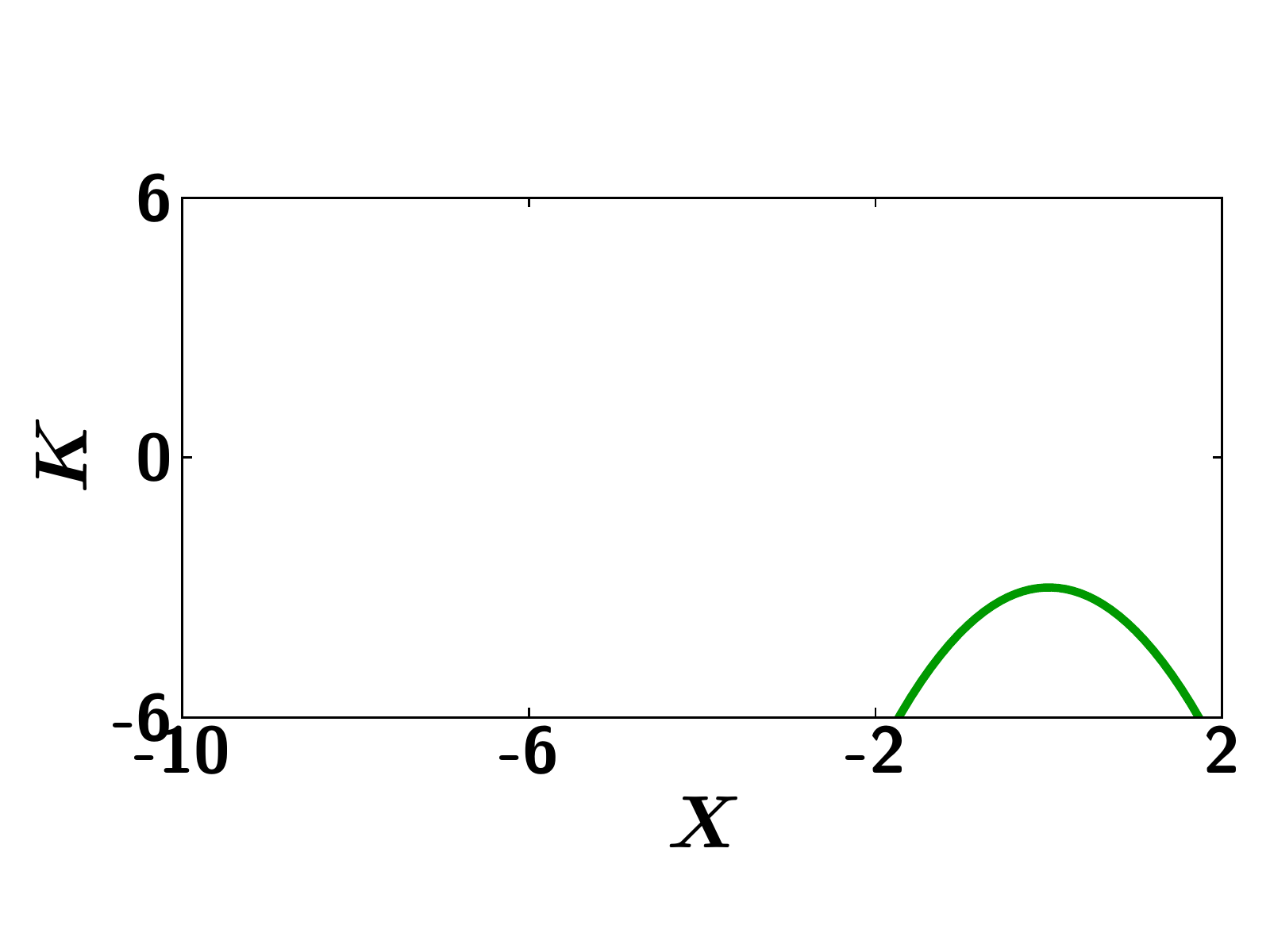}
		\put(85,15){\textbf{\color{black} \large(h)}}
	\end{overpic}
	\caption{\textbf{(a)-(d)} Comparison of the exact and WKB solution of the metaplectically-rotated Airy's equation \eq{eq:6_metAiry} for $a = -3$ and different rotation angles $t$: (a) $t = 0$, (b) $t = -\pi/4$, (c) $t = -2\pi/5$, and (d) $t = -\pi/2$. \textbf{(e)-(h)} Corresponding WKB dispersion surfaces for \Eq{eq:6_metAiry}. The caustic at $X = \tan(t)\sin(t)/4 + a\cos(t)$ coincides with the location on the dispersion surface where $\dd K/ \dd X \to \infty$.}
	\label{fig:6_rotAiry}
\end{figure}

As motivated in \Ch{ch:GO}, caustics can be resolved by rotating the phase space using metaplectic operators, an observation that constitutes the foundation for the MGO theory. Before discussing MGO in its entirety, however, it is useful to first begin with an illustrative example of this concept. Consider a wave field incident on an isolated cutoff in a $1$-D inhomogeneous medium. As is well-known, the corresponding wave field $\psi$ near the cutoff is approximately described by Airy's equation~\cite{Tracy14}
\begin{equation}
	\pd{x}^2 \psi(x) - (x - a) \psi(x) = 0 
	,
	\label{eq:6_airy}
\end{equation}

\noindent with the cutoff located at $a$. Applying the WKB approximation to \Eq{eq:6_airy} yields the dispersion surface $k(x) = \sqrt{a-x}$ on which the wave `quanta' is asymptotically confined, as well as the divergent wave envelope $\phi(x) \sim (a-x)^{-1/4}$. Thus, the caustic at $x = a$ manifests as a singularity in the WKB envelope, as illustrated in the first column of \Fig{fig:6_rotAiry}.

Let me now rotate the phase space using the MT corresponding to a phase-space rotation by (negative) angle $t$ as
\begin{align}
	x = \cos(t) X + i \sin(t) \frac{\dd}{\dd X} 
	, \quad
	\frac{\dd}{\dd x} = i \sin(t) X + \cos(t) \frac{\dd}{\dd X} 
	.
\end{align}

\noindent Then, \Eq{eq:6_airy} becomes
\begin{equation}
	\cos^2(t) \pd{X}^2 \Psi(X) + i\left[\sin(2t) X - \sin(t)\right] \pd{X} \Psi(X) - \left[ \sin^2(t) X^2 + \cos(t) X - \frac{i}{2}\sin(2t) - a\right] \Psi(X) = 0 
	,
	\label{eq:6_metAiry}
\end{equation}

\noindent where $\Psi(X)$ is the metaplectic image of $\psi(x)$. Applying the WKB approximation to \Eq{eq:6_metAiry} yields%
\begin{subequations}%
	\label{eq:6_metAiryWKB}%
	\begin{align}
		\Psi_\text{WKB}(X) &=  
		\frac{
			\alpha_+ \exp\left[ i\Theta_+(X) \right] 
			+ \alpha_- \exp\left[ i\Theta_-(X) \right]
		}{
			\left[ 
				\sin^2(t) - 4X \cos(t) + 4a\cos^2(t) 
			\right]^{1/4}
		} 
		, \\
		\Theta_\pm(X) &= \frac{\tan(t) X - \sin(t) X^2}{2 \cos(t)} \pm \frac{\left[ \sin^2(t) - 4X \cos(t) + 4a\cos^2(t) \right]^{3/2}}{12 \cos^3(t)} 
		,
	\end{align}
\end{subequations}

\noindent with $\alpha_\pm$ constants determined by boundary conditions, which should be matched on either side of the caustic separately due to Stokes phenomenon~\cite{Heading62a}. 

In \Fig{fig:6_rotAiry}, the WKB result is compared to the exact result, which can be computed in \Ch{ch:MT} as
\begin{align}
	\Psi_t(X) = \pm \frac{1}{\sqrt{\cos(t)}} \, \airyA \left[\frac{X}{\cos(t)} - \frac{\tan^2(t)}{4} - a\right] \exp\left[i\frac{\tan(t) X - \sin(t) X^2}{2 \cos(t)} - i\frac{\tan^3(t)}{12} - i\frac{a\tan(t)}{2}\right] 
	.
\end{align}

\noindent As the phase space is rotated, the caustic moves steadily towards increasing $X$. At a rotation angle of $\pi/2$, \Eq{eq:6_metAiry} becomes
\begin{equation}
	-i\pd{X} \Psi(X) + (X^2 - a)\Psi(X) = 0 
	.
	\label{eq:6_airyFT}
\end{equation}

\begin{figure}[t!]
	\centering
	\begin{overpic}[width=0.24\linewidth,trim={2mm 24mm 1mm 17mm},clip]{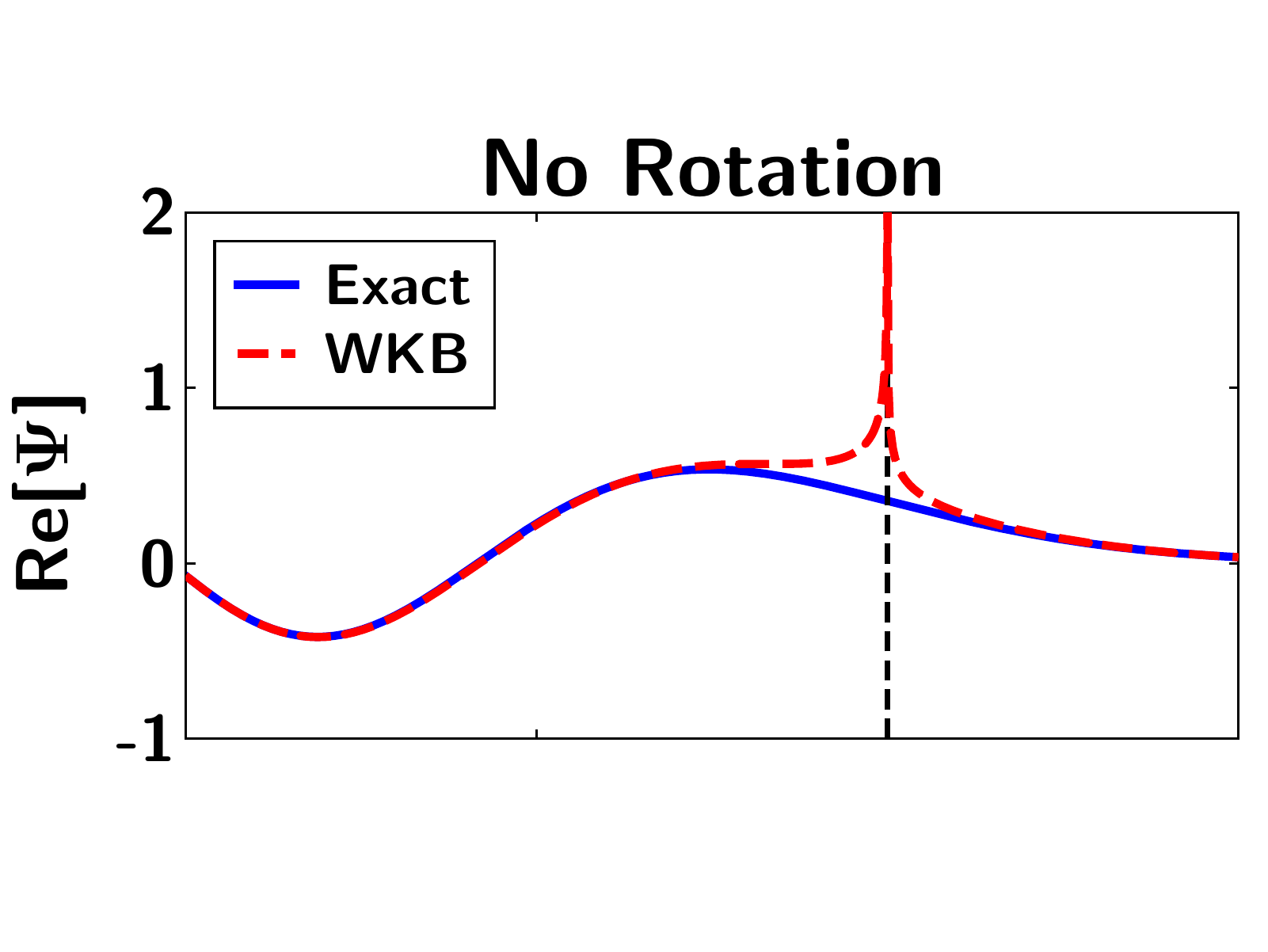}
		\put(85,5){\textbf{\color{black} \large(a)}}
	\end{overpic}
	\begin{overpic}[width=0.24\linewidth,trim={2mm 24mm 1mm 17mm},clip]{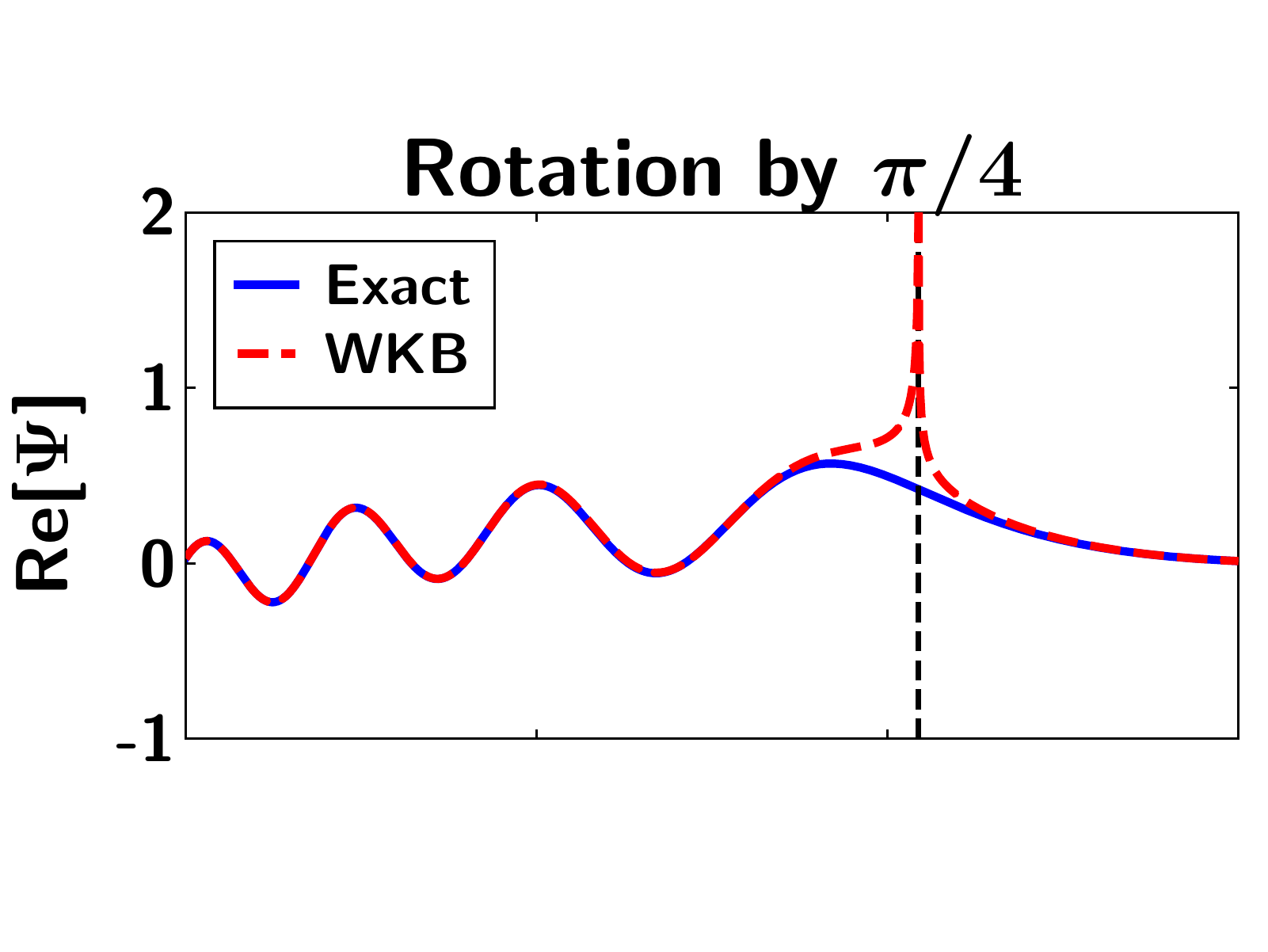}
		\put(85,5){\textbf{\color{black} \large(b)}}
	\end{overpic}
	\begin{overpic}[width=0.24\linewidth,trim={2mm 24mm 1mm 17mm},clip]{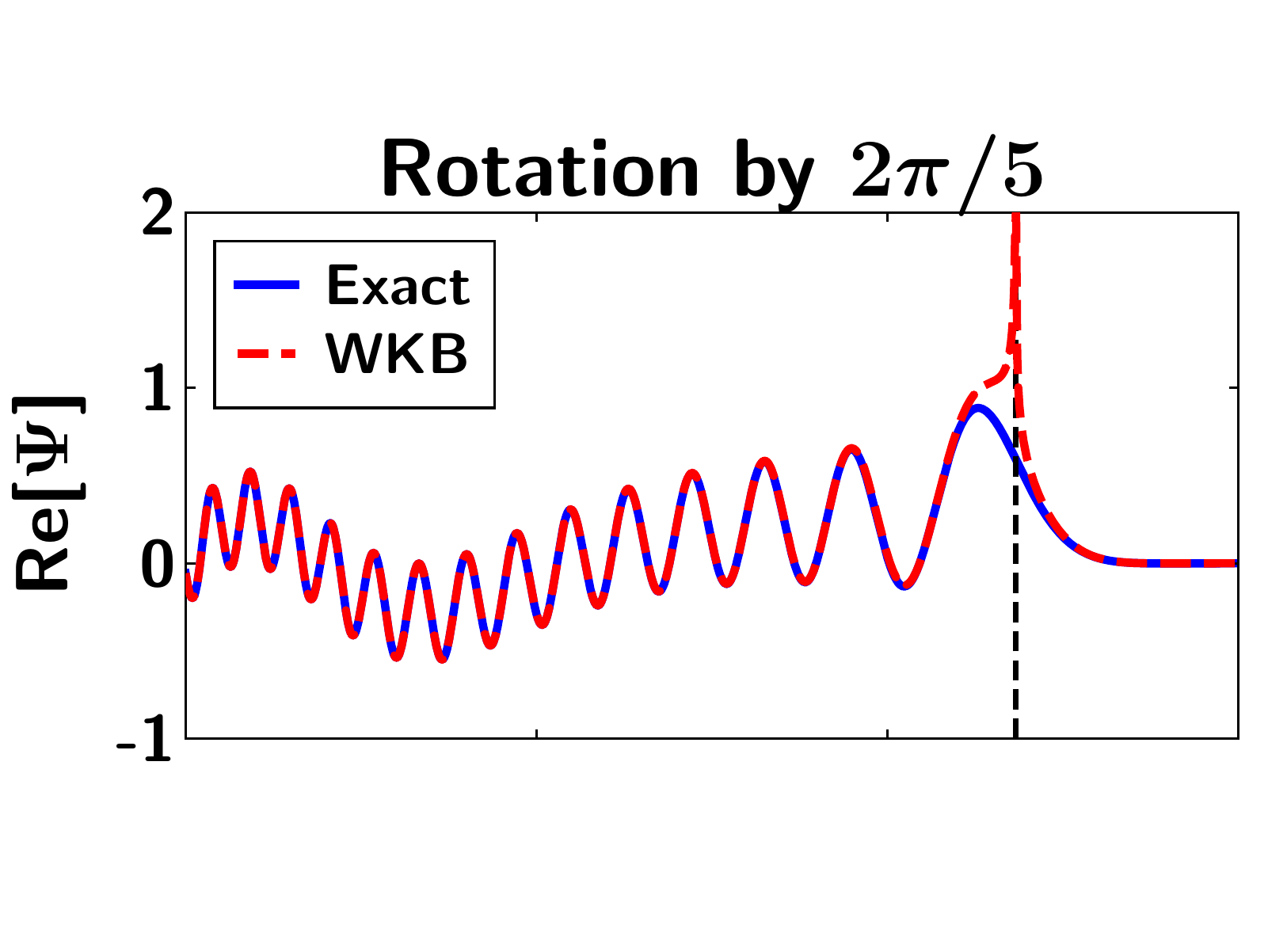}
		\put(85,5){\textbf{\color{black} \large(c)}}
	\end{overpic}
	\begin{overpic}[width=0.24\linewidth,trim={2mm 24mm 1mm 17mm},clip]{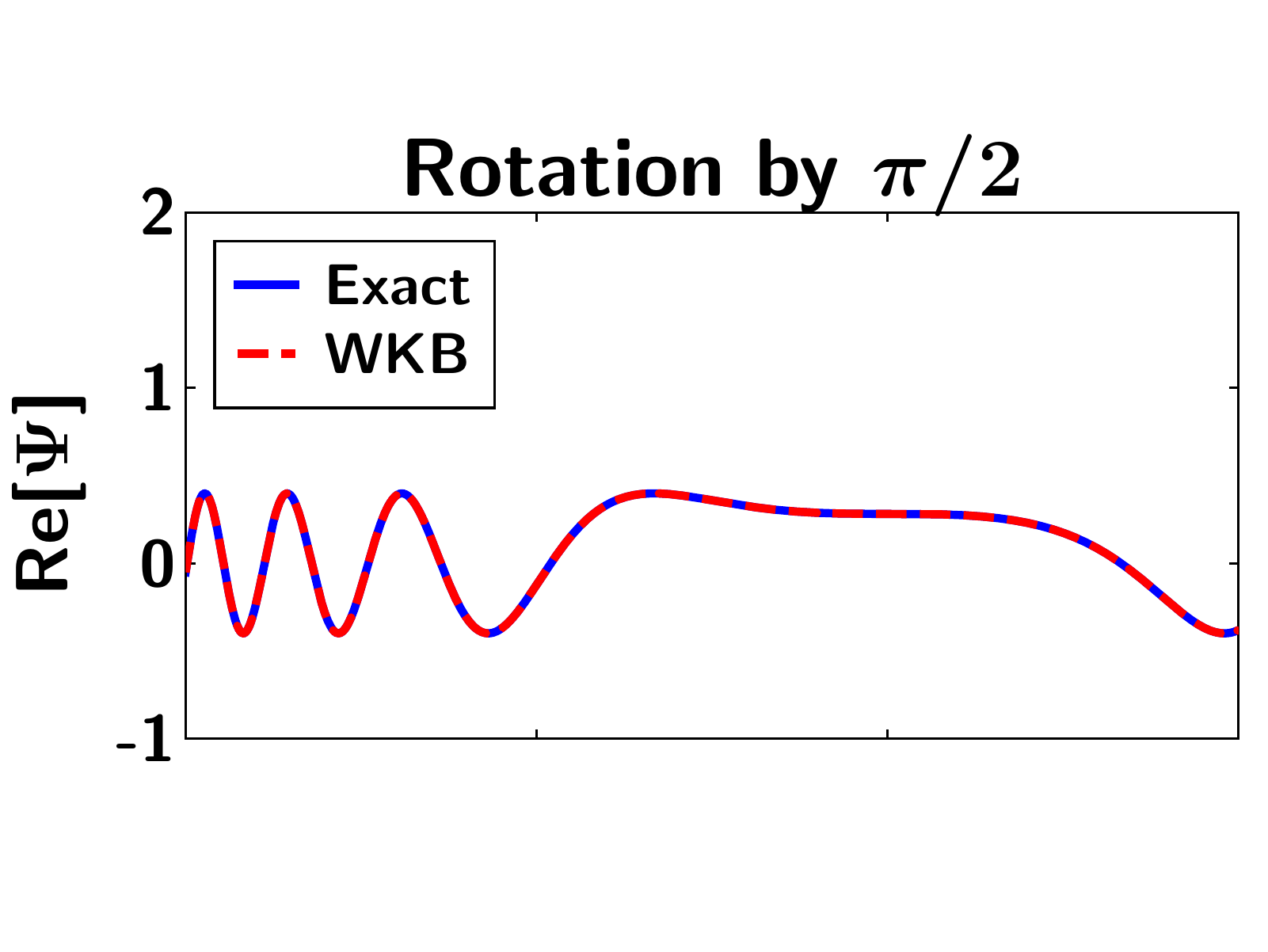}
		\put(85,5){\textbf{\color{black} \large(d)}}
	\end{overpic}

	\begin{overpic}[width=0.24\linewidth,trim={2mm 11mm 3mm 21mm},clip]{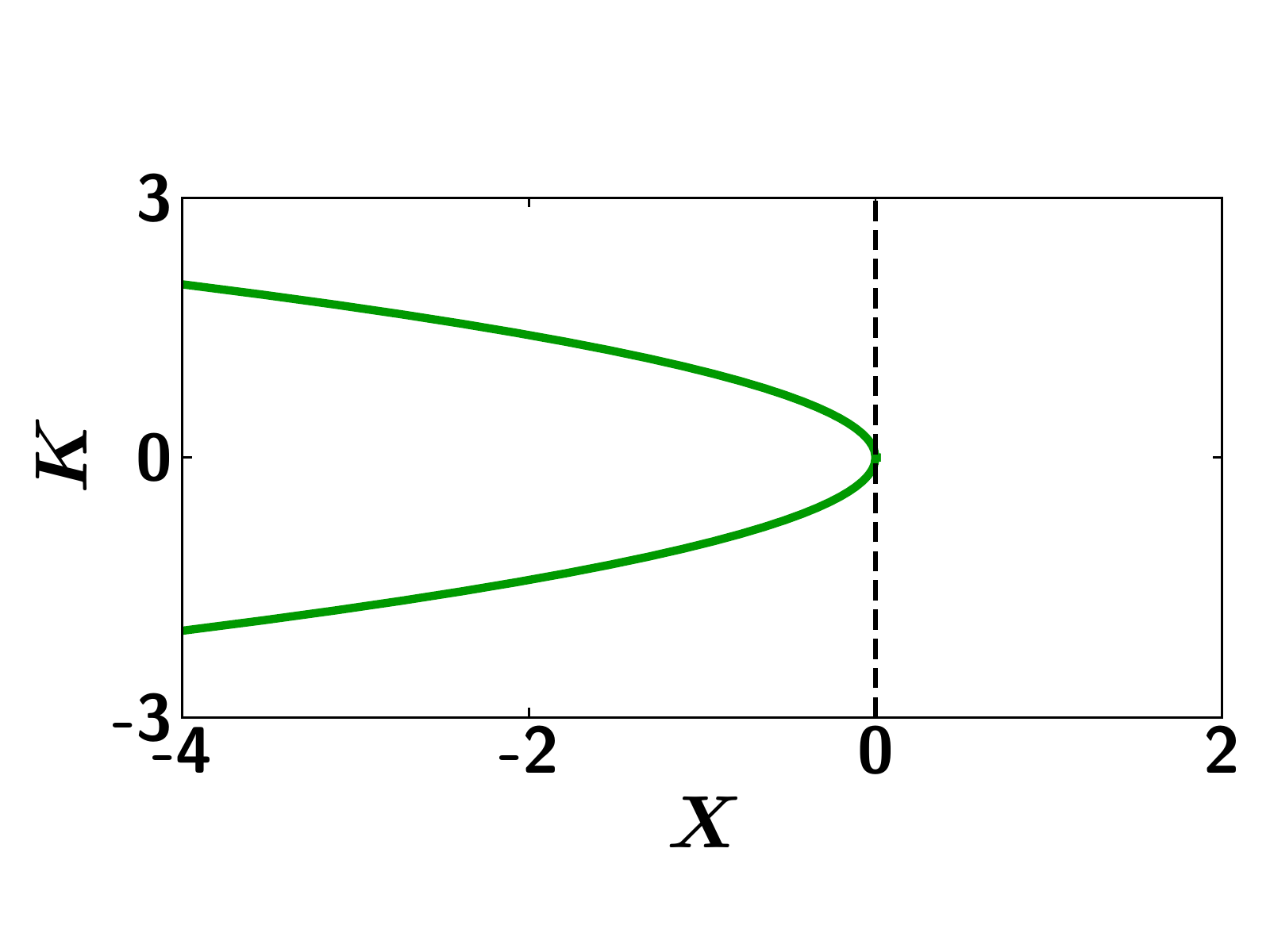}
		\put(85,15){\textbf{\color{black} \large(e)}}
	\end{overpic}
	\begin{overpic}[width=0.24\linewidth,trim={2mm 11mm 3mm 21mm},clip]{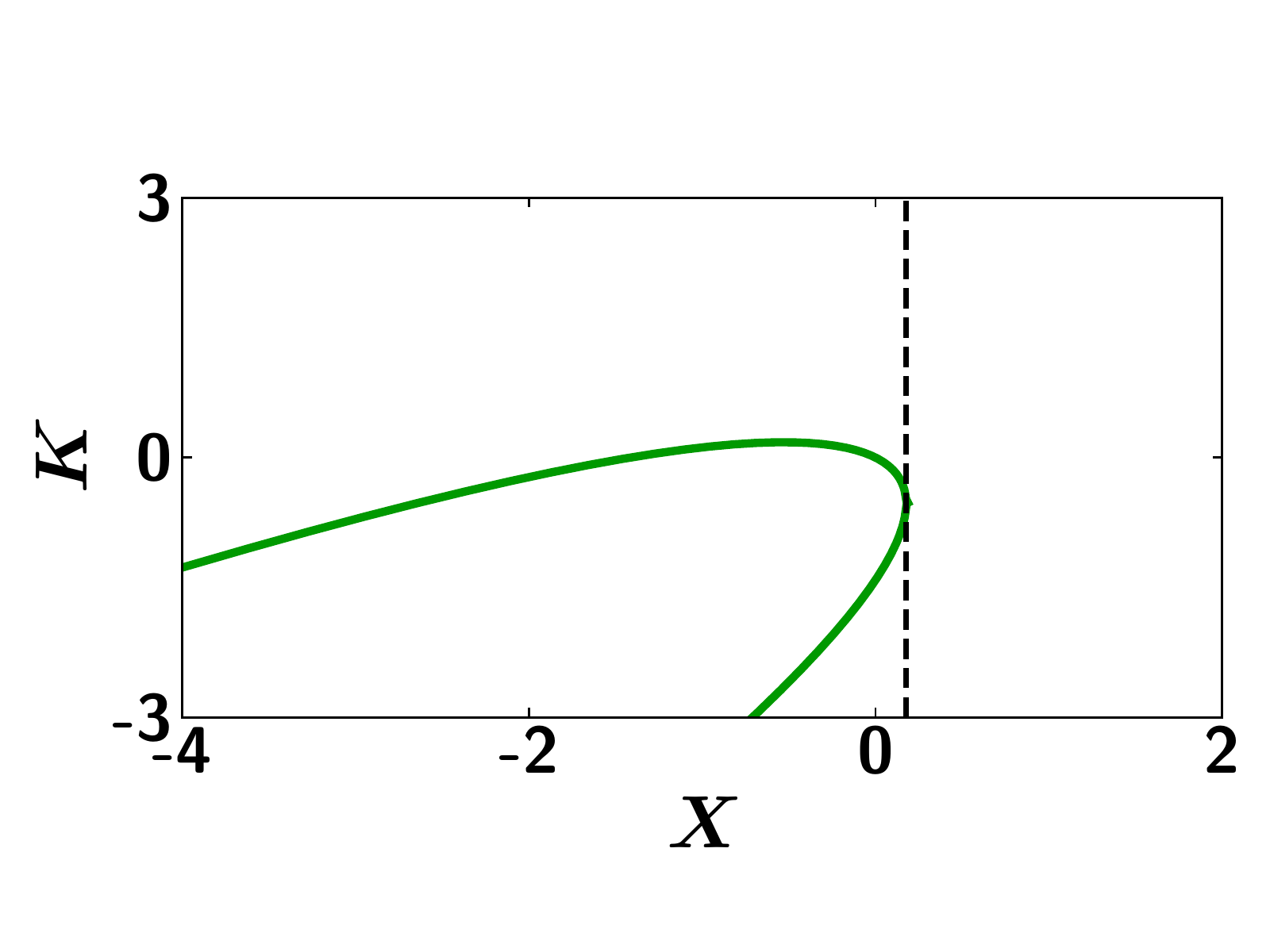}
		\put(85,15){\textbf{\color{black} \large(f)}}
	\end{overpic}
	\begin{overpic}[width=0.24\linewidth,trim={2mm 11mm 3mm 21mm},clip]{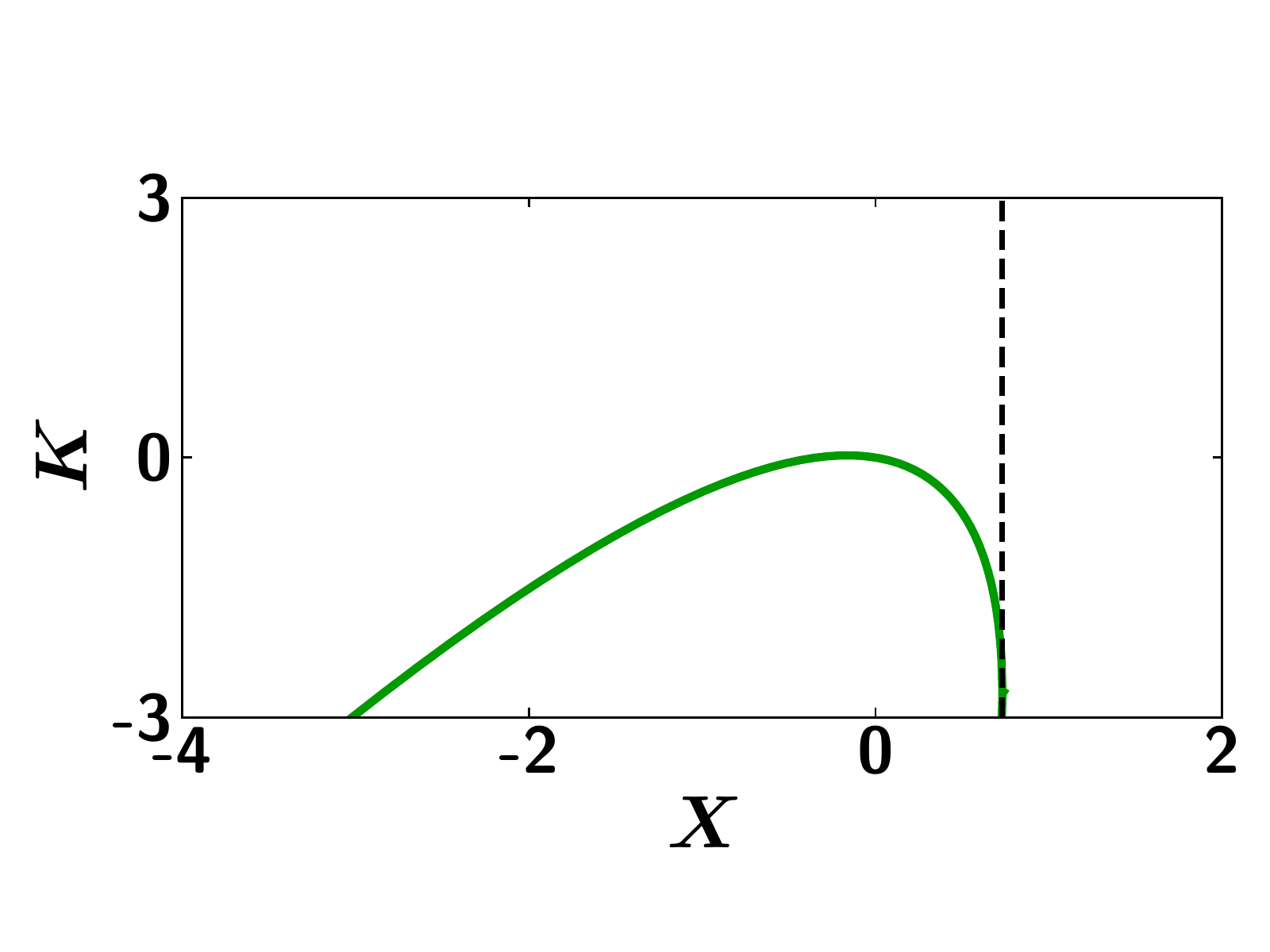}
		\put(85,15){\textbf{\color{black} \large(g)}}
	\end{overpic}
	\begin{overpic}[width=0.24\linewidth,trim={2mm 11mm 3mm 21mm},clip]{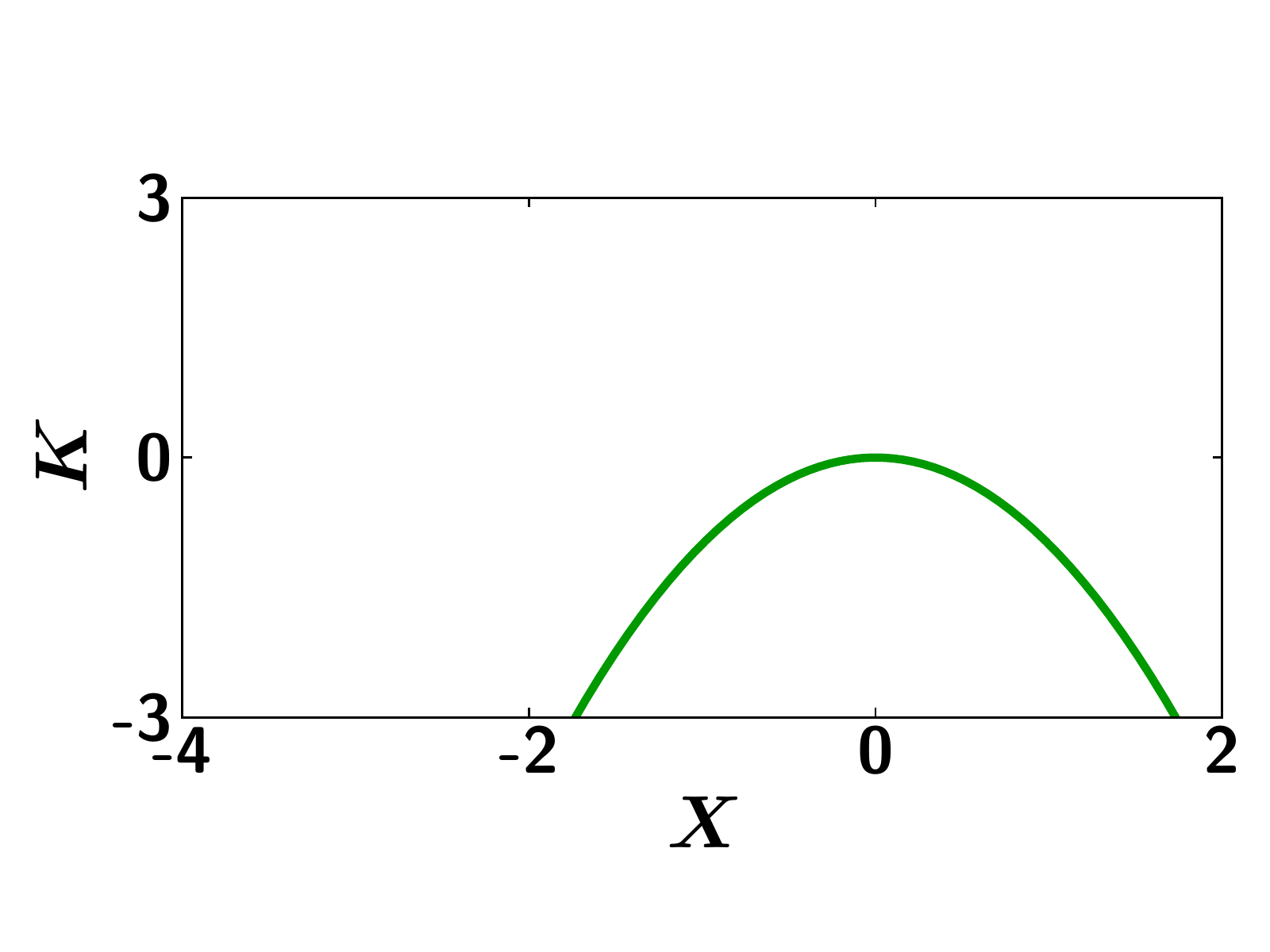}
		\put(85,15){\textbf{\color{black} \large(h)}}
	\end{overpic}
	\caption{Same as \Fig{fig:6_rotAiry}, but with $a = 0$.}
	\label{fig:6_rotAiry0}
\end{figure}

\noindent In this case, the caustic disappears entirely and the WKB approximation, obtained from the ($+$) solution to \Eqs{eq:6_metAiryWKB} as
\begin{equation}
	\Psi_\text{WKB}(X) = \alpha \exp\left( iaX - i \frac{X^3}{3} \right) 
	,
	\label{eq:6_airyFTwkb}
\end{equation}

\noindent becomes exact. Importantly, the WKB approximation to \Eq{eq:6_airyFT} holds at any $a$, even though for $a \ge 0$ there are values of $X$ at which the wavenumber $K$ approaches zero. (For example, see \Fig{fig:6_rotAiry0} for the case $a = 0$.) This is to be expected because: (i) $a$ can be removed from \Eq{eq:6_airy} by a simple variable transformation, and hence has no fundamental meaning, and (ii) it is $\dd K/ \dd X$ that determines the validity of geometrical optics, not the value of $K$ \textit{per~se} (see \Ch{ch:GO}). For wave equations that are more complicated than \Eq{eq:6_airy}, a single MT is not sufficient to reinstate geometrical optics for the entire field. However, multiple MTs applied sequentially can. Specifically, a phase space can be continually rotated using the NIMT such that $\dd K/\dd X$ always remains finite, as I shall now discuss.

\section{Geometrical optics in arbitrary projective plane}

As a first step to developing MGO, I must derive the GO equations in an arbitrarily chosen projective plane in phase space. Recall from \Ch{ch:GO} that standard GO is developed by performing an eikonal decomposition with respect to the standard coordinate variable $\Vect{x}$ and then projecting the resulting envelope operator equation onto the eigenbasis of $\VectOp{x}$; here I shall generalize this procedure to work for some rotated coordinates $\Vect{X}$. 

Generally speaking, for a given plane in phase space to be a valid choice for a GO projective plane, it must be related to $\Vect{x}$-space by a linear symplectic transformation. Therefore, let $\Stroke{\Vect{Z}}$ be the phase-space coordinates that result from a general linear symplectic transformation of the form given by \Eq{eq:3_Strans}, and let $\Stroke{\VectOp{Z}}$ be the transformed phase-space operators that result from the corresponding metaplectic transform given by \Eq{eq:3_operTRANS}. It is worth emphasizing that \Eq{eq:3_Strans} preserves the origin of phase space, \ie $\Vect{z} = \Vect{0}$ maps to $\Stroke{\Vect{Z}} = \Vect{0}$. Shifting the origin does not affect the projective properties of a plane; hence, for the purposes of developing MGO, I can identify all projective planes which differ only by a shift in the origin as equivalent. As a result, when I speak of the `tangent plane' of the dispersion manifold at $\Vect{z}(\Vect{\tau})$ later in this chapter, I really speak of the plane which is parallel to the tangent plane at $\Vect{z}(\Vect{\tau})$ and passes through the origin.

I again consider the general wave equation given in \Eq{eq:2_hilbertWAVE}, but, rather than introducing an eikonal ansatz on $\Vect{x}$-space as done in \Eq{eq:2_phiENV}, I now assume the wavefield is eikonal on the desired projective plane. Hence, I perform the unitary transformation
\begin{equation}
	\ket{\Phi} = e^{-i \Theta(\VectOp{X}) } \ket{\psi}
	\label{eq:6_PhiENV}
\end{equation}

\noindent such that \Eq{eq:2_hilbertWAVE} becomes
\begin{equation}
	e^{-i \Theta(\VectOp{X}) } 
	\oper{D}(\VectOp{z}) \, 
	e^{i \Theta(\VectOp{X}) } 
	\ket{\Phi} = \ket{0} 
	.
	\label{eq:6_wavePHI}
\end{equation}

In principle, \Eq{eq:6_wavePHI} is sufficient to develop GO; however, the simultaneous presence of $\VectOp{X}$ and $\VectOp{z}$ is inconvenient. To rectify this, let me introduce into \Eq{eq:6_wavePHI} the metaplectic operator corresponding to $\Mat{S}$ as
\begin{equation}
	e^{-i \Theta(\VectOp{X}) } 
	\oper{M}
	\oper{M}^\dagger
	\oper{D}(\VectOp{z}) \, 
	\oper{M}
	\oper{M}^\dagger
	e^{i \Theta(\VectOp{X}) } 
	\ket{\Phi} = \ket{0} 
	,
	\label{eq:6_waveMET}
\end{equation}

\noindent where I have used the unitarity of $\oper{M}$. This allows me to rigorously transform $\VectOp{z}$ to $\Stroke{\VectOp{Z}}$. As shown in \App{app:6_metWEYL},
\begin{equation}
	\Weyl\left[ 
		\oper{M}
		\oper{M}^\dagger
		\oper{D}(\VectOp{z}) \, 
		\oper{M}
		\oper{M}^\dagger 
	\right] 
	= \Symb{D}\left(
		\Mat{S}^{-1} \Stroke{\Vect{Z}} 
	\right) 
	= \Symb{D}(\Vect{z})
	,
\end{equation}

\noindent which also demonstrates the well-known `symplectic covariance' property of the Weyl symbol~\cite{Littlejohn86a,deGosson06}. In other words, the Weyl symbol of an operator at a given phase-space location does not depend on how this location is parameterized, as long as different parameterizations (here, $\Vect{z}$ and $\Vect{Z}$) are connected via symplectic transformations.

Since symplectic transformations preserve the Poisson bracket, they also preserve the Moyal star product (\Ch{ch:GO}). Hence, the GO limit of \Eq{eq:6_waveMET} can be obtained using the procedure outlined in \App{app:6_GO}, but replacing $\Symb{D}(\Vect{z})$ with $\Symb{D}(\Mat{S}^{-1} \Vect{Z})$ and $\Vect{z}$ by $\Vect{Z}$.%
\footnote{The derivation presented in \App{app:6_GO} is preferable to that presented in \Ch{ch:GO} since it does not make explicit reference to any specific basis choice; hence it is sufficiently general to accommodate the arbitrarily rotated basis $\{\ket{\Vect{X}} \}$ considered here.} %
This yields
\begin{align}
	\left\{
		\Symb{D}\left[ \Mat{S}^{-1} \Stroke{\Vect{Z}}(\VectOp{X}) \right]
		+ \Vect{V}(\VectOp{X})^\intercal \VectOp{K}
		- \frac{i}{2} \pd{\Vect{X}} \cdot \Vect{V}(\VectOp{X})
		- \frac{1}{2} \deltaQO(\Stroke{\VectOp{Z}})
	\right\} \ket{\Phi} 
	= \ket{0} 
	, 
	\label{eq:6_approxENVmet}
\end{align}

\noindent where, given recent interest~\cite{Dodin19,Yanagihara19a,Yanagihara19b,Yanagihara21a,Yanagihara21b}, I have included the quasioptical (QO) term which governs diffraction as
\begin{align}
	\deltaQO(\Stroke{\VectOp{Z}}) 
	\doteq 
	- \Mat{M}(\VectOp{X}) \dubdot \VectOp{K}\VectOp{K} 
	+ i \left[ \pd{\Vect{X}} \cdot \Mat{M}(\VectOp{X}) \right] \VectOp{K} 
	.
	\label{eq:6_deltaQO}
\end{align}

\noindent I have also defined the following quantities:
\begin{align}
	\Vect{K}(\Vect{X}) 
	\doteq \pd{\Vect{X}} \Theta(\Vect{X}) 
	, \quad
	\Vect{V}(\Vect{X}) 
	\doteq \left.\pd{\Vect{K}} \Symb{D}\left(\Mat{S}^{-1} \Stroke{\Vect{Z}} \right) \right|_{ \Vect{K} = \Vect{K}(\Vect{X}) } 
	, \quad
	\Mat{M}(\Vect{X}) \doteq \left.
		\pd{\Vect{K}\Vect{K}}^2 \Symb{D}\left( \Mat{S}^{-1} \Stroke{\Vect{Z}} \right)
	\right|_{\Vect{K} = \Vect{K}(\Vect{X})} 
	.
\end{align}

Dynamical equations that govern $\Phi(\Vect{X}) \doteq \braket{\Vect{X}}{\Phi}$ are obtained by projecting \Eq{eq:6_approxENVmet} onto $\{ \ket{\Vect{X}} \}$. Neglecting diffraction, the GO equations on this projective plane are
\begin{subequations}
	\begin{align}
		\label{eq:6_GOrayMET}
		\Symb{D}\left[ \Mat{S}^{-1} \Stroke{\Vect{Z}}(\Vect{X}) \right] &= 0 
		, \\
		\label{eq:6_GOenvMET}
		\Vect{V}(\Vect{X})^\intercal \pd{\Vect{X}} \Phi(\Vect{X}) 
		+ \frac{1}{2} \left[ \pd{\Vect{X}} \cdot \Vect{V}(\Vect{X}) \right] \Phi(\Vect{X}) &= 0 
		.
	\end{align}
\end{subequations}

\noindent As before, \Eq{eq:6_GOrayMET} is solved via ray tracing, while \Eq{eq:6_GOenvMET} can be formally solved as
\begin{equation}
	\Phi(\Vect{\tau}) = \phi_0(\Vect{\tau}_\perp) 
	\sqrt{\frac{ J(0, \Vect{\tau}_\perp) }
	{ J(\Vect{\tau}) }} 
	, \quad
	J(\Vect{\tau})
	\doteq 
	\det \pd{\Vect{\tau}} \Vect{X}(\Vect{\tau})
	.
	\label{eq:6_ENVjacobMET}
\end{equation}

However, let me make an observation regarding the rays generated by \Eq{eq:6_GOrayMET}. These rays satisfy
\begin{equation}
	\pd{\tau_1} \Stroke{\Vect{Z}} 
	= \JMat{2N} \, \pd{\Stroke{\Vect{Z}}} \Symb{D}\left( \Mat{S}^{-1} \Stroke{\Vect{Z}} \right) 
	.
\end{equation}

\noindent Since $\Mat{S}$ is constant, the chain rule yields
\begin{equation}
	\pd{\tau_1} \Stroke{\Vect{Z}} 
	= \JMat{2N} \, 
	\left( \Mat{S}^{-1} \right)^\intercal 
	\left. 
		\pd{\Vect{z}} \Symb{D}\left( \Vect{z} \right) 
	\right|_{\Vect{z} = \Mat{S}^{-1}\Stroke{\Vect{Z}}}
	.
\end{equation}

\noindent Moreover, since $\Mat{S}$ is symplectic, multiplication by $\Mat{S}^{-1}$ from the left yields
\begin{equation}
	\pd{\tau_1} \left( \Mat{S}^{-1} \Stroke{\Vect{Z}} \right)
	= \JMat{2N} \, 
	\left. 
		\pd{\Vect{z}} \Symb{D}\left( \Vect{z} \right) 
	\right|_{\Vect{z} = \Mat{S}^{-1}\Stroke{\Vect{Z}}} \, .
\end{equation}

\noindent Finally, a comparison with \Eq{eq:3_EOM} yields the relationship between the original and the transformed rays:
\begin{equation}
	\Stroke{\Vect{Z}}(\Vect{\tau}) 
	= \Mat{S} \, \Vect{z}(\Vect{\tau}) 
	.
	\label{eq:6_sRAYS}
\end{equation}

\noindent Thus, one does not need to re-trace rays every time the projection plane is changed, but rather, one can simply perform the same symplectic transformation to the rays that one applies to the ambient phase space.

Lastly, the wavefield on the projective plane is constructed as
\begin{equation}
	\Psi(\Vect{X}) \doteq \Phi(\Vect{X}) 
	\exp \left[ i \Theta(\Vect{X}) \right] 
	,
\end{equation}

\noindent summed over all branches of $\Theta(\Vect{X})$ if $\Theta(\Vect{X})$ is multivalued. The wavefield on $\Vect{x}$-space can then be obtained by taking the inverse MT of $\Psi(\Vect{X})$ using \Eq{eq:3_invMTint}. To conclude this section, let me reiterate for emphasis that I use the lowest-order GO approximation in the tangent plane [\Eqs{eq:6_GOrayMET} and \eq{eq:6_GOenvMET}] for simplicity only; it is anticipated that, after rotating into the tangent plane, one should be able to use more sophisticated wave models instead of the standard GO approximation, such as the XGO model~\cite{Ruiz17a,Ruiz15a,Ruiz15b,Ruiz15d,Ruiz17t} that includes polarization dynamics or the recently developed quasioptical extension to XGO~\cite{Dodin19,Yanagihara19a,Yanagihara19b,Yanagihara21a,Yanagihara21b} that also includes diffraction, to yield correspondingly more sophisticated MGO models that still remain free of caustics by virtue of the MGO rotation scheme.


\section{Geometrical optics in a piecewise-linear tangent space: the MGO framework}

Equation \eq{eq:6_ENVjacobMET} implies that a caustic at some position $\Vect{\tau} = \Vect{t}$ on the ray manifold can be avoided by choosing $\Vect{X}$ to be the tangent plane at $\Vect{t}$ (and subsequently denoted $\Vect{X}_\Vect{t}$), since $J(\Vect{t}) \neq 0$ is then guaranteed by definition. This is the logic of the MGO framework, which is summarized in the following three steps.
\begin{itemize}
	\item{The rotated GO equations \eq{eq:6_GOrayMET} and \eq{eq:6_GOenvMET} are solved in the tangent plane at a given ray position $\Vect{t}$.}
	\item{The obtained solution $\Psi_\Vect{t}(\Vect{X}_\Vect{t})$ is rotated into the following tangent plane at $\Vect{t} + \delta \Vect{t}$ using an infinitesimal near-identity MT (NIMT) to provide initial conditions for the corresponding next set of rotated GO equations to be solved. This step improves the continuity of the global solution.}
	\item{An inverse MT is used to map $\Psi_\Vect{t}(\Vect{X}_\Vect{t})$ back to the original $\Vect{x}$ coordinates.}
\end{itemize}

\begin{figure}[t]
	\centering
	\includegraphics[width=0.5\linewidth,trim={5mm 7mm 20mm 9mm},clip]{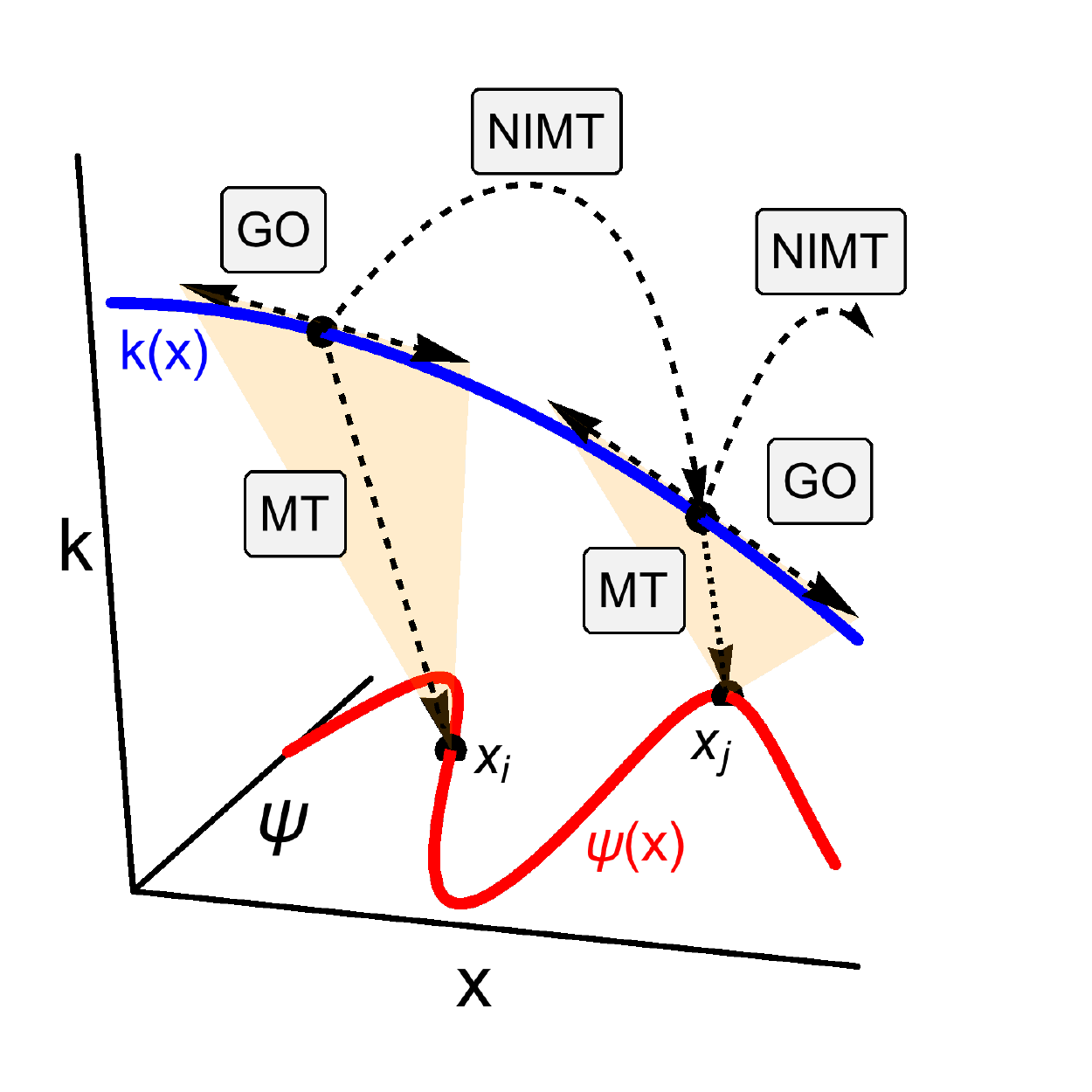}
	\caption{New framework for GO that is free from caustic singularities. This new approach consists of three steps: for a given point on the dispersion manifold, (i) the GO solution is calculated on the local tangent plane using metaplectic geometrical optics (MGO), (ii) the MGO solution is projected onto the tangent plane of the next point on the dispersion manifold using a near-identity metaplectic transform (NIMT) to initialize the next MGO calculation, (iii) the MGO solution is projected onto configuration space using a metaplectic transform (MT) subsequently evaluated using stationary phase or steepest-descent method. The process is repeated for all points on the dispersion manifold.}
	\label{fig:6_MGO}
\end{figure}

\noindent This procedure is then repeated for all points on the ray manifold, and the resulting contributions are summed over to obtain the final field. See \Fig{fig:6_MGO} for a visual summary.

\subsection{GO solution in a single tangent plane}

In the following, I shall assume that the dispersion manifold $\Vect{z}(\Vect{\tau})$ has already been obtained via \Eq{eq:3_EOM}. This is not a restrictive assumption, though, because the ray trajectories themselves are unaffected by caustics. 

Let me consider some point $\Vect{x}$ in configuration space and attempt to construct $\psi(\Vect{x})$. To do so, I shall map $\Vect{x}$ to the dispersion manifold using the ray map $\Vect{\tau}(\Vect{x})$, solve for $\Psi_\Vect{t}(\Vect{X}_\Vect{t})$ in the optimal projection plane, that is, the tangent plane of the dispersion manifold at $\Vect{\tau}$, and then map $\Psi_\Vect{t}(\Vect{X}_\Vect{t})$ to $\psi(\Vect{x})$ using an inverse MT. In general, however, $\Vect{\tau}(\Vect{x})$ will be multi-valued, and for the aforementioned scheme to work, the contributions to $\psi(\Vect{x})$ from each branch must be considered separately. Therefore, let $\Vect{t} \in \Vect{\tau}(\Vect{x})$ be a branch of $\Vect{\tau}(\Vect{x})$, and let me construct the GO wavefield in the tangent plane at $\Vect{t}$.%
\footnote{The GO approximation in the tangent plane is facilitated by the fact that $\pd{\tau_1}\Vect{K}_{\Vect{t}} = - \pd{\Vect{Q}_\Vect{t}} \fourier{\Symb{D}} = \Vect{0}$ at the tangent point by definition, where
$\fourier{\Symb{D}}(\Stroke{\Vect{Z}}_\Vect{t})= \Symb{D}( \Mat{S}_\Vect{t}^{-1} \Stroke{\Vect{Z}}_\Vect{t})$ is the Weyl symbol of the dispersion operator in the new coordinates. I also assume that $\pd{\tau_1}\Vect{K}_{\Vect{t}}$ is slowly varying in the neighborhood of the tangent point.}
Let $\Mat{S}_\Vect{t}$ be the symplectic matrix that transforms $\Vect{x}$-space to the tangent plane. [Note that by definition, $\Mat{S}_\Vect{t}$ consists of the tangent vectors and normal vectors on the dispersion manifold at position $\Vect{z}(\Vect{\tau})$. As discussed further in \Sec{sec:linking} and \App{app:6_QR}, these vectors can be calculated via the ray trajectories themselves, yielding an explicit algorithmic construction of $\Mat{S}_\Vect{t}$.] The rays are transformed into the new coordinates using \Eq{eq:6_sRAYS} as
\begin{equation}
	\Stroke{\Vect{Z}}_\Vect{t}(\Vect{\tau}) = \Mat{S}_\Vect{t} \, \Vect{z}(\Vect{\tau})  
	.
\end{equation}

\noindent Using the rays, $\Vect{K}_\Vect{t}(\Vect{X}_\Vect{t})$ can be constructed as 
\begin{equation}
	\Vect{K}_\Vect{t}(\Vect{X}_\Vect{t}) \doteq \Vect{K}_\Vect{t}\left[ \Vect{\tau}(\Vect{X}_\Vect{t}) \right] 
	,
	\label{eq:6_PFIELD}
\end{equation}

\noindent where $\Vect{\tau}(\Vect{X}_\Vect{t})$ is the function inverse of $\Vect{X}_\Vect{t}(\Vect{\tau})$, \ie the tangent-plane ray map. 

Before solving for the tangent-plane phase and envelope, it is convenient to renormalize $\Psi_\Vect{t}(\Vect{X}_\Vect{t})$ by its value at $\Vect{X}_\Vect{t}(\Vect{t})$;%
\footnote{Note the distinction between $\Vect{X}_\Vect{t}$ and $\Vect{X}_\Vect{t}(\Vect{t})$; the former denotes the general coordinates in the tangent plane at $\Vect{z}(\Vect{t})$ on the dispersion manifold, while the latter denotes the specific value of that coordinate when the ray trajectories are projected onto the tangent plane and then evaluated at $\Vect{t}$. This is analogous to the difference between $\Vect{x}$ and $\Vect{x}(\Vect{t})$.} %
that is, let
\begin{equation}
	\Psi_\Vect{t}(\Vect{X}_\Vect{t}) 
	= \alpha_\Vect{t} 
	\Phi_\Vect{t}(\Vect{X}_\Vect{t})
	\exp\left[ i \Theta_\Vect{t}(\Vect{X}_\Vect{t}) \right]
	,
	\quad
	\alpha_\Vect{t}
	\doteq 
	\Psi_\Vect{t}
	\left[
		\Vect{X}_\Vect{t}(\Vect{t})
	\right] 
	,
	\label{eq:6_wavefieldMET}
\end{equation}

\noindent and require
\begin{equation}
	\Phi_\Vect{t}
	\left[
		\Vect{X}_\Vect{t}(\Vect{t})
	\right]
	= 1
	,
	\quad
	\Theta_\Vect{t}
	\left[
		\Vect{X}_\Vect{t}(\Vect{t})
	\right]
	= 0 
	.
	\label{eq:6_MGOconstr}
\end{equation}

\noindent Then, the phase on the tangent plane is computed simply as
\begin{equation}
	\Theta_\Vect{t}(\Vect{X}_\Vect{t}) = 
	\int_{ \Vect{X}_\Vect{t}(\Vect{t}) }^{\Vect{X}_\Vect{t}} 
	\dd \Vect{X}^\intercal
	\Vect{K}_\Vect{t}\left( \Vect{X} \right) 
	,
	\label{eq:6_phase}
\end{equation}

\noindent where the line integral is taken over any path with the specified endpoints. Importantly, when $\Vect{\tau}(\Vect{x})$ is multi-valued, then $\Vect{K}_\Vect{t}(\Vect{X}_\Vect{t})$ should be restricted to the branch containing $\Vect{K}_\Vect{t}(\Vect{t})$. This is because I am only interested in the contribution near $\Vect{t}$ on the dispersion manifold. The envelope is computed using \Eq{eq:6_GOenvMET} or its QO analogue. For the lowest order GO equation, the solution is written explicitly as
\begin{equation}
	\Phi_\Vect{t}
	\left[
		\Vect{X}_\Vect{t}(\Vect{\tau})
	\right]
	= 
	\Phi_\Vect{t}
	\left[
		\Vect{X}_\Vect{t}(t_1, \Vect{\tau}_\perp)
	\right]
	\sqrt
	{
		\frac
		{
			J_\Vect{t}(t_1, \Vect{\tau}_\perp)
		}
		{
			J_\Vect{t}(\Vect{\tau})
		}
	}
	,
	\label{eq:6_MGOphi}
\end{equation}

\noindent where $\Phi_\Vect{t}[\Vect{X}_\Vect{t}(t_1, \Vect{\tau}_\perp)]$ is set by initial conditions subject to \Eq{eq:6_MGOconstr}, and
\begin{equation}
	J_\Vect{t}(\Vect{\tau})
	\doteq
	\det \pd{\Vect{\tau}} \Vect{X}_\Vect{t}(\Vect{\tau})
	\equiv
	\det \left[
		\Mat{A}_\Vect{t} 
		\pd{\Vect{\tau}} \Vect{x}(\Vect{\tau})
		+
		\Mat{B}_\Vect{t} 
		\pd{\Vect{\tau}} \Vect{k}(\Vect{\tau})
	\right]
	,
	\label{eq:6_MGOjac}
\end{equation}

\noindent represents the tangent-plane ray-map Jacobian determinant.


\subsection{Linking the tangent-plane solutions to obtain a continuous representation}
\label{sec:linking}

The overall `constant' $\alpha_\Vect{t}$ is actually a function of $\Vect{t}$ that must be properly constructed required to make $\psi(\Vect{x})$ continuous. To determine $\alpha_\Vect{t}$, note that after computing \Eq{eq:6_wavefieldMET} for all $\Vect{t} \in \Vect{\tau}(\Vect{x})$ and for all $\Vect{x}$ in the domain of interest, I will have performed an independent GO calculation at each $\Vect{\tau}$ on the dispersion manifold. This sequence of GO calculations must be `linked' together such that they each describe the same wavefield. This `linking' is governed by $\alpha_\Vect{t}$. Let me impose the following linking procedure: the arbitrary constant in $\Psi_{\Vect{t}+\Vect{h}}(\Vect{X}_{\Vect{t}+\Vect{h}})$ is obtained by projecting the neighboring $\Psi_\Vect{t}(\Vect{X}_\Vect{t})$ onto the tangent plane at $\Vect{\tau} = \Vect{t}+\Vect{h}$. Hence, $\alpha_\Vect{t}$ can be found from
\begin{equation}
	\alpha_{\Vect{t}+\Vect{h}} = 
	\alpha_\Vect{t}
	\left.\NIMT
	{
		\Mat{S}_{\Vect{t}+\Vect{h}} \Mat{S}^{-1}_\Vect{t}
	}\left\{
		\Phi_\Vect{t}(\Vect{X}_\Vect{t}) \,
		\exp\left[ i \Theta_\Vect{t}(\Vect{X}_\Vect{t}) \right]
		\nullFrac
	\right\}\right|_{\Vect{X}_{\Vect{t}+\Vect{h}}( \Vect{t}+\Vect{h} )}
	,
	\label{eq:6_nimtCONT}
\end{equation}

\noindent where $\NIMT{\Mat{S}}\left[ \psi(\Vect{x}) \right]$ denotes the NIMT of $\psi(\Vect{x})$ with respect to $\Mat{S}$ (\Ch{ch:MT}). In other words, the arbitrary constant in $\Psi_{\Vect{t}+\Vect{h}}(\Vect{X}_{\Vect{t}+\Vect{h}})$ is obtained by projecting the neighboring $\Psi_\Vect{t}(\Vect{X}_\Vect{t})$ onto the tangent plane at $\Vect{\tau} = \Vect{t}+\Vect{h}$.

Equation \eq{eq:6_nimtCONT} is sufficient for implementation in an MGO-based ray-tracing code since it can be readily performed using the NIMT algorithm discussed in \Ch{ch:NIMT}. Alternatively, after some additional manipulations, \Eq{eq:6_nimtCONT} can be solved in closed form to yield a simpler analytical expression that is more suitable for solving MGO problems analytically (as shown in \Ch{ch:Ex}) and might also be easier to implement numerically, although a detailed stability analysis (against round-off error, among other things) would need to performed before this can be stated definitively. These additional manipulations are done as follows.

In the continuous limit, \Eq{eq:6_nimtCONT} can also be written as a differential equation. Using \Eqs{eq:4_NIET} followed by \Eqs{eq:3_invPRODexpand} and \eq{eq:3_detAexpand}, one can show that
\begin{align}
	&\left.\NIMT
	{
		\Mat{S}_{\Vect{t}+\Vect{h}} \Mat{S}^{-1}_\Vect{t}
	}\left\{
		\Phi_\Vect{t}(\Vect{X}_\Vect{t}) \,
		\exp\left[ i \Theta_\Vect{t}(\Vect{X}_\Vect{t}) \right]
	\right\}\right|_{\Vect{X}_{\Vect{t}+\Vect{h}}( \Vect{t}+\Vect{h} )}
	\approx
	1 
	+ h \, \eta_\Vect{t}
	,
\end{align}

\noindent where I have defined $\eta_\Vect{t}$ as
\begin{align}
	\eta_\Vect{t}
	&\doteq
	\left[\dd_h \Vect{X}_\Vect{t}(\Vect{t}) 
		- \Mat{V}_\Vect{t}^\intercal \Vect{X}_\Vect{t}(\Vect{t}) 
		- \Mat{W}_\Vect{t} \Vect{K}_\Vect{t}(\Vect{t}) 
		\nullFrac
	\right]^\intercal
	\pd{\Vect{X}_\Vect{t}} \Phi_\Vect{t}\left[
		\Vect{X}_\Vect{t}(\Vect{t})
	\right]
	+ 
	i \Vect{K}_\Vect{t}^\intercal(\Vect{t}) \dd_h \Vect{X}_\Vect{t}(\Vect{t}) 
	- \frac{1}{2}
	\Tr\left(\Mat{V}_\Vect{t} \right) 
	\nonumber\\
	&\hspace{3mm} 
	- \frac{i}{2} \Vect{K}_\Vect{t}^\intercal(\Vect{t}) \Mat{W}_\Vect{t} \Vect{K}_\Vect{t}(\Vect{t})
	- \frac{i}{2} \Vect{X}_\Vect{t}^\intercal(\Vect{t}) \Mat{U}_\Vect{t} \Vect{X}_\Vect{t}(\Vect{t})
	- \frac{i}{2} \Vect{K}_\Vect{t}^\intercal(\Vect{t}) \Mat{V}_\Vect{t}^\intercal \Vect{X}_\Vect{t}(\Vect{t}) 
	- \frac{i}{2} \Vect{X}_\Vect{t}^\intercal(\Vect{t}) \Mat{V}_\Vect{t} \Vect{K}_\Vect{t}(\Vect{t}) 
	.
	\label{eq:6_etaDEFunsimple}
\end{align}

\noindent The $N\times N$ matrices $\Mat{U}$, $\Mat{V}$, and $\Mat{W}$ are obtained through the matrix decomposition
\begin{equation}
	\left(\dd_{h} \Mat{S}_\Vect{t}\right)\Mat{S}_\Vect{t}^{-1}
	\doteq
	\begin{pmatrix}
		\Mat{V}_\Vect{t}^\intercal & \Mat{W}_\Vect{t} \\
		- \Mat{U}_\Vect{t} & - \Mat{V}_\Vect{t}
	\end{pmatrix}
	,
	\label{eq:6_UVWdef}
\end{equation}

\noindent which is possible because $\left(\dd_{h} \Mat{S}_\Vect{t}\right)\Mat{S}_\Vect{t}^{-1}$ is a Hamiltonian matrix (\Ch{ch:MT}). Consequently, $\Mat{U}_\Vect{t}$ and $\Mat{W}_\Vect{t}$ are both symmetric. I have also defined the directional derivative as
\begin{equation}
	h \, \dd_{h} 
	\doteq 
	\Vect{h}^\intercal \pd{\Vect{t}} 
	.
\end{equation}

\noindent Note that $\dd_{h}$ is a total (directional) derivative in $\Vect{t}$, so it acts on both arguments of $\Vect{X}_\Vect{t}(\Vect{t})$, including the subscript, \ie
\begin{equation}
	\dd_{h}\Vect{X}_\Vect{t}(\Vect{t}) \equiv
	\dd_{h}\left[
		\Mat{A}_\Vect{t} \Vect{x}(\Vect{t})
		+ \Mat{B}_\Vect{t} \Vect{k}(\Vect{t})
	\right]
	=
	\pd{h} \Mat{A}_\Vect{t} \, \Vect{x}(\Vect{t})
	+\Mat{A}_\Vect{t} \, \pd{h} \Vect{x}(\Vect{t})
	+ \pd{h}\Mat{B}_\Vect{t} \, \Vect{k}(\Vect{t})
	+ \Mat{B}_\Vect{t} \, \pd{h}\Vect{k}(\Vect{t})
	.
\end{equation}

\noindent Then, \Eq{eq:6_nimtCONT} yields
\begin{equation}
	\pd{h} \log \alpha_\Vect{t} 
	= 
	\eta_\Vect{t} 
	.
	\label{eq:6_constDE}
\end{equation}

\noindent When $\alpha_\Vect{t}$ is evolved along a ray, then $\partial_h = \partial_{t_1}$, and \Eq{eq:6_constDE} is trivially integrated as
\begin{equation}
	\alpha_\Vect{t} = \alpha_{\left(0,\Vect{t}_\perp \right)} \exp\left[
		\int_0^{t_1} \dd \xi \, \eta_{\left(\xi, \Vect{t}_\perp \right)}
	\right] 
	.
	\label{eq:6_alphaINT}
\end{equation}

\noindent The remaining constant function $\alpha_{\left(0,\Vect{t}_\perp \right)}$ is determined by initial conditions.

Using the definition \eq{eq:6_UVWdef}, \Eq{eq:6_etaDEFunsimple} can be simplified considerably. First, I note that
\begin{align}
	&\Vect{K}_\Vect{t}^\intercal(\Vect{t}) \Mat{W}_\Vect{t} \Vect{K}_\Vect{t}(\Vect{t})
	+\Vect{X}_\Vect{t}^\intercal(\Vect{t}) \Mat{U}_\Vect{t} \Vect{X}_\Vect{t}(\Vect{t})
	+\Vect{K}_\Vect{t}^\intercal(\Vect{t}) \Mat{V}_\Vect{t}^\intercal \Vect{X}_\Vect{t}(\Vect{t}) 
	+\Vect{X}_\Vect{t}^\intercal(\Vect{t}) \Mat{V}_\Vect{t} \Vect{K}_\Vect{t}(\Vect{t}) 
	\nonumber\\
	&=
	- \Stroke{\Vect{Z}}^\intercal_\Vect{t}(\Vect{t}) \,
	\JMat{2N}
	\left(\dd_{h} \Mat{S}_\Vect{t}\right)\Mat{S}_\Vect{t}^{-1} \,
	\Stroke{\Vect{Z}}_\Vect{t}(\Vect{t}) 
	=
	- \Vect{z}^\intercal(\Vect{t}) \,
	\Mat{S}_\Vect{t}^\intercal
	\JMat{2N}
	\left(\dd_{h} \Mat{S}_\Vect{t}\right) \,
	\Vect{z}(\Vect{t})
	.
\end{align}

\noindent Next, I note that
\begin{align}
	\Vect{K}_\Vect{t}^\intercal(\Vect{t}) \dd_h \Vect{X}_\Vect{t}(\Vect{t}) 
	&=
	\frac{1}{2}
	\left\{
		\Vect{K}_\Vect{t}^\intercal(\Vect{t}) \dd_h \Vect{X}_\Vect{t}(\Vect{t})
		- \Vect{X}_\Vect{t}^\intercal(\Vect{t}) \dd_h \Vect{K}_\Vect{t}(\Vect{t})
		+ \dd_h \left[
			\Vect{K}_\Vect{t}^\intercal(\Vect{t}) \Vect{X}_\Vect{t}(\Vect{t})
		\right]
		\nullFrac
	\right\} 
	\nonumber\\
	&=
	\frac{1}{2}
	\left\{
		- \Stroke{\Vect{Z}}^\intercal_{\Vect{t}}(\Vect{t})
		\,
		\JMat{2N} 
		\,
		\dd_h \Stroke{\Vect{Z}}_{\Vect{t}}(\Vect{t})
		+ \dd_h \left[
			\Vect{K}_\Vect{t}^\intercal(\Vect{t}) \Vect{X}_\Vect{t}(\Vect{t})
		\right]
		\nullFrac
	\right\} 
	\nonumber\\
	&=
	\frac{1}{2}
	\left\{
		- \Vect{z}^\intercal(\Vect{t}) \, \Mat{S}^\intercal_{\Vect{t}} \JMat{2N} \left( \dd_h \Mat{S}_\Vect{t} \right) \, \Vect{z}(\Vect{t})
		-
		\Vect{z}^\intercal(\Vect{t}) \, \Mat{S}^\intercal_{\Vect{t}} \JMat{2N} \Mat{S}_\Vect{t} \, \dd_h \Vect{z}(\Vect{t})
		+ \dd_h \left[ \Vect{K}_\Vect{t}^\intercal(\Vect{t}) \Vect{X}_\Vect{t}(\Vect{t}) \right]
		\nullFrac
	\right\} 
	\nonumber\\
	&=
	\frac{1}{2}
	\left\{
		- \Vect{z}^\intercal(\Vect{t}) \, \Mat{S}^\intercal_{\Vect{t}} \JMat{2N} \left( \dd_h \Mat{S}_\Vect{t} \right) \, \Vect{z}(\Vect{t})
		-
		\Vect{z}^\intercal(\Vect{t}) \, \JMat{2N} \, \dd_h \Vect{z}(\Vect{t})
		+ \dd_h \left[ \Vect{K}_\Vect{t}^\intercal(\Vect{t}) \Vect{X}_\Vect{t}(\Vect{t}) \right]
		\nullFrac
	\right\} 
	.
\end{align}

\noindent Next, note that I can introduce the rectangular matrix and its subsequent QR decomposition
\begin{equation}
	\begin{pmatrix}
		\Mat{A}_\Vect{t}^\intercal \\
		\Mat{B}_\Vect{t}^\intercal
	\end{pmatrix}
	=
	\tilde{\Mat{Q}}_\Vect{t} \tilde{\Mat{R}}_\Vect{t}
	,
	\label{eq:6_ABqr}
\end{equation}

\noindent where $\tilde{\Mat{Q}}_\Vect{t}$ is a rectangular $2N \times N$ orthogonal matrix, and $\tilde{\Mat{R}}_\Vect{t}$ is a square $N \times N$ upper triangular matrix. The symplectic condition given by \Eq{eq:3_symplec2} implies that
\begin{equation}
	\tilde{\Mat{R}}_\Vect{t}^{-1} \tilde{\Mat{Q}}_\Vect{t}^\intercal 
	= \begin{pmatrix}
		\vspace{2mm}\Mat{D}_\Vect{t} & - \Mat{C}_\Vect{t}
	\end{pmatrix}
	.
	\label{eq:6_DCqr}
\end{equation}

\noindent I can then calculate
\begin{align}
	\Tr(\Mat{V}_\Vect{t}) \equiv
	\Tr\left(
		\Mat{D}_\Vect{t} \dd_h \Mat{A}_\Vect{t}^\intercal 
		- \Mat{C}_\Vect{t} \dd_h \Mat{B}_\Vect{t}^\intercal
	\right)
	&=
	\Tr\left[
		\begin{pmatrix}
			\vspace{2mm}\Mat{D}_\Vect{t} & - \Mat{C}_\Vect{t}
		\end{pmatrix}
		\, \dd_h
		\begin{pmatrix}
			\Mat{A}_\Vect{t}^\intercal  \\
			\Mat{B}_\Vect{t}^\intercal 
		\end{pmatrix}
	\right]
	\nonumber\\
	&=
	\Tr\left[
		\tilde{\Mat{R}}_\Vect{t}^{-1} \tilde{\Mat{Q}}_\Vect{t}^\intercal 
		\, \dd_h
		\left( \tilde{\Mat{Q}}_\Vect{t} \tilde{\Mat{R}}_\Vect{t} \right)
	\right]
	\nonumber\\
	&=
	\Tr\left[
		\tilde{\Mat{R}}_\Vect{t}^{-1} \tilde{\Mat{Q}}_\Vect{t}^\intercal 
		\tilde{\Mat{Q}}_\Vect{t} 
		\, \dd_h
		\left(\tilde{\Mat{R}}_\Vect{t} \right)
		+
		\tilde{\Mat{R}}_\Vect{t}^{-1} \tilde{\Mat{Q}}_\Vect{t}^\intercal 
		\, \dd_h
		\left( \tilde{\Mat{Q}}_\Vect{t} \right) \tilde{\Mat{R}}_\Vect{t}
	\right]
	\nonumber\\
	&=
	\Tr\left(
		\tilde{\Mat{R}}_\Vect{t}^{-1}
		\, \dd_h
		\tilde{\Mat{R}}_\Vect{t}
	\right)
	+ \Tr\left[
		\tilde{\Mat{R}}_\Vect{t}^{-1} \tilde{\Mat{Q}}_\Vect{t}^\intercal 
		\, \dd_h
		\left( \tilde{\Mat{Q}}_\Vect{t} \right) \tilde{\Mat{R}}_\Vect{t}
	\right]
	\nonumber\\
	&=
	\Tr\left(
		\tilde{\Mat{R}}_\Vect{t}^{-1}
		\, \dd_h
		\tilde{\Mat{R}}_\Vect{t}
	\right)
	+ \Tr\left(
		\tilde{\Mat{Q}}_\Vect{t}^\intercal 
		\, \dd_h
		\tilde{\Mat{Q}}_\Vect{t}
	\right)
	\nonumber\\
	&=
	\Tr\left(
		\tilde{\Mat{R}}_\Vect{t}^{-1}
		\, \dd_h
		\tilde{\Mat{R}}_\Vect{t}
	\right)
	,
\end{align}

\noindent where I have used the linearity and cyclic properties of the matrix trace, and in the final line I have used the fact that $\tilde{\Mat{Q}}_\Vect{t}^\intercal \, \dd_h \tilde{\Mat{Q}}_\Vect{t}$ is antisymmetric and thereby traceless as a direct consequence of the orthogonality relation:
\begin{equation}
	\tilde{\Mat{Q}}_\Vect{t}^\intercal \tilde{\Mat{Q}}_\Vect{t}
	= \IMat{N}
	, \quad \to \quad
	\tilde{\Mat{Q}}_\Vect{t}^\intercal \dd_h \tilde{\Mat{Q}}_\Vect{t}
	= - \left( \tilde{\Mat{Q}}_\Vect{t}^\intercal \dd_h \tilde{\Mat{Q}}_\Vect{t} \right)^\intercal
	.
\end{equation}

\noindent Hence, using Jacobi's formula for the determinant of a square invertible matrix $\Mat{M}$:
\begin{equation}
	\dd_h \log \det \Mat{M} = \Tr \left( \Mat{M}^{-1} \dd_h \Mat{M} \right)
	,
	\label{eq:6_jacobiDET}
\end{equation}

\noindent one obtains the simplification%
\footnote{As demonstrated by this calculation along with discussion presented in \Ch{ch:MT}, $\Tr(\Mat{V}_\Vect{t})$ represents the deviation of $\Mat{S}_\Vect{t}$ from orthogonality. Said differently, it represents the deviation from orthogonality of the tangent-space basis of the new coordinates $\Vect{X}_\Vect{t}$, since $( \Mat{A}_\Vect{t} \quad \Mat{B}_\Vect{t})^\intercal$ is the basis generated by $\Mat{S}_\Vect{t}$ for the tangent space spanned by $\pd{\Vect{\tau}} \Vect{z}(\Vect{t})$.}%
\begin{equation}
	\Tr(\Mat{V}_\Vect{t}) = \dd_h \log \det \tilde{\Mat{R}}_\Vect{t}
	.
\end{equation}

\noindent Lastly, since
\begin{align}
	\begin{pmatrix}
		\dd_h \Vect{X}_\Vect{t}(\Vect{t}) 
		- \Mat{V}_\Vect{t}^\intercal \Vect{X}_\Vect{t}(\Vect{t}) 
		- \Mat{W}_\Vect{t} \Vect{K}_\Vect{t}(\Vect{t}) \\
		\dd_h \Vect{K}_\Vect{t}(\Vect{t}) 
		+ \Mat{U}_\Vect{t} \Vect{X}_\Vect{t}(\Vect{t}) 
		+ \Mat{V}_\Vect{t} \Vect{K}_\Vect{t}(\Vect{t})
	\end{pmatrix}
	&=
	\dd_h \Stroke{\Vect{Z}}_\Vect{t}(\Vect{t})
	- \left( \dd_h \Mat{S}_\Vect{t} \right) \Mat{S}^{-1} 
	\Stroke{\Vect{Z}}_\Vect{t}(\Vect{t})
	\nonumber\\
	&=
	\left( \dd_h \Mat{S}_\Vect{t} \right)
	\Vect{z}(\Vect{t})
	+
	\Mat{S}_\Vect{t}
	\dd_h \Vect{z}(\Vect{t})
	- \left( \dd_h \Mat{S}_\Vect{t} \right)
	\Vect{z}(\Vect{t})
	\nonumber\\
	&=
	\Mat{S}_\Vect{t}
	\dd_h \Vect{z}(\Vect{t})
	,
\end{align}

\noindent one can simplify 
\begin{equation}
	\left[\dd_h \Vect{X}_\Vect{t}(\Vect{t}) 
		- \Mat{V}_\Vect{t}^\intercal \Vect{X}_\Vect{t}(\Vect{t}) 
		- \Mat{W}_\Vect{t} \Vect{K}_\Vect{t}(\Vect{t}) 
		\nullFrac
	\right]^\intercal
	\pd{\Vect{X}_\Vect{t}} \Phi_\Vect{t}\left[
		\Vect{X}_\Vect{t}(\Vect{t})
	\right]
	=
	\left[
		\Mat{A}_\Vect{t} \dd_h \Vect{x}(\Vect{t})
		+ \Mat{B}_\Vect{t} \dd_h \Vect{k}(\Vect{t})
	\right]^\intercal
	\pd{\Vect{X}_\Vect{t}} \Phi_\Vect{t}\left[
		\Vect{X}_\Vect{t}(\Vect{t})
	\right]
	.
\end{equation}

\noindent Hence,
\begin{subequations}
	\label{eq:6_alphaSIMPLE}
	\begin{align}
		\alpha_\Vect{t}
		&=
		\frac{\tilde{c}_0(\Vect{t}_\perp)}{ \sqrt{ \det \tilde{\Mat{R}}_\Vect{t} } }
		\exp\left[
			\frac{i}{2} \Vect{K}_\Vect{t}^\intercal(\Vect{t}) \Vect{X}_\Vect{t}(\Vect{t})
		\right]
		\exp\left[
			\int_0^{t_1} \dd \xi \, \tilde{\eta}_{(\xi, \Vect{t}_\perp)}
		\right]
		, \\
		\tilde{\eta}_\Vect{t}
		&\doteq
		- \frac{i}{2} \Vect{z}^\intercal(\Vect{t}) \, \JMat{2N} \, \dot{\Vect{z}}(\Vect{t})
		+ \left[
			\Mat{A}_\Vect{t} \dot{\Vect{x}}(\Vect{t})
			+ \Mat{B}_\Vect{t} \dot{\Vect{k}}(\Vect{t})
		\right]^\intercal
		\pd{\Vect{X}} \Phi_\Vect{t}\left[
			\Vect{X}_\Vect{t}(\Vect{t})
		\right]
		.
	\end{align}
\end{subequations}

\noindent where $\tilde{c}_0(\Vect{t}_\perp) \doteq \alpha_{\left(0,\Vect{t}_\perp \right)} \exp\left[ - \frac{i}{2} \Vect{K}_{(0, \Vect{t}_\perp)}^\intercal(0, \Vect{t}_\perp) \Vect{X}_{(0, \Vect{t}_\perp)}(0, \Vect{t}_\perp) \right] \sqrt{\det \tilde{\Mat{R}}_{(0, \Vect{t}_\perp)} }$ is another arbitrary function.

If the lowest GO model is used, \ie if $\Phi_\Vect{t}$ has the formal solution given by \Eq{eq:6_MGOphi}, then one final simplification can be made to \Eq{eq:6_alphaSIMPLE}. Indeed, since
\begin{equation}
	\Mat{A}_\Vect{t} \dot{\Vect{x}}(\Vect{t})
	+ \Mat{B}_\Vect{t} \dot{\Vect{k}}(\Vect{t})
	=
	\Vect{V}_\Vect{t}(\Vect{t})
\end{equation}

\noindent (with $\Vect{V}_\Vect{t}$ being the tangent-space group velocity, not to be confused with the Hamiltonian submatrix $\Mat{V}_\Vect{t}$), then \Eq{eq:6_GOenvMET} implies
\begin{equation}
	\left[
		\Mat{A}_\Vect{t} \dot{\Vect{x}}(\Vect{t})
		+ \Mat{B}_\Vect{t} \dot{\Vect{k}}(\Vect{t})
	\right]^\intercal
	\pd{\Vect{X}} \Phi_\Vect{t}\left[
		\Vect{X}_\Vect{t}(\Vect{t})
	\right]
	\equiv
	\Vect{V}_\Vect{t}^\intercal (\Vect{t})
	\pd{\Vect{X}} \Phi_\Vect{t}\left[
		\Vect{X}_\Vect{t}(\Vect{t})
	\right]
	=
	\left.
		- \frac{1}{2} \pd{\tau_1} \left[ \log \det \pd{\Vect{\tau}} \Vect{X}_\Vect{t}(\Vect{\tau}) \right]
	\right|_{\Vect{\tau} = \Vect{t}}
	,
\end{equation}

\noindent where I have used the fact that $\Phi_\Vect{t}[\Vect{X}_\Vect{t}(\Vect{t})] = 1$ by my normalization convention. Next, I note that
\begin{equation}
	\pd{\Vect{\tau}} \Vect{X}_\Vect{t}(\Vect{\tau})
	\equiv
	\Mat{A}_\Vect{t} 
	\pd{\Vect{\tau}} \Vect{x}(\Vect{\tau})
	+
	\Mat{B}_\Vect{t} 
	\pd{\Vect{\tau}} \Vect{k}(\Vect{\tau})
	=
	\begin{pmatrix}
		\vspace{2mm}\Mat{A}_\Vect{t} & \Mat{B}_\Vect{t}
	\end{pmatrix}
	\pd{\Vect{\tau}} \Vect{z}(\Vect{\tau})
	.
\end{equation}

\noindent Let me choose to define $\Mat{D}_\Vect{t}$ and $\Mat{C}_\Vect{t}$ such that
\begin{equation}
	\pd{\Vect{\tau}} \Vect{z}(\Vect{\tau})
	=
	\begin{pmatrix}
		\Mat{D}_\Vect{\tau}^\intercal \\
		- \Mat{C}_\Vect{\tau}^\intercal
	\end{pmatrix}
	\Mat{M}_\Vect{\tau}
	\label{eq:6_zTAUdecomp}
\end{equation}

\noindent for some square $N \times N$ invertible matrix $\Mat{M}_\Vect{\tau}$. This expression simply states that $\Mat{D}$ and $\Mat{C}$ span the same vector space as $\pd{\Vect{\tau}} \Vect{z}(\Vect{\tau})$, subject to the change of basis governed by $\Mat{M}_\Vect{\tau}$.%
\footnote{$\Mat{M}_\Vect{\tau}$ can be considered a sort of free parameter in MGO, since it represents the freedom with which one can choose a basis; for example, one can always rotate a basis about some angle within the spanning plane to obtain a new set of basis vectors, which would be effected in \Eq{eq:6_zTAUdecomp} as a change of $\Mat{M}_\Vect{t}$. Imposing additional constraints, \eg requiring $\Mat{S}_\Vect{t}$ to be orthosymplectic, can uniquely determine $\Mat{M}_\Vect{t}$, as I shall show in \Sec{sec:6_orthoMGO} and further discuss in \App{app:6_QR}} %
(Note that basis vectors transform inversely to coordinates; hence if phase-space coordinates are transformed by $\Mat{S}$, then basis vectors are transformed by $\Mat{S}^{-1}$.) Then,
\begin{align}
	\pd{\Vect{\tau}} \Vect{X}_\Vect{t}(\Vect{\tau})
	=
	\begin{pmatrix}
		\vspace{2mm}\Mat{A}_\Vect{t} & \Mat{B}_\Vect{t}
	\end{pmatrix}
	\begin{pmatrix}
		\Mat{D}_\Vect{\tau}^\intercal \\
		- \Mat{C}_\Vect{\tau}^\intercal
	\end{pmatrix}
	\Mat{M}_\Vect{\tau}
	\equiv
	\tilde{\Mat{R}}_\Vect{t}^{\intercal} \tilde{\Mat{Q}}_\Vect{t}^{\intercal}
	\tilde{\Mat{Q}}_\Vect{\tau} \tilde{\Mat{R}}_\Vect{\tau}^{-\intercal}
	\Mat{M}_\Vect{\tau}
	,
\end{align}

\noindent where I have used both \Eq{eq:6_ABqr} and \Eq{eq:6_DCqr}. Similar to the previous calculation, I invoke Jacobi's formula \eq{eq:6_jacobiDET} to conclude
\begin{align}
	\left.
		\dd_h \left[ \log \det \pd{\Vect{\tau}} \Vect{X}_\Vect{t}(\Vect{\tau}) \right]
	\right|_{\Vect{\tau} = \Vect{t}}
	&=
	\left.
		\Tr\left[
			\Mat{M}_\Vect{\tau}^{-1}
			\tilde{\Mat{R}}_\Vect{\tau}^{\intercal}
			\tilde{\Mat{Q}}_\Vect{\tau}^\intercal
			\tilde{\Mat{Q}}_\Vect{t}
			\tilde{\Mat{R}}_\Vect{t}^{-\intercal} 
			\,
			\dd_h 
			\left(
				\tilde{\Mat{R}}_\Vect{t}^{\intercal} \tilde{\Mat{Q}}_\Vect{t}^{\intercal}
				\tilde{\Mat{Q}}_\Vect{\tau} \tilde{\Mat{R}}_\Vect{\tau}^{-\intercal}
				\Mat{M}_\Vect{\tau}
			\right)
		\right]
	\right|_{\Vect{\tau} = \Vect{t}}
	\nonumber\\
	&=
	\left.
		\Tr\left[
			\Mat{M}_\Vect{t}^{-1}
			\tilde{\Mat{R}}_\Vect{t}^{\intercal} \tilde{\Mat{Q}}_\Vect{t}^{\intercal}
			\,
			\dd_h 
			\left(
				\tilde{\Mat{Q}}_\Vect{\tau} \tilde{\Mat{R}}_\Vect{\tau}^{-\intercal}
				\Mat{M}_\Vect{\tau}
			\right)
		\right]
	\right|_{\Vect{\tau} = \Vect{t}}
	\nonumber\\
	&=
	\left.
		\Tr\left[
			\Mat{M}_\Vect{t}^{-1}
			\tilde{\Mat{R}}_\Vect{t}^{\intercal} \tilde{\Mat{Q}}_\Vect{t}^{\intercal}
			\,
			\dd_h 
			\left(
				\tilde{\Mat{Q}}_\Vect{\tau}
			\right)
			\tilde{\Mat{R}}_\Vect{t}^{-\intercal}
			\Mat{M}_\Vect{t}
			+
			\Mat{M}_\Vect{t}^{-1}
			\tilde{\Mat{R}}_\Vect{t}^{\intercal}
			\dd_h 
			\left(
				\tilde{\Mat{R}}_\Vect{\tau}^{-\intercal}
			\right)
			\Mat{M}_\Vect{t}
			+
			\Mat{M}_\Vect{t}^{-1}
			\dd_h 
			\left(
				\Mat{M}_\Vect{\tau}
			\right)
		\right]
	\right|_{\Vect{\tau} = \Vect{t}}
	\nonumber\\
	&=
	\left.
		\Tr\left[
			\tilde{\Mat{R}}_\Vect{t}^{\intercal}
			\dd_h 
			\left(
				\tilde{\Mat{R}}_\Vect{\tau}^{-\intercal}
			\right)
		\right]
		+ \Tr\left[
			\Mat{M}_\Vect{t}^{-1}
			\dd_h 
			\left(
				\Mat{M}_\Vect{\tau}
			\right)
		\right]
	\right|_{\Vect{\tau} = \Vect{t}}
	\nonumber\\
	&=
	\dd_h \left[ 
		\log \det \tilde{\Mat{R}}_\Vect{t}^{-1}
		+
		\log \det \Mat{M}_\Vect{t}
	\right]
	.
\end{align}

\noindent Hence, I compute
\begin{align}
	\alpha_\Vect{t}
	=
	\frac{c_0(\Vect{t}_\perp)}{ \sqrt{ \det \Mat{M}_\Vect{t} } }
	\exp\left[
		\frac{i}{2} \Vect{K}_\Vect{t}^\intercal(\Vect{t}) \Vect{X}_\Vect{t}(\Vect{t})
	\right]
	\exp\left[
		- \frac{i}{2} \int_0^{t_1} \dd \xi \, \Vect{z}^\intercal(\xi, \Vect{t}_\perp) \, \JMat{2N} \, \dot{\Vect{z}}(\xi, \Vect{t}_\perp)
	\right]
	,
	\label{eq:6_alphaFINAL}
\end{align}

\noindent where $c_0(\Vect{t}_\perp) \doteq \alpha_{\left(0,\Vect{t}_\perp \right)} \exp\left[ - \frac{i}{2} \Vect{K}_{(0, \Vect{t}_\perp)}^\intercal(0, \Vect{t}_\perp) \Vect{X}_{(0, \Vect{t}_\perp)}(0, \Vect{t}_\perp) \right] \sqrt{\det \Mat{M}_{(0, \Vect{t}_\perp)} }$ is an arbitrary function.


\subsection{Projecting tangent-space GO solution to configuration space}

Having obtained the GO field in the optimal tangent-plane representation, and having linked all the tangent-plane representations together to obtain a continuous tangent-space GO solution, the final step in MGO is to project this solution back to the original $\Vect{x}$-space coordinates and sum over all branches of the dispersion manifold. Naively, one might attempt to calculate something of the form
\begin{equation}
	\alpha_\Vect{t} \int \dd \Vect{X}_\Vect{t} \,
	U_\Vect{t}^{-1}(\Vect{x}, \Vect{X}_\Vect{t}) \Psi_\Vect{t} (\Vect{X}_\Vect{t})
	,
	\label{eq:6_naiveINT}
\end{equation}

\noindent where $U_\Vect{t}^{-1}(\Vect{x}, \Vect{X})$ is the inverse MT kernel for $\Mat{S}_\Vect{t}$ given either by \Eq{eq:3_invMTint} when $\Mat{B}_\Vect{t}$ is invertible, or by \Eq{eq:3_invMETkernB} when $\Mat{B}_\Vect{t}$ is not invertible. There is a subtlety, however. An integral of the form given by \Eq{eq:6_naiveINT} will have contributions from all points $\Vect{X}_\Vect{t}$ that satisfy the stationary phase condition
\begin{equation}
	\Mat{D}_\Vect{t} \Vect{X}_\Vect{t} - \Mat{B}_\Vect{t}^\intercal \pd{\Vect{X}_\Vect{t}} \Theta(\Vect{X}_\Vect{t}) = \Vect{x}
	.
\end{equation} 

\noindent One of these roots, namely, $\Vect{X}_\Vect{t} = \Vect{X}_\Vect{t}(\Vect{t})$, is the desired ray contribution that the tangent-space representation has been optimized to obtain. The other roots (that lie elsewhere on the dispersion manifold) are `spurious' and undesired - they might still be caustics of $\Psi_\Vect{t}$ and must therefore be filtered out somehow. 

At this point, semiclassical methods like GO and Maslov's method traditionally evaluate \Eq{eq:6_naiveINT} using the stationary-phase approximation (SPA) about the ray contribution $\Vect{X}_\Vect{t} = \Vect{X}_\Vect{t}(\Vect{t})$~\cite{Heller77a}. However, the SPA fails when saddlepoints are close together~\cite{Chester57}, as occurs near caustics. To remedy this, note that under fairly general conditions, integrals like \Eq{eq:6_naiveINT} can be evaluated on the union of steepest-descent contours through some subset of saddlepoints in complex $\Vect{X}_\Vect{t}$-space~\cite{Bleistein86}. By integrating \Eq{eq:6_naiveINT} only along the steepest-descent contour through $\Vect{X}_\Vect{t} = \Vect{X}_\Vect{t}(\Vect{t})$ rather than the entire set, I can isolate the desired ray contribution in a manner that is asymptotically equivalent to the SPA but is also well-behaved at caustics. (In this regard, I can also define the `saddlepoint contribution' to an integral as the result of integrating along the corresponding steepest-descent contour.) Hence, the variable shift $\Vect{\epsilon} \doteq \Vect{X}_\Vect{t} - \Vect{X}_\Vect{t}(\Vect{t})$ yields
\begin{equation}
	\psi_\Vect{t}(\Vect{x})
	=
	\alpha_\Vect{t} \int_{\cont{0}} \dd \Vect{\epsilon} \,
	U_\Vect{t}^{-1}\left[
		\Vect{x}, \Vect{\epsilon} + \Vect{X}_\Vect{t}(\Vect{t}) 
	\right] \Psi_\Vect{t} \left[
		\Vect{\epsilon} + \Vect{X}_\Vect{t}(\Vect{t}) 
	\right]
	,
\end{equation}

\noindent where $\psi_\Vect{t}(\Vect{x})$ denotes the tangent-space field contribution from $\Vect{t}$ on the dispersion manifold projected back to the original $\Vect{x}$-space coordinates, and $\cont{0}$ is the steepest-descent contour through $\Vect{\epsilon} = \Vect{0}$. Note that no further approximations to this integral can be made without additional knowledge of the caustic structure - I shall discuss some of these additional approximations in \Ch{ch:Ex}.

Having isolated the field contribution from a single branch of the dispersion manifold, the final step is to sum over all such contributions to obtain the final MGO expression
\begin{equation}
	\psi(\Vect{x}) = \sum_{ \Vect{t} \in \Vect{\tau}(\Vect{x}) }
	\psi_\Vect{t}
	.
	\label{eq:6_MGOsol}
\end{equation}

\noindent As shown in \App{app:6_simple}, using the specific forms of $U_\Vect{t}^{-1}$ and $\alpha_\Vect{t}$ given by \Eq{eq:6_alphaFINAL} yields the following two simplifications: when $\det \Mat{B}_\Vect{t} \neq 0$ $\psi_\Vect{t}$ takes the form
\begin{align}
	\psi_\Vect{t}
	&=
	\frac{
		\sigma_\Vect{t} \,
		f(\Vect{t}_\perp)
	}{ 
		(- 2\pi i)^{\frac{N}{2}}
		\sqrt{ 
			\det \Mat{B}_\Vect{t} \,
			\det \Mat{M}_\Vect{t} 
		} 
	}
	\exp\left[
		i \int_0^{t_1} \dd \xi \, \Vect{k}^\intercal( \xi, \Vect{t}_\perp) \dot{\Vect{x}}(\xi, \Vect{t}_\perp)
	\right]
	\nonumber\\
	&\hspace{4mm} \times
	 \int_{\cont{0}} \dd \Vect{\epsilon} \,
	\exp\left[
		- \frac{i}{2} \Vect{\epsilon}^\intercal \Mat{D} \Mat{B}^{-1} \Vect{\epsilon}
		- i \Vect{\epsilon}^\intercal \Vect{K}_\Vect{t}(\Vect{t})
	\right]
	\Psi_\Vect{t} \left[
		\Vect{\epsilon} + \Vect{X}_\Vect{t}(\Vect{t}) 
	\right] 
	,
	\label{eq:6_MGOb}
\end{align}

\noindent while when $\det \Mat{B}_\Vect{t} = 0$ $\psi_\Vect{t}$ takes the form
\begin{align}
	\psi_\Vect{t}
	&=
	\frac{
		\sigma_\Vect{t} \,
		f(\Vect{t}_\perp)
		\exp\left[
			i \int_0^{t_1} \dd \xi \, \Vect{k}^\intercal(\xi, \Vect{t}_\perp) \dot{\Vect{x}}(\xi, \Vect{t}_\perp)
		\right]
	}{ 
		(-2\pi i)^{\rho/2}
		\sqrt
		{
			\det \Mat{\Lambda}_{\rho \rho} \,
			\det \Mat{a}_{\varsigma \varsigma}^{-1} \,
			\det \Mat{M}_\Vect{t}
		}
	}
	\nonumber\\
	&\hspace{4mm} \times
	\int_{\cont{0}} \dd \Vect{\epsilon}_\rho \,
	\exp
	\left\{
		- \frac{i}{2} \Vect{\epsilon}_\rho^\intercal \, \Mat{d}_{\rho \rho} \Mat{\Lambda}_{\rho \rho}^{-1} \, \Vect{\epsilon}_\rho
		- i \Vect{\epsilon}_\rho^\intercal \Vect{K}_\Vect{t}^\rho(\Vect{t})
	\right\}
	\, 
	\Psi_\Vect{t}
	\left[
		\Mat{L}_\textrm{s}
		\begin{pmatrix}
			\Vect{X}_\Vect{t}^\rho(\Vect{t}) + \Vect{\epsilon}_\rho \\
			\Mat{a}_{\varsigma \varsigma} \Vect{x}_\varsigma(\Vect{t})
		\end{pmatrix}
	\right]
	,
	\label{eq:6_MGObMOD}
\end{align}

\noindent where $\sigma_\Vect{t} \doteq \pm 1$ is the overall sign that remains constant unless $\det{\Mat{B}_\Vect{t}}$ crosses the branch cut for the MT. Then, $\sigma_\Vect{t}$ changes sign to ensure that $\psi_\Vect{t}$ evolves smoothly in $\Vect{t}$. Also, $f(\Vect{t}_\perp)$ is an arbitrary function that is fixed by initial conditions.


\subsection{MGO for orthosymplectic transformations}
\label{sec:6_orthoMGO}

Although I have just shown that MGO can be formulated for general symplectic matrix $\Mat{S}$, it practice it is preferable to impose the additional constraint that $\Mat{S}$ be orthosymplectic. This is because, as shown in \App{app:6_QR}, an orthosymplectic construction of $\Mat{S}$ can be done explicitly using only the local ray information $\pd{\Vect{\tau}} \Vect{z}(\Vect{t})$. It is straightforward to impose the orthosymplecticity of $\Mat{S}$ to \Eqs{eq:6_MGOb} and \eq{eq:6_MGObMOD}; the results are for $\det \Mat{B}_\Vect{t} \neq 0$,
\begin{align}
	\psi_\Vect{t}
	&=
	\frac{
		\sigma_\Vect{t} \,
		f(\Vect{t}_\perp)
	}{ 
		(- 2\pi i)^{\frac{N}{2}}
		\sqrt{ 
			\det \Mat{B}_\Vect{t} \,
			\det \Mat{R}_\Vect{t} 
		} 
	}
	\exp\left[
		i \int_0^{t_1} \dd \xi \, \Vect{k}^\intercal( \xi, \Vect{t}_\perp) \dot{\Vect{x}}(\xi, \Vect{t}_\perp)
	\right]
	\nonumber\\
	&\hspace{4mm} \times
	 \int_{\cont{0}} \dd \Vect{\epsilon} \,
	\exp\left[
		- \frac{i}{2} \Vect{\epsilon}^\intercal \Mat{A}_\Vect{t} \Mat{B}_\Vect{t}^{-1}  \Vect{\epsilon}
		- i \Vect{\epsilon}^\intercal \Vect{K}_\Vect{t}(\Vect{t})
	\right]
	\Psi_\Vect{t} \left[
		\Vect{\epsilon} + \Vect{X}_\Vect{t}(\Vect{t}) 
	\right] 
	,
	\label{eq:6_MGOb_ortho}
\end{align}

\noindent and for $\det \Mat{B}_\Vect{t} = 0$, 
\begin{align}
	\psi_\Vect{t}
	&=
	\frac{
		\sigma_\Vect{t} \,
		f(\Vect{t}_\perp)
		\exp\left[
			i \int_0^{t_1} \dd \xi \, \Vect{k}^\intercal(\xi, \Vect{t}_\perp) \dot{\Vect{x}}(\xi, \Vect{t}_\perp)
		\right]
	}{ 
		(-2\pi i)^{\rho/2}
		\sqrt
		{
			\det \Mat{\Lambda}_{\rho \rho} \,
			\det \Mat{a}_{\varsigma \varsigma}^{-1} \,
			\det \Mat{R}_\Vect{t}
		}
	}
	\nonumber\\
	&\hspace{4mm} \times
	\int_{\cont{0}} \dd \Vect{\epsilon}_\rho \,
	\exp
	\left\{
		- \frac{i}{2} \Vect{\epsilon}_\rho^\intercal \, \Mat{a}_{\rho \rho} \Mat{\Lambda}_{\rho \rho}^{-1} \, \Vect{\epsilon}_\rho
		- i \Vect{\epsilon}_\rho^\intercal \Vect{K}_\Vect{t}^\rho(\Vect{t})
	\right\}
	\, 
	\Psi_\Vect{t}
	\left[
		\Mat{L}_\textrm{s}
		\begin{pmatrix}
			\Vect{X}_\Vect{t}^\rho(\Vect{t}) + \Vect{\epsilon}_\rho \\
			\Mat{a}_{\varsigma \varsigma} \Vect{x}_\varsigma(\Vect{t})
		\end{pmatrix}
	\right]
	,
	\label{eq:6_MGObMOD_ortho}
\end{align}

\noindent where $\Mat{R}_\Vect{t}$ is the $N \times N$ square upper triangular matrix that results from a QR decomposition of $\pd{\Vect{\tau}} \Vect{z}(\Vect{t})$. Also, the orthosymplectic condition implies that $\det \Mat{a}_{\varsigma \varsigma} = \pm 1$.


\section{Comparing MGO with other known methods}

\subsection{Comparison with standard GO}

In \Ch{ch:Ex}, it will be shown numerically in a series of examples that the MGO formula \eq{eq:6_MGOsol} remains finite at caustics and agrees with the standard GO formula \eq{eq:2_GOsol} away from caustics. The fact that \eq{eq:6_MGOsol} remains finite at caustics follows immediately from the observation that all terms in \Eqs{eq:6_MGOsol}--\eq{eq:6_MGObMOD_ortho} remain finite at caustics by construction; however, the fact that MGO reproduces GO away from caustics is less immediately obvious. In this section, I shall prove this property explicitly.

Let me first consider when $\Mat{B}_\Vect{t}$ is invertible. Assuming that $\Vect{x}$ is located far from a caustic, the steepest-descent integral can be evaluated by the standard (quadratic) saddlepoint method as follows:%
\footnote{Recall that the quadratic saddlepoint method is invalid at caustics; at caustics, higher-order terms in the integrand phase must be retained since the quadratic term vanishes by definition.}
\begin{align}
	\int_{\cont{0}} \dd \Vect{\epsilon} \,
	\exp\left[
		- \frac{i}{2} \Vect{\epsilon}^\intercal \Mat{D} \Mat{B}^{-1} \Vect{\epsilon}
		- i \Vect{\epsilon}^\intercal \Vect{K}_\Vect{t}(\Vect{t})
	\right]
	\Psi_\Vect{t} \left[
		\Vect{\epsilon} + \Vect{X}_\Vect{t}(\Vect{t}) 
	\right] 
	&=
	\int_{\cont{0}} \dd \Vect{\epsilon} \,
	\exp\left( 
		- \frac{i}{2} \Vect{\epsilon}^\intercal \Mat{D}_\Vect{t} \Mat{B}_\Vect{t}^{-1} \Vect{\epsilon}
	\right)
	\nonumber\\
	&=
	\frac{(-2 \pi i)^{N/2} \sqrt{\det \Mat{B}_\Vect{t} }}{ \sqrt{\det{\Mat{D}_\Vect{t}}} }
	,
	\label{eq:6_UpsilonSADDLE}
\end{align}

\noindent where I have used the consistent normalization $\Psi_\Vect{t}\left[  \Vect{X}_\Vect{t}(\Vect{t})\right] = 1$ along with the fact that the second-order term vanishes, \ie $\pd{\Vect{X}} \Vect{K}_\Vect{t}\left[ \Vect{X}_\Vect{t}(\Vect{t}) \right] = \OMat{N}$, by definition of the tangent plane to a Lagrangian manifold. [Note that any overall sign in \Eq{eq:6_UpsilonSADDLE} that results from branch cuts can be absorbed into the overall sign $\sigma_\Vect{t}$.] Thus, \Eq{eq:6_MGOb} becomes
\begin{equation}
	\psi_\Vect{t}
	=
	\frac{
		\sigma_\Vect{t} \,
		f(\Vect{t}_\perp)
	}{ 
		\sqrt{ 
			\det \Mat{D}_\Vect{t} \,
			\det \Mat{M}_\Vect{t} 
		} 
	}
	\exp\left[
		i \int_0^{t_1} \dd \xi \, \Vect{k}^\intercal( \xi, \Vect{t}_\perp) \dot{\Vect{x}}(\xi, \Vect{t}_\perp)
	\right]
	.
\end{equation}

\noindent Next, note that \Eq{eq:6_zTAUdecomp} implies the relation
\begin{equation}
	\pd{\Vect{t}} \Vect{x}(\Vect{t}) = \Mat{D}_\Vect{t}^\intercal \Mat{M}_\Vect{t}
	.
\end{equation}

\noindent Thus, upon using the fact that
\begin{equation}
	\det{\Mat{D}_\Vect{t}} \det \Mat{M}_\Vect{t}
	= \det{\Mat{D}_\Vect{t}^\intercal } \det \Mat{M}_\Vect{t}
	= \det \Mat{D}_\Vect{t}^\intercal  \Mat{M}_\Vect{t}
	= \det{ \pd{\Vect{t}} \Vect{x}(\Vect{t}) }
	\doteq j(\Vect{t})
	,
\end{equation}

\noindent I obtain the standard GO formula for a single ray contribution:
\begin{equation}
	\psi_\Vect{t}
	=
	\frac{
		f(\Vect{t}_\perp)
	}{ 
		\sqrt{ 
			j(\Vect{t})
		} 
	}
	\exp\left[
		i \int_0^{t_1} \dd \xi \, \Vect{k}^\intercal( \xi, \Vect{t}_\perp) \dot{\Vect{x}}(\xi, \Vect{t}_\perp)
	\right]
	.
\end{equation}

Now let me consider when $\det \Mat{B}_\Vect{t} = 0$. Again, far from caustics the steepest-descent integral can be performed using the standard quadratic saddlepoint method:
\begin{align}
	&\int_{\cont{0}} \dd \Vect{\epsilon}_\rho \,
	\exp
	\left[
		- \frac{i}{2} \Vect{\epsilon}_\rho^\intercal \, \Mat{d}_{\rho \rho} \Mat{\Lambda}_{\rho \rho}^{-1} \, \Vect{\epsilon}_\rho
		- i \Vect{\epsilon}_\rho^\intercal \Vect{K}_\Vect{t}^\rho(\Vect{t})
	\right]
	\, 
	\Psi_\Vect{t}
	\left[
		\Mat{L}_\textrm{s}
		\begin{pmatrix}
			\Vect{X}_\Vect{t}^\rho(\Vect{t}) + \Vect{\epsilon}_\rho \\
			\Mat{a}_{\varsigma \varsigma} \Vect{x}_\varsigma(\Vect{t})
		\end{pmatrix}
	\right]
	\nonumber\\
	&=
	\int_{\cont{0}} \dd \Vect{\epsilon}_\rho \,
	\exp
	\left(
		- \frac{i}{2} \Vect{\epsilon}_\rho^\intercal \, \Mat{d}_{\rho \rho} \Mat{\Lambda}_{\rho \rho}^{-1} \, \Vect{\epsilon}_\rho
	\right)
	\nonumber\\
	&=
	\frac{(-2 \pi i)^{\rho/2} \sqrt{\det \Mat{\Lambda}_{\rho \rho} }}{ \sqrt{\det{\Mat{d}_{\rho \rho}}} }
	.
\end{align}

\noindent Hence, \Eq{eq:6_MGObMOD} becomes
\begin{align}
	\psi_\Vect{t}
	&=
	\frac{
		\sigma_\Vect{t} \,
		f(\Vect{t}_\perp)
		\exp\left[
			i \int_0^{t_1} \dd \xi \, \Vect{k}^\intercal(\xi, \Vect{t}_\perp) \dot{\Vect{x}}(\xi, \Vect{t}_\perp)
		\right]
	}{ 
		\sqrt
		{
			\det{\Mat{d}_{\rho \rho}} \,
			\det \Mat{a}_{\varsigma \varsigma}^{-1} \,
			\det \Mat{M}_\Vect{t}
		}
	}
	.
\end{align}

\noindent Lastly, since
\begin{equation}
	\det \Mat{D} = \det \widetilde{\Mat{D}} \equiv \det{\Mat{d}_{\rho \rho}} \, \det \Mat{a}_{\varsigma \varsigma}^{-1}
	,
\end{equation}

\noindent by the orthogonality of $\Mat{L}_\textrm{s}$ and $\Mat{R}_\textrm{s}$, the same logic as used above again yields the standard GO formula for a single ray contribution:
\begin{equation}
	\psi_\Vect{t}
	=
	\frac{
		f(\Vect{t}_\perp)
	}{ 
		\sqrt{ 
			j(\Vect{t})
		} 
	}
	\exp\left[
		i \int_0^{t_1} \dd \xi \, \Vect{k}^\intercal( \xi, \Vect{t}_\perp) \dot{\Vect{x}}(\xi, \Vect{t}_\perp)
	\right]
	.
\end{equation}

\noindent Note that the arbitrary function $f(\Vect{t}_\perp) $ encodes the initial conditions along a ray. Thus, \Eq{eq:6_MGOsol} is equivalent to the standard GO formula \eq{eq:2_GOsol} when evaluated away from caustics.


\subsection{Comparison with other semiclassical integral expressions}

Let me now relate the MGO method I have just outlined with the related caustic-removal scheme presented in \Ref{Littlejohn85}, which shares the same underlying idea with MGO of using continual phase-space rotations as the rays propagate, and has also been previously related to semiclassical methods based on wavepackets~\cite{Littlejohn86b, Kay94a, Zor96, Madhusoodanan98}. In \Ref{Littlejohn85}, the following $1$-D expression for $\psi$ was obtained~\cite[Eqs.~5 \& 7]{Littlejohn85}:
\begin{align}
	\psi(x)
	&= c 
	\int \dd \tau \,
	\exp\left[
		\frac{i}{2} K_\tau(\tau) X_\tau(\tau)
		- \frac{i}{2} \int_0^\tau \dd \xi \, \Vect{z}(\xi)^\intercal \, \JMat{2} \, \dot{\Vect{z}}(\xi)
	\right]
	\int \dd X \, 
	\frac{
		\delta\left[ X - X_\tau(\tau) \right]
	}{
		\sqrt{-2 \pi i B_\tau}
	}
	\exp\left[
		- i G_\tau(x, X)
	\right]
	\nonumber\\
	&=
	c \int \frac{  \dd \tau }{\sqrt{-2 \pi i B_\tau}} \,
	\exp\left\{
		\frac{i}{2} k(\tau) x
		+ \frac{i}{2} k(\tau) \left[x - x(\tau) \right]
		- \frac{iA_\tau}{2B_\tau} [x - x(\tau)]^2
		- \frac{i}{2}
		\int_0^\tau \dd \tau' \,
		\Vect{z}(\xi)^\intercal \JMat{2} \dot{\Vect{z}}(\xi)
	\right\}
	,
	\label{eq:6_LittlejohnEQ}
\end{align}

\noindent where I have denote the MT phase function as
\begin{equation}
	G_t(x, X) \doteq 
	\frac{D_t}{2B_t } X^2
	- \frac{x X}{B_t}
	+ \frac{A_t}{2B_t} x^2 
	.
\end{equation}

\noindent (Note that I have replaced the original notation with my notation.) Using the well-known delta-function manipulations, I can similarly express the MGO solution [\Eqs{eq:6_MGOsol} and \eq{eq:6_MGOb}] by the integral
\begin{align}
	\psi(x)
	&=
	c \int \dd t \,
	\delta \left[ t - \tau(x) \right]
	\exp
	\left[
		\frac{i}{2} K_t(t) X_t(t)
		+ \int_0^{t_1} \dd \xi \, \tilde{\eta}_{\xi}
	\right]
	\nonumber\\
	&\hspace{7mm}\times
	\int_{\cont{0}} \dd \epsilon \,
	\frac{
		\Psi_\Vect{t}\left[ 
			\epsilon + X_t(t)
		\right]
	}{
		\sqrt{ - 2 \pi i B_t}
	}
	\exp\left\{
		- i G_t[x, \epsilon + X_t(t)]
		\nullFrac
	\right\}
	.
	\label{eq:6_MGOlittlejohn}
\end{align}

\noindent After comparing \Eq{eq:6_MGOlittlejohn} with the top line of \Eq{eq:6_LittlejohnEQ}, it is clear that choosing 
\begin{equation}
	\Psi_\Vect{t}\left[\epsilon + X_t(t)\right] = \delta(\epsilon)
	\label{eq:6_LittlejohnPSI}
\end{equation}

\noindent in \Eq{eq:6_MGOlittlejohn} yields a delta-windowed formulation of \Eq{eq:6_LittlejohnEQ}:
\begin{align}
	\psi(x)
	&=
	c \int \dd \tau \, \frac{ \delta \left[ t - \tau(x) \right]}{\sqrt{-2 \pi i B_\tau}} \,
	\nonumber\\
	&\hspace{10mm}\times
	\exp\left\{
		\frac{i}{2} k(\tau) x
		+ \frac{i}{2} k(\tau) \left[x - x(\tau) \right]
		- \frac{iA_\tau}{2B_\tau} [x - x(\tau)]^2
		- \frac{i}{2}
		\int_0^\tau \dd \tau' \,
		\Vect{z}(\xi)^\intercal \JMat{2} \dot{\Vect{z}}(\xi)
	\right\}
	,
	\label{eq:6_MGOhybrid}
\end{align}

\noindent where to evaluate $\tilde{\eta}$ I have used
\begin{equation}
	\pd{X} \Phi_t\left[X_t(t) \right]
	= \delta'(0) = 0
	.
\end{equation}

Thus, one sees that there are two differences between MGO and the method of \Ref{Littlejohn85}: (i) MGO allows for an arbitrary GO solution in the tangent plane, whereas \Ref{Littlejohn85} assumes the solution is delta-shaped, and (ii) MGO filters out only the dominant contributing rays in the resulting semiclassical integral expression using a delta-shaped window function around the saddlepoints, whereas \Ref{Littlejohn85} allows for all locations to contribute. (Similar statements also hold when comparing MGO and wavepacket methods, since \Ref{Littlejohn85} is a special case.) However, I emphasize that either the pure MGO expression or the pure semiclassical integral expressions of \Refs{Littlejohn85, Littlejohn86b, Kay94a, Zor96, Madhusoodanan98} should be used; hybrid methods such as \Eq{eq:6_MGOhybrid} do not remain finite at caustics (and do not even accurately describe wave propagation in the GO limit), and are introduced here only for illustrative purposes.


\begin{subappendices}

\section{Symplectic covariance of the Weyl symbol}
\label{app:6_metWEYL}

Here I demonstrate the symplectic covariance of the Weyl symbol. Consider some operator $\oper{D}$ with symbol
\begin{equation}
	\Symb{D}(\Vect{z}) = 
	\int \frac{\dd \Vect{z}'}{(2\pi)^N} \, 
	\exp\left[
		i\left(\Vect{z}'\right)^\intercal \JMat{2N} \Vect{z}
	\right] \, 
	\Tr \left\{
		\exp\left[
			-i\left(\Vect{z}'\right)^\intercal \JMat{2N} \VectOp{z}
		\right]
		\oper{D}
		\nullFrac
	\right] 
	.
\end{equation}

\noindent Correspondingly, the symbol of $\oper{M}^\dagger \oper{D} \oper{M}$ is 
\begin{align}
	\Weyl\left[ 
		\oper{M}^\dagger 
		\oper{D}
		\oper{M} 
	\right]
	= \int \frac{\dd \Vect{z}'}{(2\pi)^N} \, 
	\exp\left[
		i\left(\Vect{z}'\right)^\intercal \JMat{2N} \Vect{z}
	\right] \,
	\Tr \left\{
		\exp\left[
			-i\left(\Vect{z}'\right)^\intercal \JMat{2N} \VectOp{z}
		\right]
		\oper{M}^\dagger \oper{D} \oper{M}
		\nullFrac
	\right\} 
	.
\end{align}

\noindent Since $\oper{M}$ is unitary,
\begin{align}
	\Tr \left\{
		\exp\left[
			-i\left(\Vect{z}'\right)^\intercal \JMat{2N} \VectOp{z}
		\right]
		\oper{M}^\dagger \oper{D} \oper{M} 
		\nullFrac
	\right\}
	&= \Tr \left\{
		\oper{M}
		\exp\left[
			-i\left(\Vect{z}'\right)^\intercal \JMat{2N} \VectOp{z}
		\right]
		\oper{M}^\dagger \oper{D} 
		\nullFrac
	\right\} \nonumber \\
	&= \Tr \left\{
		\exp\left[
			-i\left(\Vect{z}'\right)^\intercal \JMat{2N} \oper{M} \VectOp{z} \oper{M}^\dagger
		\right]
		\oper{D} 
		\nullFrac
	\right\} \nonumber \\
	&= \Tr \left\{
		\exp\left[
			-i\left(\Vect{z}'\right)^\intercal \JMat{2N} \Mat{S}^{-1} \VectOp{z}
		\right]
		\oper{D} 
		\nullFrac
	\right\}
	.
\end{align}

\noindent Since $\Mat{S}$ is symplectic,
\begin{equation}
	\left(\Vect{z}'\right)^\intercal \JMat{2N} \Mat{S}^{-1} \VectOp{z}
	= 
	\left(\Vect{z}'\right)^\intercal \Mat{S}^\intercal \Mat{S}^{-\intercal} \JMat{2N} \Mat{S}^{-1} \VectOp{z}
	= \left(\Mat{S} \Vect{z}'\right)^\intercal \JMat{2N} \, \VectOp{z} 
	.
\end{equation}

\noindent Hence, after making the variable substitution $\Vect{\zeta} \doteq \Mat{S}\Vect{z}'$,
\begin{align}
	\Weyl\left[ 
		\oper{M}^\dagger 
		\oper{D}
		\oper{M} 
	\right]
	&= \int \frac{\dd \Vect{\zeta}}{(2\pi)^N} \, 
	\exp\left[
		i\left(\Mat{S}^{-1} \Vect{\zeta}\right)^\intercal \JMat{2N} \Vect{z}
	\right] \, 
	\Tr \left[
		\exp\left(
			-i\Vect{\zeta}^\intercal \JMat{2N} \VectOp{z}
		\right)
		\oper{D}
		\nullFrac
	\right] \nonumber \\
	&= \int \frac{\dd \Vect{\zeta}}{(2\pi)^N} \, 
	\exp\left(
		i\Vect{\zeta}^\intercal 
		\JMat{2N} \,  
		\Mat{S} \, \Vect{z}
	\right) \,  
	\Tr \left[
		\exp\left(
			-i\Vect{\zeta}^\intercal \JMat{2N} \VectOp{z}
		\right)
		\oper{D}
		\nullFrac
	\right]
	,
\end{align}

\noindent where I have used the fact that $\det \Mat{S} = 1$ for any symplectic matrix. I therefore obtain
\begin{equation}
	\Weyl\left[ 
		\oper{M}^\dagger \oper{D} \oper{M} 
	\right]
	= \Symb{D}\left( \Mat{S} \Vect{z} \right)
	= \Symb{D}\left( \Stroke{\Vect{Z}} \right) 
	.
\end{equation}

\noindent Similarly, 
\begin{align}
	\Weyl\left[ 
		\oper{M}
		\oper{M}^\dagger 
		\oper{D} 
		\oper{M} 
		\oper{M}^\dagger
	\right]
	&= \int \frac{\dd \Vect{\zeta}}{(2\pi)^N} \, 
	\exp\left(
		 i\Vect{\zeta}^\intercal \JMat{2N} \, \Stroke{\Vect{Z}} 
	\right) \,
	\Tr \left[
		\exp\left(
			-i\Vect{\zeta}^\intercal \Mat{J} \VectOp{z}
		\right)
		\oper{M}
		\oper{D}
		\oper{M}^\dagger
		\nullFrac
	\right] \nonumber \\
	&= \int \frac{\dd \Vect{z}'}{(2\pi)^N} \, 
	\exp\left[
		 i\left( \Vect{z}'\right)^\intercal \JMat{2N} \Mat{S}^{-1} \, \Stroke{\Vect{Z}} 
	\right] \,
	\Tr \left\{
		\exp\left[
			-i\left(\Vect{z}' \right)^\intercal \JMat{2N} \VectOp{z}
		\right]
		\oper{D}
		\nullFrac
	\right\} 
	.
\end{align}

\noindent Therefore,
\begin{equation}
	\Weyl\left[ 
		\oper{M}
		\oper{M}^\dagger 
		\oper{D} 
		\oper{M} 
		\oper{M}^\dagger
	\right]
	= \Symb{D}\left(\Mat{S}^{-1} \Stroke{\Vect{Z}} \right) = \Symb{D}(\Vect{z}) \, .
\end{equation}


\section{Approximating the envelope equation in the GO limit via the Weyl symbol calculus}
\label{app:6_GO}

Here, I derive \Eq{eq:2_approxENV} from \Eq{eq:2_hilbertENV} using the Weyl symbol calculus. (See also \Refs{Dodin19,McDonald88a}.) To obtain the GO envelope operator, I shall (i) calculate the Weyl symbol of the envelope operator, (ii) approximate the Weyl symbol in the GO limit, and (iii) take the Weyl transform of the GO Weyl symbol.

Using \Eq{eq:2_WWTstar}, one obtains by definition
\begin{equation}
	\Weyl\left\{
		\exp\left[ -i \theta(\VectOp{x}) \right]
		\oper{D}(\VectOp{z})
		\exp\left[ i \theta(\VectOp{x}) \right]
		\nullFrac
	\right\} = 
	\exp\left[ -i \theta(\Vect{x}) \right] \star \Symb{D}(\Vect{z}) \star \exp\left[ i \theta(\Vect{x}) \right] 
	,
\end{equation}

\noindent where $\Symb{D}(\Vect{z}) \doteq \Weyl\left[\oper{D}(\VectOp{z}) \right]$. Since $\exp[i \theta(\Vect{x}) ]$ is independent of $\Vect{k}$, one readily finds using \Eq{eq:2_moyalSERIES} that
\begin{equation}
	\Symb{D}(\Vect{z}) \star \exp\left[ i\theta(\Vect{x}) \right] 
	= 
	\left.
		\sum_{s = 0}^\infty 
		\frac{1}{s!} \left(
			\frac{\pd{\Vect{k}}^\intercal \pd{\Vect{\varsigma}} }
			{2 i \kappa} 
		\right)^s 
		\Symb{D}(\Vect{z}) 
		\exp\left[
			i \kappa \theta(\Vect{\varsigma})
		\right]
	\right|_{\Vect{\varsigma} = \Vect{x}} 
	.
\end{equation}

\noindent Here, $\kappa$ is a dimensionless wavenumber that has been formally introduced to elucidate the GO ordering, where $\kappa \to \infty$. I shall keep terms up to $O(\kappa^{-2})$, rather than $O(\kappa^{-1})$ as is traditionally done, to demonstrate the ease with which `full-wave' effects such as diffraction can be included into reduced wave models.

Let me consider $1$-D for simplicity. (The $N$-D case is analogous.) Using Faa di Bruno's formula~\cite{Comtet74},
\begin{equation}
	\pd{\varsigma}^s \exp\left[ i \kappa \theta(\varsigma) \right] 
	= \exp\left[ i \kappa \theta(\varsigma) \right] 
	\sum_{j = 1}^s (i \kappa)^j B_{s,j}\left[\pd{\varsigma} \theta(\varsigma), \ldots, \pd{\varsigma}^{s - j + 1} \theta(\varsigma) \right] 
	,
\end{equation}

\noindent where $B_{s,j}$ are the incomplete, or partial, Bell polynomials. Some important Bell polynomials are
\begin{subequations}
	\begin{align}
		B_{s,s}(f_1) &= (f_1)^s 
		, \\
		B_{s, s - 1}(f_1, f_2) &= \binom{s}{2} (f_1)^{s-2}f_2 
		, \\
		B_{s, s - 2}(f_1, f_2, f_3) &= \binom{s}{3}(f_1)^{s-3} f_3 + 3\binom{s}{4} (f_1)^{s-4} (f_2)^2 
		,
	\end{align}
\end{subequations}

\noindent where $\binom{n}{m} \doteq \frac{n!}{m!(n-m)!}$. Hence, to $O(\kappa^{-2})$,
\begin{align}
	\Symb{D}(\Vect{z}) \star \exp\left[ i \theta(x) \right] 
	&\approx 
	\exp\left[ i \kappa \theta(x) \right] 
	\nonumber\\
	&\hspace{4mm}\times
	\left\{ 
		\sum_{s = 0}^\infty 
		\frac{ 
			\left[\theta'(x) /2\right]^s
		}{s!} 
		\pd{k}^s \Symb{D}(\Vect{z}) 
		- \frac{
			i \theta''(x)
		}{8\kappa} 
		\sum_{s = 2}^\infty \frac{ \left[ \theta'(x) /2 \right]^{s-2}}{(s-2)!} 
		\pd{k}^s \Symb{D}(\Vect{z}) 
		\right.\nonumber\\
		&\hspace{4mm}\left. 
		- \frac{\theta'''(x)}{48 \kappa^2} 
		\sum_{s = 3}^\infty \frac{ \left[ \theta'(x) /2\right]^{s-3}}{(s-3)!} 
		\pd{k}^s \Symb{D}(\Vect{z}) 
		- \frac{\left[\theta''(x) \right]^2}{128 \kappa^2} 
		\sum_{s = 4}^\infty \frac{ \left[ \theta'(x) /2\right]^{s-4}}{(s-4)!} 
		\pd{k}^s \Symb{D}(\Vect{z}) 
	\right\} 
	,
\end{align}

\noindent where $\theta'(x) \doteq \pd{x} \theta(x)$, since $\theta(x)$ is univariate. Upon shifting the indices back to $s = 0$, I obtain
\begin{align}
	\Symb{D}(\Vect{z}) \star 
	\exp\left[ i\theta(x) \right] 
	&\approx 
	\exp\left[ i \kappa \theta(x) \right] 
	\nonumber\\
	&\hspace{4mm}\times
	\left\{ 
		\sum_{s = 0}^\infty \frac{ \left[\theta'(x) /2\right]^s}{s!} \pd{k}^s \Symb{D}(\Vect{z}) 
		- \frac{i \theta''(x)}{8\kappa} 
		\sum_{s = 0}^\infty \frac{ \left[ \theta'(x) /2\right]^s}{s!} 
		\pd{k}^{s+2} \Symb{D}(\Vect{z}) 
		\right. \nonumber\\
		&\hspace{4mm}\left. - \frac{\theta'''(x)}{48 \kappa^2} 
		\sum_{s = 0}^\infty \frac{ \left[ \theta'(x) /2\right]^s}{s!} 
		\pd{k}^{s+3} \Symb{D}(\Vect{z}) 
		- \frac{\left[\theta''(x) \right]^2}{128 \kappa^2} \sum_{s = 0}^\infty \frac{ \left[ \theta'(x) /2\right]^s}{s!} 
		\pd{k}^{s+4} \Symb{D}(\Vect{z}) 
	\right\} 
	.
\end{align}

\noindent One recognizes these summations as the Taylor expansions of $\Symb{D}$, $\pd{k}^2 \Symb{D}$, $\pd{k}^3 \Symb{D}$, and $\pd{k}^4 \Symb{D}$ about $k + \theta'/2$. Therefore,
\begin{align}
	\Symb{D}(\Vect{z}) \star \exp\left[ i \theta(x) \right] 
	&\approx
	\exp\left[ i \kappa \theta(x) \right] 
	\nonumber\\
	&\hspace{4mm}\times
	\left\{ 
		\Symb{D}\left[x, k + \frac{\theta'(x)}{2} \right] 
		- \frac{i \theta''(x)}{8 \kappa} \pd{k}^2 \Symb{D}\left[x, k + \frac{\theta'(x)}{2} \right] 
		\right. \nonumber\\
		&\hspace{4mm}\left.- \frac{\theta'''(x)}{48 \kappa^2} \pd{k}^3 \Symb{D}\left[x, k + \frac{\theta'(x)}{2} \right] 
		- \frac{\left[\theta''(x) \right]^2}{128 \kappa^2} \pd{k}^4 \Symb{D}\left[x, k + \frac{\theta'(x)}{2} \right] 
	\right\}
	.
\end{align}

A similar calculation will show that
\begin{align}
	\exp \left[-i \theta(x) \right] \star \Symb{D}(\Vect{z}) \star \exp\left[ i \theta(x) \right]
	\approx \Symb{D}\left[x, k + \theta'(x) \right] 
	- \frac{\theta'''(x)}{24 \kappa^2} \pd{k}^3 \Symb{D}\left[x, k + \theta'(x) \right] 
	.
	\label{eq:6_symbTRANS1D}
\end{align}

\noindent In multiple dimensions, \Eq{eq:6_symbTRANS1D} readily generalizes to
\begin{equation}
	\exp\left[-i \theta(\Vect{x})\right] \star \Symb{D}(\Vect{z}) \star \exp\left[ i \theta(\Vect{x}) \right]
	\approx \QoSymb\left[ \Vect{x}, \Vect{k} + \pd{\Vect{x}} \theta(\Vect{x}) \right] 
	,
\end{equation}

\noindent where
\begin{equation}
	\QoSymb(\Vect{z}) \doteq
	\Symb{D}(\Vect{z}) - \frac{1}{24 \kappa^2} \pd{\Vect{x}\Vect{x}\Vect{x} }^3 \theta(\Vect{x}) \, \tripdot \, \pd{\Vect{k}\Vect{k}\Vect{k} }^3 \Symb{D}( \Vect{z} )
\end{equation}

\noindent is the correction to $\Symb{D}(\Vect{z})$ found in \Ref{Dodin19}. Here, $\tripdot$ denotes the triple contraction.

Since $\WeylInv\left[\Vect{k} \right] = \VectOp{k} \sim \kappa^{-1} \pd{\Vect{x}}$ on $\Vect{x}$-space, a power series expansion of $\QoSymb\left[\Vect{x}, \Vect{k} + \pd{\Vect{x}} \theta(\Vect{x}) \right]$ in $\kappa^{-1}$ is equivalent to a power series expansion in $\Vect{k}$. Hence,
\begin{align}
	\QoSymb\left[\Vect{x}, \Vect{k} + \pd{\Vect{x}} \theta(\Vect{x}) \right] 
	&\approx \QoSymb\left[\Vect{x}, \pd{\Vect{x}} \theta (\Vect{k}) \right] 
	+ \frac{1}{\kappa} \Vect{v}(\Vect{x})^\intercal \Vect{k}
	+ \frac{1}{2 \kappa^2} \Vect{k}^\intercal \Mat{m}(\Vect{x}) \Vect{k} 
	,
	\label{eq:6_approxSYMB}
\end{align}

\noindent where I have defined
\begin{align}
	\Vect{v}(\Vect{x}) \doteq \left. 
		\pd{\Vect{k}} \QoSymb( \Vect{z} )
	\right|_{\Vect{k} = \pd{\Vect{x}} \theta (\Vect{x})} 
	, \quad
	\Mat{m}(\Vect{x}) \doteq \left.
		\pd{\Vect{k} \Vect{k}}^2 \QoSymb( \Vect{z} )
	\right|_{\Vect{k} = \pd{\Vect{x}} \theta (\Vect{x})} 
	.
\end{align}

\noindent Finally, taking the WWT of \Eq{eq:6_approxSYMB} using results presented in \Ch{ch:GO} yields the reduced envelope operator
\begin{align}
	\WeylInv\left\{
		\exp\left[ -i\theta(\Vect{x}) \right] 
		\star \Symb{D}(\Vect{z}) \star 
		\exp\left[ i\theta(\Vect{x}) \right] 
		\nullFrac
	\right\}
	&\approx
	\QoSymb\left[ \VectOp{x}, \pd{\Vect{x}} \theta(\VectOp{x}) \right] 
	+ \frac{\Vect{v}(\VectOp{x})^\intercal \VectOp{k} - \frac{i}{2} \pd{\Vect{x}} \cdot \Vect{v}(\VectOp{k}) }
	{\kappa} \nonumber\\
	&\hspace{4mm}
	+ \frac{\Mat{m}(\VectOp{x}) \dubdot \VectOp{k} \VectOp{k} 
	- i \left[ \pd{\Vect{x}} \cdot \Mat{m}(\VectOp{x}) \right]^\intercal \VectOp{k}
	- \frac{1}{4} \pd{\Vect{x}\Vect{x}}^2 \dubdot \Mat{m}(\VectOp{x})}
	{2\kappa^2} \, .
	\label{eq:6_approxOP}
\end{align}

\noindent Only the lowest order GO approximation is considered in \Ch{ch:GO}; hence, \Eq{eq:2_approxENV} is obtained from \Eq{eq:6_approxOP} by dropping all $O(\kappa^{-2})$ terms. Also, $\pd{\Vect{k}\Vect{k}\Vect{k}}^3 \Symb{D}(\Vect{z})$ and $\pd{\Vect{x}\Vect{x}}^2 \dubdot \Mat{m}(\Vect{x})$ are often negligibly small. (Both are identically zero for the Helmholtz equation, for example.) I therefore drop these two terms in obtaining \Eq{eq:6_deltaQO}.


\section{Orthosymplectic construction of $\Mat{S}_\Vect{t}$}
\label{app:6_QR}

Let me now explicitly construct the symplectic matrix that maps $\Vect{x}$-space to the tangent plane of the dispersion manifold at some $\Vect{z}(\Vect{\tau})$. Recall that coordinates and coordinate axes transform oppositely (contravariantly versus covariantly). In other words, if the coordinates are transformed by $\Mat{S}$, then the coordinate axes are transformed by $\Mat{S}^{-1}$; hence, I desire $\Mat{S}^{-1}$ to map $\Vect{x}$-space to the local tangent space, rather than $\Mat{S}$. Let $\{ \unit{\Vect{T}}_j ( \Vect{t} ) \}$ and $\{ \unit{\Vect{N}}_j ( \Vect{t} ) \}$ be a symplectically dual set of $N$ tangent vectors and normal vectors to the dispersion manifold at $\Vect{\tau} = \Vect{t}$, respectively. As can be readily verified, the matrix
\begin{equation}
	\Mat{S}_\Vect{t}^{-1} = 
	\begin{pmatrix}
		\uparrow & & \uparrow 
		& \uparrow & & \uparrow\\
		\unit{\Vect{T}}_1(\Vect{t}) & \ldots & \unit{\Vect{T}}_N(\Vect{t}) 
		& \unit{\Vect{N}}_1(\Vect{t}) & \ldots & \unit{\Vect{N}}_N(\Vect{t})\\
		\downarrow & & \downarrow 
		& \downarrow & & \downarrow
	\end{pmatrix}
\end{equation}

\noindent maps $\Vect{x}$-space to the local tangent space at $\Vect{\tau} = \Vect{t}$. (The arrows emphasize that $\{ \unit{\Vect{T}}_j ( \Vect{t} ) \}$ and $\{ \unit{\Vect{N}}_j ( \Vect{t} ) \}$ form the columns of $\Mat{S}_\Vect{t}^{-1}$.) Indeed, one computes
\begin{equation}
	\begin{pmatrix}
		\uparrow & & \uparrow 
		& \uparrow & & \uparrow\\
		\unit{\Vect{T}}_1(\Vect{t}) & \ldots & \unit{\Vect{T}}_N(\Vect{t}) 
		& \unit{\Vect{N}}_1(\Vect{t}) & \ldots & \unit{\Vect{N}}_N(\Vect{t})\\
		\downarrow & & \downarrow 
		& \downarrow & & \downarrow
	\end{pmatrix}
	\unit{\Vect{e}}_j
	=
	\unit{\Vect{T}}_j(\Vect{t})
	,
	\quad j \in [1, N]
	,
\end{equation}

\noindent where $\unit{\Vect{e}}_j$ is the canonical basis vector whose $j$-th component equals $1$ and all other components equal $0$. Note also that this observation for $\Mat{S}^{-1}$ provides the underlying logic for \Eq{eq:6_zTAUdecomp}, since the tangent plane at $\Vect{t}$ is spanned by the columns of the matrix
\begin{equation}
	\pd{\Vect{\tau}} \Vect{z}(\Vect{t})
	= \begin{pmatrix}
		\uparrow & & \uparrow \\
		\pd{\tau_1} \Vect{z}(\Vect{t}) & \ldots & \pd{\tau_N} \Vect{z}(\Vect{t})\\
		\downarrow & & \downarrow 
	\end{pmatrix}
	.
\end{equation}

It is particularly convenient to impose the $\Mat{S}_\Vect{t}$ be orthosymplectic, since in that case the entirety of $\Mat{S}_\Vect{t}$ is determined by $\pd{\Vect{\tau}} \Vect{z}(\Vect{t})$. (This is because orthosymplectic matrices only require specifying two submatrices $\Mat{A}$ and $\Mat{B}$, rather than four.) For this situation, let me define the matrices $\Mat{A}_\Vect{t}$ and $\Mat{B}_\Vect{t}$ through the QR decomposition of $\pd{\Vect{\tau}} \Vect{z}(\Vect{t})$ as
\begin{equation}
	\pd{\Vect{\tau}} \Vect{z}(\Vect{t})
	=
	\Mat{Q}_\Vect{t} \Mat{R}_\Vect{t}
	, \quad
	\Mat{Q}_\Vect{t}
	=
	\begin{pmatrix}
		\Mat{A}_\Vect{t}^\intercal \\ 
		\Mat{B}_\Vect{t}^\intercal
	\end{pmatrix}
	.
	\label{eq:6_zQR}
\end{equation}

\noindent Note that this is equivalent to imposing the $\Mat{M}_\Vect{t}$ be an upper triangular matrix (specifically, $\Mat{M}_\Vect{t} = \Mat{R}_\Vect{t}$) in the more general construction provided by \Eq{eq:6_zTAUdecomp}. 

Also note that the QR decomposition is unique~\cite{Trefethen97} if the upper-triangular matrix $\Mat{R}_\Vect{t}$ is restricted to having strictly positive diagonal elements; hence, any QR decomposition algorithm can be used to construct $\Mat{S}_\Vect{t}$. One possible choice is to use Gram--Schmidt orthogonalization to obtain the set $\{ \unit{\Vect{T}}_j(\Vect{t}) \}$ from the set $\{ \pd{\tau_j} \Vect{z}(\Vect{t}) \}$. The symplectically dual normal vectors can then be obtained from the tangent vectors as
\begin{equation}
	\unit{\Vect{N}}_j(\Vect{t}) = - \JMat{2N} \unit{\Vect{T}}_j(\Vect{t})
	,
\end{equation}

\noindent as is also apparent from the block-decomposed form of an orthosymplectic matrix (\Ch{ch:MT}). I can verify this construction is indeed symplectic as follows. Orthonormality of $\{ \unit{\Vect{T}}_j(\Vect{t}) \}$ implies that
\begin{equation}
	\left[ \unit{\Vect{N}}_j (\Vect{t}) \right]^\intercal \JMat{2N} \, \unit{\Vect{T}}_{j'} (\Vect{t})
        = - \delta_{j j'} 
        .
\end{equation}

\noindent Moreover, it is a property of Lagrangian manifolds (which the dispersion manifold is) that any set of tangent vectors satisfy~\cite{Arnold89}
\begin{equation}
	\left[ \unit{\Vect{T}}_j (\Vect{t}) \right]^\intercal \JMat{2N} \, \unit{\Vect{T}}_{j'} (\Vect{t}) = 0 
	.
\end{equation}

\noindent This also implies that 
\begin{equation}
	\left[ \unit{\Vect{N}}_j (\Vect{t}) \right]^\intercal \JMat{2N} \, \unit{\Vect{N}}_{j'} (\Vect{t}) = 0
	.
\end{equation}

\noindent Hence, $\Mat{S}_\Vect{t}$ is symplectic as constructed.


\section{Curvature-dependent adaptive discretization}
\label{app:6_adapt}

Since MGO relies on evolving the tangent plane of the ray manifold as the rays propagate, it is desirable to develop a discretization of the rays that naturally congregates in regions where the tangent plane changes quickly, \ie the step size becomes smaller in regions with higher curvature. This would ensure that the angle between neighboring tangent planes is always small even when discretized, as is necessary for the accuracy of MGO.

The procedure to develop adaptive discretizations for Hamiltonian systems is actually well known~\cite{Hairer97}: simply replace the Hamiltonian with a new Hamiltonian possessing the same root structure (so the dispersion relation $\Symb{D} = 0$ is the same, and hence, so too are the generated rays) but different gradients (since ray velocities are set by $\pd{\Vect{z}} \Symb{D}$). In this manner, adaptive integration schemes can be developed which are self-supervising (do not need to be error-controlled in the traditional sense~\cite{Press07}), and amenable to symplectic methods of numerical integration~\cite{Richardson12,Hairer97}. Note that in other plasma contexts, this method of adaptive discretization has been useful in developing ray equations that naturally slow down as they approach mode-conversion regions~\cite{Tracy07,Jaun07}, analogous to what I desire here for caustics.

Let me consider a modified ray Hamiltonian
\begin{equation}
	\widetilde{\Symb{D}}(\Vect{z}) \doteq f(\Vect{z}) \Symb{D}(\Vect{z}) 
	,
\end{equation}

\noindent where $f(\Vect{z})$ is some smooth function. Let $\Vect{z}_{\Symb{D}}$ and $\Vect{z}_f$ denote the zero sets of $\Symb{D}(\Vect{z})$ and $f(\Vect{z})$ respectively, \ie
\begin{equation}
	\Vect{z}_{\Symb{D}} \doteq \left\{ \Vect{z} \, | \, \Symb{D}(\Vect{z}) = 0 \right\} \, ,
	\quad
	\Vect{z}_{f} \doteq \left\{ \Vect{z} \, | \, f(\Vect{z}) = 0 \right\} 
	.
\end{equation}

\noindent Clearly, the zero set of $\widetilde{\Symb{D}}(\Vect{z})$ is 
\begin{equation}
	\Vect{z}_{\widetilde{\Symb{D}}} \doteq \left\{ \Vect{z} \, | \, f(\Vect{z})\Symb{D}(\Vect{z}) = 0 \right\}
	= \Vect{z}_{\Symb{D}} \bigcup \Vect{z}_f 
	,
\end{equation}

\noindent where $\bigcup$ denotes the set union. For $\widetilde{\Symb{D}}(\Vect{z})$ and $\Symb{D}(\Vect{z})$ to generate the same set of rays, $f(\Vect{z})$ and $\Symb{D}(\Vect{z})$ cannot be zero simultaneously. Hence, I require
\begin{equation}
	\Vect{z}_{\Symb{D}} \bigcap \Vect{z}_f = \emptyset 
	,
	\label{eq:6_zeroINTER}
\end{equation}

\noindent where $\bigcap$ denotes the set intersection and $\emptyset$ is the empty set. Equation \eq{eq:6_zeroINTER} is most easily satisfied by requiring $f(\Vect{z})$ be sign-definite, say, positive (so $\Vect{z}_f = \emptyset$).

Let me now compute the rays generated by $\widetilde{D}(\Vect{z})$. Analogous to \Eq{eq:3_EOM}, the new rays satisfy
\begin{equation}
	\pd{\widetilde{\tau}_1} \Vect{z} = \JMat{2N} \, \pd{\Vect{z}} \widetilde{\Symb{D}}(\Vect{z})
	= f(\Vect{z}) \JMat{2N} \, \pd{\Vect{z}} \Symb{D}(\Vect{z})
	+ \Symb{D}(\Vect{z}) \JMat{2N} \, \pd{\Vect{z}} f(\Vect{z})
	\label{eq:6_modHAM}
\end{equation}

\noindent for some new parameterization $\widetilde{\tau}_1$. From \Eq{eq:6_modHAM}, rays initialized within $\Vect{z}_{\Symb{D}}$ will always remain in $\Vect{z}_{\Symb{D}}$, at least in exact arithmetic. For such rays, the second term in \Eq{eq:6_modHAM} is identically zero, making \Eq{eq:6_modHAM} simply a reparameterization of \Eq{eq:3_EOM} with
\begin{equation}
	\tau_1 = f(\Vect{z}) \widetilde{\tau}_1 
	.
	\label{eq:6_reparam}
\end{equation}

\noindent For inexact arithmetic, however, the second term in \Eq{eq:6_modHAM} is not exactly zero, and therefore must be retained to preserve the Hamiltonian structure~\cite{Hairer97,Zare75}.

By \Eq{eq:6_reparam}, $\tau_1$ is non-uniformly discretized when $\widetilde{\tau}_1$ is uniformly discretized; hence, a curvature-dependent adaptive discretization can be achieved by properly designing $f(\Vect{z})$. First, I restrict $f(\Vect{z})$ to only depend on the local curvature of the dispersion manifold, denoted $\curv(\Vect{z})$.%
\footnote{$\curv(\Vect{z})$ may be difficult to calculate numerically, since obtaining the dispersion manifold is often the \textit{result} of ray-tracing, not the prerequisite as I suggest here. An iterative approach might be possible; however, this is outside the scope of the present work.}
Next, I impose that the uniform and the adaptive discretizations are equivalent when $\curv(\Vect{z}) = 0$. Hence,
\begin{equation}
	\lim_{\curv \to 0} f(\curv) = 1 
	.
	\label{eq:6_limit}
\end{equation}

\noindent Finally, for the adaptive discretization to congregate in regions of high curvature, I require $f(\curv)$ to be a strictly decreasing function of $\curv$, that is
\begin{equation}
	f'(\curv \neq 0) < 0 
	,
	\quad
	f'(0) \le 0 
	.
	\label{eq:6_decrease}
\end{equation}

\noindent These conditions on $f$ ensure that the rays indeed slow down (but never stop entirely) in regions where $\curv$ is large and $f$ is correspondingly small. In locally flat regions where $\curv = 0$, there is no difference between the two parameterizations. Said another way, a fixed time step $\Delta \fourier{\tau}_1$ corresponds to a variable time-step $\Delta \tau_1 = f \Delta \fourier{\tau}_1$ in the original coordinates that naturally shortens where $f$ is smaller (and $\curv$ larger) without any external input or monitoring.

\begin{figure}[t]
	\centering
	\begin{overpic}[width=0.6\linewidth]{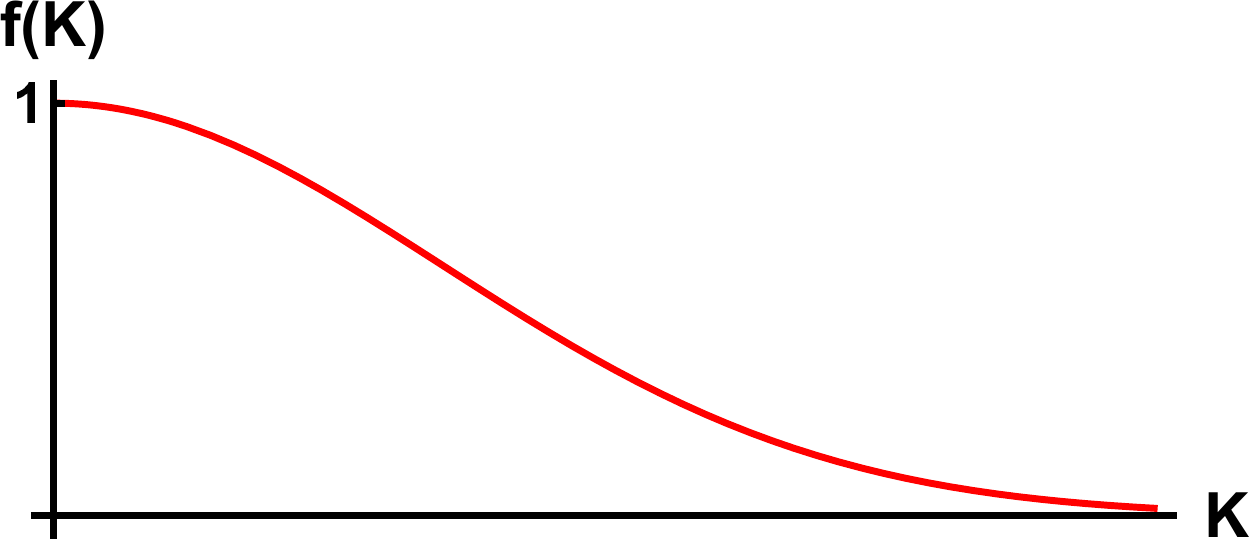}
		\put(90,9){\textbf{\large(a)}}
	\end{overpic}
	
	\begin{overpic}[width=0.6\linewidth]{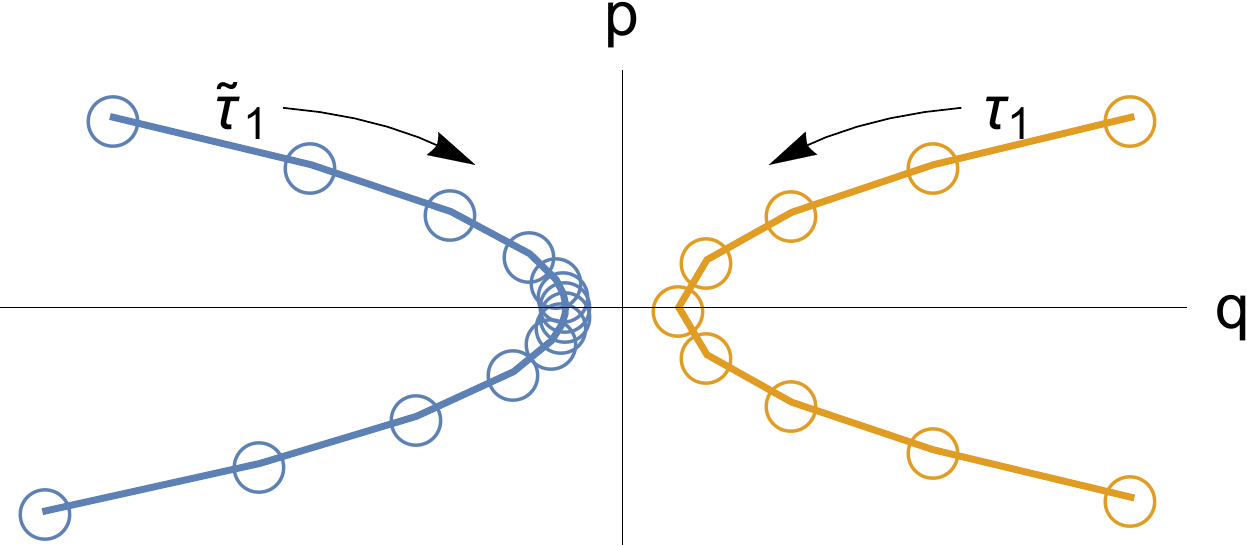}
		\put(90,9){\textbf{\large(b)}}
	\end{overpic}
	\caption{\textbf{(a)} An example $f(\curv)$, where $\curv$ is the local curvature of the dispersion manifold, which satisfies the requirements \eq{eq:6_zeroINTER}, \eq{eq:6_limit}, and \eq{eq:6_decrease}. \textbf{(b)} Comparison of the uniform discretization $\tau_1$ and the adaptive discretization $\widetilde{\tau}_1$ for the Airy dispersion manifold (\Ch{ch:Ex}) using $f(\curv)$ provided by \Eq{eq:6_fADAPT} with $\mu = 1$. The values of $\tau_1$ and $\widetilde{\tau}_1$ are uniformly discretized on the intervals $[0,4]$ and $[0,6.5]$ respectively. $\curv(\Vect{z})$ is calculated in the usual manner for a $1$-D planar curve, \ie $\curv(\Vect{z}) = \|\pd{\Vect{z}} \Symb{D}(\Vect{z}) \|^{-3} \, \left| \JMat{2N} \,  \pd{\Vect{z}} \Symb{D}(\Vect{z}) \dubdot \pd{\Vect{z}\Vect{z}}^2 \Symb{D}(\Vect{z}) \right| $~\cite{Goldman05}. For visualization purposes, the dispersion manifolds are displaced slightly from the origin, and the $\tau_1$ discretization is reflected about the $p$ axis.}
	\label{fig:6_reparam}
\end{figure}

Any function which satisfies \Eqs{eq:6_zeroINTER}, \eq{eq:6_limit}, and \eq{eq:6_decrease} will be a suitable choice, for example,
\begin{equation}
	f\left[ \curv(\Vect{z}) \right] = \frac{1}{1 + \mu \left[\curv(\Vect{z})\right]^2} 
	,
	\label{eq:6_fADAPT}
\end{equation}

\noindent with $\mu \ge 0$ a free parameter. Figure \ref{fig:6_reparam} shows an example $f(\curv)$ which satisfies these three requirements, and shows the adaptive discretization generated by \Eq{eq:6_fADAPT}. As a final remark, when reparameterized rays are used to calculate $\Phi_\Vect{t}(\Vect{X}_\Vect{t})$, additional terms related to $\pd{\Vect{z}}f(\Vect{z})$ will arise. These can be interpreted as the `gravitational' forces associated with the `time dilation' $\tau_1 \to \widetilde{\tau}_1$~\cite{Dodin19,Dodin10c}.


\section{Derivation of \Eqs{eq:6_MGOb} and \eq{eq:6_MGObMOD}}
\label{app:6_simple}

Here I shall derive the simplifications to the MGO solution \eq{eq:6_MGOsol} depending on whether $\Mat{B}_\Vect{t}$ is invertible or not. First consider the case when $\det \Mat{B}_\Vect{t} \neq 0$. Then, using \Eq{eq:3_invMTint} for $U^{-1}$, the isolated contribution from $\Vect{t}$ takes the form
\begin{align}
	\psi_\Vect{t}
	&=
	\frac{
		\sigma_\Vect{t} \, 
		c_0(\Vect{t}_\perp)
	}{ 
		(- 2\pi i)^{\frac{N}{2}}
		\sqrt{ 
			\det \Mat{B}_\Vect{t} \,
			\det \Mat{M}_\Vect{t} 
		} 
	}
	\exp\left[
		\frac{i}{2} \Vect{K}_\Vect{t}^\intercal(\Vect{t}) \Vect{X}_\Vect{t}(\Vect{t})
		- \frac{i}{2} \Vect{x}^\intercal \Mat{B}_\Vect{t}^{-1}\Mat{A}_\Vect{t} \Vect{x}
		- \frac{i}{2} \int_0^{t_1} \dd \xi \, \Vect{z}^\intercal(\Vect{t}) \, \JMat{2N} \, \dot{\Vect{z}}(\Vect{t})
	\right]
	\nonumber\\
	&\hspace{4mm} \times
	 \int_{\cont{0}} \dd \Vect{\epsilon} \,
	\exp\left\{
		- \frac{i}{2} [\Vect{\epsilon} + \Vect{X}_\Vect{t}(\Vect{t})]^\intercal \Mat{D} \Mat{B}^{-1} [\Vect{\epsilon} + \Vect{X}_\Vect{t}(\Vect{t})]
		+ i\Vect{x}^\intercal \Mat{B}^{-1} [\Vect{\epsilon} + \Vect{X}_\Vect{t}(\Vect{t})]
	\right\}
	\Psi_\Vect{t} \left[
		\Vect{\epsilon} + \Vect{X}_\Vect{t}(\Vect{t}) 
	\right] 
	\nonumber\\
	&=
	\frac{
		\sigma_\Vect{t} \,
		c_0(\Vect{t}_\perp)
	}{ 
		(- 2\pi i)^{\frac{N}{2}}
		\sqrt{ 
			\det \Mat{B}_\Vect{t} \,
			\det \Mat{M}_\Vect{t} 
		} 
	}
	\exp\left[
		- \frac{i}{2} \int_0^{t_1} \dd \xi \, \Vect{z}^\intercal(\Vect{t}) \, \JMat{2N} \, \dot{\Vect{z}}(\Vect{t})
	\right]
	\nonumber\\
	&\hspace{4mm} \times
	\exp\left[
		\frac{i}{2} \Vect{K}_\Vect{t}^\intercal(\Vect{t}) \Vect{X}_\Vect{t}(\Vect{t})
		- \frac{i}{2} \Vect{x}^\intercal \Mat{B}_\Vect{t}^{-1}\Mat{A}_\Vect{t} \Vect{x}
		- \frac{i}{2} \Vect{X}_\Vect{t}^\intercal(\Vect{t}) \Mat{D} \Mat{B}^{-1} \Vect{X}_\Vect{t}(\Vect{t})
		+ i \Vect{x}^\intercal \Mat{B}^{-1} \Vect{X}_\Vect{t}(\Vect{t})
	\right]
	\nonumber\\
	&\hspace{4mm} \times
	 \int_{\cont{0}} \dd \Vect{\epsilon} \,
	\exp\left\{
		- \frac{i}{2} \Vect{\epsilon}^\intercal \Mat{D} \Mat{B}^{-1} \Vect{\epsilon}
		- i \left[
			\Mat{D}^\intercal \Vect{X}_\Vect{t}(\Vect{t})
			- \Vect{x}
		\right]^\intercal \Mat{B}^{-1} \Vect{\epsilon}
	\right\}
	\Psi_\Vect{t} \left[
		\Vect{\epsilon} + \Vect{X}_\Vect{t}(\Vect{t}) 
	\right] 
	,
\end{align}

\noindent where $\sigma_\Vect{t} \doteq \pm 1$ is the overall sign ambiguity. Note that
\begin{align}
	\Mat{D}^\intercal \Vect{X}_\Vect{t}(\Vect{t}) - \Vect{x}
	= \Mat{D}^\intercal \Mat{A}_\Vect{t} \Vect{x}(\Vect{t}) 
	+ \Mat{D}^\intercal \Mat{B}_\Vect{t} \Vect{k}(\Vect{t})
	- \Vect{x}
	&= \Vect{x}(\Vect{t}) - \Vect{x}
	+ \Mat{B}_\Vect{t}^\intercal 
	\left[
		\Mat{C}_\Vect{t} \Vect{x}(\Vect{t})
		+ \Mat{D}_\Vect{t} \Vect{k}(\Vect{t})
	\right]
	\nonumber\\
	&=
	\Vect{x}(\Vect{t}) - \Vect{x}
	+ \Mat{B}_\Vect{t}^\intercal \Vect{K}_\Vect{t}(\Vect{t})
	.
\end{align}

\noindent Similarly, one computes
\begin{align}
	&\Vect{K}^\intercal_\Vect{t}(\Vect{t}) \Vect{X}_\Vect{t}(\Vect{t}) 
	- \Vect{x}^\intercal \Mat{B}^{-1} \Mat{A} \Vect{x}
	- \Vect{X}^\intercal_\Vect{t}(\Vect{t}) \Mat{D} \Mat{B}^{-1} \Vect{X}_\Vect{t}(\Vect{t})
	+ 2\Vect{x}^\intercal \Mat{B}^{-1} \Vect{X}_\Vect{t}(\Vect{t})
	\nonumber\\
	&=
	\left\{
		\Vect{x}^\intercal(\Vect{t})
		\left[ 
			\Mat{C}^\intercal
			- \Mat{A}^\intercal \Mat{D} \Mat{B}^{-1}
		\right]
		+ \Vect{k}^\intercal(\Vect{t}) 
		\left[
			\Mat{D}^\intercal
			- \Mat{B}^\intercal \Mat{D} \Mat{B}^{-1}
		\right]
	\right\} \Vect{X}_\Vect{t}(\Vect{t}) 
	- \Vect{x}^\intercal \Mat{B}^{-1} \Mat{A} \Vect{x}
	+ 2\Vect{x}^\intercal \Mat{B}^{-1} \Vect{X}_\Vect{t}(\Vect{t})
	\nonumber\\
	&=
	- \Vect{x}^\intercal \Mat{B}^{-1} \Mat{A} \Vect{x}
	+ \left[
		2\Vect{x}^\intercal 
		- \Vect{x}^\intercal(\Vect{t})
	\right] \Mat{B}^{-1} \Vect{X}_\Vect{t}(\Vect{t})
	\nonumber\\
	&=
	- \Vect{x}^\intercal \Mat{B}^{-1} \Mat{A} \Vect{x}
	+ 2\Vect{x}^\intercal \Mat{B}^{-1}\Mat{A} \Vect{x}(\Vect{t})
	- \Vect{x}^\intercal(\Vect{t}) \Mat{B}^{-1}\Mat{A} \Vect{x}(\Vect{t})
	+ 2\Vect{x}^\intercal \Vect{k}(\Vect{t})
	- \Vect{x}^\intercal(\Vect{t}) \Vect{k}(\Vect{t})
	\nonumber\\
	&=
	- \left[
		\Vect{x}
		- \Vect{x}(\Vect{t})
	\right]^\intercal \Mat{B}^{-1} \Mat{A} 
	\left[
		\Vect{x}
		- \Vect{x}(\Vect{t})
	\right]
	+ \left[
		2\Vect{x} 
		- \Vect{x}(\Vect{t}) 
	\right]^\intercal \Vect{k}(\Vect{t})
	.
\end{align}

\noindent Hence, when $\Vect{x} = \Vect{x}(\Vect{t})$, as is imposed by the summation in \Eq{eq:6_MGOsol}, then I obtain the simplification
\begin{align}
	\psi_\Vect{t}
	&=
	\frac{
		\sigma_\Vect{t} \,
		c_0(\Vect{t}_\perp)
	}{ 
		(- 2\pi i)^{\frac{N}{2}}
		\sqrt{ 
			\det \Mat{B}_\Vect{t} \,
			\det \Mat{M}_\Vect{t} 
		} 
	}
	\exp\left[
		\frac{i}{2} \Vect{k}^\intercal(\Vect{t}) \Vect{x}(\Vect{t})
		- \frac{i}{2} \int_0^{t_1} \dd \xi \, \Vect{z}^\intercal(\Vect{t}) \, \JMat{2N} \, \dot{\Vect{z}}(\Vect{t})
	\right]
	\nonumber\\
	&\hspace{4mm} \times
	 \int_{\cont{0}} \dd \Vect{\epsilon} \,
	\exp\left[
		- \frac{i}{2} \Vect{\epsilon}^\intercal \Mat{D} \Mat{B}^{-1} \Vect{\epsilon}
		- i \Vect{\epsilon}^\intercal \Vect{K}_\Vect{t}(\Vect{t})
	\right]
	\Psi_\Vect{t} \left[
		\Vect{\epsilon} + \Vect{X}_\Vect{t}(\Vect{t}) 
	\right] 
	.
\end{align}

\noindent Lastly, since integration by parts yields
\begin{equation}
	\frac{i}{2} \Vect{k}^\intercal(\Vect{t}) \Vect{x}(\Vect{t}) - \frac{i}{2} \int_0^{t_1} \dd \xi \, \Vect{z}^\intercal(\Vect{t}) \, \JMat{2N} \, \dot{\Vect{z}}(\Vect{t})
	=
	\frac{i}{2} \Vect{k}^\intercal(0, \Vect{t}_\perp) \Vect{x}(0, \Vect{t}_\perp)
	+ i \int_0^{t_1} \dd \xi \, \Vect{k}^\intercal(\xi, \Vect{t}_\perp) \dot{\Vect{x}}(\xi, \Vect{t}_\perp)
	,
\end{equation}

\noindent by redefining the arbitrary constant function as $f(\Vect{t}_\perp) \doteq c_0(\Vect{t}_\perp) \exp\left[ \frac{i}{2} \Vect{k}^\intercal(0, \Vect{t}_\perp) \Vect{x}(0, \Vect{t}_\perp) \right]$, one obtains \Eq{eq:6_MGOb}.

Now consider when $\det \Mat{B}_\Vect{t} = 0$. Using \Eq{eq:3_invMTint} for $U^{-1}$, the isolated contribution from $\Vect{t}$ takes the form
\begin{align}
	\psi_\Vect{t}
	&=
	\frac{
		\sigma_\Vect{t} \,
		f(\Vect{t}_\perp)
		\exp\left[
			i \int_0^{t_1} \dd \xi \, \Vect{k}^\intercal(\xi, \Vect{t}_\perp) \dot{\Vect{x}}(\xi, \Vect{t}_\perp)
		\right]
	}{ 
		(-2\pi i)^{\rho/2}
		\sqrt
		{
			\det \Mat{\Lambda}_{\rho \rho} \,
			\det \Mat{a}_{\varsigma \varsigma}^{-1} \,
			\det \Mat{M}_\Vect{t}
		}
	}
	\nonumber\\
	&\hspace{4mm} \times
	\exp\left\{
		\frac{i}{2} \Vect{K}_\Vect{t}^\intercal(\Vect{t}) \Vect{X}_\Vect{t}(\Vect{t})
		- \frac{i}{2} \Vect{k}^\intercal(\Vect{t}) \Vect{x}(\Vect{t})
		- \frac{i}{2} \Vect{X}_\rho^\intercal(\Vect{t}) \, \Mat{d}_{\rho \rho} \Mat{\Lambda}_{\rho \rho}^{-1} \, \Vect{X}_\rho(\Vect{t})
		\right.\nonumber\\
		&\left.\hspace{37mm}
		+ i \Vect{X}_\rho^\intercal(\Vect{t}) \, \Mat{M}_3 \Mat{R}_\textrm{s}^\intercal \Vect{x}
		- \frac{i}{2} \Vect{x}^\intercal \, \Mat{R}_\textrm{s} \Mat{M}_4 \Mat{R}_\textrm{s}^\intercal \, \Vect{x}
	\right\}
	\nonumber\\
	&\hspace{4mm} \times
	\int_{\cont{0}} \dd \Vect{\epsilon} \,
	\exp
	\left\{
		- \frac{i}{2} \Vect{\epsilon}_\rho^\intercal \, \Mat{d}_{\rho \rho} \Mat{\Lambda}_{\rho \rho}^{-1} \, \Vect{\epsilon}_\rho
		- i \Vect{\epsilon}_\rho^\intercal \, 
		\left[
			\Mat{d}_{\rho \rho} \Mat{\Lambda}_{\rho \rho}^{-1} \, \Vect{X}_\rho(\Vect{t})
			+ \Mat{M}_3 \Mat{R}_\textrm{s}^\intercal \Vect{x}
		\right]
	\right\}
	\, 
	\nonumber\\
	&\hspace{37mm}\times
	\delta
	\left[
		\Vect{X}_\varsigma(\Vect{t}) + \Vect{\epsilon}_\varsigma - \Mat{a}_{\varsigma \varsigma} \Vect{x}_\varsigma
	\right]
	\Psi_\Vect{t} \left[
		\Vect{\epsilon} + \Vect{X}_\Vect{t}(\Vect{t}) 
	\right]
	\nonumber\\
	&=
	\frac{
		\sigma_\Vect{t} \,
		f(\Vect{t}_\perp)
		\exp\left[
			i \int_0^{t_1} \dd \xi \, \Vect{k}^\intercal(\xi, \Vect{t}_\perp) \dot{\Vect{x}}(\xi, \Vect{t}_\perp)
		\right]
	}{ 
		(-2\pi i)^{\rho/2}
		\sqrt
		{
			\det \Mat{\Lambda}_{\rho \rho} \,
			\det \Mat{a}_{\varsigma \varsigma}^{-1} \,
			\det \Mat{M}_\Vect{t}
		}
	}
	\nonumber\\
	&\hspace{4mm} \times
	\exp\left\{
		\frac{i}{2} \Vect{K}_\Vect{t}^\intercal(\Vect{t}) \Vect{X}_\Vect{t}(\Vect{t})
		- \frac{i}{2} \Vect{k}^\intercal(\Vect{t}) \Vect{x}(\Vect{t})
		- \frac{i}{2} \Vect{X}_\rho^\intercal(\Vect{t}) \, \Mat{d}_{\rho \rho} \Mat{\Lambda}_{\rho \rho}^{-1} \, \Vect{X}_\rho(\Vect{t})
		\right.\nonumber\\
		&\left.\hspace{37mm}
		+ i \Vect{X}_\rho^\intercal(\Vect{t}) \, \Mat{M}_3 \Mat{R}_\textrm{s}^\intercal \Vect{x}
		- \frac{i}{2} \Vect{x}^\intercal \, \Mat{R}_\textrm{s} \Mat{M}_4 \Mat{R}_\textrm{s}^\intercal \, \Vect{x}
	\right\}
	\nonumber\\
	&\hspace{4mm} \times
	\int_{\cont{0}} \dd \Vect{\epsilon}_\rho \,
	\exp
	\left\{
		- \frac{i}{2} \Vect{\epsilon}_\rho^\intercal \, \Mat{d}_{\rho \rho} \Mat{\Lambda}_{\rho \rho}^{-1} \, \Vect{\epsilon}_\rho
		- i \Vect{\epsilon}_\rho^\intercal \, 
		\left[
			\Mat{d}_{\rho \rho} \Mat{\Lambda}_{\rho \rho}^{-1} \, \Vect{X}_\rho(\Vect{t})
			- \Mat{M}_3 \Mat{R}_\textrm{s}^\intercal \Vect{x}
		\right]
	\right\}
	\, 
	\Psi_\Vect{t}
	\left[
		\Mat{L}_\textrm{s}
		\begin{pmatrix}
			\Vect{X}_\Vect{t}^\rho(\Vect{t}) + \Vect{\epsilon}_\rho \\
			\Mat{a}_{\varsigma \varsigma} \Vect{x}_\varsigma(\Vect{t})
		\end{pmatrix}
	\right]
	,
\end{align}

\noindent where I have used the fact that $\Vect{X}_\varsigma(\Vect{t}) =  \Mat{a}_{\varsigma \varsigma} \Vect{x}_\varsigma(\Vect{t})$. I first compute the simplification
\begin{align}
	\Mat{d}_{\rho \rho} \Mat{\Lambda}_{\rho \rho}^{-1} \, \Vect{X}_\rho(\Vect{t}) 
	- \Mat{M}_3 \Mat{R}_\textrm{s}^\intercal \Vect{x}
	&=
	\Mat{\Lambda}_{\rho \rho}^{-1}
	\left[
		\Mat{d}_{\rho \rho}^\intercal \Vect{X}_\rho(\Vect{t}) 
		- \Vect{x}_\rho
		+ \Mat{d}_{\varsigma \rho}^\intercal \Mat{a}_{\varsigma \varsigma} \Vect{x}_\varsigma
	\right]
	\nonumber\\
	&=
	\Mat{\Lambda}_{\rho \rho}^{-1}
	\left[
		\Mat{d}_{\rho \rho}^\intercal \Mat{a}_{\rho \rho} \Vect{x}_\rho(\Vect{t})
		+ \Mat{d}_{\rho \rho}^\intercal \Mat{a}_{\rho \varsigma} \Vect{x}_\varsigma(\Vect{t})
		+ \Mat{d}_{\rho \rho}^\intercal \Mat{\Lambda}_{\rho \rho} \Vect{k}_\rho(\Vect{t})
		- \Vect{x}_\rho
		+ \Mat{d}_{\varsigma \rho}^\intercal \Mat{a}_{\varsigma \varsigma} \Vect{x}_\varsigma
	\right]
	\nonumber\\
	&=
	\left[
		\Mat{c}_{\rho \rho} \Vect{x}_\rho(\Vect{t})
		+ \Mat{c}_{\rho \varsigma} \Vect{x}_\varsigma(\Vect{t})
		+ \Mat{d}_{\rho \rho} \Vect{k}_\rho(\Vect{t})
	\right]
	+ \Mat{\Lambda}_{\rho \rho}^{-1} \left[ \Vect{x}_\rho(\Vect{t}) - \Vect{x}_\rho \right]
	\nonumber\\
	&\hspace{53mm}
	+ \Mat{\Lambda}_{\rho \rho}^{-1} \Mat{d}_{\varsigma \rho}^\intercal \Mat{a}_{\varsigma \varsigma} 
	\left[
		\Vect{x}_\varsigma
		- \Vect{x}_\varsigma(\Vect{t})
	\right]
	\nonumber\\
	&=
	\Vect{K}_\rho(\Vect{t})
	+ \Mat{\Lambda}_{\rho \rho}^{-1} \left[ \Vect{x}_\rho(\Vect{t}) - \Vect{x}_\rho \right]
	+ \Mat{\Lambda}_{\rho \rho}^{-1} \Mat{d}_{\varsigma \rho}^\intercal \Mat{a}_{\varsigma \varsigma} 
	\left[
		\Vect{x}_\varsigma
		- \Vect{x}_\varsigma(\Vect{t})
	\right]
	,
\end{align}

\noindent where I have used the definitions
\begin{subequations}
	\begin{align}
		\begin{pmatrix}
			\Vect{X}_\rho \\
			\Vect{X}_\varsigma
		\end{pmatrix}
		&=
		\begin{pmatrix}
			\Mat{a}_{\rho \rho} & \Mat{a}_{\rho \varsigma} \\
			\OMat{\varsigma \rho} & \Mat{a}_{\varsigma \varsigma}
		\end{pmatrix}
		\begin{pmatrix}
			\Vect{x}_\rho \\
			\Vect{x}_\varsigma
		\end{pmatrix}
		+
		\begin{pmatrix}
			\Mat{\Lambda}_{\rho \rho} & \OMat{ \rho \varsigma} \\
			\OMat{\varsigma \rho} & \OMat{\varsigma \varsigma}
		\end{pmatrix}
		\begin{pmatrix}
			\Vect{k}_\rho \\
			\Vect{k}_\varsigma
		\end{pmatrix}
		=
		\begin{pmatrix}
			\Mat{a}_{\rho \rho} \Vect{x}_\rho + \Mat{a}_{\rho \varsigma} \Vect{x}_\varsigma + \Mat{\Lambda}_{\rho \rho} \Vect{k}_\rho \\
			\Mat{a}_{\varsigma \varsigma} \Vect{x}_\varsigma
		\end{pmatrix}
		, \\[2mm]
		\begin{pmatrix}
			\Vect{K}_\rho \\
			\Vect{K}_\varsigma
		\end{pmatrix}
		&=
		\begin{pmatrix}
			\Mat{c}_{\rho \rho} & \Mat{c}_{\rho \varsigma} \\
			\Mat{c}_{\varsigma \rho} & \Mat{c}_{\varsigma \varsigma}
		\end{pmatrix}
		\begin{pmatrix}
			\Vect{x}_\rho \\
			\Vect{x}_\varsigma
		\end{pmatrix}
		+
		\begin{pmatrix}
			\Mat{d}_{\rho \rho} & \OMat{\rho \varsigma} \\
			\Mat{d}_{\varsigma \rho} & \Mat{a}_{\varsigma \varsigma}^{-\intercal}
		\end{pmatrix}
		\begin{pmatrix}
			\Vect{k}_\rho \\
			\Vect{k}_\varsigma
		\end{pmatrix}
		=
		\begin{pmatrix}
			\Mat{c}_{\rho \rho} \Vect{x}_\rho + \Mat{c}_{\rho \varsigma} \Vect{x}_\varsigma + \Mat{d}_{\rho \rho} \Vect{k}_\rho \\
			\Mat{c}_{\varsigma \rho} \Vect{x}_\rho + \Mat{c}_{\varsigma \varsigma} \Vect{x}_\varsigma + \Mat{d}_{\varsigma \rho} \Vect{k}_\rho + \Mat{a}_{\varsigma \varsigma}^{-\intercal} \Vect{k}_\varsigma
		\end{pmatrix}
		.
	\end{align}
\end{subequations}

\noindent Similarly, I compute
\begin{align}
	&\frac{i}{2} \Vect{K}_\Vect{t}^\intercal(\Vect{t}) \Vect{X}_\Vect{t}(\Vect{t})
	- \frac{i}{2} \Vect{k}^\intercal(\Vect{t}) \Vect{x}(\Vect{t})
	- \frac{i}{2} \Vect{X}_\rho^\intercal(\Vect{t}) \, \Mat{d}_{\rho \rho} \Mat{\Lambda}_{\rho \rho}^{-1} \, \Vect{X}_\rho(\Vect{t})
	+ i \Vect{X}_\rho^\intercal(\Vect{t}) \, \Mat{M}_3 \Mat{R}_\textrm{s}^\intercal \Vect{x}
	- \frac{i}{2} \Vect{x}^\intercal \, \Mat{R}_\textrm{s} \Mat{M}_4 \Mat{R}_\textrm{s}^\intercal \, \Vect{x}
	\nonumber\\
	&=
	\frac{i}{2} \Vect{K}_\Vect{t}^\intercal(\Vect{t}) \Vect{X}_\Vect{t}(\Vect{t})
	- \frac{i}{2} \Vect{k}^\intercal(\Vect{t}) \Vect{x}(\Vect{t})
	- \frac{i}{2} \Vect{X}_\rho^\intercal(\Vect{t}) \Mat{\Lambda}_{\rho \rho}^{-1}
	\left[
		\Mat{d}_{\rho \rho}^\intercal \, \Vect{X}_\rho(\Vect{t})
		- 2 \Vect{x}_\rho
		+ 2 \Mat{d}_{\varsigma \rho}^\intercal \Mat{a}_{\varsigma \varsigma} \Vect{x}_\varsigma
	\right]
	- \frac{i}{2} \Vect{x}^\intercal \, \Mat{R}_\textrm{s} \Mat{M}_4 \Mat{R}_\textrm{s}^\intercal \, \Vect{x}
	\nonumber\\
	&=
	\frac{i}{2} \Vect{K}_\Vect{t}^\intercal(\Vect{t}) \Vect{X}_\Vect{t}(\Vect{t})
	- \frac{i}{2} \Vect{k}^\intercal(\Vect{t}) \Vect{x}(\Vect{t})
	- \frac{i}{2} \Vect{X}_\rho^\intercal(\Vect{t}) \Vect{K}_\rho(\Vect{t})
	\nonumber\\
	&\hspace{4mm}
	+ \frac{i}{2} \Vect{X}_\rho^\intercal(\Vect{t}) \Mat{\Lambda}_{\rho \rho}^{-1}
	\left\{
		\left[ 2 \Vect{x}_\rho - \Vect{x}_\rho(\Vect{t}) \right]
		- \Mat{d}_{\varsigma \rho}^\intercal \Mat{a}_{\varsigma \varsigma} 
		\left[
			2 \Vect{x}_\varsigma
			- \Vect{x}_\varsigma(\Vect{t})
		\right]
	\right\}
	- \frac{i}{2} \Vect{x}^\intercal \, \Mat{R}_\textrm{s} \Mat{M}_4 \Mat{R}_\textrm{s}^\intercal \, \Vect{x}
	\nonumber\\
	&=
	\frac{i}{2}\Vect{x}_\rho^\intercal(\Vect{t}) \Mat{c}_{\varsigma \rho}^\intercal \Mat{a}_{\varsigma \varsigma} \Vect{x}_\varsigma(\Vect{t})
	+ \frac{i}{2} \Vect{x}_\varsigma^\intercal(\Vect{t}) \Mat{c}_{\varsigma \varsigma}^\intercal \Mat{a}_{\varsigma \varsigma} \Vect{x}_\varsigma(\Vect{t})
	+ \frac{i}{2} \Vect{k}_\rho^\intercal(\Vect{t}) \Mat{d}_{\varsigma \rho}^\intercal \Mat{a}_{\varsigma \varsigma} \Vect{x}_\varsigma(\Vect{t})
	+ \frac{i}{2} \Vect{k}_\varsigma^\intercal(\Vect{t}) \Vect{x}_\varsigma(\Vect{t})
	- \frac{i}{2} \Vect{k}^\intercal(\Vect{t}) \Vect{x}(\Vect{t})
	\nonumber\\
	&\hspace{4mm}
	+ \frac{i}{2} \left[
		\Vect{x}_\rho^\intercal (\Vect{t}) \Mat{\Lambda}_{\rho \rho}^{-1} \Mat{a}_{\rho \rho}
		+ \Vect{x}_\varsigma^\intercal (\Vect{t}) \Mat{a}_{\rho \varsigma}^\intercal \Mat{\Lambda}_{\rho \rho}^{-1}
		+ \Vect{k}_\rho^\intercal (\Vect{t})
	\right]
	\left\{
		\left[ 2 \Vect{x}_\rho - \Vect{x}_\rho(\Vect{t}) \right]
		- \Mat{d}_{\varsigma \rho}^\intercal \Mat{a}_{\varsigma \varsigma} 
		\left[
			2 \Vect{x}_\varsigma
			- \Vect{x}_\varsigma(\Vect{t})
		\right]
	\right\}
	\nonumber\\
	&\hspace{4mm}
	- \frac{i}{2} \Vect{x}_\rho^\intercal\Mat{\Lambda}_{\rho \rho}^{-1} \Mat{a}_{\rho \rho} \Vect{x}_\rho
	- \frac{i}{2} \Vect{x}_\rho^\intercal \Mat{\Lambda}_{\rho \rho}^{-1} \Mat{a}_{\rho \varsigma} \Vect{x}_\varsigma
	- \frac{i}{2} \Vect{x}_\varsigma^\intercal\Mat{a}_{\rho \varsigma}^\intercal \Mat{\Lambda}_{\rho \rho}^{-1} \Vect{x}_\rho
	- \frac{i}{2} \Vect{x}_\varsigma^\intercal\Mat{a}_{\varsigma \varsigma}^\intercal \Mat{c}_{\varsigma \varsigma}\Vect{x}_\varsigma 
	+ \frac{i}{2} \Vect{x}_\varsigma^\intercal\Mat{a}_{\varsigma \varsigma}^\intercal \Mat{d}_{\varsigma \rho} \Mat{\Lambda}_{\rho \rho}^{-1} \Mat{a}_{\rho \varsigma}\Vect{x}_\varsigma
	\nonumber\\
	&=
	\frac{i}{2} \Vect{x}_\varsigma^\intercal\Mat{n} \Vect{x}_\varsigma
	- \frac{i}{2} \Vect{x}_\varsigma^\intercal(\Vect{t}) \Mat{n} \Vect{x}_\varsigma(\Vect{t})
	+ i
	\left\{
		\Vect{k}_\rho^\intercal (\Vect{t})
		- \frac{1}{2} \left[ \Vect{x}_\rho - \Vect{x}_\rho(\Vect{t}) \right]^\intercal\Mat{\Lambda}_{\rho \rho}^{-1} \Mat{a}_{\rho \rho} 
	\right\}
	\left[ \Vect{x}_\rho - \Vect{x}_\rho(\Vect{t}) \right]
	\nonumber\\
	&\hspace{4mm}
	- i \left\{
		\Vect{k}_\rho^\intercal (\Vect{t}) \Mat{d}_{\varsigma \rho}^\intercal \Mat{a}_{\varsigma \varsigma}
		+ \Vect{x}_\varsigma^\intercal (\Vect{t}) \Mat{a}_{\rho \varsigma}^\intercal \Mat{\Lambda}_{\rho \rho}^{-1}\Mat{d}_{\varsigma \rho}^\intercal \Mat{a}_{\varsigma \varsigma}
		+ \Vect{x}_\rho^\intercal (\Vect{t}) \Mat{\Lambda}_{\rho \rho}^{-1} \Mat{a}_{\rho \rho} \Mat{d}_{\varsigma \rho}^\intercal \Mat{a}_{\varsigma \varsigma}
		+\Vect{x}_\rho^\intercal \Mat{\Lambda}_{\rho \rho}^{-1} \Mat{a}_{\rho \varsigma} 
	\right\}
	\left[ \Vect{x}_\varsigma - \Vect{x}_\varsigma(\Vect{t}) \right]
	,
\end{align}

\noindent where $\Mat{n} \doteq \Mat{a}_{\varsigma \varsigma}^\intercal \Mat{d}_{\varsigma \rho} \Mat{\Lambda}_{\rho \rho}^{-1} \Mat{a}_{\rho \varsigma} - \Mat{a}_{\varsigma \varsigma}^\intercal \Mat{c}_{\varsigma \varsigma}$ is symmetric. Hence, when $\Vect{x} = \Vect{x}(\Vect{t})$, as imposed by the summation in \Eq{eq:6_MGOsol}, I obtain \Eq{eq:6_MGObMOD}.


\section{MGO with Gaussian coherent states}

As discussed in \Ch{ch:MT}, an alternate expression for the MT kernel $U^{-1}(\Vect{x}, \Vect{X})$ that is insensitive to the invertibility of $\Mat{B}$ can be obtained by using coherent states. In this appendix I show how these coherent states can also be used to develop MGO. First, let me recall from \Eq{eq:3_gaussMT} that the coherent-state MT kernel takes the form
\begin{align}
	U^{-1}(\Vect{x}, \Vect{X})
	=
	\pm
	\int \dd \Vect{K}_0 \,
	\frac
	{
		\exp
		\left[
			\fourier{g}(\Vect{x}, \Vect{\xi})
			- |\Vect{K}_0|^2
		\right]
	}
	{
		(\sqrt{2} \pi)^N 
		\sqrt{ \det(2 \Mat{D} - i \Mat{B} ) }
	}
	,
\end{align}

\noindent where I have defined $\Vect{\xi} \doteq \Vect{X} + 2 i \Vect{K}_0$ and 
\begin{align}
	\fourier{g}(\Vect{x}, \Vect{\xi})
	\doteq
	- \frac{1}{2} \Vect{x}^\intercal 
	\left(
		2\Mat{D}
		- i \Mat{B}
	\right)^{-1}
	\left(
		\Mat{A}
		+ 2 i \Mat{C}
	\right)
	\Vect{x}
	+
	\left(
		\Vect{x}
		- \frac{1}{2} \Mat{D}^\intercal \Vect{\xi}
	\right)^\intercal
	\left(
		2 \Mat{D}
		- i \Mat{B}
	\right)^{-1} \Vect{\xi}
	.
\end{align}

Next, let me verify that the integrand phase is still stationary at the ray contribution even when coherent states are used. Indeed, the phase of 
$$
	\pm
	\int \dd \Vect{X} \, \dd \Vect{K}_0 \,
	\Phi_\Vect{t}(\Vect{X})
	\frac
	{
		\exp
		\left[
			\fourier{g}(\Vect{x}, \Vect{\xi})
			- |\Vect{K}_0|^2
			+ i \Theta_\Vect{t}(\Vect{X})
		\right]
	}
	{
		(\sqrt{2} \pi)^N 
		\sqrt{ \det(2 \Mat{D} - i \Mat{B} ) }
	}
$$
is stationary (with respect to variations in both $\Vect{X}$ and $\Vect{K}_0$ where $\Vect{X}$ and $\Vect{K}_0$ simultaneously satisfy
\begin{align}
	2i \Mat{D}_\Vect{t}^\intercal
	\left[
		\Vect{K}_\Vect{t}(\Vect{X}) 
		- \Vect{K}_0
	\right]
	+ \Vect{x}
	- \Mat{D}_\Vect{t}^\intercal \Vect{X}
	+ \Mat{B}_\Vect{t}^\intercal \Vect{K}_\Vect{t}(\Vect{X})
	= 0
	, \quad
	\Vect{x}
	- \Mat{D}_\Vect{t}^\intercal \Vect{X}
	+ \Mat{B}_\Vect{t}^\intercal \Vect{K}_0 
	= 0 .
	\label{eq:6_gaussSPA}
\end{align}

\noindent When $\Vect{x}$ is evaluated at $\Vect{x}(\Vect{t})$, a simultaneous solution to \Eqs{eq:6_gaussSPA} is $\Vect{X} = \Vect{X}_\Vect{t}(\Vect{t})$ and $\Vect{K}_0 = \Vect{K}_\Vect{t}(\Vect{t})$ as desired. Therefore, upon defining the new integration variables
\begin{align}
	\Vect{\epsilon}_r \doteq \Vect{X} - \Vect{X}_\Vect{t}(\Vect{t})
	, \quad
	\Vect{\epsilon}_i \doteq 2 \Vect{K}_0 - 2 \Vect{K}_\Vect{t}(\Vect{t})
	,
\end{align}

\noindent I obtain the following alternate representation of $\psi_\Vect{t}(\Vect{x})$:
\begin{align}
	\psi_\Vect{t}(\Vect{x})
	=
	\frac
	{
		\sigma_\Vect{t}
		\alpha_\Vect{t}
		\exp
		\left\{
			\fourier{g}_\Vect{t}[ \Vect{x}, \Vect{\xi}_\Vect{t}(\Vect{t})]
			- |\Vect{K}_\Vect{t}(\Vect{t})|^2 
		\right\}
	}
	{
		(2\sqrt{2} \, \pi)^N 
		\sqrt{ \det(2 \Mat{D}_\Vect{t} - i \Mat{B}_\Vect{t} ) }
	}
	\int_{\cont{0}} \dd \Vect{\epsilon}_r \, \dd \Vect{\epsilon}_i \,
	\Psi_\Vect{t}[\Vect{\epsilon}_r + \Vect{X}_\Vect{t}(\Vect{t})]
	\exp
	\left[
		- \fourier{\gamma}_\Vect{t}(\Vect{\epsilon}, \Vect{x}, \Vect{t})
	\right]
	,
	\label{eq:6_MGOpsiGAUSS}
\end{align}

\noindent where I have defined $\Vect{\epsilon} \doteq \Vect{\epsilon}_r + i \Vect{\epsilon}_i$ along with
\begin{subequations}%
	\begin{align}%
		\Vect{\xi}_\Vect{t}(\Vect{t}) &\doteq \Vect{X}_\Vect{t}(\Vect{t}) + 2 i \Vect{K}_\Vect{t}(\Vect{t})
		, \\
		\fourier{\gamma}_\Vect{t}(\Vect{\epsilon}, \Vect{x}, \Vect{t})
		&\doteq
		\frac{1}{2} \Vect{\epsilon}^\intercal \Mat{D}_\Vect{t}
		\left(
			2 \Mat{D}_\Vect{t}
			- i \Mat{B}_\Vect{t}
		\right)^{-1} \Vect{\epsilon}
		+ \frac{|\Vect{\epsilon}_i|^2 }{4}
		+ \Vect{\epsilon}_i^\intercal \Vect{K}_\Vect{t}(\Vect{t})
		- \Vect{\epsilon}^\intercal 
		\left(
			2 \Mat{D}_\Vect{t}
			- i \Mat{B}_\Vect{t}
		\right)^{-\intercal}
		\left[
			\Vect{x}
			- \Mat{D}_\Vect{t}^\intercal \Vect{\xi}_\Vect{t}(\Vect{t})
		\right]
		.
	\end{align}
\end{subequations}

Equation \eq{eq:6_MGOpsiGAUSS} is equivalent to \Eq{eq:6_MGObMOD} but might be advantageous since it can be applied `as is' without performing an SVD of $\Mat{B}_\Vect{t}$. That said, \Eq{eq:6_MGOpsiGAUSS} involves a $2N$-D integral, which is harder to evaluate numerically. In this sense, the representation \eq{eq:6_MGObMOD} may be more practical, especially at large $N$.

\end{subappendices}

\chapter{Summary of MGO and Examples}
\label{ch:Ex}

\section{Introduction}

Having developed the general theory of MGO and outlined its usage in \Ch{ch:MGO}, let me now demonstrate its ability to accurately and robustly model caustics on a series of analytical examples. Note that the material presented in this chapter is based on the previously published work presented in \Refs{Lopez20,Lopez21a}.



\section{Summary of MGO and Outline of an MGO-based ray-tracing algorithm}

Let me now briefly outline how the MGO formalism can be used in a ray-tracing code. For simplicity, suppose an incident eikonal wavefield is prescribed on a plane $x_1 = 0$, that is, 
\begin{equation}
	\psi_\textrm{in}(\Vect{x}_\perp) 
	= \phi_0(\Vect{x}_\perp) \exp[i \theta_0(\Vect{x}_\perp)]
	,
	\label{eq:6_psiINIT}
\end{equation}

\noindent where $\Vect{x}_\perp$ is a vector containing the remaining $N-1$ spatial coordinates besides $x_1$. [It is straightforward to generalize the following procedure for curvilinear initial surfaces, and even $\Vect{k}$-space surfaces in the event that $\psi(0,\Vect{x}_\perp)$ contains an $\Vect{x}$-space caustic.] Equation \eq{eq:6_psiINIT} provides the following initial conditions for the rays:
\begin{equation}
	\Vect{z}(0, \Vect{\tau}_\perp) = 
	\begin{pmatrix}
		0 \\
		\Vect{\tau}_\perp \\
		k_1(0, \Vect{\tau}_\perp) \\
		\nabla \theta_0(\Vect{\tau}_\perp)
	\end{pmatrix}
	,
	\label{eq:6_zINIT}
\end{equation}

\noindent where $\Vect{\tau}_\perp \equiv \Vect{x}_\perp$ are coordinates on the initial plane and $k_1(0, \Vect{\tau}_\perp)$ solves the dispersion relation \eq{eq:2_GOdisp} when $\tau_1 = 0$, \ie
\begin{equation}
	\Symb{D}[0, \Vect{\tau}_\perp, k_1(0, \Vect{\tau}_\perp), \nabla \theta_0(\Vect{x}_\perp)]
	= 0
	.
\end{equation}

\noindent The next step is to evolve the rays via \Eq{eq:3_EOM} subject to the initial conditions \eq{eq:6_zINIT}. This can be done using the adaptive time-stepping scheme presented in \App{app:6_adapt}. For each timestep $\Vect{\tau} = \Vect{t}$ along a ray, one computes the rectangular matrix $\pd{\Vect{\tau}}\Vect{z}(\Vect{t})$ (\eg via finite difference) and subsequently performs a QR decomposition to obtain $\Mat{Q}_\Vect{t}$ and $\Mat{R}_\Vect{t}$, as described in \App{app:6_QR}. Note that in practice reorthogonalization techniques may be required to ensure the norm of $\Mat{Q}_\Vect{t}$ is sufficiently close to unity. Having obtained $\Mat{Q}_\Vect{t}$, the symplectic submatrices $\Mat{A}_\Vect{t}$ and $\Mat{B}_\Vect{t}$ can be obtained via \Eq{eq:6_zQR}. Specifically, this step reads
\begin{equation}
	\pd{\Vect{\tau}} \Vect{z}(\Vect{t})
	=
	\Mat{Q}_\Vect{t} \Mat{R}_\Vect{t}
	, \quad
	\Mat{Q}_\Vect{t}
	=
	\begin{pmatrix}
		\Mat{A}_\Vect{t}^\intercal \\ 
		\Mat{B}_\Vect{t}^\intercal
	\end{pmatrix}
	.
	\label{eq:7_zQRrep}
\end{equation}

To allow for $\det \Mat{B}_\Vect{t} = 0$, the next step is to perform an SVD of $\Mat{B}_\Vect{t}$ to obtain its rank $\rho$ along with the submatrices $\Mat{\Lambda}_{\rho \rho}$, $\Mat{a}_{\rho \rho}$, and $\Mat{a}_{\varsigma \varsigma}$. Specifically, this step reads
\begin{equation}
	\Mat{B}_\Vect{t} = \Mat{L}_\textrm{s} \,
	\begin{pmatrix}
		\Mat{\Lambda}_{\rho \rho} & \OMat{ \rho \varsigma} \\
		\OMat{\varsigma \rho} & \OMat{\varsigma \varsigma}
	\end{pmatrix} \, \Mat{R}_\textrm{s}^\intercal 
	, \quad
	\Mat{A}_\Vect{t} = \Mat{L}_\textrm{s} \,
	\begin{pmatrix}
		\Mat{a}_{\rho \rho} & \OMat{\rho \varsigma} \\
		\OMat{\varsigma \rho} & \Mat{a}_{\varsigma \varsigma}
	\end{pmatrix} \, \Mat{R}_\textrm{s}^\intercal 
	\label{eq:7_SVDrep}
\end{equation}

\noindent where $\Mat{L}_\textrm{s}$ and $\Mat{R}_\textrm{s}$ are matrices containing the left and right singular vectors of $\Mat{B}_\Vect{t}$, respectively. (Importantly, one should not confuse $\Mat{R}_\textrm{s}$ with $\Mat{R}_\Vect{t}$.) One then computes the MGO prefactor function
\begin{align}
	\MTnorm
	= 
	\frac{
		\sigma_\Vect{t} \,
		\psi_\textrm{in}(\Vect{t}_\perp) \sqrt{\dot{x}_1(0, \Vect{t}_\perp)}
		\exp\left[i \int_0^{t_1} \dd \xi \, \Vect{k}(\xi, \Vect{t}_\perp)^\intercal \dot{\Vect{x}}(\xi, \Vect{t}_\perp) \right]
	}{
		(-2\pi i)^{\rho/2}
		\sqrt{\det \Mat{\Lambda}_{\rho\rho} \det \Mat{a}_{\varsigma \varsigma} \det \Mat{R}_\Vect{t} }
	}
	,
	\label{eq:6_MTnormFINAL}
\end{align}

\noindent or equivalently, if $\det \Mat{B}_\Vect{t} \neq 0$,
\begin{align}
	\MTnorm
	= 
	\frac{
		\sigma_\Vect{t} \,
		\psi_\textrm{in}(\Vect{t}_\perp) \sqrt{\dot{x}_1(0, \Vect{t}_\perp)}
		\exp\left[i \int_0^{t_1} \dd \xi \, \Vect{k}(\xi, \Vect{t}_\perp)^\intercal \dot{\Vect{x}}(\xi, \Vect{t}_\perp) \right]
	}{
		(-2\pi i)^{N/2}
		\sqrt{\det \Mat{B}_\Vect{t} \det \Mat{R}_\Vect{t} }
	}
	,
	\label{eq:6_MTnormFINAL_Bneq0}
\end{align}

\noindent Note that the phase integral has a clear incremental structure, \ie 
\begin{equation}
	\int_0^{t_1 + \Delta t} \dd \xi
	= \int_0^{t_1} \dd \xi 
	+
	\int_{t_1}^{t_1 + \Delta t} \dd \xi  
	,
\end{equation}

\noindent that can be leveraged for efficient evaluation along a ray. Also, the overall sign ambiguity $\sigma_\Vect{t}$ is chosen to maintain the continuity of $\MTnorm$ along a ray, and should be initialized such that $\MTnorm \Upsilon_\Vect{t} = \psi_\textrm{in}$ when $t_1 = 0$. Here I have also assumed that the initial surface is located far from any caustics to facilitate matching the arbitrary function to the incident wavefield. Specifically, using the quadratic saddlepoint method to evaluate the steepest-descent contour places $\psi_{(0,\Vect{t}_\perp)}$ in the approximate form
\begin{equation}
	\psi_{(0,\Vect{t}_\perp)}
	= \frac{f(\Vect{t}_\perp)}{j(0, \Vect{t}_\perp)}
	.
\end{equation}

\noindent where $f(\Vect{t}_\perp)$ is an arbitrary function of the initial conditions. Then using the fact that
\begin{equation}
	j(0, \Vect{t}_\perp)
	= \det
	\begin{pmatrix}
		\dot{x}_1(0, \Vect{t}_\perp) & \leftarrow & 0 & \rightarrow\\
		\vdots & &  \IMat{N-1} \\
		\dot{x}_N(0, \Vect{t}_\perp)
	\end{pmatrix}
	=
	\dot{x}_1(0, \Vect{t}_\perp)
	,
\end{equation}

\noindent yields \Eq{eq:6_MTnormFINAL}.

Next, one computes the tangent-space wavefield $\Psi_\Vect{t}$ by integrating $\Vect{K}_\Vect{t}$ to obtain the phase, and by using either \Eq{eq:6_GOenvMET} or \Eq{eq:6_MGOphi} to compute the envelope. One can then take the integral
\begin{align}
	\Upsilon_\Vect{t}
	=
	\int_{\cont{0}} \dd \Vect{\epsilon}_\rho \,
	\Psi_\Vect{t}
	\left[
		\Mat{L}_\textrm{s}
		\begin{pmatrix}
			\Vect{X}_\Vect{t}^\rho(\Vect{t}) + \Vect{\epsilon}_\rho \\
			\Mat{a}_{\varsigma \varsigma} \Vect{x}_\varsigma(\Vect{t})
		\end{pmatrix}
	\right]
	\exp
	\left[
		- \frac{i}{2} \Vect{\epsilon}_\rho^\intercal \, \Mat{a}_{\rho \rho} \Mat{\Lambda}_{\rho \rho}^{-1} \, \Vect{\epsilon}_\rho
		- i \Vect{\epsilon}_\rho^\intercal \Vect{K}_\Vect{t}^\rho(\Vect{t})
	\right]
	\label{eq:7_UpsilonINT}
\end{align}

\noindent over the invertible subspace of $\Mat{B}_\Vect{t}$ using the Gauss--Freud quadrature rule described in \Ch{ch:GF}. Equivalently, if $\det \Mat{B}_\Vect{t} \neq 0$ one can compute
\begin{align}
	\Upsilon_\Vect{t}
	=
	\int_{\cont{0}} \dd \Vect{\epsilon} \,
	\Psi_\Vect{t}
	\left[
		\Vect{X}_\Vect{t}(\Vect{t}) + \Vect{\epsilon}
	\right]
	\exp
	\left[
		- \frac{i}{2} \Vect{\epsilon}^\intercal \, \Mat{A}_\Vect{t} \Mat{B}_\Vect{t}^{-1} \, \Vect{\epsilon}
		- i \Vect{\epsilon}^\intercal \Vect{K}_\Vect{t}(\Vect{t})
	\right]
	\label{eq:7_UpsilonINT_Bneq0}
\end{align}

The contribution from $\Vect{t}$ is thus obtained as
\begin{equation}
	\psi_\Vect{t}
	=
	\MTnorm \Upsilon_\Vect{t}
	.
	\label{eq:7_psi_t}
\end{equation}

\noindent Finally, each branch of $\Vect{k}(\Vect{x})$ is summed over to obtain the MGO solution via \Eq{eq:6_MGOsol}:\begin{equation}
	\psi(\Vect{x})
	= \sum_{\Vect{t} \in \Vect{\tau}(\Vect{x})}
	\psi_\Vect{t}
	.
	\label{eq:7_MGOsolREP}
\end{equation} 

This final step may be difficult to perform numerically, because one must be able to identify the multivaluedness of the ray map when interpolating $\Vect{\tau}(\Vect{x})$. Essentially, ray discretization can obscure this feature through a type of aliasing, yielding a sampling of $\Vect{\tau}(\Vect{x})$ that appears single-valued but highly oscillatory as the discretization randomly samples from the different branches. Using an inverse ray-tracing framework in place of the standard forward ray-tracing framework can improve this situation by decoupling the field evaluation points from the ray discretization (sampling). Moreover, as shown by \Refs{Colaitis19a,Colaitis21}, reducing the influence of ray-discretization noise can result in large improvements in computational efficiency, even when accounting for the additional operations required by inverse ray-tracing, because less rays are needed to obtain the same field information. Inverse ray-tracing also opens the possibility for using complex rays that might allow MGO to model the evanescent fields that occur in caustic shadow regions. This is something that remains to be investigated.


\section{One-dimensional examples}

I shall first investigate a series of examples in $1$-D for simplicity.


\subsection{Plane wave}

As a first example, let us consider a plane wave propagating in a uniform medium. For simplicity, I consider $1$-D propagation governed by the one-way wave equation
\begin{equation}
	i\pd{x} \psi(x) + \psi(x) = 0
	.
	\label{eq:7_ex1Wave}
\end{equation}

\noindent There is no caustic in this case, and \Eq{eq:7_ex1Wave} is easy to integrate even without using MGO. However, this example is instructive to illustrate the MGO machinery even when $\det{\Mat{B}_\Vect{t}} = 0$ with relatively little algebra.


\subsubsection{Tracing rays}

Let us start by writing \Eq{eq:7_ex1Wave} in the integral form \eq{eq:2_intWAVE}. The corresponding kernel $D(x, x')$ can be written as
\begin{equation}
	D(x, x') 
	= 
	i \pd{x'} \delta(x - x') 
	- \delta(x - x') 
	,
\end{equation}

\noindent so the Weyl symbol \eq{eq:2_WWTkernel} is as follows:
\begin{equation}
	\Symb{D}(x, k) = k - 1 .
\end{equation}

\noindent The corresponding ray equations \eq{eq:2_GOrays} are
\begin{equation}
	\pd{\tau} x(\tau) = 1 
	, \quad
	\pd{\tau} k(\tau) = 0
	,
	\label{eq:7_ex1RayEQ}
\end{equation}

\noindent with solutions given by
\begin{equation}
	x(\tau) = \tau 
	, \quad
	k(\tau) = 1
	,
\end{equation}

\noindent where the integration constants have been chosen to satisfy $\Symb{D}[x(0), k(0)] = 0$. Since $\tau(x) = x$ is single-valued, $\psi(x)$ will be absent of caustics.


\subsubsection{Computing $\MTnorm$ via \Eq{eq:6_MTnormFINAL} or \Eq{eq:6_MTnormFINAL_Bneq0}}

I first compute the integral
\begin{equation}
	 \int_0^{t} \dd \xi \, k(\xi ) \dot{x}(\xi)
	 =
	 \int_0^{t} \dd \xi
	 =
	 t
	 .
\end{equation}

\noindent Next, since $\pd{\tau} \Vect{z} = (1, 0)^\intercal$, the QR decomposition trivially yields
\begin{equation}
	\Mat{Q}_t
	\equiv
	\begin{pmatrix}
		A_t \\
		B_t
	\end{pmatrix}
	= 
	\begin{pmatrix}
		1 \\
		0
	\end{pmatrix}
	, \quad
	R_t = 1
	.
\end{equation}

\noindent Hence, the rank of $B_t$ is zero \ie $\rho = 0$, such that $\Lambda_{\rho \rho}$ is empty and correspondingly $\det \Lambda_{\rho \rho} = 1$ by convention. Since $a_{\varsigma \varsigma} \equiv A_t = 1$, I therefore compute $\MTnorm$ via \Eq{eq:6_MTnormFINAL} to be
\begin{equation}
	\MTnorm
	= 
	\psi_\textrm{in} \sqrt{\dot{x}(0)}
	\,
	\frac{
		\sigma_t \,
		\exp\left[i \int_0^{t} \dd \xi \, k(\xi) \dot{x}(\xi) \right]
	}{
		(-2\pi i)^{\rho/2}
		\sqrt{\det \Lambda_{\rho\rho} \det a_{\varsigma \varsigma} \det R_t }
	}
	= \psi_\textrm{in}
	\exp\left(i t \right)
	,
\end{equation}

\noindent where I have used the fact that $\sigma_t = 1$ for all $t$ as there is no branch-cut crossings to consider.


\subsubsection{Computing $\Upsilon_\Vect{t}$ via \Eq{eq:7_UpsilonINT} or \Eq{eq:7_UpsilonINT_Bneq0}}

Since $\rho = 0$, the integral of $\Upsilon_\Vect{t}$ is empty and therefore equal to $1$ by convention. Hence, I trivially obtain
\begin{equation}
	\Upsilon_\Vect{t}
	= 1
	.
\end{equation}


\subsubsection{Summing over rays}

Since the ray map is the identity, \ie $\tau(x) = x$, there is only a single ray at each position $x$. As such, the summation over branches is trivially performed to yield the MGO solution
\begin{equation}
	\psi(x) = 
	\sum_{t \in \tau(x)}
	\psi_\textrm{in}
	\exp\left(i t \right)
	=
	c \, \exp\left(i q \right)
	,
	\label{eq:7_planeSOL}
\end{equation}

\noindent where $c \equiv \psi_\textrm{in}$ is an arbitrary constant. Equation \eq{eq:7_planeSOL} is an exact solution of \Eq{eq:7_ex1Wave}, which is anticipated because \eq{eq:7_ex1Wave} is a first-order equation and thus coincides with its GO approximation.


\subsection{Airy's equation}

As a second example, let us consider a simple fold caustic in $1$-D, which occurs when a wave encounters an isolated cutoff. For slowly varying media, this situation is often modeled with Airy's equation~\cite{Kravtsov93},
\begin{equation}
	\pd{x}^2 \psi(x) - x \psi(x) = 0 
	.
	\label{eq:7_airy}
\end{equation}


\subsubsection{Tracing rays}

Like with \Eq{eq:2_hilbertWAVE}, I can also write \Eq{eq:7_airy} as
\begin{equation}
	\oper{D}(\VectOp{z}) \ket{\psi} = \ket{0} 
	,
	\quad
	\oper{D}(\VectOp{z}) \doteq \oper{k}^2 + \oper{x} 
	.
\end{equation}

\noindent The Weyl symbol of $\oper{D}(\VectOp{z})$ is therefore calculated to be
\begin{equation}
	\Symb{D}(\Vect{z}) \doteq \WeylInv \left[ \oper{D}(\VectOp{z}) \right] = k^2 + x 
	.
	\label{eq:7_airyDman}
\end{equation}

\noindent In this case, the dispersion manifold $\Symb{D}(\Vect{z}) = 0$ is a parabola which opens along the negative $x$-axis.

By \Eq{eq:2_GOrays}, the rays are obtained by integrating
\begin{equation}
	\pd{\tau} x
	= 2 k
	, \quad
	\pd{\tau} k
	= -1
	.
	\label{eq:7_airyRAYeq}
\end{equation}

\noindent The solution to \Eq{eq:7_airyRAYeq} is
\begin{equation}
	x(\tau) = -(k_0 - \tau)^2
	,
	\quad
	k(\tau) = k_0 - \tau
	,
	\label{eq:7_airyRAYS}
\end{equation}

\noindent where $k_0$ is a constant determined by initial conditions. Note that the ray map
\begin{equation}
	\tau(x) =
	k_0
	\pm 
	\sqrt{- x}
	\label{eq:7_airyTAU}
\end{equation}

\noindent is double-valued when $x < 0$ and has no solution when $x > 0$. Hence, the caustic occurs at $x = 0$.


\subsubsection{Computing $\MTnorm$ via \Eq{eq:6_MTnormFINAL} or \Eq{eq:6_MTnormFINAL_Bneq0}}

I compute the integral
\begin{equation}
	\int_0^{t} \dd \xi \, k(\xi ) \dot{x}(\xi)
	=
	2 \int_0^{t} \dd \xi
	(k_0 - \xi)^2
	=
	2 k_0^2 t 
	- 2 k_0 t^2
	+ \frac{2}{3} t^3
	.
\end{equation}

\noindent Next, a QR decomposition of $\pd{\tau} \Vect{z} = (2 k(\tau), -1)^\intercal$ yields
\begin{equation}
	\Mat{Q}_t
	\equiv
	\begin{pmatrix}
		A_t \\
		B_t
	\end{pmatrix}
	= 
	\frac{1}{ \vartheta_t }
	\begin{pmatrix}
		2k(t) \\
		-1
	\end{pmatrix}
	, \quad
	R_t = \vartheta_t
	,
\end{equation}

\noindent where I have defined
\begin{equation}
	\vartheta_t \doteq \sqrt{1 + 4 k^2(t) } 
	.
\end{equation}

\noindent Hence $B_t$ is full rank for all $t$, \ie $\varsigma = 0$, such that $a_{\varsigma \varsigma}$ is empty and correspondingly $\det a_{\varsigma \varsigma} = 1$ by convention. Then, since $\Lambda_{\rho \rho} = B_t$, I compute
\begin{equation}
	\MTnorm
	= 
	c_0
	\,
	\frac{
		\sigma_t \,
		\exp\left[
			i \int_0^{t} \dd \xi \, k(\xi) \dot{x}(\xi) 
		\right]
	}{
		(-2\pi i)^{\rho/2}
		\sqrt{\det \Lambda_{\rho\rho} \det a_{\varsigma \varsigma} \det R_t }
	}
	=
	\tilde{c}_0
	\,
	\exp\left[
		2 i k_0^2 t 
		- 2 i k_0 t^2
		+ i \frac{2}{3} t^3
	\right]
	,
\end{equation}

\noindent where $c_0$ and $\tilde{c}_0 \doteq i \sigma_t c_0/ \sqrt{-2 \pi i}$ are constants that will be matched to initial conditions. I have again used the fact that $\sigma_t$ is constant due to the lack of branch-cut crossings.


\subsubsection{Computing $\Upsilon_\Vect{t}$ via \Eq{eq:7_UpsilonINT} or \Eq{eq:7_UpsilonINT_Bneq0}}

Since $B_t$ is full rank and diagonal (trivially), $\Upsilon_t$ takes the simpler form
\begin{align}
	\Upsilon_t
	&=
	\int_{\cont{0}} \dd \epsilon 
	\,
	\Psi_t
	\left[
		X_t(t) + \epsilon
	\right]
	\exp
	\left[
		- \frac{i A_t }{2 B_t} \epsilon^2
		- i K_t(t) \epsilon
	\right]
	\nonumber\\
	&=
	\int_{\cont{0}} \dd \epsilon 
	\,
	\Psi_t
	\left[
		X_t(t) + \epsilon
	\right]
	\exp
	\left[
		i k(t) \epsilon^2
		- i \frac{k^2(t)}{ \vartheta(t) } \epsilon
	\right]
	,
\end{align}

\noindent where I have used
\begin{equation}
	K_t(t)
	= - B_t x(t) + A_t k(t)
	=
	\frac{k^2(t)}{ \vartheta(t) }
	.
\end{equation}

Next, I must calculate the tangent-plane wavefield $\Psi_t$. I first focus on calculating the phase $\Theta_t$. Noting that
\begin{equation}
	\Stroke{\Vect{Z}}_t(\tau) 
	=
	\Mat{S}_t
	\Vect{z}(\tau)
	=
	\frac{1}{ \vartheta_t }
	\begin{pmatrix}
		2k(t) & -1 \\
		1 & 2k(t)
	\end{pmatrix}
	\begin{pmatrix}
		- k^2(\tau) \\
		k(\tau)
	\end{pmatrix}
	= \frac{1}{ \vartheta_t }
	\begin{pmatrix}
		- k(\tau) - 2k(t) k^2(\tau) \\
		2 k(t) k(\tau) - k^2(\tau)
	\end{pmatrix}
	,
\end{equation}

\noindent the tangent-plane ray map is given by
\begin{equation} 
	\tau(X_t)
	=
	k_0 + \frac{
		1 \pm \sqrt{1 - 8k(t)\vartheta_t X_t }
	}{
		4 k(t)
	}
	,
\end{equation}

\noindent or equivalently
\begin{equation}
	k(X_t) =
	\frac{
		- 1 \pm \sqrt{1 - 8k(t)\vartheta_t X_t }
	}{
		4 k(t)
	}
	.
\end{equation}

\noindent Hence, 
\begin{equation}
	K_t(X_t) \equiv
	K_t[\tau(X_t)]
	=
	\frac
	{
		4k(t) X_t + \vartheta_t \left[-1 \pm \sqrt{1 - 8k(t) \vartheta_t X_t} \right]
	}
	{
		8k^2(t)
	}
	.
	\label{eq:7_airyPFIELD}
\end{equation}

\noindent One therefore computes $\Theta_t$ by direct integration
\begin{equation}
	\Theta_t[\epsilon + X_t(t)]
	=
	\int_{X_t(t)}^{X_t(t) + \epsilon}
	K_t[\tau(X_t)] \dd X_t
	=
	\frac{8k^4(t) - \vartheta_t^4}{8k^2(t) \vartheta_t} \epsilon
	+ \frac{1}{4k(t)} \epsilon^2
	+ \frac{\vartheta_t^6 - \left[\vartheta_t^4 - 8k(t) \vartheta_t \epsilon \right]^{3/2}}{96k^3(t)}
	,
\end{equation}

\noindent where I have restricted $K_t(X_t)$ to the branch where $K_t[X_t(t)] = K_t(t)$ by choosing the $(+)$~sign in \Eq{eq:7_airyPFIELD}. Also as a reminder, $\epsilon \doteq X_t - X_t(t)$.

Next, I calculate the tangent-plane envelope $\Phi_t$. Since
\begin{equation}
	J_t(X_t) = 
	\frac{1 + 4k(t) k(X_t)}{\vartheta_t}
	=
	\frac{\sqrt{1 - 8k(t)\vartheta_t X_t }
	}{\vartheta_t}
	,
\end{equation}

\noindent (where again, I have chosen the $+$ sign for the ray map) I compute
\begin{equation}
	\Phi_t[\epsilon + X_t(t)]
	=
	\frac{ \vartheta_t }{
		\left[
			\vartheta_t^4 - 8k(t)\vartheta_t \epsilon
		\right]^{1/4}
	}
	.
\end{equation}

\noindent Hence, the tangent-space wavefield is
\begin{equation}
	\Psi_t[\epsilon + X_t(t)]
	=
	\frac{ \vartheta_t }{
		\left[
			\vartheta_t^4 - 8k(t)\vartheta_t \epsilon
		\right]^{1/4}
	}
	\exp
	\left\{
		i \frac{8k^4(t) - \vartheta_t^4}{8k^2(t) \vartheta_t} \epsilon
		+ i \frac{\epsilon^2}{4k(t)}
		+ i \frac{\vartheta_t^6 - \left[\vartheta_t^4 - 8k(t) \vartheta_t \epsilon \right]^{3/2}}{96k^3(t)}
	\right\}
	.
\end{equation}

\noindent As such, $\Upsilon_t$ takes the form
\begin{align}
	\Upsilon_t
	=
	\int_{\cont{0}} 
	\frac{ \dd \epsilon \, \vartheta_t }{
		\left[
			\vartheta_t^4 - 8k(t)\vartheta_t \epsilon
		\right]^{1/4}
	}
	\exp
	\left\{
		- i \frac{\vartheta_t^3}{8k^2(t)} \epsilon
		+ i \frac{\vartheta_t^2}{4k(t)} \epsilon^2
		+ i \frac{\vartheta_t^6 - \left[\vartheta_t^4 - 8k(t) \vartheta_t \epsilon \right]^{3/2}}{96k^3(t)}
	\right\}
	.
	\label{eq:7_airyUPSILON}
\end{align}

In \Ch{ch:GF}, \Eq{eq:7_airyUPSILON} is computed numerically using Gauss--Freud quadrature. Here, I shall adopt a different strategy and attempt an analytical approximation. Specifically, let us approximate the envelope by its value at $\epsilon = 0$, and let us expand the integrand phase to cubic order in $\epsilon$ (since one expects the $A_2$ normal form). One therefore obtains the approximate expression
\begin{equation}
	\Upsilon_t
	\approx
	\int_{\cont{0}} 
	\dd \epsilon \,
	\exp
	\left[
		i k(t) \epsilon^2 
		- i \frac{\epsilon^3}{3 \vartheta_t^3 }
	\right]
	.
	\label{eq:7_airyUPSILONsaddles}
\end{equation}

\noindent Since the coefficient of $\epsilon^2$ vanishes when $t = k_0$, while that of $\epsilon^3$ remains nonzero for all $t$, \Eq{eq:7_airyUPSILONsaddles} contains a pair of coalescing saddle points. By making the substitution $\varepsilon \doteq i \epsilon/ \vartheta(t) - i k(t) \vartheta^2(t)$ and using Cauchy's integral theorem, the integral of \Eq{eq:7_airyUPSILONsaddles} is placed into standard form
\begin{align}
	\int_{\cont{0}} \dd \epsilon \,
	\exp\left[
		i k(t) \epsilon^2
		- i \frac{\epsilon^3}{3 \vartheta_t^3 } 
	\right] 
	= - i \vartheta_t
	\int_{\cont{s(t)}} \dd \varepsilon \,
	\exp\left[
		i \frac{2}{3} k^3(t) \vartheta_t^6 
		+ k^2(t)\vartheta_t^4 \varepsilon 
		+ \frac{\varepsilon^3}{3 } 
	\right] 
	,
	\label{eq:7_airySTANDARD}
\end{align}

\noindent where I have introduced
\begin{equation}
	s(t) \doteq \text{sign}\left[ k(t) \right] = \text{sign}\left( k_0 - t \right) 
	.
\end{equation}

\begin{figure}
	\centering
	\includegraphics[width=0.6\linewidth,trim={0mm 0mm 0mm 0mm},clip]{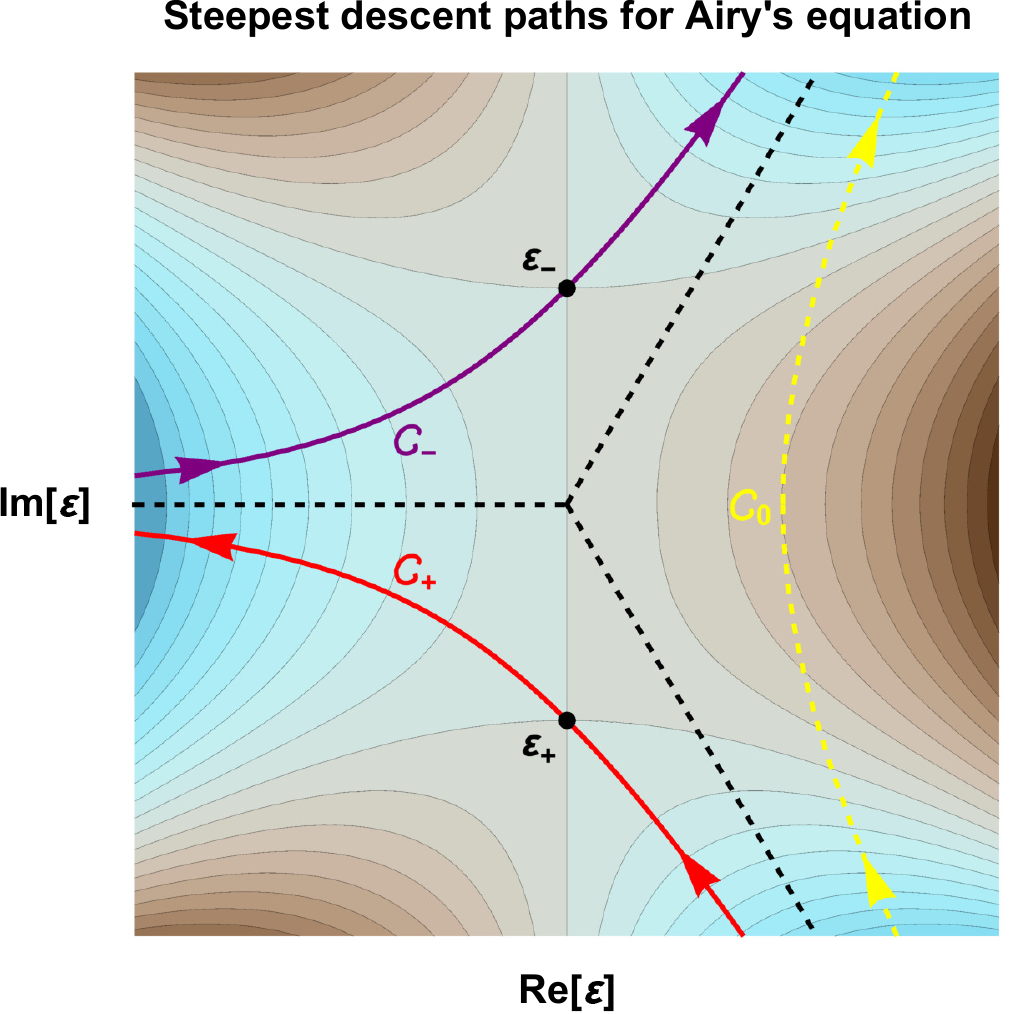}
	\caption{The function $f(\varepsilon; x) = x \varepsilon + \frac{\varepsilon^3}{3}$ in the complex $\varepsilon$ plane with $x \ge 0$, which serves as the phase in Airy's integral \eq{eq:7_airySTANDARD}. The background color depicts $\text{Re}\left[ f(\varepsilon; x) \right]$, with blue denoting negative values. The points $\varepsilon_\pm = \mp i \sqrt{|x|}$ are the saddlepoints where $\pd{\varepsilon} f(\varepsilon; x) = 0$. Correspondingly, the standard contour for Airy's integral, $\cont{0}$, can be decomposed into the two steepest descent contours $\cont{\pm}$.}
	\label{fig:7_airyCONT}
\end{figure}

\noindent The contours $\cont{+}$ and $\cont{-}$, shown in \Fig{fig:7_airyCONT}, correspond to the steepest-descent contours that pass through the saddlepoint $\varepsilon_{\pm} \doteq \mp i | k(t) \vartheta_t^2|$. (Note that their union creates the standard Airy contour, denoted $\cont{0}$ in the figure.) The determination of the desired contour for \Eq{eq:7_airySTANDARD} can be seen as follows: when $k > 0$, $\epsilon = 0$ corresponds to $\varepsilon = -ik\vartheta^2 = -i|k\vartheta^2| = \epsilon_+$, while when $k < 0$, $\epsilon = 0$ corresponds to $\varepsilon = -ik\vartheta^2 = +i|k\vartheta^2| = \epsilon_-$. 

The obtained contour integral can be evaluated analytically; indeed, it can be readily proven that
\begin{equation}
	\int_{\cont{\pm}} \hspace{-2mm}\dd \varepsilon \,
	\exp\left(
		\frac{\varepsilon^3}{3 }
		- x \varepsilon 
	\right)
	= i \pi \left[\airyA(x) \pm i \airyB(x) \nullFrac\right] 
	,
\end{equation}

\noindent where $\airyA(x)$ and $\airyB(x)$ are the Airy functions of the first and second kind, respectively~\cite{Olver10a}. Hence, I obtain
\begin{align}
	\Upsilon_t
	\approx
	\pi \vartheta_t
	\exp\left[
		i \frac{2}{3} k^3(t) \vartheta_t^6 
	\right]
	\left\{
		\airyA\left[
			- k^2(t)\vartheta_t^4
		\right]
		+ i s(t) \airyB\left[
			- k^2(t)\vartheta_t^4
		\right] \nullFrac 
	\right\}
	.
	\label{eq:7_airyUPSILONapprox}
\end{align}

\noindent As a short remark, note that including higher-order terms in the Taylor expansion of $\Phi_t(\epsilon)$ around $\epsilon = 0$ will result in additional terms in \Eq{eq:7_airyUPSILONapprox} proportional to the corresponding higher-order derivatives of $\airyA$ and $\airyB$. It would be interesting to explore the effect of such additional terms on the MGO solution; in particular, it would be enlightening to see if retaining the first-order derivative term leads to an MGO analogue of the standard uniform approximations (\ie Ludwig--Kravtsov expansion~\cite{Ludwig66,Kravtsov93}) often used in catastrophe optics.


\subsubsection{Summing over rays}

As the final step, I sum over all branches of $\tau(x)$. Using \Eq{eq:7_airyTAU}, I obtain the MGO solution
\begin{align}
	\psi(x)
	&=
	\sqrt{1 - 4 x}
	\left\{
		\airyA\left[- \varrho^2(x) \right]
		\cos[ \varpi(x)]
		- \airyB\left[ - \varrho^2(x) \right]
		\sin[ \varpi(x) ]
	\right\}
	,
	\label{eq:7_airyMGO}
\end{align}

\noindent where I have defined
\begin{align}
	\varrho(x) \doteq
	(1 - 4x) \sqrt{-x}
	, \quad
	\varpi(x) 
	\doteq 
	\frac{2}{3} \varrho^3(x) - \frac{2}{3}(-x)^{3/2}
	,
\end{align}

\begin{figure}
	\centering
	\includegraphics[width=0.6\linewidth]{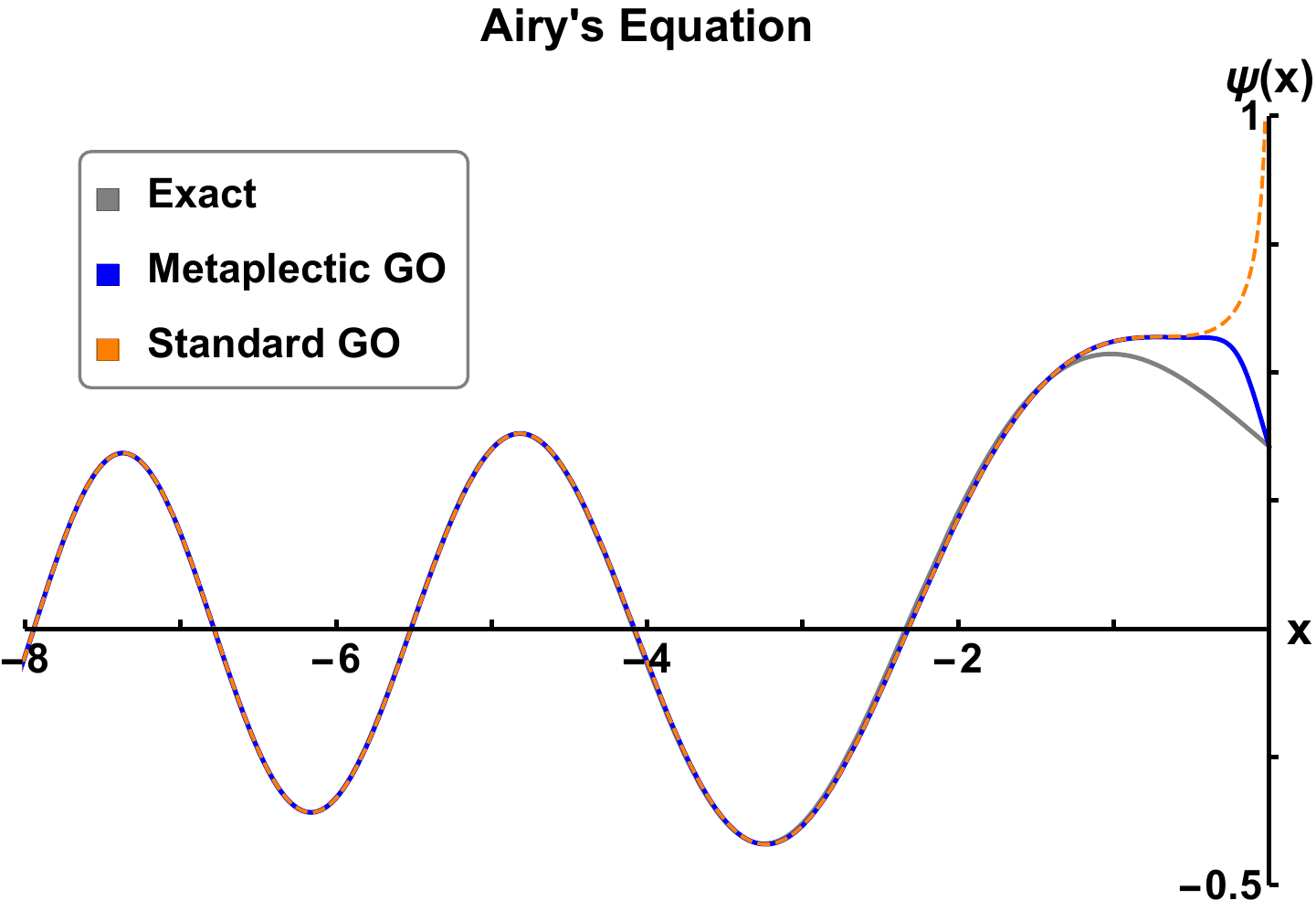}
	\caption{Comparison between the Airy function $\textrm{Ai}(x)$, its standard GO approximation [\Eq{eq:7_airyWKB}], and its MGO approximation [\Eq{eq:7_airyMGO}]. The caustic is located at $x = 0$. The error in the MGO solution is largely due to the approximations made when performing the inverse MT (which may not be needed for certain applications). Also, recall that the numerical MGO quadrature rule described in \Ch{ch:GF} agrees with the exact solution even better than the approximate analytical MGO expression derived here; see \Fig{fig:5_MGOairy}.}
	\label{fig:7_airySOL}
\end{figure}

\noindent and I have chosen $\tilde{c}_0 = \exp\left( - i \frac{2}{3} k_0^3 \right)/2\pi$ to enforce that $\psi(0) = \airyA(0)$. (Note that this is only possible because the MGO solution remains finite at caustics; it is impossible to enforce such a boundary condition on the standard GO solution because it diverges at $x = 0$.) Figure~\ref{fig:7_airySOL} compares \Eq{eq:7_airyMGO} with the exact solution,
\begin{equation}
	\psi_\text{exact}(x) = \textrm{Ai}(x) 
	,
\end{equation}

\noindent and with the standard GO approximation~\cite{Olver10a},
\begin{equation}
	\psi_\text{GO} = \pi^{-1/2} (-x)^{-1/4} 
	\sin\left[\frac{2}{3} (-x)^{3/2} + \frac{\pi}{4} \right] 
	.
	\label{eq:7_airyWKB}
\end{equation}

\noindent As can be seen, the MGO solution is almost indistinguishable from the standard GO approximation far from the caustic at $x = 0$. However, in contrast to the standard GO solution, my solution remains finite for all $x \le 0$, like the exact solution of \Eq{eq:7_airy}.


\subsection{Weber's equation}

Next, let us consider a bounded wave in a $1$-D harmonic potential, which exhibits two adjacent fold caustics. This situation is described mathematically by Weber's equation,
\begin{equation}
	\pd{x}^2 \psi(x) + \left(2\nu + 1 - x^2 \right)\psi(x) = 0 
	,
	\label{eq:7_parabEQ}
\end{equation}

\noindent which is also the Schr\"odinger equation for a quantum harmonic oscillator~\cite{Shankar94}. Equation \eq{eq:7_parabEQ} can be written as
\begin{equation}
	\oper{D}(\VectOp{z})\ket{\psi} = \ket{0} 
	,
	\quad
	\oper{D}(\VectOp{z}) \doteq \oper{k}^2 + \oper{x}^2 - 2E \, \IdentOp 
	,
\end{equation}

\noindent where $E \doteq \nu + 1/2$. The Weyl symbol is readily computed to be
\begin{equation}
	\Symb{D}(\Vect{z}) = k^2 + x^2 - 2E 
	.
\end{equation}

\noindent In this case, the dispersion manifold $\Symb{D}(\Vect{z}) = 0$ is a circle of radius $R \doteq \sqrt{2E}$.


\subsubsection{Tracing rays}

The ray equations are given as
\begin{equation}
	\pd{\tau} x = 2 k 
	,
	\quad
	\pd{\tau} k = -2 x 
	.
\end{equation}

\noindent Their solutions have the form
\begin{equation}
	x(\tau) = R \cos \left( 2 \tau \right) 
	,
	\quad
	k(\tau) = - R \sin \left( 2 \tau \right) 
	,
\end{equation}

\noindent where I have assumed the initial condition $q(0) = R$.


\subsubsection{Computing $\MTnorm$ via \Eq{eq:6_MTnormFINAL} or \Eq{eq:6_MTnormFINAL_Bneq0}}

I compute the integral
\begin{equation}
	\int_0^t \dd \xi \,
	k(\xi) \dot{x}(\xi)
	=
	2 R^2 
	\int_0^t \dd \xi \,
	\sin^2 \left( 2 \tau \right) 
	=
	t R^2
	- \frac{R^2}{4} \sin(4 t) 
	.
\end{equation}

\noindent Next, a QR decomposition of $\pd{\tau} \Vect{z} = (- 2 R \sin (2 t) , - 2 R \cos(2 t) )^\intercal$ yields

\begin{equation}
	\Mat{Q}_t
	\equiv
	\begin{pmatrix}
		A_t \\
		B_t
	\end{pmatrix}
	= 
	\begin{pmatrix}
		- \sin (2 t) \\
		- \cos(2 t)
	\end{pmatrix}
	, \quad
	R_t = 2 R
	,
\end{equation}

\noindent Unlike the previous example, $B_t$ can now change sign, and consequently, $\sigma_t$ will change sign as well. Let us choose to have $B_t$ cross the branch cut whenever $B_t$ changes from positive to negative. This is encapsulated by the phase convention
\begin{equation}
	B_t 
	= |B_t| \exp
	\left( 
		i
		\left\lfloor
			\frac{4t - \pi}{2\pi}
		\right\rfloor
		\pi
	\right) 
	,
\end{equation}

\noindent where $\lfloor \, \rfloor$ denotes the floor operation. Hence, choosing
\begin{equation}
	\sigma_t = 
	\exp
	\left( 
		-i
		\left\lfloor
			\frac{4t + \pi}{4\pi} 
		\right\rfloor
		\pi
	\right)
\end{equation}

\noindent will ensure continuity in $t$ across the branch cut. Thus, I obtain
\begin{equation}
	\MTnorm
	= 
	c_0
	\,
	\frac{
		\sigma_t \,
		\exp\left[
			i \int_0^{t} \dd \xi \, k(\xi) \dot{x}(\xi) 
		\right]
	}{
		(-2\pi i)^{\rho/2}
		\sqrt{\det \Lambda_{\rho\rho} \det a_{\varsigma \varsigma} \det R_t }
	}
	=
	c_0
	\,
	\frac{
		\exp\left[
			i t R^2
			- i \frac{R^2}{4} \sin(4 t) 
			-i
			\left\lfloor
				\frac{4t + \pi}{4\pi} 
			\right\rfloor
			\pi
		\right]
	}{
		\sqrt{-2\pi i}
		\sqrt{
			- 2 R \cos(2 t)
		}
	}
	,
\end{equation}

\noindent where I note that the square-root functions are single-valued, as the branch cut has been imposed through my choice of $\sigma_t$.


\subsubsection{Computing $\Upsilon_\Vect{t}$ via \Eq{eq:7_UpsilonINT} or \Eq{eq:7_UpsilonINT_Bneq0}}

\noindent Again, $\Upsilon_t$ takes the simpler form
\begin{align}
	\Upsilon_t
	&=
	\int_{\cont{0}} \dd \epsilon 
	\,
	\Psi_t
	\left[
		X_t(t) + \epsilon
	\right]
	\exp
	\left[
		- \frac{i A_t }{2 B_t} \epsilon^2
		- i K_t(t) \epsilon
	\right]
	\nonumber\\
	&=
	\int_{\cont{0}} \dd \epsilon 
	\,
	\Psi_t
	\left[
		X_t(t) + \epsilon
	\right]
	\exp
	\left[
		- \frac{i }{2} \tan(2 t) \epsilon^2
		- i R \epsilon
	\right]
	,
\end{align}

\noindent where I have used
\begin{equation}
	K_t(t)
	= - B_t x(t) + A_t k(t)
	=
	R
	.
\end{equation}

Next, I must calculate the tangent-plane wavefield $\Psi_t$. I first focus on calculating the phase $\Theta_t$. Noting that
\begin{equation}
	\Stroke{\Vect{Z}}_t(\tau) 
	=
	\Mat{S}_t
	\Vect{z}(\tau)
	=
	\begin{pmatrix}
		- \sin (2 t) & - \cos (2 t) \\
		\cos (2 t) & - \sin (2 t)
	\end{pmatrix}
	\begin{pmatrix}
		R \cos \left( 2 \tau \right)  \\
		- R \sin \left( 2 \tau \right) 
	\end{pmatrix}
	=
	\begin{pmatrix}
		R \sin \left( 2\tau - 2 t \right) \\
		R \cos \left( 2\tau - 2 t \right)
	\end{pmatrix}
	,
\end{equation}

\noindent the tangent-plane ray map is given by
\begin{equation} 
	\tau(X_t)
	=
	t + \frac{1}{2} \arcsin
	\left( \frac{X_t}{R} \right)
	.
\end{equation}

\noindent Hence, 
\begin{equation}
	K_t(X_t) \equiv
	K_t[\tau(X_t)]
	=
	\sqrt{R^2 - X_t^2}
	.
	\label{eq:7_parabPFIELD}
\end{equation}

\noindent One therefore computes $\Theta_t$ by direct integration
\begin{equation}
	\Theta_t[\epsilon + X_t(t)]
	=
	\int_{X_t(t)}^{X_t(t) + \epsilon}
	K_t(X_t) \dd X_t
	=
	\frac{\epsilon}{2}\sqrt{R^2 - \epsilon^2} 
	+ \frac{R^2}{2} \arcsin\left( \frac{\epsilon}{R} \right) 
	.
	\label{eq:7_parabTHETAmet}
\end{equation}

Next, I calculate the tangent-plane envelope $\Phi_t$. Since
\begin{equation}
	J_t(X_t)
	=
	2 \sqrt{R^2 - X_t^2}
	,
\end{equation}

\noindent I compute
\begin{equation}
	\Phi_t[\epsilon + X_t(t)]
	=
	\left[1 - \left( \epsilon / R \right)^2 \right]^{-1/4} 
	.
\end{equation}

\noindent Hence, the tangent-space wavefield is
\begin{equation}
	\Psi_t[\epsilon + X_t(t)]
	=
	\left[1 - \left( \epsilon / R \right)^2 \right]^{-1/4} 
	\exp
	\left[
		i \frac{\epsilon}{2}\sqrt{R^2 - \epsilon^2} 
		+ i \frac{R^2}{2} \arcsin\left( \frac{\epsilon}{R} \right) 
	\right]
	.
\end{equation}

\noindent As such, $\Upsilon_t$ takes the form
\begin{align}
	\Upsilon_t
	=
	\int_{\cont{0}} 
	\frac{ \dd \epsilon }{
		\left[
			1 - \left( \epsilon / R \right)^2
		\right]^{1/4}
	}
	\exp
	\left[
		i \frac{\epsilon}{2}\sqrt{R^2 - \epsilon^2} 
		+ i \frac{R^2}{2} \arcsin\left( \frac{\epsilon}{R} \right)
		- \frac{i }{2} \tan(2 t) \epsilon^2
		- i R \epsilon
	\right]
	.
	\label{eq:7_parabUPSILON}
\end{align}

Before proceeding, note that $\Upsilon_t$ only involves functions which are either constant ($\Phi_t$, $\Theta_t$) or $\pi$-periodic in time ($\tan 2 t$). Thus,
\begin{equation}
	\mc{N}_{t + \pi} \Upsilon_{t + \pi}
	= \MTnorm \Upsilon_{t} \exp\left[i \pi (R^2 - 1) \right] 
	.
\end{equation}

\noindent For this to be single-valued over the dispersion manifold, $R^2 - 1$ must be an even integer, which in turn requires $\nu$ to be an integer. Since $E \ge 0$ is also needed for $R$ to be real, the integer must be nonnegative. All together, this leads to the Bohr--Sommerfeld quantization of Weber's equation, more commonly known as~\cite{Shankar94}
\begin{equation}
	E = \nu + 1/2 
	,
	\quad
	\nu = 0, 1, 2, \ldots
	.
\end{equation}

Returning to $\Upsilon_t$, in \Ch{ch:GF}, \Eq{eq:7_parabUPSILON} is computed numerically using Gauss--Freud quadrature; as with the previous example, here, I shall develop an analytical approximation instead. As before, I approximate the envelope by its value at $\epsilon = 0$, and let us expand the integrand phase to cubic order in $\epsilon$ (since one expects the $A_2$ normal form). One therefore obtains the approximate expression
\begin{equation}
	\Upsilon_t
	\approx
	\int_{\cont{0}} 
	\dd \epsilon \,
	\exp
	\left[
		- i \frac{\tan \left(2 t \right)}{2} \epsilon^2 
		- i \frac{\epsilon^3}{6 R}
	\right]
	.
	\label{eq:7_parabUPSILONsaddles}
\end{equation}

\noindent Equation \eq{eq:7_parabUPSILONsaddles} is of the same basic form as \Eq{eq:7_airyUPSILONsaddles}. Therefore, I immediately conclude that
\begin{align}
	\Upsilon_t \approx 
	\pi (2R)^{1/3} \exp\left[ -i \frac{R^2}{3} \tan^3(2t) \right]
	\left\{ 
		\textrm{Ai}\left[ - \frac{\tan^2(2t)}{4} (2R)^{4/3} \right]
		+ i \, s(t) \textrm{Bi}\left[ - \frac{\tan^2(2t)}{4} (2R)^{4/3} \right]
	\right\} 
	,
\end{align}

\noindent where I have defined
\begin{equation}
	s(t) \doteq - \text{sign}\left[ \tan(2t) \right] = \text{sign}\left[ k(t) \right]
	\,
	\text{sign}\left[ x(t) \right] 
	.
\end{equation}


\subsubsection{Summing over rays}

The final step is to sum over all the rays. I specifically choose to sum over the ray interval $t \in [-\pi/2, \pi/2)$. Then, since $\cos(2 |t|) = x/R$ and $\sin(2 |t|) = \sqrt{1 - x^2/R^2}$,
\begin{align}
	\psi(x)
	&= 
	\frac{
		c_0 \, \sigma_{|t|}
	}{
		2^{2/3} R^{1/6} 
	}
	\frac{ 
		\sqrt{i \pi}
	}{
		\sqrt{ - x/R }
	}
	\exp\left[
		i R^2 |t|
		- \frac{i}{2} x \sqrt{R^2 - x^2}
		- i \frac{R^2}{3} \frac{(R^2 - x^2)^{3/2}}{x^3}
	\right]
	\nonumber\\
	&\hspace{30mm}\times
	\left\{ 
		\textrm{Ai}\left[ - \frac{R^2 - x^2}{x^2 2^{2/3}} R^{4/3} \right]
		- i \, \text{sign}(x) \textrm{Bi}\left[ - \frac{R^2 - x^2}{x^2 2^{2/3}} R^{4/3} \right]
	\right\}
	\nonumber\\
	&\hspace{4mm}+ 
	\frac{
		c_0 \, \sigma_{-|t|}
	}{
		2^{2/3} R^{1/6} 
	}
	\frac{ \sqrt{i \pi}}{\sqrt{ - x/R }}
	\exp\left[
		- i R^2 |t|
		+ \frac{i}{2} x \sqrt{R^2 - x^2}
		+ i \frac{R^2}{3} \frac{(R^2 - x^2)^{3/2}}{x^3}
	\right]
	\nonumber\\
	&\hspace{30mm}\times
	\left\{ 
		\textrm{Ai}\left[ - \frac{R^2 - x^2}{x^2 2^{2/3}} R^{4/3} \right]
		+ i \, \text{sign}(x) \textrm{Bi}\left[ - \frac{R^2 - x^2}{x^2 2^{2/3}} R^{4/3} \right]
	\right\}
	.
\end{align}

\noindent Noting that
\begin{align}
    \sigma_{|t|} = 1 
    ,
    \quad
    \sigma_{-|t|} = \exp\left[ - i \frac{\text{sign}(x) - 1}{2} \pi \right] 
    ,
    \quad
    - x/R = |x/R| \exp\left[ - i \frac{\text{sign}(x) + 1}{2} \pi \right]
    ,
\end{align}

\noindent choosing $c_0$ and using $|t| = \frac{1}{2} \cos^{-1}(x/R)$, I obtain the MGO solution
\begin{align}
	\psi(x) = \frac{ \textrm{Ai}\left[ -\varrho^2(x) \right]\cos \varpi(x)}{\sqrt{|x|}}
	- \text{sign}(x) \frac{ \textrm{Bi}\left[ -\varrho^2(x) \right] \sin \varpi(x)}{\sqrt{|x|}}
	,
	\label{eq:7_parabMGO}
\end{align}

\noindent where I have defined
\begin{subequations}
	\begin{align}
		\varrho(x) &\doteq \frac{ R^{2/3} \sqrt{R^2 - x^2} }{2^{1/3}x}
		, \\
		\varpi(x) &\doteq \frac{x\sqrt{R^2 - x^2}}{2} 
		- \frac{R^2 \cos^{-1} \left(\frac{x}{R} \right)}{2}
		+ \frac{2}{3}\varrho^3(x) 
		+ \frac{\pi}{4} \left[1 - \text{sign}(x) \right]
		.
	\end{align}
\end{subequations}

\begin{figure}
	\centering
	\begin{overpic}[width=0.46\linewidth]{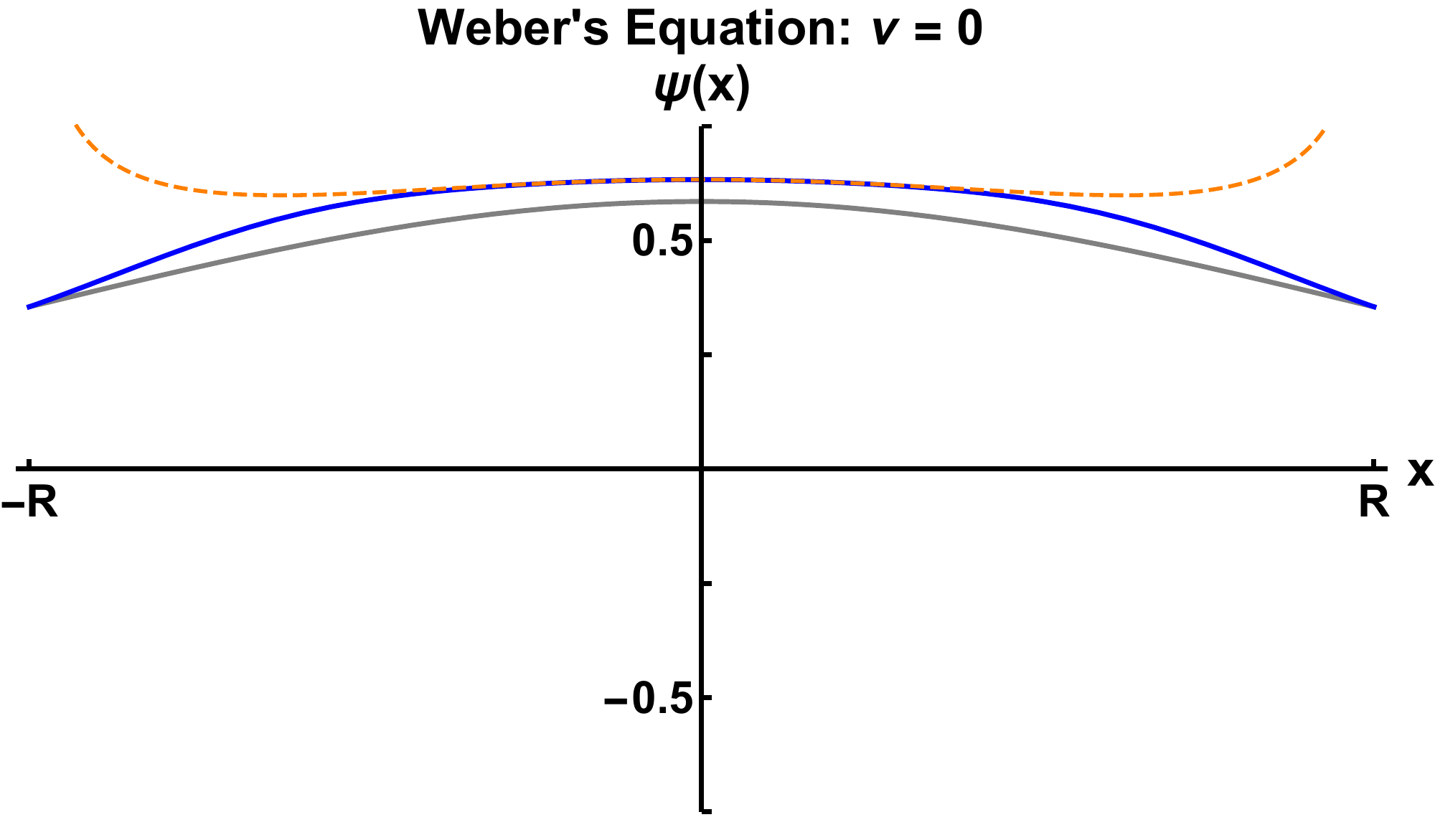}
		\put(5,5){\textbf{\large(a)}}
	\end{overpic}
	\hspace{1mm}
	\begin{overpic}[width=0.46\linewidth]{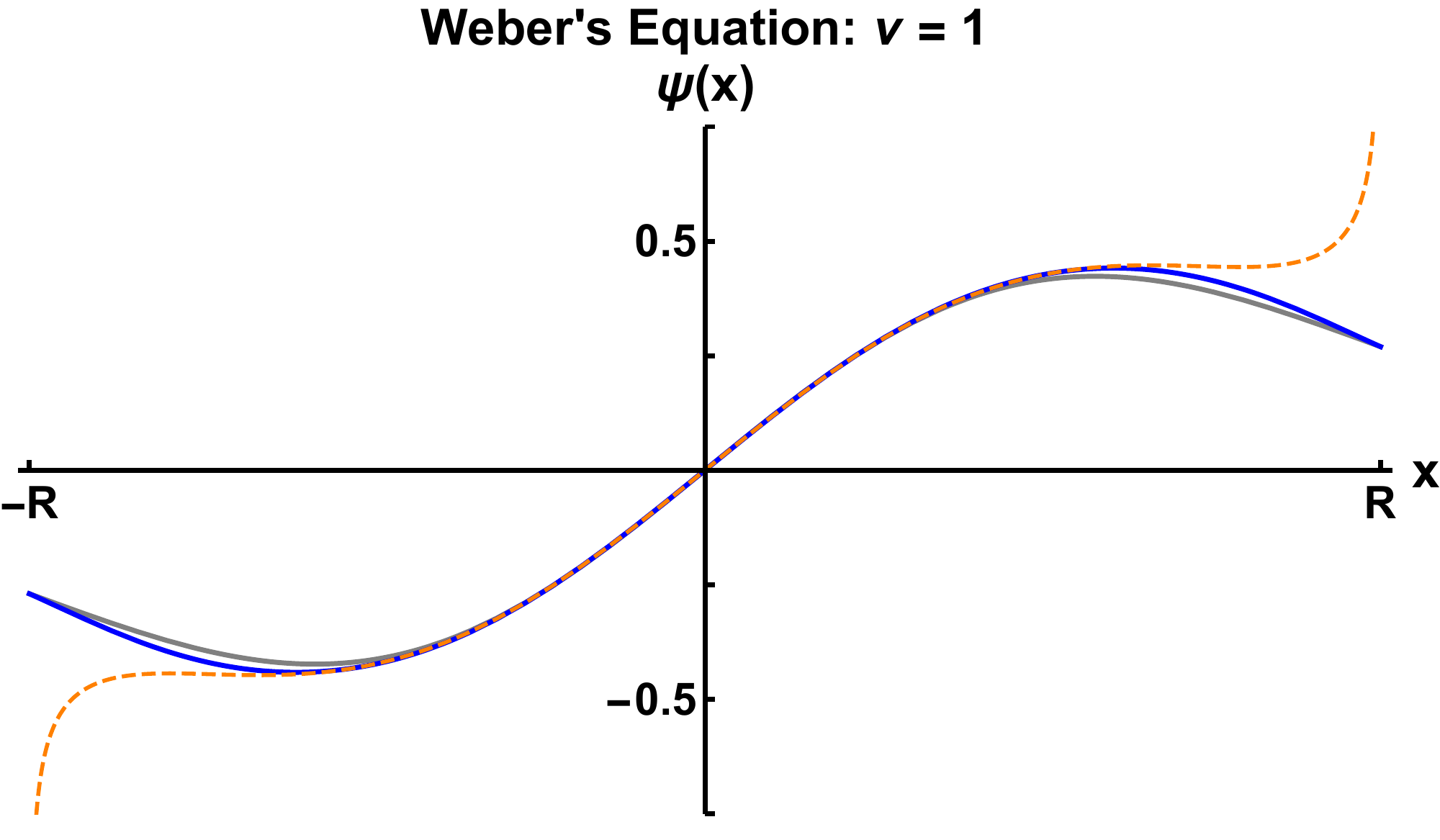}
		\put(5,5){\textbf{\large(b)}}
	\end{overpic}

	\vspace{3mm}
	\begin{overpic}[width=0.46\linewidth]{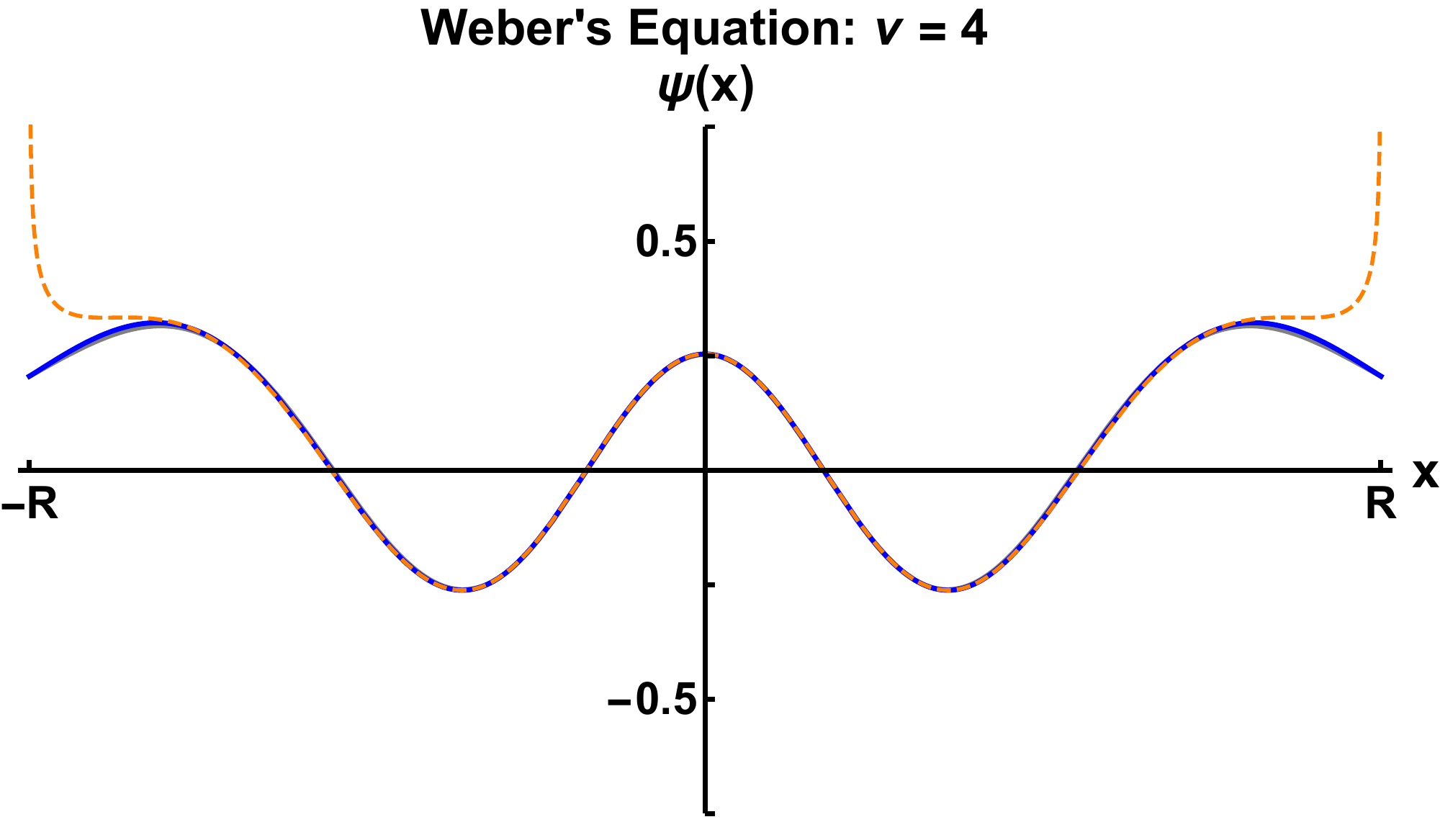}
		\put(5,5){\textbf{\large(c)}}
	\end{overpic}
	\hspace{1mm}
	\begin{overpic}[width=0.46\linewidth]{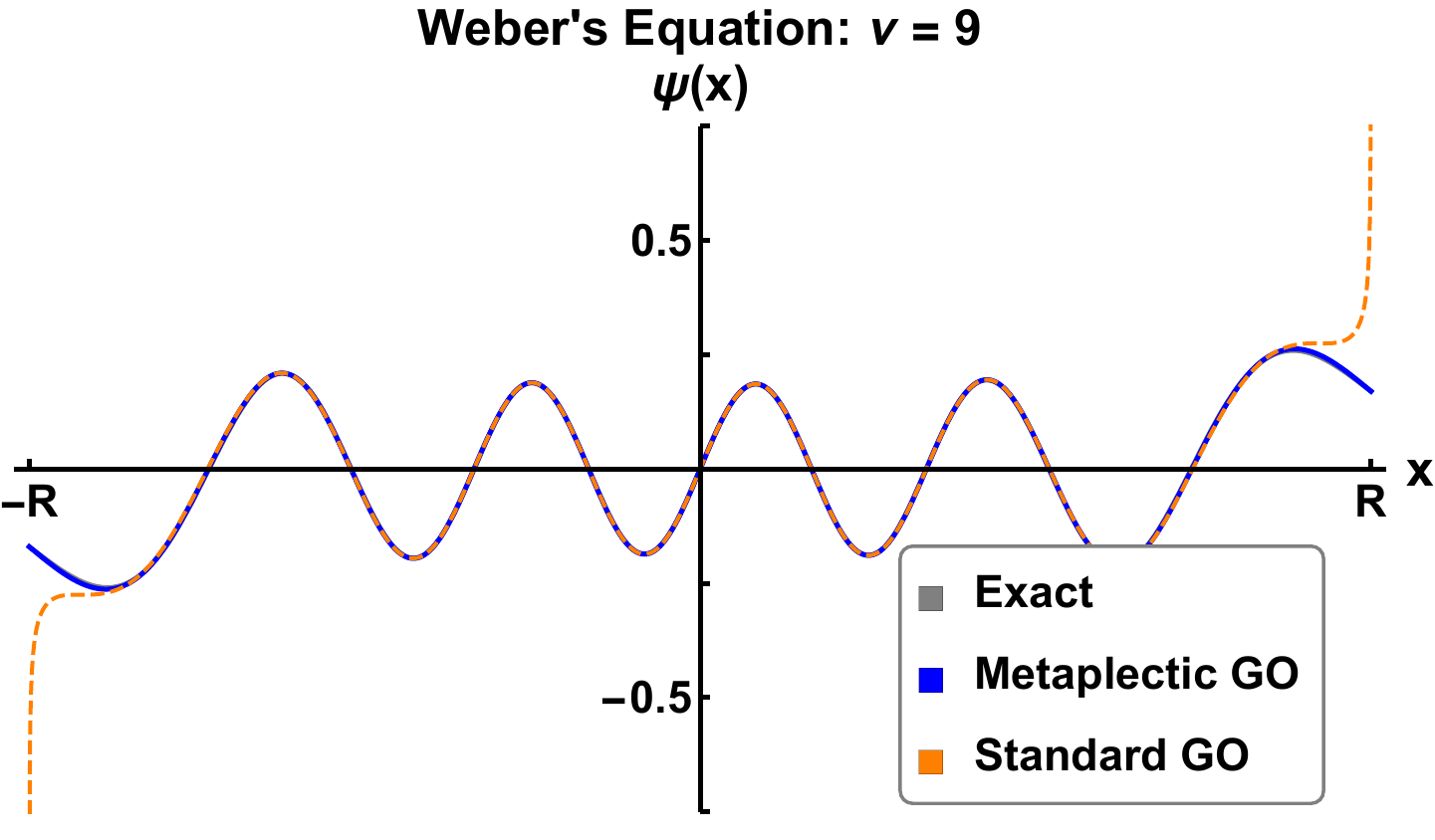}
		\put(5,5){\textbf{\large(d)}}
	\end{overpic}
	\caption{Comparison between the parabolic cylinder function [\Eq{eq:7_parabEXACT}], its standard GO approximation [\Eq{eq:7_parabWKB}], and its MGO approximation [\Eq{eq:7_parabMGO}] for $\nu = 0$, $\nu = 1$, $\nu = 4$, and $\nu = 9$. Two caustics are located at $\pm R$, where $R \doteq \sqrt{2 \nu + 1}$.}
	\label{fig:7_parabSOL}
\end{figure}

Figure~\ref{fig:7_parabSOL} compares \Eq{eq:7_parabMGO} with the exact solution for the boundary condition $\psi(R) = \textrm{Ai}(0)/\sqrt{R}$,
\begin{equation}
	\psi_\text{exact}(x) = \frac{\textrm{Ai}(0)}{\sqrt{R}} \, 
	\frac{
		\textrm{D}_{\nu} (\sqrt{2} \, x)
	}{
		\textrm{D}_\nu(\sqrt{2}\,R) 
	} \, ,
	\label{eq:7_parabEXACT}
\end{equation}

\noindent where $\textrm{D}_\nu(x)$ is Whittaker's parabolic cylinder function~\cite{Olver10a}, and with the standard GO approximation~\cite{Grunwald71},
\begin{equation}
	\psi_\text{GO} = \frac{2^{1/6} \cos\left[ \frac{x}{2}\sqrt{R^2 - x^2} - \frac{R^2 }{2}\cos^{-1}\left(\frac{x}{R} \right)  + \frac{\pi}{4} \right]}{\sqrt{\pi} \, R^{1/3} (R^2 - x^2)^{1/4}} 
	\, .
	\label{eq:7_parabWKB}
\end{equation}

\noindent Already for $\nu = 0$, my MGO solution generally captures the exact solution behavior, albeit with some small error. Beginning from $\nu = 1$, the agreement with the exact solution becomes even more remarkable, for either parity. Just like the previous example, my solution remains finite everywhere, whereas the standard GO solution becomes singular near the cutoffs at $x = \pm R$.


\section{Two-dimensional examples}

Let us also consider a series examples in $2$-D.


\subsection{Airy's equation}

As a first example, let us consider oblique propagation in a linearly stratified medium. (This is the $2$-D analogue of the $1$-D Airy problem discussed above.) Suppose that the wave is described by the Helmholtz-type equation
\begin{equation}
	\pd{\Vect{x}}^2 \psi( \Vect{x} ) 
	+ (k_0^2 - x_1) \psi( \Vect{x} ) = 0
	,
	\label{eq:7_airyEQ2D}
\end{equation}

\noindent where $k_0$ is a constant and $\pd{\Vect{x}}^2 \doteq \pd{x_1}^2 + \pd{x_2}^2$. My coordinate system is such that $x_1$ is aligned with the medium stratification and $x_2$ is transverse to $x_1$. Let us also consider the initial condition
\begin{equation}
	\psi(0, x_2) = c \exp(i k_0 x_2) ,
	\label{eq:7_airyIC2D}
\end{equation}

\noindent where $c$ is an arbitrary constant.


\subsubsection{Tracing rays}

Equation \eq{eq:7_airyEQ2D} can be equivalently written as an integral equation with integration kernel
\begin{equation}
	D(\Vect{x}, \Vect{x}') 
	= 
	- \pd{\Vect{x}'}^2 \delta(\Vect{x} - \Vect{x}') 
	+ (x_1 - k_0^2) \delta(\Vect{x} - \Vect{x}') .
\end{equation}

\noindent The corresponding Weyl symbol is
\begin{equation}
	\Symb{D}(\Vect{x}, \Vect{k}) 
	=
	k_1^2 + k_2^2 + x_1 - k_0^2
	,
\end{equation}

\noindent and the corresponding ray equations are
\begin{align}
	\pd{\tau_1} x_1(\Vect{\tau}) = 2 k_1(\Vect{\tau})
	, \quad
	\pd{\tau_1} x_2(\Vect{\tau}) = 2 k_2(\Vect{\tau})
	, \quad
	\pd{\tau_1} k_1(\Vect{\tau}) = -1
	, \quad
	\pd{\tau_1} k_2(\Vect{\tau}) = 0
	.
\end{align}

\noindent Let us define $\tau_1$ such that $x_1(0, \tau_2) = 0$. Then, the initial condition \eq{eq:7_airyIC2D} implies that $k_2(0, \tau_2) = k_0$, and the local dispersion relation $\Symb{D}[\Vect{x}(0, \tau_2), \Vect{k}(0, \tau_2)] = 0$ requires $k_1(0, \tau_2) = 0$. This leaves $x_2(0, \tau_2)$ undetermined. Since $\tau_2$ must parameterize the initial conditions for the rays, let us choose $x_2(0, \tau_2) = \tau_2$. Hence, the ray solutions are
\begin{align}
	x_1(\Vect{\tau}) 
	= - \tau_1^2
	, \quad
	x_2(\Vect{\tau}) 
	= \tau_2 + 2 k_0 \tau_1
	, \quad
	k_1( \Vect{\tau}) 
	=  - \tau_1 
	, \quad
	k_2( \Vect{\tau}) 
	=  k_0
	.
\end{align}

\noindent The inverse function $\Vect{\tau}(\Vect{x})$ is calculated as
\begin{equation}
	\tau_1(\Vect{x}) = \pm \sqrt{- x_1} 
	, \quad
	\tau_2(\Vect{x}) = x_2 \mp 2 k_0 \sqrt{- x_1}
	.
	\label{eq:7_airyBRANCH2D}
\end{equation}

\noindent Clearly, $\Vect{\tau}(\Vect{x})$ is double-valued, so there are two branches that must ultimately be summed over.


\subsubsection{Computing $\MTnorm$ via \Eq{eq:6_MTnormFINAL} or \Eq{eq:6_MTnormFINAL_Bneq0}}

I compute the integral
\begin{equation}
	\int_0^{t_1} \dd \xi \,
	\Vect{k}(\xi,t_2)^\intercal \dot{\Vect{x}}(\xi, t_2)
	=
	\int_0^{t_1} \dd \xi \,
	\left( 2 \xi^2 + 2 k_0^2 \right)
	=
	\frac{2}{3} t_1^3 + 2 k_0^2 t_1
	.
\end{equation}

\noindent Next, a QR decomposition of the matrix $\pd{\Vect{\tau}} \Vect{z}$ yields
\begin{equation}
	\pd{\Vect{\tau}} \Vect{z}(\Vect{t})
	\equiv
	\begin{pmatrix}
		- 2 t_1 & 0 \\
		2 k_0 & 1 \\
		-1 & 0 \\
		0 & 0
	\end{pmatrix}
	=
	\frac{1}{\vartheta_\Vect{t}}
	\begin{pmatrix}
		- 2 t_1 & 0 \\
		0 & \vartheta_\Vect{t} \\
		- 1 & 0 \\
		0 & 0
	\end{pmatrix}
	\begin{pmatrix}
		\vartheta_\Vect{t} & 0 \\
		2 k_0 & 1
	\end{pmatrix}
	,
\end{equation}

\noindent where I have defined
\begin{equation}
	\vartheta_\Vect{t} \doteq \sqrt{1 + 4 t_1^2}
	.
\end{equation}

\noindent Hence, I can identify the matrices
\begin{equation}
	\Mat{A}_\Vect{t}
	=
	\frac{1}{\vartheta_\Vect{t}}
	\begin{pmatrix}
		- 2 t_1 & 0 \\
		0 & \vartheta_\Vect{t}
	\end{pmatrix}
	, \quad
	\Mat{B}_\Vect{t}
	=
	\frac{1}{\vartheta_\Vect{t}}
	\begin{pmatrix}
		- 1 & 0 \\
		0 & 0
	\end{pmatrix}
	, \quad
	\Mat{R}_\Vect{t}
	=
	\begin{pmatrix}
		\vartheta_\Vect{t} & 0 \\
		2 k_0 & 1
	\end{pmatrix}
	.
\end{equation}

\noindent (Note that in this example I choose to define the QR decomposition with respect to a lower triangular matrix $\Mat{R}$ rather than an upper triangular matrix as is common convention; this is done for algebraic convenience.)

One sees that $\det \Mat{B}_\Vect{t} = 0$. Moreover, $\Mat{B}_\Vect{t}$ is already in its diagonalizing basis such that I can clearly identify the rank and corank $\rho = \varsigma = 1$, along with the submatrices
\begin{equation}
	\Lambda_{\rho \rho}
	= - \frac{1}{\vartheta_\Vect{t}}
	, \quad
	a_{\rho \rho}
	=
	- \frac{2 t_1}{\vartheta_\Vect{t}}
	, \quad
	a_{\varsigma \varsigma}
	= 1
	.
\end{equation}

\noindent Also $\Mat{L}_\textrm{s} = \Mat{R}_\textrm{s} = \IMat{2}$. Hence, I compute
\begin{equation}
	\MTnorm
	= 
	c_0(t_2)
	\,
	\frac{
		\sigma_\Vect{t} \,
		\exp\left[i \int_0^{t_1} \dd \xi \, \Vect{k}(\xi, \Vect{t}_\perp)^\intercal \dot{\Vect{x}}(\xi, \Vect{t}_\perp) \right]
	}{
		(-2\pi i)^{\rho/2}
		\sqrt{\det \Mat{\Lambda}_{\rho\rho} \det \Mat{a}_{\varsigma \varsigma} \det \Mat{R}_\Vect{t} }
	}
	=
	i c_0(t_2)
	\,
	\frac{
		\exp\left(
			i \frac{2}{3} t_1^3 
			+ 2 i k_0^2 t_1
		\right)
	}{
		\sqrt{-2\pi i}
	}
	,
\end{equation} 

\noindent where $c_0(t_2)$ is an arbitrary function that will be matched to initial conditions. I have again used the fact that $\sigma_t$ is constant due to the lack of branch-cut crossings.


\subsubsection{Computing $\Upsilon_\Vect{t}$ via \Eq{eq:7_UpsilonINT} or \Eq{eq:7_UpsilonINT_Bneq0}}

To calculate $\Upsilon_\Vect{t}$, I must calculate the tangent-space wavefield. I begin with calculating the phase. Since the rotated rays are given by
\begin{equation}
	\Stroke{\Vect{Z}}_\Vect{t}(\Vect{\tau})
	= \Mat{S}_\Vect{t} \Vect{z}(\Vect{\tau})
	=
	\begin{pmatrix}
		\frac
		{
			2 t_1 \tau_1^2 + \tau_1
		}{\vartheta_\Vect{t} }
		\\
		\tau_2 + 2 k_0 \tau_1 \\
		\frac
		{
			2 t_1 \tau_1 - \tau_1^2
		}{\vartheta_\Vect{t} }
		\\
		k_0
	\end{pmatrix}
	,
\end{equation}

\noindent one can compute the tangent-plane ray map $\Vect{\tau}(\Vect{X}_\Vect{t})$, which is double-valued. Upon restricting $\Vect{K}_\Vect{t}(\Vect{X}_\Vect{t})$ to the correct branch, I obtain
\begin{align}
	\Vect{K}_\Vect{t}(\Vect{X}_\Vect{t})
	=
	\begin{pmatrix}
		- \frac{X_{\Vect{t},1}}{2 t_1}
		- \vartheta_\Vect{t}
		\frac
		{
			1 
			- \sqrt{1 + 8  t_1 \vartheta_\Vect{t} X_{\Vect{t},1}}
		}{8 t_1^2}\\
		k_0
	\end{pmatrix}
	.
\end{align}

\noindent Then, the phase is computed via the line integral
\begin{align}
	\Theta_\Vect{t}[\Vect{\epsilon} + \Vect{X}_\Vect{t}(\Vect{t})]
	&=
	\int_{\Vect{X}_\Vect{t}(\Vect{t})}^{\Vect{\epsilon} + \Vect{X}_\Vect{t}(\Vect{t})}
	\dd \Vect{X}_\Vect{t}^\intercal
	\Vect{K}_\Vect{t}(\Vect{X}_\Vect{t})
	\nonumber\\
	&=
	\int_{\Vect{X}_\Vect{t}(\Vect{t})}^{\Vect{\epsilon} + \Vect{X}_\Vect{t}(\Vect{t})}
	\left[
		\left(
			- \frac{X_{\Vect{t},1}}{2 t_1}
			- \vartheta_\Vect{t}
			\frac
			{
				1 
				- \sqrt{1 + 8  t_1 \vartheta_\Vect{t} X_{\Vect{t},1}}
			}{8 t_1^2}
		\right)
		\dd X_{\Vect{t},1}
		+ k_0 \, \dd X_{\Vect{t},2}
	\right]
	\nonumber\\
	&=
	k_0 \epsilon_2
	+ \frac{8 t_1^4 - \vartheta^4_\Vect{t} }{8 t_1^2 \vartheta_\Vect{t}} \epsilon_1
	- \frac{1}{4 t_1} \epsilon_1^2
	+ \frac
	{
		\left(\vartheta^4_\Vect{t} + 8 t_1 \vartheta_\Vect{t} \epsilon_1\right)^{3/2}
		- \vartheta^6_\Vect{t}
	}{96 t_1^3}
	.
\end{align}

Next, I calculate the tangent-plane envelope $\Phi_t$. Since
\begin{equation}
	\pd{\Vect{\tau}} \Vect{X}_\Vect{t}(\Vect{\tau})
	=
	\begin{pmatrix}
		\frac{4 t_1 \tau_1 + 1}{\vartheta_\Vect{t}} & 0 \\
		2 k_0 & 1
	\end{pmatrix}
	,
\end{equation}

\noindent I compute
\begin{equation}
	J_\Vect{t}(\Vect{X}_\Vect{t}) = 
	\frac{\sqrt{1 + 8 t_1 \vartheta_\Vect{t} X_{\Vect{t},1} }
	}{\vartheta_\Vect{t}}
	.
\end{equation}

\noindent Hence,
\begin{equation}
	\Phi_\Vect{t}
	\left[
		\Vect{\epsilon}
		+ \Vect{X}_\Vect{t}(\Vect{t})
	\right]
	=
	\frac
	{
		\vartheta_\Vect{t}
	}
	{
		\left( 
			\vartheta^4_\Vect{t}
			+ 8 t_1 \vartheta_\Vect{t} \epsilon_1 
		\right)^{1/4}
	}
	,
\end{equation}

I can now compute $\Upsilon_\Vect{t}$. By definition,
\begin{align}
	\Upsilon_\Vect{t}
	&=
	\int_{\cont{0}} \dd \Vect{\epsilon}_\rho \,
	\Psi_\Vect{t}
	\left[
		\Mat{L}_\textrm{s}
		\begin{pmatrix}
			\Vect{X}_\Vect{t}^\rho(\Vect{t}) + \Vect{\epsilon}_\rho \\
			\Mat{a}_{\varsigma \varsigma} \Vect{x}_\varsigma
		\end{pmatrix}
	\right]
	\exp
	\left[
		- \frac{i}{2} \Vect{\epsilon}_\rho^\intercal \, \Mat{a}_{\rho \rho} \Mat{\Lambda}_{\rho \rho}^{-1} \, \Vect{\epsilon}_\rho
		- i \Vect{\epsilon}_\rho^\intercal \Vect{K}_\Vect{t}^\rho(\Vect{t})
	\right]
	\nonumber\\
	&=
	\int_{\cont{0}} 
	\frac
	{
		\dd \epsilon_1 \, \vartheta_\Vect{t}
	}
	{
		\left( 
			\vartheta^4_\Vect{t}
			+ 8 t_1 \vartheta_\Vect{t} \epsilon_1 
		\right)^{1/4}
	}
	\exp
	\left(
		- i \frac{\vartheta^3_\Vect{t} }{8 t_1^2} \epsilon_1
		- i \frac{\vartheta_\Vect{t}^2}{4 t_1} \epsilon_1^2
		+ i \frac
		{
			\left(\vartheta^4_\Vect{t} + 8 t_1 \vartheta_\Vect{t} \epsilon_1\right)^{3/2}
			- \vartheta^6_\Vect{t}
		}{96 t_1^3}
	\right)
	.
\end{align}

\noindent This is the same integral that was studied in the $1$-D Airy example above. Hence, using \Eq{eq:7_airyUPSILONapprox} I obtain
\begin{align}
	\Upsilon_\Vect{t}
	\approx
	\pi \vartheta_\Vect{t} \exp\left(- i \frac{2}{3} t_1^3 \vartheta^6_\Vect{t} \right)
	\left[
		\airyA
		\left(
			- t_1^2 \vartheta^4_\Vect{t} 
		\right) 
		- i \, \frac{t_1}{|t_1|} 
		\airyB
		\left(
			- t_1^2 \vartheta^4_\Vect{t} 
		\right) 
	\right]
	.
\end{align}


\subsubsection{Summing over rays}

Summing over both branches of $\Vect{\tau}(\Vect{x})$ yields
\begin{align}
	\psi(\Vect{x})
	&=
	\frac{
		i \sqrt{\pi}
	}{
		\sqrt{-2 i}
	}
	\sqrt{1 - 4 x_1}
	\nonumber\\
	&\hspace{4mm}\times
	\left(
		c_0(x_2 - 2 k_0 \sqrt{- x_1})
		\,
		\exp\left[
			- i \varpi(x_1)
			+ 2 i k_0^2 \sqrt{- x_1} 
		\right]
		\left\{
			\airyA
			\left[
				- \varrho^2(x_1)
			\right]
			- i 
			\airyB
			\left[
				- \varrho^2(x_1)
			\right]
			\nullFrac
		\right\}
		\nullFrac
		\right.
		\nonumber\\
		&\hspace{8mm}\left. +
		c_0(x_2 + 2 k_0 \sqrt{- x_1})
		\,
		\exp\left[
			i \varpi(x_1)
			- 2 i k_0^2 \sqrt{- x_1} 
		\right]
		\left\{
			\airyA
			\left[
				- \varrho^2(x_1)
			\right]
			+ i 
			\airyB
			\left[
				- \varrho^2(x_1)
			\right]
			\nullFrac
		\right\}
		\nullFrac
	\right)
	,
\end{align}

\noindent where I have introduced again
\begin{equation}
	\varrho(x_1) \doteq (1 - 4 x_1) \sqrt{-x_1}
	, \quad
	\varpi(x_1) \doteq \frac{2}{3} \varrho^3(x_1) - \frac{2}{3} (-x)^{3/2}
	.
\end{equation}

\noindent To satisfy the initial condition \eq{eq:7_airyIC2D} it is clear that one must choose
\begin{equation}
	c_0(t_2)
	=
	\frac{\sqrt{-2 i}}{2i \sqrt{\pi}} \exp( i k_0 t_2)
	.
\end{equation}

\noindent I thus obtain the MGO solution:
\begin{equation}
	\psi(\Vect{x}) 
	=
	\sqrt{1 - 4 x_1} \,
	\exp(i k_0 x_2)
	\left\{
		\airyA[- \varrho^2(x_1) ]
		\cos \varpi(x_1) 
		-
		\airyB[- \varrho^2(x_1) ]
		\sin \varpi(x_1) 
	\right\}
	.
	\label{eq:7_airyMGO2D}
\end{equation}

\begin{figure}
	\centering
	\includegraphics[width=0.7\linewidth,trim={2mm 16mm 2mm 14mm},clip]{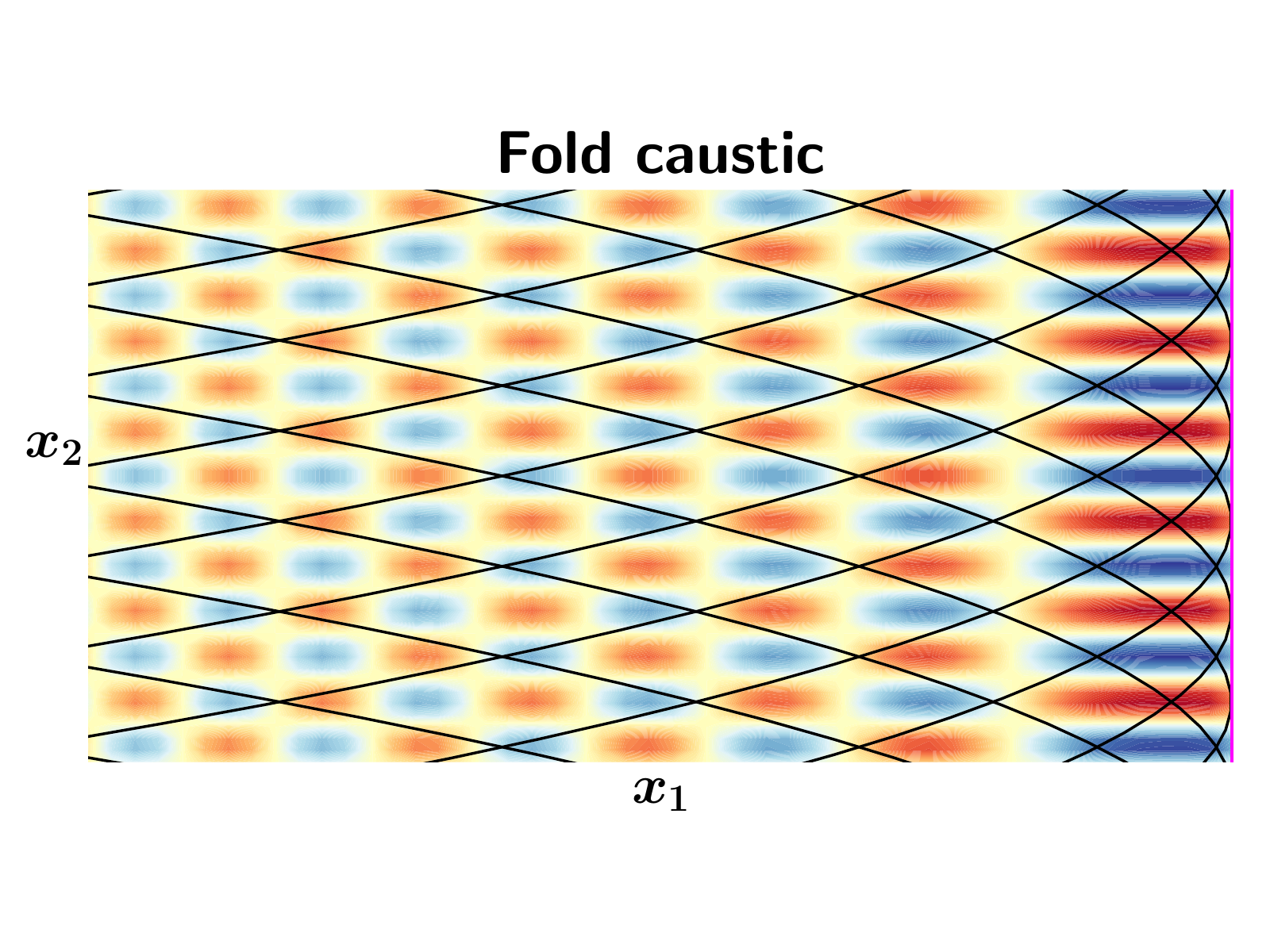}
	\caption{Contour plot showing the real part of the MGO solution \eq{eq:7_airyMGO2D}, with $k_0 = 2$, near the fold caustic (cutoff) located at $x_1 = 0$ (magenta). The ray trajectories $( x_1(\Vect{\tau}), x_2(\Vect{\tau}) )$ are shown as black curves. Note that the field remains finite along the caustic. In fact, the MGO solution is nearly indistinguishable with the exact solution \eq{eq:7_airyEXACT2D}.}
	\label{fig:7_airy2D}
\end{figure}

\noindent The MGO solution \eq{eq:7_airyMGO2D} is plotted in \Fig{fig:7_airy2D}. Notably, this solution is finite along the caustic surface (cutoff) located at $x_1 = 0$, and agrees remarkably well with the exact solution of \Eq{eq:7_airyEQ2D},
\begin{equation}
	\psi_\text{ex}(\Vect{x}) = \airyA(x_1) \exp(ik_0 x_2)
	.
	\label{eq:7_airyEXACT2D}
\end{equation}

\noindent A similar plot of \Eq{eq:7_airyEXACT2D} is not presented because it is virtually indistinguishable from \Fig{fig:7_airy2D}. It is reassuring to observe that the MGO solution \eq{eq:7_airyMGO2D} separates into a traveling wave component along the caustic surface [governed by the $\exp(i k_0 x_2)$ term] and the 1-D MGO formula for a fold caustic \eq{eq:7_airyMGO} transverse to the caustic surface, just as the exact solution \eq{eq:7_airyEXACT2D} separates into a traveling wave and a fold caustic.


\subsection{Pearcey's equation}

As a final example, let us consider a 2-D plane wave described by the paraxial wave equation~\cite{Kogelnik66}
\begin{equation}
	i \pd{x_1} \psi(\Vect{x}) + \frac{1}{2}\pd{x_2}^2 \psi(\Vect{x}) + \psi(\Vect{x}) = 0 
	,
	\label{eq:7_cuspEQ}
\end{equation}

\noindent where $x_1$ is aligned with the optical axis and $x_2$ is tranverse to it. Let us assume the initial condition 
\begin{equation}
	\psi(0 , x_2) = \sqrt{ \frac{2 \pi i }{f } } \exp\left( - \frac{i}{2f} \, x_2^2 - \frac{i \aber}{4 f^3} \, x_2^4 \right) 
	.
	\label{eq:7_cuspIC}
\end{equation}

\begin{figure}
	\centering
	\includegraphics[width=0.3\linewidth]{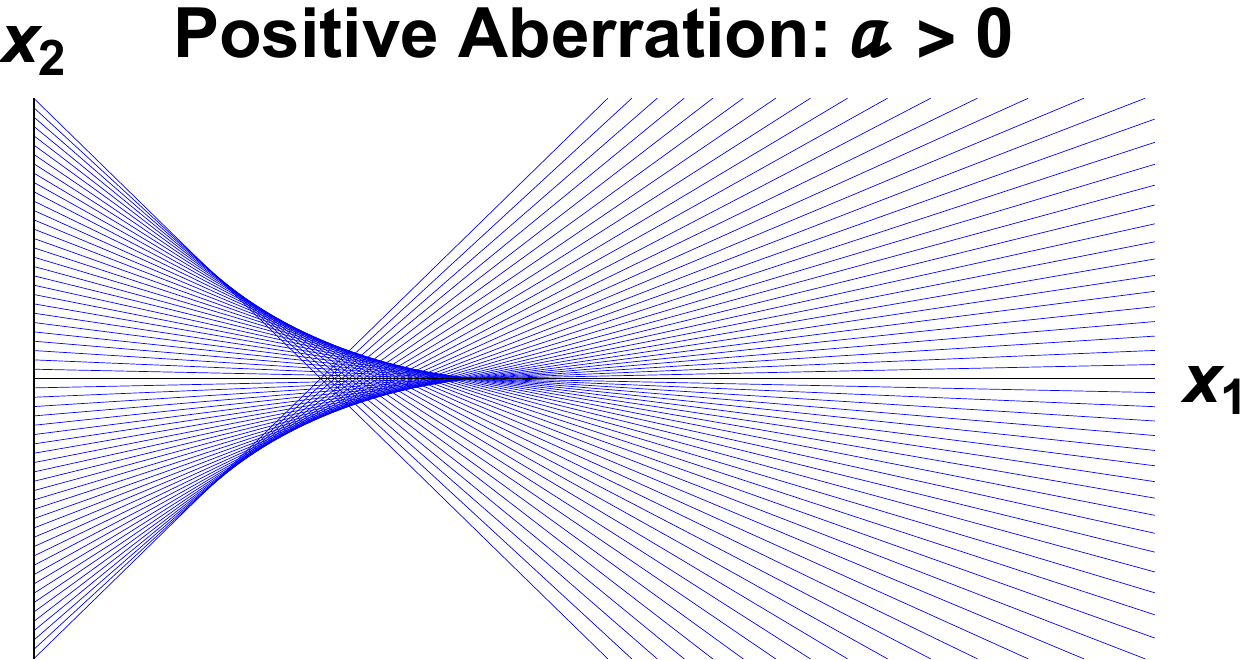}
	\hspace{3mm}
	\includegraphics[width=0.3\linewidth]{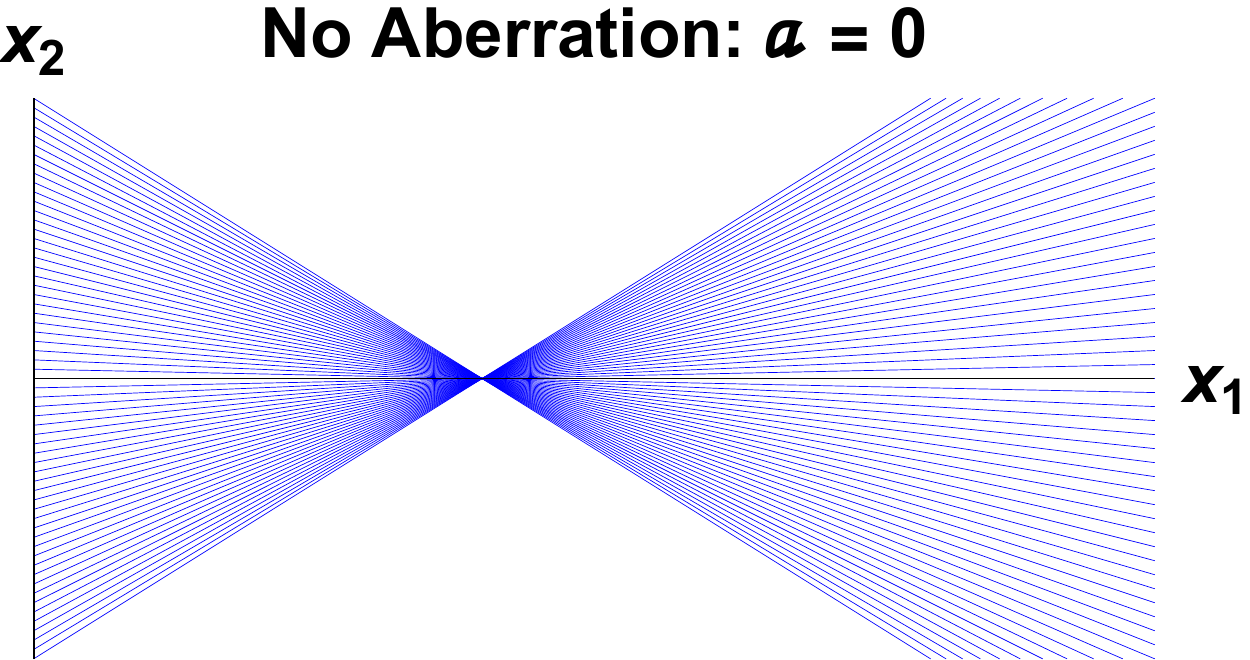}
	\hspace{3mm}
	\includegraphics[width=0.3\linewidth]{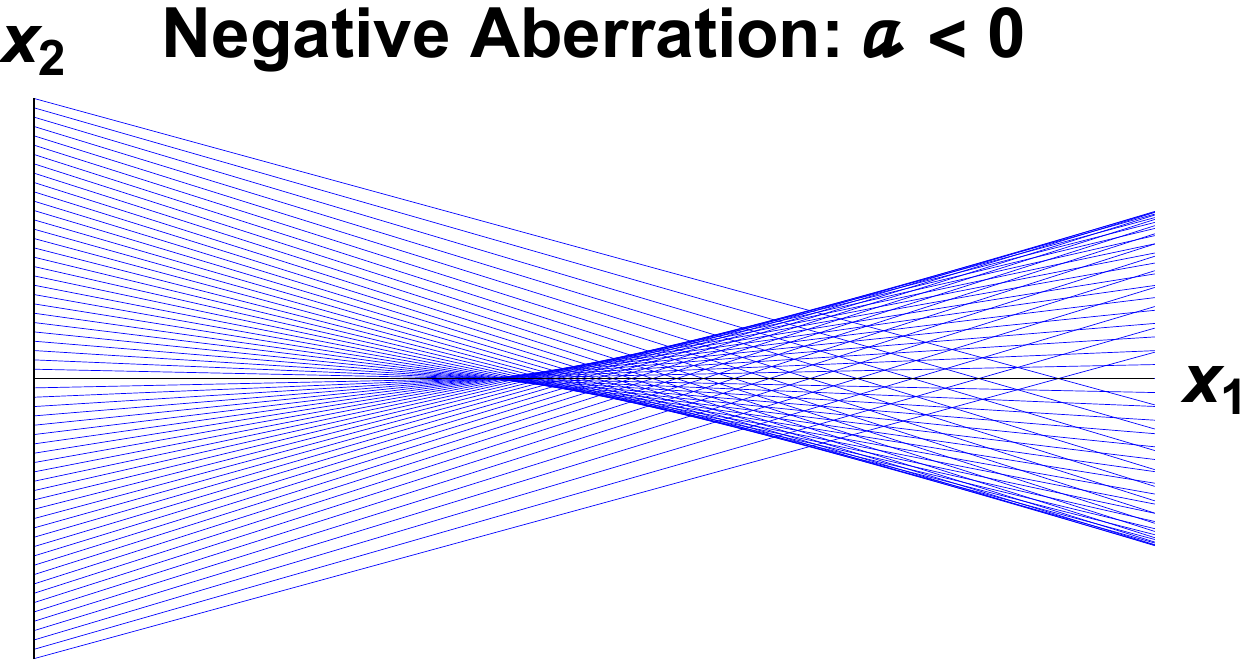}
	\caption{Ray trajectories for the paraxial equation \eq{eq:7_cuspEQ} with initial conditions given by \Eq{eq:7_cuspIC}. For no aberration ($\aber = 0$), the rays focus at $\Vect{x} = (f, 0)$. Positive aberration ($\aber > 0$) causes the outer rays to focus before the focal point, while negative aberration ($\aber < 0$) causes the outer rays to focus beyond the focal point. Both cases result in a cusped ray pattern.}
	\label{fig:7_aber}
\end{figure}

\noindent This corresponds to a wave that is being focused by an imperfect lens, with focal distance $f$ and aberration $\aber$. In the absence of aberration ($\aber = 0$), the initial field will focus at $\Vect{x} = (f, 0)$; however, as shown in \Fig{fig:7_aber}, finite aberration causes the focusing to become distorted, resulting in a cusped wavefield. For simplicity, I shall assume that $f \gg 1$ and $\aber < 0$.

\subsubsection{Tracing rays}

Equation \eq{eq:7_cuspEQ} can be equivalently written as an integral equation with integration kernel
\begin{equation}
	D(\Vect{x}, \Vect{x}') 
	= 
	i \pd{x_1'} \delta(\Vect{x} - \Vect{x}') 
	- \frac{1}{2} \pd{x_2'}^2 \delta(\Vect{x} - \Vect{x}')
	- \delta(\Vect{x} - \Vect{x}') 
	.
\end{equation}

\noindent The corresponding Weyl symbol is
\begin{equation}
	\Symb{D}(\Vect{x}, \Vect{k}) 
	=
	k_1 + \frac{k_2^2}{2} - 1
	,
\end{equation}

\noindent and the corresponding ray equations are
\begin{align}
	\pd{\tau_1} x_1(\Vect{\tau}) = 1
	, \quad
	\pd{\tau_1} x_2(\Vect{\tau}) &= k_2(\Vect{\tau})
	, \quad
	\pd{\tau_1} k_1(\Vect{\tau}) = 0
	, \quad
	\pd{\tau_1} k_2(\Vect{\tau}) = 0
	.
	\label{eq:7_cuspRAYeq}
\end{align}

Similar to the previous example, let us define $\tau_1$ and $\tau_2$ such that $x_1(0, \tau_2) = 0$ and $x_2(0, \tau_2) = f \tau_2$. Then, imposing the local dispersion relation and the initial condition \eq{eq:7_cuspIC} yields the ray trajectories
\begin{align}
	x_1(\Vect{\tau}) = \tau_1 , \quad
	x_2(\Vect{\tau}) = f \tau_2 + k_2(\Vect{\tau}) \tau_1 ,\quad
	k_1(\Vect{\tau}) =  1 - \frac{ k_2^2(\Vect{\tau})}{2} , \quad
	k_2(\Vect{\tau}) = - \, \tau_2 - \aber \, \tau_2^3 .
\end{align}

\noindent The inverse function $\Vect{\tau}(\Vect{x})$ is either single- or triple-valued, depending on the value of the discriminant
\begin{equation}
	\Delta(\Vect{x}) = 
	4 \left(\frac{x_1 - f}{ \aber x_1} \right)^3 
	+ 27 \left( \frac{x_2}{ \aber x_1} \right)^2
	.
	\label{eq:7_DeltaDEF}
\end{equation}

\noindent If $\Delta(\Vect{x}) > 0$, there is only one ray given by
\begin{equation}
	\tau_2^{(0)}(\Vect{x}) = 
	\sqrt[\textrm{\small 3}]{
		- \frac{x_2}{2 \aber x_1} 
		+ \sqrt
		{
			\frac{\Delta(\Vect{x})}{108}
		} 
	} 
	+ 
	\sqrt[\textrm{\small 3}]{ 
		- \frac{x_2}{2 \aber x_1}  
		- \sqrt
		{
			\frac{\Delta(\Vect{x})}{108}
		} 
	}
	,
\end{equation}

\noindent while if $\Delta(\Vect{x}) \le 0$, there are two additional rays given by
\begin{equation}
	\tau_2^{(\pm)}(\Vect{x}) = 
	\textrm{Re}
	\left[
		(-1 \pm i \sqrt{3})
		\sqrt[\textrm{\small 3}]{
			- \frac{x_2}{2 \aber x_1} 
			+ i \sqrt
			{
				\frac{|\Delta(\Vect{x})|}{108}
			} 
		}
	\right]
	.
\end{equation}

\noindent For all values of $\Delta(\Vect{x})$, one has $\tau_1(\Vect{x}) = x_1$.


\subsubsection{Computing $\MTnorm$ via \Eq{eq:6_MTnormFINAL} or \Eq{eq:6_MTnormFINAL_Bneq0}}

I first compute the integral
\begin{equation}
	\int_0^{t_1} \dd \xi \,
	\Vect{k}(\xi,t_2)^\intercal \dot{\Vect{x}}(\xi, t_2)
	=
	\int_0^{t_1} \dd \xi \,
	\left[
		1 + \frac{ k_2^2(\Vect{\tau})}{2}
	\right]
	=
	\left[
		1 + \frac{ k_2^2(\Vect{\tau})}{2}
	\right] t_1
	.
\end{equation}

\noindent Next, a QR decomposition of the matrix $\pd{\Vect{\tau}} \Vect{z}$ yields
\begin{equation}
	\pd{\Vect{\tau}} \Vect{z}(\Vect{t})
	\equiv
	\begin{pmatrix}
		1 & 0 \\
		k_2( \Vect{t} ) & j_\Vect{t} \\
		0 & - k_2(\Vect{t}) k_2'(\Vect{t}) \\
		0 & k_2'(\Vect{t})
	\end{pmatrix}
	=
	\begin{pmatrix}
		\frac{1}{\vartheta_\Vect{t}} & - \frac{j_\Vect{t} k_2(\Vect{t})}{\vartheta_\Vect{t} \varphi_\Vect{t}} \\
		\frac{k_2(\Vect{t})}{\vartheta_\Vect{t}} & \frac{j_\Vect{t}}{\vartheta_\Vect{t} \varphi_\Vect{t}} \\
		0 & - \frac{\vartheta_\Vect{t} k_2(\Vect{t}) k_2'(\Vect{t}) }{\varphi_\Vect{t}} \\
		0 & \frac{\vartheta_\Vect{t} k_2'(\Vect{t}) }{\varphi_\Vect{t}}
	\end{pmatrix}
	\begin{pmatrix}
		\vartheta_\Vect{t} & \frac{k_2(\Vect{t}) j_\Vect{t} }{\vartheta_\Vect{t} } \\
		0 & \frac{\varphi_\Vect{t} }{\vartheta_\Vect{t}}
	\end{pmatrix}
	,
\end{equation}

\noindent where I have defined
\begin{equation}
	j_\Vect{t}
	\doteq
	\det \pd{\Vect{\tau}} \Vect{x}(\Vect{t}) 
	= f + t_1 k_2'(\Vect{t})
	, \quad
	\vartheta_\Vect{t} \doteq \sqrt{1 + k_2^2(\Vect{t})}
	, \quad
	\varphi_\Vect{t} \doteq \sqrt{j_\Vect{t}^2 + \left[k_2'(\Vect{t}) \vartheta^2_\Vect{t} \right]^2} 
	.
\end{equation}

\noindent Hence, I can identify the matrices
\begin{equation}
	\Mat{A}_\Vect{t}
	=
	\frac{1}{\vartheta_\Vect{t} \varphi_\Vect{t}}
	\begin{pmatrix}
		\varphi_\Vect{t} & \varphi_\Vect{t} k_2(\Vect{t}) \\
		- j_\Vect{t} k_2(\Vect{t}) & j_\Vect{t}
	\end{pmatrix}
	, \quad
	\Mat{B}_\Vect{t}
	=
	\frac{\vartheta_\Vect{t} k_2'(\Vect{t}) }{\varphi_\Vect{t}}
	\begin{pmatrix}
		0 & 0 \\
		- k_2(\Vect{t}) & 1
	\end{pmatrix}
	, \quad
	\Mat{R}_\Vect{t}
	=
	\frac{1}{\vartheta_\Vect{t}}
	\begin{pmatrix}
		\vartheta_\Vect{t}^2 & k_2(\Vect{t}) j_\Vect{t} \\
		0 & \varphi_\Vect{t}
	\end{pmatrix}
	.
\end{equation}

Again, I see that $\det \Mat{B}_\Vect{t} = 0$. I perform an SVD of $\Mat{B}_\Vect{t}$ to obtain
\begin{equation}
	\Mat{L}_\textrm{s}
	=
	\begin{pmatrix}
		0 & 1 \\
		-1 & 0
	\end{pmatrix}
	, \quad
	\Mat{R}_\textrm{s}
	=
	\frac{1}{\vartheta_\Vect{t}}
	\begin{pmatrix}
		k_2(\Vect{t}) & 1 \\
		- 1 & k_2(\Vect{t})
	\end{pmatrix}
	.
\end{equation}

\noindent Hence the SVD matrix representations are given as
\begin{equation}
	\widetilde{\Mat{A}}_\Vect{t}
	=
	\Mat{L}_\textrm{s}^\intercal
	\Mat{A}_\Vect{t}
	\Mat{R}_\textrm{s}
	=
	\begin{pmatrix}
		\frac{j_\Vect{t}}{\varphi_\Vect{t}} & 0 \\
		0 & 1
	\end{pmatrix}
	, \quad
	\widetilde{\Mat{B}}_\Vect{t}
	=
	\Mat{L}_\textrm{s}^\intercal
	\Mat{B}_\Vect{t}
	\Mat{R}_\textrm{s}
	=
	\begin{pmatrix}
		\frac{\vartheta_\Vect{t}^2 k_2'(\Vect{t}) }{\varphi_\Vect{t}} & 0 \\
		0 & 0
	\end{pmatrix}
	,
\end{equation}

\noindent and consequently, I can identify the rank and corank $\rho = \varsigma = 1$, along with the submatrices
\begin{equation}
	\Lambda_{\rho \rho}
	= \frac{\vartheta_\Vect{t}^2 k_2'(\Vect{t}) }{\varphi_\Vect{t}}
	, \quad
	a_{\rho \rho}
	=
	\frac{j_\Vect{t}}{\varphi_\Vect{t}}
	, \quad
	a_{\varsigma \varsigma}
	= 1
	.
\end{equation}

\noindent Hence, I compute
\begin{equation}
	\MTnorm
	= 
	c_0(t_2)
	\,
	\frac{
		\sigma_\Vect{t} \,
		\exp\left[i \int_0^{t_1} \dd \xi \, \Vect{k}(\xi, \Vect{t}_\perp)^\intercal \dot{\Vect{x}}(\xi, \Vect{t}_\perp) \right]
	}{
		(-2\pi i)^{\rho/2}
		\sqrt{\det \Mat{\Lambda}_{\rho\rho} \det \Mat{a}_{\varsigma \varsigma} \det \Mat{R}_\Vect{t} }
	}
	=
	c_0(t_2)
	\,
	\frac{
		\sigma_\Vect{t} \,
		\exp\left(
			i \frac{ 1 + \vartheta_\Vect{t}^2}{2} t_1 
		\right]
	}{
		\sqrt{-2\pi i}
		\sqrt{ 
			\vartheta_\Vect{t}^2 k_2'(\Vect{t})
		}
	}
	,
\end{equation} 

\noindent where $c_0(t_2)$ is an arbitrary function that will be matched to initial conditions. Note that $\sigma_\Vect{t}$ is nontrivial since $k_2'(\Vect{t})$ can change sign. However, I do not need to explicitly compute $\sigma_\Vect{t}$ since it will be removed by matching to initial conditions (specifically because the branch-cut crossing only depends on $t_2$ and thus can be lumped into the arbitrary function $c_0$ that is matched to initial conditions).


\subsubsection{Computing $\Upsilon_\Vect{t}$ via \Eq{eq:7_UpsilonINT} or \Eq{eq:7_UpsilonINT_Bneq0}}

To compute $\Upsilon_\Vect{t}$ I must calculate the tangent-space wavefield. I start with the envelope. Since
\begin{equation}
	\pd{\Vect{\tau}} \Vect{X}_\Vect{t}(\Vect{\tau})
	=
	\frac{1}{\vartheta_\Vect{t} \varphi_\Vect{t}}
	\begin{pmatrix}
		\varphi_\Vect{t}[1 + k_2(\Vect{t}) k_2(\Vect{\tau})]
		&
		\varphi_\Vect{t} j_\Vect{\tau} k_2(\Vect{t})
		\\
		j_\Vect{t}[ k_2(\Vect{\tau}) - k_2(\Vect{t})]
		&
		j_\Vect{t}j_\Vect{\tau} + \vartheta_\Vect{t}^2 k_2'(\Vect{t}) k_2'(\Vect{\tau}) [1 + k_2(\Vect{t}) k_2(\Vect{\tau})]
	\end{pmatrix}
	,
\end{equation}

\noindent I compute
\begin{equation}
	J_\Vect{t}(\Vect{\tau}) = 
	\frac{
		j_\Vect{t} j_\Vect{\tau} 
		+ [1 + k_2(t_2) k_2(\tau_2)]^2 
		k_2'(\tau_2) k_2'(t_2)
	}
	{
		\varphi_\Vect{t}
	} 
	.
\end{equation}

\noindent Hence I obtain
\begin{equation}
	\Phi_\Vect{t}(\Vect{\tau})
	=
	\frac
	{
		\varphi_\Vect{t}
	}
	{
		\sqrt
		{
			j_\Vect{t} j_\Vect{\tau} 
			+ [1 + k_2(\Vect{t}) k_2(\Vect{\tau})]^2 
			k_2'(\Vect{\tau}) k_2'(\Vect{t})
		}
	}
	.
\end{equation}

\noindent In principle, one can construct $\Phi_\Vect{t}(\Vect{X}_\Vect{t})$ using the inverse ray map $\Vect{\tau}(\Vect{X}_\Vect{t})$, but this is not practical for analytical calculations.

It is also impractical to explicitly construct the tangent-space phase $\Theta_\Vect{t}(\Vect{X}_\Vect{t})$. Hence I shall adopt a different approach and attempt to approximate $\Upsilon_\Vect{t}$ directly. By definition,
\begin{align}
	\Upsilon_\Vect{t}
	&=
	\int_{\cont{0}} \dd \Vect{\epsilon}_\rho \,
	\Psi_\Vect{t}
	\left[
		\Mat{L}_\textrm{s}
		\begin{pmatrix}
			\Vect{X}_\Vect{t}^\rho(\Vect{t}) + \Vect{\epsilon}_\rho \\
			\Mat{a}_{\varsigma \varsigma} \Vect{x}_\varsigma
		\end{pmatrix}
	\right]
	\exp
	\left[
		- \frac{i}{2} \Vect{\epsilon}_\rho^\intercal \, \Mat{a}_{\rho \rho} \Mat{\Lambda}_{\rho \rho}^{-1} \, \Vect{\epsilon}_\rho
		- i \Vect{\epsilon}_\rho^\intercal \Vect{K}_\Vect{t}^\rho(\Vect{t})
	\right]
	\nonumber\\
	&=
	\int_{\cont{0}} \dd \epsilon \,
	\Psi_\Vect{t}
	\left[
		X_{\Vect{t},1}(\Vect{t}),
		X_{\Vect{t},2}(\Vect{t}) - \epsilon
	\right]
	\exp
	\left[
		- \frac{i}{2} \, \frac{j_\Vect{t}}{\vartheta_\Vect{t}^2 k_2'(\Vect{t}) } \, \epsilon^2
		+ i K_\Vect{t,2}(\Vect{t}) \epsilon
	\right]
	.
\end{align}

\noindent Since I am studying a cusp $A_3$ caustic, let us therefore attempt to fit the quartic normal form to $\Upsilon_\Vect{t}$. Indeed, making the constant envelope approximation as done in previous examples, and expanding the phase to fourth order in $\epsilon$ formally yields
\begin{equation}
	\Upsilon_\Vect{t}
	\approx
	\int_{\cont{0}} \dd \epsilon \,
	\exp
	\left[
		- \frac{i}{2} \, \frac{j_\Vect{t}}{\vartheta_\Vect{t}^2 k_2'(\Vect{t}) } \, \epsilon^2
		- \frac{i}{6} \pd{X_{\Vect{t},2}}^2 K_{\Vect{t},2}(\Vect{t}) \epsilon^3
		+ \frac{i}{24} \pd{X_{\Vect{t},2}}^3 K_{\Vect{t},2}(\Vect{t}) \epsilon^4
	\right]
	.
	\label{eq:7_upsAPPROX}
\end{equation}

One can subsequently compute the relevant derivatives of $K_{\Vect{t},2}$ as follows. If one can find the constraint that $X_{\Vect{t}, 1}(\Vect{\tau}) = X_{\Vect{t}, 1}(\Vect{t})$ imposes on $\Vect{\tau}$, then one could perform the desired derivatives of $K_{\Vect{t}, 2}$ implicitly along the constraint surface $\tau_{\Vect{t},1}(\tau_2)$ or $\tau_{\Vect{t},2}(\tau_1)$. Since
\begin{equation}
	X_{\Vect{t}, 1}(\Vect{\tau}) = 
	\frac{\tau_1[1 + k_2(t_2) k_2(\tau_2)] + f \tau_2 k_2(t_2)}{\vartheta_\Vect{t}}
	,
\end{equation}

\noindent one can invert the relation $X_{\Vect{t}, 1}(\Vect{\tau}) = Q_{\Vect{t}, 1}(\Vect{t})$ to yield the constraint
\begin{equation}
	\tau_{\Vect{t},1}(\tau_2) = \frac{t_1 \vartheta_\Vect{t}^2 + f (t_2 - \tau_2)k_2(t_2)}{1 + k_2(t_2) k_2(\tau_2)}
	.
\end{equation}

\noindent Hence, the functions $X_{\Vect{t},2}[\tau_{\Vect{t},1}(\tau_2), \tau_2]$ and $K_{\Vect{t},2}[\tau_{\Vect{t},1}(\tau_2), \tau_2]$ both lie along the surface $X_1 = X_{\Vect{t},1}(\Vect{t})$. Since both $X_{\Vect{t}, 2}$ and $K_{\Vect{t}, 2}$ can be considered univariate functions along this constraint surface, \ie
\begin{equation}
	K_{\Vect{t}, 2}(\tau_2) \doteq K_{\Vect{t},2}[\tau_{\Vect{t},1}(\tau_2), \tau_2]
	, \quad
	X_{\Vect{t}, 2}(\tau_2) \doteq Q_{\Vect{t},2}[\tau_{\Vect{t},1}(\tau_2), \tau_2]
	,
\end{equation}

\noindent taking implicit derivatives is straightforward.

Indeed, taking $\pd{X_2}$ and $\pd{X_2}^2$ of the definition
\begin{equation}
	X_{\Vect{t},2} = X_{\Vect{t},2}[\tau_{\Vect{t},1}(\tau_2), \tau_2]
\end{equation}

\noindent yield the inversion relations between derivatives of $\tau_2$ and $X_2$ at fixed $X_1 = X_{\Vect{t},1}(\Vect{t})$:%
\begin{subequations}
	\label{eq:7_tau2INV}
	\begin{align}
		1 = \dd_{\tau_2} X_{\Vect{t}, 2} \,
		\pd{X_2} \tau_2
		\quad
		&\to
		\quad
		\pd{X_2} \tau_2
		=
		\frac{1}{\dd_{\tau_2} X_{\Vect{t}, 2}}
		, \\
		0 = \dd_{\tau_2}^2 X_{\Vect{t}, 2} (\pd{X_2} \tau_2)^2
		+ \dd_{\tau_2} X_{\Vect{t}, 2} \pd{X_2}^2 \tau_2
		\quad
		&\to
		\quad
		\pd{X_2}^2 \tau_2
		= - \frac{\dd_{\tau_2}^2 X_{\Vect{t}, 2}}{(\dd_{\tau_2} X_{\Vect{t}, 2})^3}
		,
	\end{align}
\end{subequations}

\noindent where $\dd_{\tau_2} \doteq \pd{\tau_2} + \pd{\tau_2} \tau_{\Vect{t},1}(\tau_2) \pd{\tau_1}$ is the `total' derivative with respect to $\tau_2$. Note also that 
\begin{equation}
	\dd_{\tau_2} K_{\Vect{t}, 2}(\Vect{t}) = 0
	\label{eq:7_K2tang}
\end{equation}

\noindent by definition of the tangent plane. Hence, using \Eqs{eq:7_tau2INV} and \eq{eq:7_K2tang} I formally compute the second derivative
\begin{align}
	\pd{X_{\Vect{t},2}}^2 K_{\Vect{t}, 2}(\Vect{t})
	=
	\dd_{\tau_2}^2 K_{\Vect{t}, 2}(\Vect{t}) \{\pd{X_2} \tau_2[X_{\Vect{t},2}(\Vect{t})] \}^2
	=
	\frac
	{
		\dd_{\tau_2}^2 K_{\Vect{t}, 2}(\Vect{t})
	}
	{
		[\dd_{\tau_2} X_{\Vect{t}, 2}(\Vect{t})]^2
	}
	,
	\label{eq:7_2ndDER}
\end{align}

\noindent and the third derivative
\begin{align}
	\pd{X_{\Vect{t},2}}^3 K_{\Vect{t}, 2}(\Vect{t})
	&=
	\dd_{\tau_2}^3 K_{\Vect{t}, 2}(\Vect{t}) \{\pd{X_2} \tau_2[X_{\Vect{t},2}(\Vect{t})] \}^3
	+ 3 \, \dd_{\tau_2}^2 K_{\Vect{t}, 2}(\Vect{t}) \, \pd{X_2}^2 \tau_2[X_{\Vect{t},2}(\Vect{t})] \, \pd{X_2} \tau_2[X_{\Vect{t},2}(\Vect{t})] 
	\nonumber\\
	&=
	\frac
	{
		\dd_{\tau_2}^3 K_{\Vect{t}, 2}(\Vect{t})
	}
	{
		[\dd_{\tau_2} X_{\Vect{t}, 2}(\Vect{t})]^3
	}
	+ 3 
	\frac
	{
		\dd_{\tau_2}^2 K_{\Vect{t}, 2}(\Vect{t})
	}
	{
		\dd_{\tau_2} X_{\Vect{t}, 2}(\Vect{t})
	} \pd{X_2}^2 \tau_2[X_{\Vect{t},2}(\Vect{t})]
	\nonumber\\
	&=
	\frac
	{
		\dd_{\tau_2}^3 K_{\Vect{t}, 2}(\Vect{t})
	}
	{
		[\dd_{\tau_2} X_{\Vect{t}, 2}(\Vect{t})]^3
	}
	- 3 
	\frac
	{
		\dd_{\tau_2}^2 K_{\Vect{t}, 2}(\Vect{t}) \, 
		\dd_{\tau_2}^2 X_{\Vect{t}, 2}(\Vect{t})
	}
	{
		[\dd_{\tau_2} X_{\Vect{t}, 2}(\Vect{t})]^4
	}
	.
	\label{eq:7_3rdDER}
\end{align}

\noindent However, one can readily verify that bifurcation set that results from \Eqs{eq:7_2ndDER} and \eq{eq:7_3rdDER} is not equal to the exact bifurcation set of the rays given by $\Delta(\Vect{x}) = 0$ [\Eq{eq:7_DeltaDEF}].%
\footnote{This is actually to be expected, since generally a local expansion of a given function will not reproduce the global root structure.} %
Hence, it is better to adopt a hybrid local-global approximation scheme that maintains some aspects of the local expansion [\Eqs{eq:7_2ndDER} and \eq{eq:7_3rdDER}] while also reproducing the global bifurcation set such that the saddlepoint caustic of \Eq{eq:7_upsAPPROX} be exactly the same as the true caustic. This is done by choosing
\begin{subequations}
	\label{eq:7_derAPPROX}
	\begin{align}
		\pd{X_{\Vect{t},2}}^2 K_{\Vect{t}, 2}(\Vect{t})
		&\approx
		6 f |\aber| t_2 
		\left|
			\frac{t_1}{f \vartheta_\Vect{t}^2 k_2'(t_2)}
		\right|^{3/2}
		i^{3(s_\Vect{t} + 1)/2 }
		+ O(j_\Vect{t})
		, \\
		\pd{X_{\Vect{t},2}}^3 K_{\Vect{t}, 2}(\Vect{t})
		&\approx
		6 f |\aber|
		\left|
			\frac{t_1}{f \vartheta_\Vect{t}^2 k_2'(t_2) } 
		\right|^2
		+ O(|\Vect{t} - \Vect{t}_f| )
		,
	\end{align}
\end{subequations}

\noindent where I have defined $s_\Vect{t} \doteq \textrm{sgn}[k_2'(t_2)]$ and $\Vect{t}_f \doteq (f, 0)$. Also note that the ordering terms in the expansion \eq{eq:7_derAPPROX} depict the deviation from their respective critical sets.

By combining \Eqs{eq:7_upsAPPROX} and \eq{eq:7_derAPPROX}, I obtain
\begin{equation}
	\Upsilon_\Vect{t}
	\approx
	\int_{\cont{0}} \dd \epsilon \,
	\exp
	\left[
		- i \frac{j_\Vect{t}}{2 \vartheta_\Vect{t}^2 k_2'(t_2)} \epsilon^2
		- i f |\aber| t_2 
		\left|
			- \frac{f \vartheta_\Vect{t}^2 k_2'(t_2)}{t_1}
		\right|^{-3/2}
		\epsilon^3
		+ \frac{i}{4} f |\aber| 
		\left|
			\frac{t_1}{f \vartheta_\Vect{t}^2 k_2'(t_2) } 
		\right|^2 \epsilon^4
	\right]
	.
\end{equation}

\noindent I shall now transform $\Upsilon_\Vect{t}$ into the standard Pearcey form~\cite{Olver10a}. Let
\begin{align}
	\epsilon = 
	\vartheta_\Vect{t} 
	\left| \frac{2 k_2'(t_2)}{t_1} \right|^{1/2} 
	\left| 
		\frac{f}{\aber}
	\right|^{1/4} i^{- \frac{s_\Vect{t} + 1}{2}}
	\left(
		\varepsilon 
		+ \frac{t_2}{\sqrt{2} } |f \aber|^{1/4}
	\right)
	, \quad
	\dd \epsilon 
	= \vartheta_\Vect{t}
	\left| 
		\frac{f}{\aber}
	\right|^{1/4}
	\sqrt{
		\frac{- 2 k_2'(t_2)}{t_1}
	}
	\dd \varepsilon
	.
\end{align}

\noindent This puts $\Upsilon_\Vect{t}$ into the desired form:
\begin{align}
	\Upsilon_\Vect{t}
	\approx
	\vartheta_\Vect{t} 
	\left| 
		\frac{f}{\aber}
	\right|^{1/4}
	\sqrt{
		\frac{- 2 k_2'(\Vect{t})}{t_1}
	}
	\exp
	\left(
		i \frac{j_\Vect{t} + f - t_1}{4 t_1 } f t_2^2
	\right)
	\int_{\cont{t_2}} \dd \varepsilon \,
	\exp
	\left(
		i  \gamma_\Vect{t}
		\varepsilon
		+ i \chi_\Vect{t}
		\varepsilon^2
		+ i \varepsilon^4
	\right)
	,
	\label{eq:7_cuspUPSILON}
\end{align}

\noindent where I have defined
\begin{equation}
	\chi_\Vect{t} \doteq
	\left| 
		\frac{ f}{ \aber}
	\right|^{1/2}
	\frac{
		f - x_1(\Vect{t})
	}{x_1(\Vect{t})}
	, \quad
	\gamma_\Vect{t} \doteq
	\left| 
		\frac{4 f^3}{\aber} 
	\right|^{1/4}
	\frac{
		x_2(\Vect{t})
	}{x_1(\Vect{t})}
	,
\end{equation}

\noindent and $\cont{t_2}$ is the steepest-descent contour through the saddlepoint $\varepsilon = - t_2 |f \aber|^{1/4}/\sqrt{2}$.


\subsubsection{Summing over rays}

Far from the caustic, I can evaluate $\Upsilon_\Vect{t}$ in the GO limit as
\begin{align}
	\Upsilon_\Vect{t}
	\approx
	\int_{-\infty}^\infty \dd \epsilon_\rho \,
	\exp
	\left[
		- i \frac{j_\Vect{t}}{2 \vartheta_\Vect{t}^2 k_2'(t_2)} \epsilon_\rho^2
	\right]
	=
	\vartheta_\Vect{t}
	\left|
		\frac{ 2 \pi k_2'(t_2)}{ j_\Vect{t} }
	\right|^{1/2}
	\exp\left[ - i \, s_\Vect{t} \, \textrm{sign}( j_\Vect{t} ) \frac{\pi}{4} \right]
	.
\end{align}

\noindent Then, since my assumption $\aber < 0$ implies that $\Vect{\tau}(\Vect{x})$ is single-valued along the initial surface, and since my assumption that $f \gg 1$ implies that the initial surface lies sufficiently far from the caustic, the MGO solution along this initial surface takes the form
\begin{equation}
	\mc{N}_{(0,t_2)} \Upsilon_{(0,t_2)}
	=
	c_0(t_2)
	\,
	\frac{
		\sigma_{t_2} \,
	}{
		\sqrt{f}
	}
	\exp\left(
		i \frac{1 - s_\Vect{t} }{2} \pi
	\right)
	.
\end{equation}

\noindent Thus, the initial conditions \eq{eq:7_cuspIC} are satisfied by the choice
\begin{align}
	c_0(t_2) 
	= 
	\frac
	{
		\sqrt{2 \pi}
	}
	{ 
		\sigma_{t_2} 
	}
	\exp
	\left\{
		- \frac{i}{2} f t_2^2
		- \frac{i \aber}{4} f t_2^4
		+ i \frac{ 2 s_\Vect{t} - 1}{4} \pi
	\right\}
	.
\end{align} 

\begin{figure}
	\centering
	\includegraphics[width=0.7\linewidth,trim={2mm 4mm 4mm 3mm},clip]{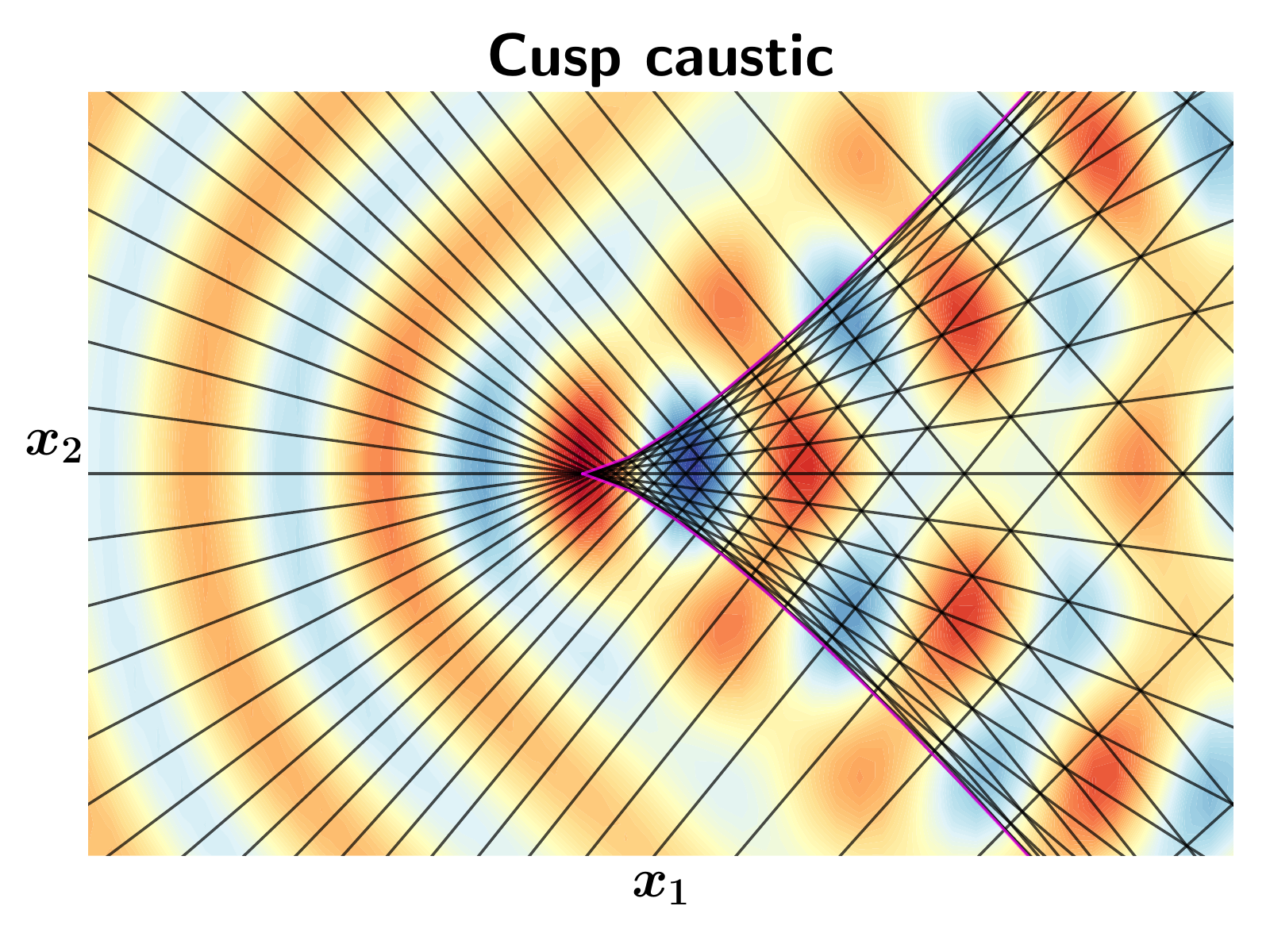}
	\caption{Contour plot showing the real part of the MGO solution \eq{eq:7_cuspMGO}, with $\aber = -4/f$, near a cusp caustic located at $x_2 = \pm \sqrt{ 4(f-x_1)^3 / 27\aber x_1 }$ (magenta). The ray trajectories $( x_1(\Vect{\tau}), x_2(\Vect{\tau}) )$ are shown as black lines. Note that the field remains finite along the caustic.}
	\label{fig:7_cusp}
\end{figure}

Hence, the MGO solution takes the form
\begin{align}
	\psi(\Vect{x}) 
	&=
	\left|
		\frac{4 f}{\aber x_1^2}
	\right|^{1/4}
	\, 
	\exp
	\left(
		i x_1
		+ i \frac{x_2^2}{2 x_1}
	\right)
	\sum_{t_2 \in \tau_2(\Vect{x})}
	\int_{\cont{t_2}} \dd \varepsilon \,
	\exp
	\left(
		i 
		\gamma_\Vect{t}
		\varepsilon
		+ i 
		\chi_\Vect{t}
		\varepsilon^2
		+ i \varepsilon^4
	\right)
	,
	\label{eq:7_cuspPSIsum}
\end{align}

\noindent where the sum is over all real saddlepoint contributions. Let us recall the Pearcey function~\cite{Paris91}, defined as
\begin{equation}
	\pearcey(x, y) \doteq 
	\int_{-\infty}^\infty \dd s \, \exp \left(i y s + i x s^2 + i s^4 \right)
	.
\end{equation}

\noindent Then, when $\Delta(\Vect{x}) \widetilde{\Delta}(\Vect{x}) \ge 0$, where $\Delta(\Vect{x})$ is defined in \Eq{eq:7_DeltaDEF} and
\begin{equation}
	\widetilde{\Delta}(\Vect{x}) \doteq 
	2 \left(
		\frac{f - x_1}{|\aber| x_1}
	\right)^3
	+ 27 (5 - \sqrt{27})
	\left(
		\frac{x_2}{|\aber| x_1}
	\right)^2
	,
\end{equation}

\noindent the summation in \Eq{eq:7_cuspPSIsum} is simplified as
\begin{equation}
	\sum_{t_2 \in \tau_2(\Vect{x})}
	\int_{\cont{t_2}} \dd \varepsilon \,
	\exp
	\left(
		i 
		\gamma_\Vect{t}
		\varepsilon
		+ i 
		\chi_\Vect{t}
		\varepsilon^2
		+ i \varepsilon^4
	\right)
	= \pearcey(\chi_\Vect{t},\gamma_\Vect{t}) 
	.
\end{equation}

\noindent When $\Delta(\Vect{x}) \widetilde{\Delta}(\Vect{x}) < 0$ (the caustic shadow), $\pearcey(x,y)$ contains an additional contribution from one of the two complex saddlepoints~\cite{Wright80}, which is not included in \Eq{eq:7_cuspPSIsum}. It does not seem possible to isolate the real saddlepoint contribution to $\pearcey(x,y)$ using a complex rotation as done in the previous examples for $\airyA(x)$. Nevertheless, the shadow contribution is asymptotically subdominant, so within the MGO accuracy I can include it such that \Eq{eq:7_cuspPSIsum} can be universally expressed through the Pearcey function as
\begin{align}
	\psi(\Vect{x})
	&=
	\left|
		\frac{4 f}{\aber x_1^2}
	\right|^{1/4}
	\, 
	\exp
	\left(
		i x_1
		+ i \frac{x_2^2}{2 x_1}
	\right)
	\pearcey
	\left(
		\left| 
			\frac{ f}{ \aber}
		\right|^{1/2}
		\frac{f - x_1}{x_1}
		, 
		\left| 
			\frac{4 f^3}{\aber} 
		\right|^{1/4}
		\frac{x_2}{x_1}
	\right)
	.
	\label{eq:7_cuspMGO}
\end{align}

\noindent As readily verified, \Eq{eq:7_cuspMGO} also happens to be the exact solution of \Eq{eq:7_cuspEQ} for the initial condition \eq{eq:7_cuspIC}. (See also \Ch{ch:GO}.) This solution is illustrated in \Fig{fig:7_cusp} for $\aber = - 4/f$.


\chapter{Conclusions}

It is common practice to employ ray-tracing codes to optimize wave-plasma interactions in nuclear fusion research. Unfortunately, these codes are based on the geometrical-optics (GO) approximation, which is not valid at caustics. This significantly limits their predictive capabilities and slows down design iteration by requiring the use of full-wave codes instead. In this thesis, I present a reformulation of GO, called `metaplectic GO' or MGO, that is well-behaved near caustics and can be applied to any linear wave equation. As a result, MGO promises to be a new paradigm for modeling caustics within ray-tracing codes. 

Rather than evolving the wavefield amplitude in the traditional $\Vect{x}$ coordinates, which leads to caustic singularities, MGO evolves the wavefield amplitude in mixed coordinate-momentum variables that are optimally chosen to avoid caustics. These mixed-variable representations are obtained using sequenced metaplectic transforms (MT) applied along the rays, after which MGO is named. (The trivial cases for mixed-variable representations, namely the pure $\Vect{x}$ or pure $\Vect{k}$ representation, are generated by the corresponding special cases of the MT: the identity and the Fourier transform.) MGO proceeds as follows: for each point on the dispersion manifold, (i) the phase space is rotated to align configuration space with the local tangent plane, (ii) GO is applied in the rotated phase space, (iii) the GO solution $\Psi_\Vect{t}(\Vect{X})$ is linked to previous and subsequent GO calculations via near-identity MTs (NIMTs) to ensure continuity, and (iv) $\Psi_\Vect{t}(\Vect{X})$ is transformed back to the original phase space using an inverse MT, summing over distinct branches of the dispersion manifold if applicable. MGO is suitable for quasioptical modeling when generalized to non-Euclidean ray-based coordinates. MGO is also `caustic-agnostic' in that it works on all types of caustics; hence, knowledge of the catastrophe-theoretical description of caustics, which is often assumed to be a prerequisite for modeling caustics, is unnecessary for MGO. This makes MGO more robust and accessible as a theory.

I present four different formulations of MGO which differ in the class of symplectic transformations and metaplectic transforms used. The first is the most basic formulation that is applicable any time $\det \Mat{B} \neq 0$. However, this formulation fails to describe `quasiuniform' ray patterns with $\det \Mat{B}_\Vect{t} = 0$ over a finite domain (an example being homogeneous media). Hence, I also present an extended formulation of MGO that can be applied to any ray pattern in both uniform and nonuniform media. This extended formulation is slightly more difficult to use in practice, since it requires one to perform an SVD of $\Mat{B}$ at each time step to determine whether the original or the extended formulation of MGO is appropriate. Therefore I have also presented a third formulation of MGO based on Gaussian coherent states that can be applied `as is' regardless the invertibility of $\Mat{B}$. Finally, since the rotation matrix $\Mat{S}$ can be easily constructed from ray information so that it is orthogonal as well as symplectic (orthosymplectic), I have also presented a formulation of MGO assuming orthosymplecticity. The formulas that result from this final formulation are considerably simpler than those published previously in \Refs{Lopez20, Lopez21a}, which allows for MGO to be related to standard GO and to other published semiclassical caustic-removal schemes in a straightforward manner. Indeed, I present here an explicit proof that MGO reduces to GO when evaluated away from caustics, which is an important result for instilling confidence in MGO but has thus far only been inferred from the results of numerical MGO calculations. I also present a new interpretation of MGO as a delta-windowed semiclassical integral that allows for arbitrary wavefunction profiles, rather than being restricted to bounded wavepackets as most semiclassical methods assume. I anticipate this observation will be the foundation for future dedicated comparison studies.
	
I demonstrate MGO analytically in five examples, namely, a plane wave propagating in uniform media (no caustic) in $1$-D, an EM wave incident on an isolated cutoff (Airy-type fold caustic) in both $1$-D and $2$-D, an EM wave bounded between two cutoffs (quantum harmonic oscillator), and an imperfectly focused plane wave in vacuum (Pearcey-type cusp caustic). In all examples, MGO provides an accurate representation of the exact solution that remains finite at the caustics, unlike traditional GO. Yet, as the final example shows, MGO does not always accurately model caustic shadows. Further extensions of MGO to incorporate caustic shadow fields by generalizing the MTs for complex-valued coordinate transformations~\cite{Wolf74} remain to be investigated. Future work should also attempt to include polarization dynamics for modeling vector waves with MGO, including spin--orbit coupling~\cite{Dodin19,Bliokh04a,Bliokh04b,Ruiz17a,Ruiz15a,Ruiz15b,Ruiz15d,Ruiz17t,Oancea20,Bliokh15}, mode conversion~\cite{Dodin19,Yanagihara19a,Yanagihara19b,Yanagihara21a,Yanagihara21b,Dodin17c}, and optical anisotropy~\cite{Ruiz17a,Kravtsov96,Kravtsov07,Bliokh07}.

Besides outlining the basic theory of MGO, this thesis also develops several algorithms for MGO. The first algorithm is a fast near-identity MT algorithm for evolving the optimal representation along a ray. This algorithm is based on a novel pseudo-differential representation of the MT in arbitrary dimensions, which can be useful for both analytical and numerical applications. An important example is the simulation of a wavepacket evolving in a quadratic potential, whose propagator is a metaplectic operator. Evolving the system by $\Delta t$ would invoke an MT that is near-identity, which is not a common consideration in MT-algorithm design. In contrast, the pseudo-differential representation that I propose here readily displays the simplicity of the MT in the near-identity limit, suggestive of a new algorithm. 

Specifically, in the near-identity limit the pseudo-differential series can be accurately truncated. Using a truncated Taylor approximation leads to what I call the `local' fast NIMT algorithm, because the correspondingly finite stencil width enables local, pointwise transformations. This is useful when transforming `incomplete' functions, \eg signals measured over finite intervals; it also leads naturally to a linear time algorithm. When applied once, this algorithm performs a fast, near-identity transformation; when iterated, the algorithm can perform an arbitrary MT by synthesizing a series of near-identity transformations. With a computational efficiency of $O(N_p N^3 K)$ (with $N_p$ the number of points, $N$ the number of dimensions, and $K$ the number of iterations), the local NIMT is potentially faster than existing MT algorithms, which often scale as $O(N_p \log N_p)$ from their similarity with the fast Fourier transform. Moreover, unlike these other algorithms, the local NIMT is the same algorithm regardless the number of dimensions and the structure of~$\Mat{S}$. Hence, the local NIMT is flexible in its application and should thereby complement the existing collection of MT algorithms. I proceed to assess the stability of the iterated local NIMT and identify two dominant instabilities: the loss of unitarity via truncation error (magnification), and the poor conditioning of discrete derivatives (d-instability). One might expect the magnification to be suppressed by reducing the transformation `step size', \ie its deviation from identity, or by increasing the number of terms retained; however, this is not true. Reducing the step size increases the number of iterations needed to perform a finite transformation, and it is not clear whether this tradeoff is beneficial in the general case. Increasing the truncation order indeed decreases the NIMT magnification, but also increases its susceptibility to the d-instability. The most robust avenue to local NIMT stability therefore appears to be the combined use of a low-order truncation with occasional smoothing, which I demonstrate in a numerical example.

Conversely, using a truncated Pad\'e approximation leads to what I call the `unitary' NIMT algorithm that (i) is exactly unitary and (ii) can be approximately computed in time that scales linearly with the number of grid points $N_p$. This is still faster than other known algorithms for the discrete MT, which scale as $O(N_p \log_2 N_p)$ due to their similarity with the fast Fourier transform. This formulation of the NIMT is a marked improvement over the local algorithm, which was not unitary and suffered from numerical instability as a consequence. Moreover, I proceed to prove that the unitary NIMT always converges to the discrete MT with a second-order accuracy when iterated along a suitable `trajectory' of near-identity symplectic matrices; hence the iterated unitary NIMT can be used to perform finite MTs in $O(KN_p)$ operations without experiencing any numerical instability. In particular, the FT can be computed with only two iterations in principle (two iterations ensure each iterate has $\det \textsf{A} \neq 0$), but $K \gg 1$ would be necessary to achieve a reasonable accuracy. I expect these results to be useful for reduced numerical modeling of wave caustics via MGO, among other possible applications.

The second key algorithm developed in this thesis is a Gauss--Freud steepest-descent quadrature rule for performing the inverse MT that reverts the optimal representation along a ray back to the original $\Vect{x}$ variables. These MT integrals are highly oscillatory and cannot be easily calculated using standard numerical methods. However, when evaluated along the steepest-descent curves the integrals become exponentially decaying and amenable to standard Gaussian quadrature. I first validate my algorithm on isolated saddlepoints (\ie cuspoid caustics) of various degeneracy to demonstrate the expected $2n-1$ polynomial accuracy of an $n$-point Gaussian quadrature formula. I then use my algorithm to simulate an EM wave propagating into an unmagnetized plasma that has a fold-type caustic at the critical cutoff density, both an isolated cutoff (\ie the Airy problem) and a pair of cutoffs (\ie the quantum harmonic oscillator problem). In both cases the numerical solutions agree remarkably well with the exact solution and significantly improve upon the analytically approximated MGO solution that are also presented in this thesis. This encouraging result provides strong evidence that MGO can be suitable for practical applications.

Lastly, other auxiliary results presented in this thesis are the GO equations for an arbitrarily rotated phase space (and more generally, a phase space obtained by an arbitrary linear symplectic transformation), an adaptive integration of the ray trajectories specifically tailored to MGO, and an orthogonalization procedure to determine the optimal MT at each point along a ray. The foundations are now set for the development of an MGO-based ray-tracing code. The orthosymplectic transformations advocated here have consequences in this regard: the orthosymplectic MGO formalism is simpler to compute and more memory efficient, since the enhanced symmetry of $\Mat{S}$ means less elements need to be stored; however, errors may develop due to any erroneous non-orthogonality of $\Mat{S}$ introduced by the chosen numerical method for performing the orthogonalization. The tradeoff between these effects can be investigated when benchmarking a future MGO-based code.

\bibliographystyle{ieeetr}
\bibliography{/Users/Nick/Documents/Biblio.bib}

\begin{thebibliography}{100}

\bibitem{Stix92}
T.~H. Stix, {\em Waves in Plasmas}.
\newblock New York: American Institute of Physics, 1992.

\bibitem{Fisch87}
N.~J. Fisch, ``Theory of current drive in plasmas,'' {\em Rev. Mod. Phys.},
  vol.~59, p.~175, 1987.

\bibitem{Prater08}
R.~Prater, D.~Farina, Y.~Gribov, R.~W. Harvey, A.~K. Ram, Y.~R. Lin-Liu,
  E.~Poli, A.~P. Smirnov, F.~Volpe, E.~Westerhof, A.~Zvonkov, and {the ITPA
  Steady State Operation Topical Group}, ``Benchmarking of codes for electron
  cyclotron heating and electron cyclotron current drive under {ITER}
  conditions,'' {\em Nucl. Fusion}, vol.~48, p.~035006, 2008.

\bibitem{Freidberg10}
J.~P. Freidberg, {\em Plasma Physics and Fusion Energy}.
\newblock Cambridge: Cambridge University Press, 2010.

\bibitem{Wesson11}
J.~Wesson, {\em Tokamaks}.
\newblock Oxford: Oxford University Press, 4~ed., 2011.

\bibitem{Farina14}
D.~Farina, M.~Henderson, L.~Figini, and G.~Saibene, ``Optimization of the
  {ITER} electron cyclotron equatorial launcher for improved heating and
  current drive functional capabilities,'' {\em Phys. Plasmas}, vol.~21,
  p.~061504, 2014.

\bibitem{Poli15}
F.~M. Poli, R.~G. Andre, N.~Bertelli, S.~P. Gerhardt, D.~Mueller, and
  G.~Taylor, ``Simulations towards the achievement of non-inductive current
  ramp-up and sustainment in the {National Spherical Torus Experiment
  Upgrade},'' {\em Nucl. Fusion}, vol.~55, p.~123011, 2015.

\bibitem{Lopez18a}
N.~A. Lopez and F.~M. Poli, ``Regarding the optimization of {O1-mode ECRH} and
  the feasibility of {EBW} startup on {NSTX-U},'' {\em Plasma Phys. Control.
  Fusion}, vol.~60, p.~065007, 2018.

\bibitem{Marinak01}
M.~M. Marinak, G.~D. Kerbel, N.~A. Gentile, O.~Jones, D.~Munro, S.~Pollaine,
  T.~R. Dittrich, and S.~W. Haan, ``Three-dimensional {HYDRA} simulations of
  {National Ignition Facility} targets,'' {\em Phys. Plasmas}, vol.~8,
  p.~2275, 2001.

\bibitem{Lindl04}
J.~D. Lindl, P.~Amendt, R.~L. Berger, S.~G. Glendinning, S.~H. Glenzer, S.~W.
  Haan, R.~L. Kauffman, O.~L. Landen, and L.~J. Suter, ``The physics basis for
  ignition using indirect-drive targets on the {National Ignition Facility},''
  {\em Phys. Plasmas}, vol.~11, p.~339, 2004.

\bibitem{Hohenberger15}
M.~Hohenberger, P.~B. Radha, J.~F. Myatt, S.~{Le Pape}, J.~A. Marozas, F.~J.
  Marshall, D.~T. Michel, S.~P. Regan, W.~Seka, A.~Shvydky, T.~C. Sangster,
  J.~W. Bates, R.~Betti, T.~R. Boehly, M.~J. Bonino, D.~T. Casey, T.~J.~B.
  Collins, R.~S. Craxton, J.~A. Delettrez, D.~H. Edgell, R.~Epstein, G.~Fiksel,
  P.~Fitzsimmons, J.~A. Frenje, D.~H. Froula, V.~N. Goncharov, D.~R. Harding,
  D.~H. Kalantar, M.~Karasik, T.~J. Kessler, J.~D. Kilkenny, J.~P. Knauer,
  C.~Kurz, M.~Lafon, K.~N. LaFortune, B.~J. MacGowan, A.~J. MacKinnon, A.~G.
  MacPhee, R.~L. McCrory, P.~W. McKenty, J.~F. Meeker, D.~D. Meyerhofer, S.~R.
  Nagel, A.~Nikroo, S.~Obenschain, R.~D. Petrasso, J.~E. Ralph, H.~G.
  Rinderknecht, M.~J. Rosenberg, A.~J. Schmitt, R.~J. Wallace, J.~Weaver,
  C.~Widmayer, S.~Skupsky, A.~A. Solodov, C.~Stoeckl, B.~Yaakobi, and J.~D.
  Zuegel, ``Polar-direct-drive experiments on the {National Ignition
  Facility},'' {\em Phys. Plasmas}, vol.~22, p.~056308, 2015.

\bibitem{Craxton15}
R.~S. Craxton, K.~S. Anderson, T.~R. Boehly, V.~N. Goncharov, D.~R. Harding,
  J.~P. Knauer, R.~L. McCrory, P.~W. McKenty, D.~D. Meyerhofer, J.~F. Myatt,
  A.~J. Schmitt, J.~D. Sethian, R.~W. Short, S.~Skupsky, W.~Theobald, W.~L.
  Kruer, K.~Tanaka, R.~Betti, T.~J.~B. Collins, J.~A. Delettrez, S.~X. Hu,
  J.~A. Marozas, A.~V. Maximov, D.~T. Michel, P.~B. Radha, S.~P. Regan, T.~C.
  Sangster, W.~Seka, A.~A. Solodov, J.~M. Soures, C.~Stoeckl, and J.~D. Zuegel,
  ``Direct-drive inertial confinement fusion: A review,'' {\em Phys. Plasmas},
  vol.~22, p.~110501, 2015.

\bibitem{Robey18}
H.~F. Robey, L.~F. {Berzak Hopkins}, J.~L. Milovich, and N.~B. Meezan, ``The
  {I-Raum}: A new shaped hohlraum for improved inner beam propagation in
  indirectly-driven {ICF} implosions on the {National Ignition Facility},''
  {\em Phys. Plasmas}, vol.~25, p.~012711, 2018.

\bibitem{Kritcher21}
A.~L. Kritcher, A.~B. Zylstra, D.~A. Callahan, O.~A. Hurricane, C.~Weber,
  J.~Ralph, D.~T. Casey, A.~Pak, K.~Baker, B.~Bachmann, S.~Bhandarkar,
  J.~Biener, R.~Bionta, T.~Braun, M.~Bruhn, C.~Choate, D.~Clark, J.~M. {Di
  Nicola}, L.~Divol, T.~Doppner, V.~Geppert-Kleinrath, S.~Haan, J.~Heebner,
  V.~Hernandez, D.~Hinkel, M.~Hohenberger, H.~Huang, C.~Kong, S.~{Le Pape},
  D.~Mariscal, E.~Marley, L.~Masse, K.~D. Meaney, M.~Millot, A.~Moore,
  K.~Newman, A.~Nikroo, P.~Patel, L.~Pelz, N.~Rice, H.~Robey, J.~S. Ross,
  M.~Rubery, J.~Salmonson, D.~Schlossberg, S.~Sepke, K.~Sequoia, M.~Stadermann,
  D.~Strozzi, R.~Tommasini, P.~Volegov, C.~Wild, S.~Yang, C.~Young, M.~J.
  Edwards, O.~L. Landen, R.~Town, and M.~Herrmann, ``Achieving record hot spot
  energies with large {HDC} implosions on {NIF} in {HYBRID-E},'' {\em Phys.
  Plasmas}, vol.~28, p.~072706, 2021.

\bibitem{Tracy14}
E.~R. Tracy, A.~J. Brizard, A.~S. Richardson, and A.~N. Kaufman, {\em Ray
  Tracing and Beyond: Phase Space Methods in Plasma Wave Theory}.
\newblock Cambridge: Cambridge University Press, 2014.

\bibitem{Kravtsov90}
Y.~A. Kravtsov and Y.~I. Orlov, {\em Geometrical Optics of Inhomogeneous
  Media}.
\newblock Berlin: Springer, 1990.

\bibitem{Poli18a}
F.~M. Poli, ``Integrated tokamak modeling: When physics informs engineering and
  research planning,'' {\em Phys. Plasmas}, vol.~25, p.~055602, 2018.

\bibitem{Poli16}
F.~M. Poli, P.~T. Bonoli, M.~Chilenski, R.~Mumgaard, S.~Shiraiwa, G.~M.
  Wallace, R.~Andre, L.~Delgado-Aparicio, S.~Scott, J.~R. Wilson, R.~W. Harvey,
  Y.~V. Petrov, M.~Reinke, I.~Faust, R.~Granetz, J.~Hughes, and J.~Rice,
  ``Experimental and modeling uncertainties in the validation of lower hybrid
  current drive,'' {\em Plasma Phys. Control. Fusion}, vol.~58,
  p.~095001, 2016.

\bibitem{Hawryluk81}
R.~J. Hawryluk, ``An empirical approach to tokamak transport,'' in {\em Physics
  of Plasmas Close to Thermonuclear Conditions} (B.~{Coppi}, G.~G. {Leotta},
  D.~{Pfirsch}, R.~{Pozzoli}, and E.~{Sindoni}, eds.), vol.~1, p.~19, Brussels:
  Commission of the European Communities, 1981.

\bibitem{Kravtsov93}
Y.~A. Kravtsov and Y.~I. Orlov, {\em Caustics, Catastrophes and Wave Fields}.
\newblock Berlin: Springer, 1993.

\bibitem{Berry80b}
M.~V. Berry and C.~Upstill, ``Catastrophe optics: Morphologies of caustics and
  their diffraction patterns,'' {\em Prog. Opt.}, vol.~18, p.~257, 1980.

\bibitem{Peng86}
Y.-K.~M. Peng and D.~J. Strickler, ``Features of spherical torus plasmas,''
  {\em Nucl. Fusion}, vol.~26, p.~769, 1986.

\bibitem{Peng00}
Y.-K.~M. Peng, ``The physics of spherical torus plasmas,'' {\em Phys. Plasmas},
  vol.~7, p.~1681, 2000.

\bibitem{Ono15a}
M.~Ono and R.~Kaita, ``Recent progress on spherical torus research,'' {\em
  Phys. Plasmas}, vol.~22, p.~040501, 2015.

\bibitem{Erckmann94}
V.~Erckmann and U.~Gasparino, ``Electron cyclotron resonance heating and
  current drive in toroidal fusion plasmas,'' {\em Plasma Phys. Control.
  Fusion}, vol.~36, p.~1869, 1994.

\bibitem{Prater04}
R.~Prater, ``Heating and current drive by electron cyclotron waves,'' {\em
  Phys. Plasmas}, vol.~11, p.~2349, 2004.

\bibitem{Ram00}
A.~K. Ram and S.~D. Schultz, ``Excitation, propagation, and damping of electron
  {Bernstein} waves in tokamaks,'' {\em Phys. Plasmas}, vol.~7,
  p.~4084, 2000.

\bibitem{Shiraiwa06}
S.~Shiraiwa, K.~Hanada, M.~Hasegawa, H.~Idei, H.~Kasahara, O.~Mitarai,
  K.~Nakamura, N.~Nishino, H.~Nozato, M.~Sakamoto, K.~Sasaki, K.~Sato,
  Y.~Takase, T.~Yamada, and H.~Zushi, ``Heating by an electron {Bernstein} wave
  in a spherical tokamak plasma via mode conversion,'' {\em Phys. Rev. Lett.},
  vol.~96, p.~185003, 2006.

\bibitem{Uchijima15}
K.~Uchijima, T.~Takemoto, J.~Morikawa, and Y.~Ogawa, ``Direct observation of
  transition to electron {Bernstein} waves from electromagnetic mode by three
  mode-conversion scenarios in the dipole confinement torus plasma,'' {\em
  Plasma Phys. Control. Fusion}, vol.~57, p.~065003, 2015.

\bibitem{Seltzman17}
A.~H. Seltzman, J.~K. Anderson, S.~J. Diem, J.~A. Goetz, and C.~B. Forest,
  ``Observation of electron {Bernstein} wave heating in a reversed field
  pinch,'' {\em Phys. Rev. Lett.}, vol.~119, p.~185001, 2017.

\bibitem{Lopez18b}
N.~A. Lopez and A.~K. Ram, ``Mode-conversion of the extraordinary wave at the
  upper hybrid resonance in the presence of small-amplitude density
  fluctuations,'' {\em Plasma Phys. Control. Fusion}, vol.~60,
  p.~125012, 2018.

\bibitem{Laqua07}
H.~P. Laqua, ``Electron {Bernstein} wave heating and diagnostic,'' {\em Plasma
  Phys. Control. Fusion}, vol.~49, p.~R1, 2007.

\bibitem{Preinhaelter73}
J.~Preinhaelter and V.~Kopecky, ``Penetration of high-frequency waves into a
  weakly inhomogeneous magnetized plasma at oblique incidence and their
  transformation to {Bernstein} modes,'' {\em J. Plasma Phys.}, vol.~10,
  p.~1, 1973.

\bibitem{Hansen85}
F.~R. Hansen, J.~P. Lynov, and P.~Michelsen, ``The {O-X-B} mode conversion
  scheme for {ECRH} of a high-density tokamak plasma,'' {\em Plasma Phys.
  Control. Fusion}, vol.~27, p.~1077, 1985.

\bibitem{Mjolhus84}
E.~Mjolhus, ``Coupling to {Z} mode near critical angle,'' {\em J. Plasma
  Phys.}, vol.~31, p.~7, 1984.

\bibitem{Laqua03}
H.~P. Laqua, H.~Maassberg, N.~B. Marushchenko, F.~Volpe, A.~Weller, and
  W.~Kasparek, ``{Electron-Bernstein-wave} current drive in an overdense plasma
  at the {Wendelstein 7-AS} stellarator,'' {\em Phys. Rev. Lett.}, vol.~90,
  p.~075003, Feb 2003.

\bibitem{Shevchenko07}
V.~F. Shevchenko, G.~Cunningham, A.~Gurchenko, E.~Gusakov, B.~Lloyd,
  M.~O'Brien, A.~N. Saveliev, A.~Surkov, F.~A. Volpe, and M.~Walsh,
  ``Development of electron {Bernstein} wave research in {MAST},'' {\em Fusion
  Sci. Technol.}, vol.~52, p.~202, 2007.

\bibitem{Tracy93}
E.~R. Tracy and A.~N. Kaufman, ``Metaplectic formulation of linear mode
  conversion,'' {\em Phys. Rev. E}, vol.~48, p.~2196, 1993.

\bibitem{Tracy01}
E.~R. Tracy, A.~N. Kaufman, and A.~Jaun, ``A ray-based algorithm for
  multi-dimensional linear conversion,'' {\em Phys. Lett. A}, vol.~290,
  p.~309, 2001.

\bibitem{Tracy03a}
E.~R. Tracy, A.~N. Kaufman, and A.~J. Brizard, ``Ray-based methods in
  multidimensional linear wave conversion,'' {\em Phys. Plasmas}, vol.~10,
  p.~2147, 2003.

\bibitem{Tracy07}
E.~R. Tracy, A.~N. Kaufman, and A.~Jaun, ``Local fields for asymptotic matching
  in multidimensional mode conversion,'' {\em Phys. Plasmas}, vol.~14,
  p.~082102, 2007.

\bibitem{Jaun07}
A.~Jaun, E.~R. Tracy, and A.~N. Kaufman, ``Eikonal waves, caustics and mode
  conversion in tokamak plasmas,'' {\em Plasma Phys. Control. Fusion}, vol.~49,
  p.~43, 2007.

\bibitem{Richardson08}
A.~S. Richardson and E.~R. Tracy, ``Quadratic corrections to the metaplectic
  formulation of resonant mode conversion,'' {\em J. Phys. A: Math. Theor.},
  vol.~41, p.~375501, 2008.

\bibitem{Ruiz15a}
D.~E. Ruiz and I.~Y. Dodin, ``Lagrangian geometrical optics of nonadiabatic
  vector waves and spin particles,'' {\em Phys. Lett. A}, vol.~379,
  p.~2337, 2015.

\bibitem{Ruiz15b}
D.~E. Ruiz and I.~Y. Dodin, ``On the correspondence between quantum and
  classical variational principles,'' {\em Phys. Lett. A}, vol.~379,
  p.~2623, 2015.

\bibitem{Ruiz17a}
D.~E. Ruiz and I.~Y. Dodin, ``Ponderomotive dynamics of waves in
  quasiperiodically modulated media,'' {\em Phys. Rev. A}, vol.~95,
  p.~032114, 2017.

\bibitem{Dodin19}
I.~Y. Dodin, D.~E. Ruiz, K.~Yanagihara, Y.~Zhou, and S.~Kubo, ``Quasioptical
  modeling of wave beams with and without mode conversion: I. {Basic} theory,''
  {\em Phys. Plasmas}, vol.~26, p.~072110, 2019.

\bibitem{Yanagihara19a}
K.~Yanagihara, I.~Y. Dodin, and S.~Kubo, ``Quasioptical modeling of wave beams
  with and without mode conversion: {II}. {Numerical} simulations of
  single-mode beams,'' {\em Phys. Plasmas}, vol.~26, p.~072111, 2019.

\bibitem{Yanagihara19b}
K.~Yanagihara, I.~Y. Dodin, and S.~Kubo, ``Quasioptical modeling of wave beams
  with and without mode conversion: {III}. {Numerical} simulations of
  mode-converting beams,'' {\em Phys. Plasmas}, vol.~26, p.~072112,
  2019.

\bibitem{Yanagihara21a}
K.~Yanagihara, S.~Kubo, I.~Dodin, and {the LHD Experiment Group},
  ``Quasioptical propagation and absorption of electron cyclotron waves:
  simulations and experiment,'' {\em Nucl. Fusion}, vol.~61, p.~106012,
  2021.

\bibitem{Yanagihara21b}
K.~Yanagihara, S.~Kubo, and I.~Y. Dodin, ``Quasioptical modeling of wave beams
  with and without mode conversion: {IV}. {Numerical} simulations of waves in
  dissipative media,'' {\em Phys. Plasmas}, vol.~28, p.~122102, 2021.

\bibitem{Berry76}
M.~V. Berry, ``Waves and {Thom's} theorem,'' {\em Adv. Phys.}, vol.~25,
  p.~1, 1976.

\bibitem{Hobbs07}
C.~A. Hobbs, J.~N.~L. Connor, and N.~P. Kirk, ``Theory and numerical evaluation
  of oddoids and evenoids: Oscillatory cuspoid integrals with odd and even
  polynomial phase functions,'' {\em J. Comput. Appl. Math.}, vol.~207, 
  p.~192, 2007.

\bibitem{Borghi16}
R.~Borghi, ``Computational optics through sequence transformations,'' {\em
  Prog. Opt.}, vol.~61, p.~1, 2016.

\bibitem{Zannotti17a}
A.~Zannotti, F.~Diebel, M.~Boguslawski, and C.~Denz, ``Optical catastrophes of
  the swallowtail and butterfly beams,'' {\em New J. Phys.}, vol.~19,
  p.~053004, 2017.

\bibitem{EspindolaRamos19}
E.~Espindola-Ramos, G.~Silva-Ortigoza, C.~T. Sosa-Sanchez, I.~Julian-Macias,
  O.~d. Cabrera-Rosas, P.~Ortega-Vidals, A.~Gonzalez-Juarez, R.~Silva-Ortigoza,
  M.~P. Velazquez-Quesada, and G.~F. {Torres del Castillo}, ``Paraxial optical
  fields whose intensity pattern skeletons are stable caustics,'' {\em J. Opt.
  Soc. Am. A}, vol.~36, p.~1820, 2019.

\bibitem{Wright09}
J.~C. Wright, P.~T. Bonoli, A.~E. Schmidt, C.~K. Phillips, E.~J. Valeo, R.~W.
  Harvey, and M.~A. Brambilla, ``An assessment of full wave effects on the
  propagation and absorption of lower hybrid waves,'' {\em Phys. Plasmas},
  vol.~16, p.~072502, 2009.

\bibitem{Shiraiwa10}
S.~Shiraiwa, O.~Meneghini, R.~Parker, P.~Bonoli, M.~Garrett, M.~C. Kaufman,
  J.~C. Wright, and S.~Wukitch, ``Plasma wave simulation based on a versatile
  finite element method solver,'' {\em Phys. Plasmas}, vol.~17,
  p.~056119, 2010.

\bibitem{Myatt17}
J.~F. Myatt, R.~K. Follett, J.~G. Shaw, D.~H. Edgell, D.~H. Froula, I.~V.
  Igumenshchev, and V.~N. Goncharov, ``A wave-based model for cross-beam energy
  transfer in direct-drive inertial confinement fusion,'' {\em Phys. Plasmas},
  vol.~24, p.~056308, 2017.

\bibitem{Ludwig66}
D.~Ludwig, ``Uniform asymptotic expansions at a caustic,'' {\em Commun. Pure
  Appl. Math.}, vol.~19, p.~215, 1966.

\bibitem{Berry72}
M.~V. Berry and K.~E. Mount, ``Semiclassical approximations in wave
  mechanics,'' {\em Rep. Prog. Phys.}, vol.~35, p.~315, 1972.

\bibitem{Jackson75}
J.~D. Jackson, {\em Classical Electrodynamics}.
\newblock New York: Wiley, 2~ed., 1975.

\bibitem{Born99}
M.~Born and E.~Wolf, {\em Principles of Optics}.
\newblock Cambridge: Cambridge University Press, 7~ed., 1999.

\bibitem{Maslov81}
V.~P. Maslov and M.~V. Fedoriuk, {\em Semiclassical Approximation in Quantum
  Mechanics}.
\newblock Dordrecht, Netherlands: Reidel, 1981.

\bibitem{Ziolkowski84}
R.~W. Ziolkowski and G.~A. Deschamps, ``Asymptotic evaluation of high-frequency
  fields near a caustic: An introduction to {Maslov's} method,'' {\em Radio
  Sci.}, vol.~19, p.~1001, 1984.

\bibitem{Thomson85}
C.~J. Thomson and C.~H. Chapman, ``An introduction to {Maslov's} asymptotic
  method,'' {\em Geophys. J. R. Astron. Soc.}, vol.~83, p.~143, 1985.

\bibitem{Littlejohn85}
R.~G. Littlejohn, ``Symplectically invariant {WKB} wave functions,'' {\em Phys.
  Rev. Lett.}, vol.~54, p.~1742, 1985.

\bibitem{Littlejohn86b}
R.~G. Littlejohn, ``Wave-packet evolution and quantization,'' {\em Phys. Rev.
  Lett.}, vol.~56, p.~2000, 1986.

\bibitem{Kay94a}
K.~G. Kay, ``Integral expressions for the semiclassical time-dependent
  propagator,'' {\em J. Chem. Phys.}, vol.~100, p.~4377, 1994.

\bibitem{Zor96}
D.~Zor and K.~G. Kay, ``Globally uniform semiclassical expressions for
  time-independent wave functions,'' {\em Phys. Rev. Lett.}, vol.~76, 
  p.~1990, 1996.

\bibitem{Madhusoodanan98}
M.~Madhusoodanan and K.~G. Kay, ``Globally uniform semiclassical wave functions
  for multidimensional systems,'' {\em J. Chem. Phys.}, vol.~109, 
  p.~2644, 1998.

\bibitem{Alonso97b}
M.~A. Alonso and G.~W. Forbes, ``Uniform asymptotic expansions for wave
  propagators via fractional transformations,'' {\em J. Opt. Soc. Am. A},
  vol.~14, p.~1279, 1997.

\bibitem{Alonso99}
M.~A. Alonso and G.~W. Forbes, ``New approach to semiclassical analysis in
  mechanics,'' {\em J. Math. Phys.}, vol.~40, p.~1699, 1999.

\bibitem{Lopez19}
N.~A. Lopez and I.~Y. Dodin, ``Pseudo-differential representation of the
  metaplectic transform and its application to fast algorithms,'' {\em J. Opt.
  Soc. Am. A}, vol.~36, p.~1846, 2019.

\bibitem{Lopez20}
N.~A. Lopez and I.~Y. Dodin, ``Restoring geometrical optics near caustics using
  sequenced metaplectic transforms,'' {\em New J. Phys.}, vol.~22, 
  p.~083078, 2020.

\bibitem{Lopez21a}
N.~A. Lopez and I.~Y. Dodin, ``Metaplectic geometrical optics for modeling
  caustics in uniform and non-uniform media,'' {\em J. Opt.}, vol.~23, 
  p.~025601, 2021.

\bibitem{Lopez21b}
N.~A. Lopez and I.~Y. Dodin, ``Exactly unitary discrete representations of the
  metaplectic transform for linear-time algorithms,'' {\em J. Opt. Soc. Am. A},
  vol.~38, p.~634, 2021.

\bibitem{Donnelly21}
S.~M. Donnelly, N.~A. Lopez, and I.~Y. Dodin, ``Steepest-descent algorithm for
  simulating plasma-wave caustics via metaplectic geometrical optics,'' {\em
  Phys. Rev. E}, vol.~104, p.~025304, 2021.

\bibitem{Lopez22}
N.~A. Lopez and I.~Y. Dodin, ``Metaplectic geometrical optics for ray-based
  modeling of caustics: Theory and algorithms,'' {\em Phys. Plasmas}, vol.~29,
  p.~052111, 2022.

\bibitem{Ozaktas96}
H.~M. Ozaktas, O.~Arikan, M.~A. Kutay, and G.~Bozdagi, ``Digital computation of
  the fractional {Fourier} transform,'' {\em IEEE Trans. Signal Processing},
  vol.~44, p.~2141, 1996.

\bibitem{Hennelly05b}
B.~M. Hennelly and J.~T. Sheridan, ``Fast numerical algorithm for the linear
  canonical transform,'' {\em J. Opt. Soc. Am. A}, vol.~22, p.~928,
  2005.

\bibitem{Healy10}
J.~J. Healy and J.~T. Sheridan, ``Fast linear canonical transforms,'' {\em J.
  Opt. Soc. Am. A}, vol.~27, p.~21, 2010.

\bibitem{Koc10a}
A.~Koc, H.~M. Ozaktas, and L.~Hesselink, ``Fast and accurate computation of
  two-dimensional non-separable quadratic-phase integrals,'' {\em J. Opt. Soc.
  Am. A}, vol.~27, p.~1288, 2010.

\bibitem{Ding12}
J.-J. Ding, S.-C. Pei, and C.-L. Liu, ``Improved implementation algorithms of
  the two-dimensional nonseparable linear canonical transform,'' {\em J. Opt.
  Soc. Am. A}, vol.~29, p.~1615, 2012.

\bibitem{Pei16}
S.-C. Pei and S.-G. Huang, ``Two-dimensional nonseparable discrete linear
  canonical transform based on {CM-CC-CM-CC} decomposition,'' {\em J. Opt. Soc.
  Am. A}, vol.~33, p.~214, 2016.

\bibitem{Sun18a}
Y.-N. Sun and B.-Z. Li, ``Segmented fast linear canonical transform,'' {\em J.
  Opt. Soc. Am. A}, vol.~35, p.~1346, 2018.

\bibitem{Healy18}
J.~J. Healy, ``Simulating first order optical systems - algorithms for and
  composition of discrete linear canonical transforms,'' {\em J. Opt.},
  vol.~20, p.~014008, 2018.

\bibitem{Deano17}
A.~Deano, D.~Huybrechs, and A.~Iserles, {\em Computing Highly Oscillatory
  Integrals}.
\newblock Philadelphia: SIAM, 2017.

\bibitem{Deano09}
A.~Deano and D.~Huybrechs, ``Complex {Gaussian} quadrature of oscillatory
  integrals,'' {\em Numer. Math.}, vol.~112, p.~197, 2009.

\bibitem{Dodin17d}
I.~Y. Dodin, D.~E. Ruiz, and S.~Kubo, ``Mode conversion in cold low-density
  plasma with a sheared magnetic field,'' {\em Phys. Plasmas}, vol.~24, 
  p.~122116, 2017.

\bibitem{Shankar94}
R.~Shankar, {\em Principles of Quantum Mechanics}.
\newblock New York: Plenum, 2~ed., 1994.

\bibitem{Stoler81}
D.~Stoler, ``Operator methods in physical optics,'' {\em J. Opt. Soc. Am.},
  vol.~71, p.~334, 1981.

\bibitem{McDonald88a}
S.~W. McDonald, ``Phase-space representations of wave equations with
  applications to the eikonal approximation for short-wavelength waves,'' {\em
  Phys. Rep.}, vol.~158, p.~337, 1988.

\bibitem{Whitham74}
G.~B. Whitham, {\em Linear and Nonlinear Waves}.
\newblock New York: Wiley, 1974.

\bibitem{Arnold89}
V.~I. Arnold, {\em Mathematical Methods of Classical Mechanics}.
\newblock New York: Springer, 1989.

\bibitem{Poston96}
T.~Poston and I.~Stewart, {\em Catastrophe Theory and Its Applications}.
\newblock New York: Dover, 1996.

\bibitem{Arnold83}
V.~I. Arnold, ``Singularities, bifurcations, and catastrophes,'' {\em Sov.
  Phys. Usp.}, vol.~26, p.~1025, 1983.

\bibitem{Olver10a}
F.~W.~J. Olver, D.~W. Lozier, R.~F. Boisvert, and C.~W. Clark, {\em NIST
  Handbook of Mathematical Functions}.
\newblock Cambridge: Cambridge University Press, 2010.

\bibitem{Arnold75}
V.~I. Arnold, ``Critical points of smooth functions and their normal forms,''
  {\em Russ. Math. Surv.}, vol.~30, p.~1, 1975.

\bibitem{Kaiser00}
T.~B. Kaiser, ``Laser ray tracing and power deposition on an unstructured
  three-dimensional grid,'' {\em Phys. Rev. E}, vol.~61, p.~895, 2000.

\bibitem{Colaitis18}
A.~Colaitis, T.~Chapman, D.~Strozzi, L.~Divol, and P.~Michel, ``A
  tesselation-based model for intensity estimation and laser plasma
  interactions calculations in three dimensions,'' {\em Phys. Plasmas},
  vol.~25, p.~033114, 2018.

\bibitem{Colaitis21}
A.~Colaitis, I.~Igumenshchev, J.~Mathiaud, and V.~Goncharov, ``Inverse ray
  tracing on icosahedral tetrahedron grids for non-linear laser plasma
  interaction coupled to {3D} radiation hydrodynamics,'' {\em J. Comput.
  Phys.}, vol.~442, p.~110537, 2021.

\bibitem{Michel09}
P.~Michel, L.~Divol, E.~A. Williams, S.~Weber, C.~A. Thomas, D.~A. Callahan,
  S.~W. Haan, J.~D. Salmonson, S.~Dixit, D.~E. Hinkel, M.~J. Edwards, B.~J.
  MacGowan, J.~D. Lindl, S.~H. Glenzer, and L.~J. Suter, ``Tuning the implosion
  symmetry of {ICF} targets via controlled crossed-beam energy transfer,'' {\em
  Phys. Rev. Lett.}, vol.~102, p.~025004, 2009.

\bibitem{Chester57}
C.~Chester, B.~Friedman, and F.~Ursell, ``An extension of the method of
  steepest descents,'' {\em Proc. Cambridge Philos. Soc.}, vol.~53, 
  p.~599, 1957.

\bibitem{Colaitis19a}
A.~Colaitis, J.~P. Palastro, R.~K. Follett, I.~V. Igumenshchev, and
  V.~Goncharov, ``Real and complex valued geometrical optics inverse
  ray-tracing for inline field calculations,'' {\em Phys. Plasmas}, vol.~26,
  p.~032301, 2019.

\bibitem{Colaitis19b}
A.~Colaitis, R.~K. Follett, J.~P. Palastro, I.~Igumenshchev, and V.~Goncharov,
  ``Adaptive inverse ray-tracing for accurate and efficient modeling of cross
  beam energy transfer in hydrodynamics simulations,'' {\em Phys. Plasmas},
  vol.~26, p.~072706, 2019.

\bibitem{Lopez21DPPb}
N.~A. Lopez, E.~Kur, T.~D. Chapman, D.~J. Strozzi, and P.~A. Michel,
  ``Influence of speckles on laser intensity profile near turning points,''
  {\em Bull. Am. Phys. Soc.}, vol.~63, p.~Abstract NP11:00063, 2021.

\bibitem{Knudson85}
S.~K. Knudson, J.~B. Delos, and B.~Bloom, ``Semiclassical calculation of
  quantum-mechanical wave functions for a two-dimensional scattering system,''
  {\em J. Chem. Phys.}, vol.~83, p.~5703, 1985.

\bibitem{Knudson86}
S.~K. Knudson, J.~B. Delos, and D.~W. Noid, ``Bound state semiclassical wave
  functions,'' {\em J. Chem. Phys.}, vol.~84, p.~6886, 1986.

\bibitem{Pool66}
J.~C.~T. Pool, ``Mathematical aspects of the {Weyl} correspondence,'' {\em J.
  Math. Phys.}, vol.~7, p.~66, 1966.

\bibitem{Littlejohn86a}
R.~G. Littlejohn, ``The semiclassical evolution of wave packets,'' {\em Phys.
  Rep.}, vol.~138, p.~193, 1986.

\bibitem{Case08}
W.~B. Case, ``Wigner functions and {Weyl} transforms for pedestrians,'' {\em
  Am. J. Phys.}, vol.~76, p.~937, 2008.

\bibitem{Alonso11}
M.~A. Alonso, ``Wigner functions in optics: describing beams as ray bundles and
  pulses as particle ensembles,'' {\em Adv. Opt. Photon.}, vol.~3,
  p.~272, 2011.

\bibitem{Ruiz17t}
D.~E. Ruiz, {\em A geometric theory of waves and its applications to plasma
  physics}.
\newblock PhD thesis, Princeton University, 2017.

\bibitem{Dodin22}
I.~Y. Dodin, ``Quasilinear theory for inhomogeneous plasma,'' {\em
  arXiv:2201.08562}, 2022.

\bibitem{Goldstein02}
H.~Goldstein, C.~P. Poole, and J.~L. Safko, {\em Classical Mechanics}.
\newblock New York: Addison-Wesley, 3~ed., 2002.

\bibitem{Luneburg64}
R.~K. Luneburg, {\em Mathematical Theory of Optics}.
\newblock Berkeley: U. California Press, 1964.

\bibitem{Press07}
W.~H. Press, S.~A. Teukolsky, W.~T. Vetterling, and B.~P. Flannery, {\em
  Numerical Recipes}.
\newblock Cambridge: Cambridge University Press, 3~ed., 2007.

\bibitem{Dragt05}
A.~J. Dragt, ``The symplectic group and classical mechanics,'' {\em Ann. N. Y.
  Acad. Sci.}, vol.~1045, p.~291, 2005.

\bibitem{Dragt82}
A.~J. Dragt, ``Lectures on nonlinear orbit dynamics,'' {\em AIP Conf. Proc.},
  vol.~87, p.~147, 1982.

\bibitem{Hall15}
B.~C. Hall, {\em Lie Groups, Lie Algebras, and Representations: An Elementary
  Introduction}.
\newblock Berlin: Springer, 2015.

\bibitem{Moshinsky71}
M.~Moshinsky and C.~Quesne, ``Linear canonical transformations and their
  unitary representations,'' {\em J. Math. Phys.}, vol.~12, p.~1772,
  1971.

\bibitem{Collins70}
S.~A. Collins, ``Lens-system diffraction integral written in terms of matrix
  optics,'' {\em J. Opt. Soc. Am.}, vol.~60, p.~1168, 1970.

\bibitem{Scully12}
M.~O. Scully and M.~S. Zubairy, {\em Quantum Optics}.
\newblock Cambridge: Cambridge University Press, 2012.

\bibitem{Kogelnik66}
H.~Kogelnik and T.~Li, ``Laser beams and resonators,'' {\em Appl. Opt.},
  vol.~5, p.~1550, 1966.

\bibitem{Wolf79}
K.~B. Wolf, {\em Integral Transforms in Science and Engineering}.
\newblock Boston: Springer, 1979.

\bibitem{Qin19}
H.~Qin, ``A necessary and sufficient condition for the stability of linear
  hamiltonian systems with periodic coefficients,'' {\em J. Math. Phys.},
  vol.~60, p.~022901, 2019.

\bibitem{Littlejohn87}
R.~G. Littlejohn and J.~M. Robbins, ``New way to compute {Maslov} indices,''
  {\em Phys. Rev. A}, vol.~36, p.~2953, 1987.

\bibitem{Zhao15}
L.~Zhao, J.~J. Healy, and J.~T. Sheridan, ``Constraints on additivity of the
  {1D} discrete linear canonical transform,'' {\em Appl. Opt.}, vol.~54,
  p.~9960, 2015.

\bibitem{Trefethen97}
L.~N. Trefethen and D.~{Bau, III}, {\em Numerical Linear Algebra}.
\newblock Philadelphia: SIAM, 1997.

\bibitem{Abramowitz70}
M.~Abramowitz and I.~A. Stegun, {\em Handbook of Mathematical Functions}.
\newblock New York: Dover, 1970.

\bibitem{Wigner32}
E.~Wigner, ``On the quantum correction for thermodynamic equilibrium,'' {\em
  Phys. Rev.}, vol.~40, p.~749, 1932.

\bibitem{deGosson06}
M.~{de Gosson}, {\em Symplectic Geometry and Quantum Mechanics}.
\newblock Basel: Birkh{\"a}user, 2006.

\bibitem{Lohmann93}
A.~W. Lohmann, ``Image rotation, {Wigner} rotation, and the fractional
  {Fourier} transform,'' {\em J. Opt. Soc. Am. A}, vol.~10, p.~2181,
  1993.

\bibitem{Cartwright76}
N.~D. Cartwright, ``A non-negative {Wigner-type} distribution,'' {\em Physica
  A}, vol.~83, p.~210, 1976.

\bibitem{OConnell81}
R.~F. {O'Connell} and E.~P. Wigner, ``Some properties of a non-negative
  quantum-mechanical distribution function,'' {\em Phys. Lett. A}, vol.~85,
  p.~121, 1981.

\bibitem{Kaminsky05}
E.~J. Kaminsky and L.~Simanjuntak, ``Chirp slope keying for underwater
  communications,'' {\em Proc. SPIE}, vol.~5778, p.~894, 2005.

\bibitem{Xiao87}
M.~Xiao, L.-A. Wu, and H.~J. Kimble, ``Precision measurement beyond the
  shot-noise limit,'' {\em Phys. Rev. Lett.}, vol.~59, p.~278, 1987.

\bibitem{Savitzky64}
A.~Savitzky and M.~J.~E. Golay, ``Smoothing and differentiation of data by
  simplified least squares procedures,'' {\em Anal. Chem.}, vol.~36,
  p.~1627, 1964.

\bibitem{Strang14}
G.~Strang and S.~MacNamara, ``Functions of difference matrices are {Toeplitz}
  plus {Hankel},'' {\em SIAM Rev.}, vol.~56, p.~525, 2014.

\bibitem{Eves80}
H.~Eves, {\em Elementary Matrix Theory}.
\newblock New York: Dover, 1980.

\bibitem{Diele98}
F.~Diele, L.~Lopez, and R.~Peluso, ``The {Cayley} transform in the numerical
  solution of unitary differential systems,'' {\em Adv. Comp. Math.}, vol.~8,
  p.~317, 1998.

\bibitem{Iserles01}
A.~Iserles, ``On {Cayley-transform} methods for the discretization of
  {Lie-group} equations,'' {\em Found. Comput. Math}, vol.~1, p.~129,
  2001.

\bibitem{Golub13}
G.~H. Golub and C.~F. {Van Loan}, {\em Matrix Computations}.
\newblock Baltimore: Johns Hopkins University Press, 4~ed., 2013.

\bibitem{Zhang20}
X.~Zhang, Y.~Fu, and H.~Qin, ``Simulating pitch angle scattering using an
  explicitly solvable energy conserving algorithm,'' {\em Phys. Rev. E},
  vol.~102, p.~033302, 2020.

\bibitem{Fu22}
Y.~Fu, X.~Zhang, and H.~Qin, ``An explicitly solvable energy-conserving
  algorithm for pitch-angle scattering in magnetized plasmas,'' {\em J. Comput.
  Phys.}, vol.~449, p.~110767, 2022.

\bibitem{Iserles00}
A.~Iserles, ``How large is the exponential of a banded matrix?,'' {\em New
  Zealand J. Math.}, vol.~29, p.~177, 2000.

\bibitem{Benzi07}
M.~Benzi and N.~Razouk, ``Decay bounds and {O(n)} algorithms for approximating
  functions of sparse matrices,'' {\em Electron. Trans. Numer. Anal.}, vol.~28,
  p.~16, 2007.

\bibitem{AlMohy09}
A.~H. {Al-Mohy} and N.~J. Higham, ``A new scaling and squaring algorithm for
  the matrix exponential,'' {\em SIAM J. Matrix Anal. Appl.}, vol.~31, 
  p.~970, 2009.

\bibitem{AlMohy11}
A.~H. {Al-Mohy} and N.~J. Higham, ``Computing the action of the matrix
  exponential, with an application to exponential integrators,'' {\em SIAM J.
  Sci. Comput.}, vol.~33, p.~488, 2011.

\bibitem{Suli03}
E.~Suli and D.~F. Mayers, {\em An Introduction to Numerical Analysis}.
\newblock Cambridge: Cambridge University Press, 2003.

\bibitem{Arsenault80b}
H.~H. Arsenault, ``Generalization of the principal plane concept in matrix
  optics,'' {\em Am. J. Phys.}, vol.~48, p.~397, 1980.

\bibitem{Nazarathy82}
M.~Nazarathy and J.~Shamir, ``First-order optics - a canonical operator
  representation: lossless systems,'' {\em J. Opt. Soc. Am.}, vol.~72, 
  p.~356, 1982.

\bibitem{Liu08}
X.~Liu and K.-H. Brenner, ``Minimal optical decomposition of ray transfer
  matrices,'' {\em Appl. Opt.}, vol.~47, p.~E88, 2008.

\bibitem{Yasir21a}
P.~A.~A. Yasir, ``Realization of general first-order optical systems using thin
  lenses of arbitrary focal length and fixed free propagation distance,'' {\em
  J. Opt. Soc. Am. A}, vol.~38, p.~42, 2021.

\bibitem{Yasir21b}
P.~A.~A. Yasir, ``Realization of general first-order optical systems using nine
  thin cylindrical lenses of arbitrary focal length and four units of free
  propagation distance,'' {\em J. Opt. Soc. Am. A}, vol.~38, p.~644,
  2021.

\bibitem{Heading62a}
J.~Heading, {\em An Introduction to Phase-Integral Methods}.
\newblock London: Methuen, 1962.

\bibitem{Rudin87}
W.~Rudin, {\em Real and Complex Analysis}.
\newblock New York: McGraw-Hill, 3~ed., 1987.

\bibitem{Gil07}
A.~Gil, J.~Segura, and N.~M. Temme, {\em Numerical Methods for Special
  Functions}.
\newblock Philadelphia: SIAM, 2007.

\bibitem{Cools97}
R.~Cools, ``Constructing cubature formulae: the science behind the art,'' {\em
  Acta Numer.}, vol.~6, p.~1, 1997.

\bibitem{Golub69}
G.~H. Golub and J.~H. Welsch, ``Calculation of {Gauss} quadrature rules,'' {\em
  Math. Comp.}, vol.~23, p.~221, 1969.

\bibitem{Gautschi16}
W.~Gautschi, {\em Orthogonal Polynomials in MATLAB: Exercises and Solutions}.
\newblock Philadelphia: SIAM, 2016.

\bibitem{Steen69}
N.~M. Steen, G.~D. Byrne, and E.~M. Gelbard, ``Gaussian quadratures for the
  integrals,'' {\em Math. Comp.}, vol.~23, p.~661, 1969.

\bibitem{Gautschi20}
W.~Gautschi, {\em Gauss quadrature and Christoffel function for halfrange Freud
  weight functions}.
\newblock Purdue University Research Repository, 2020.

\bibitem{Gautschi21}
W.~Gautschi, {\em A Software Repository for Gaussian Quadratures and
  Christoffel Functions}.
\newblock Philadelphia: SIAM, 2021.

\bibitem{Ruiz15d}
D.~E. Ruiz, C.~L. Ellison, and I.~Y. Dodin, ``Relativistic ponderomotive
  {Hamiltonian} of a {Dirac} particle in a vacuum laser field,'' {\em Phys.
  Rev. A}, vol.~92, p.~062124, 2015.

\bibitem{Heller77a}
E.~J. Heller, ``Generalized theory of semiclassical amplitudes,'' {\em J. Chem.
  Phys.}, vol.~66, p.~5777, 1977.

\bibitem{Bleistein86}
N.~Bleistein and R.~A. Handelsman, {\em Asymptotic Expansions of Integrals}.
\newblock New York: Dover, 1986.

\bibitem{Comtet74}
L.~Comtet, {\em Advanced Combinatorics}.
\newblock Dordrecht, Netherlands: Reidel, 1974.

\bibitem{Hairer97}
E.~Hairer, ``Variable time step integration with symplectic methods,'' {\em
  Appl. Numer. Math}, vol.~25, p.~219, 1997.

\bibitem{Richardson12}
A.~S. Richardson and J.~M. Finn, ``Symplectic integrators with adaptive time
  steps,'' {\em Plasma Phys. Control. Fusion}, vol.~54, p.~014004, 2012.

\bibitem{Zare75}
K.~Zare and V.~Szebehely, ``Time transformations in the extended phase-space,''
  {\em Celest. Mech.}, vol.~11, p.~469, 1975.

\bibitem{Goldman05}
R.~Goldman, ``Curvature formulas for implicit curves and surfaces,'' {\em
  Comput. Aided Geom. Des.}, vol.~22, p.~632, 2005.

\bibitem{Dodin10c}
I.~Y. Dodin and N.~J. Fisch, ``Vlasov equation and collisionless hydrodynamics
  adapted to curved spacetime,'' {\em Phys. Plasmas}, vol.~17, 
  p.~112118, 2010.

\bibitem{Grunwald71}
H.~P. Grunwald and F.~Milano, ``The {WKB} wave functions for the harmonic
  oscillator,'' {\em Am. J. Phys.}, vol.~39, p.~1394, 1971.

\bibitem{Paris91}
R.~B. Paris, ``The asymptotic behaviour of {Pearcey's} integral for complex
  variables,'' {\em Proc. R. Soc. A}, vol.~432, p.~391, 1991.

\bibitem{Wright80}
F.~J. Wright, ``The {Stokes} set of the cusp diffraction catastrophe,'' {\em J.
  Phys. A: Math. Gen.}, vol.~13, p.~2913, 1980.

\bibitem{Wolf74}
K.~B. Wolf, ``Canonical transforms {I}. complex linear transforms,'' {\em J.
  Math. Phys.}, vol.~15, p.~1295, 1974.

\bibitem{Bliokh04a}
K.~Y. Bliokh and Y.~P. Bliokh, ``Modified geometrical optics of a smoothly
  inhomogeneous isotropic medium: The anisotropy, {Berry} phase, and the
  optical {Magnus} effect,'' {\em Phys. Rev. E}, vol.~70, p.~026605,
  2004.

\bibitem{Bliokh04b}
K.~Y. Bliokh and Y.~P. Bliokh, ``Topological spin transport of photons: the
  optical {Magnus} effect and {Berry} phase,'' {\em Phys. Lett. A}, vol.~333,
  p.~181, 2004.

\bibitem{Oancea20}
M.~A. Oancea, J.~Joudioux, I.~Y. Dodin, D.~E. Ruiz, C.~F. Paganini, and
  L.~Andersson, ``The gravitational spin {Hall} effect of light,'' {\em Phys.
  Rev. D}, vol.~102, p.~024075, 2020.

\bibitem{Bliokh15}
K.~Y. Bliokh, F.~J. {Rodriguez-Fortuno}, F.~Nori, and A.~V. Zayats,
  ``Spin-orbit interactions of light,'' {\em Nat. Photon.}, vol.~9, 
  p.~796, 2015.

\bibitem{Dodin17c}
I.~Y. Dodin and A.~V. Arefiev, ``Parametric decay of plasma waves near the
  upper-hybrid resonance,'' {\em Phys. Plasmas}, vol.~24, p.~032119,
  2017.

\bibitem{Kravtsov96}
Y.~A. Kravtsov, O.~N. Naida, and A.~A. Fuki, ``Waves in weakly anisotropic {3D}
  inhomogeneous media: quasi-isotropic approximation of geometrical optics,''
  {\em Phys.-Usp.}, vol.~39, p.~129, 1996.

\bibitem{Kravtsov07}
Y.~A. Kravtsov, B.~Bieg, and K.~Y. Bliokh, ``Stokes-vector evolution in a
  weakly anisotropic inhomogeneous medium,'' {\em J. Opt. Soc. Am. A}, vol.~24,
  p.~3388, 2007.

\bibitem{Bliokh07}
K.~Y. Bliokh, D.~Y. Frolov, and Y.~A. Kravtsov, ``Non-abelian evolution of
  electromagnetic waves in a weakly anisotropic inhomogeneous medium,'' {\em
  Phys. Rev. A}, vol.~75, p.~053821, 2007.

\end{thebibliography}
\end{document}